\documentclass[12pt]{article}
\pdfoutput=1

\usepackage[dvipsnames]{xcolor} \usepackage[a4paper]{geometry}
\usepackage[english]{babel}
\usepackage{cite,enumerate,enumitem,booktabs,float,graphicx}

\usepackage[T1]{fontenc}
\usepackage[utf8]{inputenc}
\usepackage{lmodern}

\makeatletter
\g@addto@macro\bfseries{\boldmath}
\makeatother

\usepackage{bm,amsmath,amssymb,tensor,mathtools,braket}
\numberwithin{equation}{section}

\usepackage{tikz}
\usetikzlibrary{cd}

\usepackage{caption}
\usepackage{subcaption}

\usepackage[pdftex]{hyperref}
\definecolor{dark-blue}{rgb}{0.15,0.15,0.4}
\hypersetup{
pdftitle   = {The Energy-Momentum-News Complex near Future Null Infinity},
  pdfkeywords = {Carrollian geometry, celestial holography, Carrollian holography, holographic reconstruction, asymptotically flat spacetimes, holographic renormalisation},
  pdfauthor  = {Jelle Hartong, Emil Have, Vijay Nenmeli, Gerben Oling},
  colorlinks,
  linkcolor={dark-blue},
  citecolor={blue},
  linktoc=page
}

\newcommand{\RR}{\mathbb{R}}

\newcommand{\pd}{\partial}

\newcommand{\vphi}{\varphi}
\newcommand{\eps}{\varepsilon}

\newcommand{\LL}{\mathcal{L}}

\newcommand{\inv}{^{-1}}
\newcommand{\OO}{\mathcal{O}}
\newcommand{\qiq}{\quad\implies\quad}

\makeatletter
\let\save@mathaccent\mathaccent
\newcommand*\if@single[3]{%
  \setbox0\hbox{${\mathaccent"0362{#1}}^H$}%
  \setbox2\hbox{${\mathaccent"0362{\kern0pt#1}}^H$}%
  \ifdim\ht0=\ht2 #3\else #2\fi
  }
\newcommand*\rel@kern[1]{\kern#1\dimexpr\macc@kerna}
\newcommand*\widebar[1]{\@ifnextchar^{{\wide@bar{#1}{0}}}{\wide@bar{#1}{1}}}
\newcommand*\wide@bar[2]{\if@single{#1}{\wide@bar@{#1}{#2}{1}}{\wide@bar@{#1}{#2}{2}}}
\newcommand*\wide@bar@[3]{%
  \begingroup
  \def\mathaccent##1##2{%
    \let\mathaccent\save@mathaccent
    \if#32 \let\macc@nucleus\first@char \fi
    \setbox\z@\hbox{$\macc@style{\macc@nucleus}_{}$}%
    \setbox\tw@\hbox{$\macc@style{\macc@nucleus}{}_{}$}%
    \dimen@\wd\tw@
    \advance\dimen@-\wd\z@
    \divide\dimen@ 3
    \@tempdima\wd\tw@
    \advance\@tempdima-\scriptspace
    \divide\@tempdima 10
    \advance\dimen@-\@tempdima
    \ifdim\dimen@>\z@ \dimen@0pt\fi
    \rel@kern{0.6}\kern-\dimen@
    \if#31
      \overline{\rel@kern{-0.6}\kern\dimen@\macc@nucleus\rel@kern{0.4}\kern\dimen@}%
      \advance\dimen@0.4\dimexpr\macc@kerna
      \let\final@kern#2%
      \ifdim\dimen@<\z@ \let\final@kern1\fi
      \if\final@kern1 \kern-\dimen@\fi
    \else
      \overline{\rel@kern{-0.6}\kern\dimen@#1}%
    \fi
  }%
  \macc@depth\@ne
  \let\math@bgroup\@empty \let\math@egroup\macc@set@skewchar
  \mathsurround\z@ \frozen@everymath{\mathgroup\macc@group\relax}%
  \macc@set@skewchar\relax
  \let\mathaccentV\macc@nested@a
  \if#31
    \macc@nested@a\relax111{#1}%
  \else
    \def\gobble@till@marker##1\endmarker{}%
    \futurelet\first@char\gobble@till@marker#1\endmarker
    \ifcat\noexpand\first@char A\else
      \def\first@char{}%
    \fi
    \macc@nested@a\relax111{\first@char}%
  \fi
  \endgroup
}
\makeatother

\newcommand{\MM}{\mathbb{M}}
\newcommand{\scri}{\mathcal{I}}
\newcommand{\D}{{\partial}}
\newcommand{\os}[2]{\overset{(#1)}{#2}{}}

\title{\textbf{
    The Energy-Momentum-News Complex
    \\
    near Future Null Infinity
  }
}
\date{}
\author{Jelle Hartong, Emil Have, Vijay Nenmeli, Gerben Oling}

\usepackage{authblk}

\author[1]{Jelle Hartong}
\author[2,3]{Emil Have}
\author[1]{\\Vijay Nenmeli}
\author[1,4]{Gerben Oling}
\affil[1]{%
  School of Mathematics and Maxwell Institute for Mathematical Sciences,
  \protect\\
  University of Edinburgh,
  Peter Guthrie Tait Road,
  Edinburgh EH9 3FD, UK
}
\affil[2]{%
  The Mathematical Institute, University of Oxford,\protect\\ Woodstock Road, Oxford OX2 6GG, United Kingdom
}
\affil[3]{%
  Center of Gravity, Niels Bohr Institute, University of Copenhagen,\protect\\ Blegdamsvej 17, DK-2100 Copenhagen Ø, Denmark
}
\affil[4]{%
Institute for Theoretical Physics,
TU Wien,
\protect\\ 
Wiedner Hauptstrasse 8-10,
1040 Vienna,
Austria
}
\date{}
\begin{document}

\maketitle
\thispagestyle{empty}

\begin{abstract}
    We study asymptotically flat vacuum solutions of general relativity in three and four dimensions, with an emphasis on the geometric structures that emerge near null infinity.
    We construct asymptotic solutions to the three- and four-dimensional Einstein equations near future null infinity,
    which is a conformal Carroll manifold,
    starting from the most general Carroll metric data allowed by the Einstein equations.
    We use a Carroll-covariant version of Bondi--Sachs gauge,
    whose residual transformations act on the boundary Carroll geometry and shear as boundary diffeomorphisms, Weyl transformations and Carroll boosts.
    We then define a boundary energy-momentum-news complex at future null infinity by varying a suitably renormalised
    action with respect to the boundary Carroll metric data and shear.
    This involves adding boundary terms to the Einstein--Hilbert action
    on a cut-off surface near future null infinity.
    The boundary energy-momentum-news complex obeys two relations due to the boundary diffeomorphism and Weyl gauge invariance of the renormalised action.
    A third relation, due to the Carroll boost, is anomalous,
    and the corresponding anomaly is obtained from the variation of the renormalised action.
    Together, these Ward-type identities obeyed by the boundary energy-momentum-news complex lead to a Carroll-covariant generalisation of the Bondi loss equations.
\end{abstract}

\newpage
\tableofcontents

\section{Introduction}%
\label{sec:intro}
In the absence of a cosmological constant,
the energy and angular momentum radiated away by gravitational waves
are captured by the Bondi mass and angular momentum aspects.
These quantities arise in the Bondi--Sachs formalism~\cite{Bondi:1962px,Sachs:1962wk,Sachs:1962zza,Newman:1961qr},
which allows us to describe the gravitational radiation
for large classes of isolated systems,
which ends up at the future asymptotic null boundary of spacetime (denoted by $\scri^+$).
The corresponding Bondi loss equations, governing the evolution of the mass and angular momentum aspects, are a key ingredient in our theoretical understanding of gravitational radiation.

At the same time, Bondi, van der Burg, Metzner and Sachs~\cite{Bondi:1962px,Sachs:1962wk,Sachs:1962zza} showed that the asymptotic symmetry algebra in this context is not just given by the Poincaré isometries of Minkowski space.
Instead, it is enhanced to the infinite-dimensional BMS algebra,
which contains supertranslations that act independently at each point on the celestial sphere,
and these have recently played an increasingly important role in precision waveform modelling and gravitational-wave astronomy~\cite{Compere:2019gft,Mitman:2020bjf,Mitman:2024uss}.
Additionally, the seminal work by Barnich and Troessaert showed that the BMS algebra can be further extended to include superrotations~\cite{Barnich:2009se,Barnich:2010eb,Barnich:2011mi,Barnich:2010ojg},
corresponding to the infinite-dimensional conformal transformations of the celestial sphere.

Additionally, the success of the AdS/CFT correspondence naturally prompted the question if a similarly precise holographic description of (quantum) gravity in asymptotically flat spacetime can be obtained.
As we will review in more detail below,
building in particular on the `infrared triangle' developed by Strominger and collaborators~\cite{Strominger:2013jfa,He:2014laa,Strominger:2014pwa},
several related attempts to develop such holographic descriptions now exist.
The celestial holography programme~\cite{Kapec:2016jld,Pasterski:2016qvg,Pasterski:2017kqt,Pasterski:2021raf} emphasizes the Virasoro-type structure provided by the superrotations, and aims to provide a dual CFT$_2$ description of four-dimensional scattering amplitudes.
On the other hand, the Carroll holography programme builds on the conformal Carrollian nature of $\mathcal{I}^+$.
The latter can be described using so-called Carrollian geometry,
which we will describe in more detail below.

However, despite much progress, both approaches still are far from the level of precision that can be achieved in AdS through for example the $\mathcal{N}=4$ super-Yang--Mills and ABJM dualities.
One approach could be to study limits of such established top-down AdS/CFT settings, where the bulk limit sending the cosmological constant to zero maps to a Carroll limit in the dual field theory, as we will discuss below.
Though many classical Carroll limits of Lagrangian descriptions of Lorentzian field theories have been developed, the resulting field theories so far appear to be invariably deeply problematic as quantum field theories~\cite{deBoer:2023fnj,Cotler:2025npu}.

Motivated by all of the above, the first aim of the present work is to generalize the Bondi--Sachs formalism to a Carroll-covariant description of the asymptotic dynamics of general relativity near future null infinity.
In particular, this involves moving away from the coordinate choices and the restrictions on the geometric data at null infinity that come with the standard Bondi--Sachs gauge.
Using a Carroll-covariant version of Bondi--Sachs gauge, we then solve the vacuum Einstein equations in a radial expansion away from $\mathcal{I}^+$.
Next, we obtain a Carroll-covariant analogue of the dual energy-momentum tensor, both from the bulk equations of motion and by using a suitably-renormalised variational prescription.
This generalises the Bondi mass and angular momentum aspect, and we show that the associated diffeomorphism Ward identity covariantises the Bondi loss equations.
Finally, we observe that the boundary Carroll boost symmetry appears to be anomalous,
and we show that the corresponding anomaly satisfies Wess--Zumino consistency.%
\footnote{%
    At this point, it is important to emphasise that the current work does not provide a first-principles intrinsic description of Carroll quantum field theories,
    and therefore we currently use terms such as `Ward identities' and `anomalies' purely in analogy with the corresponding elements of the AdS/CFT holographic dictionary.
    In fact despite various efforts Carrollian QFTs suffer from a number of important open issues \cite{deBoer:2023fnj,Cotler:2024xhb}. Indeed, while we focus only on the dynamics near one copy of future null infinity,
    the dual description of scattering amplitudes,
     involves both future and past null infinity \cite{Kim:2023qbl,Kraus:2024gso,Isen:2026xoc}.
     However, for our present purposes of defining a boundary energy-momentum tensor and deriving the Bondi loss equations from its properties, it suffices to work with only one boundary.
}
This work expands and extends our earlier summary in~\cite{Hartong:2025jpp}.

We now provide a more detailed discussion of the background and context of this work in Section~\ref{ssec:intro-background-context}, including further references.
Next, Section~\ref{ssec:intro-aim} contains a more extensive discussion of the aims and approaches of this paper.
Finally, we give an overview of the structure of the remaining sections in Section~\ref{ssec:intro-outline}.

\subsection{Background and context}%
\label{ssec:intro-background-context}
The ground-breaking work by Bondi, van der Burg, Metzner and Sachs~\cite{Bondi:1962px,Sachs:1962wk,Sachs:1962zza} developed a framework to describe radiation in four-dimensional spacetimes.
By working in coordinates adapted to outgoing null hypersurfaces (`Bondi--Sachs gauge'), 
they were able to express Einstein's equations near $\scri^+$ as evolution equations of free `radiative data' in the asymptotic expansion of the metric.
In four bulk spacetime dimensions, which will be the focus of the majority of this paper,
such radiative data consists of the shear tensor, which encodes the asymptotic shear of the congruence of outgoing radial geodesics
along with the Bondi mass and angular momentum aspects.
As mentioned above, the components of the Einstein equations determining the evolution of the Bondi mass and angular momentum aspect 
take the form of flux-balance laws, and they are known as the Bondi loss equations.
Here, the `loss' (or the flux) is interpreted as the energy carried away by gravitational radiation.

Additionally, these authors demonstrated that the asymptotic symmetries of four-dimensional asymptotically flat spacetimes differ from the Poincaré group,
and are instead given by infinite-dimensional BMS group, which extends the Poincaré group by an infinite set of `supertranslations' consisting of smooth functions on the two-dimensional celestial sphere.
Recently,
several generalisations of the BMS$_4$ group have been proposed.
Barnich and Troessaert~\cite{Barnich:2009se,Barnich:2010eb,Barnich:2011mi,Barnich:2010ojg}
replaced the Lorentz group, which corresponds to the group of globally well-defined conformal transformations of the celestial sphere, with its local conformal transformations,
corresponding to two copies of the Witt algebra.
In this context, these local conformal symmetries of the celestial two-sphere are known as superrotations, and the resulting extension of the BMS group is known as the `extended' BMS group.
Another extension was proposed by Campiglia and Laddha in~\cite{Campiglia:2014yka,Campiglia:2015yka}, who introduced the `generalised' BMS group, which replaces the Lorentz group with the diffeomorphisms of the two-sphere. 

Shortly after the discovery of the original BMS group, Penrose's conformal approach to the study of infinity provided an elegant geometric framework for studying gravitational radiation in asymptotically flat spacetimes~\cite{Penrose:1962ij,Penrose:1964ge,Penrose:1965am}.
Conformally compactifying the spacetime by rescaling the metric brings asymptotic infinity to a finite location in a corresponding `unphysical' spacetime, where null infinity is a boundary in the proper mathematical sense.
Penrose used this to provide a precise definition of asymptotic flatness within this framework
and studied the behaviour of zero rest mass fields on~$\scri^+$, including the Weyl tensor, which encodes gravitational radiation.
In this context, the BMS symmetries arise as the residual diffeomorphisms that preserve the conformal structure of $\scri^+$.
These developments led to Ashtekar's notion of `universal structure' at $\scri^+$~\cite{Ashtekar:1978zz,Ashtekar:2023wfn}, which, as we will discuss shortly, is closely related to conformal Carrollian structures.
Ashtekar furthermore employed Hamiltonian methods to study $\scri^+$~\cite{Ashtekar:1981hw},
which was followed by work by Ashtekar and Streubel that constructed the symplectic structure of the radiative phase space in~\cite{Ashtekar:1981bq}, and in so doing demonstrated that the shear and the associated news tensor form a symplectic pair.
Later, a general framework for studying charges and their fluxes across~$\scri^+$ was established by Wald and Zoupas~\cite{Wald:1999wa},
who recognised that such charges are not generally conserved in radiating spacetimes and expressed their non-conservation as flux across $\scri^+$.

The last decade has seen a surge of interest in asymptotically flat spacetimes and their symmetries, brought about for a large part by Strominger's proposal that the four-dimensional gravitational $S$-matrix should be invariant under the diagonal subgroup of the BMS transformations at future and past null infinity~\cite{Strominger:2013jfa}.
Later papers then showed that Weinberg's soft graviton theorem is the Ward identity for BMS supertranslations~\cite{He:2014laa} and demonstrated that the change in the Bondi shear as a result of supertranslations encodes the gravitational memory effect~\cite{Strominger:2014pwa}.
This led to the notion of an `infrared triangle' relating asymptotic symmetries, soft theorems, and memory effects in many different contexts~\cite{Cachazo:2014fwa,Kapec:2014opa,Campiglia:2014yka,Campiglia:2015qka},
as reviewed in~\cite{Strominger:2017zoo}.
These developments led to the celestial holography programme~\cite{Kapec:2016jld,Pasterski:2016qvg,Pasterski:2017kqt,Pasterski:2021raf},
which aims to recast the four-dimensional gravitational $S$-matrix in terms of correlators of a putative conformal field theory living on the two-dimensional celestial sphere,
as reviewed in for example~\cite{Raclariu:2021zjz,Pasterski:2021rjz,McLoughlin:2022ljp,Donnay:2023mrd,Zhu:2026ofh}.

Additionally, as we briefly mentioned above, we can describe $\mathcal{I}^+$ using a type of non-Lorentzian geometry known as conformal Carrollian geometry.
The original notion of Carrollian symmetry was developed by L\'evy-Leblond, and, independently, Sen Gupta by taking the limit $c\to 0$, where $c$ is the speed of light~\cite{Levy1965,SenGupta1966OnAA}. The first notion of Carrollian geometry in the following sense appeared roughly a decade later in a paper by Henneaux~\cite{Henneaux:1979vn}.

Briefly, a Carrollian structure on a $(d{+}1)$-dimensional manifold $\mathcal{M}$ consists of a nowhere-vanishing vector field with components $v^\mu$ (where $\mu=0,\dots,d$), known as the Carrollian vector field, and a symmetric rank-$d$ tensor
with components $h_{\mu\nu}$ whose kernel is spanned by $v^\mu$, i.e., $h_{\mu\nu} v^\nu = 0$. The tensor $h_{\mu\nu}$ is sometimes known as the `Carrollian ruler'. The Carrollian structure $(v^\mu,h_{\mu\nu})$ is what replaces the notion of a metric in a Carrollian geometry.\footnote{Nevertheless, we will often abuse terminology and refer to the Carrollian structure as Carroll metric data (or structure), and to $h_{\mu\nu}$ as the Carroll metric.} In addition to the Carrollian structure, a Carrollian geometry also comes equipped with `inverses' $(\tau_\mu,h^{\mu\nu})$ satisfying 
\begin{equation}
    \tau_\mu v^\mu = -1\,,\qquad \tau_\mu h^{\mu\nu} = 0\,,\qquad \delta^\nu_\mu = -\tau_\mu v^\nu + h_{\mu\rho}h^{\rho\nu}\,.
\end{equation}
The pair $(\tau_\mu,h^{\mu\nu})$ is not unique, with an ambiguity captured by the Carroll boost symmetry that has the following infinitesimal action%
\footnote{%
    In the following, we will often take $\tau_\mu$ and $h_{\mu\nu}$ as our fundamental Carroll variables, as opposed to $v^\mu$ and $h_{\mu\nu}$, since $\tau_\mu$ is sensitive to the Carroll boost transformations.
    Displaying these transformations explicitly is a key aspect of our work since we will show that they are anomalous in the sense discussed above.
} 
\begin{equation}
    \delta \tau_\mu = \lambda_\mu\,,\qquad \delta h^{\mu\nu} = 2v^{(\mu}h^{\nu)\rho}\lambda_\rho\,,
\end{equation}
where $\lambda_\mu$ satisfying $\lambda_\mu v^\mu = 0$ is the parameter of the Carroll boost. The Carrollian structure itself remains inert under Carroll boost transformations. A conformal Carrollian structure, is, as discussed above, the one relevant at $\scri^+$. Such a structure transforms under infinitesimal Weyl rescalings with parameter $\Lambda_D$ as
\begin{equation}
    \delta v^\mu = -\Lambda_D v^\mu\,,\qquad \delta h_{\mu\nu} = 2\Lambda_D h_{\mu\nu}\,.
\end{equation}
In this work, we shall indeed recover a conformal Carrollian structure at $\scri^+$.

In the context of null infinity and BMS, conformal Carroll geometry was first discussed in a series of papers by Duval, Gibbons, and Horvathy~\cite{Duval:2014uva,Duval:2014uoa,Duval:2014lpa}.
It was first realised asymptotically in a fully covariant manner for three-dimensional bulk spacetimes in~\cite{Hartong:2015usd}.
The corresponding Carrollian approach to a holographic description of asymptotically flat spacetimes is complementary to the celestial approach outlined above, and relations between the two have been established in the literature~\cite{Donnay:2022aba,Bagchi:2022emh,Donnay:2022wvx}.
Recent reviews of Carrollian holography may be found in~\cite{Bagchi:2025vri,Nguyen:2025zhg,Ruzziconi:2026bix}. 
This work focuses on a better understanding of the energy-momentum sector and the associated role of the shear and news tensors in general conformal Carrollian field theories, and we will not have much to say about specific models.

\subsection{Aim of this paper}%
\label{ssec:intro-aim}
The fact that the geometric structure on $\scri^+$ is Carrollian suggests that, in the spirit of the anti-de Sitter/conformal field theory (AdS/CFT) correspondence~\cite{Maldacena:1997re}, one might be able to define a boundary energy-momentum tensor by computing the response of some appropriate action with respect to said Carroll data (along the lines of \cite{Henningson:1998gx,Balasubramanian:1999re,deBoer:1999tgo,deHaro:2000vlm,Skenderis:2000in,Bianchi:2001kw,Skenderis:2002wp} for asymptotically AdS spacetimes).

To obtain a covariant energy-momentum tensor at $\scri^+$, we will vary an appropriate action as a functional of the various sources at $\scri^+$.
To this end, we therefore must allow for the most general boundary geometry possible.
Such a construction of a boundary energy-momentum tensor was first considered in~\cite{Hartong:2015usd} for three-dimensional gravity.
Subsequent work in four bulk spacetime dimensions demonstrated that the zero-cosmological-constant limit of the anti-de Sitter fluid/gravity correspondence corresponds to a zero-speed-of-light limit in the fluid description~\cite{Ciambelli:2018wre,Mittal:2022ywl,Campoleoni:2023fug}.
The relation between these fluids and fluids with Carrollian symmetries as discussed in~\cite{deBoer:2017ing,deBoer:2021jej,deBoer:2023fnj}
was later clarified in~\cite{Armas:2023dcz}. In particular, the works~\cite{Ciambelli:2018wre,Mittal:2022ywl,Campoleoni:2023fug} argued for the existence of a nonzero energy flux at $\scri^+$, which is a consequence of the presence of the shear. 
In a theory with Carroll boost-invariance as considered in~\cite{deBoer:2017ing,deBoer:2021jej}, the energy flux must be zero,
but this statement is modified when additional sources such as the shear are present.
As already mentioned in~\cite{Hartong:2025jpp}, we show that the energy flux at future null infinity is nonzero due to two distinct effects: first, due the presence of shear as a source, and second, due to the fact that the Carroll boosts turn out to be anomalous.

The aforementioned work~\cite{Hartong:2015usd} by one of the authors considered holographic reconstruction of three-dimensional asymptotically flat spacetime. There, an energy-momentum tensor at $\scri^+$ was obtained by varying the Einstein--Hilbert action plus a certain set of boundary terms with respect to the boundary Carrollian structure, with the Bondi equations arising as Ward identities (see also \cite{Adami:2024rkr} for a similar approach). The present work builds directly on this approach, and significantly generalises the methods due to the extra complications present when the bulk spacetime dimension is four rather than three.

By applying the same null vielbein decomposition of the bulk metric of an asymptotically flat spacetime in the sense of Penrose, we use properties of the conformally compactified metric on the `unphysical spacetime' to deduce the leading-order radial behaviour of the metric components. We then fix all symmetries that leave the boundary (i.e., leading-order in the radial coordinate $r$) structure intact, and this leading-order structure then takes the form of a conformal Carrollian structure. This enlarges the solution space near $\scri^+$, and provides a generalisation of the partial Bondi gauge of ~\cite{Geiller:2022vto,Geiller:2024amx} by allowing for nonzero $d\tau$ (see also the recent paper~\cite{Geiller:2025dqe}, which, working in Newman--Penrose variables, likewise allows for a nonzero `twist' $\tau\wedge d\tau$).
See also~\cite{Campoleoni:2023fug} for related previous work aimed at providing a Carroll-covariant description of the solutions of the radial expansion of the Einstein equations.

\paragraph{Shear as boundary data.}
From the geometric description of radiative data at~$\scri^+$,
it has long been understood~\cite{Ashtekar:1981bq,Ashtekar:1981hw} 
that the shear
should be viewed as part of the boundary geometry, and we will give another perspective on this in the present work.%
\footnote{%
    Recent work that also emphasises the role of the shear in the conformal Carrollian description of null infinity includes~\cite{Herfray:2020rvq,Herfray:2021xyp,Nguyen:2022zgs,Baulieu:2025itt,Fiorucci:2025twa}.
}
As we will see, the shear arises as a subleading field in the near-boundary expansion of the metric, but nevertheless appears on par with the conformal Carrollian structure in the variational procedure, even though the latter appears at leading order in the expansion of the metric.
As we will see, the conformal Carrollian structure $(\tau_\mu,h_{\mu\nu})$ is augmented by the shear $C_{\mu\nu}$, which is symmetric, spatial, and traceless, corresponding to $v^\mu C_{\mu\nu} = 0$ and $h^{\mu\nu}C_{\mu\nu}=0$.
Physically, the shear encodes gravitational radiation, and transforms under Carroll boosts as 
\begin{equation}
\label{eq:shear-trafo-intro}
    \delta C_{\mu\nu} = h^{\rho\alpha}h^{\sigma\beta}h_{\alpha\langle\mu}h_{\nu\rangle\beta}\mathcal{L}_\lambda h_{\rho\sigma} + 2 \lambda_{\langle\mu} \mathcal{L}_v \tau_{\nu\rangle}\,,
\end{equation}
where the first Lie derivative is along $\lambda^\mu = h^{\mu\nu}\lambda_\nu$, and where $X_{\langle\mu\nu\rangle}$ denotes the symmetric traceless part with respect to $h^{\mu\nu}$ of the tensor $X_{\mu\nu}$.
The Carroll-covariant definition of the shear, along with the above transformation law (which is supplemented with boundary diffeomorphisms and Weyl transformations in the main text), is one of the main results of the first few sections of this paper.

\paragraph{Radial solution of Einstein equations.}
We can solve Einstein's equations order by order in a $1/r$-expansion. 
More specifically, we will do this using
what we call Carroll-covariant Bondi--Sachs gauge~\cite{Hartong:2025jpp}, where we have
\begin{equation}
    ds^2=-2e^\beta dr\tau_\mu dx^\mu+g_{\mu\nu}dx^\mu dx^\nu\,.
\end{equation}
Here, $\tau_\mu$ is the boundary Carroll clock one-form, which is independent of $r$, and $\beta=\mathcal{O}(r^{-1})$ depends on $r$ and $x^\mu$.
The metric $g_{\mu\nu}=r^2 h_{\mu\nu}+\mathcal{O}(r)$ is such that $v^\mu g_{\mu\nu}$ is analogous to the bulk metric in AdS/CFT, in the sense that we can algebraically solve for its subleading coefficients in a $1/r$ expansion in terms of the boundary data.
On the other hand, the STF part $h^\rho_{\langle\mu}h^\sigma_{\nu\rangle}g_{\rho\sigma}$ exhibits a different behaviour.\footnote{The $h^{\mu\nu}$ trace of $g_{\mu\nu}$ is fixed by the Carroll-covariant Bondi--Sachs gauge choice.}
Here, the Einstein equations only determine its time evolution along the $v^\mu$ vector field,
and we therefore need to fix $h^\rho_{\langle\mu}h^\sigma_{\nu\rangle}g_{\rho\sigma}$ at some instant of boundary time, for all values of $r$.
Provided we have this initial data, it can be shown that the only free data left in the solution to the Einstein equations consists of the boundary data $\{\tau_\mu\,, h_{\mu\nu}\,, C_{\mu\nu}\}$ and its responses $\{T^\mu\,, T^{\mu\nu}\,, S^{\mu\nu}\}$, which we call the energy-momentum-news complex.

\paragraph{Ward identities.}
The definition of the energy-momentum-news complex follows from the variation  with respect to $\{\tau_\mu\,, h_{\mu\nu}\,, C_{\mu\nu}\}$ of a gravitational action that is evaluated on shell at $\mathcal{I}^+$.
Just like in AdS/CFT, this leads to a boundary energy-momentum tensor that obeys a number of Ward identities,
which originate from residual gauge transformations that leave the action invariant.
For a large class of gauge choices (which include the Carroll-covariant Bondi--Sachs gauge, but also for example a Carroll-covariant Newman--Unti gauge), the residual gauge transformations consist of boundary diffeomorphisms, Weyl rescalings and Carroll boosts.
It will turn out that the Carroll boosts are anomalous. As advertised in~\cite{Hartong:2025jpp}, the totality of all these three Ward identities, i.e., the one due to boundary diffeomorphism invariance, Weyl invariance and the anomalous Carroll boosts leads to the Bondi loss equations.%
\footnote{%
    It was also shown in~\cite{Fiorucci:2025twa} that the standard Bondi loss equations can be captured by the diffeomorphism Ward identity.
    In addition, it was emphasised that a first-order approach gives a natural way to encode the shear tensor as the ambiguity encoded in solving for the connection,
    which is another perspective on its role as boundary data.
    However, the aforementioned paper does not contain a bulk derivation or a Carroll-covariant generalisation of the Bondi loss equations in the sense we develop here,
    and it is therefore most directly comparable to our results in Section~\ref{sec:variations-ward-ids} and Appendix~\ref{app:reduction-to-standard-BS}.
    Additionally, it is worth mentioning that~\cite{Fiorucci:2025twa} does not encounter an anomalous boost Ward identity, and it would be interesting to better understand the difference between the two approaches.
}
We show this in great detail in this paper.

\paragraph{Boundary constraint.}
An important difference with the AdS/CFT setting is that 
the Einstein equations now fix part of the boundary structure $\{\tau_\mu,h_{\mu\nu},C_{\mu\nu}\}$
by requiring that $\mathcal{L}_vh_{\mu\nu}$ is proportional to $h_{\mu\nu}$,
\begin{equation}
\label{eq:intro-constraint}
    \mathcal{L}_vh_{\mu\nu} \sim h_{\mu\nu}\,.
\end{equation}
In other words, the Lie derivative of the Carrollian ruler along the Carrollian vector field $v^\mu$ is pure (spatial) trace. 
This is in sharp contrast to what happens in the AdS/CFT correspondence, where the Lorentzian metric on the boundary remains unconstrained. This has the inevitable consequence that we cannot freely vary $h_{\mu\nu}$ to compute its response $T^{\mu\nu}$. Instead, we must vary within the space of solutions to the above constraint. This leads to a fundamental ambiguity in the definition of~$T^{\mu\nu}$, which affects the definition of the Bondi angular momentum aspect but does not affect the Bondi loss equations.
The ambiguity in $T^{\mu\nu}$ can be viewed as an improvement transformation of the energy-momentum tensor that leaves all the Ward identities invariant.

\paragraph{Variational setup.}
In the AdS/CFT correspondence, the on-shell gravitational action becomes a generating functional for the dual CFT, with sources provided by the boundary values of the bulk fields.
For example, the boundary metric sources the energy-momentum tensor of the dual CFT.
However, this on-shell action (and its variation) diverges as one approaches the boundary, and one must therefore add counterterms that render the variation of the on-shell action finite in a process known as holographic renormalisation~\cite{Henningson:1998gx,Balasubramanian:1999re,deBoer:1999tgo,deHaro:2000vlm,Skenderis:2000in,Bianchi:2001kw,Skenderis:2002wp}.
In practice, this works by evaluating the on-shell action at a radial `cut-off' surface, identifying terms that diverge as $r\to \infty$ and then subtracting them, while at the same time ensuring that the (Dirichlet) variational problem remains well-posed.
Varying the on-shell gravitational action with respect to the boundary value of the metric $g_{(0)\mu\nu}$ gives an explicit expression for the CFT energy-momentum tensor $T^{\mu\nu}_{\text{CFT}}$; schematically,
\begin{equation}
\label{eq:os-variation-intro}
    \delta S\big\vert_{r\to\infty} \sim \int_{\D AdS} \sqrt{-g_{(0)}}T^{\mu\nu}_{\text{CFT}}\delta g_{(0)\mu\nu}\,.
\end{equation}
This vanishes when Dirichlet boundary conditions $\delta g_{(0)\mu\nu}= 0$ are imposed, making the action stationary, which ensures that the variational problem is well-posed.
It is natural to ask whether a similar variational framework exists for asymptotically flat spacetimes,
and we summarised such a framework for three and four bulk spacetime dimensions in~\cite{Hartong:2025jpp}.
Other work towards addressing this question includes~\cite{Kraus:1999di,Mann:2005yr,Chandrasekaran:2021hxc,Freidel:2024emv,Riello:2024uvs,Bhambure:2024ftz,Ciambelli:2025mex,Campoleoni:2025bhn}.

While the overall strategy is very similar to what we described above in the case of the AdS/CFT correspondence, the technical and conceptual details are radically different. First off, the boundary structure at $\scri^+$ is given not by a Lorentzian metric, but by the conformal Carrollian structure along with the shear, which we denote collectively by $\{\tau_\mu,h_{\mu\nu},C_{\mu\nu}\}$.

Second, in the AdS case, the calculation is to vary the renormalised action (in some gauge-preserving sense) and then to evaluate this on shell.
This is often referred as the variation of the on-shell action, which suggests going about things in the other order, namely to compute an on shell action and to subsequently vary it within the space of on shell solutions.
Under appropriate circumstances, these two ways of computing commute,
but this is not automatic.

To be very precise, in the present flat case, we will vary the Einstein--Hilbert action supplemented with boundary terms that are defined at some cut-off hypersurface near $\mathcal{I}^+$ using variations that preserve our (off-shell) gauge choice, and we subsequently evaluate the result on shell.
We do not examine the on-shell action itself. This is because the on-shell action will receive contributions from past null infinity (and possibly corner terms) and, for our purposes, the above prescription is sufficient.
For this reason, we refrain from calling this computation the variation of the on-shell action.
Schematically, if $S$ denotes the sum of the bulk and boundary terms, and $\delta S$ the gauge preserving variation, then the evaluation of $\delta S$ as we approach $\mathcal{I}^+$ is required to take the following form,
\begin{equation}
\label{eq:os-variation-flat-intro}
    \delta S\big\vert_{r\to \infty} \sim \int_{\scri^+}d^3x e \left( T^\mu \delta\tau_\mu + \frac{1}{2}T^{\mu\nu}\delta h_{\mu\nu} + \frac{1}{2}S^{\mu\nu}\delta C_{\mu\nu} \right)\,,
\end{equation}
where $T^\mu$ is the energy current, $T^{\mu\nu}$ is the momentum-stress tensor, and $S^{\mu\nu}$ is the news. 
From the point of view of a well-posed variational principle,
we are not demanding that the variation written above vanishes.
We could consider Dirichlet boundary conditions for the variations of $\tau_\mu$ and $h_{\mu\nu}$,
but we do not want to do the same for the shear.
Doing so would require including past null infinity and following steps similar to the recent work~\cite{Isen:2026xoc}.
Instead, we are only demanding that the variation evaluated on shell is finite and of the form \eqref{eq:os-variation-flat-intro}.
The absence of a Dirichlet boundary condition for the variation of the shear can be viewed as a `leaky' boundary condition.\footnote{A complete (global) well-posed variational principle must take into account all boundary components, with appropriate boundary conditions imposed on each component. On its own, $\scri^+$ is an open system~\cite{Wieland:2020gno}, with gravitational radiation (encoded in the shear) carrying away energy through $\scri^+$. In the covariant phase space approach, this materialises as a non-vanishing symplectic flux through $\scri^+$, which is captured by the boundary term $\int_{\scri^+}\theta$ that appears in the on-shell variation $\delta S\vert_{\scri^+}$~\cite{Wald:1999wa}. Such boundary conditions are leaky~\cite{Donnay:2022aba,Donnay:2022wvx,McNees:2024iyu}, in that they allow for a nonzero symplectic flux through $\scri^+$. }

\paragraph{Log corrections.}
So far, logarithmic terms have played no role in the story. In the AdS/CFT correspondence, logarithmic terms play a crucial role as they encode anomalies~\cite{Henningson:1998gx,deHaro:2000vlm} when the bulk spacetime dimension is odd. Their appearance in the near-boundary expansion of the AdS metric is dictated by the Fefferman--Graham theorem, which fully determines the form of the expansion, and hence which powers and logarithms of~$r$ appear.
For asymptotically flat spacetimes, instead of anomalies, logarithmic terms in general encode information about the violation of peeling and smoothness~\cite{Polyhom_scri,friedrich2004smoothnessnullinfinitystructure}.

In this work, we will discuss the inclusion of a particular logarithmic term in the context of the asymptotic expansion,
namely a term of order $r^{-1}\log r$ in the expansion of $g_{\mu\nu}$.
We will show that this does not affect the Weyl Ward identity nor the Weyl transformation properties of the boundary energy-momentum tensor (which has a definite Weyl weight). Instead, the logarithmic term simply serves to remove a constraint on the $1/r$~correction to the shear tensor~\cite{Barnich:2010eb}.

\subsection{Outline}%
\label{ssec:intro-outline}

For a self-contained exposition of our main results, we refer the reader to~\cite{Hartong:2025jpp}.

The present manuscript is structured as follows. In Section~\ref{sec:bulk-geom}, we discuss the bulk geometry, and how the boundary of generic asymptotically flat spacetime acquires a conformal Carrollian structure. After demonstrating how this works for Minkowski space in Section~\ref{ssec:null-frame-mink}, we go through the general construction in Section~\ref{ssec:bulk-geom-general-null-frames}, where we also discuss how the local Lorentz symmetry acts on the null frame. In Section~\ref{ssec:bulk-geom-penrose-bdy}, we go through the Penrose construction, and demonstrate how this leads to fall-off conditions on the metric components after identifying the radial coordinate $r$ with the inverse of the defining function used in the conformal compactification. We also discuss how the near-boundary expansion of the bulk gauge transformations lead to gauge transformations that act on the boundary geometry. By partially fixing the gauge through setting one of the null vielbeine $V$ equal to $-\D_r$, we demonstrate in Section~\ref{ssec:bulk-geom-partial-gauge-fixing} how a conformal Carrollian structure emerges at leading order in the near-boundary expansion of the metric, while we demonstrate in Section~\ref{ssec:bulk-geom-LO-EE-in-partial-gauge} that the leading-order Einstein equations impose a constraint on this Carrollian structure.
We also demonstrate how a Carroll-covariant version of the shear tensor appears as an ambiguity in these equations of motion, and we compute its transformation under the boundary gauge transformations.

Then, in Section~\ref{sec:car-cov-BS-gauge}, we introduce our main gauge choice, which is the Carroll-covariant Bondi--Sachs gauge. There, the strategy will be to fix as much of the gauge transformations as we can achieve off shell. We begin in Section~\ref{ssec:car-cov-BS-gauge-V-geodesic} by requiring that the already-fixed radial null vector $V$ be everywhere geodesic, while in Section~\ref{ssec:car-cov-BS-gauge-fix-Sigma} we fix the null rotations that do not correspond to boundary Carroll boosts. We then fix radial diffeomorphisms in two different ways: in Section~\ref{ssec:car-cov-BS-gauge-fix-radial-diffeo}, we impose a Carroll-covariant Bondi determinant condition leading to a Carroll-covariant version of the Bondi--Sachs gauge, while in Section~\ref{ssec:Newman-Unti}, we discuss a Carroll-covariant version of Newman--Unti gauge. We discuss how this general Carroll-covariant Bondi--Sachs gauge reduces to the standard Bondi--Sachs gauge in Section~\ref{ssec:reduction-to-standard-Bondi--Sachs-gauge} by restricting the conformal Carrollian structure. Finally, we discuss different choices of connections and their associated curvatures on radial hypersurfaces in Sections~\ref{ssec:connection-choice} and~\ref{subsec:hypercurv}, respectively.

In Section~\ref{sec:rewriting-EE}, we rewrite Einstein's equations $R_{MN} = 0$, where $M = (\mu,r)$, in Carroll-covariant Bondi--Sachs gauge in general dimension. We begin in Section~\ref{ssec:rewriting-EE-strategy} by laying out the general strategy, which we carry out starting in Section~\ref{ssec:rewriting-EE-rr-equation}, where we rewrite the equation $R_{rr} = 0$. This is followed by Sections~\ref{ssec:rewriting-EE-mu-r-equation} and~\ref{ssec:rewriting-EE-mu-nu-equation}, where we rewrite the equations $R_{\mu r} = 0$ and $R_{\mu\nu} = 0$, respectively. In Sections~\ref{ssec:rewriting-EE-3d-simplifications} and~\ref{ssec:rewriting-EE-4d-simplifications}, we discuss simplifications that occur in three and four bulk spacetime dimensions, respectively.

Section~\ref{sec:radial-expansion} considers the near-boundary expansion of Einstein's equations and solves them order-by-order in $1/r$. We first demonstrate in Section~\ref{ssec:radial-expansion-bianchi} that the Bianchi identity $\nabla_M G^{M}{}_N=0$ reduces the number of independent equations. The next three Sections~\ref{ssec:radial-expansion-rr}--\ref{ssec:radial-expansion-stf} are concerned with solving these equations. In Section~\ref{ssec:radial-expansion-threedim}, we demonstrate that the radial expansion terminates in three bulk spacetime dimensions, while in Section~\ref{ssec:radial-expansion-logs} we discuss the consequences of a specific logarithmic term in the near-boundary expansion in four bulk spacetime dimensions.

In Section~\ref{sec:variations-ward-ids}, we discuss general properties of action functionals that depend on  Carrollian geometry sources, including the shear. The associated responses form the energy-momentum-news complex, with the response to the Carrollian `metric' structure forming the energy-momentum tensor, and the news being the response to the shear. We then discuss how to solve the boundary constraint $\mathcal{L}_v h_{\mu\nu} \sim h_{\mu\nu} $ in Section~\ref{ssec:variations-ward-ids-solving-constraint}, before defining the energy-momentum-news complex in Section~\ref{sec:variations-ward-ids-currents-from-constraint-surface-variation}. Next, we discuss how the boundary constraint restricts the boundary variations and how to deal with this in Section~\ref{ssec:variations-ward-ids-ambiguity-from-boundary-constraint}. This is followed in Section~\ref{ssec:variations-ward-ids-boundary-ward-identities} by a discussion of the Ward identities obeyed by the energy-momentum-news complex due to local invariance of the action functional under diffeomorphisms, Carroll boosts and Weyl transformations. We also consider the case where the Carroll boosts are anomalous. We then discuss the algebra satisfied by these transformations in Section~\ref{ssec:trafo-alg}, and we use this to derive the transformation properties of the energy-momentum tensor by considering the commutator of a general variation with a gauge transformation acting on the action functional in Section~\ref{subsec:trafoEMT}. In Section~\ref{subsec:Weylcov}, we discuss Weyl covariance, before finally deriving the transformation properties of the full energy-momentum-news complex in Section~\ref{subsec:EMTnewstrafos}. 

This is followed by Section~\ref{sec:bulk-conservation-equations}, where we derive the Carroll-covariant generalisations of the Bondi loss equations by matching the order-$r^{-d}$ contribution of a certain projection of the Einstein equations with the diffeomorphism Ward identity derived in Section~\ref{sec:variations-ward-ids}. In three bulk spacetime dimensions ($d=1$), these are actually conservation equations, and we discuss them in Section~\ref{ssec:ConservationEqnsd1}. We then discuss the Bondi loss equations in four bulk spacetime dimensions ($d=2$) in Section~\ref{ssec:bondi-loss-4d-simplified}, where we make the simplifying assumption that $d \tau = 0$. In the subsequent Section~\ref{ssec:bondi-loss-4d-general}, we lift this assumption and derive the Bondi loss equations in full generality. This is followed by Section~\ref{ssec:log-emt-news}, where we show that including the log terms discussed in Section~\ref{ssec:radial-expansion-logs} does not modify the energy-momentum-tensor-news complex, though it does lift a constraint on a subleading spatial field that (in the absence of logs) is imposed by the Einstein equations.

Then, in Section~\ref{sec:bulk-improvements}, we show that it is possible to add improvement terms to the energy-momentum-tensor-news complex derived in Section~\ref{sec:bulk-conservation-equations} such that the improved energy-momentum-tensor-news complex has definite Weyl weights. We begin in Section~\ref{subsec:Weyltrafos} by considering the transformation properties of various quantities under Weyl rescalings, followed by a discussion of Weyl improvements first in three bulk spacetime dimensions in Section~\ref{ssec:bulk-improvements-weyl-3d}, and then in four bulk spacetime dimensions in Section~\ref{ssec:bulk-improvements-weyl-4d}. Next, in Section~\ref{ssec:bulk-improvements-weyl-covariant-currents}, we consider the explicit Weyl transformations of the currents in four bulk spacetime dimensions.

In Section~\ref{sec:HoloRenormAndOn-ShellActions}, we show how the Weyl-improved energy-momentum-news complex can be obtained from the variation of a suitably renormalised gravitational action by adding appropriate counterterms on a cut-off hypersurface near $\mathcal{I}^+$. In Section~\ref{ssec:vars-and-cutoff-near-scri}, we set up a flat-space analogue of holographic renormalisation,
which is then applied to three-dimensional bulk spacetimes in Section~\ref{ssec:hol-ren-threedim-anomalies-improvements}, where the associated Weyl-improved energy-momentum tensor is derived. We then discuss the $d=1$ Carroll boost anomaly in Section~\ref{ssec:hol-ren-threedim-anomalies-improvements}. In Section~\ref{ssub:EMT-news-for-d=2}, we derive the Weyl-improved energy-momentum-news complex for four-dimensional bulk spacetimes. In Section~\ref{ssec:on-shell-action-logs}, we discuss the effects of adding logs and demonstrate that they do not change the result. Finally, the $d=2$ Carroll boost anomaly is discussed in Section~\ref{ssec:hol-ren-fourdim-anomalies-improvements}.

We end in Section~\ref{sec:discussion} with a discussion which briefly summarises our results and lays out a number of avenues for future exploration.

In addition to the main part of the paper, we have a included a number of appendices. The first of these, Appendix~\ref{app:conventions}, collects a number of useful identities and summarise our conventions.  
In Section~\ref{sapp:conventions-curvature}, we discuss the conventions we use for curvatures and list useful curvature identities. In Section~\ref{sapp:conventions-lie-derivative}, we briefly discuss Lie derivatives, before turning our attention to the special properties of spatial tensors in Section~\ref{sapp:spatial-tensors}.

In Appendix~\ref{app:carroll-geometry}, we provide a self-contained introduction to conformal Carrollian geometry. In Section~\ref{subsec:Carrolldata}, we discuss properties of the conformal Carrollian structure, before discussing the connection in Section~\ref{sapp:carroll-geometry-connection}. With the connection, we can define curvatures, which are discussed in Section~\ref{app:curvten}. When $d=2$, corresponding to four-dimensional bulk spacetimes, a number of special identities hold, which is the topic of Section~\ref{subsec:speciald=2id}. When performing the calculations described in this work, we repeatedly must commute Lie derivatives and covariant derivatives, and we collect results about this procedure in Section~\eqref{sapp:carroll-lie-covariant-commutators}. Then, in Section~\ref{ssec:bulk-improvements-weyl-cov-derivs}, we show how to construct Weyl covariant objects out of the data on a conformal Carrollian geometry. In Section~\ref{app:variations} we collect many useful results regarding variations of geometric data such as curvatures, and in Section~\ref{app:Carrollboosts} we provide a useful compendium of the behaviour of various quantities under Carroll boosts.

In Appendix~\ref{app:overview-of-expansion-results}, we provide an overview of the large-$r$ expansion of the near-boundary metric in Carroll-covariant Bondi--Sachs gauge. In Section~\ref{sapp:on-shell-expansion-fundamental-variables}, we summarise the on-shell expansion of the metric itself, while in Section~\ref{sapp:intermediate-results-expansions}, we do the same for various composite objects.

In Appendix~\ref{app:intermediate-results}, we give additional details on the $d=2$ calculation of Section~\ref{sec:bulk-conservation-equations} that allowed us to the identify the energy-momentum-tensor-news complex from the Einstein equations. In Sections~\ref{ssapp:order-r-m1-EOM} and~\ref{app:LossEqnLists}, we collect those details at orders $r^{-1}$ and $r^{-2}$, respectively.

Finally, in Appendix~\ref{app:reduction-to-standard-BS}, we demonstrate how the Carroll-covariant Bondi loss equations obtained in Sections~\ref{sec:bulk-improvements} and~\ref{sec:HoloRenormAndOn-ShellActions} reduce to the standard four-dimensional Bondi loss equations when restricted to the standard Bondi--Sachs gauge.

\newpage

\section{Bulk geometry setup}
\label{sec:bulk-geom}
In this section, our aim is to exhibit the natural Carroll metric structure associated to future null infinity in general asymptotically flat spacetimes~\cite{Duval:2014lpa,Hartong:2015usd,Herfray:2021qmp}.
(See also~\cite{Riello:2024uvs} for a related recent discussion.)
To see the emergence of this Carroll structure,
building in particular on earlier work in three spacetime dimensions~\cite{Hartong:2015usd},
we employ Penrose's construction formulated in terms of null vielbeine.
This gives us boundary conditions in terms of a polynomial expansion in the radial coordinate near null infinity.
We will consider logarithmic terms in later sections.

We first briefly illustrate our construction in terms of Minkowski spacetime in Bondi coordinates in Section~\ref{eq:minkowski-metric-bondi} below,
where a conformal compactification naturally leads to a particular null frame.
In Section~\ref{ssec:bulk-geom-general-null-frames}, we then consider null frames for general spacetime metrics and their gauge transformations.
We discuss the Penrose compactification of a general asymptotically flat spacetime in terms of null vielbeine in Section~\ref{ssec:bulk-geom-penrose-bdy}.
In order to set up a convenient asymptotic coordinate system,
we introduce a radial coordinate $r$ corresponding to the defining function in the Penrose compactification.
We will later furthermore demand that $r$ is chosen such that curves whose tangent is $\partial_r$ form a congruence of null curves (and in Section \ref{sec:car-cov-BS-gauge} we will require these curves to be geodesics).
We then choose one of the null vielbeine to be equal to the tangent vector field $\pd_r$ to this congruence of null curves.
This is a partial gauge fixing that makes the Carroll structure on $\mathcal{I}^+$ manifest,
as we show in Section~\ref{ssec:bulk-geom-partial-gauge-fixing}.

Finally, in Section~\ref{ssec:bulk-geom-LO-EE-in-partial-gauge},
we work out the implications of the leading-order Einstein equations using only this partial gauge fixing.
Here, we see that the boundary Carroll structure cannot be fully arbitrary.\footnote{We also checked explicitly that this condition is independent of the aforementioned partial gauge fixing, though details of this computation are not included in this paper.}
Instead, the Carroll metric data must satisfy a constraint which can be understood as follows.
The Carrollian geometry on any $(d+1)$-dimensional manifold such as $\mathcal{I}^+$ can be viewed as a line bundle (corresponding to boundary Carroll or retarded time)
fibred over a $d$-dimensional Riemannian base manifold, which in this case generalises the celestial sphere.
In this language, the constraint we encounter restricts how the metric on the base manifold evolves along the fibres.
As we will see in Section~\ref{sec:variations-ward-ids} later on,
this constraint can be solved covariantly (that is, without fixing boundary coordinates),
so it will not prevent us from defining a consistent variation in the space of allowed boundary metrics.
Finally, in Section~\ref{ssec:bulk-geom-LO-EE-in-partial-gauge} we also identify the Carroll-covariant version of the shear tensor. The residual bulk diffeomorphisms that respect these partial gauge choices act on the boundary Carroll data (including the shear) as a combination of arbitrary boundary diffeomorphisms, Weyl transformations and Carroll boosts.

To simplify the upcoming computations,
it is possible to restrict the gauge further whilst maintaining boundary Carroll covariance,
and this will be the subject of Section~\ref{sec:car-cov-BS-gauge}.
In particular, this will lead to the notions of a Carroll-covariant Bondi--Sachs and a Carroll-covariant Newman--Unti gauge.

\subsection{A null frame for Minkowski in Bondi coordinates}
\label{ssec:null-frame-mink}
The simplest example of the type of asymptotic geometry we will consider
is future null infinity in Minkowski spacetime.
In terms of the retarded time coordinate $u=t-r$, the metric of $(d+2)$-dimensional Minkowski spacetime $\MM_{d+2}$ reads
\begin{equation}
  \label{eq:minkowski-metric-bondi}
  ds^2 = - du^2 - 2du dr
  + r^2 \gamma_{AB} dx^A dx^B\,,
\end{equation}
where $\gamma_{AB}$ is the metric on the $d$-dimensional sphere $S^{d}$.
The latter is described by the spatial $x^A$~coordinates,
and the spacetime coordinates are
$x^M = (r,u,x^A)$.
We obtain the conformal compactification $\widebar{\MM}_{d+2}$ of Minkowski spacetime by introducing the defining function $\Omega = 1/r$ and rescaling the metric as follows,
\begin{equation}
  \label{eq:minkowski-metric-bondi-compactified}
  d\widebar{s}^2
  = \Omega^2 ds^2
  = \frac{1}{r^2} \left[
    - du^2 - 2du dr
    + r^2 \gamma_{AB} dx^A dx^B
  \right].
\end{equation}
The conformal boundary $\partial \widebar{\MM}_{d+2}$ is located at
$\Omega=0$,
corresponding to
$r\to\infty$,
and it is parametrised by the coordinates $x^\mu = (u,x^A)$.
The induced metric
$\gamma_{AB} dx^A dx^B$
on $\partial \widebar{\MM}_{d+2}$
is degenerate.
The normal one-form $d\Omega = - dr/r^2$ is null and nonvanishing at the conformal boundary.
Its contraction with the null vector $\widebar{V}^M \pd_M = -r^2 \pd_r$ is nonzero as $r\to\infty$,
which means that $\widebar{V}^M$ points away from the conformal boundary.
We can complete it into a null frame
for the metric~\eqref{eq:minkowski-metric-bondi-compactified}
by adding a null vector $\widebar{U}^M$
as well as $d$ spacelike vectors $\widebar{E}_a^M$,
which are labeled by $a=1,\ldots,d$.
Using a set of spatial vielbeine $e_a^A$ for the sphere metric $\gamma_{AB}$, we can take
\begin{subequations}
  \label{eq:minkowski-metric-bondi-unphysical-frame}
  \begin{gather}
    \widebar{U}^M \pd_M
    = \pd_r - \pd_u\,,
    \qquad
    \widebar{V}^M \pd_M
    = - r^2 \pd_r\,,
    \qquad
    \widebar{E}_a^M \pd_M
    = e_a^A \pd_A\,,
    \\
    \widebar{U}_M dx^M
    = \frac{1}{r^2} \left(dr + du\right),
    \qquad
    \widebar{V}_M dx^M
    = du\,,
    \qquad
    \widebar{E}^a_M dx^M
    = e^a_A dx^A\,.
  \end{gather}
\end{subequations}
This frame is chosen such that the rescaled metric~\eqref{eq:minkowski-metric-bondi-compactified}
can be written as follows,
\begin{equation}
  \label{eq:minkowski-metric-bondi-unphysical-frame-decomposition}
  d\widebar{s}^2
  = - 2 \widebar{U}_M \widebar{V}_N dx^M dx^N
  + \delta_{ab} \widebar{E}^a_A \widebar{E}^b_B dx^A dx^B\,.
\end{equation}
The frame~\eqref{eq:minkowski-metric-bondi-unphysical-frame} also leads to
a natural null frame
for the original Minkowski metric~\eqref{eq:minkowski-metric-bondi}
after rescaling with appropriate factors of $\Omega=1/r$,
\begin{subequations}
  \label{eq:minkowski-metric-bondi-physical-frame}
  \begin{align}
  U^M \pd_M
  &= \widebar{U}^M \pd_M
  = \pd_r - \pd_u\,,
  &
  U_M dx^M
  &= \Omega^{-2} \widebar{U} dx^M
  = dr + du\,,
  \\
  V^M \pd_M
  &= \Omega^2 \widebar{V}^M \pd_M
  = - \pd_r\,,
  \label{eq:minkowski-metric-bondi-physical-frame-V-vector}
  &
  V_M dx^M
  &= \widebar{V}_M dx^M
  = du\,,
  \\
  E_a^M \pd_M
  &= \Omega \widebar{E}_a^A \pd_A
  = r\inv e_a^M \pd_M\,,
  &
  E^a_M dx^M
  &= \Omega^{-1} \widebar{E}^a_M dx^M
  = r e^a_A dx^A\,.
  \end{align}
\end{subequations}
One of our main goals in the rest of this section is to generalise this construction away from Minkowski space to general asymptotically flat spacetimes.

\subsection{General null frames and gauge transformations}
\label{ssec:bulk-geom-general-null-frames}
Let us first consider the transformations that leave a null vielbein decomposition of a general metric invariant.
Along the lines of the decomposition in~\eqref{eq:minkowski-metric-bondi-unphysical-frame-decomposition} above, in a general
$(d+2)$-dimensional spacetime
with coordinates $x^M$
and a Lorentzian metric $g_{MN}$,
we can take
\begin{equation}
  \label{eq:gen-metric-null-frame-decomposition}
  ds^2
  = g_{MN} dx^M dx^N
  = - 2 U_M V_N dx^M dx^N
  + \delta_{ab} E^a_M E^b_N dx^M dx^N\,.
\end{equation}
Here, $U_M$ and $V_M$ are null,
while $E^a_M$ (with $a=1,\ldots,d$) are spacelike.
Together with the inverse fields $(U^M,V^M,E_a^M)$,
these frame fields satisfy
the following orthogonality relations
\begin{subequations}
  \label{eq:gen-null-bulk-frame-orthonormality}
  \begin{align}
    U_M V^M
    = V_M U^M
    &= -1\,,
    &
    U_M E_a^M
    = V_M E_a^M
    &= 0\,,
    \\
    E^a_M E_b^M
    &= \delta^a_b\,,
    &
    U^M E^a_M
    = V^M E^a_M
    &= 0\,,
  \end{align}
\end{subequations}
as well as the completeness relation
\begin{equation}
  \label{eq:gen-null-bulk-frame-completeness}
  \delta_M^N
  = - U_M V^N
  - V_M U^N
  + E^a_M E_a^N\,.
\end{equation}
Each term on the right-hand side of~\eqref{eq:gen-null-bulk-frame-completeness} corresponds to a projection operator,
which splits the (co)tangent space into two one-dimensional null subspaces and a $d$-dimensional spacelike subspace.

This null frame parametrisation of a general Lorentzian metric is invariant under local $SO(d+1,1)$ Lorentz transformations.
The generators of this gauge symmetry are split into null boosts $\Lambda$,
two sets of null rotations~$\Lambda^a$ and~$\Sigma^a$,
as well as spatial rotations $\Lambda^{ab}=-\Lambda^{ba}$.
In addition to these local Lorentz transformations, the full set of gauge symmetries include diffeomorphisms generated by vector fields~$\xi^M$.
Together, their infinitesimal action on $(U_M,V_M,E^a_M)$ is
\begin{subequations}
  \label{eq:gen-null-bulk-frame-diffeo-lor-action}
  \begin{align}
    \delta U_M
    &= \LL_\xi U_M
    + \Lambda U_M
    + \Sigma_a E^a_M\,,
    \\
    \delta V_M
    &= \LL_\xi V_M
    - \Lambda V_M
    + \Lambda_a E^a_M\,,
    \\
    \delta E^a_M
    &= \LL_\xi E^a_M
    + \Lambda^a U_M
    + \Sigma^a V_M
    + \Lambda^a{}_b E^b_M\,,
  \end{align}
\end{subequations}
whereas the infinitesimal action on the inverse vielbeine
$(U^M,V^M,E_a^M)$ is
\begin{subequations}
  \label{eq:gen-null-bulk-inverse-frame-diffeo-lor-action}
  \begin{align}
    \delta U^M
    &= \LL_\xi U^M
    + \Lambda U^M
    + \Sigma^a E_a^M\,,
    \\
    \label{eq:gen-null-bulk-inverse-frame-diffeo-lor-action-V-up}
    \delta V^M
    &= \LL_\xi V^M
    - \Lambda V^M
    + \Lambda^a E_a^M\,,
    \\
    \delta E_a^M
    &= \LL_\xi E_a^M
    + \Lambda_a U^M
    + \Sigma_a V^M
    - \Lambda^b{}_a E_b^M\,. \label{eq:inv-spatial-vielbein-trafo}
  \end{align}
\end{subequations}
Here and in the following, spatial frame indices $a,b,\cdots$ are raised and lowered with~$\delta_{ab}$,
the Kronecker delta symbol.
Together with the asymptotic boundary conditions that we will discuss momentarily,
the bulk gauge transformations in~\eqref{eq:gen-null-bulk-frame-diffeo-lor-action} and~\eqref{eq:gen-null-bulk-inverse-frame-diffeo-lor-action} are the fundamental ingredients that give rise to all asymptotic or boundary symmetries that we will encounter later on.

\subsection{Penrose boundary and null frame}
\label{ssec:bulk-geom-penrose-bdy}
Following the seminal work of Penrose~\cite{Penrose:1962ij} (see~\cite{Frauendiener:2000mk,Ashtekar:2014zsa} for reviews), we now recast the conformal compactification of a general $(d+2)$-dimensional asymptotically flat spacetime metric in terms of the null vielbein decomposition above.
We use this to define radial falloff conditions for the bulk vielbeine
in a polynomial large-radius expansion,
though we will later also consider subleading terms that are not smooth near the boundary,
such as log terms.

Recall that, in the Penrose construction,
a `physical' spacetime~$(\mathcal{M},g_{MN})$ is asymptotically flat at null infinity
if there exists an associated `unphysical' spacetime~$(\widebar{\mathcal{M}},\widebar{g}_{MN})$,
with boundary $\partial\widebar{\mathcal{M}}$ and interior $\mathcal{M}$,
and a function $\Omega$ on $\widebar{\mathcal{M}}$
such that the unphysical metric $\widebar{g}_{MN}$
is given by
\begin{equation}
  d\widebar{s}^2
  = \widebar{g}_{MN} dx^M dx^N
  = \Omega^2 g_{MN} dx^M dx^N\,.
\end{equation}
The function $\Omega$,
which is also known as the defining function in this context,
must satisfy the following conditions on $\pd\widebar{\mathcal{M}}$,
\begin{equation}
  \label{eq:general-null-vielbeine-conformal-boundary-def}
  \left.\Omega\right|_{\pd\bar{\mathcal{M}}}
  = 0\,,
  \qquad
  \left.d\Omega\right|_{\pd\bar{\mathcal{M}}}
  \neq 0\,.
\end{equation}
The unphysical spacetime $(\widebar{\mathcal{M}},\widebar{g}_{MN})$ then provides a conformal compactification of the physical spacetime,
and $\pd\widebar{\mathcal{M}}$ is the conformal boundary.

Using the Einstein equations with vanishing cosmological constant,
we can show that the one-form $d\Omega$ is null on the conformal boundary
(see for example~\cite{Riello:2024uvs}).
Because the conformal boundary $\pd\widebar{\mathcal{M}}$ has codimension one,
and because $d\Omega$ is a non-vanishing null 1-form on the boundary by~\eqref{eq:general-null-vielbeine-conformal-boundary-def},
we can choose a null frame
$(\widebar{U}^M, \widebar{V}^M, \widebar{E}^M_a)$
for the unphysical metric such that
\begin{equation}
  \label{eq:general-null-vielbeine-kernel-dOmega}
  \left.\left(
    \widebar{U}^M \pd_M \Omega
  \right)\right|_{\pd\widebar{\mathcal{M}}}
  = 0\,,
  \qquad
  \left.\left(
    \widebar{V}^M \pd_M \Omega
  \right)\right|_{\pd\widebar{\mathcal{M}}}
  \neq 0\,,
  \qquad
  \left.\left(
    \widebar{E}_a^M \pd_M \Omega
  \right)\right|_{\pd\widebar{\mathcal{M}}}
  = 0\,.
\end{equation}
We now use the defining function $\Omega$ to introduce a radial coordinate $r$ as follows%
\footnote{%
  Note that the function $\Omega$ is defined up to rescalings
  $\Omega\to\omega \Omega$, where $\omega$ is strictly positive.
  After its identification with the radial coordinate in~\eqref{eq:def-of-r}, we will see that each term in the $1/r$ expansion of the metric acquires a particular transformation under such rescalings. This will allow us to build objects with a specific Weyl weight.
  In this work, this Weyl-type symmetry  manifests itself as a residual gauge symmetry that preserves the conformal class of Carrollian structures on the boundary $\D \widebar{\mathcal{M}}$, as we discuss in Section~\ref{ssec:bulk-geom-partial-gauge-fixing}.
},
\begin{equation}
\label{eq:def-of-r}
    \Omega=1/r\,.
\end{equation}
Since $d\Omega$ is normal to the conformal boundary $\pd\widebar{\mathcal{M}}$ at $r\to\infty$,
different integral curves of $\pd_r$ reach the boundary at distinct points.
Parametrising the boundary with a $(d+1)$-dimensional set of coordinates $x^\mu$ then allows us to use
$x^M = (r,x^\mu)$ as our bulk coordinates.
In these coordinates,
the requirements~\eqref{eq:general-null-vielbeine-kernel-dOmega} are
\begin{equation}
  \label{eq:general-null-vielbeine-kernel-dOmega-in-coords}
  \widebar{U}^r
  = \OO(r)\,,
  \qquad
  \widebar{V}^r
  = \OO(r^2)
  \text{  and nonzero at that order}\,,
  \qquad
  \widebar{E}_a^r
  = \OO(r)\,.
\end{equation}
As briefly mentioned above,
we currently work with the assumption that the frame fields can be expanded in a power series in $1/r$,
though we will also consider logarithmic corrections in Sections~\ref{ssec:radial-expansion-logs} and~\ref{ssec:log-emt-news} later on. We use the Penrose compactification only to define boundary conditions for an appropriate set of vielbeine. Once this is fixed, one can let go of the Penrose construction and consider solutions that are not smooth near future null infinity due to the presence of logs.

We can then additionally orient our frame such 
that $\widebar{U}^\mu$ and $\widebar{E}_a^\mu$
span the $(d+1)$-dimensional tangent space of the conformal boundary at $r\to\infty$.
In terms of their radial falloffs,
this means that we must have
\begin{equation}
  \label{eq:general-null-vielbeine-boundary-pullback}
  \widebar{U}^\mu
  = \OO(1)\,,
  \,\,
  \widebar{E}_a^\mu
  = \OO(1)
  \text{  and both nonzero at that order.}
\end{equation}
This still leaves us with the $\widebar{V}^\mu$ components.
At this stage, we will only require that they are non-singular at the conformal boundary,
so that
\begin{equation}\label{eq:general-null-vielbeine-boundary-pullback2}
  \bar V^\mu=\mathcal{O}(1)\,.
\end{equation}
Later on, in Section~\ref{ssec:bulk-geom-partial-gauge-fixing},
we will see that we can set these components to zero using a gauge transformation,
and this will be part of our gauge fixing conditions.

\paragraph{Physical frame.}
Mirroring the rescaling~\eqref{eq:minkowski-metric-bondi-physical-frame}
between the unphysical and the physical frames that we employed in the Minkowski case,
we now define a physical null frame for our general spacetime metric $g_{MN}$ using
\begin{subequations}
  \begin{align}
    U^M
    &= \widebar{U}^M\,,
    &
    V^M
    &= \Omega^2 \widebar{V}^M\,,
    &
    E_a^M
    &= \Omega \widebar{E}_a^M\,,
    \\
    U_M
    &= \Omega^{-2} \widebar{U}_M\,,
    &
    V_M
    &= \widebar{V}_M\,,
    &
    E^a_M
    &= \Omega^{-1} \widebar{E}^a_M\,,
  \end{align}
\end{subequations}
For this physical frame, the radial falloff conditions~\eqref{eq:general-null-vielbeine-kernel-dOmega}
and~\eqref{eq:general-null-vielbeine-kernel-dOmega-in-coords} become
\begin{align}
  \label{eq:general-null-vielbeine-physical-falloffs}
  \begin{gathered}
    U^r
    = \OO(r)\,,~~ V^r = \OO(1) \text{  \& nonzero at that order}\,,~~ E_a^r
    = \OO(1)\,,
    \\
    U^\mu
    = \OO(1)\,,
    \,
    E_a^\mu
    = \OO(r^{-1})
    \text{  \& both nonzero at that order}\,,~~   
    V^\mu
    = \OO(r^{-2})\,.
  \end{gathered}
\end{align}
From these conditions on the frame components, we can determine the corresponding conditions for the coframe components using the orthogonality relations~\eqref{eq:gen-null-bulk-frame-orthonormality}.
Since we know that $V^r$ is nonzero at order $r^0$, we can use this to solve for
\begin{equation}
  \label{eq:general-null-vielbeine-physical-solve-for-r-cpts}
  U_r
  = - \frac{1}{V^r} \left(
    1 + U_\mu V^\mu
  \right),
  \qquad
  V_r
  = - \frac{1}{V^r}
  V_\mu V^\mu\,,
  \qquad
  E^a_r
  = - \frac{1}{V^r}
  E^a_\mu V^\mu\,,
\end{equation}
which subsequently implies that
\begin{subequations}
  \label{eq:general-null-vielbeine-physical-mu-contractions}
  \begin{align}
    U_\mu \left(
      U^\mu - \frac{1}{V^r}V^\mu U^r
    \right)
    &= \frac{1}{V^r} U^r\,,
    &
    U_\mu \left(
      E^\mu_a
      - \frac{1}{V^r} V^\mu E^r_a
    \right)
    &= \frac{1}{V^r} E^r_a\,,
    \\
    V_\mu \left(
      U^\mu - \frac{1}{V^r} V^\mu U^r
    \right)
    &= -1\,,
    &
    V_\mu \left(
      E^\mu_a
      - \frac{1}{V^r} V^\mu E^r_a
    \right)
    &= 0\,,
    \\
    E^a_\mu \left(
      U^\mu
      - \frac{1}{V^r} V^\mu U^r
    \right)
    &= 0\,,
    &
    E_\mu^a \left(
      E^\mu_b
      - \frac{1}{V^r} V^\mu E^r_a
    \right)
    &= \delta^a_b\,.
  \end{align}
\end{subequations}
Using the falloff conditions~\eqref{eq:general-null-vielbeine-physical-falloffs} for the frame components,
and plugging the result back into the relations~\eqref{eq:general-null-vielbeine-physical-solve-for-r-cpts},
we obtain the following falloff conditions for the coframe components,
\begin{align}
  \label{eq:general-null-inverse-vielbeine-physical-falloffs}
  \begin{split}
    V_\mu
    &= \OO(1)\,,
    \,
    E^a_\mu
    = \OO(r)
    \text{  and both nonzero at that order}\,,~~    U_\mu
    = \OO(r)\,,
    \\
    U_r
    &= \OO(1)
    \text{  and nonzero at that order}\,, ~~
    V_r
    = \OO(r^{-2})\,,~~
    E^a_r
    = \OO(r^{-1})\,.
  \end{split}
\end{align}
Following the second line of \eqref{eq:general-null-vielbeine-physical-falloffs} and the first line of \eqref{eq:general-null-inverse-vielbeine-physical-falloffs}
we can use the falloff conditions for $(U^\mu, V^\mu, E^\mu_a)$ and $(U_\mu, V_\mu, E^\mu_a)$ near the asymptotic boundary
to define a set of boundary tensors as follows,
\begin{subequations}
  \label{eq:boundary-frame-with-M-falloff-mu-parametrization}
  \begin{align}
    U^\mu
    &= v^\mu
    + \OO(r^{-1})\,,
    &
    U_\mu
    &= r b_\mu
    + \OO(1)\,,
    \\
    V^\mu
    &= r^{-2} M^\mu
    + \OO(r^{-3})\,,
    &
    V_\mu
    &= \tau_\mu
    + \OO(r^{-1})\,,
    \\
    E_a^\mu
    &= r^{-1} e^\mu_a
    + \OO(r^{-2})\,,
    &
    E^a_\mu
    &= r e^a_\mu
    + \OO(1)\,.
  \end{align}
\end{subequations}
Here, $M^\mu$ and $b_\mu$ are arbitrary boundary fields,
while $(v^\mu,e^\mu_a)$ and $(\tau_\mu,e_\mu^a)$ are nowhere-vanishing
boundary fields.
The latter two sets of fields define a boundary frame and its dual coframe,
since the leading-order terms in the expansion of the bulk orthonormality and completeness relations~\eqref{eq:gen-null-bulk-frame-orthonormality} and~\eqref{eq:gen-null-bulk-frame-completeness} imply that
\begin{subequations}
  \label{eq:boundary-frame-with-M-orthonormality-completeness}
  \begin{gather}
    \tau_\mu v^\mu
    = -1\,,
    \qquad
    \tau_\mu e_a^\mu
    = 0\,,
    \qquad
    v^\mu e^a_\mu
    = 0\,,
    \qquad
    e^a_\mu e_b^\mu
    = \delta^a_b\,,
    \\
    \delta^\mu_\nu
    = - v^\mu \tau_\nu
    + e_a^\mu e^a_\nu\,.
  \end{gather}
\end{subequations}
In particular, this means that we have the following projection operators,
\begin{equation}
  \label{eq:introduction-of-projection-operators-LO}
  - \tau_\mu v^\nu\,,
  \qquad
  h^\nu_\mu
  =e^a_\mu e_a^\nu\,.
\end{equation}
However, it is important to note that these projections are not invariant under the local boundary transformations,
which we will discuss momentarily.
(A similar caveat applies to the bulk projection operators mentioned after~\eqref{eq:gen-null-bulk-frame-completeness} above.)
It turns out that the notion of a vector field $X^\mu$ being parallel to $v^\mu$ as well as the notion of a one-form $X_\mu$ obeying $v^\mu X_\mu=0$ are invariant under local transformations.
In this sense, the `null time' vector field $v^\mu$ defines a one-dimensional fibration of the boundary manifold.

Finally, we turn to the radial components of the bulk vielbeine.
Throughout this work, we will use the notation
\begin{equation}
  \os{n}{X}
\end{equation}
to denote the term that multiplies $r^{-n}$ in the $1/r$ expansion of the object $X$. 
The identities~\eqref{eq:general-null-vielbeine-physical-solve-for-r-cpts} and~\eqref{eq:general-null-vielbeine-physical-mu-contractions} then imply that
\begin{subequations}
  \label{eq:boundary-frame-with-M-radial-parametrization}
  \begin{align}
    U^r
    &= r \os{0}{V}^r b_\mu v^\mu
    + \OO(1)\,,
    &
    U_r
    &= - \frac{1}{\os{0}{V}^r}
    + \OO(r^{-1})\,,
    \\
    V^r
    &= \os{0}{V}^r
    + \OO(r^{-1})\,,
    &
    V_r
    &= - r^{-2} \frac{\tau_\mu M^\mu}{\os{0}{V}^r}
    + \OO(r^{-3})\,,
    \\
    E_a^r
    &= \os{0}{V}^r b_\mu e^\mu_a
    + \OO(r^{-1})\,,
    &
    E^a_r
    &= - r^{-1} \frac{e^a_\mu M^\mu}{\os{0}{V}^r}
    + \OO(r^{-2})\,,
  \end{align}
\end{subequations}
where we note that $\os{0}{V}^r$ must be nonzero by~\eqref{eq:general-null-vielbeine-physical-falloffs}.

\paragraph{Gauge transformations.}
As we mentioned, the boundary fields $(\tau_\mu,e^a_\mu)$ and $(v^\mu,e_a^\mu)$ span the boundary geometry,
and they will be our fundamental geometric variables going forward.
Together with their local gauge transformations,
they define the Carroll metric data on the asymptotic null boundary.
At this point, however, we will see that there are still several additional fields and transformations on the boundary, most of which we will fix in the following.

Preserving the falloff conditions~\eqref{eq:boundary-frame-with-M-falloff-mu-parametrization}
and~\eqref{eq:boundary-frame-with-M-radial-parametrization}
constrains the radial expansion of the bulk gauge transformations~\eqref{eq:gen-null-bulk-frame-diffeo-lor-action}
and~\eqref{eq:gen-null-bulk-inverse-frame-diffeo-lor-action} in the following ways.
The diffeomorphisms $\xi^M$
must satisfy
\begin{subequations}
  \label{eq:boundary-frame-gauge-tr-expansions-diffeos}
  \begin{align}
    \xi^r
    &= r \os{-1}{\xi}^r
    + \os{0}{\xi}^r
    + \OO(r^{-1})\,,
    \\
    \xi^\mu
    &= \chi^\mu
    + r^{-1} \os{1}{\xi}^\mu
    + \OO(r^{-2})\,,
  \end{align}
\end{subequations}
while the null boosts $\Lambda$,
the two sets of null rotations $\Lambda^a$ and $\Sigma^a$
and the spatial rotations $\Lambda^{ab}$ must satisfy
\begin{subequations}
  \label{eq:boundary-frame-gauge-tr-expansions-local-gauge}
  \begin{align}
    \Lambda
    &= \lambda
    + \OO(r^{-1})\,,
    \\
    \Lambda^a
    &= r^{-1} \lambda^a
    + \OO(r^{-2})\,,
    \\
    \Sigma^a
    &= \sigma^a
    + \OO(r^{-1})\,,
    \\
    \Lambda^{ab}
    &= \lambda^{ab}
    + \OO(r^{-1})\,.
  \end{align}
\end{subequations}
Given these radial expansions of the parameters,
the boundary fields we introduced in~\eqref{eq:boundary-frame-with-M-falloff-mu-parametrization} then transform as follows,
\begin{subequations}
  \label{eq:boundary-frame-with-M-gauge-tr-full}
  \begin{align}
    \delta v^\mu
    &= \LL_\chi v^\mu
    + \lambda v^\mu\,,
    \\
    \delta e_a^\mu
    &= \LL_\chi e_a^\mu
    - \os{-1}{\xi}^r e^\mu_a
    + \lambda_a v^\mu
    - \lambda^b{}_a e^\mu_b\,,
    \\
    \delta M^\mu
    &= \LL_\chi M^\mu
    - \left(2 \os{-1}{\xi}^r + \lambda\right) M^\mu
    + \os{0}{V}^r \os{1}{\xi}^\mu
    + \lambda^a e^\mu_a\,,\label{eq:delta-M}
    \\
    \delta \tau_\mu
    &= \LL_\chi \tau_\mu
    - \lambda \tau_\mu
    + \lambda_a e^a_\mu\,,\label{eq:deltatau}
    \\
    \delta e^a_\mu
    &= \LL_\chi e^a_\mu
    + \os{-1}{\xi}^r e^a_\mu
    + \lambda^a{}_b e^b_\mu\,,\label{eq:deltae}
    \\
    \delta b_\mu
    &= \LL_\chi b_\mu
    + \left(\os{-1}{\xi}^r +\lambda\right) b_\mu
    - \frac{1}{ \os{0}{V}^r} \pd_\mu \os{-1}{\xi}^r
    + \sigma_a e^a_\mu\,,
    \label{eq:delta-b}
    \\
    \delta \os{0}{V}^r
    &= \LL_\chi \os{0}{V}^r
    - \left(\os{-1}{\xi}^r 
    + \lambda\right) \os{0}{V}^r\,.\label{eq:delta-Vr}
  \end{align}
\end{subequations}
In particular,
we see that the boundary fields transform under boundary diffeomorphisms%
\footnote{%
  \label{fn:lie-deriv-dim}
  Note that here and in the following, we will adopt the following convention concerning Lie derivatives. If we act on a $d+1$-dimensional ($d+2$-dimensional) tensor, then the Lie derivative along $X$ is to be read as a $d+1$-dimensional ($d+2$-dimensional) Lie derivative along $X^\mu$ ($X^M$).
}
generated by the $\chi^\mu$ parameters that correspond to the leading-order terms in the radial expansion~\eqref{eq:boundary-frame-gauge-tr-expansions-diffeos} of the bulk diffeomorphism parameters~$\xi^\mu$.
The boundary transformations~\eqref{eq:boundary-frame-with-M-gauge-tr-full} also include the subleading bulk diffeomorphism generators~$\os{1}{\xi}^\mu$,
as well as the leading-order radial diffeomorphisms $\os{-1}{\xi}^r$
and the leading-order local Lorentz transformation parameters
$\lambda$, $\lambda^a$, $\sigma^a$ and $\lambda^{ab}$.

\subsection{Conformal Carroll boundary from partial gauge fixing}
\label{ssec:bulk-geom-partial-gauge-fixing}
At this point,
we have imposed the boundary conditions~\eqref{eq:general-null-vielbeine-kernel-dOmega-in-coords}--\eqref{eq:general-null-vielbeine-boundary-pullback2} and chosen a set of bulk coordinates $x^M=(r,x^\mu)$ adapted to these boundary conditions,
but we have not yet imposed any gauge-fixing conditions.
However, it is easy to see from the transformations in~\eqref{eq:boundary-frame-with-M-gauge-tr-full} that several of the resulting boundary fields are superfluous.
One of our main aims in the following will be to define a gauge fixing of the bulk degrees of freedom
that allows for a description of the boundary geometry in terms of conformal Carrollian geometry.
This means that we will fix most of the bulk gauge transformations, while still allowing for arbitrary boundary diffeomorphisms, Weyl transformations and local Carroll boosts. 
We first give a brief overview of this gauge fixing procedure in terms of the boundary variables,
and we then give a detailed derivation of the corresponding bulk gauge choices.

\paragraph{Overview.}
Let us first consider the transformation of~$\os{0}{V}^r$ in~\eqref{eq:delta-Vr},
which reads
\begin{equation}
  \delta \os{0}{V}^r
  = \LL_\chi \os{0}{V}^r
  - \left(\os{-1}{\xi}^r 
  + \lambda\right) \os{0}{V}^r\,.
\end{equation}
We can interpret the last term in brackets as a shift term for
$\log\os{0}{V}^r$,
which is well-defined since $\os{0}{V}^r$ is nonzero by our boundary conditions.
As a result, we can use this gauge transformation to fix its value,
and preserving this gauge choice relates the corresponding gauge parameters as follows,
\begin{equation}
  \os{0}{V}^r
  = -1
  \qiq
  \lambda
  = - \os{-1}{\xi}^r\,.
  \label{eq:partial-gf-lo-lambda}
\end{equation}
With this,
the transformations of $\tau_\mu$ and $e_\mu^a$
in~\eqref{eq:deltatau}
and~\eqref{eq:deltae}
give
\begin{subequations}
  \begin{align}
    \delta \tau_\mu
    &= \LL_\chi \tau_\mu
    + \os{-1}{\xi}^r \tau_\mu
    + \lambda_a e^a_\mu\,,
    \\
    \delta e^a_\mu
    &= \LL_\chi e^a_\mu
    + \os{-1}{\xi}^r e^a_\mu
    + \lambda^a{}_b e^b_\mu\,.
  \end{align}
\end{subequations}
In these transformations,
the parameter $\lambda_a$ corresponds to local Carroll boosts,
and the $\lambda_{ab}$ are local spatial rotations.
Additionally,
we now see that both fields
transform under a Weyl transformation with parameter $\overset{(-1)}{\xi}{}^r$,
both with the same Weyl weight $+1$.
The transformations of $(\tau_\mu,e^a_\mu)$ also determine the transformations of the inverse fields $(v^\mu,e^\mu_a)$,
whose precise form will be given in~\eqref{eq:boundary-frame-gauge-tr-with-V-fixed-completely} below.

Next, consider the boundary vector $M^\mu$,
whose transformation~\eqref{eq:delta-M} now is
\begin{equation}
  \delta M^\mu
  = \LL_\chi M^\mu
  - \os{-1}{\xi}^r M^\mu
  - \os{1}{\xi}^\mu
  + \lambda^a e^\mu_a\,,
\end{equation}
Again, we see that the $M^\mu$ field is superfluous,
since we can use the second-to-last term in this transformation to set it to zero.
Preserving this gauge choice gives
\begin{equation}
  M^\mu
  = 0
  \qiq
  \os{1}{\xi}^\mu
  =\lambda^a e^\mu_a\,.
  \label{eq:partial-gf-lo-boosts}
\end{equation}
This means that
the boundary local Carroll boost generator $\lambda_a$ is identified
with the subleading part of the $\xi^\mu$ bulk diffeomorphism generators.
Finally,
following~\eqref{eq:delta-b},
the transformation of the boundary field $b_\mu$ now gives
\begin{equation}
    \delta b_\mu
    = \LL_\chi b_\mu
    +\pd_\mu \os{-1}{\xi}^r
    + \sigma_a e^a_\mu\,.
\end{equation}
This is the transformation of a Weyl connection,
along with an additional shift transformation with parameter $\sigma_a$.
We will fix this shift transformation later on in Section~\ref{ssec:car-cov-BS-gauge-fix-Sigma}
and we will keep the $b_\mu$ field arbitrary for now.

Together, these gauge choices therefore reduce our boundary fields and their transformations to those of a conformal Carrollian geometry.
Since $M^\mu = \os{0}{V}^\mu$,
we can summarise this gauge fixing as follows,
\begin{equation}
  \label{eq:gauge-fixing-V-vector-minimally-intro}
  V^M=-\delta^M_r+\mathcal{O}(r^{-1})\,.
\end{equation}
In fact,
as we will see in the remainder of this subsection,
we can use further bulk local Lorentz transformations and diffeomorphisms to set
\begin{equation}
  \label{eq:gauge-fixing-V-vector-completely-intro}
  V^M \pd_M
  = - \pd_r\,,
\end{equation}
to all orders in the radial expansion.
We will show that this choice fixes the null boosts generated by~$\Lambda$
and the null rotations generated by~$\Lambda^a$ to all orders.
In terms of the bulk metric, it implies that $g_{rr}=0$.
In this gauge, we also have
\begin{equation}
  \label{eq:V-asymptotically-geodesic-intro}
  V^R \nabla_R V^M=\Gamma^{M}_{rr}
  =-\Gamma^{r}_{rr}V^M+\Gamma^{\mu}_{rr} \delta^M_\mu\,,
\end{equation}
where $\Gamma^\mu_{rr}=\OO(r^{-4})$ and $\Gamma^r_{rr}=\OO(r^{-3})$. This must be contrasted with the result obtained by computing $X^R \nabla_R X^M$, where $X^M \pd_M = - \pd_r$, in a gauge where $g_{rr}=\OO(r^{-2})$, which is what we would get if we were to only use the boundary conditions. In that case,~\eqref{eq:V-asymptotically-geodesic-intro} would still hold with $V^M$ replaced by $X^M$, but we would get $\Gamma^\mu_{rr}=\OO(r^{-3})$ and $\Gamma^r_{rr}=\OO(r^{-2})$. We thus see that, in our gauge~\eqref{eq:gauge-fixing-V-vector-completely-intro}, we get a stronger fall off, because it leads to $\Gamma^\mu_{rr}=\OO(r^{-4})$ and $\Gamma^r_{rr}=\OO(r^{-3})$, with $\Gamma^r_{rr}$ the larger of the two. We thus see that the null vector $V^M \pd_M = -\pd_r$ is asymptotically geodesic.
If we also make the additional gauge choice $\Gamma^{\mu}_{rr}=0$, then $V^M$ is geodesic for all $r$, and we will do so in the next section.

Both~\eqref{eq:gauge-fixing-V-vector-minimally-intro} and~\eqref{eq:gauge-fixing-V-vector-completely-intro} are only partial gauge fixings,
since they still leave several bulk gauge transformations unfixed.
Residual gauge transformations may be physical, and the gravitational phase spaces resulting from distinct gauge choices may not be equivalent.
For example, we will consider two different ways to fix the radial diffeomorphisms
in Sections~\ref{ssec:car-cov-BS-gauge-fix-radial-diffeo} and~\ref{ssec:Newman-Unti},
which will lead to Carroll-covariant versions of the standard Bondi--Sachs and Newman--Unti gauges.
Though we will not compute the charges associated to these transformations,
they have been argued to be physical by Geiller and Zwikel~\cite{Geiller:2022vto,Geiller:2024amx} 
in a similar (though not fully Carroll-covariant) setup.

As we shall see in Section~\ref{ssec:bulk-geom-LO-EE-in-partial-gauge},
the partial gauge choice \eqref{eq:gauge-fixing-V-vector-completely-intro} is sufficient to derive a Carroll-covariant version of the shear tensor
$C_{\mu\nu}=h^\rho_{\langle\mu} h^\sigma_{\nu\rangle} \os{-1}{g}_{\rho\sigma}$ along with its transformations.
Even though the shear appears at a subleading order in the metric expansion,
we will see later on that it should be considered part of the boundary data, along with the Carroll metric data mentioned above.
In this sense, the gauge choice~\eqref{eq:gauge-fixing-V-vector-completely-intro} is sufficient to fully specify the boundary data and its Carroll-covariant transformations.

\paragraph{First bulk gauge choice.}
We now show that we can indeed always make the bulk gauge choice \eqref{eq:gauge-fixing-V-vector-completely-intro} and we work out its consequences in more detail.
We start by fixing $V^r$ to a constant to all orders in the radial expansion.
To see that this can be done,
recall from~\eqref{eq:gen-null-bulk-inverse-frame-diffeo-lor-action-V-up} that the gauge transformations of $V^r$ are
\begin{equation}
  \delta V^r
  = \xi^R \pd_R V^r
  - V^R \pd_R \xi^r
  - \Lambda V^r
  + \Lambda^a E_a^r\,.
\end{equation}
Given that the leading-order radial term in $V^r$ must be nonzero,
we can use the longitudinal boost transformation parametrised by $\Lambda$
to set
\begin{equation}
  \label{eq:boundary-frame-gauge-fixing-Vr}
  V^r
  = - 1\,.
\end{equation}
Preserving this gauge choice then fixes the null boost parameter $\Lambda$ in terms of the radial diffeomorphism parameter $\xi^r$ and the null rotation parameters $\Lambda^a$
as follows,
\begin{equation}
  \label{eq:boundary-frame-gauge-fixing-Vr-params}
  \begin{split}
  0
  = \delta V^r
  &= \pd_r \xi^r
  - V^\rho \pd_\rho \xi^r
  + \Lambda
  + \Lambda^a E_a^r\,,
  \\
  &{}
  \qiq
  \Lambda
  = - \pd_r \xi^r
  + V^\rho \pd_\rho \xi^r
  - \Lambda^a E^r_a\,.
  \end{split}
\end{equation}
The gauge choice~\eqref{eq:boundary-frame-gauge-fixing-Vr} therefore reduces the total amount of bulk gauge symmetries by one bulk function,
and this reduces to the condition in~\eqref{eq:partial-gf-lo-lambda} above
at leading order.
The latter motivates us to introduce the notation
\begin{equation}
  \label{eq:identifying-Weyl-generator-with-LO-radial-diffeo}
  \Lambda_D
  = - \lambda
  = \os{-1}{\xi}^r\,,
\end{equation}
since, as we already mentioned,
these boundary transformations now act as Weyl transformations,
with the function $\Lambda_D$ as their parameter.
Indeed, with this gauge fixing,
the transformations~\eqref{eq:boundary-frame-with-M-gauge-tr-full} of the boundary fields become
\begin{subequations}
  \label{eq:boundary-frame-gauge-tr-with-Vr-fixed}
  \begin{align}
    \delta v^\mu
    &= \LL_\chi v^\mu
    - \Lambda_D v^\mu\,,
    \\
    \label{eq:boundary-frame-gauge-tr-with-Vr-fixed-M-tr}
    \delta M^\mu
    &= \LL_\chi M^\mu
    - \Lambda_D M^\mu
    - \os{1}{\xi}^\mu
    + \lambda^a e^\mu_a\,,
    \\
    \delta e_a^\mu
    &= \LL_\chi e_a^\mu
    - \Lambda_D e^\mu_a
    + \lambda_a v^\mu
    - \lambda^b{}_a e^\mu_b\,,
    \\
    \delta \tau_\mu
    &= \LL_\chi \tau_\mu
    + \Lambda_D \tau_\mu
    + \lambda_a e^a_\mu\,,
    \\
    \delta e^a_\mu
    &= \LL_\chi e^a_\mu
    + \Lambda_D e^a_\mu
    + \lambda^a{}_b e^b_\mu\,,
    \\
    \delta b_\mu
    &= \LL_\chi b_\mu
    + \pd_\mu \Lambda_D
    + \sigma_a e^a_\mu\,.
  \end{align}
\end{subequations}
Here, we see that $\Lambda_D$ acts as a Weyl transformation
with weight $-1$
on the boundary vectors
$(v^\mu, M^\mu, e^\mu_a)$
and weight $+1$
on the boundary covectors
$(\tau_\mu, e^a_\mu)$,
and we see that $b_\mu$ is a Weyl connection.
This is close to what we expect from a conformal Carrollian geometry,
but the $M^\mu$ field does not have a natural interpretation in that context.
As we saw above,
from these transformations it is also clear that the subleading diffeomorphism parameter $\os{1}{\xi}^\mu$
can be used to set $M^\mu$ to zero.
This gauge fixing was not used in earlier work such as~\cite{Hartong:2015usd},
in an effort to retain any potentially physical effects.
However, since we do not need the $M^\mu$ field in a Carroll-covariant generalisation of Bondi gauge,
and since removing it significantly simplifies our computations,
we will gauge fix it to zero in the following.%
\footnote{
  In \cite{Hartong:2015xda,Hansen:2021fxi} it was shown that $M^\mu$ generically appears either on null hypersurfaces or when expanding around $c=0$.
  One can view $M^\mu$ as a Stückelberg field for Carroll boosts,
  and additional gauge transformations are needed to be able to remove $M^\mu$
  without fixing the Carroll boosts.
  Here, this comes via the action of subleading bulk diffeomorphisms with parameter $\os{1}{\xi}^\mu$.
}

\paragraph{Second bulk gauge choice.}
Following~\eqref{eq:gauge-fixing-V-vector-completely-intro},
we then impose the gauge choice
\begin{equation}
  \label{eq:gauge-fixing-V-vector-completely}
  V^M \pd_M
  = - \pd_r\,,
\end{equation}
which also serves to set $M^\mu$ to zero,
since the latter corresponds to the leading-order term in the radial expansion of $V^\mu$
by~\eqref{eq:boundary-frame-with-M-falloff-mu-parametrization}.
This gauge choice
aligns the null vector $V^M$ with (minus) the radial coordinate vector.
With this, the orthogonality relations~\eqref{eq:general-null-vielbeine-physical-solve-for-r-cpts} and~\eqref{eq:general-null-vielbeine-physical-mu-contractions} reduce to
\begin{equation}
  \label{eq:general-null-vielbeine-physical-mu-contractions-fixing-V-completely}
  \begin{gathered}
    U_\mu U^\mu
    = - U^r\,,
    \qquad
    V_\mu U^\mu
    = -1\,,
    \qquad
    E_\mu^a E^\mu_b
    = \delta^a_b\,,
    \\
    U_\mu E^\mu_a
    = - E^r_a
    \qquad
    V_\mu E^\mu_a
    = 0\,,
    \qquad
    E_\mu^a U^\mu
    = 0\,,
  \end{gathered}
\end{equation}
and the remaining radial components of the bulk vielbeine are
\begin{equation}
  \label{eq:general-null-vielbeine-r-components-fixing-V-completely}
  U_r
  = 1\,,
  \qquad
  V_r
  = 0\,,
  \qquad
  E_r^a
  = 0\,.
\end{equation}
Similarly,
the completeness relation~\eqref{eq:gen-null-bulk-frame-completeness} for the null vielbeine now reduces to
\begin{equation}
  \label{eq:null-bulk-frame-completeness-V-fixed}
  \delta_\mu^\nu
  = - V_\mu U^\nu + E_\mu^a E^\nu_a\,.
\end{equation}
This means in particular that that $(V_\mu, E_\mu^a)$ and $(U^\mu, E^\mu_a)$
form a complete set of vielbeine for the $(d+1)$-dimensional (co)tangent spaces
of the equal $r$ surfaces.
We note that $U_\mu$ is not contained in this set of $(d+1)$-dimensional bulk vielbeine.
Later on, in Section~\ref{ssec:car-cov-BS-gauge-fix-Sigma},
we will show that $U_\mu$ can be gauge fixed away up to a single function,
but this will not be necessary at this point.
Furthermore, we can use the vielbeine above to define the projection operators
\begin{equation}
  \label{eq:introduction-of-projection-operators-unexpanded}
  - V_\mu U^\nu\,,
  \qquad
  E^a_\mu E_a^\nu\,,
\end{equation}
which again come with the caveat that they are not invariant under all local gauge transformations.
At leading order in the large $r$ expansion,
the projectors in~\eqref{eq:introduction-of-projection-operators-unexpanded} reduce to 
the boundary projectors $-\tau_\mu v^\nu$ and $e_\mu^a e^\nu_a$
we defined in~\eqref{eq:introduction-of-projection-operators-LO}
using the leading-order boundary variables.
Since the latter are independent of $r$,
they can in fact be used as a basis on any equal $r$ surface,
and we will switch between $r$-dependent and $r$-independent frames whenever it is convenient.

Finally, to see how we can achieve the gauge choice $V^M = - \delta^M_r$,
consider
the transformation of $V^M$ from~\eqref{eq:gen-null-bulk-inverse-frame-diffeo-lor-action-V-up},
which we now write as
\begin{equation}
  \delta V^M
  = \xi^P\partial_P V^M-V^P\partial_P\xi^M
  - \Lambda V^M
  + \Lambda^a E_a^M\,.
\end{equation}
Since $V^r$ is nonzero,
this transformation allows us to set $V^\mu$ to zero
using $V_\mu \pd_r \xi^\mu$
as well as a combination of the $E_\mu^a \pd_r V^\mu$ and $\Lambda^a$ parameters.
Subsequently, we can use the $\Lambda$ transformations to set $V^r=-1$. 
Solving $\delta V^M=0$
then gives us the residual gauge transformations
and, using the orthogonality relations~\eqref{eq:general-null-vielbeine-physical-mu-contractions-fixing-V-completely},
this implies that we must have
\begin{subequations}
  \label{eq:boundary-frame-with-M-gauge-fixing-V-completely-params}
  \begin{align}
    \Lambda^a
    &= - E^a_\mu \pd_r \xi^\mu\,,\label{eq:fixedLambdaa}
    \\
    \Lambda
    &= - \pd_r \xi^r
    - U_\mu \pd_r \xi^\mu\,,
    \\
    0
    &= V_\mu \pd_r \xi^\mu\,. \label{eq:residualgaugetrafo3}
  \end{align}
\end{subequations}
The third condition is a constraint on the bulk diffeomorphisms,
which can be used to solve for $\tau_\mu \os{n}{\xi}^\mu$ for all $n\geq1$ order by order.
For example, from the first two orders of its expansion, we have
\begin{equation}
  \label{eq:tau-xi-order-one-and-two}
  \tau_\mu \os{1}{\xi}^\mu
  = 0\,,
  \qquad
  \tau_\mu \os{2}{\xi}^\mu
  = - \frac{1}{2} \os{1}{V}_\mu \os{1}{\xi}^\mu\,.
\end{equation}
On the other hand, 
the first two equations in~\eqref{eq:boundary-frame-with-M-gauge-fixing-V-completely-params} can be used to fully fix
the bulk null rotations~$\Lambda^a$
and the bulk null boosts~$\Lambda$,
at each order in their radial expansion,
in terms of the bulk diffeomorphism parameters.
At leading order, this reproduces~\eqref{eq:partial-gf-lo-boosts}
and~\eqref{eq:identifying-Weyl-generator-with-LO-radial-diffeo},
so that we have
\begin{equation}
  \label{eq:spatial-xi-order-one-boosts}
  \Lambda_D
  = - \lambda
  = \os{-1}{\xi}^r\,,
  \qquad
  \os{1}{\xi}^\mu
  = e_a^\mu \lambda^a\,.
\end{equation}
The residual gauge transformations~\eqref{eq:boundary-frame-gauge-tr-with-Vr-fixed} of the boundary fields then give
\begin{subequations}
  \label{eq:boundary-frame-gauge-tr-with-V-fixed-completely}
  \begin{align}
    \delta v^\mu
    &= \LL_\chi v^\mu
    - \Lambda_D v^\mu\,,
    \\
    \delta e_a^\mu
    &= \LL_\chi e_a^\mu
    - \Lambda_D e^\mu_a
    + \lambda_a v^\mu
    - \lambda^b{}_a e^\mu_b\,,
    \\
    \delta \tau_\mu
    &= \LL_\chi \tau_\mu
    + \Lambda_D \tau_\mu
    + \lambda_a e^a_\mu\,,
    \\
    \delta e^a_\mu
    &= \LL_\chi e^a_\mu
    + \Lambda_D e^a_\mu
    + \lambda^a{}_b e^b_\mu\,,
    \\
    \delta b_\mu
    &= \LL_\chi b_\mu
    + \pd_\mu \Lambda_D
    + \sigma_a e^a_\mu\,.
  \end{align}
\end{subequations}
The resulting boundary geometry is a conformal Carrollian manifold
with $b_\mu$~playing the role of a Weyl connection.
Its gauge symmetries include diffeomorphisms~$\chi^\mu$,
local Carroll boosts~$\lambda_a$,
local spatial rotations~$\lambda^a{}_b$,
local special conformal transformations~$\sigma^a$
and Weyl transformations~$\Lambda_D$.
In the upcoming paper~\cite{Hartong:2025WIP2}, it will be shown that the fields $(\tau_\mu, e^a_\mu, b_\mu)$ as well as $\os{0}{E}^a_\mu$ and $\os{0}{U}_\mu$ may be viewed as the components of a Cartan connection for the conformal Carroll algebra.

\paragraph{Metric variables.}
So far, we have mainly worked with vielbeine, both in the bulk and for the leading-order boundary variables.
However, we will often work with metric components instead,
which naturally leads us to the (degenerate) spatial metric $\Pi_{\mu\nu}$ and its (degenerate) projective inverse $\Pi^{\mu\nu}$,
which are defined by contracting the spatial vielbeine as follows,
\begin{equation}
  \Pi_{\mu\nu}
  = \delta_{ab} E^a_\mu E^a_\nu\,,
  \qquad
  \Pi^{\mu\nu}
  = \delta^{ab} E_a^\mu E_a^\nu\,.
\end{equation} 
One particular advantage of these variables is that they are invariant under local spatial rotations,
so we will not need to worry about those transformations in the following.
In terms of these variables, the components of the bulk metric and its inverse
after the gauge fixing~\eqref{eq:gauge-fixing-V-vector-completely} are given by
\begin{subequations}
\label{eq:metric-components-after-gauge-fixing-V-vector}
\begin{gather}
    g_{rr} = 0\,,
    \qquad
    g_{r\mu}= -V_\mu\,,
    \qquad
    g_{\mu\nu}=-U_\mu V_\nu-U_\nu V_\mu+\Pi_{\mu\nu}\,,
    \label{eq:gmunugaugefixed}
    \\ 
    g^{rr}=-2U_\mu U^\mu+\Pi^{\mu\nu}U_\mu U_\nu\,,
    \qquad
    g^{r\mu}=U^\mu-U_\nu \Pi^{\mu\nu}\,,
    \qquad
    g^{\mu\nu}=\Pi^{\mu\nu}\,.
\end{gather}
\end{subequations}
It is interesting to notice that, in this gauge, the metric takes the form of a general null reduction ansatz (see for example~\cite{Christensen:2013rfa}),
with the exception that $\partial_r$ is not a Killing vector here.
From~\eqref{eq:general-null-vielbeine-r-components-fixing-V-completely} we had
$U_r=1$, while $V_r=0$ and $E_r^a=0$,
so the bulk volume form satisfies
\begin{align}
  \sqrt{-\det g_{MN}}d^{d+2}x
  &= U\wedge V\wedge E^1\wedge\cdots \wedge E^d
  \nonumber\\
  &= dr
  \wedge \left(V_\mu dx^\mu\right)
  \wedge \left(E^1_{\nu_1}dx^{\nu_1}\right)
  \wedge \cdots 
  \wedge \left(E^d_{\nu_d}dx^{\nu_d}\right)
  \nonumber\\
  &= E\, d^{d+2}x\,,
  \label{eq:gauge-fixed-det-met}
\end{align}
where we defined $E$ as the determinant of the non-degenerate $(d+1)\times(d+1)$ matrix obtained from our metric variables as follows,
\begin{equation}
  \label{eq:bulk-r-surface-vielbein-determinant}
  E
  = \det\left(V_\mu, E^a_\mu\right)
   =\sqrt{\det\left(V_\mu V_\nu+\Pi_{\mu\nu}\right)}\,.
\end{equation}
Additionally,
we can now denote the spatial projection operator introduced in~\eqref{eq:introduction-of-projection-operators-unexpanded} using $\Pi^\mu_\nu$,
and in terms of the spatial metric variables it is given by
\begin{equation}
  \label{eq:spatial-Pi-projector}
  \Pi^\mu_\nu
  = \Pi^{\mu\rho}\Pi_{\rho\nu}
  = E^a_\nu E^\mu_a\,.
\end{equation}
Note that we have $V_\mu \Pi^{\mu\nu}=0$,
but $U_\mu \Pi^{\mu\nu} = - \delta^{ab} E^r_a E^\nu_b$
is not necessarily zero for now.
We will address this as part of our further gauge fixing.
In the radial expansion, we have 
\begin{equation}
  \label{eq:Pi-expansions}
  \Pi_{\mu\nu}
  = r^2 h_{\mu\nu}+r\os{-1}{\Pi}_{\mu\nu} + \OO(1)\,,
  \qquad
  \Pi^{\mu\nu}
  = r^{-2} h^{\mu\nu} +r^{-3}\os{3}{\Pi}^{\mu\nu}+ \OO(r^{-4})\,,
\end{equation}
where we defined
$h_{\mu\nu} = \delta_{ab} e^a_\mu e^b_\nu$
and
$h^{\mu\nu} = \delta^{ab} e_a^\mu e_b^\nu$
in terms of the leading vielbein variables we defined in~\eqref{eq:boundary-frame-with-M-falloff-mu-parametrization}.
Along the same lines, we introduce the notation
\begin{equation}
  h^\mu_\nu
  = h^{\mu\rho} h_{\rho\nu}
  = e_a^\mu e^a_\nu
\end{equation}
for the spatial projection operator that was defined in~\eqref{eq:introduction-of-projection-operators-LO}.

At this point, starting from~\eqref{eq:gen-null-bulk-frame-diffeo-lor-action} and~\eqref{eq:gen-null-bulk-inverse-frame-diffeo-lor-action} and using the conditions~\eqref{eq:boundary-frame-with-M-gauge-fixing-V-completely-params}, our remaining gauge transformations are
\begin{subequations}
  \label{eq:gauge-tr-with-pi-after-lambdas-fixing}
  \begin{align}
    \delta \Pi_{\mu\nu}
    &= \LL_\xi \Pi_{\mu\nu}+\xi^r\partial_r\Pi_{\mu\nu}
    - 2 U_{(\mu} \Pi_{\nu)\rho} \pd_r \xi^\rho
    + 2 \Sigma_a V_{(\mu} E_{\nu)}^a\,,
    \\
    \delta \Pi^{\mu\nu}
    &= \LL_\xi \Pi^{\mu\nu}
    +\xi^r\partial_r\Pi^{\mu\nu}+2U_\rho\Pi^{\rho(\mu}\partial_r\xi^{\nu)}- 2 U^{(\mu} \pd_r \xi^{\nu)}\,,
    \\
    \delta V_\mu
    &= \LL_\xi V_\mu+\xi^r\partial_r V_\mu
    + \pd_r \xi^r V_\mu
    + U_\rho \pd_r \xi^\rho V_\mu
    - \Pi_{\mu\rho} \pd_r \xi^\rho\,,\label{eq:deltaVmu-after-lambdas-fixing}
    \\
    \delta U_\mu
    &= \LL_\xi U_\mu
    +\xi^r\partial_r U_\mu+\partial_\mu\xi^r- \pd_r \xi^r U_\mu
    - U_\rho \pd_r \xi^\rho U_\mu
    + \Sigma_a E^a_\mu\,,
    \\
    \delta U^\mu
    &= \LL_\xi U^\mu
    +\xi^r\partial_r U^\mu+U_\rho U^\rho\partial_r\xi^\mu- \pd_r \xi^r U^\mu
    - U_\rho \pd_r \xi^\rho U^\mu
    + \Sigma^a E^\mu_a\,.
  \end{align}
\end{subequations}
As we already mentioned in Footnote~\ref{fn:lie-deriv-dim},
the Lie derivative $\LL_\xi$ in these expressions is a $(d+1)$-dimensional Lie derivative along $\xi^\mu\partial_\mu$.
(In the context of the analogy with null reductions mentioned immediately below \eqref{eq:gmunugaugefixed}, we see that we can think of $\Sigma^a$ as a local Galilean boost.)
Using the vielbein expressions in~\eqref{eq:boundary-frame-gauge-tr-with-V-fixed-completely},
we see that the transformations of the leading-order spatial variables are
\begin{subequations}
  \label{eq:boundary-frame-gauge-tr-with-V-fixed-completely-for-h-tensors}
  \begin{align}
  \delta h_{\mu\nu}
  &= \LL_\chi h_{\mu\nu}
  + 2 \Lambda_D h_{\mu\nu}\,,
  \\
  \delta h^{\mu\nu}
  &= \LL_\chi h^{\mu\nu}
  - 2 \Lambda_D h^{\mu\nu}
  + 2 \lambda^a e_a^{(\mu} v^{\nu)}\,,
  \end{align}
\end{subequations}
which in particular reproduces the standard Carroll boost transformation rules.

\paragraph{Inverse expansions.}
We can use the orthogonality relations~\eqref{eq:general-null-vielbeine-physical-mu-contractions-fixing-V-completely} and~\eqref{eq:null-bulk-frame-completeness-V-fixed} to fix the subleading terms in the radial expansion of
$(U^\mu, \Pi^{\mu\nu})$
in terms of the expansion of
$(V_\mu, \Pi_{\mu\nu})$.
For example, to first subleading order, this gives
\begin{equation}
  \label{eq:U-Pi-expansion-in-terms-of-V-Pi-expansion}
  \os{1}{U}^\mu
  = v^\mu v^\rho \os{1}{V}_\rho
  - h^{\mu\rho} v^\sigma  \os{-1}{\Pi}_{\rho\sigma}\,,
  \qquad
  \os{3}{\Pi}^{\mu\nu}
  = - h^{\mu\rho} h^{\nu\sigma} \os{-1}{\Pi}_{\rho\sigma}
  + 2 v^{(\mu} h^{\nu)\rho} \os{1}{V}_\rho\,,
\end{equation}
and further subleading orders can be obtained similarly.
Furthermore, from the relation $U^\mu U^\nu\Pi_{\mu\nu}=0$
we can solve for $v^\mu v^\nu\Pi_{\mu\nu}$ to all orders in $1/r$,
which gives
\begin{equation}
  \label{eq:vv-Pi-zero-at-NLO-in-partial-gauge}
    v^\mu v^\nu \os{-1}{\Pi}_{\mu\nu}=0\,.
\end{equation}
We will use such expressions at many points in the following,
and an overview of all necessary results (in the presence of additional upcoming gauge choices) can be found in Appendix~\ref{sapp:on-shell-expansion-fundamental-variables}.
For the purposes of the present section, however, the results above will be sufficient.

\subsection{Leading EOM, covariant shear and boundary constraint}
\label{ssec:bulk-geom-LO-EE-in-partial-gauge}

Before performing further gauge fixings,
let us first work out the implications of the vacuum Einstein equations $R_{MN}=0$
at leading order in the radial expansion,
using only the partial gauge choice $V^M = - \delta^M_r$ introduced in Section~\ref{ssec:bulk-geom-partial-gauge-fixing} above.

\paragraph{Expanding the equations of motion.}
Using the metric decomposition in Equation~\eqref{eq:metric-components-after-gauge-fixing-V-vector},
we find that the corresponding components of the bulk Levi-Civita connection
$\Gamma^R_{MN}$ can be written as follows,
\begin{subequations}
\label{eq:first-gauge-fixed-LCs}
    \begin{align}
    \Gamma^r_{rr}
    &=  -g^{r\sigma}\partial_r V_\sigma\,,\\
    \Gamma^\rho_{rr}
    &=  -\Pi^{\rho\sigma}\partial_r V_\sigma\,,\label{eq:further-gauge-fix-connection}\\
    \Gamma^r_{\mu r}
    &=  -\frac{1}{2}g^{r\sigma}V_{\mu\sigma}+\frac{1}{2}g^{r\sigma}\partial_r g_{\mu\sigma}\,,\\
    \Gamma^\rho_{\mu r}
    &=  -\frac{1}{2}\Pi^{\rho\sigma}V_{\mu\sigma}+\frac{1}{2}\Pi^{\rho\sigma}\partial_r g_{\mu\sigma}\,,\label{eq:Gamma-rho-mu-r-first-fixing}\\
    \begin{split}
        \Gamma^r_{\mu\nu}
    &=  \frac{1}{2}\left(\partial_\mu U_\nu+\partial_\nu U_\mu\right)-C^\rho_{\mu\nu}U_\rho-\frac{1}{2}\mathcal{L}_U\Pi_{\mu\nu}
    -\frac{1}{2}g^{rr}\partial_r g_{\mu\nu}\\  
    &{}\qquad
    -\frac{1}{2}g^{r\sigma}\left(U_\mu V_{\nu\sigma}+U_\nu V_{\mu\sigma}+V_\mu U_{\nu\sigma}+V_\nu U_{\mu\sigma}\right)\,,
    \end{split}\\
    \label{eq:rewritingGammatangential}
    \Gamma^\rho_{\mu\nu}
    &= C^\rho_{\mu\nu}
    - \frac{1}{2}\Pi^{\rho\sigma}\left(
      U_\mu V_{\nu\sigma}
      + U_\nu V_{\mu\sigma}
      +V_\mu U_{\nu\sigma}+V_\nu U_{\mu\sigma}
    \right)
    - \frac{1}{2}g^{\rho r}\partial_r g_{\mu\nu}\,,
\end{align}
\end{subequations}
where we defined
\begin{equation}
  \label{eq:Vmunu-Umunu-curly-A}
  V_{\mu\nu}
  = 2 \pd_{[\mu} V_{\nu]}\,,
  \qquad
  U_{\mu\nu}
  = 2 \pd_{[\mu} U_{\nu]}\,,
\end{equation}
as well as
\begin{equation}
  \label{eq:unHatCcon}
    C^\rho_{\mu\nu}
     =  -\frac{1}{2}U^\rho\left(\partial_\mu V_\nu+\partial_\nu V_\mu\right)
    +\frac{1}{2}\Pi^{\rho\sigma}\left(
      \partial_\mu \Pi_{\nu\sigma}+\partial_\nu \Pi_{\mu\sigma}-\partial_\sigma \Pi_{\mu\nu}
    \right).
\end{equation}
The object $C^\rho_{\mu\nu}$ transforms as a $(d+1)$-dimensional connection with respect to the coordinate transformations generated by $\xi^\mu$.
The decomposition of~$\Gamma^\rho_{\mu\nu}$ is not unique,
since we can always add a tensor to $C^\rho_{\mu\nu}$ and subtract it in the rest of the expression.
All such choices of connection have particular advantages and drawbacks
and we will discuss this in detail in Section~\ref{ssec:connection-choice} below,
but for now this will suffice.
The other objects in the expressions for the components of the Levi-Civita connection are tensorial with respect to $(d+1)$-dimensional coordinate transformations,
which are the coordinate transformations of a constant-$r$ hypersurface.

The components of the Ricci tensor of the bulk Levi-Civita connection are
\begin{equation}
\label{eq:EFE-in-terms-of-Gamma}
  R_{MN}
  = -\partial_M\Gamma^P_{PN}
  +\partial_P\Gamma^P_{MN}
  -\Gamma^P_{MQ}\Gamma^Q_{PN}
  +\Gamma^P_{PQ}\Gamma^Q_{MN}=0\,,
\end{equation}
which vanishes by the vacuum Einstein equations.
Using the metric decomposition~\eqref{eq:metric-components-after-gauge-fixing-V-vector} and the corresponding Levi-Civita decomposition~\eqref{eq:first-gauge-fixed-LCs},
together with the radial metric expansion defined in~\eqref{eq:boundary-frame-with-M-falloff-mu-parametrization},
we obtain the following radial expansion of the components of the Ricci tensor,
\begin{subequations}
\label{eq:EE-vac-LO}
    \begin{align}
    R_{\mu\nu}
    &= r \left[
      d \left(K_{\mu\nu}-\frac{1}{d}h_{\mu\nu}K\right)
      + h_{\mu\nu} \left(
        2K + d v^\rho v^\sigma \os{-1}{g}_{\rho\sigma}
      \right)
    \right]
    + \OO(1)\,,
    \label{eq:EE-vac-LO-mu-nu}
    \\
    R_{\mu r}
    &= r^{-1} \left[
      -\frac{1}{2}\tau_\mu \left(
        2K + d v^\rho v^\sigma \os{-1}{g}_{\rho\sigma}
      \right)
      +\frac{d}{2}\left(
        a_\mu-h^\rho_\mu v^\sigma \os{-1}{g}_{\rho\sigma}
      \right)
    \right]
    \\
    &{}\qquad\nonumber
    + \OO(r^{-2})\,,\label{eq:murgengauge}
    \\
    R_{rr}
    &= r^{-3} \left[
      d v^\mu \os{1}{V}_{\mu}
    \right]
    + \OO(r^{-4})\,.
\end{align}
\end{subequations}
Here, we introduced the boundary extrinsic curvature tensor $K_{\mu\nu}$
as well as the acceleration one-form $a_\mu$,
which are defined by
\begin{equation}
\label{eq:K-and-a}
    K_{\mu\nu}
    = -\frac{1}{2}\mathcal{L}_v h_{\mu\nu}\,,
    \qquad a_\mu
    = \mathcal{L}_v\tau_\mu\,.
\end{equation}
These tensors are both spatial, in the sense that $v^\mu a_\mu = 0$ and $v^\mu K_{\mu\nu} = 0$.
We denote the trace
of the extrinsic curvature by
$K = h^{\mu\nu} K_{\mu\nu}$.
Note that
the acceleration can also be written as
$a_\mu = v^\rho \tau_{\rho\mu}$,
where $\tau_{\mu\nu} = 2 \pd_{[\mu} \tau_{\nu]}$
is the leading-order term in the expansion of $V_{\mu\nu}$ defined in~\eqref{eq:Vmunu-Umunu-curly-A}.
A summary of such relevant tensors in the boundary conformal Carroll geometry can be found in Appendix~\ref{subsec:Carrolldata}.

\paragraph{Consequences of leading-order EOM.}
From~\eqref{eq:EE-vac-LO},
we see that the leading-order Einstein equations $R_{MN}=0$
allow us to solve for
part of the subleading metric variables in terms of derivatives of the leading-order variables,
\begin{subequations}
\label{eq:LOsolution}
\begin{align}
    v^\nu \os{-1}{g}_{\mu\nu}
    &= a_\mu + \frac{2}{d}K\tau_\mu\,,
    \\
    v^\mu \os{1}{V}_{\mu }
    &= 0\,.
    \label{eq:v-proj-of-V}
\end{align}
\end{subequations}
To some extent, this is similar to the familiar situation in the AdS Fefferman--Graham expansion~\cite{Henningson:1998gx,deHaro:2000vlm},
but this analogy is not perfect,
as we will discuss in more detail below.
In terms of our vielbein data and their radial expansions,
and using the decomposition
\begin{equation}
  \label{eq:Pi-NLO-decomposition-initial-partial-gauge}
    \os{-1}{\Pi}_{\mu\nu}
= - 2\tau_{(\mu} h_{\nu)}^{\rho} v^{\sigma} \os{-1}{\Pi}_{\rho\sigma}
+ h_\mu^\rho h_\nu^\sigma \os{-1}{\Pi}_{\rho\sigma}\,,
\end{equation}
the first line in~\eqref{eq:LOsolution} implies%
\footnote{
  In more detail,
  note that Equation~\eqref{eq:metric-components-after-gauge-fixing-V-vector} tells us that \begin{equation}
    \os{-1}{g}_{\mu\nu}=-\tau_\mu b_\nu-\tau_\nu b_\mu+\os{-1}{\Pi}_{\mu\nu}\,,
\end{equation}
and in~\eqref{eq:Pi-NLO-decomposition-initial-partial-gauge} we are using the fact that we know $v^\mu v^\nu\os{-1}{\Pi}_{\mu\nu}=0$ from~\eqref{eq:vv-Pi-zero-at-NLO-in-partial-gauge} above.}
\begin{subequations}
\begin{gather}
  b_\mu
  = \frac{1}{d} K \tau_\mu
  + h_\mu^\rho b_\rho\,,
  \\
  \label{eq:lo-eom-constraint-on-nlo-metric-and-spatial-weyl-conn-before-gauge-fixing}
  h_\mu^{\rho} \left(v^{\sigma} \os{-1}{\Pi}_{\rho\sigma}
  +  b_\rho \right)
  = a_\mu\,.
\end{gather}
\end{subequations}
The fact that we are not able to solve for both $h_\mu^{\rho} v^{\sigma} \os{-1}{\Pi}_{\rho\sigma}$ and $h_\mu^\rho b_\rho$ is due to the fact that we 
have not fixed the $\Sigma^a$ null rotations yet.

\paragraph{Covariant shear and its transformation.}
As is well known, the Einstein equations do not impose any restrictions on the subleading term in the expansion
of the spatial projection
$h_\mu^\rho h_\nu^\sigma g_{\rho\sigma}$
of the $g_{\mu\nu}$ components of the metric.
We split this spatial tensor into a symmetric trace-free (STF) part,
which we denote by
\begin{equation}
  \label{eq:shear-tensor-initial-definition}
  C_{\mu\nu}
  = h_{\langle\mu}^\rho h_{\nu\rangle}^\sigma \os{-1}{g}_{\rho\sigma}
  = h_{\langle\mu}^\rho h_{\nu\rangle}^\sigma \os{-1}{\Pi}_{\mu\nu}\,,
\end{equation}
and a spatial trace part,
which we denote by
\begin{equation}
  C
  = h^{\mu\nu}\os{-1}{g}_{\mu\nu}
  = h^{\mu\nu}\os{-1}{\Pi}_{\mu\nu}\,.
\end{equation}
Here, angled brackets denote the symmetric and trace-free (STF) part of a fully spatial tensor, so that for a spatial tensor $X_{\mu\nu}$ we have
\begin{equation}
\label{eq:STF-def}
    X_{\langle \mu \nu\rangle} = X_{(\mu\nu)} - \frac{1}{d} h_{\mu\nu}h^{\rho\sigma}X_{\rho\sigma}\,.
\end{equation}
(Our notations and conventions for spatial tensors are detailed in Appendix~\ref{sapp:spatial-tensors}.)

As suggested by the notation, the tensor $C_{\mu\nu}$ will turn out to be the Carroll-covariant generalisation of the shear tensor.
By expanding and projecting the residual gauge transformations of $\Pi_{\mu\nu}$ in~\eqref{eq:gauge-tr-with-pi-after-lambdas-fixing},
we find that the gauge transformations of $C_{\mu\nu}$ is given by
\begin{subequations}
  \begin{align}
    \label{eq:general-gauge-tr-of-shear}
    \delta C_{\mu\nu}
    &= \LL_\chi C_{\mu\nu}
    + \Lambda_D C_{\mu\nu}
    + h_{\langle\mu}^\rho h_{\nu\rangle}^\sigma \LL_\lambda h_{\rho\sigma}
    + 2 \lambda_{\langle\mu} h_{\nu\rangle}^\rho \left(
      b_\rho + v^\sigma \os{-1}{\Pi}_{\rho\sigma}
    \right)
    \\
    \label{eq:general-gauge-tr-of-shear-LO-on-shell}
    &= \LL_\chi C_{\mu\nu}
    + \Lambda_D C_{\mu\nu}
    + h_{\langle\mu}^\rho h_{\nu\rangle}^\sigma \LL_\lambda h_{\rho\sigma}
    + 2 \lambda_{\langle\mu} a_{\nu\rangle}\,.
  \end{align}
\end{subequations}
The second line is obtained using~\eqref{eq:lo-eom-constraint-on-nlo-metric-and-spatial-weyl-conn-before-gauge-fixing} and is therefore only valid on-shell.
The third term contains a Lie derivative with respect to $\lambda^\mu = h^{\mu\rho} \lambda_\rho$,
which is a spatial vector field determined by the Carroll boost parameter.
This Carroll-covariant expression of the shear and its transformation law is one of our main building blocks in the rest of this work.
In Section~\ref{ssec:reduction-to-standard-Bondi--Sachs-gauge},
we will discuss how further gauge choices (including a fixed boost frame) reduce our more general Carroll-covariant gauge to the standard Bondi gauge.
As we will see in~\eqref{eq:standard-bms-shear-transformation},
the expression above then reduces to the well-known BMS transformation law for the shear.

Furthermore, again using~\eqref{eq:gauge-tr-with-pi-after-lambdas-fixing}, we find that the transformation of the spatial trace $C$ which appears at the same order as the shear is given by
\begin{align}
  \label{eq:general-gauge-tr-of-NLO-spatial-trace}
  \delta C
  &= \LL_\chi C
  + h^{\mu\nu} \LL_\lambda h_{\mu\nu}
  + 2 \lambda^{\mu} \left(
    b_\mu + v^\nu \os{-1}{\Pi}_{\mu\nu}
  \right)
  + 2d \os{0}{\xi}^r
  \\
  \label{eq:general-gauge-tr-of-NLO-spatial-trace-LO-on-shell}
  &= \LL_\chi C
  + h^{\mu\nu} \LL_\lambda h_{\mu\nu}
  + 2 \lambda^{\mu} a_{\mu}
  + 2d \os{0}{\xi}^r\,,
\end{align}
where the second line once again uses~\eqref{eq:gauge-tr-with-pi-after-lambdas-fixing}.
This shows that we can use the next-to-leading-order radial diffeomorphism parameter
$\os{0}{\xi}^r$
to set the trace~$C$ to zero.
As we will see in Section~\ref{ssec:Newman-Unti}, this observation
is compatible with a Carroll-covariant version of Newman--Unti gauge,
which uses most of the radial diffeomorphisms to fix the subleading components of the $V_\mu$~vielbein but leaves its leading and next-to-leading orders undetermined.
We will also see in Section~\ref{ssec:car-cov-BS-gauge-fix-radial-diffeo} that a covariant version of the Bondi--Sachs gauge directly sets the trace~$C$ to zero by fixing the determinant of $e^\mu_a e^\nu_b\Pi_{\mu\nu}$.
Either way, we note that our definition of the Carroll-covariant shear tensor and the derivation of its transformation law only rely on the partial gauge choice $V^M = - \delta^M_r$ in~\eqref{eq:gauge-fixing-V-vector-completely}.

\paragraph{Boundary data.}
The spatial symmetric trace-free shear tensor $C_{\mu\nu}$ is
not fixed by the equations of motion.
In asymptotically AdS spacetimes, the holographic renormalisation procedure allows us to relate such arbitrary subleading metric components to energy-momentum currents in the dual CFT,
and we will similarly find generalisations of the standard Bondi mass and the Bondi angular momentum currents
at further subleading orders in the radial expansion.

However, even though $C_{\mu\nu}$ is subleading in the radial expansion,
we will see that the shear tensor in fact appears in the variation of the on-shell action at the same order as the leading-order boundary Carroll metric data specified by $(\tau_\mu, h_{\mu\nu})$.
This suggests that the shear tensor should likewise be interpreted as boundary data.%
\footnote{
  The idea of unifying the shear and the boundary metric data on null infinity is not new and goes back at least to~\cite{Ashtekar:1981hw}.
  It was recently discussed from an intrinsic Carroll perspective in~\cite{Korovin:2017xqu,Herfray:2021qmp,Baulieu:2025itt,Fiorucci:2025twa}
  and can also be understood from a gauging of the conformal Carroll algebra~\cite{Hartong:2025WIP2}.
}
While the response to varying the boundary Carroll metric data in the on-shell action will give rise to a covariant boundary energy-momentum tensor, 
the response of the Carroll-covariant shear $C_{\mu\nu}$ will provide a Carroll-covariant definition of the Bondi news.

To summarise,
to first subleading order, the on-shell asymptotic expansion of the metric~\eqref{eq:metric-components-after-gauge-fixing-V-vector} resulting from the partial gauge choice $V^M = -\delta^M_r$ is given by
\begin{subequations}
  \begin{align}
    g_{rr}
    &= 0\,,
    \\
    g_{r\mu}
    &= - \tau_\mu
    + \OO(r\inv)\,,
    \\
    g_{\mu\nu}
    &= r^2 h_{\mu\nu}
    + r \left(
      C_{\mu\nu}
      - 2 \tau_{(\mu} a_{\nu)}
      - \frac{2}{d}K \tau_\mu \tau_\nu
    \right)
    + \OO(r^0)\,.
  \end{align}
\end{subequations}
Following~\eqref{eq:boundary-frame-gauge-tr-with-Vr-fixed},
\eqref{eq:boundary-frame-gauge-tr-with-V-fixed-completely-for-h-tensors}
and~\eqref{eq:general-gauge-tr-of-shear-LO-on-shell},
the transformations of the boundary variables $\tau_\mu$, $h_{\mu\nu}$ and $C_{\mu\nu}$ under the residual boundary gauge transformations consisting of
diffeomorphisms~$\chi^\mu$,
Carroll boosts~$\lambda_\mu$ 
and Weyl transformations~$\Lambda_D$
are
\begin{subequations}
  \begin{align}
    \delta \tau_\mu
    &= \LL_\chi \tau_\mu
    + \Lambda_D \tau_\mu
    + \lambda_a e^a_\mu\,,
    \\
    \delta h_{\mu\nu}
    &= \LL_\chi h_{\mu\nu}
    + 2 \Lambda_D h_{\mu\nu}\,,
    \\
    \delta C_{\mu\nu}
    &= \LL_\chi C_{\mu\nu}
    + \Lambda_D C_{\mu\nu}
    + h_\mu^\rho h_\nu^\sigma \LL_\lambda h_{\mu\nu}
    + 2 \lambda_{(\mu} a_{\nu)}\,.
  \end{align}
\end{subequations}
As mentioned above, the latter expression uses the leading-order equation of motion in~\eqref{eq:lo-eom-constraint-on-nlo-metric-and-spatial-weyl-conn-before-gauge-fixing},
and we also recall that the spatial trace component of the metric which appears at the same order as the shear can be set to zero using additional gauge choices,
which we will discuss in Section~\ref{ssec:car-cov-BS-gauge-fix-radial-diffeo} below.

\paragraph{Constraint on boundary Carroll metric data.}
As we have seen in the above,
the leading-order Einstein equations~\eqref{eq:EE-vac-LO} can be used to solve for at least part of the subleading terms in the radial expansion in terms of the boundary data.
However, from the symmetric trace-free part of the $R_{\mu\nu}$ expression in~\eqref{eq:EE-vac-LO},
we also get the condition
\begin{equation}
    \label{eq:constraintK}
    K_{\mu\nu}=\frac{K}{d}h_{\mu\nu}\,.
\end{equation}
In this way, the leading-order Einstein equations therefore impose a constraint on the \emph{leading-order} boundary Carroll metric data,
which says its extrinsic curvature has to be pure trace.
In three bulk dimensions, this equation is trivially satisfied,
but in general dimensions it restricts the class of boundary geometries that we are allowed to consider.

The apparent necessity of this constraint is in sharp contrast to the usual Fefferman--Graham expansion of asymptotically Anti-de Sitter spacetimes,
where the Lorentzian geometry on the conformal boundary can be fully arbitrary.
As far as we know, the constraint~\eqref{eq:constraintK} cannot be avoided.
We have checked that adding log terms to the expansion does not lift the constraint.
However, as we will discuss in detail in Section~\ref{sec:variations-ward-ids},
it is perfectly possible to consider only variations within this restricted class of boundary metrics,
and such variations turn out to be sufficient to define boundary response functions which correspond to Carroll-covariant generalisations of the Bondi mass and angular momentum aspect as well as Bondi news.
Sections~\ref{sec:bulk-conservation-equations} and~\ref{sec:bulk-improvements} will then show that the covariant Bondi loss equations may be cast in the form of Ward identities for such a boundary `energy-momentum-news' complex,
and we will also recover the loss equations using an appropriate `holographic renormalisation' procedure in Section~\ref{sec:HoloRenormAndOn-ShellActions}.

\section{Carroll-covariant gauge fixing}
\label{sec:car-cov-BS-gauge}
In the previous section, we used a null vielbein parametrisation to show how  asymptotically flat spacetimes lead to conformal Carrollian geometry on future null infinity.
We defined a radial coordinate $r$ associated to the defining function of the conformal compactification, so that our bulk coordinates are $x^M=(r,x^\mu)$, with the $x^\mu$ parametrising any equal-$r$ surface.

We then introduced a limited amount of gauge fixing, setting one of the bulk null vielbein vectors $V^M \pd_M = - \pd_r$ to be equal to the radial coordinate vector.
As we saw in Equation~\eqref{eq:boundary-frame-with-M-gauge-fixing-V-completely-params},
preserving this choice led to the requirements
\begin{equation}
  \label{eq:boundary-frame-with-M-gauge-fixing-V-completely-params-repeat}
  \Lambda^a
  = - E^a_\mu \pd_r \xi^\mu\,,
  \qquad
  \Lambda
  = - \pd_r \xi^r
  - U_\mu \pd_r \xi^\mu\,,
  \qquad
  0
  = V_\mu \pd_r \xi^\mu\,.
\end{equation}
Here, the last equation imposes a constraint on the Carroll timelike $\tau_\mu$ projections of the bulk diffeomorphism generators $\xi^\mu$.
Specifically, this expression fixes $\tau_\mu \os{n}{\xi}^\mu$ for $n\geq 1$,
where $\os{n}{\xi}^\mu$ denotes all but the leading-order terms in the radial expansion.
The leading-order diffeomorphisms (which we denoted by $\chi^\mu=\os{0}{\xi}^\mu$) are still free, however,
and we saw in~\eqref{eq:boundary-frame-gauge-tr-with-V-fixed-completely} and~\eqref{eq:boundary-frame-gauge-tr-with-V-fixed-completely-for-h-tensors}
that they generate boundary diffeomorphism transformations of the boundary tensors. On the other hand, the first two equations in~\eqref{eq:boundary-frame-with-M-gauge-fixing-V-completely-params-repeat} relate the null boosts $\Lambda$ and the null rotations $\Lambda^a$ to the bulk diffeomorphism parameters.

One of the reasons we only performed a partial gauge fixing so far is to show that the results of Sections~\ref{ssec:bulk-geom-partial-gauge-fixing} and~\ref{ssec:bulk-geom-LO-EE-in-partial-gauge} are robust,
in the sense that they do not require or depend on any additional gauge choices.
Having established this, we will now proceed with our gauge fixing procedure.
Our aim is to fix the majority of the bulk gauge transformations~\eqref{eq:gen-null-bulk-frame-diffeo-lor-action} of our null vielbein parametrisation of a general asymptotically flat spacetime.
We will fix almost all of the bulk gauge transformations leaving room for a limited amount of residual transformations,
which are then responsible for the gauge transformations of the boundary Carroll structure.

Specifically, in Section~\ref{ssec:car-cov-BS-gauge-V-geodesic} we will impose the condition that
the null vector $V^M \pd_M = - \pd_r$ is geodesic, though not necessarily affinely parametrised,
which, as we will see, implies that
\begin{equation}
  \label{eq:VM-fixed-up-to-beta-intro}
  V_M(r, x) dx^M
  = e^{\beta(r,x)} \tau_\mu(x) dx^\mu\,.
\end{equation}
This means that all radial dependence of the null covector $V_M$ is encoded in the function~$\beta$,
which also parametrises the non-affinity of $V^M$ as a null geodesic vector field.
As we will see, this geometric requirement turns out to fix the remaining spacelike components of the bulk diffeomorphism generators $\xi^\mu$, which complements the timelike constraint we already had in~\eqref{eq:boundary-frame-with-M-gauge-fixing-V-completely-params-repeat} above.

Next, recall that $(V_\mu, E^a_\mu)$ and $(U^\mu, E_a^\mu)$ formed a basis of the (co)tangent spaces of equal $r$ surfaces, as we discussed around~\eqref{eq:null-bulk-frame-completeness-V-fixed}.
In Section~\ref{ssec:car-cov-BS-gauge-fix-Sigma},
we will fix the remaining null rotation generators $\Sigma^a$ by aligning the  covector $U_\mu$ with the basis covector $V_\mu$, 
reducing the number of components associated with~$U_\mu$ to a single function~$S$.

We then proceed to fix the $\xi^r$ radial diffeomorphism generators.
Our main focus will be a Carroll-covariant version of the Bondi determinant condition,
which fixes the radial dependence of the determinant of $e^\mu_a e^\nu_b\Pi_{\mu\nu}$,
as we discuss in Section~\ref{ssec:car-cov-BS-gauge-fix-radial-diffeo}.
Alternatively, we can use the radial diffeomorphisms to fix the remaining radial dependence of $V_\mu$,
which corresponds to setting $\beta=r^{-1}\os{1}{\beta}$ to all orders in~\eqref{eq:Sigma-gauge-fixing-condition}.
This gives us the Carroll-covariant equivalent of the Newman--Unti gauge, as we discuss in Section~\ref{ssec:Newman-Unti}.
However, in the remainder of this paper we will mainly use the Bondi condition.
With this, for future purposes, we note that our three gauge conditions can be conveniently summarised as
\begin{equation}
  \label{eq:additional-carroll-covariant-BS-gauge-conditions-summary}
  \Gamma^\mu_{rr}
  = - \Pi^{\mu\rho} \pd_r V_\rho
  = 0\,,
  \qquad
  U_\mu
  = \frac{1}{2} S V_\mu\,,
  \qquad
  \Gamma^\rho_{\rho r}
  = \frac{1}{2} \Pi^{\mu\nu} \pd_r \Pi_{\mu\nu}
  = r\inv d\,,
\end{equation}
where $\Gamma^R_{MN}$ is the standard bulk Levi-Civita connection.
Together, these conditions lead to a Carroll-covariant version of the standard Bondi--Sachs gauge,
and in terms of the bulk metric this corresponds to
\begin{equation}
  \label{eq:car-cov-bondi-metric-intro}
  ds^2
  = g_{MN} dx^M dx^N
  = - 2 e^\beta \tau_\mu  dx^\mu dr
  + \left(
    - e^{2\beta} S \tau_\mu \tau_\nu
    + \Pi_{\mu\nu}
  \right) dx^\mu dx^\nu\,.
\end{equation}
This gauge allows for arbitrary Carrollian boundary geometry,
up to the aforementioned constraint~\eqref{eq:constraintK} on the boundary extrinsic curvature that is imposed by the leading-order Einstein equations.

The statement that $V^M$ is geodesic is equivalent to $\Gamma^\mu_{rr}=0$ which in turn can be shown (see below) to lead to \eqref{eq:VM-fixed-up-to-beta-intro}. An equivalent way of saying the same thing is to define $V'^M=e^{-\beta}V^M$ and to require that $V'^M$ is tangent to a congruence of affinely parametrised null geodesics, so that $V'^N\nabla_N V'^M=0$.
The corresponding expansion is then equal to $\nabla_M V'^M=dr^{-1}e^{-\beta}$.
The dual 1-form obeys $V'_Mdx^M=-\tau_\mu dx^\mu$, so the evolution tensor of the null congruence is~\cite{Poisson:2009pwt}
\begin{equation}
    B_{MN} = \Pi_M^P\Pi_N^Q \nabla_P V'_Q\,.
\end{equation}
Since the radial components of $\nabla_{M}V'_N$ all vanish, we restrict our attention to $B_{\mu\nu}$, which takes the form
\begin{equation}
\label{eq:evolution-tensor}
    B_{\mu\nu} = \frac{1}{2}e^{-\beta}\mathcal{F}_{\mu\nu}-\frac{1}{2}e^{-\beta}\mathcal{G}_{\mu\nu}-r^{-1}e^{-\beta}\Pi_{\mu\nu}\,.
\end{equation}
In here, we introduced
\begin{subequations}
  \label{eq:null-congruence-data}
  \begin{align}
    \mathcal{F}_{\mu\nu}
    &= \Pi^\rho_\mu \Pi^\sigma_\nu\left(\partial_\rho V_\sigma-\partial_\sigma V_\rho\right)
    = e^\beta\Pi^\rho_\mu \Pi^\sigma_\nu\left(\partial_\rho \tau_\sigma-\partial_\sigma \tau_\rho\right)
    = F_{\mu\nu} + \mathcal{O}(r^{-1})\,,
    \label{eq:null-congruence-data-curly-F}
    \\
    \mathcal{G}_{\mu\nu}
    &=  \Pi_\mu^\rho\Pi^\sigma_\nu\left(\D_r \Pi_{\rho\sigma} - 2r^{-1}\Pi_{\rho\sigma} \right)
    = - C_{\mu\nu} + \mathcal{O}(r^{-1})
    \,,
    \label{eq:null-congruence-data-curly-G}
  \end{align}
\end{subequations}
where we defined the boundary twist tensor
\begin{equation}
  \label{eq:twist}
  F_{\mu\nu}
  =h^\rho_\mu h^\sigma_\nu
  \left(\partial_\rho\tau_\sigma-\partial_\sigma\tau_\rho\right). 
\end{equation}
We thus see that $\mathcal{F}_{\mu\nu}$, $\mathcal{G}_{\mu\nu}$ and $e^{-\beta}$ are related to the twist, shear and expansion of the null congruence
associated to $V'^M \pd_M = -  e^{-\beta}\pd_r$.
We will have more to say about these tensors in Section~\ref{ssec:rewriting-EE-strategy} below.

While general boundary spatial metrics $h_{\mu\nu}$ have been considered at several points in the literature,
with the exception of the recent paper~\cite{Geiller:2025dqe},
earlier work has mostly focused on the case of $\tau_\mu dx^\mu = du$ for $u$ a fixed null time coordinate, which corresponds in particular to a vanishing boundary twist tensor $F_{\mu\nu}=0$.
Since $\tau_\mu$ transforms under Carroll boosts, such a choice is not Carroll-covariant,
which motivates the generalisation to the Carroll-covariant Bondi gauge in~\eqref{eq:car-cov-bondi-metric-intro}.
We work out the relation between this parametrisation and the standard Bondi--Sachs gauge in Section~\ref{ssec:reduction-to-standard-Bondi--Sachs-gauge}.
Finally, we discuss our choice of connection on equal-$r$ hypersurfaces in Section~\ref{ssec:connection-choice}.

\subsection{Requiring the radial null vector to be geodesic}
\label{ssec:car-cov-BS-gauge-V-geodesic}
Around Equation~\eqref{eq:V-asymptotically-geodesic-intro} in Section~\ref{ssec:bulk-geom-partial-gauge-fixing},
we saw that $V^M \pd_M = - \pd_r$ is asymptotically geodesic.
We now go one step further and demand that it is geodesic throughout the bulk.
We allow $V^M$ to be a non-affinely parametrised geodesic vector field.
Correspondingly, if we set $V^M = f V'^M$ for some function $f$ and require~$V'^M$ to be an affinely parametrised geodesic vector field,
we have to solve
\begin{equation}
  \label{eq:general-geodesic-VF}
  V^R \nabla_R V^M
  = f\inv \left(V^R \pd_R f\right) V^M\,.
\end{equation}
Using the fact that $V^M = - \delta^M_r$,
along with the decomposition of the Levi-Civita connection in~\eqref{eq:first-gauge-fixed-LCs},
the components of this equation are given by
\begin{subequations}
  \label{eq:V-general-geodesic-VF-components}
  \begin{align}
    \label{eq:V-general-geodesic-VF-components-r}
    V^R \nabla_R V^r
    &= \Gamma^r_{rr}
    = - \left(U^\rho - \Pi^{\rho\sigma} U_\sigma\right) \pd_r V_\rho
    = f\inv \pd_r f\,,
    \\
    \label{eq:V-general-geodesic-VF-components-mu}
    V^R \nabla_R V^\mu
    &= \Gamma^\mu_{rr}
    = - \Pi^{\mu\rho} \pd_r V_\rho
    = 0\,.
  \end{align}
\end{subequations}
On a $(d+1)$-dimensional equal-$r$ surface, we know that the kernel of
$\Pi^{\mu\nu}=\delta^{ab} E^\mu_a E^\nu_b$
is one-dimensional.
Furthermore, we know that it is spanned by $V_\mu$
from the orthogonality relations~\eqref{eq:general-null-vielbeine-physical-mu-contractions-fixing-V-completely}.
Therefore, Equation~\eqref{eq:V-general-geodesic-VF-components-mu} implies that $\pd_r V_\mu$ is proportional to $V_\mu$,
which means that we must have
\begin{equation}
  \label{eq:V-general-geodesic-VF-components-mu-solution}
  V_\mu (r,x)
  = e^{\beta(r,x)} \tau_\mu (x)\,,
  \qquad
  \beta
  = \OO(r^{-1})\,,
\end{equation}
where the fall-off of $\beta$ follows from the boundary condition that $V_\mu=\mathcal{O}(1)$.
All radial dependence of $V_\mu$ is encoded by the function $\beta$ in the exponential,
and the asymptotic expansion~\eqref{eq:boundary-frame-with-M-falloff-mu-parametrization} of $V_\mu$ fixes the leading-order behaviour of this function.
Plugging this solution back into the remaining component~\eqref{eq:V-general-geodesic-VF-components-r} gives
\begin{equation}
  \label{eq:V-general-geodesic-VF-components-r-solved}
  V^R \nabla_R V^r
  = f\inv \pd_r f
  = \pd_r \beta
  \qiq
  f = e^\beta\,.
\end{equation}
We see that the function $e^\beta$
encoding the radial dependence of $V^M$
also determines the extent to which it is not affinely parametrised
as a geodesic vector field.
In other words,
$V'^M = e^{-\beta} V^M$
is an affinely parametrised geodesic vector field.

Now let us investigate the consequences of this gauge fixing on the remaining bulk gauge parameters.
As we saw just now, the only nontrivial components of the geodesic condition~\eqref{eq:general-geodesic-VF} correspond to
\begin{equation}
  \label{eq:null-geodesic-VF-constraint}
  \Pi^{\mu\rho} \pd_r V_\rho
  = 0\,,
\end{equation}
which is equivalent to $\Gamma^\mu_{rr}=0$.
Using the residual gauge transformations in~\eqref{eq:gauge-tr-with-pi-after-lambdas-fixing}
as well as all constraints imposed so far,
preserving this gauge condition implies
\begin{align}
  0
  &= \delta \left(
    - \Pi^{\mu\rho} \pd_r V_\rho
  \right)\nonumber
  \\
  &= \Pi^\mu_\rho \left(
    \pd_r^2 \xi^\rho
    - \pd_r \xi^\rho \pd_r \beta
  \right)
  + \Pi^{\mu\rho} \pd_r \xi^\sigma
  \left(
    \pd_r \Pi_{\rho\sigma}
    + e^\beta \tau_{\rho\sigma}
  \right).
  \label{eq:general-geodesic-VF-gauge-conditions}
\end{align}
We can solve this equation order by order for the spatial projection of $\pd_r\xi^\mu$.
By combining our earlier condition~\eqref{eq:residualgaugetrafo3} with \eqref{eq:V-general-geodesic-VF-components-mu-solution},
we obtain that the timelike projection 
\begin{equation}\label{eq:taudrxi}
\tau_\mu\pd_r\xi^\mu=0\,,
\end{equation}
vanishes completely.
Together, this therefore determines $\os{n}{\xi}{}^\mu$ for all $n\ge 1$.
Clearly, the leading-order terms $\os{0}{\xi}^\mu=\chi^\mu$ in the radial expansion of~$\xi^\mu$ do not appear in these equations, so they are unconstrained.
The leading-order contributions from the individual terms in Equation~\eqref{eq:general-geodesic-VF-gauge-conditions}
are at $r^{-3}$,
but they cancel, so the equation is trivial at that order. At order $r^{-4}$ in the above equation, we can then solve for
\begin{align}
  2 h^\mu_\rho \os{2}{\xi}{}^\rho
  &= 
h^{\mu\rho} \lambda^\nu \left(
    F_{\rho\nu}
    - \os{-1}{\Pi}_{\rho\nu}
  \right)
  + \lambda^\mu \os{1}{\beta}\,,
\end{align}
where we used \eqref{eq:twist} 
as well as the relations between the expansions of
$(U^\mu, \Pi^{\mu\nu})$
and the expansions of
$(V_\mu, \Pi_{\mu\nu})$
from Equation~\eqref{eq:U-Pi-expansion-in-terms-of-V-Pi-expansion}. 
This determines the spatial part of $\os{2}{\xi}^\mu$,
and we already had $\tau_\mu\os{2}{\xi}^\mu=0$. We therefore find
\begin{equation}
  \xi^\mu
  = \chi^\mu
  + r\inv \lambda^\mu
  + \frac{r^{-2}}{2} \left(
 h^{\mu\rho} \left[
      F_{\rho\nu}
      - \os{-1}{\Pi}_{\rho\nu}
    \right]
    + \os{1}{\beta}\delta^\mu_\nu
  \right) \lambda^\nu
  + \OO(r^{-3})\,.
\end{equation}
Here and in the following,
we write $\lambda^\mu = e^\mu_a \lambda^a$ for brevity,
in line with our conventions for spatial tensors that are outlined in Appendix~\ref{sapp:spatial-tensors}.
Further subleading orders in the expansion of~$\xi^\mu$ can similarly be determined fully algebraically.

\subsection{Fixing remaining null rotations}
\label{ssec:car-cov-BS-gauge-fix-Sigma}
At this point, we have gauge fixed 
the null boosts $\Lambda$ and the null rotations $\Lambda^a$ fully,
as well as essentially all of the diffeomorphisms $\xi^\mu$ to all orders in $1/r$,
except for the residual parameters $\chi^\mu$ and $\lambda^\mu$ that we need for boundary Carroll covariance.

This leaves us with the radial diffeomorphisms $\xi^r$,
which we will fix in the next step,
as well as the null rotations $\Sigma^a$. We start with the latter.
From the general transformation rules in~\eqref{eq:gen-null-bulk-frame-diffeo-lor-action},
we obtain
\begin{equation}
  \label{eq:U-mu-gauge-transformations}
  \delta U_\mu
  =  \LL_\xi U_\mu
    +\xi^r\partial_r U_\mu+\partial_\mu\xi^r- \pd_r \xi^r U_\mu
    - U_\rho \pd_r \xi^\rho U_\mu
    + \Sigma_a E^a_\mu\,.
\end{equation}
Now recall that we can use the completeness relation~\eqref{eq:null-bulk-frame-completeness-V-fixed} as well as the identities~\eqref{eq:general-null-vielbeine-physical-mu-contractions-fixing-V-completely} that we obtained after gauge fixing $V^M = - \delta^M_r$ to write
\begin{equation}
  U_\mu
  = - V_\mu U^\rho U_\rho
  + E_\mu^a E_a^\rho U_\rho
  = U^r V_\mu
  - E^r_a E^a_\mu\,.
\end{equation}
From the gauge transformation in~\eqref{eq:U-mu-gauge-transformations} above, we see that
we can use the null rotations parametrised by~$\Sigma_a$
to remove any components of $U_\mu$ proportional to the spatial $E_\mu^a$ covectors.
This leaves only the component of $U_\mu$ which is proportional to $V_\mu$ and,
in the following,
it will be convenient to parametrise it as follows,
\begin{equation}
  \label{eq:Sigma-gauge-fixing-condition}
  U_\mu
  = \frac{1}{2} S V_\mu
  = \frac{1}{2} S e^\beta \tau_\mu\,,
  \qquad
  S(r,x)
  = r \os{-1}{S}(x)
  + \OO(1)\,.
\end{equation}
where $S(r,x)$ is an arbitrary function whose radial expansion follows from the expansion of $U_\mu$ and $V_\mu$ in~\eqref{eq:boundary-frame-with-M-falloff-mu-parametrization}.
In terms of the radial components $U^r$ and $E^r_a$, this means that
\begin{equation}
  \label{eq:Sigma-gauge-fixing-implications}
  U^r
  = - U^\mu U_\mu
  = \frac{1}{2} S\,,
  \qquad
  E^r_a
  = - E^\mu_a U_\mu
  = 0\,.
\end{equation}
The last identity provides a convenient way to find the constraint on the gauge parameters that results from the gauge condition~\eqref{eq:Sigma-gauge-fixing-condition}.
Setting its variation to zero gives us
\begin{subequations}
  \label{eq:Sigma-gauge-fixing-parameter-result}
  \begin{gather}
    0
    = \delta E^r_a
    = -E^\mu_a\partial_\mu \xi^r
    - \frac{1}{2} \delta_{ab} S E^b_\mu \pd_r \xi^\mu
    - \Sigma_a
    \\
    \qiq
    \Sigma^a
    = - \frac{1}{2} S E^a_\mu \pd_r \xi^\mu
    - \delta^{ab} E^\mu_b \pd_\mu \xi^r\,.
  \end{gather}
\end{subequations}
The bulk gauge transformations then reduce to~\eqref{eq:gauge-tr-with-pi-after-lambdas-fixing} with $\Sigma^a$ now fixed by~\eqref{eq:Sigma-gauge-fixing-parameter-result}.
In particular, using that $S=2U^r$ we find
\begin{equation}
\delta S=\LL_\xi S+\xi^r\partial_r S-2S\partial_r\xi^r-2U^\mu\partial_\mu\xi^r\,.
\end{equation}
In terms of the boundary variables,
the gauge choice~\eqref{eq:Sigma-gauge-fixing-condition} implies that the boundary Weyl connection $b_\mu$ is purely timelike.%
\footnote{%
  \label{fn:alternative-sigma-gauge}%
  An alternative gauge fixing of the $\Sigma^a$ null rotations would be to set
  $E^a_\mu\pd_r U^\mu  = 0$,
  which corresponds to
  $U^\mu = e^{-\beta} v^\mu$,
  similar to how the constraint~\eqref{eq:V-general-geodesic-VF-components-mu-solution}
  simplified the radial dependence of the $V_\mu$ vielbein.
  Preserving this alternative constraint leads to a differential equation for $\Sigma^a$,
  which implies that $\sigma^a=0$ at leading order.
  Furthermore,
  we have $v^\mu E_\mu^a=0$ so that in particular $v^\mu h^\nu_\rho \os{1}{\Pi}_{\mu\nu}=0$,
  while the boundary Weyl connection is gauge fixed to
  $b_\mu = a_\mu + (1/d) K \tau_\mu$ instead.
  Note that both choices of $\Sigma^a$ gauge fixing are compatible with the leading-order equation of motion in~\eqref{eq:lo-eom-constraint-on-nlo-metric-and-spatial-weyl-conn-before-gauge-fixing},
  in that they give the same answer for $\os{-1}{g}_{\mu\nu}$.
  However, while it simplifies the radial dependence of $U^\mu$,
  this alternative gauge choice means that,
  in addition to keeping $U^r = -U_\mu U^\mu$ nonzero as above,
  the radial components
  $E^r_a=-U_\mu E^\mu_a$ are now nonzero, too.
  In this gauge, we have $V_\mu=e^\beta\tau_\mu$ and $U^\mu=e^{-\beta}v^\mu$, which makes $\Pi^\mu_\nu=h^\mu_\nu$ in particular $r$-independent.
  The corresponding gauge-fixed metric reads
  \begin{equation}
      ds^2=
      -2e^\beta\tau_\mu dx^\mu dr
      +\left(-2e^\beta\tau_{(\mu} U_{\nu)}+\Pi_{\mu\nu}\right)dx^\mu dx^\nu\,,
  \end{equation}
  where $v^\mu\Pi_{\mu\nu}=0$.
  We will not explore this gauge fixing further in the present work.
}
It fixes the special conformal transformations $\sigma^a$ in terms of the Carroll boosts $\lambda^a$ and the Weyl transformations~$\Lambda_D$,
\begin{equation}
  \label{eq:boundary-weyl-connection-timelike-gauge-fix}
  b_\mu
  = \frac{1}{2} \os{-1}{S} \tau_\mu\,,
  \qquad
  \sigma^a
  = \frac{1}{2} \os{-1}{S} \lambda^a
  - \delta^{ab} e^\mu_b \pd_\mu \Lambda_D\,.
\end{equation}
After the gauge choices~\eqref{eq:Sigma-gauge-fixing-condition}
and~\eqref{eq:V-general-geodesic-VF-components-mu-solution},
the bulk spacetime metric~\eqref{eq:metric-components-after-gauge-fixing-V-vector} now is
\begin{equation}
  \label{eq:metric-in-car-cov-bondi-gauge}
  ds^2
  = g_{MN} dx^M dx^N
  = - 2 e^\beta \tau_\mu  dx^\mu dr
  + \left(
    - e^{2\beta} S \tau_\mu \tau_\nu
    + \Pi_{\mu\nu}
  \right) dx^\mu dx^\nu.
\end{equation}
In addition,
we have
$g^{rr}= S$,
$g^{r\mu} = U^\mu$
and
$g^{\mu\nu} = \Pi^{\mu\nu}$,
so we see that the function~$S$ controls in particular the norm of the normal 1-form $\partial_M r$ to a constant $r$ hypersurface.

Notice that we have not yet fixed the determinant of the spatial metric $e^\mu_a e^\nu_b\Pi_{\mu\nu}$,
which we will do using an appropriate analogue of the Bondi gauge condition in the next subsection.
At this point, the above therefore provides a Carroll-covariant generalisation of the `partial Bondi gauge' introduced by Geiller and Zwikel in~\cite{Geiller:2022vto,Geiller:2024amx}, which incorporates both the standard Bondi--Sachs gauge as well as the Newman--Unti gauge.
We will briefly discuss the equivalent of Newman--Unti gauge in our current setup in Section~\ref{ssec:Newman-Unti} below.

\subsection{Fixing radial diffeomorphisms using Bondi condition}
\label{ssec:car-cov-BS-gauge-fix-radial-diffeo}
We then finalise our gauge choice by introducing a gauge condition to fix most of the radial diffeomorphisms.
This can be done in different ways,
and we will discuss an alternative approach in the following subsection,
but we will mainly use the equivalent of the standard Bondi gauge condition.
This fixes the determinant of the spatial bulk metric in terms of the determinant of the boundary spatial metric.
In our current language, one way to encode this condition is by setting
\begin{equation}
  \label{eq:bondi-gauge-condition}
  \frac{1}{2} \Pi^{\mu\nu} \pd_r \Pi_{\mu\nu}
  = r\inv d\,.
\end{equation}
For future purposes,
note that the left-hand side is equivalent to
$\Gamma^\rho_{\rho r}$
using the decomposition of the Christoffel symbols in~\eqref{eq:first-gauge-fixed-LCs}.
To see that this equation is equivalent to a determinant condition,
introduce the objects
\begin{equation}
  \pi_{ab}
  = e_a^\mu e_b^\nu \Pi_{\mu\nu}\,,
  \qquad
  \pi^{ab}
  = e^a_\mu e^b_\nu \Pi^{\mu\nu}\,,
\end{equation}
which are each other's inverses. Here, we have used the boundary spatial vielbeine $e^a_\mu$ and $e_a^\mu$ to create non-degenerate $d\times d$ matrices from the rank $d$ degenerate $(d+1)$-dimensional spatial bulk tensors on the right-hand side.
Since these boundary vielbeine are independent of the radial coordinate,
the condition in~\eqref{eq:bondi-gauge-condition} implies that
\begin{equation}
  \pd_r \log \det(\pi_{ab})
  = \pi^{ab} \pd_r \pi_{ab}
  = \Pi^{\mu\nu} \pd_r \Pi_{\mu\nu}
  = 2 d r\inv\,.
\end{equation}
This gives a differential equation for the determinant of $\pi_{ab}$.
Using the fact that its leading-order behavior is determined by
$e^\mu_a e^\nu_b (r^2 h_{\mu\nu}) = r^2 \delta_{ab} $,
it implies that
\begin{equation}
  \label{eq:bond-gauge-condition-as-spatial-determinant}
  \det \left(e^\mu_a e^\nu_b \Pi_{\mu\nu} \right)
  = r^{2d}\,.
\end{equation}
We can then use the gauge condition~\eqref{eq:bondi-gauge-condition} to simplify the full metric determinant.
Using also the gauge conditions~\eqref{eq:V-general-geodesic-VF-components-mu-solution} and~\eqref{eq:Sigma-gauge-fixing-condition}
we have
\begin{align}
\begin{split}
  E\, d^{d+1}x
  &= U \wedge V \wedge E^1 \wedge \cdots \wedge E^d
  \\
  &= dr \wedge e^\beta \tau \wedge E^1 \wedge \cdots \wedge E^d
  \\
  &= e^\beta \det(e^\mu_a e^\nu_b \Pi_{\mu\nu})
  dr \wedge \tau \wedge e^1 \wedge \cdots \wedge e^d\,,
\end{split}
\end{align}
where the bulk vielbein determinant $E = \det(V_\mu, E_\mu^a)$ on equal-$r$ surfaces was defined in~\eqref{eq:bulk-r-surface-vielbein-determinant},
and we now define $e = \det(\tau_\mu, e_\mu^a)$ as its boundary equivalent.
The Bondi determinant condition in the form of~\eqref{eq:bond-gauge-condition-as-spatial-determinant} is then equivalent to
\begin{equation}
  \label{eq:bond-gauge-condition-as-bulk-vielbein-determinant}
  E = r^d e^\beta e\,,
\end{equation}
which in particular means that the radial dependence of the full bulk metric is purely encoded by the function $e^\beta$ along with a simple overall factor.

By expanding either form of this gauge condition,
we can determine the trace
$h^{\mu\nu} \os{n}{\Pi}_{\mu\nu}$
at all orders $n\geq -1$.
In particular, at next-to-leading order, we find
\begin{equation}
  \label{eq:bondi-gauge-nlo-spatial-trace-zero}
    C
  = h^{\mu\nu}\os{-1}{g}_{\mu\nu}
  = h^{\mu\nu} \os{-1}{\Pi}_{\mu\nu}
  = 0\,.
\end{equation}
Due to the Carroll-covariant Bondi determinant condition, the tensor
$h_\mu^\rho h_\nu^\sigma \os{-1}{\Pi}_{\mu\nu}$
therefore reduces to the symmetric trace-free shear tensor $C_{\mu\nu}$
which,
as we saw in Section~\ref{ssec:bulk-geom-LO-EE-in-partial-gauge},
is not constrained by the equations of motion.
Of course, this is in full analogy with the standard formulation of the Bondi gauge.

Now let us determine the constraints imposed on the gauge parameters by requiring that they preserve the Bondi condition, for example in the form~\eqref{eq:bondi-gauge-condition}.
Its variation gives
\begin{align}
  0
  &=\delta \left(\Pi^{\mu\nu} \pd_r \Pi_{\mu\nu}\right)\nonumber
  \\
  &= 2dr^{-1}\left(\partial_r\xi^r
  -r^{-1}\xi^r\right)+ \Pi^{\mu\nu} \LL_{\pd_r \xi} \Pi_{\mu\nu}
   - 2 U^\mu \Pi^\nu_\rho \pd_r \xi^\rho \pd_r \Pi_{\mu\nu}\,.
  \label{eq:bondi-gauge-condition-parameter-condition}
\end{align}
Expanding this expression,
we find that all terms of order $r\inv$ cancel.
As a result, the parameter
$\os{-1}{\xi}^r = \Lambda_D$,
which generates boundary Weyl transformations,
remains unconstrained.
Next, at order $r^{-2}$, we find
\begin{align}
  2d \os{0}{\xi}^r
  &= - h^{\mu\nu} \LL_\lambda h_{\mu\nu}
  + 4 \os{1}{U}^\mu h_{\mu\nu} \lambda^\nu
  + 2 v^\mu \Pi^{(-1)}_{\mu\nu} \lambda^\nu\nonumber
  \\
  &= - h^{\mu\nu} \LL_\lambda h_{\mu\nu}
  - 2 v^\mu \os{-1}{\Pi}_{\mu\nu} \lambda^\nu\,.
\end{align}
Here, we used the relations between the expansions of
$(U^\mu, \Pi^{\mu\nu})$
and
$(V_\mu, \Pi_{\mu\nu})$
which we listed in Equation~\eqref{eq:U-Pi-expansion-in-terms-of-V-Pi-expansion}.
Using the on-shell form of the subleading metric variables obtained from the leading-order equations of motion in~\eqref{eq:LOsolution},
along with
the gauge-fixed expression for
$b_\mu = \frac{1}{2} \os{-1}{S} \tau_\mu$
that followed from~\eqref{eq:boundary-weyl-connection-timelike-gauge-fix},
we can see that
\begin{equation}
  v^\mu \os{-1}{\Pi}_{\mu\nu} \lambda^\nu
  = \lambda^\mu a_\mu\,.
\end{equation}
On shell, the expression for $\os{0}{\xi}^r$ we obtained in the above therefore reduces to
\begin{align}
  2d \os{0}{\xi}^r
  &= - h^{\mu\nu} \LL_\lambda h_{\mu\nu}
  - 2 \lambda^\mu a_\mu\nonumber
  \\
  \label{eq:nlo-radial-diffeo-from-Bondi-gauge-form2-LO-on-shell}
  &= - h^{\mu\nu} \LL_\lambda h_{\mu\nu}
  + 2 v^\mu \LL_\lambda \tau_\mu\nonumber
  \\
  &= -2 e\inv \pd_\rho \left(
    e \lambda^\rho
  \right).
\end{align}

For an alternative derivation of this result, we can also consider the variation of the vielbein determinant $E = r^d e^\beta e$ we obtained from the Bondi condition in~\eqref{eq:bond-gauge-condition-as-bulk-vielbein-determinant},
\begin{align}
  E\inv \delta E
  &= E\inv \pd_\rho \left(E \xi^\rho\right)\nonumber
  \\
  &= e\inv \pd_\rho \left(e \xi^\rho\right)
  + d r\inv \xi^r
  + \pd_r \xi^r
  + \xi^R \pd_R \beta\nonumber
  \\
  &= e\inv \pd_\rho \left(e \chi^\rho\right)
  + (d+1) \Lambda_D\nonumber
  \\
  &{}\qquad\nonumber
  + r\inv \left[
    e\inv \pd_\rho \left(e \lambda^\rho\right)
    + d \os{0}{\xi}^r
    + \xi^\rho \pd_\rho \os{1}{\beta}
    - \Lambda_D \os{1}{\beta}
  \right]
  + \OO(r^{-2})
  \\
  &= e\inv \delta e
  + \delta \beta\,.
\end{align}
The equation of motion~\eqref{eq:v-proj-of-V} implies that $\os{1}{\beta}=0$ on shell,
and in particular this means that the terms in the square brackets in the expressions above must vanish,
which reproduces our previous result for the subleading term $\os{0}{\xi}^r$ of the radial diffeomorphism parameter.

Expressions for further subleading orders in the expansion of $\xi^r$ can be obtained similarly.
Apart from the leading-order parameter $\os{-1}{\xi}^r = \Lambda_D$
which remains arbitrary and parametrises boundary Weyl transformations,
the Bondi gauge condition~\eqref{eq:bondi-gauge-condition} therefore fixes the radial diffeomorphisms completely.

Finally, recall the transformation~\eqref{eq:general-gauge-tr-of-NLO-spatial-trace-LO-on-shell} of
the next-to-leading order spatial trace
$C = h^{\mu\nu} \os{-1}{\Pi}_{\mu\nu}$,
which appears at the same order as the STF shear tensor,
\begin{align}
  \label{eq:general-gauge-tr-of-NLO-spatial-trace-LO-on-shell-repeat}
  \delta C
  &= \LL_\chi C
  + h^{\mu\nu} \LL_\lambda h_{\mu\nu}
  + 2 \lambda^{\mu} a_{\mu}
  + 2d \os{0}{\xi}^r\,.
\end{align}
In~\eqref{eq:bondi-gauge-nlo-spatial-trace-zero} above, we saw that the Bondi gauge sets this term to zero.
The Bondi gauge choice also fixes the subleading radial diffeomorphism~$\os{0}\xi^r$ as we saw in~\eqref{eq:nlo-radial-diffeo-from-Bondi-gauge-form2-LO-on-shell},
and indeed this expression precisely serves to ensure that $\delta C=0$ then holds, too.

\subsection{Fixing radial diffeomorphisms using Newman--Unti condition}
\label{ssec:Newman-Unti}
Instead of using the Bondi condition~\eqref{eq:bondi-gauge-condition}, we can also fix the radial diffeomorphisms by imposing conditions on the function $\beta(r,x^\rho)$,
which controls the radial dependence of $V_\mu(r,x^\rho)=e^{\beta(r,x^\rho)}\tau_{\mu}(x^\rho)$
following our discussion in Section~\ref{ssec:car-cov-BS-gauge-V-geodesic}.

From Equation~\eqref{eq:deltaVmu-after-lambdas-fixing} we know that $V_\mu$ transforms as
\begin{equation}
    \delta V_\mu
    = \LL_\xi V_\mu+\xi^r\partial_r V_\mu
    + \pd_r \xi^r V_\mu
    - \Pi_{\mu\rho} \pd_r \xi^\rho\,,
\end{equation}
where we used the gauge choices $U_\rho=SV_\rho/2$ and $V_\mu=e^\beta\tau_\mu$
as well as $\tau_\rho\partial_r\xi^\rho=0$. 
The $h^\mu_\nu$ projection of this equation is identically satisfied at leading order and,
as we can see after differentiation with respect to $r$,
the remaining orders are equivalent to the condition~\eqref{eq:general-geodesic-VF-gauge-conditions} we obtained from the aforementioned constraint on $V_\mu$ in Section~\ref{ssec:car-cov-BS-gauge-V-geodesic}.
On the other hand,
the $v^\mu$ projection of $\delta V_\mu=\delta \left(e^\beta\tau_\mu\right)$ leads to
\begin{equation}
  \label{eq:gaugetrafobeta}
    \delta\beta
    =\mathcal{L}_\xi\beta
    +\xi^r\partial_r\beta
    -\Lambda_D
    +\partial_r\xi^r
    +a_\rho\left(\xi^\rho-\chi^\rho\right)
    +e^{-\beta}v^\mu\Pi_{\mu\nu}\partial_r\xi^\nu\,.
\end{equation}
Note that the $-\Lambda_D$ term at order $r^0$ is cancelled by the expansion of the $\pd_r \xi^r$ term.
Likewise, at order $r^{-1}$, the terms involving $\xi^r$ drop out,
and we have
\begin{equation}
    \delta\os{1}{\beta}=\mathcal{L}_\chi\os{1}{\beta}-\Lambda_D\os{1}{\beta}+\left(a_\rho-v^\mu \os{-1}{\Pi}_{\mu\rho}\right)\lambda^\rho\,.
\end{equation}
However, after that, we see that we can use $\xi^r$ to set
\begin{equation}
  \label{eq:car-cov-NU-gauge}
    \beta=r^{-1}\os{1}{\beta}\,,
\end{equation}
to all orders,
and we refer to this as the Carroll-covariant Newman--Unti gauge.

At this point, it is useful to emphasise that this gauge fixing is entirely off shell.
For comparison with existing literature (such as~\cite{Barnich:2011ty}),
note that the leading-order equation of motion in~\eqref{eq:v-proj-of-V} implies  that $\os{1}{\beta}=0$,
which means that in the Newman--Unti gauge above $\beta$ is exactly zero on shell.
Following our discussion around Equation~\eqref{eq:V-general-geodesic-VF-components-r-solved}, having $\beta=0$ means that
\begin{equation}
\label{eq:NU-affine}
  V^N \nabla_N V^M = 0\,,
\end{equation}
so that $V^M \pd_M = - \pd_r$ corresponds to an affinely parametrised geodesic vector field.

Finally, note that the Carroll-covariant Newman--Unti gauge~\eqref{eq:car-cov-NU-gauge}
does not constrain~$\os{0}{\xi}^r$,
since it drops out of the inhomogeneous term $\partial_r\xi^r$ in the transformation of $\beta$ in~\eqref{eq:gaugetrafobeta}.
As we just mentioned in~\eqref{eq:general-gauge-tr-of-NLO-spatial-trace-LO-on-shell-repeat},
the spatial trace
$C= h^{\mu\nu} \os{-1}{\Pi}_{\mu\nu}$
that appears at the same order as the shear tensor
transforms as
\begin{equation}
  \delta C
  \label{eq:general-gauge-tr-of-NLO-spatial-trace-LO-on-shell-repeat-again}
  = \LL_\chi C
  + h^{\mu\nu} \LL_\lambda h_{\mu\nu}
  + 2 \lambda^{\mu} a_{\mu}
  + 2d \os{0}{\xi}^r\,.
\end{equation}
In the previous subsection,
we saw that the Carroll-covariant Bondi--Sachs gauge implies that the trace $C$ vanishes.
Instead, in this context, we see that we can use the residual~$\os{0}{\xi}^r$ transformation to set $C$ to zero.
In analogy with previous work on partial gauges which incorporate both Newman--Unti and Bondi--Sachs~\cite{Geiller:2022vto,Geiller:2024amx}, one may suspect that this transformation could have a nonzero charge associated to it,
but we will not investigate this question in the current paper.

\subsection{Summary and residual gauge transformations}
\label{sec:BS-summary}

At this point,
we have completed the Carroll-covariant gauge fixing of our bulk metric.
For future reference, we will now briefly summarise our results,
including the remaining variables, their radial expansion, and their residual gauge transformations.
Here and in the following,
we will use the Carroll-covariant Bondi--Sachs gauge fixing that results from the choice of radial coordinate discussed in Section~\ref{ssec:car-cov-BS-gauge-fix-radial-diffeo},
rather than the Newman--Unti gauge from Section~\ref{ssec:Newman-Unti}.
In the next subsection, we then show that our results indeed reduce to the standard Bondi--Sachs gauge if we gauge fix most of the remaining gauge freedom,
which includes boundary diffeomorphisms, Weyl rescalings and local Carroll boost transformations.

\paragraph{Metric variables.}
We have gauge fixed our bulk metric to be of the form~\eqref{eq:metric-in-car-cov-bondi-gauge},
\begin{equation}
  \label{eq:car-cov-bondi-metric-repeat}
  \begin{gathered}
    ds^2
    = - 2 e^\beta \tau_\mu dr dx^\mu
    + \left(
      - e^{2\beta} S \tau_\mu \tau_\nu
      + \Pi_{\mu\nu}
    \right) dx^\mu dx^\nu\,,
    \\
    g^{rr}
    = S\,,
    \qquad
    g^{r\mu}
    = U^\mu\,,
    \qquad
    g^{\mu\nu}
    = \Pi^{\mu\nu}\,.
  \end{gathered}
\end{equation}
This is supplemented with the determinant condition
$\Pi^{\mu\nu} \pd_r \Pi_{\mu\nu} = 2r\inv d$
from~\eqref{eq:bondi-gauge-condition}.
A concise way of formulating our gauge choices is as follows,
\begin{equation}
  \label{eq:BSgaugemetric}
  g_{rr}=0\,,\qquad \Gamma^\mu_{rr}=0\,,\qquad\Gamma^\mu_{\mu r}=dr^{-1}\,.
\end{equation}
For compactness, we will occasionally still refer to
$V_\mu(r,x)$ as opposed to $e^{\beta(r,x)} \tau_\mu(x)$,
but recall that all of the radial dependence of this covector is parametrised by the function~$\beta$
as a result of the constraint~\eqref{eq:V-general-geodesic-VF-components-mu-solution}.
The remaining variables satisfy
\begin{equation}
  \label{eq:car-cov-bondi-orthogonality-completeness-repeat}
  \begin{gathered}
    -1
    = U^\mu V_\mu\,,
    \qquad
    0
    = U^\mu E^a_\mu\,,
    \qquad
    0
    = V_\mu E_a^\mu\,,
    \qquad
    \delta^a_b
    = E^a_\mu E_b^\mu\,,
    \\
    \delta^\mu_\nu
    = - U^\mu V_\nu + \Pi^\mu_\nu\,,
  \end{gathered}
\end{equation}
where the $E^a_\mu$ and the $E_a^\mu$
are vielbeine for
$\Pi_{\mu\nu} = \delta_{ab} E^a_\mu E^b_\nu$
and
$\Pi^{\mu\nu} = \delta^{ab} E_a^\mu E_b^\nu$,
and
their contraction
$\Pi^\mu_\nu = E^a_\mu E_a^\nu$
is the spatial projection.
In particular, the relations above imply that $(U^\mu, E^\mu_a)$
and $(V_\mu, E_\mu^a)$ form a basis of the (co)tangent spaces of equal-$r$ surfaces.
Finally, the 
determinant $E = \det(V_\mu, E_\mu^a)$ on equal-$r$ surfaces is
\begin{equation}
  E = r^d e^\beta e\,,
\end{equation}
where we have $e = \det(\tau_\mu, e_\mu^a)$.

\paragraph{Radial expansion.}
The expansion of the variables
$(V_\mu, \Pi_{\mu\nu})$ and $S$ is given by
\begin{subequations}
  \label{eq:car-cov-bondi-down-variables-expansion-repeat}
  \begin{align}
    V_\mu
    =
    e^\beta \tau_\mu
    &= \tau_\mu
    \left[
      1
      + r\inv \os{1}{\beta}
      + r^{-2} \left(\frac{1}{2}
        (\os{1}{\beta})^2
        + \os{2}{\beta}
      \right)
      + \OO(r^{-3})
    \right],
    \\
    \Pi_{\mu\nu}
    &= r^2 h_{\mu\nu}
    + r \left(
      C_{\mu\nu}
      - 2 h_\mu^\rho \tau_\nu v^\sigma \os{-1}{\Pi}_{\rho\sigma}
    \right)
    + \os{0}{\Pi}_{\mu\nu}
    + \OO(r\inv)\,,
    \\
    S
    &= r \os{-1}{S}
    + \os{0}{S}
    + \OO(r\inv)\,.
  \end{align}
\end{subequations}
The spatial symmetric trace-free part
$C_{\mu\nu} = h_{\langle\mu}^\rho h_{\nu\rangle}^\sigma \os{-1}{\Pi}_{\mu\nu}$
of the subleading term in the expansion of the spatial metric is left fully arbitrary by the equations of motion,
and this is the Carroll-covariant generalisation of the shear tensor.
The Bondi determinant condition~\eqref{eq:bondi-gauge-condition} on $\Pi_{\mu\nu}$ fixes the trace $h^{\mu\nu} \os{n}{\Pi}_{\mu\nu}$ of all subleading terms $n\geq -1$ in the expansion of the spatial metric.
In particular, we saw that this sets the trace term which appears at the same order as the shear to zero.
The first few terms in the expansion of the inverse variables $(U^\mu, \Pi^{\mu\nu})$
are given by
\begin{subequations}
  \label{eq:car-cov-bondi-down-variables-expansion-inverses-repeat}
  \begin{align}
    U^\mu
    &= v^\mu
    + r^{-1} \left[
      v^\mu v^\rho \os{1}{V}_\rho
      - h^{\mu\rho} v^\sigma \os{-1}{\Pi}_{\rho\sigma}
    \right]
    + \OO(r^{-2})\,,
    \\
    \Pi^{\mu\nu}
    &=
    r^{-2} h^{\mu\nu}
    - r^{-3} h^{\mu\rho} h^{\nu\sigma} \os{-1}{\Pi}_{\rho\sigma}
    + \OO(r^{-4})\,.
  \end{align}
\end{subequations}
As we discussed around~\eqref{eq:U-Pi-expansion-in-terms-of-V-Pi-expansion},
these and further subleading orders are fixed by the terms in the expansions of
$(V_\mu, \Pi_{\mu\nu})$
using the relations~\eqref{eq:car-cov-bondi-orthogonality-completeness-repeat}.
These and other useful results on the radial expansion in the Carroll-covariant Bondi gauge are also collected in Appendix~\ref{app:overview-of-expansion-results}.

\paragraph{Gauge transformations.}
The residual transformations act on the unexpanded metric variables in~\eqref{eq:car-cov-bondi-metric-repeat} as follows
\begin{subequations}
   \label{eq:car-cov-bondi-gauge-tr-repeat}
   \begin{align}
    \delta \Pi_{\mu\nu}
    &= \LL_\xi \Pi_{\mu\nu}+\xi^r\partial_r\Pi_{\mu\nu}
    -2 S V_{(\mu} \Pi_{\nu)\rho} \pd_r \xi^\rho
  -  2V_{(\mu}\Pi^{\rho}_{\nu)} \pd_\rho \xi^r\,,
    \\
    \delta \Pi^{\mu\nu}
    &= \LL_\xi \Pi^{\mu\nu}
    +\xi^r\partial_r\Pi^{\mu\nu}- 2 U^{(\mu} \pd_r \xi^{\nu)}\,,
    \\
    \delta V_\mu
    &= \LL_\xi V_\mu+\xi^r\partial_r V_\mu
    + \pd_r \xi^r V_\mu
    - \Pi_{\mu\rho} \pd_r \xi^\rho\,,\label{eq:deltaVmu}
    \\
    \delta U^\mu
    &= \LL_\xi U^\mu
    +\xi^r\partial_r U^\mu- \pd_r \xi^r U^\mu
  -S\partial_r\xi^\mu-  \Pi^{\mu\nu} \pd_\nu \xi^r\,,\\
\delta S & =\LL_\xi S+\xi^r\partial_r S-2S\partial_r\xi^r-2U^\mu\partial_\mu\xi^r\,.
  \end{align}
\end{subequations}
This follows from Equation~\eqref{eq:gauge-tr-with-pi-after-lambdas-fixing}
using
\eqref{eq:residualgaugetrafo3}, \eqref{eq:V-general-geodesic-VF-components-mu-solution}, \eqref{eq:Sigma-gauge-fixing-condition} and
\eqref{eq:Sigma-gauge-fixing-parameter-result}.
The only parameters that enter in these transformations are the bulk diffeomorphism generators $\xi^\mu$ and $\xi^r$,
which are constrained by
the requirement that $V^\mu$ is a null geodesic, as we discussed in Section~\ref{ssec:car-cov-BS-gauge-V-geodesic},
and also by the Bondi--Sachs determinant condition,
which leads to the conditions in~\eqref{eq:general-geodesic-VF-gauge-conditions},
\eqref{eq:taudrxi}
and~\eqref{eq:bondi-gauge-condition-parameter-condition}.
These can be solved order by order in the radial expansion,
which gives
\begin{subequations}
  \begin{align}
  \begin{split}
    \xi^\mu
    &= \chi^\mu
    + r\inv \lambda^\mu
    \\
    &{}\qquad
    + \frac{r^{-2}}{2} \left(h^{\mu\rho} \lambda^\sigma \left[
        F_{\rho\sigma}
        - C_{\rho\sigma}
      \right]
      + \lambda^\mu  \os{1}{\beta}
    \right)
    + \OO(r^{-3})\,,
    \end{split}
    \\
    \xi^r
    &= r \Lambda_D
    - \frac{1}{2d} \left[
      h^{\mu\nu} \LL_\lambda h_{\mu\nu}
      + 2 v^\mu \os{-1}{\Pi}_{\mu\nu} \lambda^\nu
    \right]
    + \OO(r\inv)\,.
  \end{align}
\end{subequations}
Further subleading terms
contain no new free parameters.%
\footnote{%
  The subleading terms can also be shown to be independent of $\Lambda_D$.
  This follows from Equations~\eqref{eq:general-geodesic-VF-gauge-conditions}, \eqref{eq:taudrxi} and~\eqref{eq:bondi-gauge-condition-parameter-condition}, which define the residual diffeomorphisms that preserve \eqref{eq:BSgaugemetric}.
  The first two of these equations are independent of $\xi^r$ and constrain the radial expansion of $\xi^\mu$.
  The third equation is a $\Lambda_D$-independent equation for the radial expansion of $\xi^r-r\Lambda_D$.\label{footnote:LambdaDdep}
}
Hence,
the only remaining gauge parameters are the vector field $\chi^\mu$, which generates boundary diffeomorphisms,
the spatial vector field $\lambda^\mu = e^\mu_a \lambda^a$,
which generates boundary Carroll boosts,
and the function $\Lambda_D$,
which generates boundary Weyl transformations.

These transformations act on the leading-order Carroll metric variables in the expansions~\eqref{eq:car-cov-bondi-down-variables-expansion-repeat} and~\eqref{eq:car-cov-bondi-down-variables-expansion-inverses-repeat} as follows
\begin{subequations}
  \label{eq:car-cov-bondi-gauge-tr-LO-repeat}
  \begin{align}
    \label{eq:car-cov-bondi-gauge-tr-LO-tau-repeat}
    \delta \tau_\mu
    &= \LL_\chi \tau_\mu
    + \Lambda_D \tau_\mu
    + \lambda_\mu\,,
    \\
    \label{eq:car-cov-bondi-gauge-tr-LO-h-down-repeat}
    \delta h_{\mu\nu}
    &= \LL_\chi h_{\mu\nu}
    + 2 \Lambda_D h_{\mu\nu}\,,
    \\
    \label{eq:car-cov-bondi-gauge-tr-LO-v-repeat}
    \delta v^\mu
    &= \LL_\chi v^\mu
    - \Lambda_D v^\mu\,,
    \\
    \label{eq:car-cov-bondi-gauge-tr-LO-h-up-repeat}
    \delta h^{\mu\nu}
    &= \LL_\chi h^{\mu\nu}
    - 2 \Lambda_D h^{\mu\nu}
    + 2 \lambda^{(\mu} v^{\nu)}\,.
  \end{align}
\end{subequations}
Since it is a purely spatial vector,
the boost parameter $\lambda_\mu$ can be consistently lowered and raised with the leading-order spatial metric $h_{\mu\nu}$ and $h^{\mu\nu}$, respectively.
As we mentioned in Section~\ref{ssec:bulk-geom-LO-EE-in-partial-gauge},
the shear tensor $C_{\mu\nu}$ should also be interpreted as part of the boundary data,
and its transformation is given by
\begin{equation}
  \label{eq:car-cov-bondi-gauge-tr-of-shear}
  \delta C_{\mu\nu}
  = \LL_\chi C_{\mu\nu}
  + \Lambda_D C_{\mu\nu}
  + h_{\langle\mu}^\rho h_{\nu\rangle}^\sigma \LL_\lambda h_{\rho\sigma}
  + 2 \lambda_{\langle\mu} h_{\nu\rangle}^\rho \left(
    b_\rho + v^\sigma \os{-1}{\Pi}_{\rho\sigma}
  \right)
\end{equation}
following~\eqref{eq:general-gauge-tr-of-shear}.

\paragraph{Leading-order equations of motion.}
We already worked out the effect of the leading-order Einstein equations using only a partial gauge fixing in Section~\ref{ssec:bulk-geom-LO-EE-in-partial-gauge}.
First, we saw that the equations of motion require that the boundary extrinsic curvature
$K_{\mu\nu} = - (1/2) \LL_v h_{\mu\nu}$
satisfies
$K_{\mu\nu} = (1/d) h_{\mu\nu} K$,
so that it is purely determined by its trace.
Second, the equations allowed us to fix part of the subleading metric variables in terms of the leading ones.
In terms of the fully gauge-fixed variables in~\eqref{eq:car-cov-bondi-down-variables-expansion-repeat}, this gives
\begin{equation}
  \label{eq:car-cov-bondi-LO-eom-results}
  \os{1}{\beta}
  = 0\,,
  \qquad
  h_\mu^{\rho} v^{\sigma} \os{-1}{\Pi}_{\rho\sigma}
  = a_\mu\,,
  \qquad
  \os{-1}{S}
  = \frac{2}{d} K\,,
\end{equation}
and the transformation of the shear tensor in~\eqref{eq:car-cov-bondi-gauge-tr-of-shear} is then given by
\begin{equation}
  \delta C_{\mu\nu}
  = \LL_\chi C_{\mu\nu}
  + \Lambda_D C_{\mu\nu}
  + h^\rho_{\langle \mu} h^\sigma_{\nu\rangle} \LL_\lambda h_{\rho\sigma}
  + 2 \lambda_{\langle \mu} a_{\nu \rangle}\,.
  \label{eq:car-cov-bondi-gauge-tr-of-shear-on-shell}
\end{equation}
Here,
$a_\mu = \LL_v \tau_\mu$
is the acceleration, which is spatial.
As we will see in the following,
this expression reduces to the standard transformation law of the shear under BMS transformations
after fixing the spatial metric $h_{\mu\nu}$
and after breaking Carroll covariance
by fixing $\tau_\mu$ and going to a fixed boost frame.

\subsection{Reduction to standard Bondi--Sachs gauge}
\label{ssec:reduction-to-standard-Bondi--Sachs-gauge}
Next, we will show that our geometric setup can be used to reproduce the standard Bondi--Sachs asymptotic phase space.
Following for example~\cite{Barnich:2010eb}, our goal is to reproduce the set of metrics parametrised by
\begin{equation}
  \label{eq:standard-Bondi--Sachs-metric-form}
  ds^2
  = e^{2\tilde\beta} \frac{V}{r} du^2
  - 2 e^{\tilde\beta} du dr
  + \tilde{g}_{AB}
  \left(dx^A - \tilde{U}^A du\right)
  \left(dx^B - \tilde{U}^B du\right).
\end{equation}
The first obvious difference with respect to our metric parametrisation in~\eqref{eq:car-cov-bondi-metric-repeat} is that the above involves a coordinate split $x^\mu = (u, x^A)$ of our equal-$r$ hypersurface coordinates.%
\footnote{%
  For brevity, we will use $x$ to denote either $x^A$ or $x^\mu$
  in arguments of functions in the following
  when the context should make it clear which is meant.
}
Here, the $x^A$ with $A=1,\ldots,d$ act as spatial coordinates,
and $u$ is a retarded time coordinate.
Additionally, a function $\tilde\beta(r,u,x^A)$ appears, and we will see that it agrees with our function $\beta(r,x^\mu)$ above.
Likewise, the function $V(r,u,x^A)$ will be related to our function $S(r,x^\mu)$,
the spatial vectors $\tilde{U}^A(r,u,x^B)$ will be related to our $U^\mu(r,x^\rho)$,
and the $d$-dimensional spatial metric $\tilde{g}_{AB}(r,u,x^C)$ will be related to our $\Pi_{\mu\nu}(r,x^\rho)$.
The metric $\tilde{g}_{AB}$ is often parametrised using a conformal factor times the metric of a round $d$-sphere, but we will not do so here to retain generality.

\paragraph{Further gauge fixing.}
Our first step is to introduce the $u$ coordinate.
We will do this by identifying it with the parameter of the integral curve of the $v^\mu \pd_\mu$ vector field.
As a result, we can write
\begin{equation}
  \label{eq:gf-to-standard-bondi-v-not-spatial}
  v^\mu (x^\rho) \pd_\mu
  = v^u(u, x^B) \pd_u\,.
\end{equation}
Clearly, the requirement $v^A = 0$ fixes some of our gauge transformations.
From the transformations of $v^\mu$ in~\eqref{eq:car-cov-bondi-gauge-tr-LO-v-repeat},
we see that it means that the spatial components~$\chi^A$ of our boundary diffeomorphisms cannot depend on retarded time,
\begin{equation}
  0
  = \delta v^A
  = \LL_\chi v^A
  - \Lambda_D v^A
  = - v^u \pd_u \chi^A
  \qiq
  \pd_u \chi^A
  = 0\,.
\end{equation}
Next, we can use the boundary Weyl transformations generated by $\Lambda_D$ to set the remaining component $v^u$ to minus one,
so that we have
\begin{equation}
  \label{eq:gf-to-standard-bondi-v-fully-fixed}
  v^\mu \pd_\mu
  = - \pd_u\,.
\end{equation}
Again using~\eqref{eq:car-cov-bondi-gauge-tr-LO-v-repeat}, preserving this constraint implies that
\begin{align}
  0
  = \delta v^u
  = \LL_\chi v^u
  - \Lambda_D v^u
  = \pd_u \chi^u
  + \Lambda_D
  \qiq
  \Lambda_D
  = - \pd_u \chi^u\,.
\end{align}
This fixes the Weyl parameter in terms of the time derivative of $\chi^u$.
The orthogonality relations~\eqref{eq:car-cov-bondi-orthogonality-completeness-repeat} then imply
\begin{equation}
  \tau_\mu dx^\mu
  = du
  + \tau_A dx^A\,,
  \qquad
  h_{uu}
  = 0\,,
  \qquad
  h_{uA}
  = 0\,.
\end{equation}
Next,
note that the Carroll boost parameter
$\lambda_\mu dx^\mu
= h_{\mu\nu} \lambda^\nu dx^\mu
= h_{AB} \lambda^B dx^A$
only contains spatial $dx^A$ components.
At the price of breaking Carroll boost covariance,
we can therefore precisely use the boost transformations in~\eqref{eq:car-cov-bondi-gauge-tr-LO-tau-repeat}
to set the spatial components $\tau_A$ to zero,
so that
\begin{equation}
  \label{eq:gf-to-standard-bondi-tau-fully-fixed}
  \tau_\mu dx^\mu
  = du\,.
\end{equation}
Indeed, preserving this constraint implies that
\begin{equation}
  0
  = \delta \tau_A
  = \LL_\chi \tau_A
  + \Lambda_D \tau_A
  + \lambda_A
  = \pd_A \chi^u
  + \lambda_A
  \qiq
  \lambda_A
  = - \pd_A \chi^u\,,
\end{equation}
which fully fixes the boundary Carroll boosts.

Let us briefly consider how these gauge fixings affect the shear tensor and its transformation \eqref{eq:car-cov-bondi-gauge-tr-of-shear-on-shell}.
Since $v^\mu C_{\mu\nu}=0$ we see that in this gauge $C_{uu}=C_{uA}=0$.
Equation \eqref{eq:car-cov-bondi-gauge-tr-of-shear-on-shell} then tells us that $C_{AB}$ transforms as
\begin{equation}
  \delta C_{AB}
  = \chi^u\partial_u C_{AB}+\LL_\chi C_{AB}
  + \Lambda_D C_{AB}
  +  \LL_\lambda h_{AB}-\frac{1}{2}h_{AB}h^{CD}\LL_\lambda h_{CD}\,,
\end{equation}
where the Lie derivatives are along $\chi^A$ and $\lambda^A$.
Using that $\Lambda_D = - \pd_u \chi^u$ and $\lambda_A = - \pd_A \chi^u$
by the above,
we can also write this as
\begin{equation}\label{eq:sheartrafores}
  \delta C_{AB}
  = \chi^u\partial_u C_{AB}+\LL_\chi C_{AB}
  - \pd_u \chi^u C_{AB}
  -2D_{\langle A}\partial_{B\rangle}\chi^u\,,
\end{equation}
where $D_A$ is the 2-dimensional Levi-Civita connection with respect to the Riemannian metric $h_{AB}$ and the angle brackets now denote the symmetric trace-free~(STF) components with respect to the $h_{AB}$ metric.

\paragraph{Matching parameters.}
So far, we have mainly been concerned with the boundary variables arising from our metric degrees of freedom in~\eqref{eq:car-cov-bondi-metric-repeat}.
However, the simple form of the radial dependence in $V_\mu = e^\beta \tau_\mu$
means that our gauge fixing $\tau_\mu dx^\mu = du$ also has direct bulk consequences.
Specifically, using the orthogonality relations~\eqref{eq:car-cov-bondi-orthogonality-completeness-repeat}, it implies
\begin{equation}
\label{eq:Bondi-rels}
  \begin{gathered}
    V_\mu dx^\mu
    = e^\beta du\,,
    \qquad
    \Pi^{uu}
    = 0\,,
    \qquad
    \Pi^{uA}
    = 0\,,
    \\
    U^\mu \pd_\mu
    = - e^{-\beta} \pd_u
    + U^A \pd_A\,,
    \quad
    \Pi_{uA}
    = e^{\beta} \Pi_{AB} U^B\,,
    \quad
    \Pi_{uu}
    = e^{2\beta} \Pi_{AB} U^A U^B\,,
    \\
    \delta^A_B
    = \Pi^{AC} \Pi_{CB}\,.
  \end{gathered}
\end{equation}
Additionally,
the requirement that $v^\mu\pd_\mu = -\pd_u$
means that $U^A = \OO(r\inv)$.
Note that the independent terms in the inverse metric are now
$(\beta, S, \Pi^{AB}, U^A)$,
while the metric is specified by
$(\beta, S, \Pi_{AB}, \Pi_{AB} U^B)$.
With this, our Carroll-covariant Bondi gauge metric~\eqref{eq:car-cov-bondi-metric-repeat} reduces to
\begin{align}
\begin{split}
  ds^2
  &= - 2 e^\beta \tau_\mu dr dx^\mu
  + \left(
    - e^{2\beta} S \tau_\mu \tau_\nu
    + \Pi_{\mu\nu}
  \right) dx^\mu dx^\nu
  \\
  &= - 2 e^\beta du dr
  - e^{2\beta} S du^2
  + \Pi_{AB}
  \left(dx^A + e^\beta U^A du\right)
  \left(dx^B + e^\beta U^B du\right)
  \\
  &=
  e^{2\tilde\beta} \frac{V}{r} du^2
  - 2 e^{\tilde\beta} du dr
  + \tilde{g}_{AB}
  \left(dx^A - \tilde{U}^A du\right)
  \left(dx^B - \tilde{U}^B du\right)\,,
\end{split}
\end{align}
where we have matched the standard Bondi--Sachs parametrization~\eqref{eq:standard-Bondi--Sachs-metric-form} in the last line.
The dictionary for this identification is given by
\begin{equation}
  \label{eq:gauge-fixed-dictionary-to-standard-Bondi--Sachs-metric-form}
  \begin{gathered}
    \Pi_{AB}
    = \tilde{g}_{AB}\,,
    \qquad
    U^A
    = - e^{-2\tilde\beta} \tilde{U}^A\,,
    \qquad
    \beta
    = 2 \tilde\beta\,,
    \qquad
    S
    = - e^{-2\tilde\beta} \frac{V}{r}\,.
  \end{gathered}
\end{equation}
Likewise, the inverse metric components are given by
\begin{equation}
  \begin{gathered}
    g^{rr}
    = S
    = - e^{-2\tilde\beta} \frac{V}{r}\,,
    \qquad
    g^{ru}
    = U^u
    = - e^{-2\tilde\beta}\,,
    \qquad
    g^{rA}
    = U^A
    = - e^{-2\tilde\beta} \tilde{U}^A\,,
    \\
    g^{uu}
    = \Pi^{uu}
    = 0\,,
    \qquad
    g^{uA}
    = \Pi^{uA}
    = 0\,,
    \qquad
    g^{AB}
    = \Pi^{AB}\,,
  \end{gathered}
\end{equation}
where $g^{AB} = \Pi^{AB}$
is the inverse of the $d$-dimensional spatial metric $g_{AB} = \Pi_{AB}$.
In addition, our determinant condition~\eqref{eq:bondi-gauge-condition}
is now equivalent to
\begin{equation}
  \label{eq:bondi-gauge-condition-BS-version}
  2 r\inv d
  = \Pi^{\mu\nu} \pd_r \Pi_{\mu\nu}
  = \Pi^{AB} \pd_r \Pi_{AB}
  \qiq
  \pd_r \left[
    \det( r^{-2} g_{AB} )
  \right]
  = 0\,,
\end{equation}
which is the standard Bondi determinant condition.
Finally, based on our falloff conditions~\eqref{eq:car-cov-bondi-down-variables-expansion-repeat}
and~\eqref{eq:car-cov-bondi-down-variables-expansion-inverses-repeat},
the dictionary~\eqref{eq:gauge-fixed-dictionary-to-standard-Bondi--Sachs-metric-form} implies that
\begin{subequations}
  \begin{align}
    \Pi_{AB}
    &= r^2 h_{AB}
    + r C_{AB}
    + \OO(1)\,,\label{eq:spatial-metric-exp}
    \\
    \tilde\beta
    &= 
    \frac{r\inv}{2} \os{1}{\beta}
    + \frac{r^{-2}}{2} \os{2}{\beta}
    + \OO(r^{-3})\,,
    \\
    \frac{V}{r}
    &= - r \os{-1}{S}
    + \os{1}{\beta} \os{-1}{S}
    - \os{0}{S}
    + \OO(r\inv)\,,
    \\
    \Pi^{AB}
    &=
    r^{-2} h^{AB}
    - r^{-3} h^{AC} h^{BD} \os{-1}{\Pi}_{AB}
    + \OO(r^{-4})\,.
    \\
    \tilde{U}^A
    &=
    - r\inv \os{1}{U}^A
    + \OO(r^{-2})\,.
  \end{align}
\end{subequations}

\paragraph{Leading-order EOM.}
Now let us consider the consequences of the leading-order equations of motion.
First, note that in the gauge~\eqref{eq:gf-to-standard-bondi-v-fully-fixed} the (spatial) extrinsic curvature takes the simple form
\begin{equation}
  K_{\mu\nu} dx^\mu dx^\nu
  = - \frac{1}{2} \LL_v h_{\mu\nu} dx^\mu dx^\nu
  = \frac{1}{2} \pd_u h_{AB} dx^A dx^B\,.
\end{equation}
The leading-order constraint~\eqref{eq:constraintK} requiring the extrinsic curvature to be pure trace can then be solved as follows\footnote{See also~\cite{Baiguera:2022lsw}, where this is written as
\begin{equation}
  h_{AB}(u,x)
  = h_{AB}(0,x) \left(
    \frac{h(u,x)}{h(0,x)}
  \right)^{1/d}\,,
\end{equation}
where
$h(u,x) = \det(h_{AB}(u,x))$
is the determinant of the spatial metric.}
\begin{equation}
  \label{eq:constraintK-bs-gauge-solution}
  h_{AB}(u,x)=e^{2\varphi(u,x)}\gamma_{AB}(x)\,,
\end{equation}
where $\gamma_{AB}$ is independent of $u$. 
The extrinsic curvature is then
\begin{equation}
  K_{AB}
  = \pd_u \vphi h_{AB}
  \qiq
  \os{-2}{V}
  = - \frac{2}{d}K
  = -2 \pd_u \vphi\,,
\end{equation}
in agreement with the boundary conditions imposed in~\cite{Barnich:2010eb}.
Additionally, note that the acceleration $a_\mu = \LL_v \tau_\mu$
vanishes if $\tau_\mu dx^\mu = du$ and $v^\mu \pd_\mu = - \pd_u$
as above.

Using the dictionary in~\eqref{eq:gauge-fixed-dictionary-to-standard-Bondi--Sachs-metric-form},
the remaining parts of the leading-order equations of motion in~\eqref{eq:car-cov-bondi-LO-eom-results} then imply
\begin{equation}
  \os{1}{\beta}
  = 0\,,
  \qquad
  \os{1}{U}^A
  = 0\,,
\end{equation}
where we used the expansion of the inverse variables in~\eqref{eq:car-cov-bondi-down-variables-expansion-inverses-repeat} and
the vanishing of the acceleration
to obtain $\os{1}{U}^A= 0$ from~\eqref{eq:car-cov-bondi-LO-eom-results}.
Additionally,
the determinant condition~\eqref{eq:bondi-gauge-condition-BS-version}
implies that~$\os{-1}{\Pi}_{AB}$ is traceless.

\paragraph{Residual transformations, supertranslations and superrotations.}
So far, the gauge fixing has restricted the gauge parameters as follows,
\begin{equation}
    \lambda_A=-\partial_A\chi^u\,,\qquad\Lambda_D=-\partial_u\chi^u\,,\qquad\partial_u\chi^A=0\,.
\end{equation}
The transformation of $h_{\mu\nu}$ given in \eqref{eq:car-cov-bondi-gauge-tr-LO-h-down-repeat} then leads to the following transformation of $\varphi$ and $\gamma_{AB}$ in the parametrization~\eqref{eq:constraintK-bs-gauge-solution},
\begin{align}
\begin{split}
    \delta\gamma_{AB} & =  \mathcal{L}_\chi\gamma_{AB}-2\omega\gamma_{AB}\,,\\
    \delta\varphi & =  \mathcal{L}_\chi\varphi+\chi^u\partial_u\varphi-\partial_u\chi^u+\omega\,,
\end{split}
\end{align}
where $\partial_u\omega=0$.
In these expressions, the Lie derivative is along $\chi^A$.
Here, the additional gauge parameter $\omega$ arises from the fact that our metric only depends on the combination $h_{AB}=e^{2\varphi}\gamma_{AB}$,
and so we can shift $\varphi$ and rescale $\gamma_{AB}$ with a $u$-independent function $\omega$ while leaving this product inert. From now on, we will set $d=2$ for the remainder of this subsection.

At this stage, $\chi^A$ is any 2-dimensional coordinate transformation.
We can fix these by demanding that $\gamma_{AB}$ is a fixed metric,
which then requires us to set $\delta\gamma_{AB}=0$.
This restricts $\chi^A$ to a conformal Killing vector of $\gamma_{AB}$,
which in turn says that $\omega=\frac{1}{4}\gamma^{AB}\mathcal{L}_\chi\gamma_{AB}$.
Specifying that $\gamma_{AB}$ is the round metric on the two-sphere
thus means that the $\chi^A$ correspond to the generators of its (infinite-dimensional) conformal symmetries.

The residual transformation of the choices made thus far are generated by $\chi^\mu\partial_\mu=\chi^u(u,x)\partial_u+\chi^A(x)\partial_A$,
and these form the Newman--Unti group~\cite{Barnich:2011ty}.
We still have a completely arbitrary $\chi^u$ parameter.
This can be fixed by setting $\varphi=0$,
whose residual gauge transformations are 
\begin{equation}\label{eq:chiu}
    \chi^u=T(x)+u\frac{1}{2}D_C\chi^C\,.
\end{equation}
These are the (extended) BMS transformations.
In here,
$T(x)$ corresponds to a supertranslation
and $\chi^A$ to a superrotation.
These are simply the conformal Carrollian Killing vectors $\chi^\mu$ that solve $\delta\tau_\mu=0$ and $\delta h_{\mu\nu}=0$,
as given in~\eqref{eq:car-cov-bondi-gauge-tr-LO-tau-repeat} and~\eqref{eq:car-cov-bondi-gauge-tr-LO-h-down-repeat},
for a boundary metric of the form $\tau=du$ and $h=\gamma_{AB}(x)dx^A dx^B$.
Using \eqref{eq:chiu}, we can use \eqref{eq:sheartrafores} to find 
\begin{equation}
  \label{eq:standard-bms-shear-transformation}
  \delta C_{AB}
  = \chi^u\partial_u C_{AB}+\LL_\chi C_{AB}
  - \frac{1}{2}D_C\chi^C C_{AB}
  -2D_{\langle A}\partial_{B\rangle}\chi^u\,,
\end{equation}
where the last term is traceless with respect to $\gamma_{AB}$.
This is the standard BMS transformation law for the shear~\cite{Barnich:2010eb,Barnich:2016lyg,Donnay:2023mrd}.

\subsection{Choice of hypersurface connection}
\label{ssec:connection-choice}
To finalise our geometric setup,
we now return to the decomposition of the bulk Levi-Civita connection~$\Gamma^P_{MN}$
which we first discussed in Section~\ref{ssec:bulk-geom-LO-EE-in-partial-gauge}.
Using our full Carroll-covariant gauge,
as summarised in~\eqref{eq:car-cov-bondi-metric-repeat} above,
the bulk Levi-Civita components are now given by
\begin{subequations}
  \label{eq:main-christoffel-decomposition}
    \begin{align}
    \Gamma^r_{rr} & =  -U^\sigma\partial_r V_\sigma\,,
    \label{eq:main-christoffel-decomposition-Gamma-r-rr}
    \\
    \Gamma^\rho_{rr} & =  0\,,\\
    \Gamma^r_{\mu r} & =  \frac{1}{2}\mathcal{A}_\mu+\frac{1}{2}U^\sigma\partial_r g_{\mu\sigma}\,,\\
    \Gamma^\rho_{\mu r} & =  -\frac{1}{2}\Pi^{\rho\sigma}V_{\mu\sigma}+\frac{1}{2}\Pi^{\rho\sigma}\partial_r \Pi_{\mu\sigma}\,,\label{eq:Gamma^rho_mur}\\
    \begin{split}
    \Gamma^r_{\mu\nu} & =  -\frac{1}{2}S\partial_r g_{\mu\nu}+\frac{1}{2}V_\nu\partial_\mu S+\frac{1}{2}V_\mu\partial_\nu S-\frac{1}{2}\mathcal{L}_U\Pi_{\mu\nu}
     \\  
    &\quad\,+\frac{1}{2}S V_\mu \mathcal{A}_\nu+\frac{1}{2}S V_\nu \mathcal{A}_\mu+\frac{1}{2}V_\mu V_\nu U^\sigma\partial_\sigma S\,,
    \end{split}
    \\
    \begin{split}
    \Gamma^\rho_{\mu\nu} & =  \hat C^\rho_{\mu\nu} + \frac{1}{2}U^\rho\left(V_\mu\mathcal{A}_\nu+V_\nu\mathcal{A}_\mu\right)\\
    &\quad\,+\frac{1}{2}\Pi^{\rho\sigma}\left(SV_\mu V_{\sigma\nu}+SV_\nu V_{\sigma\mu}+V_\mu V_\nu\partial_\sigma S\right)-\frac{1}{2}U^{\rho}\partial_r g_{\mu\nu}\,.
    \label{eq:Gammafixedtangential}
     \end{split}
\end{align}
\end{subequations}
Here, the bulk acceleration $\mathcal{A}_\mu$ is defined as
\begin{equation}
  \label{eq:curly-A-def}
  \mathcal{A}_\mu
  = \LL_U V_\mu
  = a_\mu
  + \OO(r\inv)\,.
\end{equation}
This spatial tensor generalises the acceleration
$a_\mu = \LL_v \tau_\mu$
to arbitrary constant-$r$ hypersurfaces.
To obtain the results in~\eqref{eq:main-christoffel-decomposition} from \eqref{eq:first-gauge-fixed-LCs}, it is useful to note that
\begin{equation}
  U^\sigma V_{\sigma\mu}
  = \mathcal{A}_\mu\,,
  \qquad
  U^\sigma U_{\sigma\mu}
  = \frac{1}{2}S\mathcal{A}_\mu
  + \frac{1}{2}\partial_\mu S 
  + \frac{1}{2}V_\mu U^\sigma\partial_\sigma S\,.
\end{equation}
Objects such as $V_\mu$, $\partial_r\Pi_{\mu\nu}$ and $\mathcal{A}_\mu$ are tensors on equal-$r$ hypersurfaces with respect to the diffeomorphisms generated by $\xi^\mu$.
On the other hand, the object~$\Gamma^\rho_{\mu\nu}$ transforms under $\xi^\mu$ as a connection on these hypersurfaces.
As we already mentioned below~\eqref{eq:unHatCcon}, 
we can to some extent choose this connection to be what we like,
as we are always free to add and subtract tensorial terms.
In the decomposition~\eqref{eq:main-christoffel-decomposition} above,
we introduced the following choice of connection
\begin{equation}
  \begin{split}
    \hat C^\rho_{\mu\nu}
    & =  -\frac{1}{2}U^\rho\left(\partial_\mu V_\nu+\partial_\nu V_\mu\right)
    -\frac{1}{2}U^\rho\left(V_\mu\mathcal{A}_\nu+V_\nu\mathcal{A}_\mu\right)\\
    &{}\qquad
    +\frac{1}{2}\Pi^{\rho\sigma}\left(
      \partial_\mu \Pi_{\nu\sigma}+\partial_\nu \Pi_{\mu\sigma}-\partial_\sigma \Pi_{\mu\nu}
    \right).
  \end{split}
\end{equation}
While it is not unique, this choice is convenient for several reasons that we will now briefly explain.

Our main goal in the following section will be to rewrite the vacuum Einstein equations $R_{MN}=0$ into a form that is amenable to a large $r$ expansion.
To do this, we will consider $R_{rr}, R_{r\mu}, R_{\mu\nu}$ and contract the $\mu, \nu$ indices with $U^\mu$ and $\Pi^{\mu}_\rho$.
The radial expansion organises into derivatives of Carrollian structures and so it will be useful to define a notion of covariant differentiation on a constant-$r$ hypersurface
that, at leading order in $1/r$,
reduces to a convenient boundary Carroll connection. 

Such constant-$r$ hypersurfaces have a normal $\partial_M r$ whose norm is $g^{rr}=S$.
This function has no definite sign so these hypersurfaces have no definite character.
There is therefore no non-degenerate metric. Instead, we use the complete set of vielbeine $(V_\mu, E^a_\mu)$ and its inverse $(U^\mu, E^\mu_a)$.
We will now discuss the choice of connection on these constant-$r$ hypersurfaces.

\paragraph{Torsion.}
The $1/r$ expansion is a derivative expansion and so curvatures will inevitably be involved.
The Bianchi identities for these curvatures greatly simplify in the absence of torsion,
as can be seen in Appendix~\ref{sapp:conventions-curvature}.
Additionally, the relation between Lie derivatives and covariant derivatives is much simpler without torsion.
Finally, covariantisation (replacing ordinary with covariant derivatives) is much simpler without torsion, too.
For these reasons, we will require our connection to be torsion-free.
There are also some drawbacks to not using torsion that we will discuss further below.

\paragraph{Connection ansatz.}
The connection obviously needs to transform as a connection under the $\xi^\mu$ diffeomorphisms.
We can write an ansatz for such a symmetric connection in terms of our vielbeine as follows 
\begin{equation}
  \label{eq:unHatCcon-repeat}
    C^\rho_{\mu\nu}
     =  -\frac{1}{2}U^\rho\left(\partial_\mu V_\nu+\partial_\nu V_\mu\right)
    +\frac{1}{2}\Pi^{\rho\sigma}\left(
      \partial_\mu \Pi_{\nu\sigma}+\partial_\nu \Pi_{\mu\sigma}-\partial_\sigma \Pi_{\mu\nu}
    \right)+S^\rho{}_{\mu\nu}\,,
\end{equation}
where the tensor $S^\rho{}_{\mu\nu}$ is symmetric in the last two indices.
The trace of this connection is given by
\begin{equation}
  C^\rho_{\rho\mu}
  =  E\inv \pd_\mu E
  - \frac{1}{2} \mathcal{A}_\mu+S^\rho{}_{\rho\mu}\,,
\end{equation}
where the vielbein determinant $E$ is defined in Equation~\eqref{eq:bulk-r-surface-vielbein-determinant}.
The trace of the connection is important when integrating by parts,
as we have
\begin{equation}
  D_\mu X^\mu
  =\partial_\mu X^\mu+C^\rho_{\rho\mu}X^\mu
  =E^{-1}\partial_\mu\left(EX^\mu\right)
  +\left(- \frac{1}{2} \mathcal{A}_\mu+S^\rho{}_{\rho\mu}\right)X^\mu\,.
\end{equation}
Even when there is no torsion, the Ricci tensor is not guaranteed to be symmetric.
As we will see in~\eqref{eq:ricci-antisym-part-for-symmetric-connection},
the exterior derivative of the trace of the connection determines the antisymmetric part of the Ricci tensor.
Therefore, if we set $C^\rho_{\rho\mu}=E^{-1}\partial_\mu E$ by requiring
\begin{equation}\label{eq:traceS}
  S^\rho{}_{\rho\mu}=\frac{1}{2} \mathcal{A}_\mu\,,
\end{equation}
this simplifies integration by parts
and makes the Ricci tensor symmetric. 

\paragraph{Metricity.}
Next, we turn to the metricity properties of the connection~\eqref{eq:unHatCcon-repeat}.
A straightforward calculation leads to 
\begin{subequations}
  \begin{align}
    D_\rho V_\sigma
    &=  \frac{1}{2}V_{\rho\sigma}-S^\kappa{}_{\rho\sigma}V_\kappa\,,\\
    D_\rho\Pi_{\mu\nu}
    &=  \frac{1}{2}V_\mu\mathcal{L}_U\Pi_{\nu\rho}
    +\frac{1}{2}V_\nu\mathcal{L}_U\Pi_{\mu\rho}
    -S^\kappa{}_{\rho\mu}\Pi_{\kappa\nu}
    -S^\kappa{}_{\rho\nu}\Pi_{\kappa\mu}
    \label{eq:second-derivative-rel}\,,
    \\
    D_\rho U^\mu
    &=  -\frac{1}{2}U^\mu\mathcal{L}_U V_\rho
    +\frac{1}{2}\Pi^{\mu\sigma}\mathcal{L}_U\Pi_{\rho\sigma}
    +S^\mu{}_{\rho\sigma}U^\sigma\,,
    \\
    D_\rho\Pi^{\mu\nu}
    &= U^{(\mu}\Pi^{\nu)\sigma}V_{\rho\sigma}
    +S^\mu{}_{\rho\sigma} \Pi^{\sigma\nu}
    +S^\nu{}_{\rho\sigma} \Pi^{\sigma\mu}\,.
  \end{align}
\end{subequations}
There exists no symmetric tensor $S^\rho{}_{\mu\nu}$ for which the connection would be Carroll metric-compatible in the sense that $D_\rho U^\mu=D_{\rho}\Pi_{\mu\nu}=0$.
This is the drawback of using a symmetric connection. 

On the other hand, if one is willing to introduce torsion,
other metricity properties are possible.
For example, in~\cite{Hartong:2015xda,Hansen:2021fxi} a torsionful connection was used which is compatible with the Carroll boost-invariant metric data,
which in this context would mean that $D_\rho U^\mu$ and $D_\rho \Pi_{\mu\nu}$ would vanish.
Such a connection has the advantage of being easier to distribute across contractions of tensors,
but its torsion would make other manipulations more difficult,
and for the purposes of this work we favoured a symmetric connection.

However, it is important to stress that all covariant results can be rewritten in terms of any other connection, so this choice is ultimately immaterial.
Additionally, in this context, neither torsion or non-metricity introduce additional degrees of freedom since, as we can see in for example~\eqref{eq:second-derivative-rel} above, they are always expressed in terms of existing tensors and their derivatives.

\paragraph{Preferred connection.}
If we choose 
\begin{equation}
  \label{eq:connection-shift-choice}
    S^\rho{}_{\mu\nu}
    =-\frac{1}{2}U^\rho\left(V_\mu\mathcal{A}_\nu+V_\nu\mathcal{A}_\mu\right),
\end{equation}
then \eqref{eq:traceS} is obeyed.
We will call the resulting connection $\hat C^\rho_{\mu\nu}$,
and it is given by
\begin{equation}
  \label{eq:hatCcon}
  \begin{split}
    \hat C^\rho_{\mu\nu}
    & =  -\frac{1}{2}U^\rho\left(\partial_\mu V_\nu+\partial_\nu V_\mu\right)
    -\frac{1}{2}U^\rho\left(V_\mu\mathcal{A}_\nu+V_\nu\mathcal{A}_\mu\right)\\
    &{}\qquad
    +\frac{1}{2}\Pi^{\rho\sigma}\left(
      \partial_\mu \Pi_{\nu\sigma}+\partial_\nu \Pi_{\mu\sigma}-\partial_\sigma \Pi_{\mu\nu}
    \right)\,.
  \end{split}
\end{equation}
The associated covariant derivative will be called $\hat D$
and it satisfies
\begin{subequations}
  \label{eq:hatCcon-metricity-properties}
  \begin{gather}
    \hat{D}_\rho U^\mu
    = - \Pi^{\mu\alpha} \mathcal{K}_{\alpha\rho}\,,
    \qquad
    \hat{D}_\rho \Pi_{\mu\nu}
    = - 2 V_{(\mu} \mathcal{K}_{\nu)\rho}\,,
    \\
    \hat{D}_\rho V_\mu
    = \frac{1}{2} \mathcal{F}_{\rho\mu}-V_\rho\mathcal{A}_\mu\,,
    \qquad
    \hat{D}_\rho \Pi^{\mu\nu}
    = - U^{(\mu} \Pi^{\nu)\alpha} \mathcal{F}_{\alpha\rho}-2V_\rho U^{(\mu}\mathcal{A}^{\nu)}\,.
  \end{gather}
\end{subequations}
Here, following~\eqref{eq:null-congruence-data} at the start of this section,
we used the following tensors,
\begin{align}
  \label{eq:curly-K-def}
  \mathcal{K}_{\mu\nu}
  &= - \frac{1}{2} \LL_U \Pi_{\mu\nu}
  = r^2 K_{\mu\nu}
  + \OO(r)\,,
  \\
  \label{eq:curly-F-def}
  \mathcal{F}_{\mu\nu}
  &= \Pi_\mu^\rho \Pi_\nu^\sigma V_{\rho\sigma}
  = F_{\mu\nu}
  + \OO(r\inv)\,.
\end{align}
We see that the tensors above generalise the boundary spatial extrinsic curvature
$K_{\mu\nu} = - \frac{1}{2} \LL_v h_{\mu\nu}$
and the boundary twist
$F_{\mu\nu} = h_\mu^\rho h_\nu^\sigma \tau_{\mu\nu}$
from their asymptotic values to arbitrary $r$ hypersurfaces,
where
$V_{\mu\nu} = 2 \pd_{[\mu} V_{\nu]}$
and
$\tau_{\mu\nu} = 2 \pd_{[\mu} \tau_{\nu]}$.
Specifically, the twist tensor $\mathcal{F}_{\mu\nu}$ measures to what extent the clock one-form $V_\mu = e^\beta \tau_\mu$ fails to define integrable $d$-dimensional spatial hypersurfaces in the $(d+1)$-dimensional equal-$r$ hypersurfaces,
and $\mathcal{K}_{\mu\nu}$ measures the extrinsic curvature of such spatial hypersurfaces.
The choice~\eqref{eq:connection-shift-choice} has the advantage that
the non-metricity properties of the Carroll boost-invariant metric data,
corresponding to $\hat{D}_\rho U^\mu$ and $\hat{D}_\rho \Pi_{\mu\nu}$ here,
are entirely captured by the $\mathcal{K}_{\mu\nu}$ tensor.%
\footnote{%
  There is a closely related connection based on 
  \begin{equation}\label{eq:barCcon}
    \bar C^\rho_{\mu\nu}=\hat C^\rho_{\mu\nu}-V_\nu\mathcal{K}_\mu{}^\rho\,.
  \end{equation}
  This connection has torsion but is Carroll metric compatible because $\bar D_\rho U^\mu=0=\bar D_\rho\Pi_{\mu\nu}=0$.
  This connection has a non-zero intrinsic torsion in the sense of~\cite{Figueroa-OFarrill:2020gpr}.
  This connection was used in~\cite{Hansen:2021fxi}.
  What we have done is to trade this intrinsic torsion for a kind of intrinsic non-metricity.
}

To derive $\hat D_\mu V_\nu$ it is useful to note that the definitions of $\mathcal{F}_{\mu\nu}$ and $\mathcal{A}_\mu$ imply
\begin{equation}
    V_{\mu\nu}=\mathcal{F}_{\mu\nu}-V_\mu\mathcal{A}_\nu+V_\nu\mathcal{A}_\mu\,,
\end{equation}
which fixes the antisymmetric part of this covariant derivative.
Taking the Lie derivative along $U^\mu$ of this identity gives
\begin{equation}
  \partial_\mu\mathcal{A}_\nu-\partial_\nu\mathcal{A}_\mu
  =\mathcal{L}_U\mathcal{F}_{\mu\nu}
  -V_\mu\mathcal{L}_U\mathcal{A}_\nu
  +V_\nu\mathcal{L}_U\mathcal{A}_\mu\,.
\end{equation}
A few additional useful properties that follow immediately from~\eqref{eq:hatCcon-metricity-properties} are
\begin{equation}
    \hat D_\rho\Pi^{\rho\nu} = \mathcal{A}^\nu\,,\qquad \hat D_\rho U^\rho=-\mathcal{K}\,,
\end{equation}
as well as the fact that $U^\rho\hat D_\rho$ on $U^\mu$ and $\Pi_{\mu\nu}$ gives zero. Finally, the fully spatially projected covariant derivatives of $\Pi^{\mu\nu}$ and $\Pi_{\mu\nu}$ are also zero.

The right-hand side of \eqref{eq:hatCcon-metricity-properties} is written in terms of all the different kinds of single-derivative tensors that one can build out of the fields $(U^\mu, \Pi^{\mu\nu}, V_\mu, \Pi_{\mu\nu})$
which do not require a connection (so using exterior derivatives and Lie derivatives).
This can be used to show that only the covariant derivatives of spatial tensors (those for which any contraction with $U^\mu$ or $V_\mu$ gives zero) really depend on the connection $\hat C^\rho_{\mu\nu}$ for a covariant formulation.
This can be seen by repeatedly using completeness on general covariant derivatives $\hat D_\mu X_\nu$ and $\hat D_\mu X^\nu$ together with the metricity properties in~\eqref{eq:hatCcon-metricity-properties}.

\paragraph{Boundary connection.}
We will use the connection $\hat{C}^\rho_{\mu\nu}$ introduced above on general equal-$r$ hypersurfaces.
However, many of our computations will also take place at future null infinity,
corresponding to the $r\to\infty$ asymptotic boundary in these coordinates.
Both in terms of such boundary computations as well as in the context of the radial expansion of the Einstein equations,
it is essential to have an $r$-independent connection at our disposal.
For this, we can simply use the leading-order term in the expansion of the connection above,
\begin{equation}
  \hat{C}^\rho_{\mu\nu}
  = \mathcal{C}^\rho_{\mu\nu}
  + \OO(r\inv)\,,
\end{equation}
whose explicit expression is given by
\begin{equation}
  \label{eq:bdy-Ccon}
  \begin{split}
    \mathcal{C}^\rho_{\mu\nu}
    & =  -\frac{1}{2}v^\rho\left(\partial_\mu \tau_\nu+\partial_\nu \tau_\mu\right)
    -\frac{1}{2}v^\rho\left(\tau_\mu a_\nu+\tau_\nu a_\mu\right)\\
    &{}\qquad
    +\frac{1}{2}h^{\rho\sigma}\left(
      \partial_\mu h_{\nu\sigma}+\partial_\nu h_{\mu\sigma}-\partial_\sigma h_{\mu\nu}
    \right)\,.
  \end{split}
\end{equation}
We denote the corresponding covariant derivative by $\mathcal{D}_\rho$.
The main properties of this connection can be directly obtained from those of its unexpanded predecessor by simply substituting the relevant $r$-dependent objects for their boundary values.
For completeness, we list these properties in Appendix~\ref{eq:app-spatial-vielbein-def},
where we also derive many useful identities that will be essential in the upcoming sections.

\subsection{Hypersurface curvature}
\label{subsec:hypercurv}
As for any affine connection, we can associate a Riemann curvature tensor to the preferred equal-$r$ hypersurface connection that we introduced in the previous subsection.
For many of the subsequent computations, it will be very convenient to reduce this curvature tensor to as few degrees of freedom as possible.

The Riemann tensor on such hypersurfaces can be decomposed into timelike and spacelike components,
and we will see that all but the fully spacelike components can be written purely in terms of the metricity properties of the connection.
For a given connection, the latter are known tensorial expressions
(such as those in~\eqref{eq:hatCcon-metricity-properties} for the case of our preferred connection),
and therefore only the fully spacelike projection of the Riemann tensor contains new curvature data.

Furthermore, in the majority of this work we are only interested in three and four bulk spacetime dimensions, which corresponds to $d=1$ and $d=2$ spacelike directions on such hypersurfaces, respectively.
Using its symmetry properties, we show that the independent degrees of freedom in the fully spatial projection of the hypersurface Riemann tensor
either vanish identically (for $d=1$)
or reduce to a single curvature scalar (for $d=2$).
Finally, we work out the relevant expressions for our preferred hypersurface connection.

\paragraph{Conventions.}
Most of the results in this subsection are valid for an arbitrary symmetric connection,
which we denote as $C^\rho_{\mu\nu}$
and with $D_\rho$ its covariant derivative.
Such an arbitrary symmetric connection can be parametrised using the expression in~\eqref{eq:unHatCcon-repeat} if we take $S^\rho{}_{\mu\nu}$ to be a general tensor.
Following our conventions in Appendix~\ref{sapp:conventions-curvature},
its Riemann curvature tensor is defined by
\begin{subequations}
  \label{eq:curv-maintext}
  \begin{align}
    \left[D_\mu\,,D_\nu\right]X_\rho & =  R_{\mu\nu\rho}{}^\sigma X_\sigma\,,
    \label{eq:curv1}\\
    \left[D_\mu\,,D_\nu\right]X^\sigma & = -R_{\mu\nu\rho}{}^\sigma X^\rho\,,
    \label{eq:curv2}
  \end{align}
\end{subequations}
which corresponds to the usual coordinate expression for the curvature tensor,
\begin{equation}
  \label{eq:Riemann-coord-expr}
  R_{\mu\nu\rho}{}^\sigma=-\partial_\mu C_{\nu\rho}^\sigma-C_{\mu\lambda}^\sigma C^\lambda_{\nu\rho}-\left(\mu\leftrightarrow\nu\right)\,.
\end{equation}
Next, we define the Ricci tensor as follows,
\begin{equation}
    R_{\mu\rho}=R_{\mu\nu\rho}{}^\nu\,.
\end{equation}
As we remarked above,
even if our connection is symmetric,
the Ricci tensor is generically not symmetric.
Explicitly, its antisymmetric part is given by
\begin{equation}
  \label{eq:ricci-antisym-part-for-symmetric-connection}
    2R_{[\mu\rho]}
    =-\partial_\mu C^\nu_{\nu\rho}
    +\partial_\rho C^\nu_{\nu\mu}\,.
\end{equation}
Finally,
the algebraic and differential Bianchi identities are
\begin{align}
    R_{[\mu\nu\rho]}{}^\sigma
    =  0\,,
    \qquad
    D_{[\mu}R_{\nu\rho]\lambda}{}^\sigma
    =  0\,,
\end{align}
following~\eqref{eq:app-general-riemann-alg-bianchi} and~\eqref{eq:app-general-riemann-dif-bianchi}.

\paragraph{A Riemann-like tensor.}
Next, let us define the tensor
\begin{equation}
    Q_{\mu\nu\rho\alpha}=R_{\mu\nu\rho}{}^\sigma\Pi_{\sigma\alpha}+T_{\mu\nu\rho\alpha}\,,
\end{equation}
where we have also introduced the tensor
\begin{equation}\label{eq:defT}
    T_{\mu\nu\rho\alpha}=-\frac{1}{2}D_\mu\left(D_\nu\Pi_{\rho\alpha}+D_\rho\Pi_{\nu\alpha}-D_\alpha\Pi_{\nu\rho}\right)-\left(\mu\leftrightarrow\nu\right)\,.
\end{equation}
One of the main reasons for introducing the tensor $Q_{\mu\nu\rho\alpha}$ is the fact that it has the following symmetry properties,
\begin{align}
\begin{split}
    Q_{(\mu\nu)\rho\alpha} & =  0\,,\\
    Q_{[\mu\nu\rho]\alpha} & =  0\,,\\
    Q_{\mu\nu(\rho\alpha)} & =  0\,,
\end{split}
\end{align}
which are straightforward to verify.
To prove the last property, we need to use
\begin{equation}
    \left[D_\mu\,,D_\nu\right]\Pi_{\rho\alpha}=R_{\mu\nu\rho}{}^\sigma\Pi_{\sigma\alpha}+R_{\mu\nu\alpha}{}^\sigma\Pi_{\sigma\rho}\,,
\end{equation}
which follows for example from~\eqref{eq:Riemann-coord-expr} in a similar way as the identities in~\eqref{eq:curv-maintext}.
It is a standard exercise to show that the symmetry properties of $Q_{\mu\nu\rho\alpha}$ imply
\begin{align}
  \label{eq:Q-Riemann-like-symmetry-properties}
\begin{split}
    Q_{(\mu\nu)\rho\alpha} & =  0\,,\\
    Q_{\mu\nu(\rho\alpha)} & =  0\,,\\
    Q_{\mu\nu\rho\alpha}
    - Q_{\rho\alpha\mu\nu}
    & =  0\,,\\
    Q_{[\mu\nu\rho\alpha]} & =  0\,,
\end{split}
\end{align}
which are the well-known symmetries of the Levi-Civita Riemann tensor.
For this reason, we can call $Q_{\mu\nu\rho\alpha}$
a `Riemann-like' tensor,
and these symmetry properties
make it a very useful object when manipulating (contractions of) covariant derivatives.%
\footnote{%
  The constant-$r$ hypersurfaces can be viewed as a fibration with one-dimensional fibres and a $d$-dimensional Riemannian base manifold.
  It would be interesting to see if $Q_{\mu\nu\rho\alpha}$ is equal to the Riemann tensor of the base manifold.
}

Using the completeness relation in~\eqref{eq:car-cov-bondi-orthogonality-completeness-repeat} on the upper index of the curvature tensor,
we can decompose the curvature tensor as follows,
\begin{equation}\label{eq:relRandQ}
  R_{\mu\nu\rho}{}^\kappa=\Pi^{\kappa\alpha}Q_{\mu\nu\rho\alpha}-\Pi^{\kappa\alpha}T_{\mu\nu\rho\alpha}-U^\kappa\left[D_\mu\,,D_\nu\right]V_\rho\,,
\end{equation}
where we also used the identity \eqref{eq:curv1} applied to $V_\rho$.
Consequently, this implies
\begin{equation}\label{eq:relRicandQ}
    R_{\mu\rho}=\Pi^{\nu\alpha}Q_{\mu\nu\rho\alpha}-\Pi^{\nu\alpha}T_{\mu\nu\rho\alpha}-U^\nu\left[D_\mu\,,D_\nu\right]V_\rho\,.
\end{equation}
for the Ricci tensor.
Once we specialise to a particular connection,
the second and third terms on the right-hand side in both expressions can be computed fully explicitly
from the non-metricity properties of the connection.

\paragraph{Timelike and spacelike projections.}
In fact,
we will now show that all contractions of $R_{\mu\nu\rho}{}^\kappa$ with either $U^\lambda$ or $V_\kappa$ are fully determined by the non-metricity properties of the connection.
For $R_{\mu\nu\rho}{}^\kappa V_\kappa$ and for $R_{\mu\nu\rho}{}^\kappa U^\rho$ this follows  from~\eqref{eq:curv1} and~\eqref{eq:curv2}, respectively.
Since the Riemann tensor is antisymmetric in its first two indices,
we thus only need to show it for the $R_{\mu\nu\rho}{}^\kappa U^\mu$ contraction.
In the process of computing this contraction,
we encounter $Q_{\mu\nu\rho\alpha}U^\mu$
which equals $\Pi^\lambda_\nu U^\mu Q_{\rho\alpha\mu\lambda}$
by the symmetry properties of the $Q_{\mu\nu\rho\alpha}$ tensor in~\eqref{eq:Q-Riemann-like-symmetry-properties}.
We then use~\eqref{eq:relRandQ} to eliminate
$\Pi^\lambda_\nu U^\mu Q_{\rho\alpha\mu\lambda}$
in favour of
the known expression
$R_{\rho\alpha\mu}{}^\gamma U^\mu\Pi_{\gamma\nu}$,
so that we finally obtain
\begin{align}\label{eq:RUfirstindex}
  R_{\mu\nu\rho}{}^\kappa U^\mu
  &= - \Pi^{\kappa\alpha}\Pi_{\nu\gamma}
  \left[D_\rho\,,D_\alpha\right]U^\gamma
  + \Pi^{\kappa\alpha}\Pi^\lambda_\nu U^\mu
  \left(T_{\rho\alpha\mu\lambda}-T_{\mu\lambda\rho\alpha}\right)
  \\
  &{}\qquad\nonumber
  -U^\kappa U^\mu\left[D_\mu\,,D_\nu\right]V_\rho\,.
\end{align}
This shows that all projections of the curvature tensor except the purely spatial one can be written in terms of non-metricity and thus in terms of exterior and Lie derivatives.

We are thus left with the spatial projections,
which we can express as
\begin{equation}
  \label{eq:R-fully-spatial}
  \Pi^\mu_\alpha \Pi^\nu_\beta\Pi^\rho_\gamma\Pi_{\delta\kappa}
  R_{\mu\nu\rho}{}^\kappa
  =\Pi^\mu_\alpha \Pi^\nu_\beta\Pi^\rho_\gamma\Pi_{\delta}^{\sigma}
  \left(Q_{\mu\nu\rho\sigma}-T_{\mu\nu\rho\sigma}\right),
\end{equation}
using~\eqref{eq:relRandQ}.
In this expression, we can write the relevant spatial components of the $Q_{\mu\nu\rho\sigma}$ tensor as follows,
\begin{equation}
    \Pi^\mu_\alpha \Pi^\nu_\beta\Pi^\rho_\gamma\Pi_{\delta}^{\sigma}Q_{\mu\nu\rho\sigma}=E^a_\alpha E^b_\beta E^c_\gamma E^d_{\delta}Q_{abcd}\,.
\end{equation}
We remind the reader that the spatial frame indices $a,b,\cdots$ are $d$-dimensional.

One of the curvatures that we will encounter below is the Ricci tensor of the $C$-connection, i.e.,~\eqref{eq:relRicandQ}. In particular, its spatial projection, 
\begin{equation}
    \Pi^\mu_\kappa\Pi^\rho_\lambda R_{\mu\rho}=\Pi^\mu_\kappa\Pi^\rho_\lambda\Pi^{\nu\alpha}Q_{\mu\nu\rho\alpha}-\Pi^\mu_\kappa\Pi^\rho_\lambda\Pi^{\nu\alpha}T_{\mu\nu\rho\alpha}-\Pi^\mu_\kappa\Pi^\rho_\lambda U^\nu\left[D_\mu\,,D_\nu\right]V_\rho\,.
\end{equation}
We will split this in an STF part and a trace part. The latter gives
\begin{equation}
    \Pi^{\mu\rho} R_{\mu\rho}=\Pi^{\mu\rho}\Pi^{\nu\alpha}Q_{\mu\nu\rho\alpha}-\Pi^{\mu\rho}\Pi^{\nu\alpha}T_{\mu\nu\rho\alpha}-\Pi^{\mu\rho} U^\nu\left[D_\mu\,,D_\nu\right]V_\rho\,.
\end{equation}

\paragraph{Simplifications in three and four dimensions.}
The main cases of interest in this work are three and four bulk spacetime dimensions.
In the case of $d=1$ we have $Q_{abcd}=0$, since there are no tensors with the symmetry properties~\eqref{eq:Q-Riemann-like-symmetry-properties} in one dimension,
and the STF part of $\Pi^\mu_\kappa\Pi^\rho_\lambda R_{\mu\rho}$ vanishes identically.
Hence, for $d=1$ we are left with
\begin{equation}
    \Pi^{\mu\rho} R_{\mu\rho}=-\Pi^{\mu\rho}\Pi^{\nu\alpha}T_{\mu\nu\rho\alpha}-\Pi^{\mu\rho} U^\nu\left[D_\mu\,,D_\nu\right]V_\rho\,,
\end{equation}
which is once again expressed in terms of the metricity properties of the connection.

On the other hand, for $d=2$,
the symmetries in~\eqref{eq:Q-Riemann-like-symmetry-properties} mean that we have
\begin{equation}
    Q_{abcd}=\frac{1}{2}Q\left(\delta_{ac}\delta_{bd}-\delta_{bc}\delta_{ad}\right),
\end{equation}
where $Q$ is the single independent component.
This means that~\eqref{eq:R-fully-spatial} becomes
\begin{equation}
    \Pi^\mu_\alpha \Pi^\nu_\beta\Pi^\rho_\gamma\Pi_{\delta\kappa}R_{\mu\nu\rho}{}^\kappa=\frac{1}{2}Q\left(\Pi_{\alpha\gamma}\Pi_{\beta\delta}- \Pi_{\beta\gamma}\Pi_{\alpha\delta}\right)-\Pi^\mu_\alpha \Pi^\nu_\beta\Pi^\rho_\gamma\Pi_{\delta}^{\sigma}T_{\mu\nu\rho\sigma}\,.
\end{equation}
The Ricci tensor is thus
\begin{equation}
    \Pi^\mu_\kappa\Pi^\rho_\lambda R_{\mu\rho}=\frac{1}{2}Q\Pi_{\kappa\lambda}-\Pi^\mu_\kappa\Pi^\rho_\lambda\Pi^{\nu\alpha}T_{\mu\nu\rho\alpha}-\Pi^\mu_\kappa\Pi^\rho_\lambda U^\nu\left[D_\mu\,,D_\nu\right]V_\rho\,.
\end{equation}
The trace of this equation gives
\begin{equation}\label{eq:Q}
    Q=\Pi^{\mu\rho}R_{\mu\rho}+\Pi^{\mu\rho}U^\nu V_\kappa R_{\mu\nu\rho}{}^\kappa+\Pi^{\mu\rho}\Pi^{\nu\sigma}T_{\mu\nu\rho\sigma}\,,
\end{equation}
and the STF part leads to
\begin{equation}
    \Pi^\mu_{\langle\kappa}\Pi^\rho_{\lambda\rangle} R_{\mu\rho}=-\Pi^\mu_{\langle\kappa}\Pi^\rho_{\lambda\rangle}\Pi^{\nu\alpha}T_{\mu\nu\rho\alpha}-\Pi^\mu_{\langle\kappa}\Pi^\rho_{\lambda\rangle} U^\nu\left[D_\mu\,,D_\nu\right]V_\rho\,.
\end{equation}

\paragraph{Applied to preferred connection.}
Finally, let us now apply these results to the preferred equal-$r$ hypersurface connection~\eqref{eq:hatCcon} we introduced in the previous subsection.
We start with the object 
$T_{\mu\nu\rho\alpha}$ defined in \eqref{eq:defT}.
Using the metricity properties of this connection,
and in particular the second equation on the first line of \eqref{eq:hatCcon-metricity-properties}, we can show that
\begin{equation}
    \hat D_\nu\Pi_{\rho\alpha}+\hat D_\rho\Pi_{\nu\alpha}-\hat D_\alpha\Pi_{\nu\rho}=-2V_\alpha\mathcal{K}_{\nu\rho}\,,
\end{equation}
which means that we have
\begin{equation}
    T_{\mu\nu\rho\alpha}=\mathcal{K}_{\nu\rho}\hat D_\mu V_\alpha+V_\alpha\hat D_\mu\mathcal{K}_{\nu\rho}-\left(\mu\leftrightarrow\nu\right)\,.
\end{equation}
Using also the first equation on the second line of \eqref{eq:hatCcon-metricity-properties}, we can show that 
\begin{equation}\label{eq:spatialRic}
    \Pi^\mu_\kappa\Pi^\rho_\lambda\hat R_{\mu\rho}=\Pi^\mu_\kappa\Pi^\rho_\lambda\Pi^{\nu\alpha}Q_{\mu\nu\rho\alpha}-\Pi^\mu_\kappa\Pi^\rho_\lambda\left(\hat D_{(\mu}\mathcal{A}_{\rho)}+\mathcal{A}_\mu \mathcal{A}_\rho\right)\,,
\end{equation}
so that its trace is
\begin{equation}\label{eq:spatialRicscalar}
    \Pi^{\mu\rho}\hat R_{\mu\rho}=\Pi^{\mu\rho}\Pi^{\nu\alpha}Q_{\mu\nu\rho\alpha}-\hat D_\mu \mathcal{A}^\mu\,,
\end{equation}
and its STF part obeys
\begin{equation}\label{eq:spatialRicSTF}
    \Pi^\mu_{\langle\kappa}\Pi^\rho_{\lambda\rangle}\hat R_{\mu\rho}=\Pi^\mu_{\langle\kappa}\Pi^\rho_{\lambda\rangle}\Pi^{\nu\alpha}Q_{\mu\nu\rho\alpha}-\Pi^\mu_{\langle\kappa}\Pi^\rho_{\lambda\rangle}\left(\hat D_{\mu}\mathcal{A}_{\rho}+\mathcal{A}_\mu \mathcal{A}_\rho\right)\,.
\end{equation}
So far, these results hold for general dimensions.
For $d=1$, they imply
\begin{equation}\label{eq:spatialRicscalard=1}
    \Pi^{\mu\rho}\hat R_{\mu\rho}=-\hat D_\mu \mathcal{A}^\mu\,,
\end{equation}
and the hypersurface Ricci and Riemann tensor do not contain any independent components.
On the other hand, for $d=2$, the above tells us that 
\begin{align}
    \Pi^{\mu\rho}\hat R_{\mu\rho}
    &= Q-\hat D_\mu \mathcal{A}^\mu\,,
    \label{eq:spatialRicscalard=2}
    \\
    \Pi^\mu_{\langle\kappa}\Pi^\rho_{\lambda\rangle}\hat R_{\mu\rho}
    &= -\Pi^\mu_{\langle\kappa}\Pi^\rho_{\lambda\rangle}
    \left(\hat D_{\mu}\mathcal{A}_{\rho}+\mathcal{A}_\mu \mathcal{A}_\rho\right).
    \label{eq:spatialRicSTFd=2}
\end{align}
In this case, we see that all components of the equal-$r$ hypersurface Ricci and Riemann tensors are determined by the non-metricity properties of the connection,
with the exception of $\Pi^{\mu\nu}\hat R_{\mu\nu}$,
which is parametrised by a single function $Q$.

\section{Einstein equations in Carroll-covariant Bondi--Sachs gauge}
\label{sec:rewriting-EE}
In the previous sections, we developed a geometric setup to describe bulk spacetimes with arbitrary conformal Carrollian geometry at future null infinity.
We saw that the leading-order Einstein equations put a constraint on the boundary extrinsic curvature.
We then fixed the bulk gauge transformations while retaining the boundary conformal Carroll symmetries, which consist of boundary diffeomorphisms, Carroll boosts and Weyl transformations.
As summarised in Section~\ref{sec:BS-summary}, the resulting description of the bulk metric is given by
\begin{equation}
  \label{eq:car-cov-bondi-metric-repeat-again}
    ds^2
    = - 2 e^\beta \tau_\mu dr dx^\mu
    + \left(
      - e^{2\beta} S \tau_\mu \tau_\nu
      + \Pi_{\mu\nu}
    \right) dx^\mu dx^\nu\,.
\end{equation}
The functions $\beta$ and $S$ and the spatial tensor $\Pi_{\mu\nu}$ have an expansion in the radial coordinate~$r$,
while the clock one-form $\tau_\mu$ is part of the boundary metric data and does not depend on the radial coordinate.
We now want to understand the near-boundary solution by solving the bulk Einstein equations to some appropriate order in $1/r$,
and we will achieve this goal in Section~\ref{sec:radial-expansion}.

For this, we will first need to rewrite the various transverse and radial components of the bulk Einstein equations in terms of the Carroll-covariant variables introduced above.
After outlining our general strategy in Section~\ref{ssec:rewriting-EE-strategy},
we rewrite the various projections of the vacuum Einstein equations in Carroll-covariant Bondi--Sachs gauge in Sections~\ref{ssec:rewriting-EE-rr-equation},
\ref{ssec:rewriting-EE-mu-r-equation}
and~\ref{ssec:rewriting-EE-mu-nu-equation}.
As a result of the decomposition of $g_{\mu\nu}$ into $V_\mu=e^\beta\tau_\mu$, $\Pi_{\mu\nu}$ and $S$, the resulting expressions are long, but significant simplifications occur in three and (to a lesser extent) four bulk spacetime dimensions,
and we give a summary of these results in Sections~\ref{ssec:rewriting-EE-3d-simplifications}
and~\ref{ssec:rewriting-EE-4d-simplifications}. Furthermore, they are written in terms of objects that are straightforward to expand in $1/r$.

\subsection{General strategy}
\label{ssec:rewriting-EE-strategy}
Using the bulk coordinates $x^M=(r,x^\mu)$,
the vacuum Einstein equations are given
in our conventions
in terms of the bulk Levi-Civita connection $\Gamma^P_{MN}$ by
\begin{equation}
\label{eq:EFE-in-terms-of-Gamma-repeat}
  R_{MN}
  = -\partial_M\Gamma^P_{PN}
  +\partial_P\Gamma^P_{MN}
  -\Gamma^P_{MQ}\Gamma^Q_{PN}
  +\Gamma^P_{PQ}\Gamma^Q_{MN}=0\,.
\end{equation}
The decomposition of the bulk coordinates into a radial coordinate $r$ and boundary (or constant-$r$ hypersurface) coordinates $x^\mu$ leads to three classes of components of the Ricci tensor,
\begin{equation}
R_{rr}=0\,,\qquad R_{\mu r}=0\,,\qquad\text{and}\qquad R_{\mu\nu}=0\,.
\end{equation}
Using our Carroll-covariant Bondi--Sachs gauge~\eqref{eq:car-cov-bondi-metric-repeat-again},
including the corresponding inverse metric variables
$g^{rr}=S$,
$g^{r\mu}=U^\mu$
and
$g^{\mu\nu}=\Pi^{\mu\nu}$
we can further decompose these components
using $U^\mu$ and the spatial projector $\Pi^\mu_\nu$,
which gives
\begin{subequations}
\label{eq:EOMList}
\begin{gather}
  R_{rr}
  =0\,,\qquad
  U^\mu R_{\mu r}
  =0\,,\qquad
  \Pi^\mu_\kappa R_{\mu r}
  =0\,,\qquad
  \\
  \Pi^{\mu\nu}R_{\mu\nu}
  =0,\qquad
  \Pi^{\mu}_{\langle\kappa}\Pi^{\nu}_{\lambda\rangle} R_{\mu\nu}
  =0\,,\qquad
  \Pi^\mu_\kappa U^\nu R_{\mu\nu}
  =0\,,\qquad
  U^\mu U^\nu R_{\mu\nu}
  =0\,.
\end{gather}
\end{subequations}
Here, recall that
$A_{\langle\mu\nu\rangle}\equiv A_{(\mu\nu)}-(1/d)\Pi_{\mu\nu}\Pi^{\rho\sigma}A_{\rho\sigma}$
is the symmetric trace-free (STF) part of a spatial tensor.
Separating the equations of motion in this way is useful since the resulting components will play distinct roles in the radial expansion.
In particular, we will later see that the equations $\Pi^\mu_\kappa U^\nu R_{\mu\nu}
  =0$ and $U^\mu U^\nu R_{\mu\nu}=0$ (at an appropriate order in $1/r$) can be interpreted as a covariant generalisation of the Bondi loss equations in four or more bulk spacetime dimensions,
or as the conservation equations in three dimensions.

Obtaining manageable expressions from the projections in~\eqref{eq:EOMList} is a nontrivial task,
and to help with this we will use several composite variables constructed from our fundamental metric variables.
In Equations~\eqref{eq:curly-A-def},
\eqref{eq:curly-K-def}
and
\eqref{eq:curly-F-def}
above,
we already saw
the acceleration~$\mathcal{A}_\mu$,
the extrinsic curvature~$\mathcal{K}_{\mu\nu}$
and the twist~$\mathcal{F}_{\mu\nu}$ that are defined on a constant-$r$ hypersurface.
We remind the reader of their definitions and expansions,
\begin{equation}
  \begin{gathered}
    \label{eq:curly-A-K-F-def-repeat}
    \mathcal{A}_\mu
    = \LL_U V_\mu
    = a_\mu
    + \OO(r\inv)\,,
    \qquad
    \mathcal{K}_{\mu\nu}
    = - \frac{1}{2} \LL_U \Pi_{\mu\nu}
    = r^2 K_{\mu\nu}
    + \OO(r)\,,
    \\
    \mathcal{F}_{\mu\nu}
    = \Pi_\mu^\rho \Pi_\nu^\sigma V_{\rho\sigma}
    = F_{\mu\nu}
    + \OO(r\inv)\,.
  \end{gathered}
\end{equation}
Recall that
$a_\mu = \LL_v \tau_\mu$
and
$K_{\mu\nu} = - \frac{1}{2} \LL_v h_{\mu\nu}$
are the asymptotic boundary versions of the spatial acceleration and extrinsic curvature,
and 
$F_{\mu\nu} = 2h_\mu^\rho h^\nu_\sigma \pd_{[\rho} \tau_{\sigma]}$
is the boundary twist tensor,
which measures to what extent the clock one-form $\tau_\mu$ fails to define integrable spatial hypersurfaces on the boundary.
As we discussed above, the unexpanded objects have a similar interpretation.
The twist tensor $\mathcal{F}_{\mu\nu}$ measures the failure of $V_\mu$ to be orthogonal to spatial hypersurfaces inside the $(d+1)$-dimensional constant-$r$ hypersurface.
If we have $\mathcal{F}_{\mu\nu}=0$, then $\mathcal{A}_\mu$ is the acceleration and $\mathcal{K}_{\mu\nu}$ is the extrinsic curvature of these spatial hypersurface within the constant-$r$ slices.
We will not assume $\mathcal{F}_{\mu\nu}=0$ or $F_{\mu\nu}=0$ but we will still refer to $\mathcal{A}_\mu$ or $a_\mu$ and $\mathcal{K}_{\mu\nu}$ or $K_{\mu\nu}$ as the acceleration and extrinsic curvature.

In addition, following its earlier appearance in~\eqref{eq:null-congruence-data}, 
we will shortly see that it is natural to introduce the following symmetric trace-free tensor $\mathcal{G}_{\mu\nu}$,
which generalises the asymptotic shear $C_{\mu\nu}$ to arbitrary constant-$r$ surfaces,%
\begin{equation}
  \label{eq:curly-G-def}
  \mathcal{G}_{\mu\nu}
  = \Pi_{\langle\mu}^\rho \Pi_{\nu\rangle}^\sigma \pd_r \Pi_{\rho\sigma}
  = \Pi_\mu^\rho \Pi_\nu^\sigma \left(
    \pd_r \Pi_{\rho\sigma}
    - 2 r\inv \Pi_{\rho\sigma}
  \right)
  = - C_{\mu\nu}
  + \OO(r\inv)\,.
\end{equation}
In the second equality, we used our Bondi gauge condition~\eqref{eq:bondi-gauge-condition},
which can equivalently be formulated in metric variables as in~\eqref{eq:BSgaugemetric}.
As we discussed in Section~\ref{ssec:connection-choice}, the decomposition of the Levi-Civita connection leads to a choice of hypersurface connection~$\hat{C}^\rho_{\mu\nu}$ and covariant derivative~$\hat{D}_\rho$
which we will use to obtain expressions that are covariant on equal $r$ hypersurfaces.
Finally, we will also need to introduce a quantity that keeps track of the radial change of $U^\mu$ (or, equivalently, the $U^\mu$~projection of the radial change of $\Pi_{\mu\nu}$).
This leads us to define
\begin{equation}
  \label{eq:calZ}
  \mathcal{Z}_\mu
  = \Pi_{\mu\nu}\partial_r U^\nu
  - \mathcal{A}_\mu
  = r^{-1} \left(
     2h^\nu_\mu v^\rho \os{0}{\Pi}_{\nu\rho}+a^\rho F_{\rho\mu} - a^\rho C_{\rho\mu}
  \right)
  + \mathcal{O}(r^{-2})\,,
\end{equation}
where we subtracted $\mathcal{A}_\mu$ to cancel the $a_\mu$ contribution at leading order.

Note that all of the objects $\mathcal{A}_\mu$, $\mathcal{K}_{\mu\nu}$, $\mathcal{F}_{\mu\nu}$, $\mathcal{G}_{\mu\nu}$ and $\mathcal{Z}_\mu$ are spatial
in the sense that their $U^\mu$~contractions vanish.
Keeping the completeness relation~\eqref{eq:car-cov-bondi-orthogonality-completeness-repeat} in mind,
it is therefore consistent to raise and lower their indices
using the spatial objects $\Pi^{\mu\nu}$ and $\Pi_{\mu\nu}$,
which allows us to define for example
\begin{equation}
  \label{eq:calF-one-raised}
  \mathcal{F}^\mu{}_\nu
  = \Pi^{\mu\rho} \mathcal{F}_{\rho\nu}\,.
\end{equation}
We will need to do this very frequently in the following expressions,
and we will therefore often suppress explicit factors of $\Pi$ to obtain more compact expressions.
This convention is further outlined in Appendix~\ref{sapp:spatial-tensors}.
Note that raising an index adds an overall factor of $r^{-2}$,
while lowering an index adds a factor of $r^{2}$ instead,
so that for example
$\mathcal{F}^\mu{}_\nu = r^{-2} h^{\mu\rho} F_{\rho\nu} + \OO(r^{-3})$.

These composite variables such as~$\mathcal{G}_{\mu\nu}$ appear naturally as building blocks in the components of the Einstein equations.
Additionally, they are helpful in determining the correct radial falloff of the resulting expressions.
As we can see in~\eqref{ssec:rewriting-EE-4d-simplifications}, the individual terms in the definition of for example $\mathcal{G}_{\mu\nu}$ are $\OO(r)$,
but since their leading-order contributions cancel, the total expression is $\OO(1)$.
To understand the structure of the equations of motion,
it will be very useful to rewrite all resulting expressions in terms of these variables,
which will make the $1/r$~behaviour of each equation easier to parse.

Related descriptions of general hypersurface geometry can be also formulated using rigging techniques~\cite{Mars:1993mj,Gourgoulhon:2005ng,Mars:2024vai}.
One may also analyse Einstein's equations using the Newman--Penrose formalism~\cite{Newman:1961qr,Newman:1962cia,Adamo:2009vu,Madler:2016xju,Geiller:2025dqe}. 
For the purposes of this work, we prefer the present formulation,
since it is adapted to the Carrollian geometry one finds at $\mathcal{I}^+$.

\subsection{Rewriting the \texorpdfstring{$R_{rr}=0$}{rr} equation}
\label{ssec:rewriting-EE-rr-equation}
We now want to obtain explicit, covariant and compact expressions for the various projections of the Einstein equations,
using the Carroll-covariant Bondi gauge developed in Section~\ref{sec:car-cov-BS-gauge}.
The most straightforward case is the $rr$ component,
\begin{equation}
  \label{eq:Rrr1}
  R_{rr}
  = dr^{-2}-\Gamma^\rho_{r\sigma}\Gamma^\sigma_{r\rho}+dr^{-1}\Gamma^r_{rr}
  =0\,.
\end{equation}
From the decomposition of the Christoffel symbols in~\eqref{eq:main-christoffel-decomposition},
we can derive that
$\Gamma^r_{rr}$ and $\Gamma^\sigma_{r\rho}$ fall off as
$\mathcal{O}(r^{-2})$ and $\mathcal{O}(r^{-1})$, respectively,
which would suggest that the leading contribution to this equation is at order $r^{-2}$.
However, it turns out that all terms at this order cancel out.
Such a cancellation will also occur in several of the other projections of the equations of motion in~\eqref{eq:EOMList},
and,
while this particular case is still relatively straightforward to obtain by simply expanding all individual Christoffel components in~\eqref{eq:Rrr1},
such a direct approach would be very unwieldy in other cases.
For this reason, we note that we can rewrite the $rr$ component of the equations of motion \emph{prior} to the radial expansion in the following way,
\begin{align}
  \label{eq:Rrrgaugefixed}
  0
  = R_{rr}
  &= d r^{-1}\partial_r\beta
  + \frac{1}{4}\Pi^{\mu\rho}\Pi^{\nu\sigma}V_{\mu\nu}V_{\rho\sigma}
  \\
  &{}\qquad
  - \frac{1}{4}\Pi^{\mu\rho}\Pi^{\nu\sigma}
  \left(\partial_r\Pi_{\rho\sigma}-2r^{-1}\Pi_{\rho\sigma}\right)
  \left(\partial_r\Pi_{\mu\nu}-2r^{-1}\Pi_{\mu\nu}\right)\,.
  \\
  \label{eq:rr-eqCompObjForm}
  &= d r\inv \pd_r \beta
  + \frac{1}{4} \mathcal{F}^{\mu\nu} \mathcal{F}_{\mu\nu}
  - \frac{1}{4} \mathcal{G}^{\mu\nu} \mathcal{G}_{\mu\nu}\,.
\end{align}
Expressed in this way, the cancellation of the would-be leading terms is encoded in the composite variables
$\mathcal{F}_{\mu\nu}$
and
$\mathcal{G}_{\mu\nu}$
that we introduced in~\eqref{eq:curly-A-K-F-def-repeat}
and~\eqref{ssec:rewriting-EE-4d-simplifications} above,
which now
appear as natural building blocks.
Counting powers in the radial expansion using the rules for spatial tensors outlined below~\eqref{eq:calF-one-raised} above,
we can now also quickly see that both
$\mathcal{F}^{\mu\nu} \mathcal{F}_{\mu\nu}$
and
$\mathcal{G}^{\mu\nu} \mathcal{G}_{\mu\nu}$
are $\OO(r^{-4})$,
and it is then immediately clear that the leading contribution to the $rr$ equation of motion is
\begin{equation}
  \label{eq:rr-eqCompObjForm-quick-expansion}
  0
  = R_{rr}
  = -d r^{-3} \os{1}{\beta}
  + \OO(r^{-4})\,.
\end{equation}
This implies that $\os{1}{\beta}=0$,
which agrees with the result we already derived as part of our initial analysis of the leading-order equations of motion in~\eqref{eq:car-cov-bondi-LO-eom-results} above.

Again, we could have also obtained this result relatively quickly by directly expanding all of the individual terms in the original expression~\eqref{eq:Rrr1}.
However, as will become clear in the following, that approach would quickly become unfeasible at lower orders due to the large amount of terms involved.
For the remainder of this section,
we will therefore focus on rewriting all (projected) components of the equations of motion listed in~\eqref{eq:EOMList} into similarly compact forms with transparent radial falloff as~\eqref{eq:rr-eqCompObjForm} above.
Using these expressions, we will then develop the expansion of the equations of motion in Section~\ref{sec:radial-expansion}.

\subsection{Rewriting the \texorpdfstring{$R_{\mu r} = 0$}{mu r} equations}
\label{ssec:rewriting-EE-mu-r-equation}
Next, let us apply the same procedure to the $\mu r$ equation.
Before considering its vielbein projections, it is useful to first consider the full equation, which reads
\begin{equation}
  \label{eq:RmurAv1}
  \begin{split}
    0
    = R_{\mu r}
    &=  \left(
      -\D_\mu\Gamma^r_{rr}
      + \D_r \Gamma^r_{\mu r}
      + d r\inv \Gamma^r_{\mu r}
    \right)
    +
    \left(
      \D_\rho \Gamma^\rho_{\mu r}
      + \Gamma^\rho_{\rho\sigma}\Gamma^\sigma_{\mu r}
      - \Gamma^\rho_{\mu\sigma}\Gamma^\sigma_{\rho r}
    \right).
  \end{split}
\end{equation}
Again, the components of the Christoffel connection can be found in \eqref{eq:main-christoffel-decomposition}.
While the terms in the first bracket are covariant on equal-$r$ hypersurfaces,
the terms in the second bracket should be combined into a covariant derivative
of~$\Gamma^\rho_{\mu r}$, which is tensorial in a $(d+1)$-dimensional sense.
For this, recall that~\eqref{eq:main-christoffel-decomposition} included in particular the decomposition
$\Gamma^\rho_{\mu\nu} = \hat{C}^\rho_{\mu\nu} + \hat{T}^\rho{}_{\mu\nu}$,
where $\hat{C}^\rho_{\mu\nu}$ is the hypersurface connection
we discussed in Section~\ref{ssec:connection-choice}
and where we denote the remaining part by
\begin{equation}
  \label{eq:christoffel-decomposition-T-shift-tensor-def}
  \hat{T}^\rho{}_{\mu\nu}
  = U^\rho V_{(\mu}\mathcal{A}_{\nu)}
  + S \Pi^{\rho\sigma} V_{\sigma(\mu} V_{\nu)}
  + \frac{1}{2} V_\mu V_\nu \Pi^{\rho\sigma} \pd_\sigma S
  -\frac{1}{2}U^{\rho}\partial_r g_{\mu\nu}\,.
\end{equation}
Using the connection~$\hat{C}^\rho_{\mu\nu}$ to covariantise the $\mu r$ equation in~\eqref{eq:RmurAv1} then gives
\begin{align}
  \label{eq:RmurA-covariant}
  \begin{split}
        0
  &= - \partial_\mu\Gamma^r_{rr}
  + \partial_r\Gamma^r_{\mu r}
  + dr^{-1}\Gamma^r_{\mu r}
  + \hat D_\rho\Gamma^\rho_{\mu r}-\frac{1}{2}\mathcal{A}_\rho\Pi^\sigma_\mu\Gamma^\rho_{\sigma r}
  \\
  &{}\qquad
  - \frac{1}{4}\Pi^\nu_\mu\Pi^{\sigma\alpha}\mathcal{A}_\alpha \partial_r\Pi_{\nu\sigma}
  - \frac{1}{4}\mathcal{F}_{\mu\rho}\partial_r U^\rho
  + \frac{1}{4}SV_\mu\mathcal{F}^{\alpha\beta}\mathcal{F}_{\alpha\beta}\,.
  \end{split}
\end{align}
Next, we must work out the covariant derivative $\hat{D}_\rho \Gamma^\rho_{\mu r}$ explicitly
using the metricity properties listed in~\eqref{eq:hatCcon-metricity-properties}.

It is then useful to consider $\Pi^\mu_\rho R_{\mu r}$ and $U^\mu R_{\mu r}$ separately.
The latter gives
\begin{align}
\label{eq:URmur}
    0
    = U^\mu R_{\mu r}
    &= -\frac{1}{2}\left(\partial_r^2 S+dr^{-1}\partial_r S\right)
    -\frac{3}{2}\mathcal{L}_U\partial_r\beta
    -\frac{3}{2}\partial_r S \partial_r\beta
    \\
    &{}\qquad\nonumber
    -S\left(\partial^2_r\beta+
    dr^{-1}\partial_r\beta\right)-S\left(\partial_r\beta\right)^2+\frac{1}{2}\mathcal{K}\partial_r\beta
    \\
    &{}\qquad\nonumber
    +\frac{1}{2}\Pi_{\mu\nu}\partial_r U^\mu\partial_r U^\nu-\frac{1}{2}\hat D_\rho\partial_r U^\rho-\frac{1}{2}\mathcal{A}_\rho\partial_r U^\rho
    \\
    &{}\qquad\nonumber
    +\frac{1}{2}\mathcal{G}^{\mu\nu}\mathcal{K}^T_{\mu\nu}+r^{-1}\mathcal{K}
    -\frac{1}{4}S\mathcal{F}^{\alpha\beta}\mathcal{F}_{\alpha\beta}-\frac{1}{2}\hat D_\rho\mathcal{A}^\rho\,.
\end{align}
Since $\mathcal{G}_{\mu\nu}$ is traceless with respect to $\Pi^{\mu\nu}$,
only the STF part of $\mathcal{K}_{\mu\nu}$ enters this expression,
which we have denoted with\footnote{%
  We will often use the notation $\mathcal{K}^T_{\mu\nu}$
  as an alternative for the angle bracket notation $\mathcal{K}_{\langle\mu\nu\rangle}$
  since the former has some advantages in the context of index manipulations and the $1/r$ expansion.
}
\begin{equation}
  \label{eq:KTDefn}
  \mathcal{K}^T_{\mu\nu}
  = \mathcal{K}_{\mu\nu}
  - \frac{1}{d} \mathcal{K} \Pi_{\mu\nu}
  = \OO(r)\,.
\end{equation}
To see why this is not $\OO(r^2)$,
recall from our initial analysis in Section~\ref{ssec:bulk-geom-LO-EE-in-partial-gauge} that the leading-order equations of motion imply that the boundary extrinsic curvature $K_{\mu\nu}$ is pure trace,
which means that the would-be leading-order term $r^2 K_{\langle \mu\nu \rangle}$ vanishes. 
Relatedly, we used the trace $\mathcal{K}$ of $\mathcal{K}_{\mu\nu}$ which is given by
\begin{equation}
    \mathcal{K} = \Pi^{\mu\nu} \mathcal{K}_{\mu\nu}\,.
\end{equation}
This all follows our conventions for spatial tensors listed in Appendix~\ref{sapp:spatial-tensors}.
Next, we express $\partial_r U^\mu$ in terms of $\mathcal{Z}^\mu$ as defined in \eqref{eq:calZ} using
\begin{equation}
    \partial_r U^\mu=\mathcal{Z}^\mu+\mathcal{A}^\mu-U^\mu\partial_r\beta\,.
\end{equation}
Using this and reordering, we obtain
\begin{align}
\label{eq:URmur-v3prime}
    0
    = U^\mu R_{\mu r}
    &= -\frac{1}{2}\left(\partial_r^2 S
    +dr^{-1}\partial_r S-dr^{-2}S\right)+r^{-1}\bar{\mathcal{K}}
    -\mathcal{L}_U\partial_r\beta
    -\frac{3}{2}\partial_r S \partial_r\beta
    \\
    &{}\qquad\nonumber
    -S\left(\partial^2_r\beta
    +dr^{-1}\partial_r\beta\right)-S\left(\partial_r\beta\right)^2
    \\
    &{}\qquad\nonumber+\frac{1}{2}\mathcal{Z}_\mu\mathcal{Z}^\mu
    -\frac{1}{2}\left(\hat D_\mu -\mathcal{A}_\mu\right)\mathcal{Z}^\mu
    \\
    &{}\qquad\nonumber
    +\frac{1}{2}\mathcal{G}^{\mu\nu}\mathcal{K}^T_{\mu\nu}
    -\frac{1}{4}S\mathcal{F}^{\alpha\beta}\mathcal{F}_{\alpha\beta}-\hat D_\rho\mathcal{A}^\rho\,,
\end{align}
where we defined another composite object $\bar{\mathcal{K}}$ as follows,
\begin{equation}
  \label{eq:BarCalK-def}
  \bar{\mathcal{K}}
  = \mathcal{K} - \frac{d}{2r} S\,.
\end{equation}
The rationale for introducing this object is as follows.
Using our Carroll-covariant Bondi--Sachs gauge,
we saw that the boundary trace $K$
is fixed to be $(d/2) \os{-1}{S}$
by the leading-order equations of motion in~\eqref{eq:car-cov-bondi-LO-eom-results}.
We therefore see that 
\begin{equation}\label{eq:orderbarcalK}
  \bar{\mathcal{K}}
  = \OO(r\inv)\,,
\end{equation}
while $\mathcal{K}$ is only order $r^0$. This further simplifies the subsequent radial expansion.
Next, commuting $\partial_r$ and $\mathcal{L}_U$ and re-expressing the result in terms of our bulk variables such as $\mathcal{Z}_{\mu}$ we obtain the identity
\begin{align}\label{eq:partialrcalK}
    \partial_r\mathcal{K}=-\mathcal{K}\partial_r\beta-\hat D_\rho \mathcal{A}^\rho-\hat D_\rho \mathcal{Z}^\rho\,.
\end{align}
Using this, we can then obtain
\begin{align}
\label{eq:URmur-v3}
    0
    = U^\mu R_{\mu r}
    &= -\frac{1}{2}\partial_r^2 S
    +\partial_r\bar{\mathcal{K}}+r^{-1}\bar{\mathcal{K}}
    \\
    &{}\qquad\nonumber
    -S\left(\partial^2_r\beta
    +\frac{d}{2}r^{-1}\partial_r\beta\right)-S\left(\partial_r\beta\right)^2-\mathcal{L}_U\partial_r\beta
    -\frac{3}{2}\partial_r S \partial_r\beta+\bar{\mathcal{K}}\partial_r\beta
    \\
    &{}\qquad\nonumber+\frac{1}{2}\mathcal{Z}_\mu\mathcal{Z}^\mu
    +\frac{1}{2}\left(\hat D_\mu +\mathcal{A}_\mu\right)\mathcal{Z}^\mu
    +\frac{1}{2}\mathcal{G}^{\mu\nu}\mathcal{K}^T_{\mu\nu}
    -\frac{1}{4}S\mathcal{F}^{\alpha\beta}\mathcal{F}_{\alpha\beta}\,,
\end{align}
as our final expression.
With this, using the same methods as before,
we can see that the $U^\mu R_{\mu r}=0$ equation is $\OO(r^{-3})$ for any $d$.
Again, this is much more straightforward to see in terms of the composite variables than by directly expanding the original equation.

To be sure,
statements such as $\mathcal{Z}_\mu=\mathcal{O}(r^{-1})$, $\bar{\mathcal{K}}
= \OO(r\inv)$ and $\mathcal{K}^T_{\mu\nu}
= \OO(r)$ rely on using what we know after solving the leading order Einstein equations.
Since we are only interested in on-shell information at this stage, we will use these results throughout.
This completes the set of building blocks we will use to rewrite the remaining equations of motion.

Next, the remaining projection
$\Pi^\mu_\kappa R_{\mu r}=0$ of the $\mu r$ equation of motion in~\eqref{eq:RmurA-covariant}
can be rewritten in a similar way in terms of the composite variables above,
which results in
\begin{align}
\label{eq:PiRBestForm}
    0
    = \Pi^\mu_\kappa R_{\mu r}
    &= -\frac{1}{2}\Pi^\mu_\kappa\partial_r \mathcal{Z}_\mu
    -\frac{d}{2}r^{-1}\mathcal{Z}_\kappa
    +\frac{1}{2}\mathcal{F}^\rho{}_\kappa \mathcal{Z}_\rho
    -\Pi^\mu_\kappa \partial_\mu\partial_r\beta
    \\
    &{}\qquad\nonumber
    +\frac{1}{2}\Pi^\mu_\kappa\hat D_\rho \mathcal{G}^\rho{}_\mu
    -\frac{1}{2}\mathcal{A}_\rho\mathcal{G}^\rho{}_\kappa
    +\frac{1}{2}\Pi^\mu_\kappa\hat D_\rho \mathcal{F}^\rho{}_\mu
    +\frac{1}{2}\mathcal{A}_\rho\mathcal{F}^\rho{}_\kappa\,.
\end{align}  
The individual terms are all order $r^{-2}$.

\subsection{Rewriting the \texorpdfstring{$R_{\mu\nu} = 0$}{mu nu} equations}
\label{ssec:rewriting-EE-mu-nu-equation}
Finally, we turn to the $\mu\nu$ equation of motion and its various spacetime projections.
Again, we first obtain a covariant expression for the full $\mu\nu$ equation
in terms of the hypersurface connection $\hat{C}^\rho_{\mu\nu}$ and its covariant derivative $\hat{D}_\rho$ as follows,
\begin{align}
  0
  = R_{\mu\nu}
  &= -\partial_\mu\Gamma^P_{P\nu}
  +\partial_r\Gamma^r_{\mu\nu}
  +\partial_\rho\Gamma^\rho_{\mu\nu}
  -\Gamma^P_{\mu Q}\Gamma^Q_{P\nu}
  +\Gamma^P_{Pr}\Gamma^r_{\mu\nu}
  +\Gamma^P_{P\sigma}\Gamma^\sigma_{\mu\nu}\nonumber
  \\
  \label{eq:Rmunu-covariant}
  &= \hat R_{\mu\nu}
  +\hat D_\rho \hat T^\rho{}_{\mu\nu}
  -\hat T^\rho{}_{\mu\sigma}\hat T^\sigma{}_{\rho\nu}
  \\
  &{}\qquad\nonumber
  +\partial_r\Gamma^r_{\mu\nu}
  -\Gamma^r_{\mu r}\Gamma^r_{\nu r}
  -\Gamma^r_{\nu\sigma}\Gamma^\sigma_{r\mu}
  -\Gamma^r_{\mu\sigma}\Gamma^\sigma_{r\nu}
  +\Gamma^P_{Pr}\Gamma^r_{\mu\nu}\,.
\end{align}
Here,
$\hat{R}_{\mu\nu}$ is the Ricci tensor of $\hat{C}^\rho_{\mu\nu}$
and $\hat{T}^\rho{}_{\mu\nu}$ is defined in~\eqref{eq:christoffel-decomposition-T-shift-tensor-def} above.
We then want to work out the various projections of this equation.
For this, it is useful to split the above expression into two terms
$R_{\mu\nu} = X_{\mu\nu} + Y_{\mu\nu}$
where
\begin{subequations}
  \label{eq:eom-rewrite-mu-nu-building-blocks}
  \begin{align}
    X_{\mu\nu}
    &= \hat R_{\mu\nu}
    +\hat D_\rho \hat T^\rho{}_{\mu\nu}
    -\hat T^\rho{}_{\mu\sigma}\hat T^\sigma{}_{\rho\nu}
    \,,
    \\
    Y_{\mu\nu}
    &= \partial_r\Gamma^r_{\mu\nu}
    -\Gamma^r_{\mu r}\Gamma^r_{\nu r}
    -\Gamma^r_{\mu\sigma}\Gamma^\sigma_{r\nu}
    -\Gamma^r_{\nu\sigma}\Gamma^\sigma_{r\mu}
    +\Gamma^P_{Pr}\Gamma^r_{\mu\nu}\,.
  \end{align}    
\end{subequations}
To compute the spacetime projections of these terms,
we first work out the projections of $\hat{T}^\rho{}_{\mu\nu}$
which give
\begin{subequations}
  \begin{align}
    U^\mu U^\nu\hat T^\rho_{\mu\nu}
    &= S\mathcal{A}^\rho
    +\frac{1}{2}\Pi^{\rho\sigma}\partial_\sigma S
    +\frac{1}{2}U^\rho\left(\partial_r S+2S\partial_r\beta\right),
    \\
    \Pi_\kappa^\mu U^\nu\hat T^\rho_{\mu\nu}
    &= -\frac{1}{2}S\mathcal{F}^\rho{}_\kappa
    +\frac{1}{2}U^\rho\mathcal{Z}_\kappa\,,
    \\
    \Pi_\kappa^\mu \Pi_\lambda^\nu\hat T^\rho_{\mu\nu}
    &= -U^\rho\left(
      \frac{1}{2}\mathcal{G}_{\kappa\lambda}+r^{-1}\Pi_{\kappa\lambda}
    \right).
  \end{align}    
\end{subequations}
Additionally, in terms of the composite variables introduced above,
the relevant tensors arising from the decomposition of the Christoffel symbols in~\eqref{eq:main-christoffel-decomposition} are
\begin{subequations}
  \begin{align}
    \Gamma^P_{Pr}
    &= dr^{-1}+\partial_r\beta\,,
    \\
    \Gamma^r_{\mu r}
    &= -\frac{1}{2}\mathcal{Z}_\mu
    +\frac{1}{2}V_\mu\left(\partial_r S+2S\partial_r\beta\right),
    \label{eq:Gammarmur}
    \\
    \Gamma^\rho_{\mu r}
    &= r^{-1}\Pi^\rho_\mu
    +\frac{1}{2}\mathcal{F}^\rho{}_\mu
    +\frac{1}{2}\mathcal{G}^\rho{}_\mu
    +V_\mu\left(\frac{1}{2}\mathcal{Z}^\rho+\mathcal{A}^\rho\right),
    \\
    \Gamma^r_{\mu\nu}
    &= \frac{1}{2}SV_\mu V_\nu\left(\partial_r S+2S\partial_r\beta\right)
    -\frac{1}{2}V_\mu V_\nu\mathcal{L}_U S
    -SV_{(\mu}\mathcal{Z}_{\nu)}
    \\
    &{}\qquad\nonumber
    +V_{(\mu}\Pi^\rho_{\nu)}\partial_\rho S
    +\mathcal{K}_{\mu\nu}
    -\frac{1}{2}S\mathcal{G}_{\mu\nu}
    -r^{-1}S\Pi_{\mu\nu}\,.
  \end{align}
\end{subequations}
With these results in place, we can now work out the spacetime projections
of the building blocks in~\eqref{eq:eom-rewrite-mu-nu-building-blocks}.
The projections of $X_{\mu\nu}$ give
\begin{subequations}
  \begin{align}
    U^\mu U^\nu X_{\mu\nu}
    &= \hat D_\rho
    \left(S\mathcal{A}^\rho+\frac{1}{2}\Pi^{\rho\sigma}\partial_\sigma S\right)-\mathcal{Z}^\rho\left(S\mathcal{A}_\rho+\frac{1}{2}\partial_\rho S\right)+\frac{1}{4}S^2\mathcal{F}^2
    \\
    &{}\qquad\nonumber
    -\frac{1}{4}\left(\partial_r S+2S\partial_r\beta\right)^2+\frac{1}{2}\mathcal{L}_U\left(\partial_r S+2S\partial_r\beta\right)
    \\
    &{}\qquad\nonumber
    -\frac{1}{2}\mathcal{K}\left(\partial_r S+2S\partial_r\beta\right)
    -\mathcal{K}^{\rho\sigma}\mathcal{K}_{\rho\sigma}+\mathcal{L}_U\mathcal{K}\,,
    \\
    \Pi_\kappa^\mu U^\nu X_{\mu\nu}
    &= -\frac{1}{4}\mathcal{Z}_\kappa\left(\partial_r S+2S\partial_r\beta\right)
    -\frac{1}{2}\mathcal{A}_\kappa\left(\partial_r S+2S\partial_r\beta\right)
    \\
    &{}\qquad\nonumber
    -\frac{1}{2}S\mathcal{A}_\rho\mathcal{F}^\rho{}_\kappa-\frac{3}{4}\mathcal{F}^\rho{}_\kappa\partial_\rho S
    -\frac{1}{2}S\Pi^\lambda_\kappa \hat D_\rho\mathcal{F}^\rho{}_\lambda
    +\frac{1}{4}S\mathcal{Z}_\rho\mathcal{F}^\rho{}_\kappa
    \\
    &{}\qquad\nonumber
    -\frac{1}{2}\mathcal{K}\mathcal{Z}_\kappa
    +\frac{1}{2}\mathcal{K}^\rho{}_\kappa\mathcal{Z}_\rho
    +\frac{1}{2}\mathcal{L}_U\mathcal{Z}_\kappa
    \\
    &{}\qquad\nonumber
    +\frac{1}{2}\left(
      \mathcal{G}^\rho{}_\kappa+2r^{-1}\Pi^\rho_\kappa
    \right)
    \left(S\mathcal{A}_\rho+\frac{1}{2}\partial_\rho S\right)\\
    &{}\qquad\nonumber+\Pi^\mu_\kappa \partial_\mu\mathcal{K}-\Pi^\mu_\kappa\hat D_\rho\mathcal{K}^\rho{}_\mu\,,\nonumber\\
    \Pi_\kappa^\mu\Pi^\nu_\lambda X_{\mu\nu}
    &= -\frac{1}{2}\Pi^\alpha_\kappa\Pi^\beta_\lambda\mathcal{L}_U\mathcal{G}_{\alpha\beta}
    -\mathcal{G}^\sigma{}_{(\lambda}\mathcal{K}_{\kappa)\sigma}
    +\frac{1}{2}\mathcal{K}\mathcal{G}_{\kappa\lambda}
    +r^{-1}\Pi_{\kappa\lambda}\mathcal{K}
    \\
    &{}\qquad\nonumber
    -\frac{1}{2}S\mathcal{F}_{\kappa\sigma}\mathcal{F}^\sigma{}_\lambda
    -\frac{1}{2}S\mathcal{G}_{\sigma(\kappa}\mathcal{F}^\sigma{}_{\lambda)}
    -\mathcal{Z}_{(\kappa}\mathcal{A}_{\lambda)}
    -\frac{1}{4}\mathcal{Z}_\kappa\mathcal{Z}_\lambda
    +\Pi_\kappa^\mu\Pi^\nu_\lambda\hat R_{\mu\nu}\,,
  \end{align}
\end{subequations}
while the projections of $Y_{\mu\nu}$ give
\begin{subequations}
\begin{align}
    U^\mu U^\nu Y_{\mu\nu} & = -\frac{1}{2}\mathcal{A}^\rho\partial_\rho S-\frac{1}{2}S\mathcal{Z}_\mu\mathcal{Z}^\mu+S\mathcal{L}_U\partial_r\beta \nonumber\\
    &\qquad+\frac{1}{4}\left(\partial_r S+2S\partial_r\beta\right)^2-\frac{1}{2}\mathcal{L}_U\left(\partial_r S+2S\partial_r\beta\right)\nonumber\\
    &\qquad+\frac{1}{2}S\left(\partial_r+dr^{-1}+\partial_r\beta\right)\left(\partial_r S+2S\partial_r\beta\right)-\frac{1}{2}dr^{-1}\mathcal{L}_U S\,,\\
    \Pi_\kappa^\mu U^\nu Y_{\mu\nu} & =  \frac{1}{4}\left(\partial_r S+2S\partial_r\beta+2dr^{-1}S\right)\mathcal{Z}_\kappa-\frac{1}{2}(d-1)r^{-1}\Pi^\alpha_\kappa\partial_\alpha S\nonumber\\
    &\qquad-\partial_r\beta\Pi^\alpha_\kappa\partial_\alpha S+\frac{1}{4}\mathcal{G}^\alpha{}_\kappa\partial_\alpha S-\frac{1}{2}\mathcal{K}^\rho{}_\kappa\mathcal{Z}_\rho\nonumber\\
    &\qquad+\frac{1}{4}\mathcal{F}^\rho{}_\kappa\left(\partial_\rho S-S\mathcal{Z}_\rho\right)-\frac{1}{2}\Pi^\alpha_\kappa\partial_r\partial_\alpha S+\frac{1}{2}S\Pi^\rho_\kappa\partial_r\mathcal{Z}_\rho\,,\\
    \Pi_\kappa^\mu\Pi^\nu_\lambda Y_{\mu\nu} & =  \Pi^\mu_\kappa\Pi^\nu_\lambda\partial_r\left(\mathcal{K}_{\mu\nu}-r^{-1}S\Pi_{\mu\nu}-\frac{1}{2}S\mathcal{G}_{\mu\nu}\right)\nonumber\\
    &\qquad+\left((d-2)r^{-1}+\partial_r\beta\right)\left(\mathcal{K}_{\kappa\lambda}-r^{-1}S\Pi_{\kappa\lambda}-\frac{1}{2}S\mathcal{G}_{\kappa\lambda}\right)\nonumber\\
    &\qquad-\mathcal{F}^\mu{}_{(\kappa}\mathcal{K}_{\lambda)\mu}-\mathcal{G}^\mu{}_{(\kappa}\mathcal{K}_{\lambda)\mu}+\frac{1}{2}S\mathcal{F}_{\rho(\kappa}\mathcal{G}^\rho{}_{\lambda)}\nonumber\\
    &\qquad+r^{-1}S\mathcal{G}_{\kappa\lambda}+\frac{1}{2}S\mathcal{G}_{\kappa\rho}\mathcal{G}^\rho{}_{\lambda}-\frac{1}{4}\mathcal{Z}_\kappa\mathcal{Z}_\lambda\,.
\end{align}
\end{subequations}
We can then find the projections of the $\mu\nu$ equations of motion~\eqref{eq:Rmunu-covariant} by summing the relevant intermediate $X_{\mu\nu}$ and $Y_{\mu\nu}$ results.
The double $U$-projection gives
\begin{align}
  0
  &= U^\mu U^\nu R_{\mu\nu}
  \nonumber
  \\
  \label{eq:UUR}
  &= \hat D_\rho\left(S\mathcal{A}^\rho+\frac{1}{2}\Pi^{\rho\sigma}\partial_\sigma S\right)
  -\mathcal{Z}_\rho\left(S\mathcal{A}^\rho+\frac{1}{2}\Pi^{\rho\sigma}\partial_\sigma S\right)
  -\frac{1}{2}\mathcal{A}^\rho\partial_\rho S
  \\
  &{}\qquad\nonumber
  +\frac{1}{4}S^2\mathcal{F}^2 -\frac{1}{2}S\mathcal{Z}_\mu\mathcal{Z}^\mu+S\mathcal{L}_U\partial_r\beta  -\mathcal{K}^{T\rho\sigma}\mathcal{K}^T_{\rho\sigma}
  -\frac{1}{d}\bar{\mathcal{K}}^2
  +\mathcal{L}_U\bar{\mathcal{K}}
  \\
  &{}\qquad\nonumber
  +\frac{3}{2}S\partial_r S\partial_r\beta-S\bar{\mathcal{K}}\partial_r\beta+S^2(\partial_r\beta)^2+S^2\left(\partial_r^2\beta+\frac{d}{2}r^{-1}\partial_r\beta\right)
  \\
  &{}\qquad\nonumber+\frac{1}{2}S\partial_r^2 S-\frac{1}{2}\bar{\mathcal{K}}\left(\partial_rS-r^{-1}S\right)+\frac{d}{4}r^{-1}S\left(\partial_rS-r^{-1}S\right)-\frac{3}{2}r^{-1}S\bar{\mathcal{K}}
  \,.
\end{align}
Using \eqref{eq:BarCalK-def} and \eqref{eq:partialrcalK},
we can write the first line as 
\begin{align}
  &\hat D_\rho\left(S\mathcal{A}^\rho+\frac{1}{2}\Pi^{\rho\sigma}\partial_\sigma S\right)-\mathcal{Z}_\rho\left(S\mathcal{A}^\rho+\frac{1}{2}\Pi^{\rho\sigma}\partial_\sigma S\right)-\frac{1}{2}\mathcal{A}^\rho\partial_\rho S
  \nonumber
  \\
  &{}\qquad
  =\frac{1}{2}\hat D_\rho\left(\Pi^{\rho\sigma}\left(\partial_\sigma+\mathcal{A}_\sigma\right)S\right)-\frac{1}{2}S\left(\partial_r\bar{\mathcal{K}}+\frac{d}{2}r^{-1}\left(\partial_r S-r^{-1}S\right)\right)
  \\
  &{}\qquad\qquad\nonumber
  -\frac{1}{2}S\left(\hat D_\rho+\mathcal{A}_\rho\right)\mathcal{Z}^\rho-\frac{1}{2}\mathcal{Z}^\rho\left(\partial_\rho+\mathcal{A}_\rho\right)S-\frac{1}{2}S\mathcal{K}\partial_r\beta\,,
\end{align}
so that we get
\begin{align}
  \label{eq:UURnew}
  0
  = U^\mu U^\nu R_{\mu\nu}
  &= \frac{1}{2}\hat D_\rho\left(\Pi^{\rho\sigma}\left(\partial_\sigma+\mathcal{A}_\sigma\right)S\right)+\mathcal{L}_U\bar{\mathcal{K}}-r^{-1}S\bar{\mathcal{K}}
  \\
  &{}\qquad\nonumber
  -\frac{1}{2}S\left(\hat D_\rho+\mathcal{A}_\rho\right)\mathcal{Z}^\rho-\frac{1}{2}\mathcal{Z}^\rho\left(\partial_\rho+\mathcal{A}_\rho\right)S
  \\
  &{}\qquad\nonumber
  +\frac{1}{4}S^2\mathcal{F}^2 -\frac{1}{2}S\mathcal{Z}_\mu\mathcal{Z}^\mu+S\mathcal{L}_U\partial_r\beta  -\mathcal{K}^{T\rho\sigma}\mathcal{K}^T_{\rho\sigma}
  -\frac{1}{d}\bar{\mathcal{K}}^2
  \\
  &{}\qquad\nonumber
  +\frac{3}{2}S\partial_r S\partial_r\beta-\frac{3}{2}S\bar{\mathcal{K}}\partial_r\beta+S^2(\partial_r\beta)^2+S^2\left(\partial_r^2\beta+\frac{d}{4}r^{-1}\partial_r\beta\right)
  \\
  &{}\qquad\nonumber+\frac{1}{2}S\partial_r^2 S-\frac{1}{2}\bar{\mathcal{K}}\left(\partial_rS-r^{-1}S\right)-\frac{1}{2}S\left(\partial_r\bar{\mathcal{K}}+r^{-1}\bar{\mathcal{K}}\right)
  \,.
\end{align}
The terms on the first line are order~$r^{-1}$,
and the terms on the other lines are order~$r^{-2}$.
For the mixed space-time projection, we obtain
\begin{align}
  \label{eq:UPiRnew}
  0
  = \Pi_\kappa^\mu U^\nu R_{\mu\nu}
  &= 
  -\frac{1}{2}S\Pi^\mu_\kappa\hat D_\rho \mathcal{F}^\rho{}_\mu
  +\frac{1}{2}\left(\mathcal{G}^\rho{}_\kappa-\mathcal{F}^\rho{}_\kappa-2\partial_r\beta\Pi^\rho_\kappa\right)\left(\partial_\rho+\mathcal{A}_\rho\right)S
  \\
  &{}\qquad\nonumber
  -\frac{1}{2}\Pi^\alpha_\kappa\left(\partial_\alpha+\mathcal{A}_\alpha\right)\left(\partial_r S-r^{-1} S\right)
  +\frac{1}{2}S\Pi^\rho_\kappa\left(
    \partial_r\mathcal{Z}_\rho+\frac{d}{2}r^{-1}\mathcal{Z}_\rho
  \right)
  \\
  &{}\qquad\nonumber
  +\frac{1}{2}\left(\mathcal{L}_U-\bar{\mathcal{K}}\right)\mathcal{Z}_\kappa
  +\frac{d-1}{d}\Pi^\mu_\kappa \partial_\mu\bar{\mathcal{K}}
  -\Pi^\mu_\kappa\hat D_\rho \mathcal{K}^{T\rho}{}_{\mu}
  -\frac{1}{d}\mathcal{A}_\kappa\bar{\mathcal{K}}\,.
\end{align}
Finally, the spatial projection gives
\begin{align}
  \label{eq:PiPiR}
  0 =
  \Pi_\kappa^\mu\Pi^\nu_\lambda R_{\mu\nu}
  &= -\frac{1}{2}\Pi^\alpha_\kappa\Pi^\beta_\lambda\mathcal{L}_U\mathcal{G}_{\alpha\beta}
  -2\mathcal{G}^\sigma{}_{(\lambda}\mathcal{K}^T_{\kappa)\sigma}
  -\frac{2}{d}\mathcal{G}_{\kappa\lambda}
  +\frac{1}{2}\overline{\mathcal{K}}\mathcal{G}_{\kappa\lambda}
  \\
  &{}\qquad\nonumber
  +\frac{1}{2}S\mathcal{G}^\nu{}_{\lambda}\mathcal{G}_{\nu\kappa}
  -\mathcal{F}^\sigma{}_{(\lambda}\mathcal{K}^T_{\kappa)\sigma}
  -\frac{1}{2}S\mathcal{F}_{\kappa\sigma}\mathcal{F}^\sigma{}_\lambda
  -\mathcal{Z}_{(\kappa}\mathcal{A}_{\lambda)}
  -\frac{1}{2}\mathcal{Z}_\kappa\mathcal{Z}_\lambda
  \\
  &{}\qquad\nonumber
  +\Pi^\mu_\kappa\Pi^\nu_\lambda\partial_r\bar{\mathcal{K}}_{\mu\nu}
  +(d-2)r^{-1}\bar{\mathcal{K}}_{\kappa\lambda}
  +r^{-1}\Pi_{\kappa\lambda}\overline{\mathcal{K}}
  +\partial_r\beta \bar{\mathcal{K}}_{\kappa\lambda}
  \\
  &{}\qquad\nonumber
  -\frac{1}{2}\Pi^\mu_\kappa\Pi^\nu_\lambda\partial_r\left(S\mathcal{G}_{\mu\nu}\right)
  -\frac{1}{4}(d-2)r^{-1}S\mathcal{G}_{\kappa\lambda}
  -\frac{1}{2}\left(S\partial_r\beta\right)\mathcal{G}_{\kappa\lambda}
  \\
  &{}\qquad\nonumber
  -\frac{1}{2}r^{-1}\left(\partial_r S-r^{-1}S\right)\Pi_{\kappa\lambda}
  -\frac{1}{2}r^{-1}\left(S\partial_r\beta\right)\Pi_{\kappa\lambda}
  +\Pi_\kappa^\mu\Pi^\nu_\lambda\hat R_{\mu\nu}\,.
\end{align}
As we saw during our initial analysis of the leading-order equations of motion in Section~\ref{ssec:bulk-geom-LO-EE-in-partial-gauge}, it is beneficial to split the spatial projection of the $\mu\nu$ equation into
a trace equation
and a symmetric trace-free (STF) equation.
The trace can be written as
\begin{align}
    \label{eq:trace}
    0
    = \Pi^{\mu\nu} R_{\mu\nu}
    &=-\frac{1}{2}S\mathcal{F}^2
    +\partial_r\bar{\mathcal{K}}
    +2dr^{-1}\bar{\mathcal{K}}-\mathcal{Z}^\mu\mathcal{A}_\nu
    -\frac{1}{2}\mathcal{Z}_\mu\mathcal{Z}^\mu+\partial_r\beta \bar{\mathcal{K}}
    \\
    &{}\qquad\nonumber
    -\frac{d}{2}r^{-1}S\partial_r\beta
    -\frac{d}{2}r^{-1}\left(\partial_r S-r^{-1}S\right)
    +\Pi^{\mu\nu}\hat R_{\mu\nu}\,,
\end{align}
which, using \eqref{eq:partialrcalK}, can also be written as
\begin{align}
    \label{eq:trace2}
    0
    = \Pi^{\mu\nu} R_{\mu\nu}
    &=-\frac{1}{2}S\mathcal{F}^2-\left(\hat D_\rho+\mathcal{A}_\rho\right)\mathcal{Z}^\rho
    -dr^{-1}S\partial_r\beta-\frac{1}{2}\mathcal{Z}_\mu\mathcal{Z}^\mu
    \\
    &{}\qquad\nonumber
    +2dr^{-1}\bar{\mathcal{K}}
    -dr^{-1}\left(\partial_r S-r^{-1}S\right)
    +\Pi^{\mu\nu}\hat R_{\mu\nu}-\hat D_\rho \mathcal{A}^\rho\,.
\end{align}
On the other hand, the STF part of~\eqref{eq:PiPiR} can be written as
\begin{align}
    \label{eq:STF}
    0
    = \Pi_{\langle\kappa}^\mu \Pi_{\lambda\rangle}^\nu R_{\mu\nu}
    &= -\frac{1}{2}\mathcal{L}_U\mathcal{G}_{\kappa\lambda}-\frac{1}{d}\Pi_{\kappa\lambda}\mathcal{G}^{\rho\sigma}\mathcal{K}^T_{\rho\sigma}
    -2\mathcal{G}^\sigma{}_{\langle\kappa}\mathcal{K}^T_{\lambda\rangle\sigma}
    -\mathcal{F}^\sigma{}_{(\kappa}\mathcal{K}^T_{\lambda)\sigma}
    \\
    &{}\qquad\nonumber
    -\frac{1}{2}\left(
      \partial_rS+\frac{1}{2}(d-2)r^{-1}S+S\partial_r\beta-\frac{d-2}{d}\bar{\mathcal{K}}
    \right)\mathcal{G}_{\kappa\lambda}
    \\
    &{}\qquad\nonumber
    -\frac{1}{2}S\Pi^\mu_\kappa\Pi^\nu_\lambda\partial_r\mathcal{G}_{\mu\nu}
    +\frac{1}{2}S\mathcal{G}_{\kappa\sigma}\mathcal{G}^\sigma{}_{\lambda}
    +\frac{1}{2}S\mathcal{F}_{\sigma\langle\kappa}\mathcal{F}^\sigma{}_{\lambda\rangle}
    \\
    &{}\qquad\nonumber
    +\Pi^\mu_{\langle\kappa}\Pi^\nu_{\lambda\rangle}\partial_r\mathcal{K}^T_{\mu\nu}
    +(d-2)r^{-1}\mathcal{K}^T_{\kappa\lambda}
    +\partial_r\beta \mathcal{K}^T_{\kappa\lambda}
    \\
    &{}\qquad\nonumber
    -\mathcal{Z}_{\langle\kappa}\mathcal{A}_{\lambda\rangle}
    -\frac{1}{2}\mathcal{Z}_{\langle\kappa}\mathcal{Z}_{\lambda\rangle}
    +\Pi_{\langle\kappa}^\mu\Pi^\nu_{\lambda\rangle}\hat R_{\mu\nu}\,.
\end{align}
The trace and STF part of the spatial projection of the Einstein equation $R_{\mu\nu}=0$ contain $\Pi^{\mu\nu}\hat R_{\mu\nu}$ and $\Pi_{\langle\kappa}^\mu\Pi^\nu_{\lambda\rangle}\hat R_{\mu\nu}$. Here, we can use the identity~\eqref{eq:spatialRicSTF}, but this we will do once we consider the cases $d=1,2$ later on.

The result for $\Pi_{\langle\kappa}^\mu \Pi_{\lambda\rangle}^\nu R_{\mu\nu}$ can be rewritten as follows.
By commuting $\partial_r$ and $\mathcal{L}_U$,
using the identity~\eqref{eq:partialrcalK},
and re-expressing the result in terms of our bulk variables such as $\mathcal{G}_{\mu\nu}$,
we obtain the identity
\begin{align}\label{eq:idpartialrKT}
    \Pi^\mu_{\langle\kappa}\Pi^\nu_{\lambda\rangle}\partial_r\mathcal{K}^T_{\mu\nu} 
    &=  -\frac{1}{2}\mathcal{L}_U\mathcal{G}_{\kappa\lambda}-\frac{1}{d}\Pi_{\kappa\lambda}\mathcal{G}^{\rho\sigma}\mathcal{K}^T_{\rho\sigma}-\Pi_{\langle\kappa}^\mu\Pi^\nu_{\lambda\rangle}\left(
    \hat D_\mu\mathcal{A}_\nu+\mathcal{A}_\mu\mathcal{A}_\nu\right)
    \\
    &{}\qquad\nonumber
    +2r^{-1}\mathcal{K}^T_{\kappa\lambda}-\frac{1}{d}\mathcal{K}\mathcal{G}_{\kappa\lambda}-\partial_r\beta \mathcal{K}^T_{\kappa\lambda}-\Pi_{\langle\kappa}^\mu\Pi^\nu_{\lambda\rangle}\left(
    \hat D_\mu\mathcal{Z}_\nu+\mathcal{A}_\mu\mathcal{Z}_\nu\right).
\end{align}
We use this identity to eliminate
$-\frac{1}{2}\mathcal{L}_U\mathcal{G}_{\kappa\lambda}$
from~\eqref{eq:STF},
leading to
\begin{align}
    \label{eq:STFnew}
    0
    = \Pi_{\langle\kappa}^\mu \Pi_{\lambda\rangle}^\nu R_{\mu\nu}
    &= \Pi_{\langle\kappa}^\mu\Pi^\nu_{\lambda\rangle}
    \hat D_\mu\mathcal{Z}_\nu
    -2\mathcal{G}^\sigma{}_{\langle\lambda}\mathcal{K}^T_{\kappa\rangle\sigma}
    -\mathcal{F}^\mu{}_{(\kappa}\mathcal{K}^T_{\lambda)\mu}
    \\
    &{}\qquad\nonumber
    -\frac{1}{2}\left(
      \partial_rS+\frac{1}{2}(d-4)r^{-1}S+S\partial_r\beta-\bar{\mathcal{K}}
    \right)\mathcal{G}_{\kappa\lambda}
    \\
    &{}\qquad\nonumber
    -\frac{1}{2}S\Pi^\mu_\kappa\Pi^\nu_\lambda\partial_r\mathcal{G}_{\mu\nu}
    +\frac{1}{2}S\mathcal{G}_{\kappa\sigma}\mathcal{G}^\sigma{}_{\lambda}
    +\frac{1}{2}S\mathcal{F}_{\sigma\langle\kappa}\mathcal{F}^\sigma{}_{\lambda\rangle}
    \\
    &{}\qquad\nonumber
    +2\Pi^\mu_{\langle\kappa}\Pi^\nu_{\lambda\rangle}\partial_r\mathcal{K}^T_{\mu\nu}
    +(d-4)r^{-1}\mathcal{K}^T_{\kappa\lambda}
    +2\partial_r\beta \mathcal{K}^T_{\kappa\lambda}
    \\
    &{}\qquad\nonumber
    -\frac{1}{2}\mathcal{Z}_{\langle\kappa}\mathcal{Z}_{\lambda\rangle}
    +\Pi_{\langle\kappa}^\mu\Pi^\nu_{\lambda\rangle}\left(\hat R_{\mu\nu}+
    \hat D_\mu\mathcal{A}_\nu+\mathcal{A}_\mu\mathcal{A}_\nu\right)\,.
\end{align}
This completes our rewriting of the equations of motion in general dimensions.
While the expressions we obtained above are rather involved,
there are significant simplifications in three and four bulk spacetime dimensions.
For example, the spatial STF components above vanish identically in three dimensions,
while they can be simplified significantly in four dimensions.
We will work this out in full detail in the remaining subsections.

\subsection{Simplifications in three bulk dimensions}
\label{ssec:rewriting-EE-3d-simplifications}
In three bulk spacetime dimensions,
corresponding to $d=1$,
the spatial tensors on a constant-$r$ hypersurface are one-dimensional in tangent space.
As a result, there are no antisymmetric or symmetric trace-free
spatial tensors,
and therefore the tensors
$\mathcal{F}_{\mu\nu}$,
$\mathcal{G}_{\mu\nu}$
and
$\mathcal{K}^T_{\mu\nu}$
are all identically zero.
The $rr$ equation of motion we derived previously in~\eqref{eq:rr-eqCompObjForm}
therefore reduces to
\begin{equation}
\label{eq:deq1rr}
    0
    = R_{rr}
    = \partial_r\beta,
\end{equation}  
which simply sets $\beta$ to zero.
Similarly, the spacetime projections~\eqref{eq:URmur-v3} and~\eqref{eq:PiRBestForm} of the $\mu r$~equations reduce to
\begin{align}
  \label{eq:UR-r-eq-d1}
    0
    = U^\mu R_{\mu r}
    &= -\frac{1}{2}\partial_r^2 S
    +\partial_r\bar{\mathcal{K}}+r^{-1}\bar{\mathcal{K}}
    \\
    &{}\qquad\nonumber+\frac{1}{2}\mathcal{Z}_\mu\mathcal{Z}^\mu
    +\frac{1}{2}\left(\hat D_\mu +\mathcal{A}_\mu\right)\mathcal{Z}^\mu\,,
    \\
    \label{eq:PiR-r-eq-d1}
    0
    = \Pi^\mu_\kappa R_{\mu r}
    &= -\frac{1}{2}\Pi^\mu_\kappa\left(\partial_r \mathcal{Z}_\mu+r^{-1}\mathcal{Z}_\mu\right).
\end{align}
As we mentioned, the symmetric spatial trace-free part of the $\mu\nu$ equation in~\eqref{eq:STFnew}
now vanishes identically since in $d=1$ such tensors are necessarily zero.
The trace of the $\mu\nu$ equation in~\eqref{eq:trace2} reduces to
\begin{align}
    \label{eq:trace2-eq-d1}
    0
    = \Pi^{\mu\nu} R_{\mu\nu}
    &=-\left(\hat D_\rho+\mathcal{A}_\rho\right)\mathcal{Z}^\rho
    -\frac{1}{2}\mathcal{Z}_\mu\mathcal{Z}^\mu-r^{-1}\left(\partial_r S-r^{-1}S\right)
    \\
    &{}\qquad\nonumber
    +2r^{-1}\bar{\mathcal{K}}
    -2\hat D_\rho \mathcal{A}^\rho\,,
\end{align}
where we also used the expression in~\eqref{eq:spatialRicscalard=1}
for the spatial Ricci scalar,
which in this case does not encode any independent curvature components.
Finally, the remaining projections of the $\mu\nu$ equations are
\begin{align}
  \label{eq:U-U-R-d1}
  0
  = U^\mu U^\nu R_{\mu\nu}
  &= \frac{1}{2}\hat D_\rho\left(\Pi^{\rho\sigma}\left(\partial_\sigma+\mathcal{A}_\sigma\right)S\right)+\mathcal{L}_U\bar{\mathcal{K}}-r^{-1}S\bar{\mathcal{K}}
  \\
  &{}\qquad\nonumber
  -\frac{1}{2}S\left(\hat D_\rho+\mathcal{A}_\rho\right)\mathcal{Z}^\rho-\frac{1}{2}\mathcal{Z}^\rho\left(\partial_\rho+\mathcal{A}_\rho\right)S
   -\frac{1}{2}S\mathcal{Z}_\mu\mathcal{Z}^\mu 
  -\bar{\mathcal{K}}^2
  \\
  &{}\qquad\nonumber
  +\frac{1}{2}S\partial_r^2 S-\frac{1}{2}\bar{\mathcal{K}}\left(\partial_rS-r^{-1}S\right)-\frac{1}{2}S\left(\partial_r\bar{\mathcal{K}}+r^{-1}\bar{\mathcal{K}}\right)
  \,.
\end{align}
\begin{align}
  \label{eq:U-Pi-R-d1-2}
  0
  = \Pi_\kappa^\mu U^\nu R_{\mu\nu}
  &= 
  -\frac{1}{2}\Pi^\alpha_\kappa\left(\partial_\alpha+\mathcal{A}_\alpha\right)\left(\partial_r S-r^{-1} S\right)
  \\
  &{}\qquad\nonumber
  +\frac{1}{2}S\Pi^\rho_\kappa\left(
    \partial_r\mathcal{Z}_\rho+\frac{1}{2}r^{-1}\mathcal{Z}_\rho
  \right)
  +\frac{1}{2}\left(\mathcal{L}_U-\bar{\mathcal{K}}\right)\mathcal{Z}_\kappa
  -\mathcal{A}_\kappa\bar{\mathcal{K}}\,.
\end{align}
We will study the expansion of these equations in Sections~\ref{ssec:radial-expansion-threedim} and~\ref{ssec:ConservationEqnsd1}.
There, we will see that the latter two equations give rise to conservation equations,
while the previous equations can be used to solve all subleading components of the radial expansion that are not associated to conserved quantities.

\subsection{Simplifications in four bulk dimensions}
\label{ssec:rewriting-EE-4d-simplifications}
In the case of four bulk spacetime dimensions, where $d=2$,
antisymmetric and spatial symmetric trace-free tensors such as the shear exist,
and the equations of motion are significantly more complicated.
However, some simplifications still occur due to the special properties of two-dimensional spatial STF tensors and,
following our discussion in Section~\ref{subsec:hypercurv},
due to the limited degrees of freedom in the curvature tensors on such constant-$r$ hypersurfaces.

The most important simplification for $d=2$ happens in the STF equation~\eqref{eq:STFnew}.
Due to various $d=2$ identities, in particular \eqref{eq:spatialRicSTFd=2} for the curvature as well as the general tensorial identities \eqref{eq:app-d2-spatial-STF-antisym-product} and \eqref{eq:app-d2-spatial-STF-STF-product},
this reduces to
\begin{align}
    \label{eq:STFnewd=2}
    0
    = \Pi_{\langle\kappa}^\mu \Pi_{\lambda\rangle}^\nu R_{\mu\nu}
    &= 
    2\Pi^\mu_{\langle\kappa}\Pi^\nu_{\lambda\rangle}\left(\partial_r\mathcal{K}^T_{\mu\nu}
    -r^{-1}\mathcal{K}^T_{\mu\nu}\right)
    +2\partial_r\beta \mathcal{K}^T_{\kappa\lambda}-\mathcal{F}^\mu{}_{(\kappa}\mathcal{K}^T_{\lambda)\mu}
    \\
    &{}\qquad\nonumber
    -\frac{1}{2}S\Pi^\mu_{\langle\kappa}\Pi^\nu_{\lambda\rangle}\partial_r\mathcal{G}_{\mu\nu}-\frac{1}{2}\left(
      \partial_rS-r^{-1}S-\bar{\mathcal{K}}+S\partial_r\beta
    \right)\mathcal{G}_{\kappa\lambda}
    \\
    &{}\qquad\nonumber
    -\frac{1}{2}\mathcal{Z}_{\langle\kappa}\mathcal{Z}_{\lambda\rangle}+\Pi_{\langle\kappa}^\mu\Pi^\nu_{\lambda\rangle}
    \hat D_\mu\mathcal{Z}_\nu
    \,.
\end{align}
We see that the STF equation is now manifestly order $r^{-1}$ for $d=2$. Incidentally, this can also be written as
\begin{align}
    0=\Pi_{\langle\kappa}^\mu \Pi_{\lambda\rangle}^\nu R_{\mu\nu} & =  2\Pi^\mu_{\langle\kappa}\Pi^\nu_{\lambda\rangle}\left(\partial_r-r^{-1}+\partial_r\beta\right)\left[\mathcal{K}^T_{\mu\nu}-\frac{1}{4}S\mathcal{G}_{\mu\nu}\right]-\mathcal{F}^\mu{}_{(\kappa}\mathcal{K}^T_{\lambda)\mu}\nonumber\\
    &\qquad+\frac{1}{2}\bar{\mathcal{K}}\mathcal{G}_{\kappa\lambda}-\frac{1}{2}\mathcal{Z}_{\langle\kappa}\mathcal{Z}_{\lambda\rangle}+\Pi_{\langle\kappa}^\mu\Pi^\nu_{\lambda\rangle}
    \hat D_\mu\mathcal{Z}_\nu\,.
\end{align}

Though most of the remaining equations of motion admit no significant further simplifications,
we will repeat them here specifically for $d=2$ for completeness.
First, the $rr$ equation gives
\begin{equation}
    \label{eq:rr-eq-d2}
    0
    = R_{rr}
    = -\frac{1}{4}\mathcal{G}^{\mu\nu} \mathcal{G}_{\mu\nu}
    +\frac{1}{4}\mathcal{F}^{\mu\nu} \mathcal{F}_{\mu\nu}
    +2r^{-1}\partial_r\beta\,,
\end{equation}  
while the projections of the $\mu r$ equation are
\begin{align}
  \label{eq:UR-r-eq-2d}
  0
  =
  U^\mu R_{\mu r}
  &= \frac{1}{2}\mathcal{Z}^\mu \mathcal{Z}_\mu
  +\frac{1}{2}\left(\hat D_\mu+\mathcal{A}_\mu\right)\mathcal{Z}^\mu
  +\partial_r\bar{\mathcal{K}}+r^{-1}\bar{\mathcal{K}}-\frac{1}{2}\partial_r^2 S
  \\
  &{}\qquad\nonumber
  +\frac{1}{2}\mathcal{G}^{\mu\nu}\mathcal{K}^T_{\mu\nu}
  -\frac{1}{4}S\mathcal{F}^{\mu\nu} \mathcal{F}_{\mu\nu}
  -\mathcal{L}_U\partial_r\beta+\bar{\mathcal{K}}\partial_r\beta\nonumber 
  \\
  &{}\qquad\nonumber
  -S\left(\partial^2_r\beta+r^{-1}\partial_r\beta\right)
  -\frac{3}{2}\partial_r S \partial_r\beta
  -S\left(\partial_r\beta\right)^2,
  \\
  \label{eq:PiR-r-eq-d2}
  0
  = \Pi^\mu_\kappa R_{\mu r}
  &= -\frac{1}{2}\Pi^\mu_\kappa\partial_r \mathcal{Z}_\mu
  -r^{-1}\mathcal{Z}_\kappa
  +\frac{1}{2}\mathcal{F}^\rho{}_\kappa \mathcal{Z}_\rho
  -\Pi^\mu_\kappa \partial_\mu\partial_r\beta
  \\
  &{}\qquad\nonumber
  +\frac{1}{2}\Pi^\mu_\kappa\hat D_\rho \mathcal{G}^\rho{}_\mu
  -\frac{1}{2}\mathcal{A}_\rho\mathcal{G}^\rho{}_\kappa
  +\frac{1}{2}\Pi^\mu_\kappa\hat D_\rho \mathcal{F}^\rho{}_\mu
  +\frac{1}{2}\mathcal{A}_\rho\mathcal{F}^\rho{}_\kappa\,.
\end{align}
The trace equation for $d=2$ is given by
\begin{align}
\label{eq:trace-eq-d=2}
    0
    = \Pi^{\mu\nu} R_{\mu\nu}
    &=-\frac{1}{2}S\mathcal{F}^2
    -\left(\hat D_\rho+\mathcal{A}_\rho\right) \mathcal{Z}^\rho-\frac{1}{2}\mathcal{Z}_\mu\mathcal{Z}^\mu
    -2r^{-1}S\partial_r\beta
    \\
    &{}\qquad\nonumber
    -2r^{-1}\left(\partial_r S-r^{-1}S-2\bar{\mathcal{K}}\right)
    +\Pi^{\mu\nu}\hat R_{\mu\nu}-\hat D_\rho \mathcal{A}^\rho\,.
\end{align}
We have used \eqref{eq:partialrcalK} to rewrite
both the $U^\mu R_{\mu r}$ and the $\Pi^{\mu\nu} R_{\mu\nu}$ equations.
Finally, the remaining $\mu\nu$ projections are given by
\begin{align}
  \label{eq:U-U-R-d2}
  0
  = U^\mu U^\nu R_{\mu\nu}
  &= \hat D_\rho\left(S\mathcal{A}^\rho+\frac{1}{2}\Pi^{\rho\sigma}\partial_\sigma S\right)-\frac{1}{2}\mathcal{A}^\rho\partial_\rho S-\mathcal{Z}^\rho\left(S\mathcal{A}_\rho+\frac{1}{2}\partial_\rho S\right)
  \\
  &{}\qquad\nonumber
  +\frac{1}{4}S^2\mathcal{F}^2 -\frac{1}{2}S\mathcal{Z}_\mu\mathcal{Z}^\mu+S\mathcal{L}_U\partial_r\beta
  +S\left(\partial_r+r^{-1}\right)\left(S\partial_r\beta\right)
  \\
  &{}\qquad\nonumber
  +\frac{1}{2}S\left(\partial_r^2S+r^{-1}\partial_r S-r^{-2}S\right)
  +\frac{1}{2}S\partial_r\beta\left(\partial_r S+2S\partial_r\beta\right)
  \\
  &{}\qquad\nonumber
  -\frac{1}{2}\bar{\mathcal{K}}\left(
    \partial_r S+2r^{-1}S+2S\partial_r\beta
  \right)
  -\mathcal{K}^{T\rho\sigma}\mathcal{K}^T_{\rho\sigma}
  -\frac{1}{2}\bar{\mathcal{K}}^2
  +\mathcal{L}_U\bar{\mathcal{K}}\,,
\end{align}
as well as
\begin{align}
  \label{eq:Pi-U-R-d2}
  0
  = \Pi_\kappa^\mu U^\nu R_{\mu\nu}
  &= -\frac{1}{2}\mathcal{F}^\rho{}_\kappa\left(\partial_\rho +\mathcal{A}_\rho \right)S+\frac{1}{2}\mathcal{G}^\rho{}_\kappa\left(\partial_\rho +\mathcal{A}_\rho \right)S
  \\
  &{}\qquad\nonumber 
  -\partial_r\beta\Pi^\alpha_\kappa\left(\partial_\alpha +\mathcal{A}_\alpha \right)S-\frac{1}{2}\Pi^\alpha_\kappa\left(\partial_\alpha+\mathcal{A}_\alpha\right)\left(\partial_r S-r^{-1} S\right)
  \\
  &{}\qquad\nonumber
  -\frac{1}{2}S\Pi^\mu_\kappa\hat D_\rho \mathcal{F}^\rho{}_\mu+\frac{1}{2}S\Pi^\rho_\kappa\left(
    \partial_r\mathcal{Z}_\rho+r^{-1}\mathcal{Z}_\rho
  \right)
  \\
  &{}\qquad\nonumber
  +\frac{1}{2}\left(\mathcal{L}_U-\bar{\mathcal{K}}\right)\mathcal{Z}_\kappa
  +\frac{1}{2}\Pi^\mu_\kappa \left(\partial_\mu-\mathcal{A}_\mu\right)\bar{\mathcal{K}}
  -\Pi^\mu_\kappa\hat D_\rho \mathcal{K}^{T\rho}{}_{\mu}\,.
\end{align}
Again using~\eqref{eq:partialrcalK}, we can also rewrite $U^\mu U^\nu R_{\mu\nu}=0$ as
\begin{align}
  0
  = U^\mu U^\nu R_{\mu\nu}
  &= \frac{1}{2}\hat D_\rho\left(\Pi^{\rho\sigma}\left(\partial_\sigma+\mathcal{A}_\sigma\right)S\right)-r^{-1}S\bar{\mathcal{K}}+\mathcal{L}_U\bar{\mathcal{K}}\label{eq:newformUUR}\\
  &{}\qquad\nonumber-\frac{1}{2}\partial_r\left(S\bar{\mathcal{K}}\right)-\frac{1}{2}\bar{\mathcal{K}}^2
  -\frac{1}{2}S\left(\hat D_\rho+\mathcal{A}_\rho\right)\mathcal{Z}^\rho-\frac{1}{2}\mathcal{Z}^\rho\left(\partial_\rho+\mathcal{A}_\rho\right)S
  \\
  &{}\qquad\nonumber
  +\frac{1}{4}S^2\mathcal{F}^2 -\frac{1}{2}S\mathcal{Z}_\mu\mathcal{Z}^\mu+\frac{1}{2}S\partial_r^2S+S\mathcal{L}_U\partial_r\beta
  +S^2\partial^2_r\beta
  \\
  &{}\qquad\nonumber
  +\frac{3}{2}S\partial_r S\partial_r\beta+S^2(\partial_r\beta)^2-\frac{3}{2}S\bar{\mathcal{K}}\partial_r\beta+\frac{1}{2}r^{-1}S^2\partial_r\beta
  \\
  &{}\qquad\nonumber
  -\mathcal{K}^{T\rho\sigma}\mathcal{K}^T_{\rho\sigma}
  \,.
\end{align}
The first line is order $r^{-1}$ and the rest is order $r^{-2}$. It could be interesting to see if differential Bianchi identities for the curvature tensor $\hat R_{\mu\nu\rho}{}^\sigma$ can help to simplify the $U^\mu U^\nu R_{\mu\nu}=0$ and $\Pi^\mu_\kappa U^\nu R_{\mu\nu}=0$ equations.

As we will show in Section~\ref{ssec:radial-expansion-bianchi},
except at order $r^{-d}$,
the $U^\mu U^\nu R_{\mu\nu}=0$ and $\Pi^\mu_\kappa U^\nu R_{\mu\nu}=0$ equations are a consequence of all the other Einstein equations
due to the bulk Bianchi identities.
Likewise, the only nontrivial part of the
$\Pi^{\mu\nu} R_{\mu\nu}=0$
trace equation is at order $r^{-2}$,
and the subsequent terms in its expansion are then automatically satisfied to all order in $1/r$.
For this reason, it is of limited interest to further simplify these equations at this point.

We will study the expansion of these $d=1$ and $d=2$ equations of motion starting in Section~\ref{sec:radial-expansion} below.
We will see that the expansion of the $R_{rr}=0$, $R_{\mu r}=0$, trace and STF equations is rather straightforward and not too much work (to the orders that we need them).
The expansion of $U^\mu U^\nu R_{\mu\nu}=0$ and $\Pi^\mu_\kappa U^\nu R_{\mu\nu}=0$ to their relevant orders will be worked out in Section \ref{sec:bulk-conservation-equations}.
There, we will see that the last two equations encode Carroll-covariant Bondi loss equations,
generalising the usual Bondi mass and angular momentum loss equations.

Finally, as a curiosity, we note that in all of the above Einstein equations we could replace $\hat D_\mu$ with $\bar D_\mu$, which is the covariant derivative with respect to the connection \eqref{eq:barCcon} that has torsion but is Carroll metric compatible.
The difference between these two tensors does not contribute in the above equations due to various projections with $\Pi^\mu_\nu$.
This is also true for the curvature scalar, as we have $\Pi^{\mu\nu}\hat R_{\mu\nu}=\Pi^{\mu\nu}\bar R_{\mu\nu}$ where $\bar R_{\mu\nu}$ is the Ricci tensor associated with \eqref{eq:barCcon}.

It would be interesting to find clearer geometrical interpretations for the building blocks used in these equations, or indeed to find a better set of building blocks that allows us to make more geometrical structures manifest.
For now, the usefulness of these buildings blocks is in their $1/r$ expansions,
as we will discuss next.

\section{Radial expansion of the Einstein equations}
\label{sec:radial-expansion}
Having rewritten the various projections of the bulk Einstein equations in terms of our Carroll-covariant Bondi--Sachs variables,
we now turn to the task of systematically solving their large $r$ expansion.
For this, we first briefly review the relevant elements of our geometric setup.
We then discuss how we can use the twice contracted differential Bianchi identities, $\nabla_M G^M{}_N=0$ where $G_{MN}$ is the Einstein tensor, to obtain a reduced set of independent equations of motion in Section~\ref{ssec:radial-expansion-bianchi}.
Next, we show how the $rr$, the $r\mu$ and the spatial projections of the $\mu\nu$ equations can be used to solve for most of the subleading components in the radial expansion of our metric variables.
In the special case of three bulk spacetime dimensions,
this expansion terminates,
as we show in Section~\ref{ssec:radial-expansion-threedim}.
Finally, we discuss the effect of subleading logarithmic corrections in Section~\ref{ssec:radial-expansion-logs}.

Recall from~\eqref{eq:car-cov-bondi-metric-repeat} that our Carroll-covariant Bondi--Sachs gauge-fixed parametrisation of the $(d+2)$-dimensional bulk metric is
\begin{equation}
  \label{eq:car-cov-bondi-metric-repeat-yet-again}
    ds^2
    = - 2 e^\beta \tau_\mu dr dx^\mu
    + \left(
      - e^{2\beta} S \tau_\mu \tau_\nu
      + \Pi_{\mu\nu}
    \right) dx^\mu dx^\nu\,,
\end{equation}
The radial expansions of these bulk variables are given in~\eqref{eq:car-cov-bondi-down-variables-expansion-repeat} and read
\begin{subequations}
  \label{eq:radial-expansion-metric-variables-repeat}
  \begin{align}
    \beta
    &= r^{-2} \os{2}{\beta}
    + \OO(r^{-3})\,,
    \\
    S
    &= r \os{-1}{S}
    + \os{0}{S}
    + \OO(r\inv)\,,
    \\
    \Pi_{\mu\nu}
    &= r^2 h_{\mu\nu}
    + r \os{-1}{\Pi}_{\mu\nu}
    + \OO(r\inv)\,,
  \end{align}
\end{subequations}
where we already used $\os{1}{\beta}=0$. 
The inverse objects $U^\mu$ and $\Pi^{\mu\nu}$ are expanded as in~\eqref{eq:car-cov-bondi-down-variables-expansion-inverses-repeat} which, in our gauge and after using the leading-order equations~\eqref{eq:car-cov-bondi-LO-eom-results}, is
\begin{subequations}
  \label{eq:radial-expansion-metric-variables-inverse-repeat}
  \begin{align}
    U^\mu
    &= v^\mu
    - r^{-1} 
       a^\mu   
    + \OO(r^{-2})\,,
    \\
    \Pi^{\mu\nu}
    &=
    r^{-2} h^{\mu\nu}
    - r^{-3} 
       h^{\mu\rho} h^{\nu\sigma} C_{\rho\sigma}
    + \OO(r^{-4})\,,
  \end{align}
\end{subequations}
where we remind the reader that on the boundary we raise and lower indices on spatial tensors such as $a_\mu$ with $h^{\mu\nu}$ and $h_{\mu\nu}$
as per our conventions in Appendix~\ref{sapp:spatial-tensors}.
The expansion of $(U^\mu,\Pi^{\mu\nu})$ follows from
the expansion of $(V_\mu,\Pi_{\mu\nu})$
using the orthonormality relations \eqref{eq:car-cov-bondi-orthogonality-completeness-repeat},
as further detailed in Appendix~\ref{sapp:on-shell-expansion-fundamental-variables}.
These relations also imply
$U^\mu U^\nu \Pi_{\mu\nu}=0$,
and this allows us to solve for
$v^\mu v^\nu \os{n}{\Pi}_{\mu\nu}$
for all orders $n\geq-1$.
Likewise, by expanding the Bondi determinant condition, 
we can solve for the trace $h^{\mu\nu} \os{n}{\Pi}_{\mu\nu}$ for any order $n\geq-1$.
At the first orders, this gives
\begin{equation}
  \label{eq:vvantrace-projection-of-Pi-at-NLO-and-NNLO}
 \begin{array}{ccccccc}
  v^\mu v^\nu \os{-1}{\Pi}_{\mu\nu}
  & = & 0\,,
  & \qquad &
  v^\mu v^\nu \os{0}{\Pi}_{\mu\nu}
  & = & a^\mu a_\mu\,,
  \\
  h^{\mu\nu} \os{-1}{\Pi}_{\mu\nu}
  & = & 0\,,
  & \qquad &
  h^{\mu\nu} \os{0}{\Pi}_{\mu\nu}
  & = & \frac{1}{2} C^{\mu\nu}C_{\mu\nu}\,.
  \end{array}
\end{equation}
In the remainder of this section we will solve for or constrain the remaining undetermined coefficients in the $1/r$ expansion,
which are
\begin{equation}
  \os{n+2}{\beta},
  \quad
  \os{n}{S}\,,
  \quad
  v^\rho h_\mu^\sigma \os{n}{\Pi}_{\rho\sigma}\,,
  \quad
  h_{\langle \mu}^\rho h_{\nu \rangle}^\sigma \os{n}{\Pi}_{\rho\sigma}\,,
  \qquad
  \text{for } n\geq0\,.
\end{equation}
We will do this to the order where the Bondi mass and angular momentum aspect appear, namely $n=d-1$.
We restrict our attention to $d=1,2$,
corresponding to three and four bulk spacetime dimensions.
We will then derive the generalisations of the Bondi loss equations to null infinity equipped with the most general allowed Carrollian geometry in Section~\ref{sec:bulk-conservation-equations}.

\subsection{Using the contracted Bianchi identities}
\label{ssec:radial-expansion-bianchi}
The components of the vacuum Einstein equations
$G_{MN} = R_{MN} - R g_{MN}/2=0$
are not all independent.
Instead, they are related by the Bianchi identities
\begin{equation}
  \label{eq:BianchiId}
  \nabla_M G^M{}_N=0\,.
\end{equation}
In particular, let us assume that
$R_{rr}$,
$R_{\mu r}$
and
$\Pi^\rho_{\langle \mu}\Pi^\sigma_{\nu\rangle}R_{\rho\sigma}$
are all zero.
The radial component of the Bianchi identities then tell us that 
\begin{equation}
    \partial_r\left(r^2\Pi^{\mu\nu}R_{\mu\nu}\right)
    =0\,.
\end{equation}
Expanding this identity tells us the trace $\Pi^{\mu\nu}R_{\mu\nu}$ is then automatically zero at all orders,
except possibly at the leading order $r^{-2}$.
Assuming that
$R_{rr}$,
$R_{\mu r}$
and
$\Pi^\rho_{\langle \mu}\Pi^\sigma_{\nu\rangle}R_{\rho\sigma}$
vanish,
this means
that we only need to compute the trace $\Pi^{\mu\nu}R_{\mu\nu}$ at the leading order $r^{-2}$
to be able to conclude that it is zero at all orders.

Next, once we have shown in this way that
$R_{r r}$,
$R_{\mu r}$
and
the full spatial projection
$\Pi_\mu^\rho \Pi_\nu^\sigma R_{\rho\sigma}$
all vanish,
the $\nu$ components of the Bianchi identity~\eqref{eq:BianchiId} imply
\begin{equation}\label{eq:radialevoW}
  \partial_r W_\nu
  -\Pi^{\rho\sigma}V_{\sigma\nu}W_\rho
  -V_\nu e^\beta E^{-1}\partial_\mu\left(E e^{-\beta}\Pi^{\mu\rho}W_\rho\right)
  =0\,,
\end{equation}
where we defined
$W_\mu = r^d e^\beta U^\rho R_{\rho\mu}$
for brevity.

Let us consider $d=2$.
Equations \eqref{eq:U-U-R-d1} and \eqref{eq:U-Pi-R-d1-2} tell us that $U^\rho R_{\rho\mu}=\mathcal{O}(r^{-1})$ and hence that $W_\mu=\mathcal{O}(r)$.
The second and third terms in~\eqref{eq:radialevoW} are always lower order in $1/r$ than $\partial_r W_\mu$ so this equation tells us that $W_\mu$ at order $r$ must be zero.
This tells us that $W_\mu$ is $\OO(r^0)$
but the equation does not tell us that $W_\mu$ at order~$r^0$ must be zero, so we need to impose this by hand.
We conclude that $W_\mu$~is guaranteed to vanish to all orders if it vanishes at order $r^0$.
A similar argument works in the case of $d=1$ too,
and we thus only need to show that
\begin{equation}
  \left.
  U^\rho R_{\rho\mu}
  \right|_{r^{-d}}
  =0
\end{equation}
to conclude that all components of the Einstein equations are satisfied.

To summarise, the Bianchi identities~\eqref{eq:BianchiId} ensure that the reduced set of equations below is equivalent to the full set of equations of motion,
\begin{subequations}
  \label{eq:EOMList2}
  \begin{gather}
    \label{eq:EOMList2a}
    R_{rr}
    =0\,,
    \qquad
    U^\mu R_{\mu r}
    =0\,,
    \qquad
    \Pi^\mu_\nu R_{\mu r}
    =0\,,
    \qquad\Pi^{\mu}_{\langle\kappa}\Pi^{\nu}_{\lambda\rangle} R_{\mu\nu}
    =0
    \,,
    \\
    \label{eq:EOMList2b}
    \left.
      \Pi^{\mu\nu}R_{\mu\nu}
    \right|_{r^{-2}}
    = 0\,,
    \qquad
    \left.
    U^\mu U^\nu R_{\mu\nu}
    \right|_{r^{-d}}
    =0\,.
    \qquad
    \left.
    \Pi^\mu_\kappa U^\nu R_{\mu\nu}
    \right|_{r^{-d}}
    =0\,.
  \end{gather}
\end{subequations}
We will now see how the first three equations in~\eqref{eq:EOMList2a} and the first equation in \eqref{eq:EOMList2b} can be used to solve for most of the metric data at subleading order.
Next, in Section~\ref{ssec:radial-expansion-stf}, we will show that the symmetric trace-free equation gives rise to an evolution equation.
The remaining two equations in~\eqref{eq:EOMList2b} will play the role of the Bondi loss equations,
and we will return to them later on in Section~\ref{sec:bulk-conservation-equations}.

\subsection{Solving the \texorpdfstring{$rr$}{rr} equation}
\label{ssec:radial-expansion-rr}
Recall that we introduced the following antisymmetric and symmetric trace-free (STF) spatial tensors
as part of the rewriting of the equations of motion in Section~\ref{sec:rewriting-EE},
\begin{subequations}
  \label{eq:curly-F-G-def-repeat}
  \begin{align}
    \mathcal{F}_{\mu\nu}
    &= \Pi_\mu^\rho \Pi_\nu^\sigma V_{\rho\sigma}
    = F_{\mu\nu}
    + \OO(r\inv)\,,
    \\
    \mathcal{G}_{\mu\nu}
    &= \Pi_\mu^\rho \Pi_\nu^\sigma \left(
      \pd_r \Pi_{\rho\sigma}
      - 2 r\inv \Pi_{\rho\sigma}
    \right)
    = - C_{\mu\nu}
    + \OO(r\inv)\,.
  \end{align}
\end{subequations}
These generalise the asymptotic twist and shear tensors to arbitrary $r$ hypersurfaces, as we also discussed around~\eqref{eq:null-congruence-data}.
In terms of these variables, we showed in~\eqref{eq:rr-eqCompObjForm} that the $rr$ component of the Ricci tensor is given by
\begin{equation}
  \label{eq:rr-eqCompObjForm-repeat}
  0
  = R_{rr}
  = d r\inv \pd_r \beta
  + \frac{1}{4} \mathcal{F}^{\mu\nu} \mathcal{F}_{\mu\nu}
  - \frac{1}{4} \mathcal{G}^{\mu\nu} \mathcal{G}_{\mu\nu}\,.
\end{equation}
As we already remarked in Section~\ref{ssec:rewriting-EE-3d-simplifications},
no spatial antisymmetric or STF tensors exist in three bulk spacetime dimensions (where $d=1$),
so in that case the $rr$ equation implies that the function $\beta$ vanishes entirely.
This vanishing of $\beta$ is analogous to the statement that the Bondi--Sachs gauge is equivalent to the Newman-Unti gauge in $d=1$.
On the other hand, in general dimensions,
expanding the $rr$ equation allows us to solve for $\beta$ order by order,
so that for example
\begin{equation}
  \label{eq:beta-up-to-order-2}
   \os{1}{\beta}
   = 0\,,
   \qquad
   \os{2}{\beta}
   = -\frac{1}{8d} C^{\mu\nu} C_{\mu\nu}
   + \frac{1}{8d} F^{\mu\nu} F_{\mu\nu}\,.
\end{equation}
Spatial indices are raised and lowered with $h^{\mu\nu}$ and $h_{\mu\nu}$
following our conventions in Appendix~\ref{sapp:spatial-tensors},
so that for example
$C^{\mu\nu} = h^{\mu\rho} h^{\nu\sigma} C_{\rho\sigma}$.
More generally, it is easy to see that
the vanishing of $R_{rr}$ at order $r^{-n}$ gives us an algebraic equation for $\overset{(n-2)}{\beta}$ for $n\geq3$
in terms of objects that appear at previous orders.
This equation therefore allows us to determine
the function $\beta(r,x)$ to all orders in a $1/r$ expansion. 

Finally, a comment on notation.
From now on, to keep expressions compact, we will denote fully  contracted products of spatial tensors as follows,
\begin{equation}
    X^2=X_{\mu\nu}X^{\mu\nu}\,,\qquad X\cdot Y=X_{\mu\nu}Y^{\mu\nu}\,.
\end{equation}
For example, we will write~\eqref{eq:beta-up-to-order-2} as
\begin{equation}
   \os{2}{\beta}
   = \frac{1}{8d}\left(F^2-C^2\right)\,.
\end{equation}
See Appendix~\ref{sapp:spatial-tensors} for more details. 

\subsection{Solving the \texorpdfstring{$\mu r$}{mu r} and trace equations}
\label{ssec:radial-expansion-rmu-trace}
Now let us turn to the two projections $U^\mu R_{\mu r}=0$ and $\Pi^\mu_\kappa R_{\mu r}=0$
of the $\mu r$ equation.
We obtained their hypersurface-covariant expressions in~\eqref{eq:URmur-v3}
and~\eqref{eq:PiRBestForm} as part of Section~\ref{ssec:rewriting-EE-mu-r-equation} above.
The resulting expressions are rather involved,
so we will not repeat them here.
We can solve them perturbatively as follows.

\paragraph{Solution strategy.}
Let us start with $U^\mu R_{\mu r}=0$ in~\eqref{eq:URmur-v3}.
Every term in this equation is order~$r^{-3}$.
At order $r^{-n-2}$,
this equation receives contributions from~$r^{-n}\os{n}{S}$ in the expansion of $S$.
The coefficient of $\os{n}{S}$ is $n(d-1-n)$ and so when $n\neq 0$ and $n\neq d-1$ we can solve \eqref{eq:URmur-v3} algebraically for $\os{n}{S}$.
When $d=1$, this means we cannot solve for~$\os{0}{S}$,
and we will later see that this coefficient is related to the Bondi mass for a three-dimensional bulk.
When $d=2$, the equation $U^\mu R_{\mu r}=0$ cannot provide us with either $\os{0}{S}$ or $\os{1}{S}$,
and we will see that the Bondi mass for $d=2$ is related to~$\os{1}{S}$.
On the other hand, we can solve for $\os{0}{S}$ algebraically using the trace equation~$\Pi^{\mu\nu}R_{\mu\nu}$ at order $r^{-2}$.
Indeed, from \eqref{eq:trace} it is easy to see that at order $r^{-2}$ the coefficient of $\os{0}{S}$ is proportional to $(d-1)$.

Next, consider the $\Pi^\mu_\kappa R_{\mu r}=0$ equation in \eqref{eq:PiRBestForm} at order $r^{-n-2}$.
This receives contributions from the composite variable $\mathcal{Z}_\mu$ defined in~\eqref{eq:calZ} at order $r^{-n-1}$.
The coefficient of $\os{n+1}{\mathcal{Z}}_\mu$ is proportional to $n+1-d$.
From the definition of $\mathcal{Z}_\mu$,
we can see%
\footnote{%
  Roughly speaking,
  we have $\mathcal{Z}_\mu=-U^\nu\partial_r\Pi_{\mu\nu}+\cdots$
  so that, at order $r^{-n-1}$, we get
  \begin{equation}
    \os{n+1}{\mathcal{Z}}_\mu=-2h_{\mu\nu}\os{n+2}{U}^\nu+nv^\nu\os{n}{\Pi}_{\mu\nu}+\cdots\,.
  \end{equation}
  Using the orthogonality relation $U^\nu \Pi_{\mu\nu}=0$,
  we furthermore get
  \begin{equation}
    h_{\mu\nu}\os{n+2}{U}^\nu=-v^\nu\os{n}{\Pi}_{\mu\nu}+\cdots\,,
  \end{equation}
  so that we see
  \begin{equation}
    \os{n+1}{\mathcal{Z}}_\mu=(n+2)v^\nu\os{n}{\Pi}_{\mu\nu}+\cdots\,.
  \end{equation}
  Since $n+2\neq 0$,
  we can always solve for $\os{n}{\Pi}_{\mu\nu}$ whenever $n\neq d-1$.
  In the above expressions, the dots are products of lower-order terms. The relation $U^\nu \Pi_{\mu\nu}=0$ allows us to solve for $v^\mu v^\nu\os{n}{\Pi}_{\mu\nu}$ so we are interested in the $h^\rho_\mu v^\sigma \os{n}{\Pi}_{\rho\sigma}$ part of $\os{n+1}{\mathcal{Z}}_\mu$.
}
that $\os{n+1}{\mathcal{Z}}_\mu$ contains terms proportional to $h^\rho_\mu v^\sigma\os{n}{\Pi}_{\rho\sigma}$.
In other words we can solve the $\Pi^\mu_\kappa R_{\mu r}=0$ equation at order $r^{-n-2}$ for 
$h^\rho_\mu v^\sigma\os{n}{\Pi}_{\rho\sigma}$ whenever $n\neq d-1$. We will see that $h^\rho_\mu v^\sigma\os{d-1}{\Pi}_{\rho\sigma}$ corresponds to the Bondi angular momentum.

We still have to consider what $U^\mu R_{\mu r}=0$ and $\Pi^\mu_\kappa R_{\mu r}=0$ tell us when $n=d-1$, i.e., at order $r^{-d-1}$.
For $d=1$, the equation $U^\mu R_{\mu r}=0$ (see \eqref{eq:UR-r-eq-d1}) is identically satisfied at order $r^{-2}$,
as we already concluded above.
We will see below that $U^\mu R_{\mu r}=0$ is identically satisfied for $d=2$,
at order $r^{-3}$, too.
For $d=1$, the equation $\Pi^\mu_\kappa R_{\mu r}=0$ will be identically satisfied at order $r^{-2}$.
However, as we will see, for $d=2$ the equation $\Pi^\mu_\kappa R_{\mu r}=0$ will not be identically satisfied at order $r^{-3}$.
This breaks the pattern so far,
and it is associated with the appearance of a log term in the radial expansion,
as we will discuss in Section~\ref{ssec:radial-expansion-logs}.
When including the appropriate log term, the equation $\Pi^\mu_\kappa R_{\mu r}=0$ can be algebraically solved at order $r^{-3}$ for the coefficient of the log term (whilst it is identically satisfied at the log order).

\paragraph{Explicit solutions for $d=2$.}
Now that we know what the $\mu r$ equations can be solved for, we can discuss the details.
We will only explicitly solve for the coefficients in the $1/r$ expansion to the order where the Bondi mass and angular momentum appear.
For $d=2$, this means we need to know $\os{0}{S}$ and $h^\rho_\mu v^\sigma\os{0}{\Pi}_{\rho\sigma}$.

Consider the trace equation \eqref{eq:trace} for general $d$.
At order $r^{-2}$, the terms that contribute are
\begin{align}\label{eq:traceorder-2}
   0 &=
   \left( \partial_r\bar{\mathcal{K}}
    +2dr^{-1}\bar{\mathcal{K}}
    -\frac{d}{2}r^{-1}\left(\partial_r S-r^{-1}S\right)
    +\Pi^{\mu\nu}\hat R_{\mu\nu}\right)\Big\vert_{@r^{-2}}\,.
\end{align}
In order to continue, we need to use the leading-order connection~$\mathcal{C}^\rho_{\mu\nu}$ that we obtained from the hypersurface connection~$\hat C^\rho_{\mu\nu}$ in~\eqref{eq:bdy-Ccon},
\begin{equation}
  \label{eq:calCcon}
  \begin{split}
    \mathcal{C}^\rho_{\mu\nu}
     = & -\frac{1}{2}v^\rho\left(\partial_\mu \tau_\nu+\partial_\nu \tau_\mu\right)
    -\frac{1}{2}v^\rho\left(\tau_\mu a_\nu+\tau_\nu a_\mu\right)\\
    &
    +\frac{1}{2}h^{\rho\sigma}\left(
      \partial_\mu h_{\nu\sigma}+\partial_\nu h_{\mu\sigma}-\partial_\sigma h_{\mu\nu}
    \right)\,.
  \end{split}
\end{equation}
Recall that we denote its associated covariant derivative and curvature by $\mathcal{D}_\mu$ and~$\mathcal{R}_{\mu\nu\rho}{}^\sigma$.
These objects obey the same properties as the unexpanded connection and its curvature, which we discussed in Sections~\ref{ssec:connection-choice} and~\ref{subsec:hypercurv},
and we only have to replace all tensors in the relevant expressions by their leading-order boundary values. See Appendix \ref{sapp:carroll-geometry-connection} for details.
At the boundary, we also have $K_{\mu\nu}=\frac{1}{d}Kh_{\mu\nu}$ due to the leading-order equations of motion,
which gives further simplifications.

Using the identity $\os{1}{\mathcal{K}}=\mathcal{D}_\mu a^\mu$ from~\eqref{eq:app-curly-K-expansion-d2-r-1},
Equation~\eqref{eq:traceorder-2} then gives us
\begin{equation}
\label{eq:TraceEqnO1Int}
  d(d-1)\overset{(0)}S
  =h^{\mu\nu}\mathcal{R}_{\mu\nu}
  +(2d-1)\mathcal{D}_\mu a^\mu\,.
\end{equation}
Just like we defined a $Q$-tensor in Section \ref{subsec:hypercurv} for the hypersurface connection $\hat C^\rho_{\mu\nu}$ we can define a similar object for the boundary connection $\mathcal{C}^\rho_{\mu\nu}$ (see Appendix \ref{app:curvten} for details).
The equivalent of~\eqref{eq:spatialRicscalar} for the boundary objects is
\begin{equation}
    h^{\mu\nu}\mathcal{R}_{\mu\nu}=\mathcal{Q}-\mathcal{D}_\mu a^\mu\,,
\end{equation}
where $\mathcal{Q}$ is defined in~\ref{eq:Qscalar-boundary-d=2}.
In terms of this $\mathcal{Q}$ scalar,
which encodes the single independent degree of freedom of the boundary curvature,
we thus get
\begin{equation}
  \label{eq:TraceRmunuO-2}
  d(d-1)\overset{(0)}S
  =\mathcal{Q}
  +2(d-1)\mathcal{D}_\mu a^\mu\,.
\end{equation}
In fact, it can be shown that $\mathcal{Q}$ is the Ricci scalar of the 2-dimensional Riemannian base manifold~\cite{Hartong:2025WIP2}.
For $d=2$, we thus obtain
\begin{equation}\label{eq:S0d=2}
    2\os{0}{S}=h^{\mu\nu}\mathcal{R}_{\mu\nu}+3\mathcal{D}_\mu a^\mu=\mathcal{Q}+2\mathcal{D}_\mu a^\mu\,.
\end{equation}

In a similar fashion, we can solve $\Pi^\mu_\kappa R_{\mu r}=0$ in~\eqref{eq:PiRBestForm} at order $r^{-2}$.
Keeping only the terms that are relevant at order $r^{-2}$, we find
\begin{align}
    0
    &= \left(-\frac{1}{2}\Pi^\mu_\kappa\partial_r \mathcal{Z}_\mu
    -\frac{d}{2}r^{-1}\mathcal{Z}_\kappa+\frac{1}{2}\Pi^\mu_\kappa\hat D_\rho \mathcal{G}^\rho{}_\mu\right.
    \\
    &{}\qquad\nonumber\left.
    -\frac{1}{2}\mathcal{A}_\rho\mathcal{G}^\rho{}_\kappa
    +\frac{1}{2}\Pi^\mu_\kappa\hat D_\rho \mathcal{F}^\rho{}_\mu
    +\frac{1}{2}\mathcal{A}_\rho\mathcal{F}^\rho{}_\kappa\right)\Big\vert_{@r^{-2}}\,.
\end{align} 
Each of these terms is individually order $r^{-2}$ and so we can simply replace each of them with their leading-order term.
This leads to an equation for~$\os{1}{\mathcal{Z}}_\kappa$ that reads
\begin{equation}
\label{eq:Z-eq}
\begin{split}
    (d-1) \os{1}{\mathcal{Z}}_\kappa
    &= - h^\mu_\kappa \mathcal{D}_\rho C^\rho{_\mu} + h^\mu_\kappa  \mathcal{D}_\rho  F^\rho{_\mu} + a_\rho C^\rho{_\kappa} + a_\rho F^\rho{_\kappa}\,.
\end{split}
\end{equation}
As we already saw in~\eqref{eq:calZ},
the terms in $\mathcal{Z}_\mu$ at order $r\inv$ are given by
\begin{equation}
\label{eq:Z-1}
    \os{1}{\mathcal{Z}}_\kappa =  2h^\nu_\kappa v^\rho \os{0}{\Pi}_{\nu\rho}+a^\rho F_{\rho\kappa}  - a^\rho C_{\rho\kappa}\,,
\end{equation}
so that we obtain an expression of the form
\begin{align}\label{eq:Pvg0}
    (d-1)h^{\rho}_\kappa v^\sigma \os{0}{\Pi}_{\rho\sigma} 
    &= -\frac{1}{2}h_\kappa^\rho\mathcal{D}{}_\sigma C^\sigma{}_\rho
    +\frac{1}{2}h_\kappa^\rho\mathcal{D}_\sigma F^\sigma{}_\rho
    \\
    &{}\qquad\nonumber
    +\frac{d}{2}a_\sigma C^\sigma{}_\kappa
    +\left(1-\frac{d}{2}\right)a_\sigma F^\sigma{}_\kappa\,.
\end{align}
Hence, when $d=2$, we can solve this algebraically for $h^{\rho}_\kappa v^\sigma \os{0}{\Pi}_{\rho\sigma}$.

\paragraph{Explicit solutions for $d=1$.}
In the case of three bulk dimensions,
both of the equations~\eqref{eq:TraceRmunuO-2} and~\eqref{eq:Pvg0}
vanish identically,
since in that case we have no spatial curvature ($\mathcal{Q}=0$)
and the spatial STF and antisymmetric tensors $C_{\mu\nu}$ and $F_{\mu\nu}$ both vanish.
As a result,
the equations of motion at this order do not constrain the radial expansion in any way.
In particular, we cannot determine $\os{0}{S}$ and $h^\mu_\rho v^\nu\os{0}{\Pi}_{\mu\nu}$ in terms of the boundary data in three dimensions.
Instead, these variables are free data in our radial expansion.
We will see in Section~\ref{ssec:ConservationEqnsd1} that they parametrise the Bondi mass aspect and the Bondi angular momentum aspect.

\paragraph{Possible constraints for $d=2$.}
Based on our analysis at the beginning of this subsection,
we expect that,
at order $r^{-3}$,
we should encounter $(d-2)\os{1}{S}$ in the $U^\mu R_{\mu r}=0$ equation
and $(d-2)h^\rho_\mu v^\sigma\os{1}{\Pi}_{\rho\sigma}$ in the $\Pi^\mu_\kappa R_{\mu r}=0$ equation.
This means that, for $d=2$, these equation cannot be solved for $\os{1}{S}$ and $h^\rho_\mu v^\sigma\os{1}{\Pi}_{\rho\sigma}$,
which are therefore free data.
We thus need to make sure if the corresponding equations are identically satisfied or if they lead to constraints on the other data.

Consider $U^\mu R_{\mu r}=0$ in~\eqref{eq:URmur-v3} which, at order $r^{-3}$,
becomes
\begin{align}
  \label{eq:UmuRmur-order-r-m3}
  0 & =  \frac{d-2}{2}\os{1}{S}-\os{2}{\mathcal{K}}+\left(2-\frac{6}{d}\right)K\os{2}{\beta}+2\mathcal{L}_v\os{2}{\beta}-\frac{1}{2d}KF^2
  \\
  &\qquad\nonumber
  +\frac{1}{2}\mathcal{D}_\mu\left(h^{\mu\nu}\os{1}{\mathcal{Z}}_\nu\right)+\frac{1}{2}a^\mu\os{1}{\mathcal{Z}}_\mu-\frac{1}{2}C^{\mu\nu}\os{-1}{\mathcal{K}}^T_{\mu\nu}\,,
\end{align}
where $\os{2}{\beta}$ is given in \eqref{eq:beta-up-to-order-2} and $\os{1}{\mathcal{Z}}_\mu$ is given in \eqref{eq:Z-1}.
It is now straightforward to compute the subleading orders $\os{2}{\mathcal{K}}$ as well as $\os{-1}{\mathcal{K}}^T_{\mu\nu}$ of these composite objects from the on-shell expansion,
and we summarise such results in Appendix~\ref{eq:app-car-cov-S-beta-Pi-dd-expansion}.
Specifically,
in~\eqref{eq:app-curly-KT-expansion-d2-r1}
and~\eqref{eq:app-curly-K-expansion-d2-r-2},
we obtain
\begin{align}
    \os{-1}{\mathcal{K}}^T_{\mu\nu}
    &=  -\frac{1}{2}\mathcal{L}_vC_{\mu\nu}-\frac{1}{2}KC_{\mu\nu}+A_{\mu\nu}\,,
    \\
    \os{2}{\mathcal{K}}
    &=  \frac{1}{4}\mathcal{L}_v F^2-K\os{2}{\beta}-\frac{1}{2}KF^2-\frac{1}{2}a^\rho\mathcal{D}_\mu C^\mu{}_\rho
    \\
    &{}\qquad\nonumber
    +\frac{1}{2}a^\rho\mathcal{D}_\mu F^\mu{}_\rho-\frac{1}{2}C^{\mu\rho}A_{\mu\rho}+\frac{1}{2}\mathcal{D}_\mu\left(h^{\mu\nu}\os{1}{\mathcal{Z}}_\nu\right)\,,
\end{align}
where it proves convenient to define the tensor
\begin{equation}
  \label{eq:A-first-def}
    A_{\mu\nu}
    =h^\rho_{\langle\mu}h^\sigma_{\nu\rangle}
    \left(\mathcal{D}_\rho a_\sigma+a_\rho a_\sigma\right)
\end{equation}
following~\eqref{eq:app-Amunu-def}.
In the end, we find that~\eqref{eq:UmuRmur-order-r-m3} becomes
\begin{align}
  \label{eq:UmuRmur-order-r-m3-final}
  0
  &=\frac{1}{2}(d-2)\os{1}{S}
  +\frac{3}{8d^2}(d-2)K\left(F^2-C^2\right)
  +\frac{d-2}{8d}\mathcal{L}_v\left(C^2-F^2\right)
  \\
  &{}\qquad\nonumber
  +h^{\nu\beta}a_\nu\left[
    v^\alpha \os{0}{\Pi}_{\alpha\beta}
    -a_\rho C^\rho{}_{\beta}+\frac{1}{4}a_\rho F^\rho{}_\beta
    + \frac{1}{2}\mathcal{D}{}_\rho C^\rho{}_\beta-\frac{1}{2}\mathcal{D}{}_\rho F^\rho{}_\beta
  \right]\,.
\end{align}
Indeed, the first line (and in particular the coefficient of $\os{1}{S}$) vanishes for $d=2$.
Using the on-shell result in~\eqref{eq:Pvg0},
we see that the second line also vanishes in that case.
Hence, we conclude that this equation is identically satisfied for $d=2$.

Next, let us consider $\Pi^\mu_\kappa R_{\mu r}=0$ at order $r^{-3}$.
Since computing at such subleading orders gets quite involved,
and since we are only interested in the $d=2$ case,
we will use $d=2$ from the beginning.
If we consider~\eqref{eq:PiR-r-eq-d2} we see that,
with the exception of the last two terms on the first line,
all terms are individually order $r^{-2}$.
We therefore need to expand most of the terms to the next order,
which makes this calculation a fair bit more involved than the previous calculations.
Because we already solved the equation at order $r^{-2}$,
and because we get zero if we contract it with $U^\kappa$,
the $v^\kappa$ projection of the equation at order $r^{-3}$ will vanish on account of results we already obtained.
We can thus contract~\eqref{eq:PiR-r-eq-d2} with $h^\kappa_\alpha$ before we expand it.
A quick calculation shows that the first two terms in that equation, which are
$-\frac{1}{2}\Pi^\mu_\kappa\partial_r \mathcal{Z}_\mu
  -r^{-1}\mathcal{Z}_\kappa$,
do not contribute at order $r^{-3}$
when contracted with $h^\kappa_\alpha$.
Hence, we find that
\begin{align}
  0
  &= 
  \frac{1}{2}F^\rho{}_\alpha \os{1}{\mathcal{Z}}_\rho+2
  h^\mu_\alpha \partial_\mu\os{2}{\beta}
  \\
  &{}\qquad\nonumber
  +\frac{1}{2}h^\kappa_\alpha\left(\Pi^\mu_\kappa\hat D_\rho \mathcal{G}^\rho{}_\mu
  -\mathcal{A}_\rho\mathcal{G}^\rho{}_\kappa
  +\Pi^\mu_\kappa\hat D_\rho \mathcal{F}^\rho{}_\mu
  +\mathcal{A}_\rho\mathcal{F}^\rho{}_\kappa\right)\Big\vert_{@r^{-3}}\,,
\end{align}
where $\os{1}{\mathcal{Z}}_\rho$ is given in \eqref{eq:Z-1}.
To make progress,
it is useful to note that
\begin{equation}
    h^\kappa_\alpha\Pi^\mu_\kappa\hat D_\rho\mathcal{G}^\rho{}_\mu\Big\vert_{@r^{-3}}=h^\mu_\alpha\mathcal{D}_\rho\mathcal{G}^\rho{}_\mu\Big\vert_{@r^{-3}}+h^\mu_\alpha C^\rho{}_\sigma\hat C^\sigma_{\rho\mu}\Big\vert_{@r^{-1}}\,.
\end{equation}
There is a similar expression for $h^\kappa_\alpha\Pi^\mu_\kappa\hat D_\rho\mathcal{F}^\rho{}_\mu$ at order $r^{-3}$.
It is now simply a matter of working out the expansions of various objects.
For example, we have
\begin{align}
  \label{eq:C1}
  \hat C^\rho_{\mu\nu}\Big\vert_{r^{-1}}
  &= -\frac{1}{2}v^\rho\left(\tau_\mu\os{1}{\mathcal{A}}_\nu+\tau_\nu\os{1}{\mathcal{A}}_\mu\right)
  \\
  &{}\qquad\nonumber
  +\frac{1}{2}h^{\rho\sigma}\left(\mathcal{D}_\mu\os{-1}{\Pi}_{\nu\sigma}+\mathcal{D}_\nu\os{-1}{\Pi}_{\mu\sigma}-\mathcal{D}_\sigma\os{-1}{\Pi}_{\mu\nu}\right)\,.
\end{align}
The end result for $\Pi^\mu_\kappa R_{r\mu}=0$ for $d=2$ at order $r^{-3}$ can be written as 
\begin{equation}\label{eq:PDZ}
    h^{\mu\nu}\mathcal{D}_\mu D_{\nu\kappa}=0\,,
\end{equation}
using also the tensor identity in~\eqref{eq:app-d2-spatial-STF-STF-product},
and where we defined
\begin{equation}
\label{eq:Dmunu-def}
    D_{\mu\nu}=h^{\rho}_\mu h^\sigma_\nu\left(\os{0}{\Pi}_{\rho\sigma}-\frac{1}{4}C^2h_{\rho\sigma}\right)-\frac{1}{2}F_{\mu\sigma} C^\sigma{}_\nu\,.
\end{equation}
The Bondi--Sachs gauge condition ensures that this $D_{\mu\nu}$ tensor is traceless.
It is manifestly spatial and symmetric,
and so $D_{\mu\nu}$ is an STF tensor that obeys \eqref{eq:PDZ}.
To see that $F_{\mu\sigma} C^\sigma{}_\nu$ is STF, we can use the identity~\eqref{eq:app-d2-spatial-STF-antisym-product}.

We thus see that $\Pi^\mu_\kappa R_{\mu r}=0$ is not identically satisfied
at order $r^{-3}$ for $d=2$.
Instead, it imposes a constraint on the part of $\os{0}{\Pi}_{\mu\nu}$ parametrised by the tensor~\eqref{eq:Dmunu-def}.
This breaks the pattern that $\Pi^\mu_\kappa R_{\mu r}=0$ is either identically satisfied or allows us to solve for a coefficient in the metric expansion.
Following~\cite{Barnich:2010eb},
this pattern can be restored by introducing a $r^{-1}\log r$ term to the expansion of $\Pi_{\mu\nu}$,
as we will discuss in Section \ref{ssec:radial-expansion-logs}.
With that addition, we will see that $\Pi^\mu_\kappa R_{\mu r}=0$ is identically satisfied at order $r^{-3}\log r$
and at order $r^{-3}$ allows us to solve for the coefficient of the log term.

\subsection{Solving the symmetric trace-free equation}
\label{ssec:radial-expansion-stf}
So far, we have seen that the $rr$, $\mu r$ and trace equations allowed us to solve for all terms in the expansion of $\beta$,
as well as most of the terms in the expansion of~$S$
and~$h^\mu_\kappa v^\nu\Pi_{\mu\nu}$,
except for those at order $r^{-d+1}$, which are free.
Our next goal will be to understand what the Einstein equations have to say about the spatial symmetric trace-free (STF) terms $h^\mu_{\langle\rho}h^\nu_{\sigma\rangle}\Pi_{\mu\nu}$
by studying the STF equation of motion \eqref{eq:STF}.
This equation simplifies considerably for $d=2$, where it reduces to~\eqref{eq:STFnewd=2}. 

We already saw in Section~\ref{ssec:bulk-geom-LO-EE-in-partial-gauge} above
that the STF equation at its leading order~$r$ implies
the boundary extrinsic curvature $K_{\mu\nu} = - \frac{1}{2} \LL_v h_{\mu\nu}$ is pure trace,
\begin{equation}
  \label{eq:constraintK-repeat}
  K^T_{\mu\nu}
  = K_{\mu\nu}-\frac{1}{d}Kh_{\mu\nu}
  = 0\,.
\end{equation}
This tells us that $\mathcal{K}^T_{\mu\nu}=\mathcal{O}(r)$.
The condition \eqref{eq:constraintK-repeat} is only non-trivial for $d\ge 2$.
In fact, the 
STF equation is trivial in three bulk dimensions,
since no spatial STF tensors exist for $d=1$.
In general dimensions, however,
we will see that the STF equation dictates the time evolution along $v^\mu$ of the $h^\mu_{\langle\rho}h^\nu_{\sigma\rangle}\os{n}{\Pi}_{\mu\nu}$ components.

We will study the $1/r$ expansion of the STF equation only for $d=2$. 
If we consider \eqref{eq:STFnewd=2} at order $r^{-1}$ we get
\begin{align}
  0
  & =  -2h^\mu_{\langle\kappa}h^\nu_{\lambda\rangle}
  \os{0}{\mathcal{K}}^T_{\mu\nu}
  -F^\mu{}_{(\kappa}\os{-1}{\mathcal{K}}^T_{\lambda)\mu}
  +\frac{1}{2}Kh^{\mu}_{\langle\kappa}h^\nu_{\lambda\rangle}
  \os{1}{\mathcal{G}}_{\mu\nu}\\
  &\qquad\nonumber
  -\frac{1}{2}\os{1}{\mathcal{K}}C_{\kappa\lambda}
  +h^\mu_{\langle\kappa}h^\nu_{\lambda\rangle}
  \mathcal{D}_\mu\os{1}{\mathcal{Z}}_\nu\,.
\end{align}
In here,
expanding the definitions in for example~\eqref{eq:app-curly-G-def} and~\eqref{eq:app-KT-BarCalK-def},
we have
\begin{subequations}
\begin{align}
    h^{\mu}_{\langle\kappa}h^\nu_{\lambda\rangle}\os{1}{\mathcal{G}}_{\mu\nu} & =  -2h^{\mu}_{\langle\kappa}h^\nu_{\lambda\rangle}\os{0}{\Pi}_{\mu\nu}\,,\\
    h^\mu_{\langle\kappa}h^\nu_{\lambda\rangle}\os{0}{\mathcal{K}}^T_{\mu\nu} & =  h^\mu_{\langle\kappa}h^\nu_{\lambda\rangle}\os{0}{\mathcal{K}}_{\mu\nu}-\frac{1}{2}Kh^{\mu}_{\langle\kappa}h^\nu_{\lambda\rangle}\os{0}{\Pi}_{\mu\nu}-\frac{1}{2}\os{1}{\mathcal{K}}C_{\kappa\lambda}\,.
\end{align}
\end{subequations}
This leads to
\begin{eqnarray}
    0  =  -2h^\mu_{\langle\kappa}h^\nu_{\lambda\rangle}\os{0}{\mathcal{K}}_{\mu\nu}-F^\mu{}_{(\kappa}\os{-1}{\mathcal{K}}^T_{\lambda)\mu}+\frac{1}{2}\os{1}{\mathcal{K}}C_{\kappa\lambda}+h^\mu_{\langle\kappa}h^\nu_{\lambda\rangle}\mathcal{D}_\mu\os{1}{\mathcal{Z}}_\nu\,,\label{eq:STFr-1}
\end{eqnarray}
where 
\begin{equation}
    \os{0}{\mathcal{K}}_{\mu\nu}=-\frac{1}{2}\mathcal{L}_v\os{0}{\Pi}_{\mu\nu}+\frac{1}{2}\mathcal{L}_a\os{-1}{\Pi}_{\mu\nu}-\frac{1}{2}\mathcal{L}_{\os{2}{U}}h_{\mu\nu}\,,
\end{equation}
so that 
\begin{equation}
\begin{split}
    -2h^\mu_{\langle\kappa}h^\nu_{\lambda\rangle}\os{0}{\mathcal{K}}_{\mu\nu} & =  \mathcal{L}_v\left(h^\mu_{\langle\kappa}h^\nu_{\lambda\rangle}\os{0}{\Pi}_{\mu\nu}\right)-2h^\mu_{\langle\kappa}h^\nu_{\lambda\rangle}a_\nu v^\rho\os{0}{\Pi}_{\mu\rho}\\
    &\qquad-h^\mu_{\langle\kappa}h^\nu_{\lambda\rangle}\mathcal{L}_a C_{\mu\nu}+2a_\rho F^\rho{}_{(\kappa}a_{\lambda)}+2h^\mu_{\langle\kappa}h^\nu_{\lambda\rangle}\mathcal{D}_\mu \left(h_{\rho\nu}\os{2}{U}^\rho\right)\,.\label{eq:calKT0}
\end{split}
\end{equation}
From the definition of $\mathcal{Z}_\mu$ in~\eqref{eq:app-calZ}, we learn that
\begin{equation}
    \os{1}{\mathcal{Z}}_\mu=-2h_{\mu\rho}\os{2}{U}^\rho+a_\rho C^\rho{}_\mu+a_\rho F^\rho{}_\mu\,.
\end{equation}
We use this to solve for $h_{\mu\rho}\os{2}{U}^\rho$ and substitute the result into \eqref{eq:calKT0}.

Next, we use the on-shell result~\eqref{eq:Pvg0} for $d=2$ and express the $\mathcal{L}_a C_{\mu\nu}$ Lie derivative in terms of covariant derivatives.
To make progress, it is useful to use
\begin{align}
    h^\mu_{\langle\kappa}h^\nu_{\lambda\rangle} F^\rho{}_\nu\mathcal{D}_\mu a_\rho & =  h^\mu_{\langle\kappa}h^\nu_{\lambda\rangle} F^\rho{}_\nu\left(\frac{1}{2}\mathcal{L}_v F_{\mu\rho}+A_{\mu\rho}-a_\mu a_\rho+\frac{1}{2}h_{\mu\rho}\mathcal{D}_\sigma a^\sigma\right)\nonumber\\
    & =  F^\rho{}_\lambda A_{\kappa\rho}-a_\rho F^\rho{}_{(\kappa}a_{\lambda)}\,,\\
    h^\mu_{\langle\kappa}h^\nu_{\lambda\rangle} C^\rho{}_\nu\mathcal{D}_\mu a_\rho & =  h^\mu_{\langle\kappa}h^\nu_{\lambda\rangle} C^\rho{}_\nu\left(\frac{1}{2}\mathcal{L}_v F_{\mu\rho}+A_{\mu\rho}-a_\mu a_\rho+\frac{1}{2}h_{\mu\rho}\mathcal{D}_\sigma a^\sigma\right)\nonumber\\
    & =  \frac{1}{2}C^\rho{}_\lambda\mathcal{L}_v F_{\kappa\rho}+\frac{1}{2}C_{\kappa\lambda}\mathcal{D}_\sigma a^\sigma-h^\mu_{\langle\kappa}h^\nu_{\lambda\rangle}a_\mu a_\rho C^\rho{}_\nu\,,\label{eq:CDa}
\end{align}
where we used the identities in~\eqref{eq:app-d2-spatial-STF-and-antisym-product-rules}
for the multiplication of (anti)symmetric $d=2$ spatial tensors
as well as the decomposition of~$\mathcal{D}_\mu a_\nu$ from~\eqref{eq:app-boundary-cov-der-of-a-decomposition},
which parametrises its STF part using the field $A_{\mu\nu}$ we introduced in~\eqref{eq:A-first-def},
\begin{equation}
  A_{\mu\nu}
  =h^\rho_{\langle\mu}h^\sigma_{\nu\rangle}\left(\mathcal{D}_{\rho}a_{\sigma}+a_\rho a_\sigma\right).
\end{equation}
After some rewriting, this allows us to express the identity~\eqref{eq:calKT0} as
\begin{align}
  \label{eq:calKT02}
  -2h^\mu_{\langle\kappa}h^\nu_{\lambda\rangle}\os{0}{\mathcal{K}}_{\mu\nu}
  & =  -h^\mu_{\langle\kappa}h^\nu_{\lambda\rangle}\mathcal{D}_\mu\os{1}{\mathcal{Z}}_\nu-\frac{1}{2}C_{\kappa\lambda}\os{1}{\mathcal{K}}+F^\rho{}_{(\kappa}\os{-1}{\mathcal{K}}^T_{\lambda)\rho}
  \\
  &\qquad\nonumber
  +\mathcal{L}_v\left(h^\mu_{\langle\kappa}h^\nu_{\lambda\rangle}\os{0}{\Pi}_{\mu\nu}-\frac{1}{2}C_\kappa{}^\rho F_{\lambda\rho}\right)
  \\
  &\qquad\nonumber
  +h^\mu_{\langle\kappa}h^\nu_{\lambda\rangle}a^\rho\left(\mathcal{D}_\mu C_{\rho\nu}-\mathcal{D}_\rho C_{\mu\nu}+h_{\rho\nu}h^{\alpha\beta}\mathcal{D}_\alpha C_{\beta\mu}\right)
  \\
  &\qquad\nonumber
  +h^\mu_{\langle\kappa}h^\nu_{\lambda\rangle}\left[a^\rho\mathcal{D}_\mu F_{\rho\nu}-a_\nu\mathcal{D}_\sigma F^\sigma{}_\mu+a_\nu a_\rho F^\rho{}_\mu\right]\,.
\end{align}
Finally, we apply the $d=2$ identities \eqref{eq:DCident} and \eqref{eq:calDFid} which make the last two lines vanish.
The STF equation at order $r^{-1}$, corresponding to~\eqref{eq:STFr-1}, can then be seen to reduce simply to 
\begin{equation}
  \label{eq:STF1/rSimple}
  \mathcal{L}_v D_{\mu\nu}=0\,,
\end{equation}
where we used the $D_{\mu\nu}$ tensor defined in~\eqref{eq:Dmunu-def}.

Recalling this definition of $D_{\mu\nu}$ above,
we see that this equation gives a constraint on the evolution of the STF part of $\os{0}{\Pi}_{\mu\nu}$.

From the all-orders expression of the STF equation of motion in~\eqref{eq:STFnewd=2} it is easy to see that a similar structure arises at higher orders.
Equation~\eqref{eq:STFnewd=2} contains radial derivatives of $\mathcal{K}^T_{\mu\nu}$ and so we can solve it algebraically for coefficients of $\mathcal{K}^T_{\mu\nu}$.
These solutions will always contain a term of the form $\mathcal{L}_v \left(
    h^\mu_{\langle\rho}h^\nu_{\sigma\rangle}\overset{(n)}\Pi_{\mu\nu}
  \right)$. We conclude that 
at order $r^{-n-1}$ in the expansion of the STF equation, we will obtain an equation of the form
\begin{equation}
  \label{eq:further-subleading-stf-lie-derivatives}
  \mathcal{L}_v \left(
    h^\mu_{\langle\rho}h^\nu_{\sigma\rangle}\overset{(n)}\Pi_{\mu\nu}
  \right)
  = T^{(hh\Pi,n)}_{\rho\sigma}\,,
\end{equation}
where the tensorial expression on the right-hand side is comprised of products of objects that appear at lower orders in the radial expansion.
Equations of this form therefore give first-order differential equations for the evolution along  $v^\mu$ of the symmetric trace-free parts of all subleading orders of the spatial metric.

For $d=2$, the STF equation \eqref{eq:STFnewd=2} at its leading order~$r$ puts a constraint on the boundary geometry.
At the next order, it is identically satisfied, which is related to the appearance of the arbitrary shear tensor.
Then, once we get to order~$r^{-1}$ and beyond,
we find new initial data in the expansion of 
$h^\mu_{\langle\rho}h^\nu_{\sigma\rangle}\Pi_{\mu\nu}$
at each order in the $1/r$ expansion.

It is this aspect of the Einstein equations with zero cosmological constant that sets the solutions apart from asymptotically locally AdS solutions,
where the Fefferman--Graham theorem
tells us that the near-boundary expansion is entirely fixed
once we specify the boundary metric and the boundary energy-momentum tensor~\cite{FeffermanGraham,Graham:1999jg,Balasubramanian:1999re,deHaro:2000vlm, Skenderis:2002wp}.
One of the main motivations of this project is to examine to what extent this is still true for asymptotically flat spacetimes.

What we have seen is that for the variables
$\beta$, $S$ and $h^{\rho}_\mu v^\sigma\Pi_{\rho\sigma}$,
the radial expansion is AdS-like in the sense that we can solve algebraically for the coefficients in the expansion,
with the exception of $\os{d-1}{S}$ and $h^{\rho}_\mu v^\sigma\os{d-1}{\Pi}_{\rho\sigma}$,
which we will relate to components of a boundary energy-momentum tensor in Section~\ref{sec:bulk-conservation-equations}.
On the other hand, for 
$h^\mu_{\langle\rho}h^\nu_{\sigma\rangle}\Pi_{\mu\nu}$, the expansion is not AdS-like.
Instead, as is well-known, it takes the form of an initial-value problem,
where we need to specify 
$h^\mu_{\langle\rho}h^\nu_{\sigma\rangle}\Pi_{\mu\nu}$
everywhere along some cut of $\mathcal{I}^+$.
In standard Bondi coordinates an example of a cut of $\mathcal{I}^+$ is to take $u=\text{cst}$,
and a Carroll-covariant definition of a cut will be discussed in~\cite{Hartong:2025WIP2}.
We thus see that,
given initial data for $h^\mu_{\langle\rho}h^\nu_{\sigma\rangle}\Pi_{\mu\nu}(r,x)$ along some cut of~$\mathcal{I}^+$,
the near-boundary expansion of the metric is fully determined in terms of the boundary metric, the shear and the Bondi mass and angular momentum aspect (which we define in this language in Section~\ref{sec:bulk-conservation-equations} below).

The radial expansion of the initial data of 
$h^\mu_{\langle\rho}h^\nu_{\sigma\rangle}\Pi_{\mu\nu}(r,x)$ is an important open problem. 
In four-dimensional asymptotically flat spacetimes, polyhomogeneous expansions near null infinity have been studied in~\cite{winicour_1985,Chrusciel:1993hx,ValienteKroon:2002gb}, and recent work has provided further evidence that smooth null infinity is generically too restrictive~\cite{Kehrberger:2021uvf,Kehrberger:2021vhp,Kehrberger:2024clh,Kadar:2025xmo}.
In higher dimensions, the situation is even more subtle: in higher odd bulk spacetime dimensions, smooth null infinity is obstructed by radiation, and half-integer powers generally appear in Bondi-type expansions, and so it is no longer sufficient to assume that the ansatz only contains integer powers of $1/r$ and logs~\cite{Hollands:2003ie,Hollands:2004ac,Tanabe:2011es,Godazgar:2012zq}.
We briefly discuss the effects of adding certain subleading log terms
in Section~\ref{ssec:radial-expansion-logs} below.

\subsection{Three-dimensional bulk spacetimes}
\label{ssec:radial-expansion-threedim}
For $d=1$, which corresponds to three bulk spacetime dimensions,
several significant simplifications occur.
First, as we already saw in Section~\ref{ssec:rewriting-EE-3d-simplifications},
the $rr$ equation of motion implies that the function $\beta(r,x)$ vanishes entirely.
Additionally, since no spatial STF or antisymmetric tensors exist in $d=1$,
we also saw that the remaining equations of motion simplify significantly.
In fact, as we will show in more detail later on in this subsection,
it turns out that
\begin{equation}
  \label{eq:three-dim-radial-expansion-terminates-preview}
  \pd_r^3 g_{\mu\nu}
  = 0\,,
\end{equation}
so that the radial expansion of the metric truncates at order $r^0$.
Together with $\beta=0$, the results from the leading-order equations of motion in~\eqref{eq:car-cov-bondi-LO-eom-results} as well as the expansions~\eqref{eq:car-cov-bondi-down-variables-expansion-repeat}
almost completely fix the total metric,
which now reads
\begin{align}
  ds^2
  &= - 2\tau_\mu dr dx^\mu
  + \left(- S\tau_\mu \tau_\nu + \Pi_{\mu\nu}\right) dx^\mu dx^\nu\nonumber
  \\
  \label{eq:3Dsol}
  &= - 2\tau_\mu dr dx^\mu
  + r^2 h_{\mu\nu}dx^\mu dx^\nu
  -2r \left(K \tau_\mu \tau_\nu + \tau_{(\mu} a_{\nu)}\right) dx^\mu dx^\nu
  \\
  &{}\qquad\nonumber
  - \left(\os{0}{S} - a^2\right)\tau_\mu \tau_\nu dx^\mu dx^\nu
  - 2 \tau_{(\mu} v^\rho h^\sigma_{\nu)} \os{0}{\Pi}_{\rho\sigma} dx^\mu dx^\nu\,.
\end{align}
The individual variables $S$ and $\Pi_{\mu\nu}$ may have further subleading terms beyond order $r^0$,
but the result~\eqref{eq:three-dim-radial-expansion-terminates-preview} tells us that they must cancel out in the $g_{\mu\nu}$ components of the metric.
As we will see in Section~\ref{ssec:bulk-improvements-weyl-3d}, the combination $\os{0}{S}-a^2$ parametrises a Carroll-covariant generalisation of the Bondi mass.
Likewise, let us introduce the notation for the spatial vector which will turn out to parametrise the equivalent of the Bondi momentum current,
\begin{equation}
  \label{eq:3Dsol-momentum-current-def}
  P_\mu
  = v^\rho h_\mu^\sigma \os{0}{\Pi}_{\rho\sigma}\,.
\end{equation}
The fact that the expansion of the metric in~\eqref{eq:3Dsol} terminates allows us to use a shortcut to determine the full expansion of our individual variables.
Computing the inverse metric $g^{MN}$ gives
\begin{subequations}
  \label{eq:3Dsol-inverse-components}
  \begin{align}
    \label{eq:3Dsol-inverse-components-S}
    S
    &= g^{rr}
    = 2rK
    + \os{0}{S}
    + 2r\inv a^\mu P_\mu
    + r^{-2} P^2
    \\
    \label{eq:3Dsol-inverse-components-U}
    U^\mu
    &= g^{r \mu}
    = v^\mu - r\inv a^\mu - r^{-2} P^\mu\,,
    \\
    \label{eq:3Dsol-inverse-components-Pi}
    \Pi^{\mu\nu}
    &= g^{\mu\nu}
    = r^{-2} h^{\mu\nu}\,,
  \end{align}
\end{subequations}
where $P^2= h^{\mu\nu} P_\mu P_\nu$ as per our conventions for spatial tensors discussed in Appendix~\ref{sapp:spatial-tensors}.
The form of the total metric in~\eqref{eq:3Dsol} then allows us to deduce that
\begin{equation}
  \label{eq:3Dsol-components-Pi}
  \Pi_{\mu\nu}
  =  r^2 h_{\mu\nu}
  - 2r \tau_{(\mu}a_{\nu)}
  + a^2 \tau_\mu \tau_\mu
  - 2 \tau_{(\mu}P_{\nu)}
  + 2r\inv a^\rho P_\rho \tau_\mu \tau_\nu
  + r^{-2} P^2 \tau_\mu \tau_\nu\,.
\end{equation}
This determines the full on-shell radial expansion of the three-dimensional metric in Carroll-covariant Bondi gauge.
Note that also the inverse metric terminates.

\paragraph{Comparison to standard Bondi--Sachs gauge.}
For now, we can use the dictionary in Section~\ref{ssec:reduction-to-standard-Bondi--Sachs-gauge} to compare~\eqref{eq:3Dsol} to the standard three-dimensional Bondi--Sachs result in~\cite{Barnich:2010eb}.
After using the boundary diffeomorphisms, Weyl transformations and local Carroll boosts to gauge fix $v^\mu \pd_\mu = - \pd_u$ and $\tau_\mu dx^\mu = du$,
so that in particular $a_\mu = \LL_v \tau_\mu = 0$,
and using
$h_{\mu\nu} dx^\mu dx^\nu = r^2 e^{2\vphi} d\phi^2$
as the boundary spatial metric
so $K = \pd_u \vphi$,
we find that the general solution~\eqref{eq:3Dsol} reduces to%
\footnote{%
  Additionally, note that the solution for $S$ in~\eqref{eq:3Dsol-inverse-components-S} above is consistent with the corresponding expression obtained in~\cite{Barnich:2010eb},
  where $S=-V/r$ contains terms at order $r^{-2}$ which are related to the momentum current.
}
\begin{equation}
  ds^2
  = -2 du dr
  - \left(2r\pd_u \vphi + \os{0}{S}\right) du^2
  + r^2 e^{2\vphi} d\phi^2
  - 2 P_\phi du d\phi\,.
\end{equation}
This expression is to be compared to the three-dimensional metric found by Barnich and Troessaert in~\cite{Barnich:2010eb} after solving the $rr$, $ru$ and $r\phi$ equations of motion,
which reads
\begin{equation}
  ds^2
  = -2dudr
  + \left(-2r \pd_u \vphi + M(u,\phi)\right) du^2
  + r^2 e^{2\vphi} d\phi^2
  + 2N(u,\phi) du d\phi\,.
\end{equation}
In that setting, the Bianchi identities likewise imply that only the $uu$ and $u\phi$ equations of motion at order $1/r$ still need to be imposed,
which correspond to the three-dimensional Bondi conservation equations.
The functions $M$ and $N$ then give rise to currents whose charges generate the BMS$_3$ algebra,
and their zero modes parametrise the mass and angular momentum of the so-called flat-space cosmology solutions~\cite{Barnich:2006av,Barnich:2010eb,Barnich:2012aw}.
Our general three-dimensional metric~\eqref{eq:3Dsol} clearly incorporates these solutions,
and we will show in later sections that its free parameters similarly give rise to conserved currents.

\paragraph{Proof of truncation.}
We will now show that $\pd_r^3 g_{MN}=0$ for a three-dimensional vacuum Einstein metric in the Carroll-covariant Bondi gauge~\eqref{eq:car-cov-bondi-metric-repeat-yet-again}{}\footnote{Since $\beta=0$, this is also a Carroll-covariant Newman--Unti gauge.}, as claimed in~\eqref{eq:three-dim-radial-expansion-terminates-preview} above.
Since we already showed in Section~\ref{eq:EFE-in-terms-of-Gamma-repeat} that the function $\beta$ vanishes in three dimensions by the $rr$~component of the Ricci tensor,
the gauge-fixed metric~\eqref{eq:car-cov-bondi-metric-repeat-yet-again} reduces to
\begin{equation}
  \label{eq:car-cov-bondi-metric-repeat-yet-again-threedim}
    ds^2
    = - 2 \tau_\mu dr dx^\mu
    + \left(
      - S \tau_\mu \tau_\nu
      + \Pi_{\mu\nu}
    \right) dx^\mu dx^\nu\,,
\end{equation}
The boundary metric variable $\tau_\mu(x)$ does not depend on the radial coordinate,
so we only need to show that the third radial derivative of
$g_{\mu\nu}=-S \tau_\mu \tau_\nu + \Pi_{\mu\nu}$
vanishes.
For this, it will be useful to split $g_{\mu\nu}$ using the boundary space and time projectors,
and we denote the result using
\begin{equation}
  \begin{gathered}
    \mathcal{Q}_{\mu\nu}
    = h_\mu^\rho h_\nu^\sigma g_{\rho\sigma}
    = h^\rho_\mu h^\sigma_\nu \Pi_{\rho \sigma}\,,
    \qquad
    \mathcal{P}_\mu
    = -h_\mu^\rho v^\sigma g_{\rho\sigma}
    = -h^\rho_\mu v^\sigma\Pi_{\rho \sigma}\,,
    \\
    \mathcal{U}
    = v^\rho v^\sigma g_{\rho\sigma}
    = -S + v^\mu v^\nu \Pi_{\mu\nu}\,.
  \end{gathered}
\end{equation}

The Weyl tensor vanishes in three dimensions,
so the vanishing of the Ricci tensor is equivalent to the vanishing of the Riemann tensor.
Following the strategy for asymptotically AdS$_3$ spacetimes in~\cite{Skenderis:1999nb}
we focus on the $R_{r\mu r\nu}=0$ equations, which we decompose using the boundary space and time projectors.
Just like we did for the Ricci tensor, we express $R_{r\mu r\nu}$ in terms of the composite objects defined in Section \ref{sec:rewriting-EE},
which leads to
\begin{subequations}
\begin{align}
        \Pi^{\mu\nu}R_{r\mu r\nu} & =  0\,,\\
    U^\rho\Pi^\sigma_\mu R_{r\rho r\sigma} & =  \frac{1}{2}\Pi^\rho_\mu\left(\partial_r+r^{-1}\right)\mathcal{Z}_\rho\,,\\
    \begin{split}
    U^\rho U^\sigma R_{r\rho r\sigma} & =  \frac{1}{2}\partial_r^2 S-\frac{1}{2}U^\rho U^\sigma \partial_r^2\Pi_{\rho\sigma}\\
    &\qquad +\frac{1}{4}\Pi_{\rho\sigma}\left(\partial_r U^\rho+\Pi^{\rho\kappa}\mathcal{A}_\kappa\right)\left(\partial_r U^\sigma+\Pi^{\sigma\lambda}\mathcal{A}_\lambda\right)\,.
    \end{split}
\end{align}
\end{subequations}
It will be useful to note that
\begin{align}
  \partial_r\Pi^{\mu\nu}
  &=\partial_r\left(\Pi_{\rho\sigma} \Pi^{\rho\mu} \Pi^{\nu\sigma}\right)\nonumber
  \\
  &= 2 \Pi_{\rho\sigma}\pd_r \Pi^{\rho(\mu} \Pi^{\nu)\sigma}
  + \Pi^{\mu\rho} \Pi^{\nu\sigma} \pd_r \Pi_{\rho\sigma}\nonumber
  \\
  &= -\Pi^{\mu\rho} \Pi^{\nu\sigma} \pd_r \Pi_{\rho\sigma}\,,
\end{align}
where we have used
$\Pi_{\rho\sigma}\Pi^{\rho\mu} = \delta_\sigma^\mu + \tau_\sigma U^\mu$
as well as 
$\tau_\sigma\Pi^{\nu\sigma} = 0$
to get to the last line.
Likewise, using the fact that $h_\mu^\rho = \delta_\mu^\rho + \tau_\mu v^\rho$,
we see that
\begin{equation}
  \Pi_\mu^\rho \Pi_\nu^\sigma \pd_r \mathcal{Q}_{\rho\sigma}
  = \Pi_\mu^\rho \Pi_\nu^\sigma h_\rho^\alpha h_\sigma^\beta \pd_r \Pi_{\alpha\beta}
  = \Pi_\mu^\rho \Pi_\nu^\sigma \pd_r \Pi_{\rho\sigma}\,.
\end{equation}
Because there are no STF tensors for $d=1$,
the tensor $\mathcal{Q}_{\mu\nu}$ is fully determined by the Bondi determinant condition~\eqref{eq:bondi-gauge-condition},
which for $d=1$ reads $\Pi^{\mu\nu}\partial_r\Pi_{\mu\nu}=2r^{-1}$.
In terms of the variables above, we can write this condition as
\begin{equation}
    \partial_r\mathcal{Q}_{\mu\nu}=2r^{-1}\mathcal{Q}_{\mu\nu}\,.
\end{equation}
Differentiating this equation twice and using the intermediate results then gives
\begin{equation}
    \partial_r^3\mathcal{Q}_{\mu\nu}=0\,.
\end{equation}

Next, the $U^\rho\Pi^\sigma_\mu R_{r\rho r\sigma}=0$ equation leads to
\begin{equation}
  \label{eq:CalQEqn}
  0
  =\pd_r^2 \mathcal{P}_\mu
  -\frac{1}{2}\pd_r\mathcal{Q}_{\mu\nu}\Pi^{\nu\rho}\left(\pd_r \mathcal{P}_\rho+a_\rho\right)\,.
\end{equation}
and differentiating this similarly implies
$\pd_r^3 \mathcal{P}_\mu=0$.
Finally, $U^\rho U^\sigma R_{r\rho r\sigma}=0$ gives
\begin{equation}
  0 =
  \pd_r^2\mathcal{U}
  -\frac{1}{2}\left(\pd_r \mathcal{P}_\mu+a_\mu\right)\Pi^{\mu\nu}\left(\pd_r \mathcal{P}_\nu+a_\nu\right)\,,
\end{equation}
which leads to $\pd_r^3 \mathcal{U} = 0$.
Then the third radial derivative of all projections of $g_{\mu\nu}$ vanishes,
and from this we conclude that $\pd_r^3 g_{MN}=0$ as claimed above.
In particular, note that this result rules out any subleading logarithmic corrections in three bulk dimensions.

\subsection{Adding log terms in \texorpdfstring{$d=2$}{d=2}}
\label{ssec:radial-expansion-logs}
As we saw in Section~\ref{ssec:radial-expansion-rmu-trace},
in four dimensions,
the equation $\Pi^\mu_\kappa R_{\mu r}=0$
gives rise to following constraint~\eqref{eq:PDZ}
at order $r^{-3}$,
\begin{equation}\label{eq:PDZ-repeat}
    h^{\mu\nu}\mathcal{D}_\mu D_{\nu\kappa}=0\,.
\end{equation}
Here, $D_{\mu\nu}$ is the STF tensor defined in~\eqref{eq:Dmunu-def}
which encodes the
$h_\mu^\rho h_\nu^\sigma \os{0}{\Pi}_{\rho\sigma}$ coefficient in the metric expansion.
As was briefly mentioned at the end of that subsection,
this constraint can be removed by adding an appropriate log term.
Specifically, by including a log term at order $r^{-1}\log r$ in the expansion of $\Pi_{\mu\nu}$,
the equation above instead produces an algebraic equation that determines the coefficient of that log term.
The purpose of this subsection is to work through this procedure explicitly. 

Note that we do not derive the most general polyhomogeneous structure of the metric near $\scri^+$.
Instead, we restrict our attention to the log term mentioned above,
which serves to restores the recursive structure of the spatial $\mu r$ equation at order $r^{-3}$.
Additionally, we saw in Section~\ref{eq:vvantrace-projection-of-Pi-at-NLO-and-NNLO} that
the STF equation at order~$r^{-1}$ produces another
constraint~\eqref{eq:STF1/rSimple} on the evolution of $D_{\mu\nu}$ which reads
\begin{equation}
  \label{eq:STF1/rSimple-repeat}
  \mathcal{L}_v D_{\mu\nu}=0\,,
\end{equation}
Unlike the $R_{\mu r}=0$ equation,
the STF equation does not allow us to solve algebraically for metric components,
and therefore this constraint is of a different nature than~\eqref{eq:PDZ-repeat}.

\paragraph{Modified metric expansion.}
We will consider the following log terms in the radial expansions of the metric variables~\eqref{eq:radial-expansion-metric-variables-repeat},
\begin{subequations}
  \label{eq:first-log-block}
  \begin{align}
    \beta
    &=  r^{-2}\overset{(2)}{\beta}+r^{-3}\log r\overset{(3,1)}{\beta}+\OO(r^{-3})\,,
    \\
    S
    &=  r\overset{(-1)}{S}+\overset{(0)}{S}+r^{-1}\log r\overset{(1,1)}{S}+\OO(r^{-1})\,,
    \\
    \Pi_{\mu\nu}
    &= r^2 h_{\mu\nu}+r\overset{(-1)}{\Pi}_{\mu\nu}+\overset{(0)}{\Pi}_{\mu\nu}+r^{-1}\log r\overset{(1,1)}{\Pi}_{\mu\nu}+\OO(r^{-1})\,.
  \end{align}
\end{subequations}
With this, the expansions of the inverse metric variables~\eqref{eq:radial-expansion-metric-variables-inverse-repeat} now read
\begin{subequations}
\label{eq:second-log-block}
\begin{align}
    U^\rho & =  v^\rho+r^{-1}\os{1}{U}^\rho+r^{-2}\os{2}{U}^\rho+r^{-3}\log r \os{3,1}{U}^\rho+\OO(r^{-3})\,,\\
    \Pi^{\rho\sigma} & = r^{-2}h^{\rho\sigma}+r^{-3}\os{3}{\Pi}^{\rho\sigma}+r^{-4}\os{4}{\Pi}^{\rho\sigma}+r^{-5}\log r \os{5,1}{\Pi}^{\rho\sigma}+\OO(r^{-5})\,,
\end{align}
\end{subequations}
where the coefficients of the log terms are determined by
\begin{subequations}
\begin{align}
    \os{3,1}{U}^\rho & =  -\os{3,1}{\beta} v^\rho-h^{\rho\mu}v^\nu\os{1,1}{\Pi}_{\mu\nu}\,,\label{eq:U-Pi-rel-logs}\\
    \os{5,1}{\Pi}^{\rho\sigma} & =  -h^{\rho\mu}h^{\sigma\nu} \os{1,1}{\Pi}_{\mu\nu}\,.
\end{align}
\end{subequations}
Here, we use the notation $\os{n,1}{X}$ for the coefficient of $r^{-n}\log r$ in the radial expansion of the object $X$.
From the orthogonality condition $U^\mu\Pi_{\mu\nu}=0$
and the $d=2$ Bondi determinant condition $\Pi^{\mu\nu} \pd_r \Pi_{\mu\nu}= 4r\inv$
we learn that
\begin{align}
    v^\mu v^\nu\os{1,1}{\Pi}_{\mu\nu} =  0\,,\qquad
    h^{\mu\nu}\os{1,1}{\Pi}_{\mu\nu}  =  0\,.
\end{align}
Likewise, following their definitions in Section~\ref{ssec:rewriting-EE-strategy}
along with the relations above,
we find that the composite objects that appear in the Einstein equations acquire the following log terms in their expansions,
\begin{subequations}
\label{eq:logexpansionsNNLO}
\begin{align}
    \mathcal{Z}_\mu & = r^{-1}\overset{(1)}{\mathcal{Z}}_\mu+r^{-2}\log r\overset{(2,1)}{\mathcal{Z}}_\mu+\mathcal{O}(r^{-2})\,,\label{eq:NNLOinZ}\\
    \mathcal{A}_\mu & = a_\mu+\cdots +r^{-3}\log r \overset{(3,1)}{\mathcal{A}}_\mu+\mathcal{O}(r^{-3})\,,\\
    \mathcal{G}^\mu{}_\nu & = -r^{-2}C^\mu{}_\nu+\cdots +r^{-4}\log r\overset{(4,1)}{\mathcal{G}}{}^\mu{}_\nu+\mathcal{O}(r^{-4})\,,\\
    \mathcal{F}^\mu{}_\nu & = r^{-2}F^\mu{}_\nu+\cdots +r^{-5}\log r\overset{(5,1)}{\mathcal{F}}{}^\mu{}_\nu+\mathcal{O}(r^{-5})\,,\\
    \mathcal{K}_{\mu\nu} & = r^2\frac{1}{2}K h_{\mu\nu}+\cdots+r^{-1}\log r\overset{(1,1)}{\mathcal{K}}_{\mu\nu}+\mathcal{O}(r^{-1})\,,\\
    \mathcal{K} & = K+\cdots+r^{-3}\log r \overset{(3,1)}{\mathcal{K}}+\mathcal{O}(r^{-3})\,.
\end{align}
\end{subequations}
Here, we explicitly listed the leading-order contributions and the first log contributions that appears in the expansion, and we suppressed all other terms.

\paragraph{Consequences in the equations of motion.}
We begin by isolating the first logarithmic contributions to the Einstein equations and working out their consequences, starting with the $rr$ equation in~\eqref{eq:rr-eqCompObjForm}.
Here, the first log term arises from the piece $2r^{-1}\D_r\beta$ and appears at $\OO(r^{-5}\log r)$,
which leads to
\begin{equation}
  \label{eq:beta31-zero}
  \os{3,1}{\beta} = 0\,.
\end{equation}
Next, we consider the equation $\Pi^\mu_\kappa R_{\mu r} = 0$ in~\eqref{eq:PiRBestForm}.
At first glance, the two terms 
\begin{equation}
    -\frac{1}{2}\Pi^\mu_\kappa\partial_r \mathcal{Z}_\mu-r^{-1}\mathcal{Z}_\kappa\,,
\end{equation}
appear to give rise to terms of order $r^{-3}\log r$, but they cancel at that order. 
As a result, this does not fix the~$\os{2,1}{\mathcal{Z}}_\mu$ coefficient,
and we will see shortly that it is instead fixed by another equation of motion.

Next, we consider the $U^\mu R_{\mu r} = 0$ equation in~\eqref{eq:URmur-v3}.
In this case, the first log terms appear at order $r^{-3}\log r$ and arise from the terms
\begin{equation}
    -\frac{1}{2}\left(\partial_r^2 S+2r^{-1}\partial_r S-2r^{-2}S\right)+r^{-1}\bar{\mathcal{K}}\,.
\end{equation}
Again, however, all log terms at that order cancel,
and this equation therefore does not restrict the leading log terms either.

The log terms do not affect the trace equation~\eqref{eq:trace-eq-d=2} at the order $r^{-2}$, which is its only relevant order due to the Bianchi identities.
We are therefore left with the STF equation.
At order $r^{-1}$ it is unchanged, and it therefore still produces the constraint~\eqref{eq:STF1/rSimple-repeat} on the Lie derivative of the $D_{\mu\nu}$ tensor.
The first log terms appear at order $r^{-2}\log r$, where the STF equation now produces an equation of the form~\eqref{eq:further-subleading-stf-lie-derivatives} for
$\mathcal{L}_v\left(
  h^\mu_{\langle\rho}h^\nu_{\sigma\rangle}\overset{(1,1)}\Pi_{\mu\nu}
\right)$.
This is an evolution equation for the STF part of the log term coefficient $\overset{(1,1)}\Pi_{\mu\nu}$ along the $v^\mu$ Carroll time vector field.

We now turn our attention to what happens at the non-log order immediately following the log terms discussed above. In this context, only the $\mu r$ components of Einstein's equations at $\OO(r^{-3})$ are relevant.
We do not need to consider the $rr$ equation since we do not need to know $\beta$ beyond $\OO(r^{-2})$
while, as mentioned above,
the trace equation is unaffected to this order.
As we saw in Section~\ref{ssec:radial-expansion-rmu-trace}, the equation $U^\mu R_{\mu r} = 0$ is identically satisfied at order $r^{-3}$ in the absence of logs.
Therefore, only terms explicitly involving radial derivatives $\D_r$ can contribute. Using the $d=2$ form of this equation in~\eqref{eq:URmur-v3prime}, we see that the only term that contributes is
\begin{equation}
\label{eq:has-S-log-term}
    -\frac{1}{2}\left(\partial_r^2 S+2r^{-1}\partial_r S\right)=\cdots +\frac{1}{2} r^{-3}\overset{(1,1)}{S}+\cdots\,.
\end{equation}
Note that both $\mathcal{G}^\mu{}_\nu$ and $\mathcal{Z}_\mu$ contain explicit factors of $\D_r$,
and so they receive log corrections at some appropriate non-log order in $1/r$. However these log corrections to $\mathcal{G}^\mu{}_\nu$ and $\mathcal{Z}_\mu$ do not contribute to $U^\mu R_{\mu r}=0$ at order $r^{-3}$.
Thus, we conclude that this equation implies
\begin{equation}
\label{eq:S11van}
    \overset{(1,1)}{S}=0\,.
\end{equation}

Similarly, the only term that can give rise to a contribution from these log terms in the $\Pi^\mu_\kappa R_{\mu r}=0$ equation in~\eqref{eq:PiR-r-eq-d2} at order $r^{-3}$ is the term
\begin{equation}
\label{eq:Z-contribution}
    -\frac{1}{2}\Pi^\mu_\kappa\partial_r \mathcal{Z}_\mu = \cdots -r^{-3}\frac{1}{2}\overset{(2,1)}{\mathcal{Z}}_\kappa + \cdots\,,
\end{equation}
where, as in~\eqref{eq:has-S-log-term}, the right-hand side isolates the relevant order $r^{-3}$ term that appears in the expansion of the left-hand side.
As we just mentioned,
$\Pi^\mu_\kappa R_{\mu r}=0$ at order $r^{-3}$ leads to the constraint on $D_{\mu\nu}$ in~\eqref{eq:PDZ-repeat} in the absence of logs.
With the addition of the contribution from~\eqref{eq:Z-contribution},
the spatial projection of the $\mu r$ equation at order $r^{-3}$ therefore becomes
\begin{equation}
\label{eq:first-lifted-constraint}
     h^{\mu\nu}\mathcal{D}_\mu D_{\nu\kappa}=-\frac{1}{2}\overset{(2,1)}{\mathcal{Z}}_\kappa\,.
\end{equation}
Using the definition of $\mathcal{Z}_\mu$~in~\eqref{eq:calZ} and the relation~\eqref{eq:U-Pi-rel-logs} as well as the equation of motion~\eqref{eq:beta31-zero},
we find that the corresponding coefficient in the expansion of the metric variables in~\eqref{eq:first-log-block} is
\begin{equation}
  \overset{(2,1)}{\mathcal{Z}}_\mu
  = - 3h_{\mu\nu}\os{3,1}{U}^\nu
  = 3 h_\mu^\rho v^\sigma\os{1,1}{\Pi}_{\rho\sigma}\,.
\end{equation}
Therefore, the modified constraint equation~\eqref{eq:first-lifted-constraint} implies that
\begin{equation}\label{eq:U31}
   h_\kappa^\mu v^\nu\os{1,1}{\Pi}_{\mu\nu}=-\frac{2}{3}h^{\mu\rho}\mathcal{D}_\mu D_{\rho\kappa}\,.
\end{equation}
In this way, we see that $\Pi^\mu_\kappa R_{\mu r}=0$ at order $r^{-3}$ no longer imposes a constraint on the $D_{\mu\nu}$~tensor.
Instead, it now allows us to solve for the mixed spatial and temporal projection of the logarithmic coefficient $\os{1,1}{\Pi}_{\mu\nu}$ in the metric expansion~\eqref{eq:first-log-block}.

\section{Variations and Ward identities}
\label{sec:variations-ward-ids}
In the previous section, we gave a procedure to systematically solve the Einstein equations in our Carroll-covariant Bondi--Sachs gauge.
Due to the Bianchi identities, the only remaining non-trivial equations that we have not studied yet correspond to $U^\mu U^\nu R_{\mu\nu}=0$ and $\Pi^\mu_\kappa U^\nu R_{\mu\nu}$ at order $r^{-d}$. We claimed
that these remaining equations correspond to covariant versions of the Bondi loss equations
and that they correspond to boundary diffeomorphism Ward identities.

To substantiate these claims, it is useful to first gain some insight into the general form of the response functions and the associated Ward identities obtained by varying our boundary data.
This is subtle for two reasons.
First, even though the shear $C_{\mu\nu}$ appears only at the first subleading order in the radial expansion, we will see in Section~\ref{sec:HoloRenormAndOn-ShellActions} that it nevertheless still contributes to the variation of the action evaluated on shell.
Since it is a fully independent variable,
this means that we must incorporate the shear in our boundary data, along with the standard Carroll metric data.
This metric data is parametrised by the clock one-form~$\tau_\mu$ and the degenerate spatial metric~$h_{\mu\nu}$,
so that the total variation of some putative action with sources will be of the form
\begin{equation}
  \label{eq:on-shell-action-leading-order-intro}
  \delta S
  = \int d^{d+1} x\, e \left(
    T^\mu \delta \tau_\mu
    + \frac{1}{2} T^{\mu\nu} \delta h_{\mu\nu}
    + \frac{1}{2} S^{\mu\nu} \delta C_{\mu\nu}
  \right).
\end{equation}
As usual in non-Lorentzian geometry, the currents $T^\mu$ and $T^{\mu\nu}$ can be assembled into an energy-momentum tensor~\cite{Hartong:2015usd,deBoer:2020xlc}.
On the other hand, we will see that~$S^{\mu\nu}$ can be related to a covariant version of the Bondi news tensor.

While this first subtlety is straightforward to deal with on a technical level,
its conceptual consequence is significant,
as it implies that the shear can be viewed as a boundary source on par with the metric data.%
\footnote{%
  In addition to serving as a boundary source,
  the shear also describes the gravitational radiation via the news tensor,
  which, roughly speaking, corresponds to its (boundary/retarded) time derivative.
  This `dual' nature of the shear allows for an equivalent alternative viewpoint.
  In that approach, we define a different boundary energy-momentum tensor,
  which obeys all the standard properties associated with a Carroll theory that does not include the shear.
  (Such a `standard' Carroll energy-momentum tensor has vanishing energy flux $T^\mu h_{\mu\nu}=0$, which follows from Carroll boost invariance  ~\cite{Hartong:2015usd,deBoer:2017ing}.)
  However, in this case, the energy-momentum tensor is not conserved,
  and the non-conservation is captured by the shear and the news tensors.
  For more details, we refer the reader to~\cite{Hartong:2025WIP2}.
  In the present paper, we adopt the first viewpoint above.
}
In this approach, we obtain a triplet of responses $(T^\mu, T^{\mu\nu}, S^{\mu\nu})$,
which we will refer to as the boundary \emph{energy-momentum-news complex}.
For the majority of the current section,
we will assume that the generic actions we study are invariant under diffeomorphisms, Carroll boosts and Weyl transformations,
though we will also discuss boost anomalies in Sections~\ref{subsec:trafoEMT} and~\ref{subsec:EMTnewstrafos}.
This allows us to derive a notion of conservation from the diffeomorphism Ward identity.
In Section~\ref{sec:bulk-conservation-equations},
we then demonstrate explicitly that the aforementioned remaining bulk equations of motion can be written in the form of this diffeomorphism Ward identity.

The second subtlety is that we must account for the fact that we cannot consider arbitrary degenerate spatial metrics $h_{\mu\nu}$ in the action.
Instead, as we already saw in Subsection~\ref{ssec:bulk-geom-LO-EE-in-partial-gauge}, the leading-order equations of motion require the boundary extrinsic curvature $K_{\mu\nu}$ associated to this metric to be pure trace.%
\footnote{%
  \label{fn:shear-subleading}%
  Due to this constraint on the boundary geometry,
  it is actually debatable to what extent the shear is truly subleading to a higher-order term in the radial expansion.
  For $d=2$, the tensor $h_{\mu\nu}$ has five independent components, since it is a symmetric tensor with rank 2 whose determinant vanishes.
  The constraint $K_{\langle\mu\nu\rangle}=0$ removes two of those five components.
  In a sense, these two components `resurface' as the two independent components of the shear tensor.
  We know that the constraint on the boundary geometry cannot be lifted by adding logs that respect the boundary conditions. In fact, they are an immediate consequence of the boundary conditions.
  The fact that the shear appears to take over these `missing' boundary metric components makes it natural that the boundary constraint is robust.
}
Our first option is to find a covariant way to describe the set of all solutions to this constraint equation,
and we will pursue this in Section~\ref{ssec:variations-ward-ids-solving-constraint}.
There, we will see that
the metric
\begin{equation}
  \label{eq:fourdim-constraint-solution-intro}
  h_{\mu\nu}
  dx^\mu dx^\nu
  = e^{2\varphi} \left((dX^1)^2 + (dX^2)^2\right),
\end{equation}
with $\varphi$, $X^1$ and $X^2$ arbitrary functions,
provides the most general solution to the extrinsic curvature constraint.
We can then vary within the space of allowed boundary data by replacing variations with respect to $h_{\mu\nu}$ by variations with respect to the functions $\varphi$, $X^1$ and $X^2$.
This will result in a set of response functions,
and we then use these to parametrise $T^{\mu\nu}$. As we will see, this leaves a part of $T^{\mu\nu}$ undetermined.

Alternatively, we can impose the constraint $K_{\langle\mu\nu\rangle}=0$ using a Lagrange multiplier $\zeta^{\mu\nu}$.
In this case, we replace~\eqref{eq:on-shell-action-leading-order-intro} with an action of the form
\begin{equation}
  \label{eq:on-shell-action-leading-order-with-constraint-intro}
  S_\text{L}
  =
  S[\tau_\mu, h_{\mu\nu}, C_{\mu\nu}]
  + \frac{1}{2} \int d^{d+1}x\,e\, \zeta^{\mu\nu} K_{\langle\mu\nu\rangle}\,.
\end{equation}
Adding this term to the action ensures that we stay on the constraint surface without having to use a specific parametrisation for our $h_{\mu\nu}$ variations.
It also means that $\zeta^{\mu\nu}$ will contribute to $T^{\mu\nu}$,
and we will show these contributions are precisely the components of~$T^{\mu\nu}$ that we could not determine when only varying in the $\varphi, X^1, X^2$ solution space above.
This only concerns the response $T^{\mu\nu}$ associated to $h_{\mu\nu}$ variations,
as it turns out that the response $T^\mu$ from $\tau_\mu$ variations is not affected by the Lagrange multiplier.
We therefore have
\begin{equation}
  \label{eq:on-shell-action-leading-order-with-constraint-variation-intro}
  \delta S_L
  = \int d^{d+1} x\, e \left(
    T^\mu \delta \tau_\mu
    + \frac{1}{2} \left(T^{\mu\nu} + t^{\mu\nu}\right) \delta h_{\mu\nu}
    + \frac{1}{2} S^{\mu\nu} \delta C_{\mu\nu}
    + \frac{1}{2} K_{\langle\mu\nu\rangle} \delta\zeta^{\mu\nu}
  \right),
\end{equation}
where we used $t^{\mu\nu}$ to denote the contributions from the Lagrange multiplier term.
Since the Lagrange multiplier~$\zeta^{\mu\nu}$ is by definition arbitrary,
the tensor $t^{\mu\nu}$ corresponds to a fundamental ambiguity in the~$T^{\mu\nu}$ tensor in our energy-momentum-news complex.
It will turn out that this does not affect the Bondi loss equations.
Indeed, since the constraint term does not break any of the symmetries of the original action,
the $t^{\mu\nu}$ ambiguity should drop out of
the corresponding Ward identities.
We will demonstrate all of this explicitly in Section~\ref{ssec:variations-ward-ids-boundary-ward-identities} below.
Additionally, we will later see that the presence of this ambiguity implies that there is a fundamental ambiguity in the definition of the Bondi angular momentum aspect. We will come back to this ambiguity in Section \ref{ssec:bulk-improvements-weyl-covariant-currents}.

\subsection{Solving the boundary constraint}
\label{ssec:variations-ward-ids-solving-constraint}

As we already saw in Subsection~\ref{ssec:bulk-geom-LO-EE-in-partial-gauge}, the leading-order equations of motion require that the extrinsic curvature of the boundary Carrollian geometry is pure trace.
While this is trivially true for $d=1$,
it imposes a constraint on the allowed boundary Carroll metric data in higher dimensions.
We now want to find a parametrization of the space of allowed metric data by explicitly and covariantly solving this constraint.
Specifically, we focus on~$d=2$, where the constraint is given by
\begin{equation}
  \label{eq:extrinsic-curvature-constraint-d=2-repeat}
  K_{\mu\nu} = \frac{1}{2} K h_{\mu\nu}\,.
\end{equation}
It will be useful to split $h_{\mu\nu} = \delta_{ab} e^a_\mu e^b_\nu$
using the boundary spatial frame that we originally introduced in Section~\ref{sec:bulk-geom}.
Projected onto this frame using the inverse vielbeine $e^\mu_a$,
the extrinsic curvature tensor and its trace are then given by
\begin{equation}
  K_{ab}
  = e^\mu_a e^\nu_b K_{\mu\nu}
  = \delta_{ac} e^\mu_b v^\nu \pd_{[\mu} e^c_{\nu]}
  + \delta_{bc} e^\mu_a v^\nu \pd_{[\mu} e^c_{\nu]}\,,
  \qquad
  K
  = 2e^\mu_a v^\nu \pd_{[\mu} e^a_{\nu]}\,.
\end{equation}
Using the two-dimensional Levi-Civita symbol $\eps_{ab}$
with $\eps_{12}=+1$, we can write this as
\begin{equation}
  K_{ab} \tau \wedge e^1 \wedge e^2
  = \delta_{c(a} \eps_{b)d} de^c \wedge e^d\,.
\end{equation}
The constraint~\eqref{eq:extrinsic-curvature-constraint-d=2-repeat} requiring the extrinsic curvature to be pure trace then amounts to setting
$K_{11} = K_{22} = K/2$
as well as
$K_{12}=0$,
which is equivalent to
\begin{equation}\label{eq:constraintinforms}
  de^1 \wedge e^2
  = - de^2 \wedge e^1
  = \frac{K}{2} \tau \wedge e^1 \wedge e^2\,,
  \qquad
  de^1 \wedge e^1
  = de^2 \wedge e^2\,.
\end{equation}
Decomposing the two-forms $de^a$ using the basis provided by $\tau$ and the spatial frame $e^a$,
this constraint implies that we can write the exterior derivatives as
\begin{subequations}
  \label{eq:fourdim-constraint-spatial-exterior-derivatives-alpha-beta}
  \begin{align}
    de^1
    &= \alpha \wedge e^1
    + \beta \wedge e^2\,,
    \\
    de^2
    &= \alpha \wedge e^2
    - \beta \wedge e^1\,.
  \end{align}
\end{subequations}
Here, the two one-forms $\alpha$ and $\beta$ are 1-forms
such that the component of $\alpha$ proportional to $\tau$ is equal to $K/2$
and the exterior derivative of the right-hand side of~\eqref{eq:fourdim-constraint-spatial-exterior-derivatives-alpha-beta} vanishes.
We now want to rotate our spatial frame so that \eqref{eq:fourdim-constraint-spatial-exterior-derivatives-alpha-beta}~allows us to use Frobenius' theorem.
For this, consider an infinitesimal rotation generated by $\lambda_{ab} = \epsilon_{ab} \lambda$,
which transforms our frame $e^a$ into a new frame $w^a$,
\begin{subequations}
  \begin{align}
    w^1
    &= e^1 + \lambda e^2+\mathcal{O}(\lambda^2)\,,
    \\
    w^2
    &= e^2 - \lambda e^1+\mathcal{O}(\lambda^2)\,.
  \end{align}
\end{subequations}
After this transformation, the exterior derivatives in~\eqref{eq:fourdim-constraint-spatial-exterior-derivatives-alpha-beta} become
\begin{subequations}
  \label{eq:fourdim-constraint-spatial-exterior-derivatives-after-rotation}
  \begin{align}
    dw^1
    &= \alpha \wedge w^1
    + \left(d\lambda + \beta\right) \wedge w^2
    = \tilde\alpha \wedge w^1\,,
    \\
    dw^2
    &= \alpha \wedge w^2
    - \left(d\lambda + \beta\right) \wedge w^1
    = \tilde\alpha \wedge w^2\,,
  \end{align}
\end{subequations}
to first order in $\lambda$.
To obtain the last equality on each line, we assumed
that the rotation parameter satisfies
\begin{equation}
  \label{eq:fourdim-constraint-rotation-condition}
  d\lambda = -\beta + A w^1 + B w^2\,.
\end{equation}
Here, $A$ and $B$ can be any functions,
since we can define
$\tilde\alpha = \alpha - A w^2 + B w^1$
so that the right-hand side of~\eqref{eq:fourdim-constraint-spatial-exterior-derivatives-after-rotation} holds.
Since $A$ and $B$ are arbitrary, the only nontrivial part of the condition on the derivative of the rotation parameter in~\eqref{eq:fourdim-constraint-rotation-condition} is its $\tau$ component,
which gives $v^\rho \pd_\rho \lambda = - v^\rho\beta_\rho$.
In adapted coordinates, this is a differential equation for $\lambda$ which can always be solved locally.

We have shown that we can always rotate our vielbeine so they obey~\eqref{eq:fourdim-constraint-spatial-exterior-derivatives-after-rotation}.
We will now go back to our notation of denoting vielbeine with $e^1$ and $e^2$,
and then the above result implies we can set $\beta=0$ in \eqref{eq:fourdim-constraint-spatial-exterior-derivatives-alpha-beta} without loss of generality,
\begin{equation}
  \label{eq:fourdim-constraint-spatial-exterior-derivatives-alpha-only}
  de^1
  = \alpha \wedge e^1\,,
  \qquad
  de^2
  = \alpha \wedge e^2\,.
\end{equation}
This equation can be solved with the help of the Frobenius theorem,
which gives
\begin{equation}
  \label{eq:fourdim-constraint-frobenius-inital}
  e^1 = M dX^1\,,
  \qquad
  e^2 = N dX^2\,,
\end{equation}
where $M$, $N$, $X^1$ and $X^2$ are all smooth functions on our boundary manifold.
Note that we now have $de^1 = M\inv dM \wedge e^1$
and $de^2 = N\inv dN \wedge e^2$.
Since the same one-form $\alpha$ appears on the right-hand side on both lines in~\eqref{eq:fourdim-constraint-spatial-exterior-derivatives-alpha-only},
we see that
\begin{equation}
  N\inv dN \wedge e^1 \wedge e^2
  = \alpha \wedge e^1 \wedge e^2
  = M\inv dM \wedge e^1 \wedge e^2\,.
\end{equation}
Only the components of $N\inv dN$ and $M\inv dM$ that are proportional to $\tau$ survive in this equation, so it implies that
$
N\inv v^\rho \pd_\rho N
= M\inv v^\rho \pd_\rho M
= - K/2
$, where we used $v^\mu\alpha_\mu=-K/2$. Hence, we obtain
\begin{equation}
  \label{eq:fourdim-constraint-fix-time-dependence-of-Frobenius-functions}
  N = fM\,,
  \qquad
  v^\rho \pd_\rho f = 0\,,
  \qquad
  M\inv v^\rho \pd_\rho M = - \frac{K}{2}\,,
\end{equation}
where $f$ is any function obeying $v^\rho \pd_\rho f=0$.
We see that the time dependence of the function $N$ equals that of the function $M$,
and this also parametrises the trace of the extrinsic curvature. We can easily verify that \eqref{eq:fourdim-constraint-frobenius-inital} with $N=fM$ and $v^\mu\partial_\mu f=0$ obeys \eqref{eq:constraintinforms} and thus solves the constraint.

The most general set of solutions to \eqref{eq:extrinsic-curvature-constraint-d=2-repeat} is therefore parametrised by metrics of the form
\begin{equation}
  h_{\mu\nu} dx^\mu dx^\nu
  = \delta_{ab} e^a e^b
  = M^2 \left((dX^1)^2 + f^2 (dX^2)^2\right)\,.
\end{equation}
Note that
$v^\rho \pd_\rho X=0$
and
$v^\rho \pd_\rho Y=0$
since $v^\mu$ and the spatial vielbeine are orthogonal. Hence, $v^\rho \pd_\rho f=0$ is solved by $f=F(X,Y)$. We can then take one more step to simplify this parametrisation.
We can get rid of the function $f$ by redefining the functions
$\tilde{X} = \tilde{X}(X,Y)$
and
$\tilde{Y} = \tilde{Y}(X,Y)$
in such a way that,
dropping the tildes for simplicity,
we have
\begin{equation}
  \label{eq:fourdim-constraint-solution}
  h_{\mu\nu} dx^\mu dx^\nu
  = e^{2\varphi} \left((dX^1)^2 + (dX^2)^2\right)\,,
\end{equation}
where $e^{2\varphi}$ contains $M^2$ and some overall factor due to the redefinitions of $X^1$ and~$X^2$.
Such a transformation always exists by the same arguments that establish the existence of isothermal coordinates on Riemann surfaces.
For this, it is crucial that we have $v^\rho \pd_\rho f = 0$,
so that the field redefinition can be viewed as a 2D diffeomorphism in field space with coordinates $(X^1, X^2)$.
All time dependence is contained in the overall function~$e^{2\varphi}$,
and following~\eqref{eq:fourdim-constraint-fix-time-dependence-of-Frobenius-functions} this is directly related to the trace $K$ of the boundary extrinsic curvature.

To summarise,
we showed that the expression in~\eqref{eq:fourdim-constraint-solution}
parametrises the most general spatial tensor~$h_{\mu\nu}$ that satisfies the constraint~\eqref{eq:extrinsic-curvature-constraint-d=2-repeat} on the boundary extrinsic curvature using three arbitrary functions.
We can always choose vielbeine such that $e^a=e^\varphi dX^a$.
The inverse metric obeys
\begin{equation}
    e^{2\varphi}h^{\mu\nu}\partial_\mu X^a\partial_\nu X^b=\delta^{ab}\,.
\end{equation}
We can vary within the space of all allowed boundary $h_{\mu\nu}$ tensors by varying
the functions $\varphi$ and $X^a$ with $a=1,2$ above.

Finally, recall that $h_{\mu\nu}$ and the clock one-form $\tau_\mu$ together fully determine the boundary Carroll metric data,
and they uniquely fix the inverse variables $(v^\mu,h^{\mu\nu})$ using the standard orthogonality relations~\eqref{eq:boundary-frame-with-M-orthonormality-completeness}.
With the parametrisation of~$h_{\mu\nu}$ in terms of the functions $\varphi$ and $X^a$ above, these orthogonality conditions in particular imply the following equations,
\begin{subequations}
\begin{align}
  v^\mu \pd_\mu X^a
  & =  0\,,\label{eq:vMXY}
  \qquad
  \\
  h^{\mu\nu} e^{2\varphi} \pd_\mu X^a \pd_\nu X^b
  & =  \delta^{ab}\,.\label{eq:invhMXY}
\end{align}
\end{subequations}
These equations should be solved for $v^\mu$ and $h^{\mu\nu}$. We can use \eqref{eq:vMXY} to find that
\begin{equation}
    K_{\mu\nu}=-\frac{1}{2}\mathcal{L}_v h_{\mu\nu}=-\mathcal{L}_v\varphi h_{\mu\nu}=\frac{1}{2}K h_{\mu\nu}\,,
\end{equation}
so that
\begin{equation}
    K=-2\mathcal{L}_v \varphi\,.
\end{equation}
To wrap things up, we work out the boundary connection $\mathcal{C}^\rho_{\mu\nu}$ in~\eqref{eq:bdy-Ccon} for the boundary geometries that obey the constraint.
Upon inserting \eqref{eq:fourdim-constraint-solution}, we obtain
\begin{align}
 \label{eq:GammaMXY}
   \mathcal{C}^\rho_{\mu\nu} & =  -\frac{1}{2}v^\rho\left(\partial_\mu\tau_\nu+\partial_\nu\tau_\mu+\tau_\mu a_\nu+\tau_\nu a_\mu\right)+\partial_\mu\varphi h^\rho_\nu+\partial_\nu\varphi h^\rho_\mu\nonumber\\
   &\qquad-h_{\mu\nu}h^{\rho\sigma}\partial_\sigma\varphi+e^{2\varphi}h^{\rho\sigma}\partial_\sigma X^a\partial_\mu\partial_\nu X^a\,.
\end{align}
Two useful properties that follow from this are
\begin{subequations}
\begin{align}
    h^\rho_\mu h^\sigma_\nu\mathcal{D}_\rho\partial_\sigma X^a & =  -2h^{\rho}_{\langle\mu}h^\sigma_{\nu\rangle}\partial_\rho\varphi\partial_\sigma X^a\,,
    \\
    v^\mu\mathcal{D}_\mu\partial_\nu X^a & =  \frac{1}{2}K\partial_\nu X^a\,,
    \label{eq:vcovderX}
\end{align}
\end{subequations}
which also imply
\begin{subequations}
\begin{align}
    \mathcal{D}_{(\mu}\left(e^{2\varphi}\partial_{\nu)} X^a\right) & = h_{\mu\nu}e^{2\varphi}h^{\rho\sigma}\partial_\rho\varphi\partial_\sigma X^a\,,\label{eq:symcovderX}\\
    h^{\mu\nu}\mathcal{D}_\mu\partial_\nu X^a & =  0\,.
\end{align}
\end{subequations}
This concludes our discussion of the solutions of the $d=2$ boundary constraint.

\subsection{The energy-momentum-news complex}
\label{sec:variations-ward-ids-currents-from-constraint-surface-variation}
Let us now return to the general variation of the action in~\eqref{eq:on-shell-action-leading-order-intro}, which reads
\begin{equation}
  \label{eq:on-shell-action-leading-order}
  \delta S
  = \int d^{3} x\, e \left(
    T^\mu \delta \tau_\mu
    + \frac{1}{2} T^{\mu\nu} \delta h_{\mu\nu}
    + \frac{1}{2} S^{\mu\nu} \delta C_{\mu\nu}
  \right).
\end{equation}
First, it is useful to note that we can assume without loss of generality
that $S^{\mu\nu}$ is a spatial symmetric trace-free (STF) tensor,
since the same is true for the shear.
To see this explicitly, we can decompose a general tensor $S^{\mu\nu}$ into time, mixed space-time, spatial trace and spatial STF components as follows,
\begin{equation}
  \label{eq:general-S-tensor-decomposition}
  S^{\mu\nu}
  = S^{\langle\mu\nu\rangle}
  + \frac{1}{2} h^{\mu\nu} h_{\rho\sigma} S^{\rho\sigma}
  - 2 v^{(\mu} h^{\nu)}_\rho \tau_\sigma S^{\rho\sigma}
  + v^\mu v^\nu \tau_\rho \tau_\sigma S^{\rho\sigma}\,.
\end{equation}
This follows from using completeness.
Using the fact that $C_{\mu\nu}$ is spatial and STF,
which in particular implies $v^\mu v^\nu \delta C_{\mu\nu} = 0$,
we can then rewrite the contribution to the total variation~\eqref{eq:on-shell-action-leading-order} as follows,
\begin{equation}
  \frac{1}{2} S^{\mu\nu} \delta C_{\mu\nu}
  = \frac{1}{2} S^{\langle\mu\nu\rangle} \delta C_{\mu\nu}
  + \frac{1}{2} \left(
    \frac{1}{2} C^{\mu\nu} h_{\rho\sigma} S^{\rho\sigma}
    - 2v^{(\mu} C^{\nu)}{}_\rho \tau_\sigma S^{\rho\sigma}
  \right) \delta h_{\mu\nu}\,.
\end{equation}
The final terms in this expression can be absorbed in the $T^{\mu\nu}$ current,
so we see that we can indeed assume $S^{\langle\mu\nu\rangle}$ is STF without loss of generality. In deriving this, we used the identities~\eqref{eq:app-variation-relations-up-to-down}.

The current $T^\mu$ is the energy current.
The response $T^{\mu\nu}$ is the momentum stress tensor with the momentum $\tau_\mu h_{\nu\rho}T^{\mu\nu}$ and the stress $h^\rho_\mu h^\sigma_\nu T^{\mu\nu}$.
Finally, the STF response $S^{\mu\nu}$ is the news tensor,
and will refer to the triplet $(T^\mu, T^{\mu\nu}, S^{\mu\nu})$ as the energy-momentum-news complex.
In the remainder of this section, we will derive its properties by assuming that the action $S$ has a number of symmetry properties that we will later confirm for the variational problem near $\mathcal{I}^+$.

\subsection{Variations and the boundary constraint}
\label{ssec:variations-ward-ids-ambiguity-from-boundary-constraint}
Having parametrised the solutions of the extrinsic curvature constraint on the boundary Carroll geometries that we are allowed to consider,
we now work out its consequences for the energy-momentum currents.
We first study the variation within the solution space established above,
and we find the precise form of the ambiguity $t^{\mu\nu}$ in the $T^{\mu\nu}$ response that results from the constraint.
We then consider implementing the geometric constraint using a Lagrange multiplier term, and we see that the same ambiguity in $T^{\mu\nu}$ arises,
which now is parametrised by the Lagrange multiplier.
Finally, we show that this ambiguity does not affect the Ward identities associated to diffeomorphisms, Carroll boosts and Weyl transformations,
which follows because these transformations leave the constraint invariant.

\paragraph{Variation within solution space.}
We start by rewriting the variations of $h_{\mu\nu}$
in terms of the variations of the functions $\varphi$ and $X^a$ in the space of solutions~\eqref{eq:fourdim-constraint-solution} to the constraint,
which gives
\begin{equation}
  \frac{1}{2} T^{\mu\nu} \delta h_{\mu\nu}
  = h_{\mu\nu} T^{\mu\nu} \delta \varphi
  + e^{2\varphi} T^{\mu\nu} \pd_\mu X^a \pd_\nu \delta X^a\,.
\end{equation}
Integrating by parts and dropping total derivatives, the total variation is then
\begin{equation}
  \label{eq:on-shell-action-variation-on-constraint-surface}
  \delta S
  = \int d^{3} x\, e \left(
    T^\mu \delta \tau_\mu
    + T^{\mu\nu}h_{\mu\nu} \delta\varphi
    + T^a \delta X^a
    + \frac{1}{2} S^{\mu\nu} \delta C_{\mu\nu}
  \right),
\end{equation}
where the currents associated to the variations with respect to $X^a$ are
\begin{align}
  \label{eq:on-shell-action-variation-on-constraint-surface-MXY-currents}
  T^a
  &= - e\inv \pd_\mu \left(
    e e^{2\varphi} T^{\mu\nu} \pd_\nu X^a
  \right)\,.
\end{align}
These functions capture only part of the full response $T^{\mu\nu}$ that one obtains from varying a general degenerate spatial Carroll metric.

In general,
the symmetric $T^{\mu\nu}$ tensor has $5=6-1$ components,
since it satisfies the relation $v^\mu v^\nu\delta h_{\mu\nu}=0$.
Of those 5 components, our variations give us access to the trace $T^{\mu\nu}h_{\mu\nu}$ and the $T^a$ defined above.
Our ignorance of $T^{\mu\nu}$ can be formulated as follows:
consider a different response, say $T'^{\mu\nu}=T^{\mu\nu}+t^{\mu\nu}$,
where $h_{\mu\nu}t^{\mu\nu}=0$,
so that the $\varphi$ response is the same for both $T^{\mu\nu}$ and $T'^{\mu\nu}$.
Then our variations cannot distinguish between $T^{\mu\nu}$ and $T'^{\mu\nu}$ whenever $t^{\mu\nu}$ also satisfies
\begin{equation}\label{eq:condt}
      e^{-1}\pd_\mu \left(
      e e^{2\varphi} t^{\mu\nu} \pd_\nu X^a
    \right)=0\,,
\end{equation}
which implies that we must have
\begin{equation}\label{eq:defU}
    e e^{2\varphi} t^{\mu\nu} \pd_\nu X^a=\partial_\rho\left(e e^{2\varphi}U^{\rho\mu\nu}\partial_\nu X^a\right)\,,
\end{equation}
for some $U^{\rho\mu\nu}$ satisfying $U^{\rho\mu\nu}=-U^{\mu\rho\nu}$.
The object of interest is $t^{\mu\nu}$ so without loss of generality we can restrict our attention to $U^{\rho\mu\nu}$ tensors for which the right-hand side is non-zero.
Using the connection \eqref{eq:GammaMXY}, we can also write this as
\begin{equation}\label{eq:tmunuinU}
    e^{2\varphi} t^{\mu\nu} \pd_\nu X^a=\mathcal{D}_\rho \left(e^{2\varphi}\partial_\nu X^aU^{\rho\mu\nu}\right)\,.
\end{equation}
Contracting this equation with $\partial_\mu X^b$ leads to
\begin{equation}\label{eq:spatt}
    e^{2\varphi} t^{\mu\nu}\partial_\mu X^b \pd_\nu X^a=\mathcal{D}_\rho \left(e^{2\varphi}\partial_\mu X^b\partial_\nu X^aU^{\rho\mu\nu}\right)\,,
\end{equation}
where we used $U^{\rho\mu\nu}=-U^{\mu\rho\nu}$.
The left-hand side is STF in the $a$ and $b$ indices and so $\partial_\mu X^b\partial_\nu X^aU^{\rho\mu\nu}$ must be, too.
Since the spatial vielbeine for~\eqref{eq:fourdim-constraint-solution} are proportional to $\partial_\mu X^a$,
we conclude from this that if we take the spatial projection
$h^{\alpha}_\rho h^\beta_\mu h^\gamma_\nu U^{\rho\mu\nu}$
of all indices of $U^{\rho\mu\nu}$,
then this object needs to be antisymmetric in its first two indices and symmetric in its last two indices.
Such a tensor is necessarily zero, and so we conclude that $h^{\alpha}_\rho h^\beta_\mu h^\gamma_\nu U^{\rho\mu\nu}=0$.
It follows that we must have
\begin{equation}
    U^{\rho\mu\nu}\partial_\nu X^a=\left(v^\rho Y^{\mu\nu}-v^\mu Y^{\rho\nu}\right)\partial_\nu X^a\,.
\end{equation}
Because $U^{\rho\mu\nu}$ is always contracted with $\partial_\nu X^a$, we can take it to be 
\begin{equation}
    U^{\rho\mu\nu}=v^\rho Y^{\mu\nu}-v^\mu Y^{\rho\nu}\,,
\end{equation}
so that it does not have a $v^\nu$ component.
We can take $Y^{\mu\nu}$ to be spatial because any component along $v^\mu$ or $v^\nu$ will drop out of the equation.
Furthermore, for $\partial_\mu X^b\partial_\nu X^aU^{\rho\mu\nu}$ to be STF in $a$ and $b$, we need $Y^{\mu\nu}$ to be STF as well.
If we now go back to \eqref{eq:tmunuinU} and compute the right-hand side using that $Y^{\mu\nu}$ is STF as well as Equations \eqref{eq:vcovderX} and \eqref{eq:symcovderX}, we obtain
\begin{equation}
    e^{2\varphi} t^{\mu\nu} \pd_\nu X^a=e^{2\varphi}\left(\mathcal{D}_\rho U^{\rho\mu\nu}\right)\partial_\nu X^a-\frac{1}{2}K e^{2\varphi} Y^{\mu\nu}\partial_\nu X^a\,.
\end{equation}
This tells us that 
\begin{equation}\label{eq:finalt}
    t^{\mu\nu}=\mathcal{D}_\rho U^{\rho\mu\nu}-\frac{1}{2}K Y^{\mu\nu}+B^\mu v^\nu\,,
\end{equation}
where we can determine $B^\mu$ by demanding that $t^{\mu\nu}$ is symmetric, which leads to $B^\mu=-\mathcal{D}_\rho Y^{\rho\mu}$.
We therefore finally obtain
\begin{equation}\label{eq:tmunu}
    t^{\mu\nu}=\mathcal{D}_\rho\left(v^\rho Y^{\mu\nu}-v^\mu Y^{\nu\rho}-v^\nu Y^{\mu\rho}\right)-KY^{\mu\nu}\,.
\end{equation}
The part of $T^{\mu\nu}$ that is not determined as a result of the constraint is thus parametrised by an STF tensor $Y^{\mu\nu}$.
Since the latter has two components, these degrees of freedom indeed account for the remaining components of $T^{\mu\nu}$ that are not one of the responses $h_{\mu\nu}T^{\mu\nu}$ or the $T^a$ in~\eqref{eq:on-shell-action-variation-on-constraint-surface-MXY-currents}.
We have thus established that, when varying the action, we cannot distinguish between $T^{\mu\nu}$ and $T^{\mu\nu}+t^{\mu\nu}$, where $t^{\mu\nu}$ is given by \eqref{eq:tmunu}.
For later purposes, we observe that
we can write~\eqref{eq:condt} as
\begin{equation}\label{eq:divt}
  \pd_\nu X^a\mathcal{D}_\mu t^{\mu\nu} =0\,,
\end{equation}
using $h_{\mu\nu}t^{\mu\nu}=0$ as well as \eqref{eq:symcovderX}.

\paragraph{Varying with Lagrange multiplier term.}
We now show that the $Y^{\mu\nu}$~parameter in~\eqref{eq:tmunu} can be interpreted as the Lagrange multiplier enforcing the constraint. 
Instead of explicitly parametrising the solutions to the $K_{\langle\mu\nu\rangle}=0$ constraint as we did above,
we can impose it on the level of the action using a Lagrange multiplier,
\begin{equation}
  \label{eq:on-shell-action-leading-order-with-constraint}
  S_L
  =
  S[\tau_\mu, h_{\mu\nu}, C_{\mu\nu}]
  + \frac{1}{2} \int d^{d+1}x\,e\, \zeta^{\mu\nu} K_{\langle\mu\nu\rangle}\,.
\end{equation}
We can take the Lagrange multiplier $\zeta^{\mu\nu}$ to be spacelike and STF.
As we already briefly mentioned after~\eqref{eq:on-shell-action-leading-order-with-constraint-intro},
varying this Lagrange multiplier term leads to
additional contributions to the $\delta h_{\mu\nu}$ variations,
\begin{align}
  \delta\left(\frac{1}{2} \int d^{d+1}x\,e\zeta^{\mu\nu} K_{\langle\mu\nu\rangle}\right)
  & = 
  \frac{1}{2}\int d^{d+1}x\,e t^{\mu\nu} \delta h_{\mu\nu}
  + \frac{1}{2} \int d^{d+1}x\,eK_{\langle\mu\nu\rangle} \delta \zeta^{\mu\nu}
  \label{eq:varLMterm}
  \\
  &{}\qquad\nonumber
  -\frac{1}{4}\int d^{d+1}x \partial_\rho\left(e\left[v^\rho\zeta^{\mu\nu}-\zeta^{\rho\mu}v^\nu-\zeta^{\rho\nu}v^\mu\right]\delta h_{\mu\nu}\right).
\end{align}
Note that the variations with respect to $\tau_\mu$ cancel out.
The second line is a total derivative, and the last term on the first line simply serves to impose the constraint.
On the other hand,
the variations of the constraint with respect to $h_{\mu\nu}$ give rise to
\begin{align}
  \label{eq:constraint-response-term}
  t^{\mu\nu}
  &= \frac{1}{2}h^{\mu\rho}h^{\nu\sigma}\mathcal{L}_v\zeta_{\rho\sigma}
  -v^{(\mu}h^{\nu)\sigma}\left(
    \mathcal{D}_\rho-a_\rho
  \right) \zeta^\rho{}_\sigma
  \\
  \label{eq:constraint-response-term-alt}
  &= \frac{1}{2}\mathcal{D}_\rho
  \left(
    v^\rho\zeta^{\mu\nu}
    -v^\mu\zeta^{\nu\rho}
    -v^\nu\zeta^{\mu\rho}
  \right)
  -\frac{1}{2}K\zeta^{\mu\nu}\,.
\end{align}
Comparing this with \eqref{eq:finalt},
we see that we have $Y^{\mu\nu}=\frac{1}{2}\zeta^{\mu\nu}$.
Equation~\eqref{eq:varLMterm} also tells us that for variations $\delta h_{\mu\nu}$ that respect the constraint,
we have
\begin{equation}\label{eq:idtmunuosvar}
  \frac{1}{2}\int d^{d+1}x\,e t^{\mu\nu} \delta h_{\mu\nu}
  = \frac{1}{4}\int d^{d+1}x \partial_\rho\left(e\left[v^\rho\zeta^{\mu\nu}-\zeta^{\rho\mu}v^\nu-\zeta^{\rho\nu}v^\mu\right]\delta h_{\mu\nu}\right)\,,
\end{equation}
which will be useful later on.
On the constraint surface, we then have
\begin{equation}
  \label{eq:on-shell-action-leading-order-with-constraint-variation}
  \delta S_L
  = \int d^{d+1} x\, e \left(
    T^\mu \delta \tau_\mu
    + \frac{1}{2} \left(T^{\mu\nu} + t^{\mu\nu}\right) \delta h_{\mu\nu}
    + \frac{1}{2} S^{\mu\nu} \delta C_{\mu\nu}
  \right),
\end{equation}
which gives rise to the same ambiguity $t^{\mu\nu}$ in $T^{\mu\nu}$ as we deduced above from directly varying inside the constraint surface.
Finally, we want to stress that the fact that we do not have access to all of $T^{\mu\nu}$ should not be viewed as a problem,
but rather as a characteristic feature of the physics at $\mathcal{I}^+$.

\paragraph{Consequences for Ward identities.}
The ambiguity tensor $t^{\mu\nu}$ will drop out of all Ward identities we derive from the symmetries of the boundary action. This follows simply from the fact that the constraint respects our three boundary gauge transformations: boundary diffeomorphisms, Weyl and local Carroll boosts.
These transformations are given in~\eqref{eq:car-cov-bondi-gauge-tr-LO-repeat}
which are repeated here for convenience,
\begin{subequations}
  \label{eq:car-gauge-tr-repeat}
  \begin{align}
    \label{eq:car-gauge-tr-tau-h-repeat}
    \delta \tau_\mu
    &= \LL_\chi \tau_\mu
    + \Lambda_D \tau_\mu
    + \lambda_\mu\,,
    &
    \delta h_{\mu\nu}
    &= \LL_\chi h_{\mu\nu}
    + 2 \Lambda_D h_{\mu\nu}\,,
    \\
    \label{eq:car-gauge-tr-v-h-repeat}
    \delta v^\mu
    &= \LL_\chi v^\mu
    - \Lambda_D v^\mu\,,
    &
    \delta h^{\mu\nu}
    &= \LL_\chi h^{\mu\nu}
    - 2 \Lambda_D h^{\mu\nu}
    + 2 \lambda^{(\mu} v^{\nu)}\,.
  \end{align}
\end{subequations}
From this, it is easy to see that,
under boundary diffeomorphisms, Carroll boosts and Weyl transformations,
we have
\begin{align}
  \delta K_{\mu\nu}
  &= \LL_\chi K_{\mu\nu}
  + \Lambda_D K_{\mu\nu}
  - h_{\mu\nu} v^\rho \pd_\rho \Lambda_D\,,
  \\
  \delta K_{\langle\mu\nu\rangle}
  &= \LL_\chi K_{\langle\mu\nu\rangle}
  + \Lambda_D K_{\langle\mu\nu\rangle}\,.
\end{align}
If we set
$\delta \zeta^{\mu\nu} = \LL_\chi \zeta^{\mu\nu} - 4\Lambda_D \zeta^{\mu\nu}$,
the Lagrange multiplier term $e\zeta^{\mu\nu} K_{\langle\mu\nu\rangle}/2$ is clearly diffeomorphism-invariant,
boost-invariant and Weyl-invariant,
so that none of the boundary gauge symmetries in~\eqref{eq:car-gauge-tr-repeat} are broken by the presence of the constraint.
Consequently, the form of the associated Ward identities must also be unchanged.
If we derive such a Ward identity from the variation of the total action 
on the constraint surface in~\eqref{eq:on-shell-action-leading-order-with-constraint-variation},
the only contribution from the Lagrange multiplier term
comes from
\begin{equation}
  \frac{1}{2} et^{\mu\nu} \delta h_{\mu\nu}\,.
\end{equation}
This term can at most contribute a total derivative if we consider transformations $\delta h_{\mu\nu}$ that leave both the initial action and the Lagrange multiplier term invariant,
and therefore it cannot affect the form of the Ward identities.

We can explicitly check this for the boundary gauge transformations of $\delta h_{\mu\nu}$ in~\eqref{eq:car-gauge-tr-tau-h-repeat} as follows.
First,
there are no $\lambda_\mu$ contributions,
since $h_{\mu\nu}$ is invariant under boosts.
The Weyl term $\Lambda_D$ vanishes since $t^{\mu\nu}$~is traceless.
Finally, to see that the diffeomorphisms generated by $\chi^\mu$ do not contribute either,
we can use the STF nature of $t^{\mu\nu}$ to derive
\begin{equation}\label{eq:tlieh}
  \frac{1}{2} et^{\mu\nu} \mathcal{L}_\chi h_{\mu\nu}=et^{\mu\nu}\mathcal{D}_\mu\left(\chi^\rho h_{\rho\nu}\right),
\end{equation}
and, using~\eqref{eq:divt}, we see that this is a total derivative term.

This establishes that the form of the Ward identities is not affected by the presence of the extrinsic curvature constraint.
We will now explicitly work out the corresponding Ward identities for our gauge transformations of interest.

\subsection{Boundary Ward identities}
\label{ssec:variations-ward-ids-boundary-ward-identities}

The energy-momentum-news complex defined in \eqref{eq:on-shell-action-leading-order} will obey various properties as a result of the symmetries of the action,
and we will refer to these properties as Ward identities.%
\footnote{%
  Though, strictly speaking, it is a priori not clear that they have an interpretation in terms of a full-fledged QFT.
  As we already mentioned in the introduction,
  it seems likely that such an interpretation, if it exists, would have to involve both future and past null infinity.
}
We will be concerned with Ward identities arising from invariance under diffeomorphisms, local Carroll boosts and Weyl transformations.
In fact, we will see later on in Sections~\ref{ssec:hol-ren-threedim-anomalies-improvements} and~\ref{ssec:hol-ren-fourdim-anomalies-improvements} that the action near $\mathcal{I}^+$ is not Carroll boost-invariant, in a manner that is reminiscent of an anomaly in the boost Ward identity,
and we will therefore refer to this as a Carroll boost anomaly.%
\footnote{%
  The same caveats from the previous footnote apply.
}
However, to first derive the non-anomalous form of the Ward identities,
we will consider in this subsection an abstract action $S$ that is fully invariant under all the gauge transformations, including local Carroll boosts.

Recall from~\eqref{eq:car-cov-bondi-gauge-tr-of-shear-on-shell} that the on-shell transformation of the shear $C_{\mu\nu}$ is
\begin{equation}
  \label{eq:car-cov-bondi-gauge-tr-of-shear-on-shell-repeat}
  \delta C_{\mu\nu}
  = \LL_\chi C_{\mu\nu}
  + \Lambda_D C_{\mu\nu}
  + \Sigma_{\mu\nu}\,,
\end{equation}
where the inhomogeneous shift $\Sigma_{\mu\nu}$ is given by
\begin{equation}\label{eq:Sigma}
    \Sigma_{\mu\nu}=h^\rho_{\langle \mu} h^\sigma_{\nu\rangle} \LL_\lambda h_{\rho\sigma}
  + 2 \lambda_{\langle \mu} a_{\nu \rangle}=2h^\rho_{\langle \mu} h^\sigma_{\nu\rangle} \left(\mathcal{D}_\rho+a_\rho\right)\lambda_\sigma\,.
\end{equation}
In here, $\chi^\mu$ are (boundary) diffeomorphisms,
$\Lambda_D$ are Weyl transformations
and $\lambda_\mu=\lambda^a e^a_\mu$ are local Carroll boosts. 
Additionally, see~\eqref{eq:car-gauge-tr-repeat} above for the transformation of the Carroll metric data.
In the following, we will again mainly work with $d=2$,
corresponding to four bulk and three boundary spacetime dimensions,
but we will also indicate the equivalent $d=1$ results.

\paragraph{Weyl Ward identity.}
First,
let us consider the Weyl transformations.
Substituting the Weyl transformations with parameter $\Lambda_D$ from~\eqref{eq:car-gauge-tr-repeat} and~\eqref{eq:car-cov-bondi-gauge-tr-of-shear-on-shell-repeat} into the variation~\eqref{eq:on-shell-action-leading-order-with-constraint-variation} of the action,
we get
\begin{align}
  0
  = \delta_{\Lambda_D}
  S
  &= \int d^{3} x\, e \Lambda_D \left(
    T^\mu \tau_\mu
    + \left(T^{\mu\nu} + t^{\mu\nu}\right) h_{\mu\nu}
    + \frac{1}{2} S^{\mu\nu} C_{\mu\nu}
  \right).
\end{align}
As we already noted above, the contributions due to the $t^{\mu\nu}$ ambiguity drop out
since the latter is traceless.
The resulting Ward identity is
\begin{equation}
  \label{eq:boundary-Weyl-WI}
  0
  = T^\mu \tau_\mu
  + T^{\mu\nu} h_{\mu\nu}
  + \frac{1}{2} S^{\mu\nu} C_{\mu\nu}\,.
\end{equation}
In three bulk dimensions, the shear term vanishes.

\paragraph{Boost Ward identity.}
Next, we consider the local Carroll boosts with parameter
$\lambda_\mu$, which obeys $v^\mu\lambda_\mu=0$.
Substituting the boost transformations into the variation~\eqref{eq:on-shell-action-leading-order-with-constraint-variation} gives
\begin{equation}
  0
  = \delta_\lambda S
  = \int d^{3} x\, e \left(
    T^\mu \lambda_\mu
    + \frac{1}{2} S^{\mu\nu} \LL_\lambda h_{\mu\nu}
    + S^{\mu\nu} \lambda_{\mu} a_{\nu}
  \right)\,.
\end{equation}
Since $h_{\mu\nu}$ is boost-invariant,
both $T^{\mu\nu}$ and the $t^{\mu\nu}$ ambiguity do not appear.
Recall from our arguments around~\eqref{eq:general-S-tensor-decomposition} that the shear response tensor $S^{\mu\nu}$ is spatial, symmetric and trace-free (STF).
As a result, we can use Equation~\eqref{eq:tlieh} to rewrite the $\frac{1}{2} S^{\mu\nu} \LL_\lambda h_{\mu\nu}$ term above.
Integrating by parts,
this leads to the Ward identity
\begin{equation}
  \label{eq:boundary-boost-WI}
  0
  = h^\mu_\rho T^\rho
  -\left(\mathcal{D}_\rho-a_\rho\right) S^{\rho\mu}\,.
\end{equation}
In three bulk dimensions, the latter term vanishes.
As we will see in Section~\ref{sec:HoloRenormAndOn-ShellActions},
both for $d=1$ and for $d=2$
this equation will have a non-vanishing anomaly on the left-hand side
when we derive it from the bulk gravitational action near future null infinity.

\paragraph{Diffeomorphism Ward identity.}
Finally, let us consider the boundary diffeomorphisms with parameter
$\chi^\mu$.
Their transformation of the action~\eqref{eq:on-shell-action-leading-order-with-constraint-variation} gives
\begin{equation}
  0
  = \delta_\chi S
  = \int d^{3} x\, e \left(
    T^\mu \LL_\chi \tau_\mu
    + \frac{1}{2} \left(T^{\mu\nu} + t^{\mu\nu}\right) \LL_\chi h_{\mu\nu}
    + \frac{1}{2} S^{\mu\nu} \LL_\chi C_{\mu\nu}
  \right).
\end{equation}
The term involving $t^{\mu\nu}$ vanishes,
as we showed explicitly around \eqref{eq:tlieh}.
This variation leads to the following diffeomorphism Ward identity,
\begin{align}
  \label{eq:boundary-diffeo-WI-not-covariant}
  0
  = -e^{-1}\partial_\mu \left(e \left[T^\mu{}_\nu+S^{\mu\rho}C_{\rho\nu}\right]\right)
  +T^\mu\partial_\nu\tau_\mu
  +\frac{1}{2}T^{\mu\rho}\partial_\nu h_{\mu\rho}
  +\frac{1}{2}S^{\mu\rho}\partial_\nu C_{\mu\rho}\,,
\end{align}
where we defined the total energy-momentum tensor
\begin{equation}
    T^\mu{}_\nu=T^\mu\tau_\nu+T^{\mu\rho}h_{\rho\nu}\,,
\end{equation}
following for example~\cite{deBoer:2020xlc}.

The expression in~\eqref{eq:boundary-diffeo-WI-not-covariant} is covariant,
though not manifestly so.
In the next section,
we will show for $d=1$ and $d=2$ that the Einstein equations $U^\mu U^\nu R_{\mu\nu}=0$ and $\Pi_\kappa^\mu U^\nu R_{\mu\nu}=0$ at order $r^{-d}$,
which correspond to the Bondi loss equations,
can be recast in the form of \eqref{eq:boundary-diffeo-WI-not-covariant}.
For this, it is useful to work out the covariant forms of the $v^\nu$ and spatial $h^\nu_\kappa$ projections
of the Ward identity~\eqref{eq:boundary-diffeo-WI-not-covariant},
which are
\begin{align}
    \label{eq:energyeq}
    0
    &= -\mathcal{L}_v\left(\tau_\rho T^\rho\right)
    +K\left(
      \tau_\rho T^\rho
      -\frac{1}{d}h_{\rho\sigma}T^{\rho\sigma}
      -\frac{1}{4}C_{\rho\sigma}S^{\rho\sigma}
    \right)
    \\
    &{}\qquad\nonumber
    -\frac{1}{2}S^{\rho\sigma}N_{\rho\sigma}
    +\left(\mathcal{D}_\mu+a_\mu\right)\left(T^\rho h^\mu_\rho\right)
    \,,
    \\
    \label{eq:diffeoWIspatialproj}
    0
    &= -\left(\mathcal{L}_v-K\right) P_\kappa
    +h^\lambda_\kappa\mathcal{D}_\mu\left(\tilde T^{\mu\sigma}h_{\sigma\lambda}\right)
    +\frac{1}{d}h^\mu_\kappa\partial_\mu\left(T^{\rho\sigma}h_{\rho\sigma}\right)
    \\
    &{}\qquad\nonumber
    +a_\kappa\left(\frac{1}{d}T^{\rho\sigma}h_{\rho\sigma}-\tau_\rho T^\rho\right)
    +T^\rho F_{\rho\kappa}
    -\frac{1}{2}S^{\mu\rho}h^\nu_\kappa\mathcal{D}_\nu C_{\mu\rho}
    \\
    &{}\qquad\nonumber
    +h^{\nu}_\kappa\mathcal{D}_\mu S^{\mu\rho}C_{\rho\nu}\,.
\end{align}
We remind the reader that $S^{\mu\nu}$ is by definition spatial and STF.
Here, for $d=2$, we have defined the \emph{news tensor} $N_{\mu\nu}$ as follows,
\begin{equation}
  \label{eq:NewsDefn}
  N_{\mu\nu}
  = -\left(\mathcal{L}_v +\frac{1}{2}K\right) C_{\mu\nu}\,.
\end{equation}

This is a spatial symmetric trace-free tensor,
which gives a covariant generalization of the Bondi news in standard Bondi--Sachs coordinates.
The significance of this object has to do with the fact that,
as we will see in Section~\ref{sec:bulk-improvements},
its on-shell value can be taken to be such that $S^{\mu\nu}=\frac{1}{2}N^{\mu\nu}$.
Additionally, we have defined the momentum $P_\mu$ and the STF part of the stress tensor $\tilde T^{\rho\sigma}$
as follows,
\begin{equation}
    P_\kappa=T^{\mu\nu}\tau_\mu h_{\nu\kappa}\,,\qquad\tilde T^{\rho\sigma}=T^{\mu\nu}h^{\langle\rho}_\mu h^{\sigma\rangle}_\nu\,,
\end{equation}
In the $d=1$ case,
the equations~\eqref{eq:energyeq} and~\eqref{eq:diffeoWIspatialproj} simplify to
\begin{subequations}
  \label{eq:diffeo-WI-threedim}
  \begin{align}
    \label{eq:diffeo-WI-time-proj-threedim}
    0
    &= -\left(\mathcal{L}_v-2K\right)\left(\tau_\rho T^\rho\right)
    +\left(\mathcal{D}_\mu+a_\mu\right)\left(T^\rho h^\mu_\rho\right)-K\left(\tau_\rho T^\rho+h_{\mu\nu}T^{\mu\nu}\right),
    \\
    \label{eq:diffeo-WI-spatial-proj-threedim}
    0
    &= -\left(\mathcal{L}_v-K\right) P_\kappa +h^\mu_\kappa\left(\partial_\mu+2a_\mu\right)\left(T^{\rho\sigma}h_{\rho\sigma} \right)-a_\kappa \left(
    \tau_\rho T^\rho
    +h_{\mu\nu}T^{\mu\nu}
  \right).
  \end{align}
\end{subequations}
As we will see in the following,
$\tau_\mu T^\mu$ and $P_\kappa$ can be identified with $v^\mu\os{d-1}{g}_{\mu\nu}$ in the expansion of the metric.

We can use the Weyl Ward identity~\eqref{eq:boundary-Weyl-WI} to remove $T^{\mu\nu} h_{\mu\nu}$ from Equation~\eqref{eq:energyeq} and $T^\mu \tau_\mu$ from Equation~\eqref{eq:diffeoWIspatialproj}.
This results in\footnote{%
  In writing \eqref{eq:energyeqplusWeyl},
  we dropped a term
  $+K\frac{1}{2}\left(\frac{1}{d}-\frac{1}{2}\right) S^{\mu\nu} C_{\mu\nu}$
  because we are only interested in $d=1$ and $d=2$,
  and in those cases this term vanishes.
}
\begin{align}
    \label{eq:energyeqplusWeyl}
    0
    &= -\left(\mathcal{L}_v-\frac{d+1}{d}K\right)\left(\tau_\rho T^\rho\right)
    -\frac{1}{2}S^{\rho\sigma}N_{\rho\sigma}
    +\left(\mathcal{D}_\mu+a_\mu\right)\left(T^\rho h^\mu_\rho\right)
    \,,
    \\
    \label{eq:diffeoWIspatialprojplusWeyl}
    0
    &= -\left(\mathcal{L}_v-K\right) P_\kappa
    +h^\lambda_\kappa\mathcal{D}_\mu\left(\tilde T^{\mu\sigma}h_{\sigma\lambda}\right)
    +\frac{1}{d}h^\mu_\kappa\left(\partial_\mu+(d+1)a_\mu\right)\left(T^{\rho\sigma}h_{\rho\sigma}\right)
    \\
    &{}\qquad\nonumber
    +T^\rho F_{\rho\kappa}
    -\frac{1}{2}S^{\mu\rho}h^\nu_\kappa\left(\mathcal{D}_\nu-a_\nu\right) C_{\mu\rho}
    +h^{\nu}_\kappa\mathcal{D}_\mu S^{\mu\rho}C_{\rho\nu}\,.
\end{align}
We will later see that this version of the diffeomorphism Ward identities has nice properties with respect to Weyl transformations. 

\subsection{Algebra of transformations}
\label{ssec:trafo-alg}
Equation \eqref{eq:on-shell-action-leading-order} defines the energy-momentum-news complex.
We know how the residual gauge transformations act on $\tau_\mu$, $h_{\mu\nu}$ and $C_{\mu\nu}$.
These residual gauge transformation form an algebra, which we will derive in this subsection.
Once we know the commutator $\left[\delta_1\,,\delta_2\right]=-\delta_{[1,2]}$, where $\delta_{[1,2]} $ is a specific gauge transformation given below in terms of the parameters appearing in $\delta_1$ and $\delta_2$,
we can
demand that the algebra is realised on the action
by requiring that
\begin{equation}
    \label{eq:WZ-intro}
    \left(\left[\delta_1\,,\delta_2\right]+\delta_{[1,2]}\right)S=0\,.
\end{equation}
This condition simply states that the fields appear in the action in a manner consistent with the algebra of residual gauge transformations.
If we assume that all transformations are actual invariances of the action $S$,
this allows us to derive the transformation properties of the energy-momentum-news complex.
We will also discuss what changes in the case of a Carroll boost anomaly. 

In order to compute the commutator $\left[\delta_1\,,\delta_2\right]$ on the fields $h_{\mu\nu}$ and $v^\mu$,
we simply act successively with $\delta_1$ and $\delta_2$ (and subtract $1\leftrightarrow 2$) on the fields $h_{\mu\nu}$ and $v^\mu$, but not on the diffeomorphism and Weyl transformation parameters.%
\footnote{%
Note that we have not fixed a gauge on the boundary that restricts the Carroll metric data, and so the parameters here are not field-dependent, i.e., they are independent of $h_{\mu\nu}$ and $v^\mu$.
}
This leads to
\begin{align}
    \left[\delta_1\,,\delta_2\right]h_{\mu\nu} & = -\mathcal{L}_{[\chi_1\,,\chi_2]}h_{\mu\nu}-2\left(\mathcal{L}_{\chi_1}\Lambda^2_D-\mathcal{L}_{\chi_2}\Lambda^1_D\right)h_{\mu\nu}=-\delta_{[1,2]} h_{\mu\nu}\,,\\
    \left[\delta_1\,,\delta_2\right]v^\mu & = -\mathcal{L}_{[\chi_1\,,\chi_2]}v^\mu+\left(\mathcal{L}_{\chi_1}\Lambda^2_D-\mathcal{L}_{\chi_2}\Lambda^1_D\right)v^\mu=-\delta_{[1,2]} v^\mu\,,
\end{align}
where $\delta_{[1,2]}$ when acting on $h_{\mu\nu}$ and $v^\mu$ is a transformation with parameters
\begin{align}
  \label{eq:chi-3}
  \chi_{[1,2]}^\mu
  &= \left[\chi_1\,,\chi_2\right]^\mu\,,
  \\
  \label{eq:LambdaD-3}
  \Lambda_D^{[1,2]}
  &= \mathcal{L}_{\chi_1}\Lambda^2_D
  -\mathcal{L}_{\chi_2}\Lambda^1_D\,.
\end{align}
Note the minus sign%
\footnote{%
  For a diffeomorphism generated by $\chi^\mu$,
  a tensor is taken to transforms via the pullback map,
  which is why $\delta_\chi$ is an anti-homomorphism from the Lie algebra of generators to differential operators acting on the fields.
} 
in front of $\mathcal{L}_{[\chi_1\,,\chi_2]}$ in the expressions for $[\delta_1\,,\delta_2]$.
When it comes to the transformation of $\tau_\mu$, however,
we cannot treat $\lambda_\mu$ as field-independent,
because it obeys the constraint $v^\mu\lambda_\mu=0$.
Equivalently, $\lambda_\mu$ is of the form $\lambda_\mu=\lambda^a e^a_\mu$,
where $\lambda^a$ is independent of the metric.
For the commutator on $\tau_\mu$,
we get
\begin{align}
    \left[\delta_1\,,\delta_2\right]\tau_\mu & =  -\mathcal{L}_{[\chi_1\,,\chi_2]}\tau_\mu-\left(\mathcal{L}_{\chi_1}\Lambda^2_D-\mathcal{L}_{\chi_2}\Lambda^1_D\right)\tau_\mu\nonumber\\
    &{}\qquad+\delta_1\lambda^2_\mu-\mathcal{L}_{\chi_1}\lambda^2_\mu-\Lambda_D^1\lambda_\mu^2+\mathcal{L}_{\chi_2}\lambda^1_\mu+\Lambda_D^2\lambda_\mu^1-\delta_2\lambda^1_\mu\nonumber\\
    &=-\delta_{[1,2]}\tau_\mu=-\mathcal{L}_{\chi_{[1,2]}}\tau_\mu-\Lambda^{[1,2]}_D\tau_\mu-\lambda^{[1,2]}_\mu\,,\label{eq:comgaugetrafo}
\end{align}
where
\begin{equation}\label{eq:lambda3}
     \lambda^{[1,2]}_\mu = -\delta_1\lambda^2_\mu+\mathcal{L}_{\chi_1}\lambda^2_\mu+\Lambda_D^1\lambda_\mu^2-\left(1\leftrightarrow 2\right)\,.
\end{equation}
In here, we have
\begin{equation}
    \delta_1\lambda^2_\mu-\mathcal{L}_{\chi_1}\lambda^2_\mu-\Lambda_D^1\lambda_\mu^2=-e^a_\mu\mathcal{L}_{\chi_1}\lambda^2_a+\lambda^2_a\lambda_1^{ab}e^b_\mu\,,
\end{equation}
so that $v^\mu\lambda_\mu^{[1,2]}=0$, as it should.
We used that $\delta_1\lambda^2_\mu=\lambda^2_a\delta_1 e^a_\mu$,
where $\delta_1 e^a_\mu$ involves a local rotation with parameter $\lambda^{ab}$ on top of the usual general coordinate and Weyl transformations.
With a bit more work, we can also show that
\begin{align}
    \left[\delta_1\,,\delta_2\right]C_{\mu\nu} & =  -\mathcal{L}_{[\chi_1\,,\chi_2]}C_{\mu\nu}-\left(\mathcal{L}_{\chi_1}\Lambda^2_D-\mathcal{L}_{\chi_2}\Lambda^1_D\right)C_{\mu\nu}\nonumber\\
    &\qquad-h^\rho_{\langle\mu}h^\sigma_{\nu\rangle}\left(\mathcal{L}_{\lambda^{[1,2]}}h_{\rho\sigma}+2a_\rho\lambda^{[1,2]}_\sigma\right)=-\delta_{[1,2]} C_{\mu\nu}\,,
\end{align}
where again $\lambda^{[1,2]}_\mu$ is given by \eqref{eq:lambda3}.
Together with $\chi^\mu_{[1,2]}$ and $\Lambda_D^{[1,2]}$ in~\eqref{eq:chi-3} and~\eqref{eq:LambdaD-3},
this completes the transformation under $\delta_{[1,2]}$ of all our boundary variables,
which fixes the algebra of transformations.

\subsection{Transformation properties of the energy-momentum tensor}\label{subsec:trafoEMT}
Let us now return to the definition of the energy-momentum-news complex in~\eqref{eq:on-shell-action-leading-order-intro}.
We will consider general~$d$,
but for simplicity we will suppress any shear terms until Section~\ref{subsec:EMTnewstrafos}.
We therefore have
\begin{equation}
  \label{eq:general-d-variation-no-shear}
  \delta S
  = \int d^{d+1} x\, e \left(
    T^\mu \delta \tau_\mu
    + \frac{1}{2} T^{\mu\nu} \delta h_{\mu\nu}
  \right).
\end{equation}
The variations in this expression are arbitrary.
For now, we will also assume that the action is invariant under gauge transformations,
so that when we take $\delta=\delta_g$ we have~$\delta_g S=0$.
This leads to the Ward identities for the energy-momentum tensor that we derived earlier in Section~\ref{eq:on-shell-action-leading-order-with-constraint-variation-intro} (though with $S^{\mu\nu}=0=C_{\mu\nu}$).

Let us first consider $\left[\delta_1\,,\delta_2\right]S$,
where both $\delta_1$ and $\delta_2$ are arbitrary variations.
In this case, we have
\begin{eqnarray}
    \left[\delta_1\,,\delta_2\right]S &=&\nonumber\\
   &&\hspace{-2cm}\int d^{d+1}x \,\Big(\delta_1\left(e T^\mu\right)\delta_2\tau_\mu-\delta_2\left(e T^\mu\right)\delta_1\tau_\mu+\frac{1}{2}\delta_1\left(e T^{\mu\nu}\right)\delta_2 h_{\mu\nu}-\frac{1}{2}\delta_2\left(e T^{\mu\nu}\right)\delta_1 h_{\mu\nu}\Big)\nonumber\\
   &&\hspace{-2cm}+\int d^{d+1}x e\Big(T^\mu\left[\delta_1\,,\delta_2\right]\tau_\mu+\frac{1}{2}T^{\mu\nu}\left[\delta_1\,,\delta_2\right]h_{\mu\nu}\Big)\,.
\end{eqnarray}
Now consider the case where $\delta_1=\delta_g$ is a gauge transformation and $\delta_2=\delta$ is an arbitrary transformation.
Using the `variation by parts' identity
\begin{align}
  &-\delta(eT^\mu)\delta_g\tau_\mu-\frac{1}{2}\delta(eT^{\mu\nu})\delta_g h_{\mu\nu}
  \\
  &{}\qquad\nonumber
  =
  \delta\left[-eT^\mu\delta_g\tau_\mu-\frac{1}{2}eT^{\mu\nu}\delta_g h_{\mu\nu}\right]+eT^\mu\delta\delta_g\tau_\mu-\frac{1}{2}eT^{\mu\nu}\delta\delta_g h_{\mu\nu}\,,
\end{align}
we can write the above as follows,
\begin{align}
    \left[\delta_g\,,\delta\right]S & = \int d^{d+1}x \,\Big(\delta_g\left(e T^\mu\right)\delta\tau_\mu+\frac{1}{2}\delta_g\left(e T^{\mu\nu}\right)\delta h_{\mu\nu}+eT^\mu\delta\delta_g\tau_\mu-\frac{1}{2}eT^{\mu\nu}\delta\delta_g h_{\mu\nu}\Big)\nonumber\\
   &\qquad+\int d^{d+1}x \delta\left[-eT^\mu\delta_g\tau_\mu-\frac{1}{2}eT^{\mu\nu}\delta_g h_{\mu\nu}\right]\nonumber\\
   &\qquad+\int d^{d+1}x e\Big(T^\mu\left[\delta_g\,,\delta\right]\tau_\mu+\frac{1}{2}T^{\mu\nu}\left[\delta_g\,,\delta\right]h_{\mu\nu}\Big)\,.
\end{align}
Up to a total derivative (or boundary term), the middle line is zero by virtue of the fact that we assumed that $S$ is gauge-invariant.
For the first line,
we use $\delta_g e=\partial_\rho\left(e\chi^\rho\right)+(d+1)\Lambda_D e$
as well as integration by parts on the $\delta\delta_g\tau_\mu$ and $\delta\delta_g h_{\mu\nu}$ terms,
which give us
\begin{align}
    \delta\delta_g\tau_\mu & =  \mathcal{L}_\chi\delta\tau_\mu+\Lambda_D\delta\tau_\mu+\delta\lambda_\mu\,,\\
    \delta\delta_g h_{\mu\nu} & =  \mathcal{L}_\chi \delta h_{\mu\nu}+\Lambda_D \delta h_{\mu\nu}\,.
\end{align}
In the end,
we obtain
\begin{align}
  \label{eq:gauge-gen-tr-comm-without-shear}
    \left[\delta_g\,,\delta\right]S & =  \int d^{d+1}x e\,\Big(\delta_g T^\mu-\mathcal{L}_\chi T^\mu+(d+2)\Lambda_DT^\mu\Big)\delta\tau_\mu\\
   &\qquad+\frac{1}{2}\int d^{d+1}x e\,\Big(\delta_g T^{\mu\nu}-\mathcal{L}_\chi T^{\mu\nu}+(d+3)\Lambda_DT^{\mu\nu}+2T^{(\mu}\lambda^{\nu)}\Big)\delta h_{\mu\nu}\nonumber\\
   &\qquad+\int d^{d+1}x e T^\mu h_{\mu\nu}\delta\lambda^\nu+\int d^{d+1}x e\Big(T^\mu\left[\delta_g\,,\delta\right]\tau_\mu+\frac{1}{2}T^{\mu\nu}\left[\delta_g\,,\delta\right]h_{\mu\nu}\Big)\,.\nonumber
\end{align}
The first term on the third line is zero by virtue of the assumed Carroll boost-invariance of $S$,
which, with zero shear, implies $T^{\mu}h_{\mu\nu}=0$ by~\eqref{eq:boundary-boost-WI}.
We will demand that $\left[\delta_g\,,\delta\right]$ is just an arbitrary variation, so that
\begin{equation}\label{eq:symmetryEMT}
    \left[\delta_g\,,\delta\right]S = \int d^{d+1}x e\Big(T^\mu\left[\delta_g\,,\delta\right]\tau_\mu+\frac{1}{2}T^{\mu\nu}\left[\delta_g\,,\delta\right]h_{\mu\nu}\Big)\,,
\end{equation}
following~\eqref{eq:general-d-variation-no-shear}.
Comparing to~\eqref{eq:gauge-gen-tr-comm-without-shear} we see that $T^\mu$ and $T^{\mu\nu}$ must transform as
\begin{align}
    \delta_g T^\mu & =  \mathcal{L}_\chi T^\mu-(d+2)\Lambda_DT^\mu\,,\label{eq:classtrafoTmu}\\
    \delta_g T^{\mu\nu} & = \mathcal{L}_\chi T^{\mu\nu}-(d+3)\Lambda_DT^{\mu\nu}-2T^{(\mu}\lambda^{\nu)}\,.\label{eq:classtrafoTmunu}
\end{align}
In particular, this implies that the energy-momentum tensor $T^\mu{}_\nu=T^\mu\tau_\nu+T^{\mu\rho}h_{\rho\nu}$ is Carroll boost-invariant and has Weyl weight $-(d+1)$.
The requirement \eqref{eq:symmetryEMT} is equivalent to \eqref{eq:classtrafoTmu} and \eqref{eq:classtrafoTmunu}.

\paragraph{Including a boost anomaly.}
We will now revisit the previous calculation with the modified assumption that the Carroll boosts are allowed to be anomalous.
Explicitly, what we mean by a boost anomaly is the following,
\begin{equation}
  \label{eq:variation-with-boost-anomaly-no-shear}
    \delta_g S =\int d^{d+1}xe\mathcal{A}_B^\mu\lambda_\mu\,,
\end{equation}
where $\mathcal{A}_B^\mu$ is an object that is constructed from the boundary geometry and derivatives.%
\footnote{%
  For $\mathcal{A}_B^\mu$ to be an anomaly,
  we also need to require that it cannot be cancelled by adding a local counterterm to the action.
  We will check this explicitly for the anomalies we will encounter from the bulk gravitational action in Section~\ref{sec:HoloRenormAndOn-ShellActions} later on.
}
This modifies the boost Ward identity~\eqref{eq:boundary-boost-WI} to
\begin{equation}
  T^\rho h^\mu_\rho
  =\mathcal{A}_B^\mu\,.
\end{equation}
We again demand that~\eqref{eq:symmetryEMT} holds.
Following very similar steps as before, we now find
\begin{align}
    0 & =  -\int d^{d+1}x\left(\delta\left(e\mathcal{A}_B^\mu\lambda_\mu\right)-eT^\mu\delta\lambda_\mu\right)+\int d^{d+1}x e\left[\delta_g T^\mu-\mathcal{L}_\chi T^\mu+(d+2)\Lambda_D T^\mu\right]\delta\tau_\mu\nonumber\\
    &\qquad +\frac{1}{2}\int d^{d+1}x e\left[\delta_g T^{\mu\nu}-\mathcal{L}_\chi T^{\mu\nu}+(d+3)\Lambda_D T^{\mu\nu}\right]\delta h_{\mu\nu}\,.
\end{align}
If we split the energy current in time and space components as follows,
\begin{equation}
    T^\mu=-\tau_\rho T^\rho v^\mu+\mathcal{A}_B^\mu\,,
\end{equation}
we can write the variation above as 
\begin{align}
    0 & =  -\int d^{d+1}x\lambda_\mu\delta\left(e\mathcal{A}_B^\mu\right)+\int d^{d+1}x e\left[\delta_g T^\mu-\mathcal{L}_\chi T^\mu+(d+2)\Lambda_D T^\mu\right]\delta\tau_\mu
    \\
    &\qquad\nonumber
    +\frac{1}{2}\int d^{d+1}x e\left[\delta_g T^{\mu\nu}-\mathcal{L}_\chi T^{\mu\nu}+(d+3)\Lambda_D T^{\mu\nu}-2\tau_\rho T^\rho v^{(\mu}\lambda^{\nu)}\right]\delta h_{\mu\nu}\,.
\end{align}
When $\mathcal{A}_B^\mu=0$ we see that we recover the transformations in~\eqref{eq:classtrafoTmu} and~\eqref{eq:classtrafoTmunu}.
To make progress, let us define
\begin{equation}\label{eq:anomalyfunc}
    A_B=\int d^{d+1}xe\mathcal{A}_B^\mu\lambda_\mu\,.
\end{equation}
We can then formally write
\begin{equation}
    \int d^{d+1}x\lambda_\mu\delta\left(e\mathcal{A}_B^\mu\right)=\int d^{d+1}xe\left(\frac{\delta A_B}{\delta\tau_\mu}\delta\tau_\mu+\frac{\delta A_B}{\delta h_{\mu\nu}}\delta h_{\mu\nu}\right)\,,
\end{equation}
where on the right-hand side we keep $\lambda_\mu$ fixed when computing the functional derivatives.
This leads to the following anomalous transformation rules,
\begin{align}
    \delta_g T^\mu & =  \mathcal{L}_\chi T^\mu-(d+2)\Lambda_DT^\mu+\frac{\delta A_B}{\delta\tau_\mu}\,,\label{eq:anomtrafoTmu}\\
    \delta_g T^{\mu\nu} & =  \mathcal{L}_\chi T^{\mu\nu}-(d+3)\Lambda_DT^{\mu\nu}+2\tau_\rho T^\rho v^{(\mu}\lambda^{\nu)}+2\frac{\delta A_B}{\delta h_{\mu\nu}}\,.\label{eq:aontrafoTmunu}
\end{align}
Equation~\eqref{eq:anomtrafoTmu} implies that the anomaly itself transforms as
\begin{equation}\label{eq:trafoanomaly}
    \delta_g\mathcal{A}^\mu_B=\mathcal{L}_\chi\mathcal{A}_B^\mu-(d+2)\Lambda_D\mathcal{A}_B^\mu+h^\mu_\rho\frac{\delta A_B}{\delta\tau_\rho}+v^\mu\lambda_\rho\mathcal{A}^\rho_B\,.
\end{equation}
In particular, it is a tensor with respect to general coordinate transformations~$\chi^\mu$ and it has a definite Weyl weight of $-(d+2)$
under $\Lambda_D$.

In the presence of an anomaly,
we need to check that the algebra of (gauge) transformations still closes on the action,
which is also known as Wess--Zumino consistency.
This follows immediately from~\eqref{eq:symmetryEMT}.
Following the notation of Section~\ref{eq:extrinsic-curvature-constraint-d=2-repeat}, we use $\delta_1$ and $\delta_2$ to denote two arbitrary gauge transformations.
Equation \eqref{eq:symmetryEMT} tells us that
\begin{equation}\label{eq:d=1WZ}
    \left(\left[\delta_1\,,\delta_2\right]+\delta_{[1,2]}\right)S = 0\,,
\end{equation}
even in the presence of a boost anomaly. The condition \eqref{eq:d=1WZ} can be shown to follow from \eqref{eq:trafoanomaly}.
We will see in Section~\ref{ssec:hol-ren-threedim-anomalies-improvements} that for $d=1$ 
a boost anomaly with the properties above indeed exists.

\subsection{Weyl transformations}\label{subsec:Weylcov}
The fact that $T^\mu$ and $T^{\mu\nu}$ transform with definite Weyl weights
suggests that we should be able to find a Weyl-covariant formulation of the Ward identities.
In fact, we can see that this is what the modified form of the diffeomorphism Ward identities in~\eqref{eq:energyeqplusWeyl} and~\eqref{eq:diffeoWIspatialprojplusWeyl} achieves.
The subject of Weyl covariance is discussed in more detail in Appendix~\ref{ssec:bulk-improvements-weyl-cov-derivs},
but we will now mention a few key results which will allow us to see the Weyl covariance of the aforementioned equations.

As we show in~\eqref{eq:b-tilde-def-and-weyl-tr},
we can interpret the following object as a Weyl connection
\begin{equation}
    \tilde b_\mu=a_\mu+\frac{1}{d}K\tau_\mu\,,
    \qquad
    \label{eq:Wtrafotildeb}
    \delta_{\Lambda_D}\tilde b_\mu=\partial_\mu\Lambda_D\,.    
\end{equation}
Additionally, the affine connection transforms $\mathcal{C}^\rho_{\mu\nu}$ transforms as
\begin{equation}
    \delta_{\Lambda_D}\mathcal{C}^\rho_{\mu\nu}=-v^\rho\tau_\mu\tau_\nu\mathcal{L}_v\Lambda_D-h_{\mu\nu}h^{\rho\sigma}\partial_\sigma\Lambda_D+\delta^\rho_\mu\partial_\nu\Lambda_D+\delta^\rho_\nu\partial_\mu\Lambda_D\,.
\end{equation}
We can use these observations to build Weyl-covariant derivatives of tensors with a definite Weyl weight. 
For example, let $X^{\mu\nu}$ be a spatial STF tensor and assume that it has Weyl weight $w$,
so that
\begin{equation}
    \delta_{\Lambda_D}X^{\mu\nu}=w\Lambda_D X^{\mu\nu}\,.
\end{equation}
Then we can see that the following combination also has Weyl weight $w$,
\begin{equation}\label{eq:Weylcovdiv}
    \left(\mathcal{D}_\mu-(w+d+3)a_\mu\right)X^{\mu\nu}\,.
\end{equation}
This can be seen either by direct computation using the properties above
or using the general result in~\eqref{eq:weyl-homog-covariant-deriv}.
Additionally,
note that $X^{\mu}{}_{\rho}=X^{\mu\nu}h_{\nu\rho}$ has Weyl weight $w+2$
and
\begin{equation}
   h_{\nu\kappa}\left(\mathcal{D}_\mu-(w+d+3)a_\mu\right)X^{\mu\nu}=h^\rho_{\kappa}\left(\mathcal{D}_\mu-(w+d+3)a_\mu\right)X^{\mu}{}_{\rho}\,,
\end{equation}
likewise transforms covariantly with weight $w+2$.
As a second example, let $X^\mu$ be any vector with Weyl weight $w$.
Then we can check that
\begin{equation}\label{eq:Weylcovdivvector}
    \left(\mathcal{D}_\mu-(w+d+1)\tilde b_\mu\right)X^{\mu}
\end{equation}
has Weyl weight $w$.
This latter statement generalises straightforwardly to any antisymmetric $(p,0)$-tensor of Weyl weight $w$.
On the other hand, if we have a spatial one-form $X_\nu$ of Weyl weight $w$,
then the precise form of the Weyl-covariant derivative depends on whether it involves the divergence, exterior derivative or some STF projection of the $\mathcal{D}_\mu X_\nu$ covariant derivative.
Concretely, we find that
\begin{gather}
    h^{\mu\nu}\left(\mathcal{D}_\mu-(w+d-2)a_\mu\right)X_\nu\,,
    \\
    \left(\partial_\mu-w\tilde b_\mu\right)X_\nu-\left(\partial_\nu-w\tilde b_\nu\right)X_\mu\,,
    \\
    h^{\rho}_{\langle\mu}h^\sigma_{\nu\rangle}\left(\mathcal{D}_\rho-(w-2)a_\rho\right)X_\sigma\,.\label{eq:STFprojWcovD1form}
\end{gather}
are Weyl-covariant combinations.

Similar observations can be made when we have a Lie derivative with respect to $v^\mu$ acting on some scalar, one-form or another tensor.
For example, if $f$ is a scalar field with Weyl weight $w$, then 
\begin{equation}\label{eq:Lievscalar}
    \left(\mathcal{L}_v+\frac{w}{d}K\right)f\,,
\end{equation}
has Weyl weight $w-1$, keeping in mind that $v^\mu$ has weight $-1$.
This extends to the case where $f_\kappa$ is a spatial 1-form of weight $w$ or in fact to any spatial $(0,q)$ tensor.
In particular, since the shear~$C_{\mu\nu}$ is spatial and has Weyl weight $+1$ by~\eqref{eq:car-cov-bondi-gauge-tr-of-shear-on-shell-repeat},
this shows that the news tensor~$N_{\mu\nu}$ we defined in~\eqref{eq:NewsDefn},
\begin{equation}
  \label{eq:NewsDefn-firstrepeat}
  N_{\mu\nu}
  = -\left(\mathcal{L}_v +\frac{1}{d}K\right) C_{\mu\nu}\,,
\end{equation}
is Weyl-covariant with weight zero,
which is one of our main motivations for using this particular definition.
On the other hand, when $\mathcal{L}_v$ acts on a vector (or a higher-rank $(p,0)$ tensor), things are a bit different.
Let $X^\rho$ be a spatial vector of weight $w$. Then we have
\begin{equation}
  \delta_{\Lambda_D}
  \left(\LL_v X^\rho\right)
  = (w-1) \Lambda_D \LL_v X^\rho
  + w \LL_v \Lambda_D X^\rho
  + v^\rho X^\sigma \pd_\sigma \Lambda_D\,.
\end{equation}
If we project this with $h^\mu_\rho$, we see that 
\begin{equation}\label{eq:WeylcovLievvector}
    h^\mu_\rho \left(
    \LL_v
    + \frac{w}{d} K
  \right) X^\rho\,,
\end{equation}
is Weyl-covariant with weight $w-1$.
With this, we now have enough data to check that the form of the diffeomorphism Ward identities in~\eqref{eq:energyeqplusWeyl} and~\eqref{eq:diffeoWIspatialprojplusWeyl} is entirely composed of Weyl-covariant building blocks.

It is also useful to note that
the exterior derivative of the Weyl connection $\tilde{b}_\mu$ is Weyl-invariant.
Its space and time projections therefore give us Weyl-covariant objects, too,
and they can be written as follows,
\begin{subequations}
  \label{eq:weyl-connection-curvature-projections}
  \begin{align}
    v^\rho h^\sigma_\mu
    \left(\pd_\rho\tilde b_\sigma-\pd_\sigma\tilde b_\rho\right)
    &=\LL_v a_\mu
    +\frac{1}{d}K a_\mu
    +\frac{1}{d} h^\rho_\mu \pd_\rho K\,,
    \\
    h^\rho_\mu h^\sigma_\nu
    \left(\pd_\rho\tilde b_\sigma-\pd_\sigma\tilde b_\rho\right)
    &=\left(\LL_v+\frac{1}{d}K\right)F_{\mu\nu}\,.
  \end{align}
\end{subequations}
The former is a weight $-1$ covector,
though this is not obvious from the expression on the right-hand side.
The latter is again Weyl invariant.

\subsection{Transformation properties of the energy-momentum-news complex}\label{subsec:EMTnewstrafos}
We now generalize the discussion of Section~\ref{subsec:trafoEMT} to include the shear and the news, and for this we will specialize to $d=2$.
The variation of the action is then
\begin{equation}\label{eq:varactionnoshear}
  \delta S
  = \int d^{3} x\, e \left(
    T^\mu \delta \tau_\mu
    + \frac{1}{2} T^{\mu\nu} \delta h_{\mu\nu}+\frac{1}{2}S^{\mu\nu}\delta C_{\mu\nu}
  \right)\,.
\end{equation}
We will assume that $S$ is invariant under diffeomorphisms and Weyl transformations.
Under a Carroll boost, we have
\begin{equation}
    \delta_g S=\int d^{3} x\, e\lambda_\mu\left(T^\mu-\mathcal{D}_\nu S^{\mu\nu}+a_\nu S^{\mu\nu}\right)\,.
\end{equation}
If there is no boost anomaly, the left-hand side is zero,
and this gives us the boost Ward identity we derived in~\eqref{eq:boundary-boost-WI},
\begin{equation}
    T^\rho h^\mu_\rho=\left(\mathcal{D}_\nu -a_\nu\right) S^{\mu\nu}\,.
\end{equation}
Instead, we will now allow this Ward identity to be anomalous,
\begin{equation}\label{eq:defanom}
    T^\rho h^\mu_\rho=\left(\mathcal{D}_\nu -a_\nu\right) S^{\mu\nu}+\mathcal{A}_B^\mu\,,
\end{equation}
where the boost anomaly $\mathcal{A}_B^\mu$ is a local expression built from the $(\tau_\mu, h_{\mu\nu}, C_{\mu\nu})$ boundary variables and derivatives thereof.
In terms of the action, this means
\begin{equation}\label{eq:gaugetrafoSos}
    \delta_g S=\int d^{3} x\, e\lambda_\mu\mathcal{A}_B^\mu\,.
\end{equation}
Including the news and shear in the key condition~\eqref{eq:symmetryEMT} means we now demand
\begin{equation}\label{eq:comdeltadeltaggeneral}
    \left[\delta_g\,,\delta\right]S = \int d^{3}x e\Big(T^\mu\left[\delta_g\,,\delta\right]\tau_\mu+\frac{1}{2}T^{\mu\nu}\left[\delta_g\,,\delta\right]h_{\mu\nu}+\frac{1}{2}S^{\mu\nu}\left[\delta_g\,,\delta\right]C_{\mu\nu}\Big)\,.
\end{equation}
This leads to the requirement
\begin{align}
    0 & = \int d^{3}xe\left[\left(\delta_g T^\mu-\mathcal{L}_\chi T^\mu+4\Lambda_D T^\mu\right)\delta\tau_\mu+\frac{1}{2}\left(\delta_g T^{\mu\nu}-\mathcal{L}_\chi T^{\mu\nu}+5\Lambda_D T^{\mu\nu}\right)\delta h_{\mu\nu}\right.\nonumber\\
    &\quad\left.+\frac{1}{2}\left(\delta_g S^{\mu\nu}-\mathcal{L}_\chi S^{\mu\nu}+4\Lambda_D S^{\mu\nu}\right)\delta C_{\mu\nu}+\frac{1}{2}S^{\mu\nu}\delta\Sigma_{\mu\nu}\right]-\delta\int d^{3}xe\lambda_\mu\mathcal{A}_B^\mu\,,
\end{align}
where the inhomogeneous shear transformation $\Sigma_{\mu\nu}$ is defined in \eqref{eq:Sigma}.
A straightforward but somewhat lengthy computation gives
\begin{align}
    \frac{1}{2}S^{\mu\nu}\delta\Sigma_{\mu\nu} & =  \Big(v^\mu\mathcal{D}_\rho\left(S^{\rho\sigma}\lambda_\sigma\right)-h^{\mu\rho}\lambda^\sigma\mathcal{L}_v S_{\rho\sigma}\Big)\delta\tau_\mu-\left(\mathcal{D}_\nu-a_\nu\right)S^{\mu\nu}\delta\lambda_\mu\\
    &\qquad +\frac{1}{2}\left(-\mathcal{L}_\lambda S^{\mu\nu}-2S^{\mu\nu}\mathcal{D}_\rho\lambda^\rho+2\lambda^\nu\mathcal{D}_\rho S^{\mu\rho}-2\lambda^\rho v^\mu F^{\nu\sigma}S_{\rho\sigma}\right)\delta h_{\mu\nu}\,.\nonumber
\end{align}
Using \eqref{eq:defanom}, we then end up with 
\begin{align}
    0 & =  \int d^{3}xe\left[\left(\delta_g T^\mu-\mathcal{L}_\chi T^\mu+4\Lambda_D T^\mu+v^\mu\mathcal{D}_\rho\left(S^{\rho\sigma}\lambda_\sigma\right)-h^{\mu\rho}\lambda^\sigma\mathcal{L}_v S_{\rho\sigma}\right)\delta\tau_\mu\right.\nonumber\\
    &\qquad \left.+\frac{1}{2}\left(\delta_g T^{\mu\nu}-\mathcal{L}_\chi T^{\mu\nu}+5\Lambda_D T^{\mu\nu}-2\lambda^{(\mu}v^{\nu)}\tau_\rho T^\rho-\mathcal{L}_\lambda S^{\mu\nu}-2S^{\mu\nu}\mathcal{D}_\rho\lambda^\rho\right.\right.\nonumber\\
    &\qquad\left.\left.+2\lambda^{(\nu}\mathcal{D}_\rho S^{\mu)\rho}-2\lambda^\rho v^{(\mu} F^{\nu)\sigma}S_{\rho\sigma}\right)\delta h_{\mu\nu}\right.\nonumber\\
    &\qquad\left.+\frac{1}{2}\left(\delta_g S^{\mu\nu}-\mathcal{L}_\chi S^{\mu\nu}+4\Lambda_D S^{\mu\nu}\right)\delta C_{\mu\nu}\right]-\int d^{3}x\lambda_\mu\delta\left(e\mathcal{A}_B^\mu\right)\,.
\end{align}
Defining again the functional $A_B$ as in~\eqref{eq:anomalyfunc}, which now also depends on $C_{\mu\nu}$
(and where we keep $\lambda_\mu$ fixed in the variations below),
we finally obtain the following transformation rules,
\begin{align}
    \delta_g T^\mu & = \mathcal{L}_\chi T^\mu-4\Lambda_D T^\mu-v^\mu\mathcal{D}_\rho\left(S^{\rho\sigma}\lambda_\sigma\right)+h^{\mu\rho}\lambda^\sigma\mathcal{L}_v S_{\rho\sigma}+\frac{\delta A_B}{\delta\tau_\mu}\,,\label{eq:gaugetrafoTshear}\\
    \delta_g T^{\mu\nu} & =  \mathcal{L}_\chi T^{\mu\nu}-5\Lambda_D T^{\mu\nu}-2\lambda^{(\mu}T^{\nu)}+2\lambda^{(\mu}\mathcal{A}_B^{\nu)}+2\lambda^\rho v^{(\mu} F^{\nu)\sigma}S_{\rho\sigma}
    \nonumber\\
    &\qquad-2\lambda^{(\nu}a_\rho S^{\mu)\rho}+\mathcal{L}_\lambda S^{\mu\nu}+2S^{\mu\nu}\mathcal{D}_\rho\lambda^\rho+\frac{\delta A_B}{\delta h_{\mu\nu}}\,,\label{eq:gaugetrafoTshear2}\\
    \delta_g S^{\mu\nu} & =  \mathcal{L}_\chi S^{\mu\nu}-4\Lambda_D S^{\mu\nu}+\frac{\delta A_B}{\delta C_{\mu\nu}}\,,\label{eq:gaugetrafoTshear3}
\end{align}
where in the expression for $\delta_g T^{\mu\nu}$ we used \eqref{eq:defanom}.
From \eqref{eq:gaugetrafoTshear}, we can work out the transformation of $h^\mu_\rho T^\rho$, and by subtracting the gauge transformation of $\left(\mathcal{D}_\nu-a_\nu\right)S^{\mu\nu}$ it follows using~\eqref{eq:defanom} that
\begin{equation}\label{eq:anomtrafo}
    \delta_g\mathcal{A}^\mu_B=\mathcal{L}_\chi\mathcal{A}^\mu_B-4\Lambda_D\mathcal{A}^\mu_B+\delta_\lambda\mathcal{A}^\mu_B\,.
\end{equation}
Here, $\delta_\lambda\mathcal{A}^\mu_B$ is the transformation of the anomaly with respect to Carroll boosts which, as we will not need this in general, we did not work out.
We conclude that the anomaly transforms as a tensor with Weyl weight -4.
In Section~\ref{ssec:hol-ren-fourdim-anomalies-improvements}, we will show that general relativity near asymptotic null infinity in four dimensions gives an explicit realisation of such an object.
Finally, equation \eqref{eq:comdeltadeltaggeneral} implies that Wess--Zumino consistency is obeyed since it implies
\begin{equation}\label{eq:WZcon}
    \left(\left[\delta_1\,,\delta_2\right]+\delta_{[1,2]}\right)S = 0\,,
\end{equation}
for two gauge transformations $\delta_1$ and $\delta_2$.
Wess--Zumino consistency can also be obtained directly from the gauge transformations of the anomaly.
This requires computing the transformation of $\mathcal{A}_B^\mu$ under Carroll boosts. We will do this explicitly in Section~\ref{ssec:hol-ren-fourdim-anomalies-improvements}, where we work with an explicit form of the boost anomaly.

\section{Bondi loss as diffeomorphism Ward identities}
\label{sec:bulk-conservation-equations}
As we saw in Section~\ref{sec:radial-expansion}, the bulk Bianchi identities imply that not all orders in the radial expansion of the bulk equations of motion are independent.
In particular, we saw that the double time projection $U^\mu U^\nu R_{\mu\nu}=0$
and the space-time projection $\Pi^\mu_\rho U^\nu R_{\mu\nu}=0$
are only non-trivial at order $r^{-d}$ for $d+2$ bulk spacetime dimensions.
Furthermore, we expect that these equations should provide us with covariant generalisations of the Bondi loss equations.

In Section~\ref{sec:variations-ward-ids}, we subsequently derived the general form of the classical Ward identities in a field theory coupled to a background conformal Carrollian geometry consisting of metric and shear tensors.
Varying these background fields leads to an energy-momentum tensor (EMT) and a news tensor,
which together satisfy diffeomorphism, Weyl and Carroll boost Ward identities.
Our goal in this section is to rewrite the aforementioned components of the bulk equations, which generalise the Bondi loss equations, to be manifestly of the form of the boundary diffeomorphism Ward identities.
In doing so, we will derive a EMT-news complex from the diffeomorphism Ward identity formulation of the bulk equations of motion.

Of course, in this way, we can only determine these currents up to terms that leave the diffeomorphism Ward identity unchanged, which correspond to possible improvements of the EMT-news complex.
We will subsequently explore such improvement terms in Section~\ref{sec:bulk-improvements}.

We start this section with the case of $d=1$, corresponding to three-dimensional bulk spacetime.
As most boundary geometric invariants vanish in this case, the computation is straightforward.
We rewrite the bulk equations into the form of the boundary diffeomorphism Ward identity and extract the corresponding EMT-news currents in Section~\ref{ssec:ConservationEqnsd1}.
The case of $d=2$ is significantly more complicated, and we will initially consider a simplified setting with $d\tau=0$ in Section~\ref{ssec:bondi-loss-4d-simplified}.
We then turn to the general case in Section~\ref{ssec:bondi-loss-4d-general}.
Finally, we consider the effect of including logarithmic terms in the radial expansion in Section~\ref{ssec:log-emt-news}.

\subsection{Bondi conservation equations in three dimensions}
\label{ssec:ConservationEqnsd1}
Our starting point is the diffeomorphism Ward identity we derived in Section~\ref{ssec:variations-ward-ids-boundary-ward-identities}
from varying a boundary action coupled to a conformal Carroll background.
In three bulk dimensions,
its time and space projections as given in~\eqref{eq:diffeo-WI-threedim} are
\begin{subequations}
\begin{align}
  0
  &=
  -\left(\mathcal{L}_v-2K\right)\left(\tau_\rho T^\rho\right)
+e^{-1}\left(\partial_\mu+a_\mu\right)\left(eT^\rho h^\mu_\rho\right)-K\left(\tau_\rho T^\rho+h_{\mu\nu}T^{\mu\nu}\right)
  \,,
  \label{eq:d=1DiffWI-1R1}
  \\
  0
  &=
  -\left(\mathcal{L}_v-K\right) P_\kappa
  +h^\mu_\kappa\left(\partial_\mu+2a_\mu\right)\left( T^{\rho\sigma}h_{\rho\sigma} \right)
  -a_\kappa \left(
    \tau_\rho T^\rho
    +h_{\mu\nu}T^{\mu\nu}
  \right)\,.
  \label{eq:d=1DiffWI-2R2}
\end{align}
\end{subequations}
We now want to match these equations to the order $r^{-1}$ contributions to the $U^\mu U^\nu R_{\mu\nu}=0$ and the $U^\mu \Pi^\nu_\kappa R_{\mu\nu}=0$ equations of motion.
The expressions we get for these equations of motion are significantly simpler in three bulk dimensions,
and we saw in~\eqref{eq:U-U-R-d1} that the double time projection reduces to
\begin{align} \label{eq:U-U-R-d1-repeat}
  0
  = U^\mu U^\nu R_{\mu\nu}
  &= \frac{1}{2}\hat D_\rho\left(\Pi^{\rho\sigma}\left(\partial_\sigma+\mathcal{A}_\sigma\right)S\right)+\mathcal{L}_U\bar{\mathcal{K}}-r^{-1}S\bar{\mathcal{K}}\nonumber\\
  &{}\qquad-\frac{1}{2}S\left(\hat D_\rho+\mathcal{A}_\rho\right)\mathcal{Z}^\rho-\frac{1}{2}\mathcal{Z}^\rho\left(\partial_\rho+\mathcal{A}_\rho\right)S
   -\frac{1}{2}S\mathcal{Z}_\mu\mathcal{Z}^\mu 
  -\bar{\mathcal{K}}^2\nonumber
  \\
  &{}\qquad+\frac{1}{2}S\partial_r^2 S-\frac{1}{2}\bar{\mathcal{K}}\left(\partial_rS-r^{-1}S\right)-\frac{1}{2}S\left(\partial_r\bar{\mathcal{K}}+r^{-1}\bar{\mathcal{K}}\right)
  \,.
\end{align}
Recall that we have
$\mathcal{A}_\mu = \LL_U V_\mu$
and
$\mathcal{K} = - \frac{1}{2} \Pi^{\mu\nu} \LL_U \Pi_{\mu\nu}$,
while in~\eqref{eq:calZ}, in~\eqref{eq:BarCalK-def} and in~\eqref{eq:orderbarcalK}
we defined the composite objects
\begin{gather}
  \label{eq:calZ-def-repeat}
  \mathcal{Z}_\mu
  = \Pi_{\mu\nu}\partial_r U^\nu
  - \mathcal{A}_\mu
  = \mathcal{O}(r^{-1})\,,
  \\
  \label{eq:BarCalK-def-repeat}
  \overline{\mathcal{K}}
  = \mathcal{K} - \frac{d}{2r} S
  = \OO(r\inv)\,.
\end{gather}
Expanding~\eqref{eq:U-U-R-d1-repeat} and collecting all terms at order $r^{-1}$,
we find that only the terms on the first line contribute.
The result is
\begin{equation}
  \label{eq:d=1constrainteq}
  0
  =
  \left(\mathcal{L}_v-2K\right)\left(\overset{(1)}{\mathcal{K}}-\frac{1}{2}\overset{(0)}{S}\right)+e^{-1}\partial_\mu\left(e \left[
      h^{\mu\nu}\partial_\nu K+K a^\mu
  \right]\right)
  \,,
\end{equation}
where we remind the reader that $\overset{(1)}{\mathcal{K}}=e^{-1}\partial_\mu\left(e h^{\mu\nu}a_\nu\right)$ following~\eqref{eq:app-curly-K-expansion-d2-r-1}.
This is an evolution equation for the free parameter $\os{0}{S}$,
which in three dimensions parametrises the Bondi mass aspect.

Next, we saw in~\eqref{eq:U-Pi-R-d1-2} that the $U^\mu \Pi^\nu_\kappa R_{\mu\nu}=0$ equation in three dimensions is
\begin{align}\label{eq:U-Pi-R-d1-2-repeat}
  0
  = \Pi_\kappa^\mu U^\nu R_{\mu\nu}
  &= 
  -\frac{1}{2}\Pi^\alpha_\kappa\left(\partial_\alpha+\mathcal{A}_\alpha\right)\left(\partial_r S-r^{-1} S\right)
  +\frac{1}{2}S\Pi^\rho_\kappa\left(
    \partial_r\mathcal{Z}_\rho+\frac{1}{2}r^{-1}\mathcal{Z}_\rho\nonumber
  \right)
  \\
  &{}\qquad
  +\frac{1}{2}\left(\mathcal{L}_U-\bar{\mathcal{K}}\right)\mathcal{Z}_\kappa
  -\mathcal{A}_\kappa\bar{\mathcal{K}}\,.
\end{align}
At order $r^{-1}$, the resulting expression can be written as follows,
\begin{equation}
  \label{eq:constraintd=1spatial}
  0
  = \frac{1}{2}\left(\mathcal{L}_v-K\right)\overset{(1)}{\mathcal{Z}}_\kappa
  +\frac{1}{2}h^\mu_\kappa\left(\partial_\mu+2a_\mu\right)\overset{(0)}{S}
  -a_\kappa\overset{(1)}{\mathcal{K}}\,.
\end{equation}
As we can already see from~\eqref{eq:calZ-def-repeat} above,
or alternatively from~\eqref{eq:app-curly-Z-expansion-d2-r1-with-U2-or-hvPi0},
the spatial tensor $\os{1}{\mathcal{Z}}_\mu$ contains the arbitrary spatial vector
$h^\mu_\rho v^\nu \os{0}{\Pi}_{\mu\nu}$,
which parametrises the Bondi angular momentum aspect.

Indeed, comparing~\eqref{eq:d=1constrainteq} with~\eqref{eq:d=1DiffWI-1R1} and~\eqref{eq:constraintd=1spatial} with~\eqref{eq:d=1DiffWI-2R2},
we see that the three-dimensional bulk conservation equations immediately match the expected form of the boundary Ward identities once we identify the boundary current components with the following combinations of bulk metric variables,
\begin{subequations}
  \label{eq:SETd1}
  \begin{align}
    \tau_\rho T^\rho
    &= \frac{1}{2}\overset{(0)}{S}
    -e^{-1}\partial_\mu\left(e h^{\mu\nu}a_\nu\right),
    \label{eq:3DbdryEMT-1}
    \\
    h_{\rho\sigma}T^{\rho\sigma}
    &= -\frac{1}{2}\overset{(0)}{S}\,,
    \\
    T^\rho h_\rho^\mu
    &= h^{\mu\nu}\partial_\nu K\,,
    \\
    P_\kappa
    = T^{\mu\nu}\tau_\mu h_{\nu\kappa}
    &= \frac{1}{2}\overset{(1)}{\mathcal{Z}}_\kappa\,.
    \label{eq:3DbdryEMT-4}
  \end{align}
\end{subequations}
However, this matching procedure does not uniquely determine the boundary energy-momentum tensor,
since it can only fix it up to the addition of `improvement' terms that drop out of the diffeomorphism Ward identities~\eqref{eq:d=1DiffWI-1R1} and~\eqref{eq:d=1DiffWI-2R2}.
Using the energy-momentum tensor we obtained in~\eqref{eq:SETd1} above, we get
\begin{align}
    \tau_\rho T^\rho+h_{\rho\sigma}T^{\rho\sigma}
    &=-e^{-1}\partial_\mu\left(e h^{\mu\nu}a_\nu\right),
    \label{eq:traceT}
    \\
    T^\rho h_\rho^\mu
    &=  h^{\mu\nu}\partial_\nu K\,.
\end{align}
This does not agree immediately with the $d=1$ Weyl and Carroll boost Ward identities
\eqref{eq:boundary-Weyl-WI} and \eqref{eq:boundary-boost-WI} (which correspond to setting the shear terms in the linked expressions to zero).
However, this could simply mean that we have to perform an improvement transformation.

As we will see in Section~\ref{ssec:bulk-improvements-weyl-3d} later on, the right-hand side of \eqref{eq:traceT} can indeed be removed using such improvement terms.
After re-deriving the above current components from holographic renormalisation in Section~\ref{subsec:holrend=1},
we will discuss improvement terms again in Section~\ref{ssec:hol-ren-threedim-anomalies-improvements},
where we will also see that no improvements exist which can set $T^\rho h_\rho^\mu$ equal to zero,
resulting in a Carroll boost anomaly.

\subsection{Simplified Bondi loss equations in four dimensions}
\label{ssec:bondi-loss-4d-simplified}
Having obtained a boundary energy-momentum tensor by matching the order $r^{-1}$ contributions to the $d=1$ versions of the $U^\mu U^\nu R_{\mu\nu}=0$ and $U^\mu \Pi^\nu_\rho R_{\mu\nu}=0$ bulk equations of motion with the boundary diffeomorphism Ward identity, we now want to do the same thing for $d=2$.

This computation is significantly more difficult for several reasons.
First, the non-trivial contributions to the equations mentioned above now sit at order $r^{-2}$ in the radial expansion.
Extracting the Bondi-type evolution equations from the expansion is therefore more cumbersome, and we delegate most of these computations to Appendix~\ref{app:LossEqnLists}.
The general $d=2$ boundary geometry also contains several geometric objects that vanish identically in the $d=1$ case.
It includes the shear tensor $C_{\mu\nu}$ and the associated news current $S^{\mu\nu}$, which enter in the expected form of the diffeomorphism Ward identity we derived in~\eqref{eq:energyeq} and~\eqref{eq:diffeoWIspatialproj}.
The boundary twist
$F_{\mu\nu} = 2h_\mu^\rho h_\nu^\sigma \pd_{[\rho} \tau_{\sigma]}$
is another significant source of complication.

As a test case, we can therefore simplify the computation ahead
by setting the exterior derivative of the boundary clock form $\tau_\mu$ to zero.
We can do this using the boundary Weyl transformations and boundary Carroll boosts,
following similar arguments as the ones we used in Section~\ref{ssec:reduction-to-standard-Bondi--Sachs-gauge} to reduce our general covariant gauge to the standard Bondi--Sachs gauge.
In this case,
the twist~$F_{\mu\nu}$
and the acceleration~$a_\mu = 2 v^\rho \pd_{[\rho} \tau_{\mu]}$
both vanish,
and the resulting expressions simplify significantly.%
\footnote{%
    There is a hierarchy of simplifications here.
    The simplest option is to set $a_\mu=F_{\mu\nu}=0$. 
    Then, one can consider $F_{\mu\nu}=0$ but $a_\mu\neq 0$, and of course the general case where $\tau_\mu$ is arbitrary.
    On top of this, one can consider cases with either $K=0$ or $K\neq 0$.
    In this section, we will move directly from the simplest to the most difficult of these cases.
    As a result, a large part of the computational strategy will be quite different,
    but we decided to include the simplest case since it is easier to digest.
}
Of course, our main goal is to obtain Carroll-covariant results where $d\tau$ is arbitrary,
so the purpose of the simplified results we obtain here is to lead to the full computation, which we undertake in Section~\ref{ssec:bondi-loss-4d-general} below.

Another motivation for studying the $d\tau=0$ case separately is that it relates naturally to the standard Bondi--Sachs phase space.
In Appendix~\ref{app:reduction-to-standard-BS} we show how our notion of the boundary energy-momentum-news complex relates to the standard expressions in the literature for the Bondi mass and angular momentum.

\subsubsection{Simplified Bondi mass loss equation}
\label{sssec:bondi-loss-4d-simplified-mass}
With $d\tau$ and hence also $F_{\mu\nu}$ and $a_\mu$ set to zero,
the terms in the $U^\mu U^\nu R_{\mu\nu}=0$ equation of motion~\eqref{eq:U-U-R-d2} at order $r^{-2}$,
which we derive in~\ref{app:LossEqnLists-mass},
reduce to
\begin{align}
  \label{eq:UURd=2simplified}
  0
  &= -\left(\mathcal{L}_v-\frac{3}{2}K\right)\left(\overset{(1)}{S}-\overset{(2)}{\mathcal{K}}\right)
  +\frac{1}{2}\mathcal{D}_\mu\left(h^{\mu\nu} \partial_\nu 
  \os{0}{S}\right)
  \\
  &{}\qquad\nonumber
  -\frac{1}{2}C^{\mu\nu}\mathcal{D}_\mu \partial_\nu K
  -\frac{1}{4}N^2\,.
\end{align}
This expression contains the covariant news tensor
$N_{\mu\nu} = -\left(\mathcal{L}_v +\frac{1}{2}K\right)C_{\mu\nu}$ that we defined in~\eqref{eq:NewsDefn}.
We also use our conventions for spatial tensors, such as $N_{\mu\nu}$,
as outlined in Appendix~\ref{sapp:spatial-tensors}.
They include raising and lowering their indices with $h^{\mu\nu}$ and $h_{\mu\nu}$
and defining for example $N^2 = N^{\mu\nu} N_{\mu\nu}$ for brevity.

We now want to match~\eqref{eq:UURd=2simplified} to the time projection of the boundary diffeomorphism Ward identity~\eqref{eq:energyeq}.
With our simplifications, the latter is
\begin{align}
  \label{eq:DiffWardvProjR1}
  0
  &= -\left(\mathcal{L}_v-\frac{3}{2}K\right)\left(\tau_\rho T^\rho\right)
  -\frac{1}{2}K\left(
    \tau_\rho T^\rho
    +h_{\rho\sigma}T^{\rho\sigma}
    +\frac{1}{2}C_{\rho\sigma}S^{\rho\sigma}
  \right)
  \\
  &{}\qquad\nonumber
  -\frac{1}{2}S^{\rho\sigma}N_{\rho\sigma}
  +\mathcal{D}_\mu\left(T^\rho h^\mu_\rho\right)\,.
\end{align}
The structure of this expression is already quite similar to what we get from the equations of motion in~\eqref{eq:UURd=2simplified} above.
In particular, it seems likely that we will be able to identify at least part of the $\tau_\rho T^\rho$ term inside the Lie derivative in~\eqref{eq:DiffWardvProjR1} with the $\os{1}{S}-\os{2}{\mathcal{K}}$ terms inside the Lie derivative in~\eqref{eq:UURd=2simplified}.
Similarly, up to a factor $(1/2)$, the tensor $S^{\mu\nu}$ associated to the variation of the shear will likely be related to the news tensor $N^{\mu\nu}$, which is promising.
Additionally, 
$h^{\mu\nu}\partial_\nu\os{0}{S}$
is a spatial vector, and we can  incorporate the last term on the first line of~\eqref{eq:UURd=2simplified}
into the term involving $T^\rho h^\mu_\rho$ in~\eqref{eq:DiffWardvProjR1}.

The only apparent obstruction is the term $-(1/2) C^{\mu\nu}\mathcal{D}_\mu \partial_\nu K$ in~\eqref{eq:UURd=2simplified}, which, at this point, does not seem to fit in with any of the structures present in~\eqref{eq:DiffWardvProjR1}. 
We can move the covariant derivatives away from the $K$ factor,
so that we have
\begin{align}
  -\frac{1}{2} C^{\mu\nu} \mathcal{D}_\mu \partial_\nu K
  &= -\frac{1}{2} \mathcal{D}_\mu \mathcal{D}_\nu \left(C^{\mu\nu}K\right)
  + \mathcal{D}_\mu C^{\mu\nu} \mathcal{D}_\nu K
  + \frac{1}{2} K \mathcal{D}_\mu \mathcal{D}_\nu C^{\mu\nu}
  \\
  \label{eq:bondi-loss-4d-simplified-mass-rewrite-product-rule}
  &= \mathcal{D}_\mu \mathcal{D}_\nu \left(
    N^{\mu\nu}
    + h^{\mu\rho} h^{\nu\sigma} \LL_v C_{\rho\sigma}
  \right)
  \\
  &{}\qquad\nonumber
  + \mathcal{D}_\mu C^{\mu\nu} \pd_\nu K
  + \frac{1}{2} K \mathcal{D}_\mu \mathcal{D}_\nu C^{\mu\nu}\,.
\end{align}
The final term can now be absorbed in the terms in~\eqref{eq:DiffWardvProjR1} proportional to $K$,
while the double covariant derivative of $N^{\mu\nu}$ can be absorbed in the $\mathcal{D}_\mu\left(T^\rho h_\rho^\mu\right)$ term. Note that $\mathcal{D}_\nu N^{\mu\nu}$ is spatial.

With $d\tau=0$, the identity in~\eqref{eq:comLiecovSTF} also holds without the $h_{\nu\gamma}$ projection it is written with there.
Applying the resulting expression to the shear tensor, we get
\begin{equation}
    \left(\mathcal{L}_v-2K\right)\mathcal{D}_\mu C^{\mu\nu}=\mathcal{D}_\mu \left(\mathcal{L}_v-2K\right)C^{\mu\nu}=\mathcal{D}_\mu\left(h^{\mu\alpha}h^{\nu\beta}\mathcal{L}_v C_{\alpha\beta}\right)\,,
\end{equation}
as the $\tau_\nu$ projection is zero on both sides and where the second equality used that $a_\mu=0$.
We can now take the divergence of this equation and use the identity~\eqref{eq:vectorcase},
which results in
\begin{equation}
  \label{eq:bondi-loss-4d-simplified-mass-rewrite-lie-double-cov}
  \mathcal{D}_\mu \mathcal{D}_\nu \left(h^{\mu\rho}h^{\nu\sigma}\mathcal{L}_v C_{\rho\sigma}\right)
  = \mathcal{L}_v\left(\mathcal{D}_\mu \mathcal{D}_\nu C^{\mu\nu}\right)
  -2K\mathcal{D}_\mu \mathcal{D}_\nu C^{\mu\nu}
  -\mathcal{D}_\mu C^{\mu\nu}\partial_\nu K\,.
\end{equation}
Combining this with~\eqref{eq:bondi-loss-4d-simplified-mass-rewrite-product-rule}, we can then write the expression~\eqref{eq:UURd=2simplified} as
\begin{align}
  \label{eq:UURdtau0Ward}
  0
  &=  -\left(\mathcal{L}_v-\frac{3}{2}K\right)\left(
    \overset{(1)}{S}
    -\overset{(2)}{\mathcal{K}}
    -\mathcal{D}_\mu\mathcal{D}_\nu C^{\mu\nu}
  \right)
  \\
  &{}\qquad\nonumber
  +\mathcal{D}_\mu \left(
    \frac{1}{2}h^{\mu\nu}\partial_\nu\overset{(0)}{S}
    +\mathcal{D}_\nu N^{\mu\nu}
  \right)
  -\frac{1}{4}N^2\,,
\end{align}
which is manifestly of the same form as the time projection of the diffeomorphism Ward identity in~\eqref{eq:DiffWardvProjR1}.
This identification allows us to extract the following ansätze for several components of the energy-momentum-news complex,
\begin{subequations}
  \label{eq:bondi-loss-4d-simplified-rewrite-currents-from-mass-eqn}
  \begin{align}
    \tau_\rho T^\rho
    &= \os{1}{S}
    -\os{2}{\mathcal{K}}
    -\mathcal{D}_\mu\mathcal{D}_\mu C^{\mu\nu}\,,
    \\
    T^\rho h^\mu_\rho
    &= 
      \frac{1}{2}h^{\mu\nu}\partial_\nu\overset{(0)}{S}
      +\mathcal{D}_\nu N^{\mu\nu}
    \,,
    \\
    \label{eq:bondi-loss-4d-simplified-rewrite-currents-from-mass-eqn-trace}
    T^{\rho\sigma} h_{\rho\sigma}
    &= -\os{1}{S}
    +\os{2}{\mathcal{K}}
    +\mathcal{D}_\mu\mathcal{D}_\mu C^{\mu\nu}
    -\frac{1}{4}C\cdot N,
    \\
    \label{eq:bondi-loss-4d-simplified-rewrite-currents-from-mass-eqn-news}
    S^{\rho\sigma}
    &= \frac{1}{2}N^{\rho\sigma}\,,
  \end{align}
\end{subequations}
where we have used the abbreviation~$C\cdot N = C^{\mu\nu} N_{\mu\nu}$.
Some of these current components also appear in the spatial projection of the diffeomorphism Ward identity,
and we can therefore not be sure these ansätze are correct until we can align them with a similar rewriting of the bulk angular momentum loss equation.

\subsubsection{Simplified Bondi angular momentum loss equation}
\label{sssec:bondi-loss-4d-simplified-angular momentum}
We now turn to the spatial evolution equations,
starting from the order $r^{-2}$ contributions of the $\Pi^\mu_\kappa U^\nu R_{\mu\nu}=0$ equations of motion.
The general expression is derived in Appendix~\ref{app:LossEqnLists-momentum}.
With $d\tau$ and hence $a_\mu$ and $F_{\mu\nu}$ set to zero,
it reduces to
\begin{align}
  \label{eq:UPiRsimplified2}
  0
  &= -\frac{1}{2}\left(\mathcal{L}_v-K\right)\left(h^\rho_\kappa\overset{(2)}{\mathcal{Z}}_\rho\right)
  +\frac{1}{2}h^\rho_\kappa\partial_\rho\left(
    -\overset{(1)}{S}
    -2\overset{(2)}{\mathcal{K}}
    +\frac{1}{8}KC^2
  \right)
  \\
  &{}\qquad\nonumber
  + h^\rho_\kappa\mathcal{D}_\sigma\left(
    \frac{1}{2}\overset{(0)}{S}C^\sigma{}_\rho
    +h^{\sigma\alpha}h^\beta_\rho\overset{(0)}{\mathcal{K}}_{\alpha\beta}
    +\frac{1}{2}KD^\sigma{}_\rho
  \right)
  \\
  &{}\qquad\nonumber
  +\frac{1}{16}C^2h^\rho_\kappa\partial_\rho K
  -\frac{1}{4}N^{\mu\nu}h^\rho_\kappa \mathcal{D}_\rho C_{\mu\nu}
  -\frac{1}{2}h^\rho_\kappa \mathcal{D}_\sigma\left(C^{\sigma\lambda}N_{\lambda\rho}\right).
\end{align}
We now want to rewrite this equation to be manifestly of the form of the spatial projection of the diffeomorphism Ward identity~\eqref{eq:diffeoWIspatialproj}.
Here, the latter reduces to
\begin{align}
  \label{eq:hDiffIdentity}
  0
  &= -\left(\mathcal{L}_v-K\right) P_\kappa
  +\frac{1}{2}h^\mu_\kappa\partial_\mu\left(T^{\rho\sigma}h_{\rho\sigma}\right)
  +h^\lambda_\kappa\mathcal{D}_\mu\left(\tilde T^{\mu\sigma}h_{\sigma\lambda}\right)
  \\
  &{}\qquad\nonumber
  -\frac{1}{2}S^{\mu\rho}h^\nu_\kappa\mathcal{D}_\nu C_{\mu\rho}
  +h^{\nu}_\kappa\mathcal{D}_\mu\left(S^{\mu\rho}C_{\rho\nu}\right).
\end{align}
Again, there is already some similarity between the equations,
and in particular the first line of~\eqref{eq:UPiRsimplified2} maps well onto the first two terms of~\eqref{eq:hDiffIdentity}. 
The second line of~\eqref{eq:UPiRsimplified2} is almost of the form of the third term in~\eqref{eq:hDiffIdentity}
but,
since $\tilde{T}^{\mu\nu}$ needs to be trace-free,
we must split off the trace of the middle term,
which is
\begin{equation}
  \label{eq:curly-K-trace-order0}
  h^{\alpha\beta}\overset{(0)}{\mathcal{K}}_{\alpha\beta}
  = \overset{(2)}{\mathcal{K}}
  + \frac{1}{2} C\cdot N\,,
\end{equation}
as follows from expanding $\mathcal{K}=\Pi^{\mu\nu}\mathcal{K}_{\mu\nu}$.
This trace contribution can then be absorbed into the second term of~\eqref{eq:hDiffIdentity}.
The last term in~\eqref{eq:UPiRsimplified2} resembles the last term in~\eqref{eq:hDiffIdentity},
but the order of the shear and the news tensor in the contraction must be reversed.
For this, we can use the $d=2$ identity
\begin{align}
  \label{eq:UPiRsimplified2-CN-to-NC}
  C^{\lambda\sigma}N_{\lambda\rho}
  &= -N^{\lambda\sigma}C_{\lambda\rho}
  +h^\sigma_\rho C\cdot N\,,
\end{align}
which follows from applying~\eqref{eq:app-d2-spatial-STF-STF-product} to the news tensor and the shear tensor.
The additional term can be written as follows,
\begin{align}
  - \frac{1}{2} h^\rho_\kappa \mathcal{D}_\sigma \left(
    h^\sigma_\rho C\cdot N
  \right)
  &= - \frac{1}{2} h^\rho_\kappa \pd_\rho \left(C\cdot N\right)\nonumber
  \\
  &= \frac{1}{4} h^\rho_\kappa \pd_\rho \left(\LL_v C^2\right)
  - \frac{1}{4} h^\rho_\kappa \pd_\rho \left(K C^2\right)\nonumber
  \\
  \label{eq:STFProdAC-result}
  &= \frac{1}{4} \left(\LL_v -K\right)\left(h^\rho_\kappa \pd_\rho C^2\right)
  - \frac{1}{4} h^\rho_\kappa C^2 \pd_\rho K.
\end{align}
With this, we can then rewrite~\eqref{eq:UPiRsimplified2}
as follows,
\begin{align}
  \label{eq:UPiRsimplified4a}
  0
  &= -\left(\mathcal{L}_v-K\right)\left(
    \frac{1}{2}h^\rho_\kappa\overset{(2)}{\mathcal{Z}}_\rho
    -\frac{1}{4}h^\rho_\kappa\partial_\rho C^2
  \right)
  \\
  &{}\qquad\nonumber
  +\frac{1}{2}h^\rho_\kappa\partial_\rho\left(
    -\overset{(1)}{S}
    -\overset{(2)}{\mathcal{K}}
    +\frac{1}{8}KC^2
    +\frac{1}{2} C\cdot N
  \right)
  \\
  &{}\qquad\nonumber
  +h^\rho_\kappa\mathcal{D}_\sigma\left(
    \frac{1}{2}\overset{(0)}{S}C^\sigma{}_\rho
    +h^{\sigma\alpha}h^\beta_\rho\overset{(0)}{\mathcal{K}}_{\langle\alpha\beta\rangle}
    +\frac{1}{2}KD^\sigma{}_\rho
  \right)
  \\
  &{}\qquad\nonumber
  -\frac{3}{16}C^2h^\rho_\kappa\partial_\rho K
  -\frac{1}{4}N^{\mu\nu}h^\rho_\kappa \mathcal{D}_\rho C_{\mu\nu}
  +\frac{1}{2}h^\rho_\kappa \mathcal{D}_\sigma\left(N^{\sigma\lambda}C_{\lambda\rho}\right).
\end{align}
To get rid of the first term on the final line,
we can use the identity~\eqref{eq:STFProdAC-result} again.
The resulting expression then manifestly fits with form of the spatial projection of the diffeomorphism Ward identity in~\eqref{eq:hDiffIdentity},
\begin{align}
  \label{eq:UPiRsimplified5}
  0
  &= -\left(\mathcal{L}_v-K\right)\left(
    \frac{1}{2}h^\rho_\kappa\overset{(2)}{\mathcal{Z}}_\rho
    -\frac{1}{16}h^\rho_\kappa\partial_\rho C^2
  \right)
  \\
  &{}\qquad\nonumber
  +\frac{1}{2}h^\rho_\kappa\partial_\rho\left(
    -\overset{(1)}{S}-\overset{(2)}{\mathcal{K}}+\frac{1}{8}KC^2-\frac{1}{4}C\cdot N
  \right)
  \\
  &{}\qquad\nonumber
  +h^\rho_\kappa\mathcal{D}_\sigma\left(
    \frac{1}{2}\overset{(0)}{S}C^\sigma{}_\rho
    +h^{\sigma\alpha}h^\beta_\rho\overset{(0)}{\mathcal{K}}_{\langle\alpha\beta\rangle}
    +\frac{1}{2}KD^\sigma{}_\rho
  \right)
  \\
  &{}\qquad\nonumber
  -\frac{1}{4}N^{\mu\nu}h^\rho_\kappa \mathcal{D}_\rho C_{\mu\nu}
  +\frac{1}{2}h^\rho_\kappa \mathcal{D}_\sigma\left(N^{\sigma\lambda}C_{\lambda\rho}\right).
\end{align}
As before, this expression now allows us to read off ansätze for several components of the energy-momentum-news complex,
\begin{subequations}
  \label{eq:bondi-loss-4d-simplified-rewrite-currents-from-ang-mom-eqn}
  \begin{align}
  \label{eq:bondi-loss-4d-simplified-rewrite-currents-from-ang-mom-eqn-trace}
    T^{\rho\sigma} h_{\rho\sigma}
    &= -\overset{(1)}{S}
    - \os{2}{\mathcal{K}}
    +\frac{1}{8}KC^2
    -\frac{1}{4}C\cdot N\,,
    \\
    \tilde T^{\rho\sigma}
    &= \frac{1}{2}\overset{(0)}{S}C^{\rho\sigma}
    +h^{\rho\alpha}h^{\sigma\beta}\overset{(0)}{\mathcal{K}}_{\langle\alpha\beta\rangle}
    +\frac{1}{2}KD^{\rho\sigma}\,,
    \\
    P_\kappa
    &= \frac{1}{2}h^\rho_\kappa\overset{(2)}{\mathcal{Z}}_\rho
    -\frac{1}{16}h^\rho_\kappa\partial_\rho C^2\,,
    \\
    \label{eq:bondi-loss-4d-simplified-rewrite-currents-from-ang-mom-eqn-trace-news}
    S^{\rho\sigma}
    &= \frac{1}{2}N^{\rho\sigma}\,.
  \end{align}
\end{subequations}
We see that the expression for the trace $T^{\rho\sigma} h_{\rho\sigma}$ we obtain here
agrees with the expression we obtained in~\eqref{eq:bondi-loss-4d-simplified-rewrite-currents-from-mass-eqn-trace} from the mass loss equation.
Likewise, the news tensor $S^{\rho\sigma}$ we find here agrees with the one in~\eqref{eq:bondi-loss-4d-simplified-rewrite-currents-from-mass-eqn-news}.
In the present simplified case, the ansätze for the energy-momentum-news components that enter in both of the projections of the diffeomorphism Ward identity are therefore immediately compatible, and no further modifications are necessary.

\subsubsection{Resulting simplified energy-momentum-news complex}
\label{sssec:bondi-loss-4d-simplified-emt-news-complex}
In the previous two subsections we established that with $d\tau=0$ both of the bulk loss equations can be rewritten as boundary diffeomorphism Ward identities.
Combining the results in~\eqref{eq:bondi-loss-4d-simplified-rewrite-currents-from-mass-eqn} and~\eqref{eq:bondi-loss-4d-simplified-rewrite-currents-from-ang-mom-eqn},
this matching identifies the following components of the energy-momentum-news complex (up to improvements),
\begin{subequations}
  \label{eq:bondi-loss-4d-simplified-currents}
  \begin{align}
    \tau_\rho T^\rho
    &= \os{1}{S}
    -\os{2}{\mathcal{K}}
    -\mathcal{D}_\mu\mathcal{D}_\mu C^{\mu\nu}\,,
    \\
    T^\rho h^\mu_\rho
    &= 
      \frac{1}{2}h^{\mu\nu}\partial_\nu\overset{(0)}{S}
      +\mathcal{D}_\nu N^{\mu\nu}
    \,,
    \\
    T^{\rho\sigma} h_{\rho\sigma}
    &= -\os{1}{S}
    +\os{2}{\mathcal{K}}
    +\mathcal{D}_\mu\mathcal{D}_\mu C^{\mu\nu}
    -\frac{1}{4}C\cdot N,
    \\
    \tilde T^{\rho\sigma}
    &= \frac{1}{2}\overset{(0)}{S}C^{\rho\sigma}
    +h^{\alpha\langle\rho}h^{\sigma\rangle\beta}\overset{(0)}{\mathcal{K}}_{\alpha\beta}
    +\frac{1}{2}KD^{\rho\sigma}\,,
    \\
    P_\kappa
    &= \frac{1}{2}h^\rho_\kappa\overset{(2)}{\mathcal{Z}}_\rho
    -\frac{1}{16}h^\rho_\kappa\partial_\rho C^2\,,
    \\
    S^{\rho\sigma}
    &= \frac{1}{2}N^{\rho\sigma}\,.
  \end{align}
\end{subequations}
In here, we have
\begin{equation}
  \label{eq:curly-Z-order2}
  \os{2}{\mathcal{Z}}_\mu
  = 3 v^\rho h^\sigma_\mu \os{1}{\Pi}_{\rho\sigma}
  + \frac{1}{2} C_{\mu\rho} \mathcal{D}_\sigma C^{\rho\sigma}
  + \frac{1}{16} h_\mu^\rho \pd_\rho C^2\,,
\end{equation}
which is worked out in \eqref{eq:genZ2} (where we set $a_\mu=0=F_{\mu\nu}$). 

In terms of the components of the bulk metric, we have
\begin{subequations}
  \begin{align}
    \os{1}{S}
    &= - v^\mu v^\nu \os{1}{g}_{\mu\nu}
    + \frac{1}{8} K C^2\,,
    \\
    v^\mu h^\nu_\rho \os{1}{\Pi}_{\rho\sigma}
    &= v^\mu h^\nu_\rho \os{1}{g}_{\rho\sigma}\,.
  \end{align}
\end{subequations}
We can thus solve the above expressions for $\tau_\rho T^\rho$ and $P_\kappa$ for $v^\mu v^\nu \os{1}{g}_{\mu\nu}$ and $v^\mu h^\nu_\rho \os{1}{g}_{\rho\sigma}$.

The STF part of the stress tensor, $\tilde{T}^{\rho\sigma}$, features the spatial $D_{\mu\nu}$ tensor defined in~\eqref{eq:Dmunu-def}.
Its time evolution is fixed by the STF equation as per~\eqref{eq:STF1/rSimple}
but this tensor is otherwise arbitrary\footnote{One can repeat the calculation above in the presence of the log terms discussed in Section~\ref{ssec:radial-expansion-logs}, which remove the condition \eqref{eq:PDZ} on $D_{\mu\nu}$. When doing this, we again find the same form of the EMT as above, but where $D_{\mu\nu}$ now only obeys \eqref{eq:STF1/rSimple}.}.
Following~\eqref{eq:app-4Dsol-S0},
\eqref{eq:app-curly-K-expansion-d2-r-2}
and~\eqref{eq:STFpartofcalK0},
the other objects in~\eqref{eq:bondi-loss-4d-simplified-currents} are fully fixed in terms of the boundary Carroll metric data and the shear tensor,
\begin{align}
  \label{eq:S-order0}
  \os{0}{S}
  &= \frac{1}{2} h^{\mu\nu}\mathcal{R}_{\mu\nu}=\frac{1}{2}\mathcal{Q}\,,
  \\
  \label{eq:curly-K-order2}
  \overset{(2)}{\mathcal{K}}
  &=-\frac{1}{2}\mathcal{D}_\mu\mathcal{D}_\nu C^{\mu\nu}
  +\frac{1}{16}KC^2\,,
  \\
  \label{eq:curly-K-stf-order0}
  h^\mu_{\langle\alpha} h^\nu_{\beta\rangle}\overset{(0)}{\mathcal{K}}_{\mu\nu}
  &=-\frac{1}{2}h^\mu_{\langle\alpha}h_{\beta\rangle}^\nu h_{\nu\sigma}
  \mathcal{D}_\mu\mathcal{D}_\rho C^{\rho\sigma}\,.
\end{align}
Note that we cannot solve for the momentum stress current $\tilde{T}^{\mu\nu}$ in terms of a particular component in the radial expansion of the metric.
As such, this current appears to be less fundamental. We refer to \cite{Hartong:2025WIP2} for an alternative viewpoint on the boundary energy-momentum tensor that explains why this is the case.

Having confirmed that the bulk equations of motion imply the boundary diffeomorphism Ward identities for the EMT-news complex above,
we can now investigate the Weyl and Carroll boost Ward identities.
Following the expressions in~\eqref{eq:boundary-Weyl-WI} and~\eqref{eq:boundary-boost-WI} and keeping in mind that $a_\mu$ is set to zero, the EMT-news complex in~\eqref{eq:bondi-loss-4d-simplified-currents} leads to
\begin{align}
  T^\mu\tau_\mu
  +T^{\mu\nu}h_{\mu\nu}
  +\frac{1}{2}S^{\mu\nu}C_{\mu\nu}
  &= 0 \,,
  \\
  h_\sigma^\rho T^\sigma-\mathcal{D}_\mu S^{\mu\rho}
  &= 
    \frac{1}{2}h^{\mu\rho}\partial_\mu\overset{(0)}{S}
    +\frac{1}{2} \mathcal{D}_\mu N^{\mu\rho}
  \,.\label{eq:spatTmusimplified}
\end{align}
We see that the Weyl Ward identity is satisfied identically,
while the Carroll boost Ward identity is not.
We will show in Section~\ref{ssec:hol-ren-fourdim-anomalies-improvements}
that there exists no improvement of the EMT-news complex that make the right-hand side of \eqref{eq:spatTmusimplified} zero and so the Carroll boosts are anomalous. 
However, before we get to that, we first derive the general energy-momentum-news complex without the present simplifications.

In Appendix \ref{app:reduction-to-standard-BS}, we relate our present results for the Bondi mass and angular momentum aspects with the usual expressions for the metric in the standard Bondi--Sachs phase space.

\subsection{General Bondi loss equations in four dimensions}
\label{ssec:bondi-loss-4d-general}
We now repeat the above computation without any additional simplifying restrictions on the boundary Carrollian geometry.
This reintroduces many terms proportional to
$a_\mu = 2v^\rho \pd_{[\rho} \tau_{\mu]}$
and/or
$F_{\mu\nu} = 2 h_\mu^\rho h_\nu^\sigma \pd_{[\rho} \tau_{\sigma]}$,
which make rewriting the relevant equations of motion into the form of the diffeomorphism Ward identity significantly more complicated.

Since it has more structure than the Bondi mass loss equation coming from the $U^\mu U^\nu R_{\mu\nu}=0$ equation of motion, it will be convenient to start by rewriting the momentum loss equation coming from
the $U^\mu \Pi_\kappa^\nu R_{\mu\nu}=0$ equations of motion.
We do this in Section~\ref{sssec:d2AngLoss}.
To further simplify the rewriting of the mass loss equation in the form of the relevant diffeomorphism Ward identity,
we assume that the ansatz for the ${S}^{\mu\nu}$ news current that we will obtain from the momentum loss equation can also be used in the mass loss equation,
and this turns out to be consistent,
as we will see in Section~\ref{sssec:d2MassLoss}.
We then obtain an energy-momentum-news complex consistent with both sets of equations in Section~\ref{sssec:d2EOMCurrents},
confirming that the bulk equations of motion corresponding to the Bondi loss equations are equivalent to the boundary diffeomorphism Ward identity.

\subsubsection{General Bondi angular momentum loss equation}
\label{sssec:d2AngLoss}
Here, our starting point is the spatial projection of the boundary diffeomorphism Ward identity derived in~\eqref{eq:diffeoWIspatialproj},
which we can write as follows,
\begin{align}
  \label{eq:diffeoWIspatialproj-repeat}
  0
  &= - \left(\mathcal{L}_v - K\right) P_\kappa
  +h^\mu_\kappa \mathcal{D}_\nu
  \left(\tilde T^{\nu\sigma}h_{\sigma\mu}\right)
  +\frac{1}{2}h^\mu_\kappa\left(\partial_\mu+3a_\mu\right)\left(T^{\rho\sigma}h_{\rho\sigma}\right)
  \\
  &{}\qquad\nonumber
  +h^{\nu}_\kappa\mathcal{D}_\mu
  \left(S^{\mu\rho}C_{\rho\nu}\right)
  -\frac{1}{2}S^{\mu\rho}h^\nu_\kappa\left(\mathcal{D}_\nu-a_\nu\right) C_{\mu\rho}
  \\
  &{}\qquad\nonumber
  -a_\kappa\left(T^{\rho\sigma}h_{\rho\sigma}+\tau_\rho T^\rho+\frac{1}{2}S^{\rho\sigma}C_{\rho\sigma}\right)
  +T^\mu F_{\mu\kappa}\,.
\end{align}
Note that the two terms on the final line are distinguished by having a factor proportional to $a_\kappa$ or $F_{\rho\kappa}$ with a free $\kappa$ index.
This is one of the factors that makes it easier to start our present computation with the angular momentum loss equation.
The $\Pi^\mu_\kappa U^\nu R_{\mu\nu}=0$ equation of motion was given in~\eqref{eq:Pi-U-R-d2}
and reads
\begin{align}
  0
  = \Pi_\kappa^\mu U^\nu R_{\mu\nu}
  &= -\frac{1}{2}\mathcal{F}^\rho{}_\kappa\left(\partial_\rho +\mathcal{A}_\rho \right)S+\frac{1}{2}\mathcal{G}^\rho{}_\kappa\left(\partial_\rho +\mathcal{A}_\rho \right)S
  \\
  &{}\qquad\nonumber 
  -\partial_r\beta\Pi^\alpha_\kappa\left(\partial_\alpha +\mathcal{A}_\alpha \right)S-\frac{1}{2}\Pi^\alpha_\kappa\left(\partial_\alpha+\mathcal{A}_\alpha\right)\left(\partial_r S-r^{-1} S\right)
  \\
  &{}\qquad\nonumber
  -\frac{1}{2}S\Pi^\mu_\kappa\hat D_\rho \mathcal{F}^\rho{}_\mu+\frac{1}{2}S\Pi^\rho_\kappa\left(
    \partial_r\mathcal{Z}_\rho+r^{-1}\mathcal{Z}_\rho
  \right)
  \\
  &{}\qquad\nonumber
  +\frac{1}{2}\left(\mathcal{L}_U-\bar{\mathcal{K}}\right)\mathcal{Z}_\kappa
  +\frac{1}{2}\Pi^\mu_\kappa \left(\partial_\mu-\mathcal{A}_\mu\right)\bar{\mathcal{K}}
  -\Pi^\mu_\kappa\hat D_\rho \mathcal{K}^{T\rho}{}_{\mu}\,.
\end{align}
We want to match the diffeomorphism Ward identity~\eqref{eq:diffeoWIspatialproj-repeat}
with a suitable rewriting of
the order $r^{-2}$ contributions to this equation of motion,
which are worked out in Appendix~\ref{app:LossEqnLists-momentum}.
To organise their rewriting, it is useful to split these contributions into six different groups,
which can individually be written in the form of particular terms entering in the diffeomorphism Ward identity expression~\eqref{eq:diffeoWIspatialproj-repeat}.
This splitting is clearly not unique, but our choice below is motivated by counting the number of derivatives involved.
With this, we write the expression for the order $r^{-2}$ contributions to~\eqref{eq:Pi-U-R-d2}
obtained in~\eqref{eq:UPiRinterm2} from Appendix~\ref{app:LossEqnLists-momentum} as\footnote{This calculation is done without including the log terms of Section~\ref{ssec:radial-expansion-logs}. We will show in Section~\ref{ssec:log-emt-news} that including those log terms does not change the form of the expression for the EMT-news complex obtained from the equations of motion.}
\begin{align}
  \label{eq:UPiRinterm}
  0 
  &= B^1_\kappa + B^2_\kappa + B^3_\kappa + B^4_\kappa + B^5_\kappa + B^6_\kappa\,,
\end{align}
where the explicit expressions for the six `blocks' of terms are given by
\begin{subequations}
  \begin{align}
    B^1_\kappa
    &=
    \frac{1}{2}\left(\mathcal{L}_v-K\right)\left(h^\mu_\kappa\overset{(2)}{\mathcal{Z}}_\mu\right)
    + \frac{1}{2}h^\mu_\kappa\left(\partial_\mu+3a_\mu\right)\left(\overset{(2)}{\mathcal{K}}+\overset{(1)}{S}\right)
    -2a_\kappa\overset{(2)}{\mathcal{K}}
    \,,
    \\
    B^2_\kappa
    &=
    -Ka_\mu D^\mu{}_\kappa
    -D^\mu{}_\kappa\partial_\mu K\,,\label{eq:block-2}
    \\
    B^3_\kappa
    &=
    -\frac{3}{2}a^\mu F_{\mu\kappa}\overset{(0)}{S}
    -\frac{1}{2}\partial_\mu\overset{(0)}{S}F^\mu{}_\kappa
    -\frac{1}{2}\overset{(0)}{S}h_{\kappa\nu}\mathcal{D}_\mu F^{\mu\nu}
    -\frac{1}{2}\overset{(0)}{S}a_\mu C^\mu{}_\kappa
    \\
    &{}\qquad\nonumber
    -\frac{1}{2}C^\mu{}_\kappa\partial_\mu\overset{(0)}{S}
    +\frac{1}{2}\overset{(0)}{S}\overset{(1)}{\mathcal{Z}}_\kappa\,,
    \\
    B^4_\kappa
    &=
    -\frac{1}{2}\overset{(1)}{\mathcal{K}}\overset{(1)}{\mathcal{Z}}_\kappa
    -\frac{1}{2}a_\kappa a^\mu\overset{(1)}{\mathcal{Z}}_\mu
    -\frac{1}{2}h^\mu_\kappa\mathcal{L}_a\overset{(1)}{\mathcal{Z}}_\mu\,,
    \\
    B^5_\kappa
    &=
    a_\mu \overset{(-1)}{\mathcal{K}}{}^{T\,\mu\nu}F_{\nu\kappa}
    +\frac{1}{2}h^\mu_\kappa\overset{(-1)}{\mathcal{K}}{}^{T}{}^{\nu\rho}
    \mathcal{D}_\mu C_{\nu\rho}
    \\
    &{}\qquad\nonumber
    -h^\nu_\kappa\mathcal{D}_\rho
    \left(
      h^{\rho\sigma}h^\mu_\nu\overset{(0)}{\mathcal{K}}{}^T_{\sigma\mu}
      -C^{\rho\sigma}\overset{(-1)}{\mathcal{K}}{}^T_{\sigma\nu}
    \right),
    \\
    B^6_\kappa
    &=
    - \frac{1}{8}K\left(3F^2+C^2\right)a_\kappa
    + \partial_\mu K C^{\mu\nu}F_{\nu\kappa}
    - \frac{1}{2}KC_{\kappa\nu}\left(
      \mathcal{D}_\mu F^{\mu\nu}
      + a_\mu F^{\mu\nu}
    \right)
    \\
    &{}\qquad\nonumber
    + \frac{1}{8}\left(F^2-C^2\right)h^\mu_\kappa\partial_\mu K
    + \frac{1}{2}\overset{(1)}{\mathcal{K}}a^\mu F_{\mu\kappa}\,.
  \end{align}
\end{subequations}
Here, recall we have $\os{1}{\mathcal{K}}=e^{-1}\partial_\mu (ea^\mu)$ from~\eqref{eq:app-curly-K-expansion-d2-r-1},
and $\os{1}{\mathcal{Z}}_\kappa$ is given in~\eqref{eq:CalZ1new}.

In addition to the number of derivatives involved,
these terms are distinguished by the independent fields they contain.
For example,
since $D_{\mu\nu}$ is an independent field in the radial expansion of the metric,
the terms in $B^2_\kappa$ cannot combine with any of the other terms
and must fit into the form of the spatial diffeomorphism Ward identity~\eqref{eq:diffeoWIspatialproj-repeat} on their own.
We now rewrite the above `blocks' one by one and extract their contributions to the energy-momentum-news complex.

\paragraph{Block 1.}
The terms in this block already fit with the form of the diffeomorphism Ward identity~\eqref{eq:diffeoWIspatialproj-repeat}.
It contributes the following contributions to the current components:
\begin{subequations}
  \label{eq:gen-mom-eq-WI-matching-block-1-results}
  \begin{align}
    P_\kappa:
    &{}\qquad
    -\frac{1}{2} h^\mu_\kappa\overset{(2)}{\mathcal{Z}}_\mu\,,
    \\
    T^{\rho\sigma} h_{\rho\sigma}:
    &{}\qquad
    \overset{(2)}{\mathcal{K}}
    +\overset{(1)}{S}\,,
    \\
    T^\rho \tau_\rho:
    &{}\qquad
    \overset{(2)}{\mathcal{K}}
    -\overset{(1)}{S}\,.
  \end{align}
\end{subequations}
As before, recall that any expression for the energy-momentum-news components is not final until we have processed both the angular momentum and mass loss equations.
We then need to reach an agreement between the anzätze we extract for the components that appear in both equations\footnote{The issue is that there can be redefinitions of the components of the EMT-news complex that cancel out in one of the equations but not the other one.}.
At this point, we are therefore merely collecting contributions to a preliminary result.

\paragraph{Block 2.}
Next, we have the contributions coming from the $D_{\mu\nu}$ field,
which arises at order $r^0$ in the expansion of the
$\Pi_{\mu\nu}$ tensor following for example~\eqref{eq:Dmunu-def}.
The covariant derivative of this field is constrained by~\eqref{eq:PDZ},
which is the order $r^{-3}$ term in the
$\Pi^\mu_\rho R_{\mu r}=0$ equations of motion.
Using this constraint, we have
\begin{align}
  \label{eq:gen-mom-eq-WI-matching-block-2-results}
  B^2_\kappa
  &= -Ka_\mu D^\mu{}_\kappa
  -D^\mu{}_\kappa\partial_\mu K\nonumber
  \\
  &= - h^\mu_\kappa
  \mathcal{D}_\nu
  \left(KD^\nu{}_\mu\right)\,.
\end{align}
which corresponds to a contribution of
$- K D^\nu{}_\kappa$
to the
$\tilde{T}^{\nu\sigma} h_{\sigma\kappa}$
stress tensor. We note that $\mathcal{D}_\nu
  \left(KD^\nu{}_\mu\right)$ is already spatial, but we keep the $h^\mu_\kappa$ for ease of comparison.

\paragraph{Block 3.}
Next, we consider all terms containing $\os{0}{S}$,
which give
\begin{align}
\label{eq:Block3}
B^3_\kappa
    &=
    -\frac{3}{2}a^\mu F_{\mu\kappa}\overset{(0)}{S}
    -\frac{1}{2}\partial_\mu\overset{(0)}{S}F^\mu{}_\kappa
    -\frac{1}{2}\overset{(0)}{S}h_{\kappa\nu}\mathcal{D}_\mu F^{\mu\nu}
    -\frac{1}{2}\overset{(0)}{S}a_\mu C^\mu{}_\kappa
    \\
    &{}\qquad\nonumber
    -\frac{1}{2}C^\mu{}_\kappa\partial_\mu\overset{(0)}{S}
    +\frac{1}{2}\overset{(0)}{S}\overset{(1)}{\mathcal{Z}}_\kappa\,.
\end{align}
Since $\os{0}{S}$ is essentially  the boundary Ricci scalar, as can be seen in \eqref{eq:S0d=2},
it is unlikely that these terms would have to be combined with any other terms.
Furthermore, since the twist $F_{\mu\nu}$ and the shear $C_{\mu\nu}$ are independent variables,
the terms in this block that contain either of these factors will have to fit into the diffeomorphism Ward identity~\eqref{eq:diffeoWIspatialproj-repeat} separately.
Indeed, using~\eqref{eq:CalZ1new}, which says
\begin{equation}
\label{eq:CalZFC}
    \overset{(1)}{\mathcal{Z}}{}^\rho=-\mathcal{D}_\sigma C^{\sigma\rho}+a_\sigma C^{\sigma\rho}+\mathcal{D}_\sigma F^{\sigma\rho}-\frac{1}{2}v^\rho F^2+a_\sigma F^{\sigma\rho}\,,
\end{equation}
we find that
we can write this block as
\begin{equation}
  B^3_\kappa
  =
  -\frac{1}{2} h^\mu_\kappa\mathcal{D}_\nu
  \left(\overset{(0)}{S}C^\nu{}_\mu\right)
  -\left(\overset{(0)}{S}a^\mu+\frac{1}{2}h^{\mu\nu}\partial_\nu\overset{(0)}{S}\right)
  F_{\mu\kappa}\,.
\end{equation}
This corresponds to the following contributions
\begin{align}
  \label{eq:gen-mom-eq-WI-matching-block-3-results}
  \tilde{T}^{\nu\rho} h_{\rho\mu}:
  &{}\qquad
  -\frac{1}{2}\overset{(0)}{S}C^\nu{}_\mu\,,
  \\
  T^\mu:
  &{}\qquad
  -\overset{(0)}{S}a^\mu
  -\frac{1}{2}h^{\mu\nu}\partial_\nu\overset{(0)}{S}\,.
\end{align}
to the current components.

\paragraph{Block 4.}
We now collect the remaining terms
containing a factor of $\os{1}{\mathcal{Z}}_\mu$,
\begin{align}
  \label{eq:ZTerms}
  B^4_\kappa
  &=
  -\frac{1}{2}\overset{(1)}{\mathcal{K}}\overset{(1)}{\mathcal{Z}}_\kappa
  -\frac{1}{2}a_\kappa a^\mu\overset{(1)}{\mathcal{Z}}_\mu
  -\frac{1}{2}h^\mu_\kappa\mathcal{L}_a\overset{(1)}{\mathcal{Z}}_\mu\,,
  \\
  \nonumber
  &= - \frac{1}{2} \os{1}{\mathcal{Z}}_\kappa \mathcal{D}_\mu a^\mu
  - \frac{1}{2} h^\mu_\kappa a^\rho \mathcal{D}_\rho \os{1}{\mathcal{Z}}_\mu
  - \frac{1}{2} h^\mu_\kappa \os{1}{\mathcal{Z}}^\rho \mathcal{D}_\mu a_\rho
  - \frac{1}{2} a_\kappa a^\mu \os{1}{\mathcal{Z}}_\mu\,.
\end{align}
It turns out that we can rewrite this block into a suitable form
without using the explicit expression for~$\os{1}{\mathcal{Z}}_\mu$ given in~\eqref{eq:CalZFC} above.
In the second line of~\eqref{eq:ZTerms}, we rewrote the Lie derivative with respect to $a^\mu$ to a covariant derivative
and we substituted the expression $\os{1}{\mathcal{K}}
  =\mathcal{D}_\mu a^\mu$. 
The covariant derivative terms in~\eqref{eq:ZTerms} suggest matching it with the $\tilde{T}^{\mu\nu}$
STF momentum stress tensor contributions to the diffeomorphism Ward identity~\eqref{eq:diffeoWIspatialproj-repeat}.
Given the terms entering in~\eqref{eq:ZTerms}, a natural guess for such a contribution is
\begin{align}
  &h^\mu_\kappa\mathcal{D}_\nu
  \left(
    2\os{1}{\mathcal{Z}}^{\langle \nu} a^{\rho \rangle} h_{\rho\mu}
  \right)\nonumber
  \\
  &{}\quad
  = h^\mu_\kappa \mathcal{D}_\nu
  \left(
    \os{1}{\mathcal{Z}}^\nu a_\mu
    + a^\nu \os{1}{\mathcal{Z}}_\mu
    - h^\nu_\mu a_\rho \os{1}{\mathcal{Z}}^\rho
  \right)\nonumber
  \\
  &{}\quad
  = a_\kappa \mathcal{D}_\nu \os{1}{\mathcal{Z}}^\nu
  + \os{1}{\mathcal{Z}}^\nu h^\mu_\kappa \mathcal{D}_\nu a_\mu
  + \os{1}{\mathcal{Z}}_\kappa \mathcal{D}_\nu a^\nu
  + h^\mu_\kappa a^\nu \mathcal{D}_\nu \os{1}{\mathcal{Z}}_\mu\nonumber
  \\
  &{}\qquad
  -a_\kappa a_\rho \os{1}{\mathcal{Z}}^\rho
  - h^\mu_\kappa \pd_\mu \left(a_\rho \os{1}{\mathcal{Z}}^\rho\right)\,.
\end{align}
Using this, together with the identity~\eqref{eq:app-da-decomposition},
which implies
\begin{equation}
  \label{eq:commute-indices-of-derivative-on-acceleration}
  2h^\mu_\rho h^\nu_\sigma
  \left(\mathcal{D}_{[\mu}a_{\nu]}\right)
  =\mathcal{L}_vF_{\rho\sigma}\,,
\end{equation}
we then see that~\eqref{eq:ZTerms} can be written as
\begin{align}
  \label{eq:Z1iden}
  B^4_\kappa
  &=\frac{1}{2}\left(\mathcal{L}_v-K\right)
  \left(\os{1}{\mathcal{Z}}^\mu F_{\mu\kappa} \right)
  -\frac{1}{2}F_{\mu\kappa}
    \left(\mathcal{L}_v-K\right)\overset{(1)}{\mathcal{Z}}{}^\mu
  \\
  &{}\qquad\nonumber
  -\frac{1}{2}h^\mu_\kappa\mathcal{D}_\nu
  \left(
    a^\nu\overset{(1)}{\mathcal{Z}}_\mu
    +\overset{(1)}{\mathcal{Z}}{}^\nu a_\mu
    -h^\nu_\mu a^\rho \overset{(1)}{\mathcal{Z}}_\rho
  \right)
  \\
  &{}\qquad\nonumber
  -\frac{1}{2}h^\mu_\kappa\left(\partial_\mu+3a_\mu\right)
  \left(a^\nu \overset{(1)}{\mathcal{Z}}_\nu\right)
  +\frac{1}{2}a_\kappa\left(
    \mathcal{D}_\mu \overset{(1)}{\mathcal{Z}}{}^\mu+a_\mu \overset{(1)}{\mathcal{Z}}{}^\mu
  \right),
\end{align}
which corresponds to the following contributions
\begin{subequations}
  \label{eq:gen-mom-eq-WI-matching-block-4-results}
  \begin{align}
    P_\kappa:
    &{}\qquad
    - \frac{1}{2}\os{1}{\mathcal{Z}}^\rho F_{\rho\kappa}\,,
    \\
    \tilde{T}^{\nu\rho} h_{\rho\mu}:
    &{}\qquad
    -\frac{1}{2}a^\nu\overset{(1)}{\mathcal{Z}}_\mu
    -\frac{1}{2}\overset{(1)}{\mathcal{Z}}{}^\nu a_\mu
    +\frac{1}{2}h^\nu_\mu a_\rho \overset{(1)}{\mathcal{Z}}{}^\rho\,,
    \\
    T^{\rho\sigma} h_{\rho\sigma}:
    &{}\qquad
    - a_\rho \overset{(1)}{\mathcal{Z}}{}^\rho\,,
    \\
    T^\rho \tau_\rho:
    &{}\qquad
    - \frac{1}{2}\mathcal{D}_\rho \overset{(1)}{\mathcal{Z}}{}^\rho+\frac{1}{2} a_\rho \overset{(1)}{\mathcal{Z}}{}^\rho\,,
    \\
    T^\mu:
    &{}\qquad
    -\frac{1}{2} h^\mu_\rho
    \left(
      \mathcal{L}_v-K\right)\overset{(1)}{\mathcal{Z}}{}^\rho\,,
  \end{align}
\end{subequations}
to the components of the energy-momentum-news currents.

\paragraph{Block 5.}
We now collect all terms involving
$\os{-1}{\mathcal{K}}^T$ and $\os{0}{\mathcal{K}}^T$,
\begin{align}
  \label{eq:Block5}
  B^5_\kappa
  &=
  a_\mu \overset{(-1)}{\mathcal{K}}{}^{T\,\mu\nu}F_{\nu\kappa}
  +\frac{1}{2}h^\mu_\kappa\overset{(-1)}{\mathcal{K}}{}^{T}{}^{\nu\rho}
  \mathcal{D}_\mu C_{\nu\rho}
  \\
  &{}\qquad\nonumber
  -h^\nu_\kappa\mathcal{D}_\rho
  \left(
    h^{\rho\sigma}h^\mu_\nu\overset{(0)}{\mathcal{K}}{}^T_{\sigma\mu}
    -C^{\rho\sigma}\overset{(-1)}{\mathcal{K}}{}^T_{\sigma\nu}
  \right).
\end{align}
The terms on the second line look like a mix of contributions to
$\tilde{T}^{\rho\sigma}h_{\sigma\nu}$
and
$S^{\rho\sigma}C_{\sigma\nu}$,
but the former of these needs to be trace-free,
and the index structure of the latter does not yet match either.
To remedy this,
it is useful to write out the spatial projection of the $r^0$ contribution to $\mathcal{K}^T_{\mu\nu}$ following its definition in~\eqref{eq:app-KT-BarCalK-def} as follows,
\begin{align}
  h^\rho_\mu h^\sigma_\nu\os{0}{\mathcal{K}}^T_{\rho\sigma}
  &= h^\rho_\mu h^\sigma_\nu\os{0}{\mathcal{K}}_{\rho\sigma}
  - \frac{1}{2} h^\rho_\mu h^\sigma_\nu \left[
     \os{2}{\mathcal{K}}h_{\rho\sigma}
    + \os{1}{\mathcal{K}} \os{-1}{\Pi}_{\rho\sigma}
    +K \os{0}{\Pi}_{\rho\sigma} 
  \right]\nonumber
  \\
  &= \os{0}{\mathcal{K}}_{\langle\mu\nu\rangle}
  - \frac{1}{2} \os{1}{\mathcal{K}} C_{\mu\nu}
  - \frac{1}{2} K D_{\mu\nu}
  + \frac{1}{4} K F_{\mu\rho} C^\rho{}_\nu\label{eq:calKT0v2}
  \\
  &{}\qquad\nonumber
  - \frac{1}{2} h_{\mu\nu} \os{2}{\mathcal{K}}
  + \frac{1}{2} h_{\mu\nu} h^{\rho\sigma} \os{0}{\mathcal{K}}_{\rho\sigma}
  - \frac{1}{8} h_{\mu\nu} K C^2\,,
\end{align}
where we defined
\begin{equation}
    \os{0}{\mathcal{K}}_{\langle\mu\nu\rangle}=h^\rho_\mu h^\sigma_\nu\os{0}{\mathcal{K}}_{\rho\sigma}-\frac{1}{2}h_{\mu\nu}h^{\rho\sigma}\os{0}{\mathcal{K}}_{\rho\sigma}\,.
\end{equation}
Here, we used \eqref{eq:Dmunu-def} and we collected all the trace terms in the last line.
The first line of the second equality is spatial and symmetric trace-free with respect to~$h^{\mu\nu}$.
We can then write the $r^{-2}$ part of $\mathcal{K}$, which enters in the trace part of this expression, as
\begin{align}
  \os{2}{\mathcal{K}}
  &= h^{\mu\nu} \os{0}{\mathcal{K}}_{\mu\nu}
  + \os{3}{\Pi}^{\mu\nu} \os{-1}{\mathcal{K}}_{\mu\nu}
  + \frac{1}{2} \os{4}{\Pi}^{\mu\nu} K h_{\mu\nu}\nonumber
  \\
  &= h^{\mu\nu} \os{0}{\mathcal{K}}_{\mu\nu}
  - C \cdot \os{-1}{\mathcal{K}}
  + \frac{1}{4} K C^2.
\end{align}
These expressions allow us to rewrite the terms entering into the derivative on the last line of~\eqref{eq:Block5} as follows,
\begin{align}
  h^{\rho\sigma}h^\mu_\nu\overset{(0)}{\mathcal{K}}{}^T_{\sigma\mu}
  -C^{\rho\sigma}\overset{(-1)}{\mathcal{K}}{}^T_{\sigma\nu}
  &= h^{\rho\mu}\left( \os{0}{\mathcal{K}}_{\langle\mu\nu\rangle}
  - \frac{1}{2} \os{1}{\mathcal{K}} C_{\mu\nu}
  - \frac{1}{2} K D_{\mu\nu}
  + \frac{1}{4} K F_{\mu\sigma} C^\sigma{}_\nu\right)\nonumber
  \\
  &{}\qquad\nonumber
  + \frac{1}{2} h^\rho_\nu C\cdot\os{-1}{\mathcal{K}}
  - \frac{1}{4} h^\rho_\nu K C^2
  - C^{\rho\sigma}\overset{(-1)}{\mathcal{K}}{}^T_{\sigma\nu}
  \\
  &= h^{\rho\mu}\left( \os{0}{\mathcal{K}}_{\langle\mu\nu\rangle}
  - \frac{1}{2} \os{1}{\mathcal{K}} C_{\mu\nu}
  - \frac{1}{2} K D_{\mu\nu}
  + \frac{1}{4} K F_{\mu\sigma} C^\sigma{}_\nu\right)
  \\
  &{}\qquad\nonumber
  - \frac{1}{2} h^\rho_\nu C\cdot\os{-1}{\mathcal{K}}^T
  + \os{-1}{\mathcal{K}}{}^T{}^{\rho\sigma} C_{\sigma\nu}\,,
\end{align}
where we used the identity~\eqref{eq:app-d2-spatial-STF-STF-product} to modify the index structure on $- C^{\rho\sigma}\overset{(-1)}{\mathcal{K}}{}^T_{\sigma\nu}$.
To get to the slightly more compact expression on the last line,
we also used
\begin{equation}
  C\cdot \os{-1}{\mathcal{K}}
  = C\cdot \os{-1}{\mathcal{K}}^T
  + \frac{1}{2} K C^2\,,
\end{equation}
which follows from the fact that
$\os{-1}{\mathcal{K}}^T_{\mu\nu}=\os{-1}{\mathcal{K}}_{\mu\nu}-\frac{1}{2}h_{\mu\nu}\os{1}{\mathcal{K}}-\frac{1}{2}K\os{-1}{\Pi}_{\mu\nu}$.
In total, this allows us to write the fifth block of terms in~\eqref{eq:Block5} as follows,
\begin{align}
  B^5_\kappa
  &=
  -h^\mu_\kappa\mathcal{D}_\nu
  \left(\overset{(-1)}{\mathcal{K}}{}^{T\,\nu\rho} C_{\rho\mu}\right)
  + \frac{1}{2}\overset{(-1)}{\mathcal{K}}{}^{T}{}^{\nu\rho}h^\mu_\kappa\mathcal{D}_\mu C_{\nu\rho}
  \\
  &{}\qquad\nonumber
  +\frac{1}{2}h^\mu_\kappa\left(\partial_\mu+3a_\mu\right)\left(C\cdot\overset{(-1)}{\mathcal{K}}{}^T\right)
  -a_\kappa C\cdot\overset{(-1)}{\mathcal{K}}{}^T
  +a_\rho \overset{(-1)}{\mathcal{K}}{}^{T\,\mu\rho}F_{\mu\kappa}
  \\
  &{}\qquad\nonumber
  -h^\mu_\kappa\mathcal{D}_\nu
  \left[h^{\nu\rho}\left(
      \overset{(0)}{\mathcal{K}}_{\langle\rho\mu\rangle}
      -\frac{1}{2}KD_{\rho\mu}
      -\frac{1}{4}KF_\rho{}^\sigma C_{\sigma\mu}
      -\frac{1}{2}\overset{(1)}{\mathcal{K}}C_{\rho\mu}
  \right)\right]\,.
\end{align}
This matches the form of the diffeomorphism Ward identity~\eqref{eq:diffeoWIspatialproj-repeat} and gives rise to the following contributions
\begin{subequations}
  \label{eq:gen-mom-eq-WI-matching-block-5-results}
  \begin{align}
    S^{\mu\nu}:
    &{}\qquad
    -\overset{(-1)}{\mathcal{K}}{}^{T\,\mu\nu}\,,
    \\
    T^{\rho\sigma} h_{\rho\sigma}:
    &{}\qquad
    C\cdot\overset{(-1)}{\mathcal{K}}{}^T\,,
    \\
    T^\mu:
    &{}\qquad
    a_\rho \overset{(-1)}{\mathcal{K}}{}^{T\,\mu\rho}\,,
    \\
    \tilde{T}^{\nu\rho} h_{\rho\mu}:
    &{}\qquad
    -h^{\nu\rho}\left(
      \overset{(0)}{\mathcal{K}}_{\langle\rho\mu\rangle}
      -\frac{1}{2}KD_{\rho\mu}
      -\frac{1}{4}KF_\rho{}^\sigma C_{\sigma\mu}
      -\frac{1}{2}\overset{(1)}{\mathcal{K}}C_{\rho\mu}
  \right)\,,
  \end{align}
\end{subequations}
to the current components.

\paragraph{Block 6.}
Finally, the remaining terms are given by
\begin{align}
  \label{eq:block-six}
  B^6_\kappa
  &=
  - \frac{1}{8}K\left(3F^2+C^2\right)a_\kappa
  + \partial_\mu K C^{\mu\nu}F_{\nu\kappa}
  - \frac{1}{2}KC_{\kappa\nu}\left(
    \mathcal{D}_\mu F^{\mu\nu}
    + a_\mu F^{\mu\nu}
  \right)
  \\
  &{}\qquad\nonumber
  + \frac{1}{8}\left(F^2-C^2\right)h^\mu_\kappa\partial_\mu K
  + \frac{1}{2}\overset{(1)}{\mathcal{K}}a^\mu F_{\mu\kappa}\,.
\end{align}
All terms in this expression are proportional to the twist~$F_{\mu\nu}$ and/or the shear~$C_{\mu\nu}$,
and these are independent tensors.
We can therefore separate off the terms that only depend on the shear or only on the twist.
Using the definition of the news tensor~$N_{\mu\nu}$ in for example~\eqref{eq:NewsDefn},
it is easy to check that we have
\begin{align}
  &-\frac{1}{8}\left(\mathcal{L}_v-K\right)\left(h^\rho_\kappa\partial_\rho C^2\right)
  -\frac{1}{4}\left(h^\rho_\kappa\partial_\rho + 3a_\kappa\right)\left(C\cdot N\right)+\frac{1}{2}a_\kappa C\cdot N
  \nonumber
  \\
  \label{eq:gen-mom-loss-rewriting-C-terms}
  &{}\qquad
  =-\frac{1}{8}C^2 \left(
    h^\mu_\kappa\partial_\mu K + K a_\kappa
  \right).
\end{align}
This allows us to interpret the twist-independent terms in~\eqref{eq:block-six} as contributions to the $P_\kappa$ and $T^{\rho\sigma}h_{\rho\sigma}$ current components.
We can similarly check that the terms involving only the twist can be recovered from the following expression,
\begin{align}
  &\label{eq:gen-mom-loss-rewriting-F-terms}
  \frac{1}{8}\left(\mathcal{L}_v-K\right)\left(h^\mu_\kappa\partial_\mu F^2\right)
  -\frac{1}{8}h^\mu_\kappa\left(\partial_\mu+3a_\mu\right)\left(\mathcal{L}_v-K\right)F^2
  \\
  &{}\qquad\qquad\nonumber
  +a_\kappa\left(\frac{1}{4}(\mathcal{L}_v-K) F^2- \frac{1}{2}KF^2 \right)
  +\frac{1}{2}\overset{(1)}{\mathcal{K}}a^\mu F_{\mu\kappa}
  \\
  &{}\qquad
  =-\frac{3}{8}KF^2 a_\kappa
  +\frac{1}{8}F^2 h^\mu_\kappa\partial_\mu K
  +\frac{1}{2}\overset{(1)}{\mathcal{K}}a^\mu F_{\mu\kappa}
  \nonumber
\end{align}
which corresponds to $P_\kappa$,
$T^{\rho\sigma}h_{\rho\sigma}$
and $T^\rho$ contributions.

This leaves the terms in~\eqref{eq:block-six} that involve both the shear and the twist tensors.
To rewrite these into a suitable form,
we first rewrite the term involving a covariant derivative using the product rule,
which gives
\begin{align}
  \label{eq:IntId5}
  -\frac{1}{2}KC_{\kappa\nu}\mathcal{D}_\mu F^{\mu\nu}
  &= - \mathcal{D}_\mu \left(
    \frac{1}{2} K C_{\kappa\nu} F^{\mu\nu}
  \right)
  + \frac{1}{2} \pd_\mu K C_{\kappa\nu} F^{\mu\nu}
  + \frac{1}{2} K F^{\mu\nu} \mathcal{D}_\mu C_{\kappa\nu}\nonumber
  \\
  &= \mathcal{D}_\mu \left(
    \frac{1}{2} K C^{\mu\nu} F_{\nu\kappa}
  \right)
  - \frac{1}{2} \pd_\mu K C^{\mu\nu} F_{\nu\kappa}
  + \frac{1}{2} K F^{\mu\nu} \mathcal{D}_\mu C_{\kappa\nu}\nonumber
  \\
  &= h^\rho_\kappa \mathcal{D}_\mu
  \left(
    \frac{1}{2} K C^{\mu\nu} F_{\nu\rho}
  \right)
  \\
  &{}\qquad\nonumber
  - \frac{1}{2} \left(
    \pd_\mu K C^{\mu\nu}
    + K \mathcal{D}_\mu C^{\mu\nu}
  -K a_\mu C^{\mu\nu}\right) F_{\nu\kappa}
\end{align}
In the second equality,
we used~\eqref{eq:app-d2-spatial-STF-antisym-product} to rearrange the indices of the first two terms.
To get to the third line,
we pulled a spatial projector through the covariant derivative term,
and we used the identity
\begin{equation}
  \label{eq:FDCid}
  F^{\mu\nu}\mathcal{D}_\mu C_{\nu\kappa}
  =-\mathcal{D}_\mu C^{\mu\nu}F_{\nu\kappa}
  +a_\mu C^{\mu\nu}F_{\nu\kappa}\,,
\end{equation}
which follows from applying~\eqref{eq:DCident}
to the shear tensor
and contracting the resulting expression with the twist tensor.

We can then rewrite the mixed twist and shear terms
in terms of contributions to
$T^\sigma h^\rho_\sigma$
and $\tilde{T}^{\mu\sigma} h_{\sigma\lambda}$
as follows,
\begin{align}
  \label{eq:gen-mom-loss-rewriting-mixed-FC-terms}
  &\partial_\mu K C^{\mu\rho}F_{\rho\kappa}
  -\frac{1}{2}KC_{\kappa\nu}\left(
    \mathcal{D}_\mu F^{\mu\nu}+a_\mu F^{\mu\nu}
  \right)
  \\
  &{}\quad\nonumber
  = h_\kappa^\rho\mathcal{D}_\nu
  \left(\frac{1}{2}KC^{\nu\mu}F_{\mu\rho}\right)
  +\left(
    \frac{1}{2}\partial_\rho K C^{\rho\mu}
    -\frac{1}{2}K\mathcal{D}_\rho C^{\rho\mu}+Ka_\rho C^{\rho\mu}
  \right)F_{\mu\kappa}\,.
\end{align}
Together with the previous results in~\eqref{eq:gen-mom-loss-rewriting-C-terms}
and~\eqref{eq:gen-mom-loss-rewriting-F-terms},
we then get
\begin{subequations}
  \label{eq:gen-mom-eq-WI-matching-block-6-results}
  \begin{align}
    P_\kappa:
    &{}\qquad
    \frac{1}{8}h^\mu_\kappa\partial_\mu \left(C^2 - F^2\right),
    \\
    T^{\rho\sigma} h_{\rho\sigma}:
    &{}\qquad
    - \frac{1}{2} C\cdot N
    - \frac{1}{4}\left(\mathcal{L}_v - K\right)F^2\,,
    \\
    T^\rho \tau_\rho:
    &{}\qquad
    \frac{1}{2}KF^2\,,
    \\
    T^\mu:
    &{}\qquad
    \frac{1}{2}\os{1}{\mathcal{K}} a^\mu
    + \frac{1}{2}\partial_\rho K C^{\mu\rho}
    +Ka_\rho C^{\mu\rho}
    -\frac{1}{2}K\mathcal{D}_\rho C^{\mu\rho}\,,
    \\
    \tilde{T}^{\nu\rho} h_{\rho\mu}:
    &{}\qquad
    \frac{1}{2}KC^{\nu\rho}F_{\rho\mu}\,,
  \end{align}
\end{subequations}
as total contributions to the current components.

\paragraph{Intermediate conclusion.}
At this point,
we have confirmed that all of the six blocks entering in the decomposition~\eqref{eq:UPiRinterm}
of the order $r^{-2}$ contributions to the
$\Pi^\mu_\kappa U^\nu R_{\mu\nu}=0$
equation of motion can be rewritten in the form of the spatial projection of the boundary diffeomorphism Ward identity~\eqref{eq:diffeoWIspatialproj-repeat}.
From this identification,
we have extracted contributions to the various components of the energy-momentum-news tensor complex.
We have so far only considered the spatial part of the matching between bulk equations of motion and boundary diffeomorphism Ward identities,
and we will consider the timelike part in the following subsection. The latter equation contains $S^{\mu\nu}$, $\tau_\mu T^\mu$, $h_{\mu\nu}T^{\mu\nu}$ and $h^\mu_\rho T^\rho$ and we need the $U^\mu U^\nu R_{\mu\nu}=0$ equation to be obeyed at order $r^{-2}$ for the same $S^{\mu\nu}$, $\tau_\mu T^\mu$, $h_{\mu\nu}T^{\mu\nu}$ as we encountered in the rewriting of the $\Pi^\mu_\kappa U^\nu R_{\mu\nu}=0$
equation at order $r^{-2}$.
In general, the EMT-news complex is subject to improvements that do not affect the diffeomorphism Ward identities,
but such improvements affect the components of the EMT-news complex in both spatial and timelike projections of the diffeomorphism Ward identity in the same way.
There can, however,
exist certain identities that allow us to change the components of the EMT-news complex in for example only the spatial projection of the diffeomorphism Ward identity but not in the timelike projection, and vice versa.
Such identities are crucial in order to find agreement between $\tau_\mu T^\mu$, $h_{\mu\nu}T^{\mu\nu}$ across both equations. 

It will turn out that the following identity pertaining to the spatial projection of the diffeomorphism Ward identity in~\eqref{eq:diffeoWIspatialproj-repeat}, is sufficient for our purposes,
\begin{equation}
  \label{eq:spatial-diffeo-ward-id-currents-ambiguity-one}
  0
  = - \left(\LL_v - K\right) h^\mu_\kappa \pd_\mu F^2
  + h^\mu_\kappa \pd_\mu \left(
    \LL_v F^2 - K F^2
  \right)
  + a_\kappa \LL_v F^2
  - 2 F^{\rho\mu} \pd_\rho K F_{\mu\kappa}\,.
\end{equation}
This expression matches the form of the $P_\kappa$,
$h_{\mu\nu} T^{\mu\nu}$
and $T^\rho h^\mu_\rho$ terms
in~\eqref{eq:diffeoWIspatialproj-repeat},
and it therefore encodes an ambiguity in these current components.

Let us now collect the contributions to the components of the energy-momentum-news complex
which we derived in~\eqref{eq:gen-mom-eq-WI-matching-block-1-results},
\eqref{eq:gen-mom-eq-WI-matching-block-2-results},
\eqref{eq:gen-mom-eq-WI-matching-block-3-results},
\eqref{eq:gen-mom-eq-WI-matching-block-4-results},
\eqref{eq:gen-mom-eq-WI-matching-block-5-results}
and~\eqref{eq:gen-mom-eq-WI-matching-block-6-results}
above.
To ensure that, for example, the energy density has the expected sign,
we multiply all expressions with an overall minus sign,\footnote{We fix the sign by demanding that $\tau_\mu T^\mu$ is negative for the Schwarzschild solution. The EMT is $T^{\mu}{}_\nu=T^\mu\tau_\nu+T^{\mu\rho}h_{\rho\nu}$ and the time-time component is minus the energy density.}
which we are free to do since the Ward identity~\eqref{eq:diffeoWIspatialproj-repeat} is linear in the currents.
This gives
\begin{subequations}
  \label{eq:UPiREMTNewsComplex-part1}
  \begin{align}
    P_\kappa^{\text{U$\Pi$R}}
    &= \frac{1}{2}h^\kappa_\alpha\overset{(2)}{\mathcal{Z}}_\kappa
    +\frac{1}{2}\overset{(1)}{\mathcal{Z}}{}^\lambda  F_{\lambda\alpha}
    -\frac{1}{8}h^\rho_\alpha\partial_\rho C^2
    +\frac{1}{8}h^\gamma_\alpha\partial_\gamma F^2
    +b_1h^\gamma_\alpha\partial_\gamma F^2\,,
    \\
    \tilde T^{\rho\sigma}_{\text{U$\Pi$R}}
    &= \frac{1}{2}\overset{(0)}{S}C^{\rho\sigma}
    +\frac{1}{2}a^\rho\overset{(1)}{\mathcal{Z}}{}^\sigma
    +\frac{1}{2}\overset{(1)}{\mathcal{Z}}{}^\rho a^\sigma
    -\frac{1}{2}h^{\rho\sigma} a^\lambda \overset{(1)}{\mathcal{Z}}_\lambda
    +\frac{1}{2}KD^{\rho\sigma}
    \\
    &{}\qquad\nonumber
    -\frac{1}{4}KC^{\rho}{}_\nu F^{\nu\sigma}
    +h^{\rho\mu}h^{\sigma\nu}\overset{(0)}{\mathcal{K}}_{\langle\mu\nu\rangle}
    -\frac{1}{2}\overset{(1)}{\mathcal{K}}C^{\rho\sigma}\,,
    \\
    S^{\mu\nu}_{\text{U$\Pi$R}}
    &=\os{-1}{\mathcal{K}}^{T\mu\nu}
    = \frac{1}{2}N^{\mu\nu}
    -\frac{1}{4}KC^{\mu\nu}
    +A^{\mu\nu}\,,
    \label{eq:tildeSUPiR2}
  \end{align}
\end{subequations}
as well as
\begin{subequations}
  \label{eq:UPiREMTNewsComplex-part2}
  \begin{align}
    h_{\mu\nu}T^{\mu\nu}_{\text{U$\Pi$R}}
    &= -\overset{(2)}{\mathcal{K}}
    -\overset{(1)}{S}
    +a^\lambda \overset{(1)}{\mathcal{Z}}_\lambda
    -C\cdot\overset{(-1)}{\mathcal{K}}{}^T
    +\frac{1}{2}C\cdot N
    +\frac{1}{4}\mathcal{L}_vF^2
    -\frac{1}{4}KF^2\nonumber
    \\
    &{}\qquad\nonumber
    +2b_1\left(\mathcal{L}_v F^2-KF^2\right)\,,
    \\
    &= -\overset{(2)}{\mathcal{K}}
    -\overset{(1)}{S}
    +a^\lambda \overset{(1)}{\mathcal{Z}}_\lambda
    -C\cdot A+\frac{1}{4}KC^2
    +\frac{1}{4}\mathcal{L}_vF^2
    -\frac{1}{4}KF^2
    \\
    &{}\qquad\nonumber
    +2b_1\left(\mathcal{L}_v F^2-KF^2\right)\,,
    \\
    \tau_\mu T^\mu_{\text{U$\Pi$R}}
    &= \overset{(1)}{S}
    -\overset{(2)}{\mathcal{K}}
    -\frac{1}{2}a_\mu\overset{(1)}{\mathcal{Z}}{}^\mu
    +\frac{1}{2}\mathcal{D}_\mu\overset{(1)}{\mathcal{Z}}{}^\mu
    -\frac{1}{2}KF^2
    -b_1KF^2\,,
    \\
    h^\mu_\rho T^\rho_{\text{U$\Pi$R}}
    &= \overset{(0)}{S}a^\mu
    +\frac{1}{2}h^{\mu\nu}\partial_\nu\overset{(0)}{S}
    +\frac{1}{2}h^\mu_\rho\mathcal{L}_v\overset{(1)}{\mathcal{Z}}{}^\rho
    -\frac{1}{2}K\overset{(1)}{\mathcal{Z}}{}^\mu
    -a_\nu \overset{(-1)}{\mathcal{K}}{}^{T\,\mu\nu}
    \\
    &{}\qquad\nonumber
    -\frac{1}{2}\partial_\nu K C^{\mu\nu}
    -Ka_\nu C^{\mu\nu}
    +\frac{1}{2}K\mathcal{D}_\nu C^{\mu\nu}
    -\frac{1}{2}a^\mu\overset{(1)}{\mathcal{K}}
    \\
    &{}\qquad\nonumber
    +2b_1 F^{\mu\sigma}\partial_\sigma K\,,
  \end{align}
\end{subequations}
where we multiplied \eqref{eq:spatial-diffeo-ward-id-currents-ambiguity-one} with $b_1$, and where in the second equality for $h_{\mu\nu}T^{\mu\nu}_{\text{U$\Pi$R}}$ we used \eqref{eq:tildeSUPiR2}.
Having completed the spatial part of the identification with the boundary diffeomorphism Ward identity,
we now move on to the temporal part.

\subsubsection{General Bondi mass loss equation}
\label{sssec:d2MassLoss}
We now turn to the mass loss equation,
which arises at order $r^{-2}$ in the $U^\mu U^\nu R_{\mu\nu}=0$ equation of motion.
Our goal is to identify this equation with the timelike projection of the diffeomorphism Ward identity in~\eqref{eq:energyeq}, which reads
\begin{align}
  \label{eq:energyeq2}
  0
  &= - \left(\mathcal{L}_v - \frac{3}{2}K\right) \left(\tau_\rho T^\rho\right)
  - \frac{1}{2}K\left(\tau_\rho T^\rho+
    h_{\rho\sigma}T^{\rho\sigma}
    +\frac{1}{2}C_{\rho\sigma}S^{\rho\sigma}
  \right)
  \\
  &{}\qquad\nonumber
  -\frac{1}{2}S^{\rho\sigma}N_{\rho\sigma}
  +\left(\mathcal{D}_\mu + a_\mu\right)
  \left(T^\rho h^\mu_\rho\right).
\end{align}
Note that
this equation is independent of the first half of the current components we obtained in~\eqref{eq:UPiREMTNewsComplex-part1} in the previous subsection,
except for the $S^{\mu\nu}$ news current.

As we briefly mentioned at the start of Subsection~\ref{eq:calZ-def-repeat},
we now make the assumption
that we can take~$S^{\mu\nu}$ to be the same in both equations, i.e., we will assume that $S^{\mu\nu}$ in the Bondi mass loss equation is given by \eqref{eq:tildeSUPiR2},
so that
\begin{equation}
  \label{eq:news-current-from-eom-ansatz}
  S^{\mu\nu}
  = S^{\mu\nu}_{\text{U$\Pi$R}}
  =\os{-1}{\mathcal{K}}^{T\mu\nu}\,.
\end{equation}
As we will see, this assumption indeed turns out to be consistent,
as it will still allow us to match the time projection of the diffeomorphism Ward identity to the relevant bulk equation of motion.

The $U^\mu U^\nu R_{\mu\nu}=0$ equation of motion was given in~\eqref{eq:U-U-R-d2},
and its expansion is worked out in Appendix~\ref{app:intermediate-results}. 
In particular, we find in~\eqref{eq:UUR2R1} of Appendix~\ref{app:LossEqnLists-mass} that its $r^{-2}$ contributions are given by
\begin{align}
  \label{eq:R1UUR2}
  0
  &= \mathcal{D}_\rho\left[
      \frac{1}{2}h^{\rho\sigma}\partial_\sigma\overset{(0)}{S}
      +\overset{(0)}{S}a^\rho
      -\frac{1}{2}C^{\rho\sigma}\partial_\sigma K
      -KC^{\rho\sigma}a_\sigma+Ka^\sigma F^\rho{}_\sigma
  \right]
  \\
  &{}\qquad\nonumber
  -\overset{(1)}{\mathcal{Z}}_\rho\left(K a^\rho+\frac{1}{2}h^{\rho\sigma}\partial_\sigma K\right)
  +\frac{1}{4}K^2 F^2
  +\frac{1}{2}C^{\rho\sigma}a_\rho\partial_\sigma K
  +\frac{1}{2}a^\rho\partial_\rho\overset{(0)}{S}
  \\
  &{}\qquad\nonumber
  -\frac{1}{2}F^\rho{}_\sigma a^\sigma\partial_\rho K
  +\frac{1}{8}K\mathcal{L}_v C^2
  -\frac{1}{8}K\mathcal{L}_v F^2
  +\frac{1}{16}K^2\left(F^2-C^2\right)
  \\
  &{}\qquad\nonumber
  +\frac{3}{2}K\left(\overset{(1)}{S}-\overset{(2)}{\mathcal{K}}\right)
  -\mathcal{L}_v\left(\overset{(1)}{S}-\overset{(2)}{\mathcal{K}}\right)
  -\mathcal{L}_a\overset{(1)}{\mathcal{K}}
  -\frac{1}{2}\left(\overset{(1)}{\mathcal{K}}\right)^2\nonumber
  -\overset{(-1)}{\mathcal{K}}{}^{T\,\mu\nu}\overset{(-1)}{\mathcal{K}}{}^{T}_{\mu\nu}\,.
\end{align}
The ansatz~\eqref{eq:news-current-from-eom-ansatz} now strongly suggests that all contributions to $-\frac{1}{2}S^{\rho\sigma}N_{\rho\sigma}$ in~\eqref{eq:energyeq2} should come from the last term above.
Indeed, we can use the expression for the order $r$ contributions to $\mathcal{K}^T_{\mu\nu}$ in~\eqref{eq:app-curly-KT-expansion-d2-r1} to expand this term as follows,
\begin{align}
  -\overset{(-1)}{\mathcal{K}}{}^{T\,\mu\nu}\overset{(-1)}{\mathcal{K}}{}^{T}_{\mu\nu}
  \nonumber
  &=-\overset{(-1)}{\mathcal{K}}{}^{T\,\mu\nu} \left(
    \frac{1}{2}N_{\mu\nu}
    -\frac{1}{4}KC_{\mu\nu}
    +h^{\rho}_{\langle\mu}h^\sigma_{\nu\rangle}\left[
      \mathcal{D}_\rho a_\sigma+a_\rho a_\sigma
    \right]
  \right)
  \\
  \label{eq:R1IntId11}
  &= -\frac{1}{2}\overset{(-1)}{\mathcal{K}}{}^{T\,\mu\nu}N_{\mu\nu}
  +\frac{1}{4}K\overset{(-1)}{\mathcal{K}}{}^{T}\cdot C
  \\
  &{}\qquad\nonumber
  -\left(\mathcal{D}_\mu +a_\mu\right)
  \left(\overset{(-1)}{\mathcal{K}}{}^{T\,\mu\nu}a_\nu\right)
  +a_\nu\mathcal{D}_\mu\overset{(-1)}{\mathcal{K}}{}^{T\,\mu\nu}\,.
\end{align}
It turns out that the final term in this expression can be rewritten into a more suitable form using an identity that can be derived from the order $r^{-1}$ terms in the $U^\mu\Pi^\nu_\rho R_{\mu\nu}=0$ equation of motion.
Following Appendix~\ref{ssapp:order-r-m1-Pi-U-R-EOM}, this gives\footnote{See Equation \eqref{eq:BId2R1}, where we use \eqref{eq:S0d=2} to eliminate $\mathcal{Q}$ in favour of $\os{0}{S}$ here.
}
\begin{align}
  \label{eq:UPiRO1r}
  0
  &= \overset{(0)}{S}a_\kappa
  -\frac{1}{2}\partial_\rho K F^\rho{}_\kappa
  -\frac{1}{2}K\left(\mathcal{D}_\rho F^\rho{}_\kappa+a_\rho F^\rho{}_\kappa\right)
  -\frac{1}{2}K a_\rho C^\rho{}_\kappa
  \\
  &{}\qquad\nonumber
  -\frac{1}{2}\partial_\rho K C^\rho{}_\kappa
  +\frac{1}{2}h^\rho_\kappa\partial_\rho\overset{(1)}{\mathcal{K}}
  -\frac{1}{2}a_\kappa \overset{(1)}{\mathcal{K}}
  +\frac{1}{2}\mathcal{L}_v\overset{(1)}{\mathcal{Z}}_\kappa
  -\mathcal{D}_\rho
  \left(h^{\rho\sigma}\overset{(-1)}{\mathcal{K}}{}^T_{\sigma\kappa}\right)\,.
\end{align}
At the end of Appendix~\ref{app:curvten}, we show that this equation is identically satisfied by virtue of a boundary Bianchi identity. 
After contracting it with $a^\kappa$, we obtain the following expression for the final term in~\eqref{eq:R1IntId11},
\begin{align}
  \label{eq:DifBI2}
  a_\nu\mathcal{D}_\mu\overset{(-1)}{\mathcal{K}}{}^{T\,\mu\nu} 
  &= \overset{(0)}{S}a^2
  -\frac{1}{2}a_\sigma\partial_\rho K F^{\rho\sigma}
  -\frac{1}{2}Ka_\sigma\mathcal{D}_\rho F^{\rho\sigma}
  \\
  &{}\qquad\nonumber
  -\frac{1}{2}K C^{\rho\sigma} a_\rho a_\sigma
  -\frac{1}{2}a_\sigma\partial_\rho K C^{\rho\sigma}
  \\
  &{}\qquad\nonumber
  +\frac{1}{2}\mathcal{L}_a\overset{(1)}{\mathcal{K}}
  -\frac{1}{2}a^2 \overset{(1)}{\mathcal{K}}
  +\frac{1}{2}a^\kappa\mathcal{L}_v\overset{(1)}{\mathcal{Z}}_\kappa\,.
\end{align}
Using \eqref{eq:R1IntId11} and \eqref{eq:DifBI2} we can rewrite the expression for the $U^\mu U^\nu R_{\mu\nu}=0$ equation of motion at order $r^{-2}$ in~\eqref{eq:R1UUR2} as follows,
\begin{align}
  \label{eq:UURIntQ1}
  0 &=  \mathcal{D}_\rho\left[
      \frac{1}{2}h^{\rho\sigma}\partial_\sigma\overset{(0)}{S}
      +\overset{(0)}{S}a^\rho
      -\frac{1}{2}C^{\rho\sigma}\partial_\sigma K
      -KC^{\rho\sigma}a_\sigma
      +Ka^\sigma F^\rho{}_\sigma
  \right]
  \\
  &{}\qquad\nonumber
  -\overset{(1)}{\mathcal{Z}}_\rho\left(K a^\rho+\frac{1}{2}h^{\rho\sigma}\partial_\sigma K\right)
  +\frac{1}{4}K^2 F^2
  +\frac{1}{2}C^{\rho\sigma}a_\rho\partial_\sigma K
  +\frac{1}{2}a^\rho\partial_\rho\overset{(0)}{S}
  -\frac{1}{2}F^\rho{}_\sigma a^\sigma\partial_\rho K
  \\
  &{}\qquad\nonumber
  +\frac{1}{8}K\mathcal{L}_v C^2
  -\frac{1}{8}K\mathcal{L}_v F^2
  +\frac{1}{16}K^2\left(F^2-C^2\right)
  \\
  &{}\qquad\nonumber
  +\frac{3}{2}K\left(\overset{(1)}{S}-\overset{(2)}{\mathcal{K}}\right)
  -\mathcal{L}_v\left(\overset{(1)}{S}-\overset{(2)}{\mathcal{K}}\right)
  -\mathcal{L}_a\overset{(1)}{\mathcal{K}}
  -\frac{1}{2}\left(\overset{(1)}{\mathcal{K}}\right)^2
  \\
  &{}\qquad\nonumber
  -\frac{1}{2}\overset{(-1)}{\mathcal{K}}{}^{T\,\mu\nu}N_{\mu\nu}
  +\frac{1}{4}K\overset{(-1)}{\mathcal{K}}{}^{T}\cdot C
  -\left(\mathcal{D}_\mu +a_\mu\right)
  \left(\overset{(-1)}{\mathcal{K}}{}^{T\,\mu\nu}a_\nu\right)
  \\
  &{}\qquad\nonumber
  +\overset{(0)}{S}a^2
  -\frac{1}{2}a_\sigma\partial_\rho K F^{\rho\sigma}
  -\frac{1}{2}Ka_\sigma\mathcal{D}_\rho F^{\rho\sigma}
  -\frac{1}{2}K C^{\rho\sigma} a_\rho a_\sigma
  -\frac{1}{2}a_\sigma\partial_\rho K C^{\rho\sigma}
  \\
  &{}\qquad\nonumber
  +\frac{1}{2}\mathcal{L}_a\overset{(1)}{\mathcal{K}}
  -\frac{1}{2}a^2 \overset{(1)}{\mathcal{K}}
  +\frac{1}{2}a^\kappa\mathcal{L}_v\overset{(1)}{\mathcal{Z}}_\kappa\,.
\end{align}
Using that $\os{1}{\mathcal{K}}=\mathcal{D}_\rho a^\rho$,
we can write the corresponding terms as follows,
\begin{equation}
  - \frac{1}{2} \LL_a \os{1}{\mathcal{K}}
  - \frac{1}{2} a^2 \os{1}{\mathcal{K}}
  - \frac{1}{2} \os{1}{\mathcal{K}}^2
  = \left(
    \mathcal{D}_\rho
    +  a_\rho
  \right) \left(
    - \frac{1}{2} a^\rho \os{1}{\mathcal{K}}
  \right).
\end{equation}
Using the product rule in reverse on terms with derivatives of $K$, and rearranging terms inspired by the form of~\eqref{eq:energyeq2} we obtain,
\begin{align}
  \label{eq:UURQ2}
  0 &=  \left(\mathcal{D}_\rho+a_\rho\right)
  \left(
    \frac{1}{2}h^{\rho\sigma}\partial_\sigma\overset{(0)}{S}
    +\overset{(0)}{S}a^\rho
    -\frac{1}{2}C^{\rho\sigma}\partial_\sigma K
    -KC^{\rho\sigma}a_\sigma
    +Ka^\sigma F^\rho{}_\sigma
  \right)
  \\
  &\quad\,\nonumber
  +\left(\mathcal{D}_\rho+a_\rho\right)
  \left(
    -\frac{1}{2}a^\rho\overset{(1)}{\mathcal{K}}
    -\overset{(-1)}{\mathcal{K}}{}^{T\,\mu\nu}a_\nu
    -Ka^\sigma F^\rho{}_\sigma
    +\frac{1}{2}Ka_\sigma C^{\rho\sigma}
    -\frac{1}{2}K\overset{(1)}{\mathcal{Z}}{}^\rho
  \right)
  \\
  &\quad\,\nonumber
  -\frac{1}{2}Ka_\rho\overset{(1)}{\mathcal{Z}}{}^\rho
  +K\mathcal{D}_\rho\left(a_\sigma F^{\rho\sigma}\right)
  -\frac{1}{2}K\mathcal{D}_\rho\left(a_\sigma C^{\rho\sigma}\right)
  +\frac{1}{2}K\mathcal{D}_\rho\overset{(1)}{\mathcal{Z}}{}^\rho
  \\
  &\quad\,\nonumber
  -\frac{1}{2}Ka_\sigma\mathcal{D}_\rho F^{\rho\sigma}
  +\frac{1}{4}K^2 F^2
  +\frac{1}{8}K\mathcal{L}_v C^2
  -\frac{1}{8}K\mathcal{L}_v F^2
  \\
  &\quad\,\nonumber
  +\frac{1}{4}K\overset{(-1)}{\mathcal{K}}{}^{T}\cdot C
  +\frac{1}{16}K^2\left(F^2-C^2\right)
  \\
  &\quad\,\nonumber
  -\left(\mathcal{L}_v-\frac{3}{2}K\right)\left(\overset{(1)}{S}-\overset{(2)}{\mathcal{K}}\right)
  -\frac{1}{2}\overset{(-1)}{\mathcal{K}}{}^{T\,\mu\nu}N_{\mu\nu}
  +\frac{1}{2}a^\kappa\mathcal{L}_v\overset{(1)}{\mathcal{Z}}_\kappa\,.
\end{align}
To rewrite the final term in the above expression,
we first use the expression~\eqref{eq:vectorcase} for the commutator of Lie and covariant derivatives
applied to $\os{1}{\mathcal{Z}}_\rho$ to get
\begin{equation}
  \mathcal{D}_\rho\mathcal{L}_v\overset{(1)}{\mathcal{Z}}{}^\rho
  -\mathcal{L}_v\mathcal{D}_\rho\overset{(1)}{\mathcal{Z}}{}^\rho
  =\overset{(1)}{\mathcal{Z}}{}^\rho\partial_\rho K
  \,.
\end{equation}
With this identity, we can show that
\begin{align}
  &\left(\mathcal{D}_\rho+a_\rho\right)
  \left(
    \frac{1}{2}h^\rho_\sigma\mathcal{L}_v\overset{(1)}{\mathcal{Z}}{}^\sigma
    -\frac{1}{2}K\overset{(1)}{\mathcal{Z}}{}^\rho
  \right)
  -\frac{1}{2}\left(\mathcal{L}_v - K\right)
  \left(
    \mathcal{D}_\rho \overset{(1)}{\mathcal{Z}}{}^\rho
    -a_\rho \overset{(1)}{\mathcal{Z}}{}^\rho
  \right)
  \\
  &{}\qquad\nonumber
  =-\frac{1}{2}Ka_\rho\overset{(1)}{\mathcal{Z}}{}^\rho+\frac{1}{2}a_\rho\mathcal{L}_v\overset{(1)}{\mathcal{Z}}{}^\rho
  = \frac{1}{2}a^\rho\mathcal{L}_v\overset{(1)}{\mathcal{Z}}_\rho\,,
\end{align}
where we lowered the index inside the Lie derivative in the final equality.
Additionally,
we can rewrite $\LL_v C^2$
in terms of the news tensor,
which can subsequently be expressed in terms of the $r^{-1}$ terms in $\mathcal{K}^T_{\mu\nu}$
using its expression in~\eqref{eq:app-curly-KT-expansion-d2-r1}.
All together, this allows us to write~\eqref{eq:UURQ2} as follows,
\begin{align}
  \label{eq:UURQ3}
  0
  &=  \left(\mathcal{D}_\rho+a_\rho\right)
  \left(
    \frac{1}{2}h^{\rho\sigma}\partial_\sigma\overset{(0)}{S}
    +\overset{(0)}{S}a^\rho
    -\frac{1}{2}C^{\rho\sigma}\partial_\sigma K
    -\frac{1}{2}KC^{\rho\sigma}a_\sigma
  \right)
  \\
  &{}\qquad\nonumber
  +\left(\mathcal{D}_\rho+a_\rho\right)
  \left(
    -\frac{1}{2}a^\rho\overset{(1)}{\mathcal{K}}
    -\overset{(-1)}{\mathcal{K}}{}^{T\,\rho\nu}a_\nu
    -K\overset{(1)}{\mathcal{Z}}{}^\rho
    +\frac{1}{2}h^\rho_\sigma\mathcal{L}_v\overset{(1)}{\mathcal{Z}}{}^\sigma
  \right)
  \\
  &{}\qquad\nonumber
  -\frac{1}{2}K\left[
    \frac{1}{8}KC^2
    -\frac{1}{2}\mathcal{D}_\rho\overset{(1)}{\mathcal{Z}}{}^\rho-\frac{1}{2}a_\rho\overset{(1)}{\mathcal{Z}}{}^\rho
    -\frac{1}{4}\mathcal{L}_v F^2
    +\frac{3}{8}KF^2+\frac{1}{2}\overset{(-1)}{\mathcal{K}}{}^{T}\cdot C
  \right]
  \\
  &{}\qquad\nonumber
  -\left(\mathcal{L}_v-\frac{3}{2}K\right)\left(
    \overset{(1)}{S}
    -\overset{(2)}{\mathcal{K}}
    +\frac{1}{2}\mathcal{D}_\rho \overset{(1)}{\mathcal{Z}}{}^\rho
    -\frac{1}{2}a_\rho \overset{(1)}{\mathcal{Z}}{}^\rho
  \right)
  -\frac{1}{2}\overset{(-1)}{\mathcal{K}}{}^{T\,\mu\nu}N_{\mu\nu}\,.
\end{align}
At this point, we have succeeded in writing the Bondi mass loss equation
in the form of the time projection of the diffeomorphism Ward identity~\eqref{eq:energyeq2}.

However, as in the previous subsection concerning the angular momentum loss equation,
extracting the components of the energy-momentum-news complex from this expression is ambiguous.
In particular, we can show that
\begin{align}
  \label{eq:UUR-currents-ambiguity}
  0 &=  \left(\mathcal{D}_\rho+a_\rho\right)
  \left(
    F^{\rho\sigma}\partial_\sigma K
    -Kh^\rho_\mu\left(\mathcal{D}_\sigma+a_\sigma\right) F^{\sigma\mu}
  \right)
  \\
  &{}\qquad\nonumber
  - \frac{1}{4}K \left(
    \mathcal{L}_v-K\right)F^2
  -  \left(\mathcal{L}_v - \frac{3}{2}K\right)
  \left(\frac{1}{2} KF^2\right)\,.
\end{align}
This takes the form of the temporal projection of the diffeomorphism Ward identity. This is 
similar to the identity~\eqref{eq:spatial-diffeo-ward-id-currents-ambiguity-one} in the previous subsection.
Incorporating this ambiguity, the current components we extract from matching the diffeomorphism Ward identity with~\eqref{eq:UURQ3} are given by
\begin{subequations}
\label{eq:UUREMTNewsComplex}
\begin{align}
    \tau_\mu T^\mu_{\text{UUR}}
    &= \overset{(1)}{S}-\overset{(2)}{\mathcal{K}}+\frac{1}{2}\mathcal{D}_\rho \overset{(1)}{\mathcal{Z}}{}^\rho-\frac{1}{2}a_\rho \overset{(1)}{\mathcal{Z}}{}^\rho+\frac{1}{2}b_2KF^2\,,
    \\
    h^\rho_\mu T^\mu_{\text{UUR}}
    &= \frac{1}{2}h^{\rho\sigma}\partial_\sigma\overset{(0)}{S}
    +\overset{(0)}{S}a^\rho
    -\frac{1}{2}C^{\rho\sigma}\partial_\sigma K
    -\frac{1}{2}KC^{\rho\sigma}a_\sigma
    -\frac{1}{2}a^\rho\overset{(1)}{\mathcal{K}}
    -K\overset{(1)}{\mathcal{Z}}{}^\rho
    \\
    &{}\qquad\nonumber
    -\overset{(-1)}{\mathcal{K}}{}^{T\,\mu\nu}a_\nu
    +\frac{1}{2}h^\rho_\sigma\mathcal{L}_v\overset{(1)}{\mathcal{Z}}{}^\sigma
    +b_2F^{\rho\sigma}\partial_\sigma K-b_2Kh^\rho_\mu\left(\mathcal{D}_\sigma+a_\sigma\right)F^{\sigma\mu}\,,\\
    h_{\mu\nu}T^{\mu\nu}_{\text{UUR}}
    &= -\overset{(1)}{S}+\overset{(2)}{\mathcal{K}}+\frac{1}{8}KC^2-\mathcal{D}_\rho\overset{(1)}{\mathcal{Z}}{}^\rho
    -\frac{1}{4}\mathcal{L}_v F^2
    +\frac{3}{8}KF^2
    \\
    &{}\qquad\nonumber
    +\frac{1}{2}b_2\mathcal{L}_vF^2
    -b_2KF^2\,,
\end{align}
\end{subequations}
along with
$S^{\mu\nu} = \overset{(-1)}{\mathcal{K}}{}^{T\,\mu\nu}$,
which we already imposed in~\eqref{eq:news-current-from-eom-ansatz}. The terms proportional to $b_2$ stem from the identity \eqref{eq:UUR-currents-ambiguity}.
Our next goal will be to match these results with the current components~\eqref{eq:UPiREMTNewsComplex-part2} we obtained from the angular momentum loss equation in the previous subsection. There are many other ambiguities, but 
\eqref{eq:UUR-currents-ambiguity} and \eqref{eq:spatial-diffeo-ward-id-currents-ambiguity-one} will prove to be sufficient to perform the matching of the current components.

\subsubsection{EMT-News complex from bulk equations of motion}
\label{sssec:d2EOMCurrents}
In the previous subsections, we rewrote the relevant bulk equations of motion to be manifestly of the form of the boundary diffeomorphism Ward identities.
We subsequently read off expressions for the components of the energy-momentum-news currents.
This does not uniquely fix the currents, and we indeed obtained different answers in~\eqref{eq:UPiREMTNewsComplex-part2} and~\eqref{eq:UUREMTNewsComplex} for the currents which enter in both the spacelike and the timelike projections.

To be precise, while we started the computation in Section~\ref{sssec:d2MassLoss} by (correctly) guessing that we would be able to require the expression for the shear current ${S}^{\mu\nu}$ to be identical to the one obtained in~\ref{sssec:d2AngLoss},
the expressions we obtained in each subsection for
$\tau_\mu T^\mu$,
$h^\rho_\mu T^\mu$ and
$h_{\mu\nu} T^{\mu\nu}$
do not immediately agree.
Thankfully, we can resolve this initial discrepancy
fixing the $b_1$ and $b_2$ parameters entering in the current components~\eqref{eq:UPiREMTNewsComplex-part2} and~\eqref{eq:UUREMTNewsComplex}.

In comparing \eqref{eq:UPiREMTNewsComplex-part2} and~\eqref{eq:UUREMTNewsComplex}, it is useful to use the following results
\begin{align}
  \label{eq:Z1Simp}
  \overset{(1)}{\mathcal{Z}}{}^\rho
  &=-\mathcal{D}_\sigma C^{\sigma\rho}
  +a_\sigma C^{\sigma\rho}
  +h_\nu^\rho\mathcal{D}_\sigma F^{\sigma\nu}
  +a_\sigma F^{\sigma\rho}\nonumber\\
  &=-\mathcal{D}_\sigma C^{\sigma\rho}
  +a_\sigma C^{\sigma\rho}
  +\mathcal{D}_\sigma F^{\sigma\rho}
  -\frac{1}{2}v^\rho F^2
  +a_\sigma F^{\sigma\rho}\,,
  \\
  \label{eq:DZ1Simp}
  \mathcal{D}_\rho\overset{(1)}{\mathcal{Z}}{}^\rho
  &= -\mathcal{D}_\rho\mathcal{D}_\sigma C^{\rho\sigma}
  +C\cdot A
  -C^{\rho\sigma}a_\rho a_\sigma
  +a_\sigma\mathcal{D}_\rho C^{\rho\sigma}
  \\
  &{}\qquad\nonumber
  -\frac{3}{4}\mathcal{L}_v F^2
  +KF^2
  -a_\sigma\mathcal{D}_\rho F^{\rho\sigma}\,,
  \\
  \label{eq:calK2}
   \overset{(2)}{\mathcal{K}}
  &= -\frac{1}{2}\mathcal{D}_\mu\mathcal{D}_\nu C^{\mu\nu}+\frac{1}{16}KC^2+\frac{3}{16}KF^2-\frac{1}{4}\mathcal{L}_v F^2\,,
\end{align}
where, in obtaining the expression for $\mathcal{D}_\rho\overset{(1)}{\mathcal{Z}}{}^\rho$, we used
\begin{equation}
    \mathcal{D}_\rho\mathcal{D}_\sigma F^{\rho\sigma}=\frac{1}{2}\left[\mathcal{D}_\rho\,,\mathcal{D}_\sigma\right] F^{\rho\sigma}=0\,.
\end{equation}
The expressions for $\overset{(1)}{\mathcal{Z}}{}^\rho$ and $\overset{(2)}{\mathcal{K}}$ are obtained in~\eqref{eq:CalZ1new} and \eqref{eq:app-curly-K-expansion-d2-r-2}.
Using these results, we can then work out that the differences between the aforementioned current components,
as derived in~\eqref{eq:UPiREMTNewsComplex-part2} and~\eqref{eq:UUREMTNewsComplex} from the matching of spacelike and timelike projection of the diffeomorphism Ward identity,
are given by
\begin{subequations}
  \label{eq:bFix}
  \begin{align}
    \tau_\mu T^\mu_{\text{UUR}}
    - \tau_\mu T^\mu_{\text{U$\Pi$R}}
    &= \left(\frac{1}{2}+b_1+\frac{1}{2}b_2\right)KF^2\,,
    \\
    h^\mu_\rho T^\rho_{\text{UUR}}
    - h^\mu_\rho T^\rho_{\text{U$\Pi$R}}
    &= -\left(\frac{1}{2}+b_2\right)Kh^\mu_\sigma
    \left(\mathcal{D}_\rho+a_\rho\right) F^{\rho\sigma}
    \\
    &{}\qquad\nonumber
    +(b_2-2b_1)F^{\mu\sigma}\partial_\sigma K\,,\\
    h_{\mu\nu}T^{\mu\nu}_{\text{UUR}}
    - h_{\mu\nu}T^{\mu\nu}_{\text{U$\Pi$R}}
    &=(2b_1-b_2)KF^2
    +\left(-\frac{1}{4}+\frac{1}{2}b_2-2b_1\right)\mathcal{L}_v F^2\,.
  \end{align}
\end{subequations}
From this, we can see that the current components agree if we take
\begin{equation}
  b_1=-\frac{1}{4}\,,
  \qquad
  b_2=-\frac{1}{2}\,.
\end{equation}
This finally confirms that we can indeed write the relevant bulk equations of motion
in the form of the boundary diffeomorphism Ward identity.
After substituting in the above values for the $b_1, b_2$ parameters,
and combining~\eqref{eq:UPiREMTNewsComplex-part1},
\eqref{eq:UPiREMTNewsComplex-part2}
and~\eqref{eq:UUREMTNewsComplex},
the total energy-momentum-news complex we obtain is given by
\begin{subequations}
  \label{eq:EMTNewsComplexSec7}
  \begin{align}
    \tau_\mu T^\mu_{\text{EOM}}
    =& \overset{(1)}{S}-\overset{(2)}{\mathcal{K}}
    -\frac{1}{2}a_\mu\overset{(1)}{\mathcal{Z}}{}^\mu
    +\frac{1}{2}\mathcal{D}_\mu\overset{(1)}{\mathcal{Z}}{}^\mu
    -\frac{1}{4}KF^2
    \\
    =&\os{1}{S}-\frac{1}{16}KC^2+\frac{1}{2}C\cdot A+a_\sigma\mathcal{D}_\rho C^{\rho\sigma}-a_\rho a_\sigma C^{\rho\sigma}\nonumber\\
    &-a_\sigma\mathcal{D}_\rho F^{\rho\sigma}+\frac{1}{16}KF^2-\frac{1}{8}\mathcal{L}_v F^2\,,\\
    h^\mu_\rho T^\rho_{\text{EOM}}
    =& \overset{(0)}{S}a^\mu
    +\frac{1}{2}h^{\mu\nu}\partial_\nu\overset{(0)}{S}
    +\frac{1}{2}h^\mu_\rho\mathcal{L}_v\overset{(1)}{\mathcal{Z}}{}^\rho
    -\frac{1}{2}K\overset{(1)}{\mathcal{Z}}{}^\mu
    -a_\nu \overset{(-1)}{\mathcal{K}}{}^{T\,\mu\nu}
    \\
    &\nonumber
    -\frac{1}{2}\partial_\nu K C^{\mu\nu}
    -Ka_\nu C^{\mu\nu}
    +\frac{1}{2}K\mathcal{D}_\nu C^{\mu\nu}
    -\frac{1}{2}a^\mu\overset{(1)}{\mathcal{K}}
    -\frac{1}{2} F^{\mu\sigma}\partial_\sigma K\,,\label{eq:energyfluxEOM}
    \\
    h_{\mu\nu}T^{\mu\nu}_{\text{EOM}}
    =& -\overset{(2)}{\mathcal{K}}-\overset{(1)}{S}
    +a^\lambda \overset{(1)}{\mathcal{Z}}_\lambda
    -C\cdot A
    +\frac{1}{4}KC^2
    -\frac{1}{4}\mathcal{L}_vF^2
    +\frac{1}{4}KF^2
    \\
    =&-\os{1}{S}+\frac{3}{16}KC^2+\frac{1}{2}\mathcal{D}_\rho\left(\mathcal{D}_\sigma-2a_\sigma\right) C^{\rho\sigma}\nonumber\\
    &+a_\sigma\mathcal{D}_\rho F^{\rho\sigma}+\frac{1}{16}KF^2\,,\\
    P_\kappa^{\text{EOM}}
    =& \frac{1}{2}h_\kappa^\mu\overset{(2)}{\mathcal{Z}}_\mu
    +\frac{1}{2}\overset{(1)}{\mathcal{Z}}{}^\lambda  F_{\lambda\kappa}
    -\frac{1}{8}h^\mu_\kappa\partial_\mu C^2
    -\frac{1}{8}h^\mu_\kappa\partial_\mu F^2\,,\label{eq:PEOM}
    \\
    \tilde T^{\rho\sigma}_{\text{EOM}}
    =& \frac{1}{2}\overset{(0)}{S}C^{\rho\sigma}
    +\frac{1}{2}a^\rho\overset{(1)}{\mathcal{Z}}{}^\sigma
    +\frac{1}{2}\overset{(1)}{\mathcal{Z}}{}^\rho a^\sigma
    -\frac{1}{2}h^{\rho\sigma} a^\lambda \overset{(1)}{\mathcal{Z}}_\lambda
    +\frac{1}{2}KD^{\rho\sigma}\label{eq:tildeTEOM}
    \\
    &\nonumber
    -\frac{1}{4}KC^{\rho}{}_\nu F^{\nu\sigma}
    +h^{\rho\mu}h^{\sigma\nu}\overset{(0)}{\mathcal{K}}_{\langle\mu\nu\rangle}
    -\frac{1}{2}\overset{(1)}{\mathcal{K}}C^{\rho\sigma}\,,
    \\
    S^{\mu\nu}_{\text{EOM}}
    =&\os{-1}{\mathcal{K}}^{T\mu\nu}
    = \frac{1}{2}N^{\mu\nu}
    -\frac{1}{4}KC^{\mu\nu}
    +A^{\mu\nu}\,,
    \label{eq:tildeSUPiR22}
  \end{align}
\end{subequations}
where in the expressions for $h_{\mu\nu}T^{\mu\nu}_{\text{EOM}}$ and $\tau_\mu T^\mu_{\text{EOM}}$ we used \eqref{eq:Z1Simp}, \eqref{eq:DZ1Simp} and \eqref{eq:calK2}.

Next, we rewrite the energy flux \eqref{eq:energyfluxEOM} as follows 
\begin{eqnarray}
    h^\mu_\rho T^\rho_{\text{EOM}}
    &=& \overset{(0)}{S}a^\mu
    +\frac{1}{2}h^{\mu\nu}\partial_\nu\overset{(0)}{S}
    +\frac{1}{2}h^\mu_\rho\left(\mathcal{L}_v-K\right)\overset{(1)}{\mathcal{Z}}{}^\rho
    -\frac{1}{2}a_\nu N^{\mu\nu}-a_\nu A^{\mu\nu}\nonumber\\
    &&
    -\frac{1}{2}\partial_\nu K C^{\mu\nu}
    -\frac{3}{4}Ka_\nu C^{\mu\nu}
    +\frac{1}{2}K\mathcal{D}_\nu C^{\mu\nu}
    -\frac{1}{2}a^\mu\mathcal{D}_\rho a^\rho
    -\frac{1}{2} F^{\mu\sigma}\partial_\sigma K\nonumber
    \\
    &=& \frac{1}{2}h^{\mu\nu}\left(\partial_\nu+2a_\nu\right)\left(\overset{(0)}{S}-a^2\right)
    +\frac{1}{2}h^\mu_\rho\left(\mathcal{L}_v-K\right)\overset{(1)}{\mathcal{Z}}{}^\rho
    -\frac{1}{2}a_\nu N^{\mu\nu}\nonumber\\
    &&
    -\frac{1}{2}\partial_\nu K C^{\mu\nu}
    -\frac{3}{4}Ka_\nu C^{\mu\nu}
    +\frac{1}{2}K\mathcal{D}_\nu C^{\mu\nu}
    -\frac{1}{2} F^{\mu\sigma}\partial_\sigma K+\frac{1}{2}h^{\mu\rho}a^\sigma\mathcal{L}_v F_{\rho\sigma}\nonumber
    \\
    &=& \frac{1}{2}h^{\mu\nu}\left(\partial_\nu+2a_\nu\right)\left(\overset{(0)}{S}-a^2\right)
    +h^\mu_\rho\left(\mathcal{L}_v-K\right)\os{0}{P}^\rho
    \nonumber\\
    &&
    -\frac{1}{2}v^\rho\left(\partial_\rho\tilde b_\sigma-\partial_\sigma\tilde b_\rho\right)\left(C^{\sigma\mu}-F^{\sigma\mu}\right)+\frac{1}{4}Ka_\sigma F^{\sigma\mu}\nonumber\\
    &&
    -\frac{1}{4}\partial_\nu K C^{\mu\nu}
    -\frac{3}{4}Ka_\nu C^{\mu\nu}
    +\frac{1}{2}K\mathcal{D}_\nu C^{\mu\nu}
    +\frac{1}{4} F^{\sigma\mu}\partial_\sigma K\,,\label{eq:energyfluxEOMv2}
\end{eqnarray}
where in the first equality we used \eqref{eq:app-curly-KT-expansion-d2-r1} and in the second equality we used
\begin{equation}\label{eq:aA}
    -a_\nu A^{\mu\nu}=\frac{1}{2}a^\mu\mathcal{D}_\rho a^\rho-a^2 a^\mu-\frac{1}{2}h^{\rho\mu}\partial_\rho a^2+\frac{1}{2}h^{\mu\rho}a^\sigma\mathcal{L}_v F_{\rho\sigma}\,.
\end{equation}
Finally, in the third equality we used
\begin{align}
    &\frac{1}{2}h^\mu_\rho\left(\mathcal{L}_v-K\right)\overset{(1)}{\mathcal{Z}}{}^\rho
    -\frac{1}{2}a_\nu N^{\mu\nu}+\frac{1}{2}h^{\mu\rho}a^\sigma\mathcal{L}_v F_{\rho\sigma}\nonumber\\
    =& h^\mu_\rho\left(\mathcal{L}_v-K\right)\os{0}{P}^\rho-\frac{1}{2}v^\rho\left(\partial_\rho\tilde b_\sigma-\partial_\sigma\tilde b_\rho\right)\left(C^{\sigma\mu}-F^{\sigma\mu}\right)\nonumber\\
    &+\frac{1}{4}\partial_\sigma K C^{\sigma\mu}-\frac{1}{4}\partial_\sigma K F^{\sigma\mu}+\frac{1}{4}Ka_\sigma F^{\sigma\mu}\,,
\end{align}
where we used
\begin{equation}\label{eq:Z1toP0}
    \os{1}{\mathcal{Z}}^\mu=2\os{0}{P}^\mu-a_\rho C^{\rho\mu}+a_\rho F^{\rho\mu}\,,
\end{equation}
and, following~\eqref{eq:app-4Dsol-P0-def}, we defined
\begin{equation}\label{eq:P0up}
    \os{0}{P}^\mu=-\frac{1}{2}\left(\mathcal{D}_\rho-2a_\rho\right)C^{\rho\mu}+\frac{1}{2}h^\mu_\sigma\mathcal{D}_\rho F^{\rho\sigma}\,.
\end{equation}
Then consider the momentum density $P^{\text{EOM}}_\kappa$ in \eqref{eq:PEOM}. We can write this as
\begin{eqnarray}
    P_\kappa^{\text{EOM}}
    &=&\frac{3}{2} h^\mu_\kappa v^\nu\os{1}{\Pi}_{\mu\nu}-\frac{1}{2}h^\mu_\kappa\left(\partial_\mu+a_\mu\right)\os{2}{\beta}-a^\mu D_{\mu\kappa}+\frac{1}{2}\os{0}{P}^\mu\left(F_{\mu\kappa}-C_{\mu\kappa}\right)\nonumber\\
    &&+\left(\os{0}{P}^\mu-\frac{1}{2}a_\rho C^{\rho\mu}+\frac{1}{2}a_\rho F^{\rho\mu}\right)F_{\mu\kappa}-\frac{1}{8}h^\mu_\kappa\partial_\rho C^2
    -\frac{1}{8}h^\mu_\kappa\partial_\mu F^2\nonumber\\
    &=&\frac{3}{2}\left(h^\mu_\kappa v^\nu\os{1}{g}_{\mu\nu}-\os{2}{\beta}a_\kappa\right)+\frac{3}{2}\os{0}{P}^\mu F_{\mu\kappa}-\frac{1}{2}\os{0}{P}^\mu C_{\mu\kappa}-\frac{1}{2}a^\rho C_{\rho\mu}F^\mu{}_\kappa\nonumber\\
    &&-\frac{3}{32}h^\mu_\kappa\partial_\mu C^2-\frac{5}{32}h^\mu_\kappa\partial_\mu F^2-\frac{3}{16}a_\kappa F^2-\frac{1}{16}a_\kappa C^2-a^\mu D_{\mu\kappa}\,,\label{eq:momdensityEOM}
\end{eqnarray}
where we used \eqref{eq:Z1toP0} and \eqref{eq:genZ2} as well as $h^\mu_\kappa v^\nu\os{1}{\Pi}_{\mu\nu}=h^\mu_\kappa v^\nu\os{1}{g}_{\mu\nu}$.
The combination $h^\mu_\kappa v^\nu\os{1}{g}_{\mu\nu}-\os{2}{\beta}a_\kappa$ has nice Weyl transformation properties as we will see in Section~\ref{subsec:Weyltrafos}.
We will postpone rewriting $\tilde T^{\rho\sigma}_{\text{EOM}}$ in \eqref{eq:tildeTEOM} until the next section.

Of course, we are still free to redefine the energy-momentum-news complex in a way that leaves the boundary diffeomorphism Ward identity invariant.
In the following section, we will consider such improvement transformations in the context of the boundary Weyl and Carroll boost Ward identities.
In particular, we will perform improvements such that the components of the EMT-news complex have a definite Weyl weight.
In other words, the current form is not yet the final form,
and so we will not try to write it in a nicer form yet.
Finally, we cannot directly compare with the results obtained in Section~\eqref{sssec:bondi-loss-4d-simplified-emt-news-complex} where we used $d\tau=0$ because this requires an improvement transformation that makes $S^{\mu\nu}=\frac{1}{2}N^{\mu\nu}$.
Such transformations will be the subject of the next section.

\subsection{The EMT-news complex when including logs}
\label{ssec:log-emt-news}

In the calculations above, we did not include the log terms discussed in Section~\ref{ssec:radial-expansion-logs}.
In this subsection, we will show that the EMT-news complex takes the same form when the log terms of that section are included.
The only difference with the results above is that the $D_{\mu\nu}$~tensor now obeys one fewer constraint, and therefore is a different contribution to the EMT-news complex.

We begin with the $\Pi_\kappa^\mu U^\nu R_{\mu\nu} = 0$ equation, which is given explicitly in~\eqref{eq:Pi-U-R-d2}. We are interested in this equation at order $r^{-2}$, and so log terms only enter via radial derivatives, since
\begin{equation}
    \D_r (r^n \log r) = n r^{n-1}\log r + r^{n-1}\,.
\end{equation}
Using the on-shell versions of the expansions in~\eqref{eq:first-log-block},~\eqref{eq:second-log-block} and~\eqref{eq:logexpansionsNNLO},
we see that only the term 
\begin{equation}
    \frac{1}{2}S\Pi^\rho_\kappa\D_r \mathcal{Z}_\rho\,,
\end{equation}
gives rise to a logarithmic contribution at order $r^{-2}$, namely
\begin{equation}
\label{eq:log-terms-D}
    \frac{1}{2}\os{-1}{S}h^\rho_\kappa \os{2,1}{\mathcal{Z}}_\rho = -K h^\rho_\kappa h^{\mu\nu}\mathcal{D}_\mu D_{\nu\rho} = -K h^\rho_\kappa (\mathcal{D}_\mu - a_\mu)D^\mu{_\rho}\,,
\end{equation}
where we used~\eqref{eq:first-lifted-constraint} as well as the first identity in~\eqref{eq:app-boundary-cov-der-h-up-v-divergence}. 
This extra contribution is proportional to the left-hand side of the constraint~\eqref{eq:PDZ}, which is no longer set to zero when logs are included. All terms involving the $D_{\mu\nu}$ tensor appear in block~$2$, corresponding to~\eqref{eq:block-2}, which was rewritten in Equation~\eqref{eq:gen-mom-eq-WI-matching-block-2-results} by imposing the constraint~\eqref{eq:PDZ}. In the presence of logs, this constraint is replaced by~\eqref{eq:first-lifted-constraint}. Combining the terms of block $2$ with those in~\eqref{eq:log-terms-D} gives
\begin{equation}
    B^2_\kappa -K h^\rho_\kappa (\mathcal{D}_\mu - a_\mu)D^\mu{_\rho} = -D^\mu{_\kappa}\D_\mu K - K h_\kappa^\rho\mathcal{D}_\mu D^\mu{_\rho} = -h_\kappa^\rho \mathcal{D}_\mu(KD^\mu{_\rho})\,,
\end{equation}
which is the same as what we obtained in~\eqref{eq:gen-mom-eq-WI-matching-block-2-results}. Thus, the constraint~\eqref{eq:PDZ} is replaced by the logarithmic contributions in such a way that the final result takes the same form. The currents obtained from the $U^\mu U^\nu R_{\mu\nu} = 0$ equation,
which is given in~\eqref{eq:U-U-R-d2},
are unaffected by the log terms of Section~\ref{ssec:radial-expansion-logs}.
This implies that the EMT-news complex maintains the same form when including these log terms,
though of course the logs change the nature of the $D_{\mu\nu}$ tensor.

We thus recover the same form for the expression of $P^{\text{EOM}}_\kappa$ given in \eqref{eq:PEOM}. 
However, the expression for $\os{2}{\mathcal{Z}}_\kappa$ in \eqref{eq:genZ2} receives a contribution from  $\log r$ terms.
This is because $\mathcal{Z}_\kappa=\Pi_{\kappa\rho}\partial_r U^\rho-\mathcal{A}_\kappa$ and so the $r^{-3}\log r$ term in $U^\rho$ contributes to $\os{2}{\mathcal{Z}}_\kappa$.
We thus get $\os{2}{\mathcal{Z}}_\kappa=\cdots+\frac{2}{3}h^{\rho\sigma}\mathcal{D}_\rho D_{\sigma\mu}$, where the dots denote the terms given in \eqref{eq:genZ2}.
This contributes a term $\frac{1}{3}h^{\rho\sigma}\mathcal{D}_\rho D_{\sigma\mu}$ to $P^{\text{EOM}}_\kappa$.
However, we will momentarily show that 
\begin{equation}\label{eq:IdforD}
    \left(\mathcal{L}_v-K\right)\left(h^{\rho\sigma}\mathcal{D}_\rho D_{\sigma\kappa}\right)=0\,,
\end{equation}
on account of the fact that $\mathcal{L}_v D_{\mu\nu}=0$ and so we can ignore the $h^{\rho\sigma}\mathcal{D}_\rho D_{\sigma\kappa}$ term in $P^{\text{EOM}}_\kappa$ as it is of the form of the $t^{\mu\nu}$ improvement term \eqref{eq:constraint-response-term} where $\zeta_{\mu\nu}$ is proportional to $D_{\mu\nu}$. Since this fundamental ambiguity will be added in Section~\ref{ssec:bulk-improvements-weyl-covariant-currents} later on anyway, we can ignore the $h^{\rho\sigma}\mathcal{D}_\rho D_{\sigma\kappa}$ term in $P^{\text{EOM}}_\kappa$.

What about logarithmic orders in the equations of motion themselves?
For both $\Pi_\kappa^\mu U^\nu R_{\mu\nu} = 0$ and $U^\mu U^\nu R_{\mu\nu} = 0$, there is no contribution at order $r^{-1}\log r$.
The first of these equations, listed in~\eqref{eq:Pi-U-R-d2},
gets a contribution at order $r^{-2}\log r$ from the terms $\frac{1}{2}S\Pi^\rho_\kappa(\D_r\mathcal{Z}_\rho + r^{-1}\mathcal{Z}_\rho)$ and $\frac{1}{2}\mathcal{L}_U \mathcal{Z}_\kappa$, namely
\begin{equation}
    \Pi_\kappa^\mu U^\nu R_{\mu\nu}\big\vert_{@r^{-2}\log r} = \frac{1}{2}\left(\mathcal{L}_v - K \right)\os{2,1}{\mathcal{Z}}_\kappa = -\left(\mathcal{L}_v - K \right)( h^\rho_\kappa h^{\mu\nu}\mathcal{D}_\mu D_{\nu\rho} )\,.
\end{equation}
The tensor $D_{\mu\nu}$ is STF and spatial, implying that
\begin{equation}
\label{eq:identity-for-DD}
    h^\rho_\kappa h^{\mu\nu}\mathcal{D}_\mu D_{\nu\rho} = h_{\kappa\nu}\mathcal{D}_\mu D^{\mu\nu} - a_\rho D^\rho{_\kappa}\,,
\end{equation}
while the condition $\mathcal{L}_v D_{\mu\nu} = 0$ from~\eqref{eq:STF1/rSimple} implies that 
\begin{equation}
\label{eq:lemma-for-D}
    (\mathcal{L}_v - 2K)D^{\mu\nu} = 2v^{(\mu}a_\rho D^{\nu)\rho}\,,
\end{equation}
where we used the identity $\mathcal{L}_v h^{\mu\nu} = 2a^{(\mu} v^{\nu)} + Kh^{\mu\nu}$. Now, using the identity~\eqref{eq:comLiecovSTF} with $X^{\mu\nu} = D^{\mu\nu}$, the left-hand side of that identity becomes
\begin{equation}
    h_{\kappa\nu}(\mathcal{L}_v - 2K)\mathcal{D}_\mu D^{\mu\nu} = (\mathcal{L}_v - K)(h_{\kappa\nu}\mathcal{D}_\mu D^{\mu\nu})\,,
\end{equation}
while the right-hand side involves~\eqref{eq:lemma-for-D} and reduces to
\begin{equation}
    2h_{\kappa\nu}\mathcal{D}_\mu \left(v^{(\mu}a_\rho D^{\nu)\rho}\right) = (\mathcal{L}_v - K)(a_\rho D^\rho{_\kappa})\,,
\end{equation}
where we used that, for any spatial $X_\mu$, we have $\mathcal{L}_v X_\mu = v^\rho\mathcal{D}_\rho X_\mu - \frac{K}{2}X_\mu$. Subtracting one side from the other in~\eqref{eq:comLiecovSTF} with $X^{\mu\nu} = D^{\mu\nu}$ therefore gives
\begin{equation}
    0 = -(\mathcal{L}_v - K)\left(h_{\kappa\nu}\mathcal{D}_\mu D^{\mu\nu} -  a_\rho D^\rho{_\kappa}\right) = -\left(\mathcal{L}_v - K \right)(h^{\mu\nu}\mathcal{D}_\mu D_{\nu\kappa})\,,
\end{equation}
where we used~\eqref{eq:identity-for-DD}, and hence 
\begin{equation}
    \Pi_\kappa^\mu U^\nu R_{\mu\nu}\big\vert_{@r^{-2}\log r} = 0\,.
\end{equation}
Finally, there are no contributions at order $r^{-2}\log r$ for the equation $U^\mu U^\nu R_{\mu\nu} =0$. This shows that all terms in the equations of motion that appear at logarithmic order up to $\OO(r^{-2})$ vanish when using lower-order equations of motion, so they impose no new constraints or evolution-type equations. 

\section{Improvements and the EMT-news complex}
\label{sec:bulk-improvements}
In the previous section, we demonstrated that the bulk equations of motion
corresponding to the Bondi loss equations
could be recast in the form of a (so far putative) boundary diffeomorphism Ward identity.
Along the way,
we were able to read off the components of a corresponding (putative) boundary energy-momentum-news tensor complex, up to improvements.

Recall from Section~\ref{sec:variations-ward-ids} that if our currents are given as the response to varying a (suitably renormalised)
action with respect to the boundary conformal Carroll metric and shear data as follows,%
\footnote{%
    On the left-hand side, we have written an action whose variation is evaluated on shell.
    This is the calculation we will do in the context of gravity near future null infinity in Section~\ref{sec:HoloRenormAndOn-ShellActions} below.
}
\begin{equation}
  \left(\delta S\right)_\text{os}
  = \int d^{d+1}x\,e \left(
    T^\mu \delta \tau_\mu
    + \frac{1}{2} T^{\mu\nu} \delta h_{\mu\nu}
    + \frac{1}{2} S^{\mu\nu} \delta C_{\mu\nu}
  \right)\,,
\end{equation}
and if the action is Weyl-invariant,
then there must exist an improvement of our EMT-news complex such that it obeys the Weyl Ward identity~\eqref{eq:boundary-Weyl-WI}.
The improved EMT-news complex should then also have definite Weyl weights.
In this section, we will show that such an improvement indeed exists,
and we will construct it explicitly.
The situation with Carroll boost invariance is more complicated and will be further discussed in Section~\ref{sec:HoloRenormAndOn-ShellActions}.
There, we will also show that the EMT-news complex can be derived from the variation of the gravitational action, supplemented with appropriate counterterms.

Before we discuss improvement transformations of the EMT-news complex,
we first continue the discussion of Weyl-covariant objects that we started in Section~\ref{subsec:Weylcov} (see also Appendix \ref{ssec:bulk-improvements-weyl-cov-derivs}).

\subsection{Weyl transformations of various objects}\label{subsec:Weyltrafos}
Since our goal is to improve the EMT-news complex in such a way that its components have definite Weyl weights,
it is useful to first gain some knowledge about Weyl covariance of the various pieces that make up to the EMT-news complex.
By an improvement, we mean a redefinition of $T^\mu, T^{\mu\nu}, S^{\mu\nu}$ that leaves the diffeomorphism Ward identity form-invariant.
The Weyl weights of the (suitably improved) EMT-news complex are given in equations \eqref{eq:gaugetrafoTshear}, \eqref{eq:gaugetrafoTshear2} and \eqref{eq:gaugetrafoTshear3}.

The shear $C_{\mu\nu}$ is a spatial STF tensor
with Weyl weight $+1$ by~\eqref{eq:weyl-tr-shear-twist}. Then by~\eqref{eq:weyl-homog-lie-deriv-special-cases}
the news tensor,
\begin{equation}
  \label{eq:NewsDefn-repeat}
  N_{\mu\nu}
  = - \left(\LL_v 
  + \frac{1}{2} K\right) C_{\mu\nu}\,,
\end{equation}
is Weyl-invariant (so Weyl weight 0).
This is the reason we originally introduced this particular definition of~$N_{\mu\nu}$ in Section~\ref{eq:on-shell-action-leading-order-with-constraint-variation-intro}.
Since $C^{\mu\nu}$ and $N^{\mu\nu}$ are weight $-3$ and $-4$ spatial STF tensors,
it is easy to check, using \eqref{eq:Weylcovdiv} or \eqref{eq:weyl-homog-covariant-deriv}, that 
\begin{equation}
   \left(
    \mathcal{D}_\rho - 2 a_\rho
  \right) C^{\rho\mu}\,,\qquad\left(
    \mathcal{D}_\rho - a_\rho
  \right) N^{\rho\mu}\,,  \label{eq:weyl-covariant-divergence-shear-news}
\end{equation}
are Weyl-covariant with weights $-3$, and  $-4$, respectively.
Next, the tensor $F^{\mu\nu}$ has Weyl weight $-3$ and, because it is antisymmetric, it follows from \eqref{eq:Weylcovdivvector} and the comment immediately below it that
\begin{equation}
    \mathcal{D}_\rho F^{\rho\mu}\,,
\end{equation}
is Weyl-covariant with weight $-3$.
Note that the result of the Weyl covariant derivative depends on the symmetry properties of the tensor it acts on.
Since $\left(
    \mathcal{D}_\rho - 2 a_\rho
  \right) C^{\rho\mu}$ is a vector of Weyl weight $-3$, Equation~\eqref{eq:Weylcovdivvector}  tells us that its Weyl-covariant divergence is given by
\begin{gather}
  \label{eq:weyl-double-covariant-divergence-shear}
  \mathcal{D}_\mu
  \left(
    \mathcal{D}_\rho - 2 a_\rho
  \right) C^{\rho\mu}\,.
\end{gather}

The EMT-news complex~\eqref{eq:EMTNewsComplexSec7} we derived in the previous Section also contains several objects appearing in the radial expansion of our bulk metric variables $\beta$, $\Pi_{\mu\nu}$, $U^\mu$, $\Pi^{\mu\nu}$ and $S$.
Even though we can derive explicit expressions in terms of boundary variables for most of these objects,
an easy way to derive their Weyl transformation is as follows.
Consider the residual transformations~\eqref{eq:car-cov-bondi-gauge-tr-repeat} and~\eqref{eq:gaugetrafobeta}
of the unexpanded variables. We can restrict the bulk diffeomorphism to $\xi^r=r\Lambda_D$
and $\xi^\mu=0$,
which corresponds to considering only Weyl transformations on the boundary.
This gives (for $d=1,2$)
\begin{subequations}
  \label{eq:S-Pi-beta-Weyl-transformations}
  \begin{align}
    \delta_{\Lambda_D}
    \beta
    &= \Lambda_D r \pd_r \beta\,,
    \\
    \delta_{\Lambda_D}
    \Pi_{\mu\nu}
    &= \Lambda_D r \pd_r \Pi_{\mu\nu}
    - 2 r e^\beta \tau_{(\mu} \Pi^\rho_{\nu)} \pd_\rho \Lambda_D\,,
    \\
    \delta_{\Lambda_D}
    U^\mu
    &= \Lambda_D \left(r \pd_r - 1\right) U^\mu
    - r \Pi^{\mu\nu} \pd_\nu \Lambda_D\,,
    \\
    \delta_{\Lambda_D}
    \Pi^{\mu\nu}
    &= \Lambda_D r \pd_r \Pi^{\mu\nu}\,,
    \\
    \delta_{\Lambda_D}
    S
    &= \Lambda_D \left(r \pd_r - 2\right) S
    - 2 r U^\mu \pd_\mu \Lambda_D\,.
  \end{align}
\end{subequations}
In deriving these transformations, we used that $\Lambda_D$ only appears at leading order in $\xi^r$ and not in the subleading orders of $\xi^r$ nor in the expansion of $\xi^\mu$,
which is discussed in footnote \ref{footnote:LambdaDdep}. 
This means that $\os{k}{\beta}$ has Weyl weight $-k$. The terms in the expansion of $\Pi_{\mu\nu}$ and $S$ are generically inhomogeneous.
For example,
we have
\begin{equation}\label{eq:Wtrafoorder0}
  \delta_{\Lambda_D}
  \os{0}{S}
  = - 2 \Lambda_D \os{0}{S}
  + 2 a^\mu \pd_\mu \Lambda_D\,,
  \qquad
  \delta_{\Lambda_D}
  \os{0}{\Pi}_{\mu\nu}
  = 2 \tau_\mu \tau_\nu a^\rho \pd_\rho \Lambda_D\,,
\end{equation}
where we used the on-shell results
as summarised in Appendix~\ref{sapp:on-shell-expansion-fundamental-variables}.
From this, it is easy to see that the combinations
\begin{equation}
  \label{eq:S-Pi-expansion-Weyl-covariant-projections}
  \os{0}{S}
  - a^2\,,
  \qquad
  \os{0}{\Pi}_{\mu\nu}-\tau_\mu\tau_\nu a^2\,,
\end{equation}
where $a^2=h^{\mu\nu}a_\mu a_\nu$,
are Weyl-covariant with weights $-2$ and $0$, respectively. 

We furthermore note that
\begin{align}
  \delta_{\Lambda_D}
  g_{\mu\nu}
  =\delta_{\Lambda_D}
  \left(
    - e^{2\beta} S \tau_\mu\tau_\nu
    + \Pi_{\mu\nu}
  \right)
  \label{eq:weyl-transformation-of-gmunu}
  = \Lambda_D r \pd_r g_{\mu\nu}
  - 2r e^\beta \tau_{(\mu} \pd_{\nu)} \Lambda_D\,.
\end{align}
We can use this to show that the following combination is Weyl-covariant at every order in its radial expansion,
\begin{equation}
    \label{eq:g-tau-b-weyl-cov-combination}
    g_{\mu\nu}+re^\beta\left(\tau_\mu\tilde b_\nu+\tau_\nu\tilde b_\mu\right)\,.
\end{equation}
To show this, use the transformation 
\begin{equation}
    \delta_{\Lambda_D}\left(r e^\beta\right)=\Lambda_D\left(r\partial_r-1\right)\left( re^\beta\right)\,,
\end{equation}
so that we have
\begin{equation}
    \delta_{\Lambda_D}
  \left(g_{\mu\nu}+re^\beta\left(\tau_\mu\tilde b_\nu+\tau_\nu\tilde b_\mu\right)\right)=\Lambda_D r\partial_r\left(g_{\mu\nu}+re^\beta\left(\tau_\mu\tilde b_\nu+\tau_\nu\tilde b_\mu\right)\right)\,.
\end{equation}
We can use~\eqref{eq:g-tau-b-weyl-cov-combination} to construct further Weyl-covariant expressions.
For example,
\begin{equation}
  \delta_{\Lambda_D} \left(
    h^\mu_\rho v^\nu g_{\mu\nu}
    - r e^\beta a_\rho
  \right)
  = \Lambda_D \left(r \pd_r - 1\right) \left(
    h^\mu_\rho v^\nu g_{\mu\nu}
    - r e^\beta a_\rho
  \right)\,,
\end{equation}
as well as
\begin{align}
  \delta_{\Lambda_D}
  \left(
    v^\mu v^\nu g_{\mu\nu}
    + \frac{2}{d} r e^\beta K
  \right)
  = \Lambda_D \left(r \pd_r - 2\right)
  \left(
    v^\mu v^\nu g_{\mu\nu}
    + \frac{2}{d} r e^\beta K
  \right)\,.
\end{align}
For $d=2$ and at order $r^{-1}$,
the following combination has Weyl weight $-2$,
\begin{equation}
  \label{eq:beta-2-Weyl-covariant-combination}
  h^\mu_\rho v^\nu \os{1}{g}_{\mu\nu}
  - \os{2}{\beta} a_\rho=h^\mu_\rho v^\nu \os{1}{\Pi}_{\mu\nu}
  - \os{2}{\beta} a_\rho\,.
\end{equation}
Furthermore, we also conclude that 
\begin{align}
  \label{eq:S-1-Weyl-covariant-combination}
  v^\mu v^\nu \os{1}{g}_{\mu\nu}
  + \os{2}{\beta}K
  &= - \os{1}{S}
  - a_\sigma\mathcal{D}_\rho C^{\rho\sigma}
  + a_\rho a_\sigma C^{\rho\sigma}
  \\
  &{}\qquad\nonumber
  + a_\sigma\mathcal{D}_\rho F^{\rho\sigma}
  - \frac{1}{16}K\left(F^2-C^2\right)\,,
\end{align}
has Weyl weight $-3$.
We used~\eqref{eq:vvPione} and~\eqref{eq:app-4Dsol-beta2},
which says that $\os{2}{\beta}=\frac{1}{16}\left(F^2-C^2\right)$.
It is also useful to note that
\begin{equation}
    v^\mu v^\nu \os{1}{g}_{\mu\nu}=v^\mu v^\nu \os{1}{\Pi}_{\mu\nu}-\os{1}{S}-2K\os{2}{\beta}\,.
\end{equation}
This gives all Weyl-covariant data we will need in the following computations.

\subsection{Weyl improvements in three dimensions}
\label{ssec:bulk-improvements-weyl-3d}
Let us first consider the three-dimensional case,
and recall that the news, shear and extrinsic curvature all vanish here.
In Equation~\eqref{eq:SETd1} of Section~\ref{ssec:ConservationEqnsd1},
we obtained the following energy-momentum tensor components,
\begin{subequations}
  \label{eq:SETd1-repeat}
  \begin{align}
    \tau_\rho T^\rho
    &= \frac{1}{2}\overset{(0)}{S}
    -\mathcal{D}_\mu a^\mu\,,
    \\
    T^\rho h_\rho^\mu
    &= h^{\mu\nu}\partial_\nu K\,,
    \\
    h_{\rho\sigma}T^{\rho\sigma}
    &= -\frac{1}{2}\overset{(0)}{S}\,,
    \\
    P_\kappa
    = T^{\mu\nu}\tau_\mu h_{\nu\kappa}
    &= \frac{1}{2}\overset{(1)}{\mathcal{Z}}_\kappa=h^\rho_\kappa v^\sigma\os{0}{g}_{\rho\sigma}\,,\label{eq:calZ1d=1}
  \end{align}
\end{subequations}
where, in the last equality, we used that for $d=1$ we have
\begin{equation}
    \os{1}{\mathcal{Z}}_\kappa=2h^\rho_\kappa v^\sigma\os{0}{g}_{\rho\sigma}\,.
\end{equation}
This can be derived using the results of Appendix\ref{sapp:on-shell-expansion-fundamental-variables}.

This particular EMT does not obey the Weyl Ward identity,
since we get
\begin{equation}
  \label{eq:seemingly-anomalous-Weyl-WI-threedim}
  \tau_\rho T^\rho+h_{\rho\sigma}T^{\rho\sigma}
  =-e^{-1}\partial_\mu\left(e h^{\mu\nu}a_\nu\right)\,,
\end{equation}
which is non-zero on the right-hand side.
At the same time, using \eqref{eq:Wtrafoorder0}, we also see that this EMT has no definite Weyl weight. 
We will now construct an improvement that sets the right-hand side of~\eqref{eq:seemingly-anomalous-Weyl-WI-threedim} to zero.
We will consider improvements that are variational.
By this, we mean that we write down a diffeomorphism-invariant action,
which we then vary with respect to $(\tau_\mu,h_{\mu\nu})$ to generate the improvements.
This guarantees that the diffeomorphism Ward identity is form-invariant.

Based on 
\eqref{eq:S-Pi-expansion-Weyl-covariant-projections}, it is easy to guess that
\begin{equation}
   \int d^2 x \, e \, h^{\mu\nu} a_\mu a_\nu\,,
\end{equation}
will be the `generator' of our improvement transformation. To compute the variation we use \eqref{eq:vara} which in particular implies that for $d=1$ we have
\begin{equation}
    a^\mu\delta a_\mu=-Ka^\mu\delta\tau_\mu-h^{\mu\nu}\mathcal{L}_v a_\nu\delta \tau_\mu+\mathcal{D}_\rho a^\rho v^\mu\delta\tau_\mu\,,
\end{equation}
up to total derivatives.
This leads to the following improvement,
\begin{eqnarray}
    T'^\mu & = & T^\mu -2cKa^\mu-2ch^{\mu\nu}\mathcal{L}_v a_\nu-cv^\mu a^2 +2c v^\mu\mathcal{D}_\rho a^\rho\,,\\
    T'^{\mu\nu} & = & T^{\mu\nu}-ch^{\mu\nu}a^2\,,
\end{eqnarray}
where $c$ is some real constant. If we take $c=-1/2$ we end up with 
\begin{subequations}
  \label{eq:SETd1-impr}
  \begin{align}
    \tau_\rho T'^\rho
    &= \frac{1}{2}\left(\overset{(0)}{S}-a^2\right)\,,
    \\
    T'^\rho h_\rho^\mu
    &= h^{\mu\nu}\left(\partial_\nu K+Ka_\nu+\mathcal{L}_v a_\nu\right)=h^{\mu\nu}v^\rho\left(\partial_\rho\tilde b_\nu-\partial_\nu\tilde b_\rho\right)\,,
    \\
    h_{\rho\sigma}T^{\rho\sigma}
    &= -\frac{1}{2}\left(\overset{(0)}{S}-a^2\right)\,,
    \\
    P_\kappa
    = T^{\mu\nu}\tau_\mu h_{\nu\kappa}
    &=h^\rho_\kappa v^\sigma\os{0}{g}_{\rho\sigma}\,,
  \end{align}
\end{subequations}
where, in the third equality, we used that for $d=1$ we have $\tilde b_\mu=a_\mu+K\tau_\mu$
following~\eqref{eq:b-tilde-def-and-weyl-tr}.
We conclude that the improved EMT obeys 
\begin{equation}
  \tau_\rho T'^\rho+h_{\rho\sigma}T'^{\rho\sigma}
  =0\,,
\end{equation}
and that $T'^\mu$ and $T'^{\mu\nu}$ now have Weyl weights $-3$ and $-4$, respectively. Finally we see that 
\begin{equation}
    T'^\rho h_\rho^\mu=h^{\mu\nu}v^\rho\left(\partial_\rho\tilde b_\nu-\partial_\nu\tilde b_\rho\right)\,,
\end{equation}
is non-zero. In Section~\ref{ssec:hol-ren-threedim-anomalies-improvements} we will see this is related to a Carroll boost anomaly. 

\subsection{Weyl improvements in four dimensions}
\label{ssec:bulk-improvements-weyl-4d}
In the four-dimensional case,
the current components we derived in~\eqref{eq:EMTNewsComplexSec7} give
\begin{align}
  \label{eq:seemingly-anomalous-Weyl-WI-fourdim}
  T^\mu \tau_\mu
  + T^{\mu\nu} h_{\mu\nu}
  + \frac{1}{2} {S}^{\mu\nu} C_{\mu\nu}
  &=
  \frac{1}{4}C\cdot N
  +\frac{1}{2}\mathcal{D}_\rho\mathcal{D}_\sigma C^{\rho\sigma}
  +\frac{1}{8}KF^2
  -\frac{1}{8}\mathcal{L}_v F^2\,,
\end{align}
where we dropped the subscript EOM.
The first term on the right-hand side of~\eqref{eq:seemingly-anomalous-Weyl-WI-fourdim} is quadratic in the shear and first order in derivatives.
Up to total derivatives,
the only diffeomorphism-invariant action we can write that is quadratic in shear and first order in derivatives is 
\begin{equation}
  \label{eq:fourdim-weyl-improvement-term-1}
   a_1 \int d^3x\,e\, KC^2\,,
\end{equation}
where $a_1$ is some real parameter. The next term on the right-hand side of \eqref{eq:seemingly-anomalous-Weyl-WI-fourdim} is linear in the shear and second order in derivatives. For $d=2$, the most general diffeomorphism invariant action with these properties is of the form 
\begin{equation}
  \label{eq:fourdim-weyl-improvement-term-23}
  \int d^3x\, e \left(
    a_2 C^{\mu\nu}a_\mu a_\nu
    + a_3 C^{\mu\nu}\mathcal{D}_\mu a_\nu
  \right)\,,
\end{equation}
with $a_2$ and $a_3$ arbitrary parameters. Finally, the last two terms in \eqref{eq:seemingly-anomalous-Weyl-WI-fourdim} are quadratic in the twist tensor $F_{\mu\nu}$ and cubic in derivatives. The only action with these properties is 
\begin{equation}
  \label{eq:fourdim-weyl-improvement-term-4}
   a_4 \int d^3x\,e\, KF^2\,.
\end{equation}
In total, we therefore obtain
\begin{equation}\label{eq:improvgen}
    \int d^3x\, e \left(a_1 KC^2+
    a_2 C^{\mu\nu}a_\mu a_\nu
    + a_3 C^{\mu\nu}\mathcal{D}_\mu a_\nu+a_4 KF^2
  \right)\,.
\end{equation}
as the generating action for our improvement transformation.

Consider first the expression for the news current $S^{\mu\nu}$ in~\eqref{eq:tildeSUPiR22}.
We would like to improve this until it becomes Weyl-covariant.
The action~\eqref{eq:fourdim-weyl-improvement-term-23} leads to the following improvement of $S^{\mu\nu}$,
\begin{equation}\label{eq:imprS}
    S'^{\mu\nu}=S^{\mu\nu}+4a_1KC^{\mu\nu}+h^{\rho\langle\mu}h^{\nu\rangle\sigma}\left(2a_3A_{\rho\sigma}+2(a_2-a_3)a_\rho a_\sigma\right)\,.
\end{equation}
Comparing with \eqref{eq:tildeSUPiR22} we see that we need to set $a_2=a_3$,
since we would otherwise generate a term proportional to $a^{\langle\mu}a^{\nu\rangle}$ which we don't want.
If we furthermore take $a_1=1/16$ and $a_2=a_3=-1/2$ then we get $S'^{\mu\nu}=\frac{1}{2}N^{\mu\nu}$,
which is Weyl-covariant,
as we can see from the discussion around~\eqref{eq:NewsDefn-repeat}.
However, for now we will only set $a_2=a_3$ and work out how this improvement affects the EMT.

To this end,
we vary \eqref{eq:improvgen} with $a_3=a_2$ with respect to $\tau_\mu$ and $h_{\mu\nu}$,
using~\eqref{eq:app-variation-relations-up-to-down} and~\eqref{eq:app-integration-measure-variation} as well as~\eqref{eq:varconn}, \eqref{eq:vara} and~\eqref{eq:varK}.
This leads to the following improvement transformation,
\begin{subequations}
  \label{eq:fourdim-weyl-improvement-term-23-currents}
  \begin{align}
    T'^\mu &=  T^\mu
    -a_2\left(\mathcal{D}_\rho-a_\rho\right)N^{\rho\mu}-\frac{1}{2}a_2\partial_\rho K C^{\rho\mu}\nonumber
    \\
    &\qquad\nonumber
    +\frac{1}{2}a_2 K\mathcal{D}_\rho C^{\rho\mu}
    -\frac{1}{2}a_2 K C^{\mu\rho}a_\rho-a_2v^\mu\mathcal{D}_\rho\mathcal{D}_\sigma C^{\rho\sigma}
    \\
    &\qquad\nonumber
    +a_2 v^\mu C\cdot A-2a_2 v^\mu C^{\rho\sigma}a_\rho a_\sigma+2a_2v^\mu a_\sigma\mathcal{D}_\rho C^{\rho\sigma}
    \\
    &\qquad -2a_4Kv^\mu F^2-4a_4\partial_\rho K F^{\rho\mu}-4a_4K h^\mu_\sigma\mathcal{D}_\rho F^{\rho\sigma}+4a_4 K a_\nu F^{\mu\nu}\,,\\
    T'^{\mu\nu}  &=  T^{\mu\nu}
    -a_1KC^2h^{\mu\nu}-2a_1 C\cdot N h^{\mu\nu}-a_1\partial_\rho C^2\left(h^{\rho\mu}v^\nu+h^{\rho\nu}v^\mu\right)\nonumber\\
    &\qquad -a_2h^{\mu\nu}C\cdot A
    +2a_2a^{(\mu}\mathcal{D}_\rho C^{\nu)\rho}-a_2\mathcal{L}_a C^{\mu\nu}-2a_2C^{\mu\nu}\mathcal{D}_\rho a^\rho\nonumber
    \\
    &\qquad \nonumber
    +2a_2 F^{(\mu}{}_\sigma v^{\nu)}\left(\mathcal{D}_\rho-2a_\rho\right) C^{\rho\sigma}
    \\
    &\qquad  -2a_4 KF^2 h^{\mu\nu}+a_4 h^{\mu\nu}\mathcal{L}_v F^2-a_4\left(h^{\rho\mu}v^\nu+h^{\rho\nu}v^\mu\right)\partial_\rho F^2\,,
  \end{align}
\end{subequations}
where we used the identity in~\eqref{eq:app-d2-spatial-STF-STF-product} for the symmetric product of $C_{\mu\nu}$ and $A_{\mu\nu}$, which are both STF tensors. This implies that
\begin{subequations}
\label{eq:improve-eom-to-weyl-d=2}
    \begin{align}
    \tau_\mu T'^\mu & =  \tau_\mu T^\mu+a_2\mathcal{D}_\rho\mathcal{D}_\sigma C^{\rho\sigma}-a_2  C\cdot A\nonumber\\
    &\qquad-2a_2\left(a_\sigma\mathcal{D}_\rho C^{\rho\sigma}-C^{\rho\sigma}a_\rho a_\sigma\right)\label{eq:improvtauTmu}
     +2a_4K F^2\,,\\
     h^\mu_\rho T'^\rho & =  h^\mu_\rho T^\rho-a_2\left(\mathcal{D}_\rho-a_\rho\right)N^{\rho\mu}-\frac{1}{2}a_2\partial_\rho K C^{\rho\mu}\nonumber
    \\
    &\qquad\nonumber
    +\frac{1}{2}a_2 K\mathcal{D}_\rho C^{\rho\mu}
    -\frac{1}{2}a_2 K a_\rho C^{\rho\mu}
    \\
    &\qquad -4a_4\partial_\rho K F^{\rho\mu}-4a_4K h^\mu_\sigma\mathcal{D}_\rho F^{\rho\sigma}-4a_4 K a_\rho F^{\rho\mu}\,,\\
    h_{\mu\nu}T'^{\mu\nu} & =  h_{\mu\nu}T^{\mu\nu}-2a_1 KC^2-4a_1 C\cdot N+2a_2\left(a_\sigma\mathcal{D}_\rho C^{\rho\sigma}-C^{\rho\sigma}a_\rho a_\sigma\right)\nonumber\\
    &\qquad-4a_4 KF^2+2a_4\mathcal{L}_v F^2\,,\\
    P'_\kappa & = P_\kappa+a_1 h^\rho_\kappa\partial_\rho C^2-a_2 a^\rho C_\rho{}^\sigma F_{\sigma\kappa}-a_2 F_{\kappa\sigma}\left(\mathcal{D}_\rho-2a_\rho\right)C^{\rho\sigma}\nonumber\\
    &\qquad+a_4 h^\rho_\kappa\partial_\rho F^2\,,\\
    \tilde T'^{\mu\nu} & = \tilde T^{\mu\nu}+a_2 h^{\langle\mu}_\rho h^{\nu\rangle}_\sigma\left(2a^{\rho}\mathcal{D}_\alpha C^{\sigma\alpha}-\mathcal{L}_a C^{\rho\sigma}-2C^{\rho\sigma}\mathcal{D}_\alpha a^\alpha\right)\,,
\end{align}
\end{subequations}
where we used
\begin{align}
\begin{split}
    h_{\mu\nu}\mathcal{L}_a C^{\mu\nu} & = -2C\cdot A+2 C^{\rho\sigma}a_\rho a_\sigma\,,\\
\tau_\mu h_{\nu\kappa}\mathcal{L}_a C^{\mu\nu} & = -a^\rho F_{\rho\sigma}C^\sigma{}_\kappa\,.
\end{split}
\end{align}
Together with \eqref{eq:imprS}, this implies that the Weyl Ward identity for these improved currents gives
\begin{align}
    T'^\mu \tau_\mu
  + T'^{\mu\nu} h_{\mu\nu}
  + \frac{1}{2} {S'}^{\mu\nu} C_{\mu\nu}= &T^\mu \tau_\mu
  + T^{\mu\nu} h_{\mu\nu}+ \frac{1}{2} {S}^{\mu\nu} C_{\mu\nu}\label{eq:imprtracerel}\\
  &
  -4a_1 N\cdot C+a_2\mathcal{D}_\rho\mathcal{D}_\sigma C^{\rho\sigma}-2a_4 KF^2+2a_4\mathcal{L}_v F^2\,.\nonumber
\end{align}
Using the initial result in~\eqref{eq:seemingly-anomalous-Weyl-WI-fourdim} we see that if we take
\begin{equation}
    a_1=\frac{1}{16}\,,\qquad a_2=-\frac{1}{2}\,,\qquad a_4=\frac{1}{16}\,,
\end{equation}
the right-hand side of \eqref{eq:imprtracerel} vanishes,
and the Weyl Ward identity holds.

For these values, we get the following EMT-news complex:
\begin{subequations}
  \label{eq:WeylImprovedEMTNewsComplex}
  \begin{align}S^{\mu\nu}_W
    =& \frac{1}{2}N^{\mu\nu}\,,
    \\
    \label{eq:WeylImprovedEMTNewsComplex-energy-density}
    \tau_\mu T^\mu_{\text{W}}
    =&-\frac{1}{2}\mathcal{D}_\rho\left(\mathcal{D}_\sigma-2a_\sigma\right) C^{\rho\sigma}-\frac{1}{8}\left(\mathcal{L}_v-K\right) F^2\nonumber\\
    &-v^\mu v^\nu \os{1}{g}_{\mu\nu}
    -\frac{1}{16}K\left(F^2-C^2\right)\,,\\
    h^\mu_\rho T^\rho_{\text{W}}
    =& \frac{1}{2}h^{\mu\nu}\left(\partial_\nu+2a_\nu\right)\left(\overset{(0)}{S}-a^2\right)
    +h^\mu_\rho\left(\mathcal{L}_v-\frac{3}{2}K\right)\os{0}{P}^\rho
    \nonumber\\
    &
    -\frac{1}{2}v^\rho\left(\partial_\rho\tilde b_\sigma-\partial_\sigma\tilde b_\rho\right)\left(C^{\sigma\mu}-F^{\sigma\mu}\right)
    +\frac{1}{2}\left(\mathcal{D}_\rho-a_\rho\right)N^{\rho\mu}\,,\\
    h_{\mu\nu}T^{\mu\nu}_{\text{W}}
    =&\frac{1}{2}\mathcal{D}_\rho\left(\mathcal{D}_\sigma-2a_\sigma\right) C^{\rho\sigma}+\frac{1}{8}\left(\mathcal{L}_v-K\right) F^2-\frac{1}{4} C\cdot N\nonumber\\
    &+v^\mu v^\nu \os{1}{g}_{\mu\nu}
  + \frac{1}{16}K\left(F^2-C^2\right)\,,\\
    P^{\text{W}}_\kappa  = & \frac{3}{2}\left(h^\mu_\kappa v^\nu\os{1}{g}_{\mu\nu}-\os{2}{\beta}a_\kappa\right)+\frac{5}{2}\os{0}{P}^\mu F_{\mu\kappa}-\frac{1}{2}\os{0}{P}^\mu C_{\mu\kappa}\nonumber\\
    &-\frac{1}{32}h^\mu_\kappa(\partial_\mu+2a_\mu) C^2+\frac{1}{32}h^\mu_\kappa(\partial_\mu+2a_\mu) F^2-a^\mu D_{\mu\kappa}\,,\label{eq:WeylImprovedEMTNewsComplex-momentum-density}
    \\
    \tilde T^{\mu\nu}_{\text{W}}
    =& \frac{1}{2}\left(\overset{(0)}{S}-a^2\right)C^{\mu\nu}
    +\frac{1}{2}KD^{\mu\nu}\label{eq:tildeTW}\\
    &+h^{\rho\langle\mu} h^{\nu\rangle\sigma}\left(\left(\mathcal{D}_\rho+3a_\rho\right)\os{0}{P}_\sigma-\frac{1}{4}\mathcal{L}_v\left(F_{\rho\alpha}C^\alpha{}_\sigma\right)+\frac{1}{2}C^\alpha{}_\sigma\left(\mathcal{L}_v +\frac{1}{2}K\right)F_{\rho\alpha}\right)\nonumber\,,
  \end{align}
\end{subequations}
where we used \eqref{eq:S-1-Weyl-covariant-combination} in the expression for $h_{\mu\nu}T^{\mu\nu}_{\text{W}}$ and $\tau_{\mu}T^{\mu}_{\text{W}}$ above,%
\footnote{%
  The expression for~$\tau_\mu T^\mu_\text{W}$ can be further rewritten using 
  \begin{equation}
    \mathcal{D}_\mu\os{0}{P}^\mu+\frac{1}{8}\left(\mathcal{L}_v-K\right)F^2=-\frac{1}{2}\mathcal{D}_\mu\left(\mathcal{D}_\rho-2a_\rho\right)C^{\rho\mu}-\frac{1}{8}\left(\mathcal{L}_v-K\right)F^2\,,
  \end{equation}
  which is how we write it in \cite{Hartong:2025jpp}.
  Here, recall that we have
  $
  \os{0}{P}^\mu=-\frac{1}{2}\left(\mathcal{D}_\rho-2a_\rho\right)C^{\rho\mu}+\frac{1}{2}h^\mu_\sigma\mathcal{D}_\rho F^{\rho\sigma}
  $
  by~\eqref{eq:P0up}.
}
and where in the expression for $h^\mu_\rho T^\rho_{\text{W}}$ we used \eqref{eq:energyfluxEOMv2} and \eqref{eq:P0up}.
In the expression for $P^{\text{W}}_\kappa$, we started from \eqref{eq:momdensityEOM} and used \eqref{eq:P0up} and \eqref{eq:FDFid}.
The subscript $W$ is used to indicate that the currents in this particular improved version of the EMT-news complex have definite Weyl weight.

The derivation of the above expression for $\tilde T^{\mu\nu}_{\text{W}}$ goes as follows. We start with the expression in \eqref{eq:tildeTEOM}, which we repeat for convenience,
\begin{align}
    \tilde T^{\mu\nu}_{\text{W}}
    =& \frac{1}{2}\overset{(0)}{S}C^{\mu\nu}
    +\frac{1}{2}a^\mu\overset{(1)}{\mathcal{Z}}{}^\nu
    +\frac{1}{2}a^\nu\overset{(1)}{\mathcal{Z}}{}^\mu 
    -\frac{1}{2}h^{\mu\nu} a^\lambda \overset{(1)}{\mathcal{Z}}_\lambda
    +\frac{1}{2}KD^{\mu\nu}
    \\
    &
    -\frac{1}{4}KC^{\mu}{}_\rho F^{\rho\nu}
    +h^{\rho\langle\mu}h^{\nu\rangle\sigma}\overset{(0)}{\mathcal{K}}_{\rho\sigma}
    -\frac{1}{2}\overset{(1)}{\mathcal{K}}C^{\mu\nu}\nonumber\\
    &-\frac{1}{2} h^{\langle\mu}_\rho h^{\nu\rangle}_\sigma\left(2a^{\rho}\mathcal{D}_\alpha C^{\sigma\alpha}-\mathcal{L}_a C^{\rho\sigma}-2C^{\rho\sigma}\mathcal{D}_\alpha a^\alpha\right)\,.
\end{align}
This can alternatively be written as
\begin{eqnarray}
    \tilde T^{\mu\nu}_{\text{W}}
    &=& \frac{1}{2}\overset{(0)}{S}C^{\mu\nu}
    +\frac{1}{2}KD^{\mu\nu}\nonumber
    \\
    &&
    -\frac{1}{4}KC^{\mu}{}_\rho F^{\rho\nu}
    +h^{\rho\langle\mu}h^{\nu\rangle\sigma}\overset{(0)}{\mathcal{K}}_{\rho\sigma}
    +\frac{1}{2}C^{\mu\nu}\mathcal{D}_\rho a^\rho \nonumber\\
    &&+h^{\rho\langle\mu} h^{\nu\rangle\sigma}\left(a_\rho\os{1}{\mathcal{Z}}_\sigma-a_{\rho}\mathcal{D}_\alpha C^{\alpha}{}_{\sigma}+\frac{1}{2}h_{\rho\alpha}h_{\sigma\beta} \mathcal{L}_a C^{\alpha\beta}\right)\,.
\end{eqnarray}
Next, we use that
\begin{align}
    &h^{\rho\langle\mu} h^{\nu\rangle\sigma}\left(a_\rho\os{1}{\mathcal{Z}}_\sigma-a_{\rho}\mathcal{D}_\alpha C^{\alpha}{}_{\sigma}+\frac{1}{2}h_{\rho\alpha}h_{\sigma\beta} \mathcal{L}_a C^{\alpha\beta}\right)\\
    &=h^{\rho\langle\mu} h^{\nu\rangle\sigma}\left(2a_\rho\os{0}{P}_\sigma+a_\rho a^\alpha F_{\alpha\sigma}+a_\rho a_\alpha C^\alpha{}_\sigma-a_{\rho}\mathcal{D}_\alpha C^{\alpha}{}_{\sigma}+\frac{1}{2}\mathcal{L}_a C_{\rho\sigma}-C_{\rho\sigma}\mathcal{D}_\alpha a^\alpha\right)
    \nonumber\\
    &=h^{\rho\langle\mu} h^{\nu\rangle\sigma}\left(2a_\rho\os{0}{P}_\sigma+a_\rho a^\alpha F_{\alpha\sigma}-a_{\rho}\mathcal{D}_\alpha C^{\alpha}{}_{\sigma}+\frac{1}{2}a^\alpha\mathcal{D}_\alpha C_{\rho\sigma}-\frac{1}{2}C_{\rho\sigma}\mathcal{D}_\alpha a^\alpha+\frac{1}{2}C^\alpha{}_\sigma\mathcal{L}_v F_{\rho\alpha}\right)\,,
    \nonumber
\end{align}
where we used 
\begin{equation}
    h^{\rho\langle\mu} h^{\nu\rangle\sigma} C_{\rho}{}^{\alpha}A_{\alpha\sigma}=0\,,
\end{equation}
as well as
\begin{equation}
    h^\rho_\mu h^\sigma_\nu\mathcal{D}_\rho a_\sigma=h^\rho_\mu h^\sigma_\nu\left(\frac{1}{2}\mathcal{L}_v F_{\rho\sigma}-a_\rho a_\sigma+\frac{1}{2}h_{\rho\sigma}\mathcal{D}_\alpha a^\alpha+A_{\rho\sigma}\right)\,.
\end{equation}
We know from \eqref{eq:STFpartofcalK0} that
\begin{eqnarray}
    h^\rho_{\langle\mu}h^\sigma_{\nu\rangle}\os{0}{\mathcal{K}}_{\rho\sigma} & = & h^\rho_{\langle\mu}h^\sigma_{\nu\rangle}\left(\left(\mathcal{D}_\rho+a_\rho\right)\os{0}{P}_\sigma-\frac{1}{4}\mathcal{L}_v\left(F_{\rho\alpha}C^\alpha{}_\sigma\right)-a^\alpha F_{\alpha\rho}a_\sigma\right.\nonumber\\
    &&\left.+\frac{1}{2}a^\alpha\left(\mathcal{D}_\alpha C_{\rho\sigma}-\mathcal{D}_\rho C_{\sigma\alpha}-\mathcal{D}_\sigma C_{\rho\alpha}\right)\right)\,.
\end{eqnarray}
Combining this with the previous result leads to
\begin{align}
  &h^{\rho\langle\mu} h^{\nu\rangle\sigma}\left(\os{0}{\mathcal{K}}_{\rho\sigma}+a_\rho\os{1}{\mathcal{Z}}_\sigma-a_{\rho}\mathcal{D}_\alpha C^{\alpha}{}_{\sigma}+\frac{1}{2}h_{\rho\alpha}h_{\sigma\beta} \mathcal{L}_a C^{\alpha\beta}\right)\nonumber\\
  &{}\qquad
  =h^{\rho\langle\mu} h^{\nu\rangle\sigma}\left(\left(\mathcal{D}_\rho
    +3a_\rho\right)\os{0}{P}_\sigma
    +a^\alpha\left(\mathcal{D}_\alpha C_{\rho\sigma}
    -\mathcal{D}_\rho C_{\sigma\alpha}\right)
  \right.
  \nonumber\\
  &{}\qquad\qquad\qquad\qquad\left.
    -a_{\rho}\mathcal{D}_\alpha C^{\alpha}{}_{\sigma}
    -\frac{1}{2}C_{\rho\sigma}\mathcal{D}_\alpha a^\alpha
    -\frac{1}{4}\mathcal{L}_v\left(F_{\rho\alpha}C^\alpha{}_\sigma\right)
  +\frac{1}{2}C^\alpha{}_\sigma\mathcal{L}_v F_{\rho\alpha}\right)
  \nonumber\\
  &=h^{\rho\langle\mu} h^{\nu\rangle\sigma}\left(\left(\mathcal{D}_\rho+3a_\rho\right)\os{0}{P}_\sigma
    -a_{\sigma}a_\alpha C^{\alpha}{}_{\rho}
  \right.
  \nonumber\\
  &{}\qquad\qquad\qquad\qquad\left.
    -\frac{1}{2}C_{\rho\sigma}\mathcal{D}_\alpha a^\alpha
  -\frac{1}{4}\mathcal{L}_v\left(F_{\rho\alpha}C^\alpha{}_\sigma\right)+\frac{1}{2}C^\alpha{}_\sigma\mathcal{L}_v F_{\rho\alpha}\right)
  \nonumber
  \\
  &=h^{\rho\langle\mu} h^{\nu\rangle\sigma}\left(\left(\mathcal{D}_\rho+3a_\rho\right)\os{0}{P}_\sigma
    -\frac{1}{2}a^2 C_{\rho\sigma}
  \right.
  \\
  \nonumber
  &{}\qquad\qquad\qquad\qquad\left.
    -\frac{1}{2}C_{\rho\sigma}\mathcal{D}_\alpha a^\alpha
  -\frac{1}{4}\mathcal{L}_v\left(F_{\rho\alpha}C^\alpha{}_\sigma\right)+\frac{1}{2}C^\alpha{}_\sigma\mathcal{L}_v F_{\rho\alpha}\right)\,.
\end{align}
Combining everything, we end up with 
\begin{align}
  \tilde T^{\mu\nu}_{\text{W}}
  &= \frac{1}{2}\left(\overset{(0)}{S}-a^2\right)C^{\mu\nu}
  +\frac{1}{2}KD^{\mu\nu}\\
  &{}\qquad\nonumber
  +h^{\rho\langle\mu} h^{\nu\rangle\sigma}\left(
    \left(\mathcal{D}_\rho+3a_\rho\right)\os{0}{P}_\sigma
    -\frac{1}{4}\mathcal{L}_v\left(F_{\rho\alpha}C^\alpha{}_\sigma\right)
    +\frac{1}{2}C^\alpha{}_\sigma
    \left(\mathcal{L}_v +\frac{1}{2}K\right)F_{\rho\alpha}
  \right)\,,
\end{align}
which is the expression that is written in~\eqref{eq:tildeTW}.

\subsection{Weyl-covariant currents in four dimensions}
\label{ssec:bulk-improvements-weyl-covariant-currents}
As we already mentioned at the start of this section,
the energy-momentum and news tensors
should transform homogeneously under Weyl transformations
if the corresponding on-shell action is Weyl-invariant.
Furthermore, if the on shell action is Weyl-invariant,
the EMT-news complex satisfies the Weyl Ward identity.

In the previous subsection, we have shown that the improved EMT-news complex \eqref{eq:EMTNewsComplexSec7} indeed obeys the Weyl Ward identity.
Using the results of Sections~\ref{subsec:Weylcov} and~\ref{subsec:Weyltrafos},
we can also check that $(T^\mu_{\text{W}}, T^{\mu\nu}_{\text{W}}, S^{\mu\nu}_{\text{W}})$ have Weyl weights $(-4, -5, -4)$, respectively.
We will now discuss the different components of the energy-momentum tensor in turn.

\paragraph{The news.}
The news tensor $S^{\mu\nu}_W = (1/2) N^{\mu\nu}$, which is spatial and STF, is Weyl-covariant with Weyl weight $-4$.
This is obvious,
since we already saw around~\eqref{eq:NewsDefn-repeat}
that the news tensor $N_{\mu\nu}$ is Weyl-invariant,
and raising both the indices with $h^{\mu\nu}$ changes the Weyl weight to $-4$. The news tensor just like the energy-momentum tensor is subject to improvement transformations.

\paragraph{Energy density.}
Now let us consider the energy density~\eqref{eq:WeylImprovedEMTNewsComplex-energy-density},
which reads
\begin{align}
  \label{eq:WeylImprovedEMTNewsComplex-energy-density-repeat}
  \tau_\mu T^\mu_{\text{W}}
    =&-\frac{1}{2}\mathcal{D}_\rho\left(\mathcal{D}_\sigma-2a_\sigma\right) C^{\rho\sigma}-\frac{1}{8}\left(\mathcal{L}_v-K\right) F^2\nonumber\\
    &-v^\mu v^\nu \os{1}{g}_{\mu\nu}
    -\frac{1}{16}K\left(F^2-C^2\right)\,.
\end{align}
Consider the first term on the right-hand side.
$C^{\rho\sigma}$ has Weyl weight $-3$, so Equation~\eqref{eq:Weylcovdiv} tells us that $\left(\mathcal{D}_\sigma-2a_\sigma\right) C^{\rho\sigma}$ is Weyl-covariant with weight $-3$ and Equation~\eqref{eq:Weylcovdivvector} then tells us that $\mathcal{D}_\rho\left(\mathcal{D}_\sigma-2a_\sigma\right) C^{\rho\sigma}$ is also Weyl-covariant with weight -3.
For the second term, we use that $F^2$ has Weyl weight $-2$.
Equation~\eqref{eq:Lievscalar} then tells us that $\left(\mathcal{L}_v-K\right) F^2$ has Weyl weight $-3$.
Note that $v^\mu$ has Weyl weight $-1$.
The second line also has Weyl weight $-3$,
as follows from~\eqref{eq:S-1-Weyl-covariant-combination}.
We conclude that $\tau_\mu T^\mu_{\text{W}}$ has weight $-3$.
Note that this expression can be used to solve for the metric component $v^\mu v^\nu \os{1}{g}_{\mu\nu}$ in terms of $\tau_\mu T^\mu_{\text{W}}$.

\paragraph{Energy flux.}
The energy flux is given by
\begin{align}
  \label{eq:PTWeylcov}
  h^\mu_\rho T^\rho_{\text{W}}
    =& \frac{1}{2}h^{\mu\nu}\left(\partial_\nu+2a_\nu\right)\left(\overset{(0)}{S}-a^2\right)
    +h^\mu_\rho\left(\mathcal{L}_v-\frac{3}{2}K\right)\os{0}{P}^\rho
    \nonumber\\
    &
    -\frac{1}{2}v^\rho\left(\partial_\rho\tilde b_\sigma-\partial_\sigma\tilde b_\rho\right)\left(C^{\sigma\mu}-F^{\sigma\mu}\right)
    +\frac{1}{2}\left(\mathcal{D}_\rho-a_\rho\right)N^{\rho\mu}\,.
\end{align}
In Equation~\eqref{eq:S-Pi-expansion-Weyl-covariant-projections},
we saw that $\overset{(0)}{S}-a^2$ has weight $-2$, and hence the first term on the right-hand side has weight $-4$. 
The second term contains $\os{0}{P}^\rho$,
which is defined in~\eqref{eq:P0up}.
This has weight $-3$, as follows from our previous comments regarding the Weyl covariant divergence of the shear, as well as from the fact that \eqref{eq:P0up} can be straightforwardly generalised to any antisymmetric tensor so that the $F$~term in \eqref{eq:Weylcovdivvector} also has a definite Weyl weight of $-3$.
Finally, using \eqref{eq:WeylcovLievvector},
we conclude that the second term on the first line is also Weyl-covariant, with weight $-4$. 
The first term on the second line has weight $-4$, as follows from the fact that $\partial_\rho\tilde b_\sigma-\partial_\sigma\tilde b_\rho$ is Weyl-invariant,
see for example below \eqref{eq:b-tilde-def-and-weyl-tr}.
Finally, the last term on the second line also has weight $-4$,
as can be seen by using~\eqref{eq:Weylcovdiv} and the fact that $N^{\rho\mu}$ has weight $-4$.
We have thus established that $h^\mu_\rho T^\rho_{\text{W}}$ has weight $-4$.
Combining this with the previous result concerning the energy density,
we see that $T^\mu$ is a weight $-4$ object, the energy current.

In Section~\ref{ssec:variations-ward-ids-boundary-ward-identities}, we saw that if the EMT comes from the variation of an action and if said action is Carroll boost-invariant,
then we should be able to find an improved EMT such that the Carroll boost Ward identity~\eqref{eq:boundary-boost-WI} is obeyed.
In our case, we instead find
\begin{equation}\label{eq:anomCboostWI}
  h^\mu_\rho T^\rho_{\text{W}}
  -\left(\mathcal{D}_\rho-a_\rho\right) S^{\rho\mu}_{\text{W}}=h^\mu_\rho T^\rho_{\text{W}}
  -\frac{1}{2}\left(\mathcal{D}_\rho-a_\rho\right) N^{\rho\mu}_{\text{W}} = \mathcal{A}_\text{B}^\mu\,,
\end{equation}
where we defined
\begin{align}
    \label{eq:Cboostanom}
    \mathcal{A}_\text{B}^\mu
    &= \frac{1}{2}h^{\mu\nu}\left(\partial_\nu+2a_\nu\right)\left(\overset{(0)}{S}-a^2\right)
    + h^\mu_\rho\left(\mathcal{L}_v-\frac{3}{2}K\right)\os{0}{P}^\rho\\
    &{}\qquad\nonumber
    -\frac{1}{2}v^\rho\left(\partial_\rho\tilde b_\sigma-\partial_\sigma\tilde b_\rho\right)
    \left(C^{\sigma\mu}-F^{\sigma\mu}\right)\,.
\end{align}
Clearly, the right-hand side of \eqref{eq:anomCboostWI} is non-zero.
In Section~\ref{sec:HoloRenormAndOn-ShellActions},
we will see that the boundary energy-momentum tensor we obtain by constructing an appropriate set of counterterms in GR leads to the same result.
Moreover, we will see in Section~\ref{ssub:EMT-news-for-d=2} that there are in fact no local counterterms that can remove the terms on the right-hand side,
even when we consider giving up Weyl invariance,
and so we will see that the Carroll boosts are anomalous,
with the anomaly given by~\eqref{eq:Cboostanom}.

There is another useful way of writing the anomaly.
This follows from using the following identity,
\begin{eqnarray}
    h^\mu_\rho\left(\mathcal{L}_v-\frac{3}{2}K\right)\os{0}{P}^\rho &=&\frac{1}{2}\left(\mathcal{D}_\rho-a_\rho\right)N^{\rho\mu}+\frac{1}{2}v^\rho\left(\partial_\rho\tilde b_\sigma-\partial_\sigma\tilde b_\rho\right)C^{\sigma\mu}\nonumber\\
    &&+\frac{1}{2}h^\mu_\rho\left(\mathcal{L}_v-\frac{3}{2}K\right)\mathcal{D}_\sigma F^{\sigma\rho}\,,
    \label{eq:idLievP0}
\end{eqnarray}
which follows from the definition of $\os{0}{P}^\rho$ and \eqref{eq:comLiecovSTF}. This leads to 
\begin{eqnarray}
    \mathcal{A}_\text{B}^\mu & = & \frac{1}{2}h^{\mu\nu}\left(\partial_\nu+2a_\nu\right)\left(\overset{(0)}{S}-a^2\right)
    +\frac{1}{2}\left(\mathcal{D}_\rho-a_\rho\right)N^{\rho\mu}\nonumber\\
    &&
    +\frac{1}{2}v^\rho\left(\partial_\rho\tilde b_\sigma-\partial_\sigma\tilde b_\rho\right)F^{\sigma\mu}+\frac{1}{2}h^\mu_\rho\left(\mathcal{L}_v-\frac{3}{2}K\right)\mathcal{D}_\sigma F^{\sigma\rho}\,.\label{eq:Cboostanomv2}
\end{eqnarray}
The divergence of the news appears both in the anomaly and on the left-hand side of the Carroll boost Ward identity. 

Finally, we remark that the energy flux does not correspond to a particular component of the metric expansion like we saw with the energy density, which we could use to solve for $v^\mu v^\nu \os{1}{g}_{\mu\nu}$.
Instead, the energy flux is entirely fixed by the (anomalous) Carroll boost Ward identity.

\paragraph{Trace of the stress tensor.}
The stress tensor is the spatial projection of $T^{\mu\nu}$,
and so its trace is $h_{\mu\nu}T^{\mu\nu}$.
The result for $h_{\mu\nu}T^{\mu\nu}_{\text{W}}$ is entirely fixed by the Weyl Ward identity and thus we can write
\begin{equation}
    h_{\mu\nu}T^{\mu\nu}_{\text{W}}=-\tau_{\mu}T^{\mu}_{\text{W}}-\frac{1}{4}C\cdot N\,.
\end{equation}

\paragraph{Momentum density.}
Now let's turn to the momentum density, which is
\begin{align}
  P^{\text{W}}_\kappa  = & \frac{3}{2}\left(h^\mu_\kappa v^\nu\os{1}{\Pi}_{\mu\nu}-\os{2}{\beta}a_\kappa\right)+\frac{5}{2}\os{0}{P}^\mu F_{\mu\kappa}-\frac{1}{2}\os{0}{P}^\mu C_{\mu\kappa}\nonumber\\
    &-\frac{1}{32}h^\mu_\kappa(\partial_\mu+2a_\mu) C^2+\frac{1}{32}h^\mu_\kappa(\partial_\mu+2a_\mu) F^2-a^\mu D_{\mu\kappa}\,.
\end{align}
From Equation~\eqref{eq:beta-2-Weyl-covariant-combination},
we see that the first term in parentheses on the right-hand side has Weyl weight $-2$.
It is straightforward to check that all the other terms,
with the exception of the last term on the second line, also all have weight $-2$.

This only leaves the term which involves the $D_{\mu\nu}$ tensor.
Recall that the latter parametrises the $r^0$ contributions to $\Pi_{\mu\nu}$ following for example~\eqref{eq:app-Dmunu-def-and-evolution}.
We saw in Section~\ref{ssec:log-emt-news} that the presence of log terms relax the properties the $D_{\mu\nu}$ tensor has to obey.
When including appropriate logs, it only has to obey $\mathcal{L}_v D_{\mu\nu}=0$.
Furthermore, the form of the energy-momentum tensor is the same as when we do not include any logs.

In Section \ref{ssec:variations-ward-ids-ambiguity-from-boundary-constraint},
we saw that energy-momentum tensors obtained from the variation of an action where the source $h_{\mu\nu}$ obeys the constraint~\eqref{eq:extrinsic-curvature-constraint-d=2-repeat}
have a fundamental ambiguity in the expression for $T^{\mu\nu}$.
Specifically, we saw that we cannot distinguish between $T^{\mu\nu}$ and $T'^{\mu\nu}=T^{\mu\nu}+t^{\mu\nu}$, where $t^{\mu\nu}$ is given by~\eqref{eq:constraint-response-term-alt} as
\begin{equation}
    t^{\mu\nu}=\frac{1}{2}\mathcal{D}_\rho
  \left(
    v^\rho\zeta^{\mu\nu}
    -v^\mu\zeta^{\nu\rho}
    -v^\nu\zeta^{\mu\rho}
  \right)
  -\frac{1}{2}K\zeta^{\mu\nu}\,.
\end{equation}
Here, $\zeta^{\mu\nu}$ is an arbitrary STF tensor.
This fundamental ambiguity is of course of the form of an improvement transformation,
as we also discussed in Section~\ref{ssec:variations-ward-ids-boundary-ward-identities}.
We can therefore include this ambiguity in the expression for the momentum density, where it enters as $h_{\mu\rho} \tau_\sigma t^{\rho\sigma}$, which leads to 
\begin{align}
  P'^{\text{W}}_\kappa  = & \frac{3}{2}\left(h^\mu_\kappa v^\nu\os{1}{g}_{\mu\nu}-\os{2}{\beta}a_\kappa\right)+\frac{5}{2}\os{0}{P}^\mu F_{\mu\kappa}-\frac{1}{2}\os{0}{P}^\mu C_{\mu\kappa}\label{eq:finalWeylP}\\
    &-\frac{1}{32}h^\mu_\kappa(\partial_\mu+2a_\mu) C^2+\frac{1}{32}h^\mu_\kappa(\partial_\mu+2a_\mu) F^2-a^\mu D_{\mu\kappa}+\frac{1}{2}h_{\nu\kappa}\left(\mathcal{D}_\mu-a_\mu\right)\zeta^{\mu\nu}\,.\nonumber
\end{align}

In the presence of logs, as we derived in Section \ref{ssec:radial-expansion-logs}, we have
\begin{subequations}
\label{eq:first-log-block-repeat}
\begin{align}
    \beta & =  r^{-2}\overset{(2)}{\beta}+\OO(r^{-3})\,,\\
    S & =  r\overset{(-1)}{S}+\overset{(0)}{S}+\OO(r^{-1})\,,\\
    \Pi_{\mu\nu} & = r^2 h_{\mu\nu}+r\overset{(-1)}{\Pi}_{\mu\nu}+\overset{(0)}{\Pi}_{\mu\nu}+r^{-1}\log r\overset{(1,1)}{\Pi}_{\mu\nu}+\OO(r^{-1})\,,
\end{align}
\end{subequations}
where
\begin{equation}
    \overset{(1,1)}{\Pi}_{\mu\nu}=h^\rho_{\langle\mu}h^\sigma_{\nu\rangle}\overset{(1,1)}{\Pi}_{\rho\sigma}+\frac{2}{3}\tau_\mu h^{\rho\sigma}\mathcal{D}_\rho D_{\sigma\nu}+\frac{2}{3}\tau_\nu h^{\rho\sigma}\mathcal{D}_\rho D_{\sigma\mu}\,.
\end{equation}
Next, Equation~\eqref{eq:weyl-transformation-of-gmunu} tells us that the expression in~\eqref{eq:beta-2-Weyl-covariant-combination} picks up an inhomogeneous term under Weyl transformations,
\begin{equation}\label{eq:modifiedtrafohvg1}
    \delta_{\Lambda_D}\left(h^\mu_\kappa v^\nu \os{1}{g}_{\mu\nu}
  - \os{2}{\beta} a_\kappa\right)=-2\Lambda_D\left(h^\mu_\kappa v^\nu \os{1}{g}_{\mu\nu}
  - \os{2}{\beta} a_\kappa\right)-\frac{2}{3}\Lambda_D h^{\mu\nu}\mathcal{D}_\mu D_{\nu\kappa}\,.
\end{equation}
It follows from the second equation in \eqref{eq:S-Pi-expansion-Weyl-covariant-projections} combined with~\eqref{eq:app-Dmunu-def-and-evolution} that the aforementioned $D_{\mu\nu}$ tensor has Weyl weight $0$. If we assign $\zeta^{\mu\nu}$ the Weyl transformation 
\begin{equation}\label{eq:Weyltrafozeta}
    \delta_{\Lambda_D}\zeta^{\mu\nu}=-4\Lambda_D\zeta^{\mu\nu}+2\Lambda_D D^{\mu\nu}\,,
\end{equation}
then we find that 
\begin{align}
    &\delta_{\Lambda_D}\left(-a^\mu D_{\mu\kappa}+\frac{1}{2}h_{\nu\kappa}\left(\mathcal{D}_\mu-a_\mu\right)\zeta^{\mu\nu}\right)\nonumber\\
    =&-2\Lambda_D\left(-a^\mu D_{\mu\kappa}+\frac{1}{2}h_{\nu\kappa}\left(\mathcal{D}_\mu-a_\mu\right)\zeta^{\mu\nu}\right)+\Lambda_D h^{\mu\nu}\mathcal{D}_\mu D_{\nu\kappa}\,,\label{eq:trafoDzeta}
\end{align}
so that, in the end, we obtain the transformation law
\begin{equation}
    \delta_{\Lambda_D}P'^{\text{W}}_\kappa =-2\Lambda_D P'^{\text{W}}_\kappa\,.
\end{equation}
The inhomogeneous terms in \eqref{eq:modifiedtrafohvg1} and in \eqref{eq:trafoDzeta} cancel each other out,
and we thus see that the $P'^{\text{W}}_\kappa$ has a definite Weyl weight of $-2$.

Finally, we can solve $P'^{\text{W}}_\kappa$ for $h^\mu_\kappa v^\nu\os{1}{g}_{\mu\nu}$,
but the solution for $h^\mu_\kappa v^\nu\os{1}{g}_{\mu\nu}$ in terms of $P'_\kappa$ will contain the ambiguity parametrised by $\zeta^{\mu\nu}$.
Again, recall that by our arguments in Section~\ref{ssec:variations-ward-ids-ambiguity-from-boundary-constraint},
this ambiguity drops out of all Ward identities,
including the diffeomorphism Ward identity, which encodes the Bondi loss equations.
It is interesting to speculate that this may be why there are so many different expressions in the literature for the relation between $h^\mu_\kappa v^\nu\os{1}{g}_{\mu\nu}$ and the Bondi angular momentum aspect.

\paragraph{Stress tensor.}
Finally, we turn to the STF part of the stress tensor,
which is
\begin{eqnarray}
    \tilde T^{\mu\nu}_{\text{W}}
    &=& \frac{1}{2}\left(\overset{(0)}{S}-a^2\right)C^{\mu\nu}
    +\frac{1}{2}KD^{\mu\nu}\label{eq:repeattildeTW}\\
    &&+h^{\rho\langle\mu} h^{\nu\rangle\sigma}\left(\left(\mathcal{D}_\rho+3a_\rho\right)\os{0}{P}_\sigma-\frac{1}{4}\mathcal{L}_v\left(F_{\rho\alpha}C^\alpha{}_\sigma\right)+\frac{1}{2}C^\alpha{}_\sigma\left(\mathcal{L}_v +\frac{1}{2}K\right)F_{\rho\alpha}\right)\nonumber\,.
\end{eqnarray}
If we ignore the $KD^{\mu\nu}$ term in \eqref{eq:repeattildeTW},
we can use the results of Section \ref{subsec:Weylcov} to see that all the terms have Weyl weight $-5$.
For the first term in parenthesis on the second line,
this follows from \eqref{eq:STFprojWcovD1form} and the fact that $\os{0}{P}_\sigma$ has weight $-1$.
However, the $KD^{\mu\nu}$ term in \eqref{eq:repeattildeTW} is not Weyl covariant. 
Again, as with the momentum density, we can modify $\tilde T^{\mu\nu}_{\text{W}}$ by adding the STF part of $t^{\mu\nu}$, leading to 
\begin{eqnarray}
    \tilde T'^{\mu\nu}_{\text{W}}
    &=& \frac{1}{2}\left(\overset{(0)}{S}-a^2\right)C^{\mu\nu}
    +\frac{1}{2}KD^{\mu\nu}+\frac{1}{2}h^{\mu\rho}h^{\nu\sigma}\mathcal{L}_v\zeta_{\rho\sigma}\\
    &&+h^{\rho\langle\mu} h^{\nu\rangle\sigma}\left(\left(\mathcal{D}_\rho+3a_\rho\right)\os{0}{P}_\sigma-\frac{1}{4}\mathcal{L}_v\left(F_{\rho\alpha}C^\alpha{}_\sigma\right)+\frac{1}{2}C^\alpha{}_\sigma\left(\mathcal{L}_v +\frac{1}{2}K\right)F_{\rho\alpha}\right)\nonumber\,.
\end{eqnarray}
The variation of~$\zeta^{\mu\nu}$ we imposed in~\eqref{eq:Weyltrafozeta} implies that 
\begin{equation}
    \delta_{\Lambda_D}\zeta_{\mu\nu}=2\Lambda_D D_{\mu\nu}\,,
\end{equation}
so that the combination
\begin{equation}
\frac{1}{2}KD^{\mu\nu}+\frac{1}{2}h^{\mu\rho}h^{\nu\sigma}\mathcal{L}_v\zeta_{\rho\sigma}\,,    
\end{equation}
has Weyl weight $-5$.
Here, we used the fact that that, even in the presence of log terms,
we have $\mathcal{L}_v D_{\mu\nu}=0$, as derived in Section~\ref{ssec:radial-expansion-logs}).
Hence, we conclude that the modified stress tensor $\tilde T'^{\mu\nu}_{\text{W}}$ has Weyl weight $-5$.
Equivalently, one can say that the $\tilde T^{\mu\nu}_{\text{W}}$ stress tensor has weight $-5$ up to improvement terms.

We see that $\tilde T^{\mu\nu}_{\text{W}}$ does not contain a metric component for which we can solve the expression.%
\footnote{%
  This should be contrasted with the situation in AdS/CFT,
  where the whole boundary energy-momentum tensor is in a one-to-one relation with the metric (in Fefferman--Graham gauge) at some appropriate order in the $1/r$ expansion.
}
We also saw this earlier for the energy flux.
These observations play an important role in a rewriting of the Bondi loss equations in terms of a different boundary energy-momentum tensor,
whose non-conservation is determined by certain flux terms that measure how radiative the spacetime is~\cite{Hartong:2025WIP2}.

Finally, it is interesting to observe that, even in the presence of $\log r$ terms in the radial expansion, the EMT-news complex has a definite Weyl weight and the Weyl Ward identity is non-anomalous.

\section{Action and boundary terms near future null infinity}\label{sec:HoloRenormAndOn-ShellActions}
The goal of this section is to derive the Weyl-improved energy-momentum-news complex we obtained from the equations of motion in~\eqref{ssec:bulk-improvements-weyl-covariant-currents} from the action of general relativity, supplemented with suitable boundary and renormalisation terms.
To this end, we will first study the properties of the Einstein--Hilbert action near future null infinity.
This will lead to a discussion of counterterms to be added on a $r=\text{cst}$ cut-off hypersurface.
As usual, we will study the $d=1,2$ cases,
corresponding to three and four bulk spacetime dimensions.

\subsection{Variations and cut-off surface near future null infinity}
\label{ssec:vars-and-cutoff-near-scri}
Our starting point is the Einstein--Hilbert action,
\begin{equation}
    S_{\text{EH}}=\int_{\mathcal{M}}d^{d+2}x\sqrt{-g}R\,.
\end{equation}
It is well known that its variation is given by
\begin{equation}
    \delta S_{\text{EH}} = \int_{\mathcal{M}}d^{d+2}x\sqrt{-g} G_{MN}\delta g^{MN}+\int_{\mathcal{M}}d^{d+2}x\partial_P\left(\sqrt{-g}J^P\right)\,,
\end{equation}
where we defined
\begin{equation}
    J^P=g^{MN}\delta\Gamma^P_{MN}-g^{PN}\delta\Gamma^M_{MN}\,.
\end{equation}
We will take $\mathcal{M}$ to be an asymptotic region that contains $r=\text{cst}$ slices for large~$r$,
and we consider the properties of $\delta S_{\text{EH}}$ near future null infinity $\mathcal{I}^+$.
In other words, we will ignore any boundary terms that do not pertain to future null infinity.
Additionally, we will not consider corner terms,
which arise from boundary terms on $\mathcal{I}^+$ itself.

\paragraph{Generalised GHY boundary term.}
We will consider a cut-off boundary denoted by $\Sigma$,
which is defined as a level set $F(r,x)=\text{cst}$ for some appropriate function $F$.
There thus exists a normal 1-form given by $N_M=\partial_M F$.
Let~$V^M$ be any null vector that obeys $V\cdot N=-1$.
We can write the metric as
\begin{equation}\label{eq:metricinVNPi}
    ds^2=-N^2 V_M V_Ndx^M dx^N-2V_M N_N dx^M dx^N+\Pi_{MN}dx^M dx^N\,.
\end{equation}
This maps onto the double null vielbein decomposition we used starting from~\eqref{eq:gen-metric-null-frame-decomposition} in Section~\ref{sec:bulk-geom} if we take $U_M=N_M+\frac{1}{2}S V_M$ and $S=N^2$ (and $\Pi_{MN}=\delta_{ab}E^a_M E^b_N$).
In our Carroll-covariant Bondi--Sachs gauge,
following for example~\eqref{eq:car-cov-bondi-metric-repeat},
we have
\begin{equation}
\begin{gathered}
    \label{eq:down-vielbeine-map-to-BS}
    V_M dx^M
    = e^\beta \tau_\mu dx^\mu\,,
    \quad
    U_M dx^M
    = dr
    + \frac{1}{2} S e^\beta \tau_\mu dx^\mu\,,
    \quad
    N_M dx^M
    = dr\,,
    \\
    V^M \pd_M
    = - \pd_r\,,
    \quad
    \frac{1}{2} \Pi^{\mu\nu} \pd_r \Pi_{\mu\nu}
    = r\inv d\,.
\end{gathered}
\end{equation}
We will use the general expressions above as much as possible in the variational computations below.
Using this parametrisation~\eqref{eq:metricinVNPi} in the Einstein--Hilbert boundary term near $\mathcal{I}^+$,
we then have
\begin{equation}
    \int_{\mathcal{M}}d^{d+2}x\partial_P\left(\sqrt{-\tilde g}J^P\right)=\int_\Sigma d^{d+1}\xi \sqrt{-\tilde g}N_P J^P+\cdots\,,
\end{equation}
where the dots denote contributions from other components of the boundary of~$\mathcal{M}$,
and where $\xi^\mu$ are coordinates on $\Sigma$.
Finally, $\tilde g_{MN}$ is the metric in the coordinates~$x^M=(F, \xi^\mu)$.
Since everything in this section will take place on the cut-off surface~$\Sigma$
we will drop the tilde on the metric from now on.

The cut-off surface $\Sigma$ does not have a definite signature as the norm of the normal, which corresponds to $S$, does not have a fixed sign.
This means that we cannot add the usual Gibbons--Hawking--York (GHY) boundary term to ensure that there are no normal derivatives of the variation of the metric on $\Sigma$.
We can address this by adding the counterterm derived in~\cite{Parattu:2016trq}, which generalises the GHY term to surfaces of arbitrary signature.
This counterterm is given by
\begin{equation}
  \label{eq:ParattuPadmanabhanCT}
  S_{\text{ext}}=a\int_\Sigma d^{d+1}\xi \sqrt{-g}\left(\delta^M_P+V^M N_P\right)\nabla_M N^P\,,
\end{equation}
where $a$ is some real parameter, and we will see it has to be equal to $2$.
If we vary $S_{\text{EH}}+S_{\text{ext}}$ keeping $V^M$ and $N_M$ fixed,
we obtain
\begin{align}   \delta\left(S_{\text{EH}}+S_{\text{ext}}\right) & =  \cdots+\int_{\Sigma}d^{d+1}\xi\sqrt{-g}\left[(1-a)\mathcal{P}^Q{}_P\nabla_Q\left(N^M\mathcal{P}^{PN}\delta g_{MN}\right)\right.\nonumber\\
    &\qquad\left.+(a-1)\left(\mathcal{P}^Q{}_P\nabla_Q V^P\right) N^M N^N\delta g_{MN}-(a-1)N^M V^P\nabla_P N^N\delta g_{MN}\right.\nonumber\\
    &\qquad\left.-g^{MP}\nabla_P N^N\delta g_{MN}+\frac{a}{2}\left(\mathcal{P}^Q{}_P\nabla_Q N^P\right)g^{MN}\delta g_{MN}\right.\nonumber\\
    &\qquad\left.+\frac{1}{2}(a-2)\left(g^{MN}N^Q\nabla_Q\delta g_{MN}+\mathcal{L}_V\left(N^M N^N\delta g_{MN}\right)\right)\right]\,,
\end{align}
where we defined the projector
$\mathcal{P}^Q{}_P=\delta^Q_P+V^Q N_P$,
which annihilates $V^P$ and $N_Q$,
and where $\mathcal{P}^{PN}=\mathcal{P}^P{}_Q g^{QN}$.
We need to pick $a=2$ in order to remove normal derivatives of variations, which arise in the first term on the last line. 

We next evaluate (for $a=2$) the variation $\delta\left(S_{\text{EH}}+S_{\text{ext}}\right)$ in our Carroll-covariant Bondi--Sachs gauge, using for example~\eqref{eq:down-vielbeine-map-to-BS} and~\eqref{eq:BSgaugemetric}.
This leads to
\begin{align}
  \label{eq:varSEH+ext}
  &\delta\left(S_{\text{EH}}+S_{\text{ext}}\right)\big\vert_{\text{Car-cov BS gauge}}
  \\
  \nonumber
  &=  \cdots+\int_{r=\Lambda}d^{d+1}x E\left[-E^{-1}\partial_\mu\left(E X^\mu\right)+\mathcal{T}^\mu\delta V_\mu
  +\frac{1}{2}\mathcal{T}^{\mu\nu}\delta\Pi_{\mu\nu}+dr^{-1}E^{-1}\delta (ES)\right],
\end{align}
where $E$ is defined in \eqref{eq:gauge-fixed-det-met}.
In this gauge, the cut-off boundary is $r=\Lambda$ for some large constant $\Lambda$,
and using \eqref{eq:down-vielbeine-map-to-BS} we have $V^M=-\delta^M_r$ and $N_M=\partial_M r$.
We furthermore defined
\begin{equation}
    X^\mu = \mathcal{P}^{\mu M}N^N\delta g_{MN}=-U^\mu U^\rho\delta V_\rho+\Pi^{\mu\rho}U^\sigma \delta\Pi_{\rho\sigma}\,,
\end{equation}
as well as
\begin{subequations}
  \begin{align}
    \mathcal{T}^\mu
    &= 2U^\mu\bar{\mathcal{K}}-S\mathcal{Z}^\mu+\Pi^{\mu\nu}\partial_\nu S\,,\label{eq:calTmu}
    \\
    \mathcal{T}^{\mu\nu}
    &= \Pi^{\mu\nu}\left(-\frac{2}{d}(d-1)\bar{\mathcal{K}}+\partial_r S-r^{-1}S+2S\partial_r\beta\right)\label{eq:calTmunu}
    \\
    &{}\qquad\nonumber
    +2\mathcal{K}^{T\,\mu\nu}-S\mathcal{G}^{\mu\nu}-2\mathcal{Z}^{(\mu}U^{\nu)}\,.
  \end{align}
\end{subequations}
We raised indices on spatial tensors such as $\mathcal{Z}_\rho$ with $\Pi^{\mu\rho}$.
We remind the reader that some of the objects used here were defined in Sections
\ref{ssec:connection-choice} and \ref{sec:rewriting-EE}.

\paragraph{Norm counterterm.}
Ultimately, we will be interested in the on-shell value of~\eqref{eq:varSEH+ext},
augmented with appropriate boundary terms,
and we require that as $r\to\infty$ this on shell result takes the form of the variation in~\eqref{eq:on-shell-action-leading-order},
\begin{equation}
  \label{eq:model-boundary-variation}
  \delta S
  = \int d^{d+1} x\, e \left(
    T^\mu \delta \tau_\mu
    + \frac{1}{2} T^{\mu\nu} \delta h_{\mu\nu}
    + \frac{1}{2} S^{\mu\nu} \delta C_{\mu\nu}
  \right)\,.
\end{equation}
This applies to $d=2$, while for $d=1$ the shear and news terms are absent.
However, the last term in~\eqref{eq:varSEH+ext} will always give rise to an on-shell variation of the Bondi news $\os{d-1}{S}$,
which we do not want and
which cannot be cancelled by terms coming from the $\delta V_\mu$ and $\delta \Pi_{\mu\nu}$ variations.
We thus need to add a second boundary term that removes the last term in~\eqref{eq:varSEH+ext}.
Since the term in question is already a total variation, it is easy to find the appropriate boundary term.
We refer to this as the `norm' counterterm, as it involves the norm $S=N^2$ of the normal $N_M$.
In a general gauge, we thus add 
\begin{equation}
  \label{eq:normCT}
  S_{\text{norm}}=b\int_\Sigma d^{d+1}\xi \sqrt{-g} N^2\left(\delta^M_P+V^M N_P\right)\nabla_M V^P\,,
\end{equation}
where we will see that we must have $b=1$.
The variation of this counterterm is
\begin{align}
  \delta S_{\text{norm}} & =  b\int_\Sigma d^{d+1}\xi \sqrt{-g}\left[ \frac{1}{2}N^2\left(\mathcal{P}^P{}_Q\nabla_P V^Q\right) g^{MN}\delta g_{MN}\right.\nonumber\\
  &\qquad\left.-\left(\mathcal{P}^P{}_Q\nabla_P V^Q\right)N^M N^N\delta g_{MN}+N^2\mathcal{P}^P{}_Q V^R\delta\Gamma^Q_{PR} \right]\,.
\end{align}
In our Carroll-covariant Bondi--Sachs gauge~\eqref{eq:BSgaugemetric}, this leads to
\begin{eqnarray}
  \delta S_{\text{norm}}\big\vert_{\text{Car-cov BS gauge}}  =  -b\int_{r=\Lambda} d^{d+1}x E\left[dr^{-1}E^{-1}\delta(ES)+S\delta\Gamma^\mu_{\mu r} \right]\,,
\end{eqnarray}
which indeed cancels the offending term in~\eqref{eq:varSEH+ext} for $b=1$.
When restricting to on-shell variations that respect our Bondi--Sachs gauge choice,
as we will do, the last term containing $\delta\Gamma^\mu_{\mu r}$ will vanish,
since our gauge choice~\eqref{eq:BSgaugemetric} involves $\Gamma^\mu_{\mu r}=dr^{-1}$. We notice that the norm counterterm \eqref{eq:normCT} in our gauge is equal to
\begin{equation}
  \label{eq:normCTgf}
  S_{\text{norm}}=-b\int_{r=\Lambda} d^{d+1}x E S dr^{-1}\,.
\end{equation}
When restricting to on-shell gauge-preserving variations,
we can construct the counterterms directly within our gauge choice.
We will do this from now on.
We now have (with $a=2$ and $b=1$),
\begin{align}
&\delta\left(S_{\text{EH}}+S_{\text{ext}}+S_{\text{norm}}\right)\big\vert_{\text{Car-cov BS gauge}}
  \nonumber
  \\
  &{}\qquad
  \label{eq:varSEH+ext+norm}
  = \cdots+\int_{r=\Lambda}d^{d+1}x E\left[-E^{-1}\partial_\mu\left(E X^\mu\right)
  +\mathcal{T}^\mu\delta V_\mu+\frac{1}{2}\mathcal{T}^{\mu\nu}\delta\Pi_{\mu\nu}\right]\,,
\end{align}
where we consider only gauge-preserving variations.

\paragraph{Finite counterterm.}
For future purposes, it is also useful to consider the following finite counterterm,
which again involves the norm $N^2=S$ but now equals
\begin{equation}
    S_{\text{fin,norm}}=-\tilde b\int_\Sigma d^{d+1}\xi\sqrt{-g}N^2 N_P V^M\nabla_M V^P\,.
\end{equation}
In our gauge, this is equal to the following term,
which is finite as $r\to\infty$,
\begin{equation}\label{eq:finnormct}
    S_{\text{fin,norm}}=-\tilde b\int_{r=\Lambda} d^{d+1}x E S\partial_r\beta\,.
\end{equation}
This leads to variations of the type $\delta S$ and $\partial_r\delta\beta$.
However, when evaluated on shell, as $r\to\infty$, 
these variations also turn out to be of the required form~\eqref{eq:model-boundary-variation},
which contains variations of $(\tau_\mu, h_{\mu\nu}, C_{\mu\nu})$.
When $\tilde b=b$, we get
\begin{equation}
 S_{\text{norm}}+S_{\text{fin,norm}}=b\int_\Sigma d^{d+1}\xi\sqrt{-g}N^2\nabla_M V^M\,.
\end{equation}
Because $V^M\pd_M = - \pd_r$ is kept fixed in our gauge, we have that $\partial_r\delta\beta=-\mathcal{L}_V\delta\beta$ is the same as $-\delta\mathcal{L}_V\beta$.
Hence, if $\delta\beta$ is zero for sufficiently large $r$, then so is $\mathcal{L}_V\beta$. 
The variation is given by
\begin{align}
  \delta S_{\text{fin,norm}}
  &=  -\tilde b\int_\Sigma d^{d+1}\xi\sqrt{-g}\left(
    \frac{1}{2}N^2 N_P V^Q\nabla_Q V^P g^{MN}\delta g_{MN}
  \right.\nonumber\\
  &\qquad\nonumber
    -N_P V^Q\nabla_Q V^P N^M N^N\delta g_{MN}
  \left.
    +N^2 N^M V^N\mathcal{L}_V\delta g_{MN}\right.\\
   &\qquad\left. -N^2N^N V^P\nabla_P V^M\delta g_{MN}
  \right)\,.
\end{align}
This obeys
\begin{align}
\label{eq:varSnormfin}  
\delta S_{\text{fin,norm}}\big\vert_{\text{Car-cov BS gauge}} & = -\tilde b \int_{r=\Lambda}d^{d+1}x E\left(S\partial_r\beta E^{-1}\delta E+\partial_r\beta\delta S+S\partial_r\delta\beta\right)\nonumber\\
&=-\tilde b\delta \int_{r=\Lambda}d^{d+1}x ES\partial_r\beta\,.
\end{align}
This is as it should be. The variations preserving our gauge choice commute with the gauge choice.
We will determine the contributions to the EMT-news complex coming from this counterterm in the following.

\paragraph{Intermediate result.}
Consider the responses to the variation of $V_\mu$ and $\Pi_{\mu\nu}$ in~\eqref{eq:calTmu} and \eqref{eq:calTmunu}.
Their leading-order nonzero contributions are
\begin{align}
  \overset{(1)}{\mathcal{T}}{}^\mu
  =& 2v^\mu\left(\overset{(1)}{\mathcal{K}}-\frac{d}{2}\overset{(0)}{S}\right)
  +\frac{2}{d}h^{\mu\nu}\partial_\nu K\,,\label{eq:calT1mu}
  \\
  \overset{(3)}{\mathcal{T}}{}^{\mu\nu}
  =& h^{\mu\nu}\left(
    -\frac{2}{d}(d-1)\overset{(1)}{\mathcal{K}}
  +(d-2)\overset{(0)}{S}\right)
  -2\overset{(1)}{\mathcal{Z}}{}^{(\mu}v^{\nu)}
  \nonumber
  \\
  &+\left(
    -\mathcal{L}_v C_{\rho\sigma}
    +2A_{\rho\sigma}\right)
  h^{\mu\rho}h^{\nu\sigma}\,.\label{eq:calT3munu}
\end{align}
This means that $\mathcal{T}^\mu$ is order $r^{-1}$ and $\mathcal{T}^{\mu\nu}$ is order $r^{-3}$. Then, because for on-shell variations we have
\begin{equation}
    \delta V_\mu=\delta\tau_\mu+\mathcal{O}(r^{-2})\,,\qquad\delta\Pi_{\mu\nu}=r^2\delta h_{\mu\nu}+ \mathcal{O}(r)\,,
\end{equation}
we see that 
\begin{equation}\label{eq:integrandorder}
    E\left(\mathcal{T}^\mu\delta V_\mu+\frac{1}{2}\mathcal{T}^{\mu\nu}\delta\Pi_{\mu\nu}\right)=\mathcal{O}(r^{d-1})\,.
\end{equation}
Here, we used Equation~\eqref{eq:bond-gauge-condition-as-bulk-vielbein-determinant},
which tells us that $E = r^d e^\beta e$.
We now work out the three-dimensional $d=1$ case and the four-dimensional $d=2$ case separately.

\subsection{Boundary energy-momentum tensor for \texorpdfstring{$d=1$}{d=1}}\label{subsec:holrend=1}
For $d=1$, the integrand \eqref{eq:integrandorder} is order $r^0$,
and so we see that we do not need to add any counterterms to make it finite.
In this case, the STF tensors $\mathcal{K}^T_{\mu\nu}$ and $\mathcal{G}_{\mu\nu}$ are identically zero.
We also know that $\beta=0$ on shell,
so that we have $E= re$.

With that, the finite contributions to the boundary currents come from the leading-order terms $\overset{(1)}{\mathcal{T}}{}^\mu$ and $\overset{(3)}{\mathcal{T}}{}^{\mu\nu}$ above, and we have
\begin{equation}
  \delta \left(S_{\text{EH}}+S_{\text{ext}}+S_\text{norm}\right)\big\vert_{\text{os}}
  =\cdots+2\int_{r = \Lambda} d^{2}x\, e\left(
    T^\mu_{\text{ren}} \delta \tau_\mu
    + \frac{1}{2} T^{\mu\nu}_{\text{ren}} \delta h_{\mu\nu}
  \right)\,,\label{eq:varaction}
\end{equation}
where the left-hand side is the total variation evaluated on shell.
On the right-hand side, the renormalised boundary currents are given by
\begin{align}
  T^\mu_{\text{ren}}
  = \frac{1}{2}\overset{(1)}{\mathcal{T}}{}^\mu
  &= v^\mu\left(e\inv\pd_\rho(e a^\rho)-\frac{1}{2}\overset{(0)}{S}\right)
  +h^{\mu\rho}\partial_\rho K\,,
  \\
  T^{\mu\nu}_{\text{ren}}
  = \frac{1}{2}\overset{(3)}{\mathcal{T}}{}^{\mu\nu}
  &= - \frac{1}{2}h^{\mu\nu} \overset{(0)}{S}
  -2 v^{(\mu} h^{\nu)\rho} P_\rho\,,
\end{align}
where we used \eqref{eq:calT1mu} and \eqref{eq:calT3munu} for $d=1$ as well as the expression for $P_\kappa$ in~\eqref{eq:calZ1d=1}.
Note the factor of 2 in~\eqref{eq:varaction}, which is there for normalisation purposes.

These currents agree with the ones we obtained from the bulk equations of motion in~\eqref{eq:SETd1}.
Since we are evaluating the variation on shell,
and given that the equations of motion in particular imply that
$T^\mu$ and $T^{\mu\nu}$ obey the equations in~\eqref{eq:d=1constrainteq} and~\eqref{eq:constraintd=1spatial},
which we know correspond to the diffeomorphism Ward identity,
we conclude that~\eqref{eq:varaction} is invariant under boundary diffeomorphisms. 

As we already knew,
these currents do not satisfy the Weyl Ward identity,
which here takes the form $\tau_\mu T^\mu+h_{\mu\nu}T^{\mu\nu}=0$.
To remedy this, inspired by the related discussion in Section~\ref{ssec:bulk-improvements-weyl-3d}, we add the finite and boundary-covariant counterterm
\begin{equation}
  \label{eq:finite-d=1-counterterm}
  S_\text{fin}
  = - \int_{r=\Lambda} d^2x\,E\, r \Pi^{\mu\nu} \mathcal{A}_\mu \mathcal{A}_\nu\,.
\end{equation}
We wrote this counterterm directly in the Carroll-covariant Bondi--Sachs gauge. 
We can now define
\begin{align}
  \label{eq:vartotactiond=1}
  &\delta \left(S_{\text{EH}}+S_{\text{ext}}+S_\text{norm}+S_\text{fin}\right)\big\vert_{\text{os}}
  \nonumber
  \\
  &{}\qquad
  =\cdots
  +2\int_{r = \Lambda} d^{2}x\, e\left(
    T'^\mu_{\text{ren}} \delta \tau_\mu
    + \frac{1}{2} T'^{\mu\nu}_{\text{ren}} \delta h_{\mu\nu}
  \right)\,,
\end{align}
where the improved renormalised currents are given by
\begin{subequations}\label{eq:imprcurrentsd=1}
  \begin{align}
    T'^\mu_{\text{ren}}
    &= - \frac{1}{2}\left(\os{0}{S} - a^2\right)  v^\mu
    + h^{\mu\sigma}v^\rho\left(\pd_\rho\tilde b_\sigma-\pd_\sigma \tilde b_\rho\right)\,,
    \\
    T'^{\mu\nu}_{\text{ren}}
    &= -2P^{(\mu}v^{\nu)} - \frac{1}{2}\left(\os{0}{S} - a^2\right) h^{\mu\nu}\,.
  \end{align}
\end{subequations}
These are precisely the improved currents we found in \eqref{eq:SETd1-impr},
which now satisfy the Weyl Ward identity.
Here,
$\os{0}{S}-a^2$
and
$P_\mu = h_\mu^\rho v^\sigma \os{0}{\Pi}_{\rho\sigma}=h_\mu^\rho v^\sigma \os{0}{g}_{\rho\sigma}$
parametrise the Bondi mass and angular momentum aspects.
Additionally, we know that the currents above are Weyl-covariant,
since they are made up of the Weyl-covariant building blocks we identified previously in Section~\ref{ssec:bulk-improvements-weyl-3d}.

\subsection{Carroll boost anomaly for \texorpdfstring{$d=1$}{d=1}}
\label{ssec:hol-ren-threedim-anomalies-improvements}
As we saw,
we can construct a boundary energy-momentum tensor in $d=1$ by varying a suitably-renormalised on-shell action with respect to the boundary geometry,
and the resulting EMT in~\eqref{eq:imprcurrentsd=1} obeys the diffeomorphism and Weyl Ward identities.
We will drop the prime and subscript for brevity.

However, this EMT fails to satisfy the Carroll boost Ward identity,
which for $d=1$ states that
$h_{\mu\nu}T^\mu=0$.
Using \eqref{eq:vartotactiond=1},
we find that, under a Carroll boost,
\begin{equation}
  \delta \left(S_{\text{EH}}+S_{\text{ext}}+S_\text{norm}+S_\text{fin}\right)\big\vert_{\text{os}}
  =\cdots+2\int_{r = \Lambda} d^{2}x\, e v^\rho\lambda^\sigma \left(\pd_\rho\tilde b_\sigma-\pd_\sigma \tilde b_\rho\right)\,.
\end{equation}
The question now is if there exists a local finite counterterm whose variation under Carroll boosts can cancel the non-invariance above.
Such a local and diffeomorphism-invariant counterterm, whose variation under Carroll boosts should be proportional to 
\begin{equation}\label{eq:Cboostnoninva}
    \int_{r = \Lambda} d^{2}x\, e v^\rho\lambda^\sigma \left(\pd_\rho\tilde b_\sigma-\pd_\sigma \tilde b_\rho\right)=\int_{r = \Lambda} d^{2}x\, e\lambda e^\mu \left(
     \LL_v a_\mu
    + K  a_\mu
    +  \pd_\mu K
  \right)\,,
\end{equation}
would have to be second order in derivatives.
For $d=1$, this means it has to be formed out of products of $a_\mu$ and $K$. 
Hence, any such counterterm would have to be of the form%
\footnote{%
  This is already taking into account the freedom to perform integration by parts,
  so that terms such as $\mathcal{L}_v \left(e^\mu a_\mu\right)$ are equivalent to the ones already written above.
}
\begin{equation}\label{eq:potentialcts}
    \int_{r = \Lambda} d^{2}x\, e\left(aK^2+bK e^\mu a_\mu+ch^{\mu\nu}a_\mu a_\nu\right)\,.
\end{equation}
Since we are in $d=1$, we used the fact that we can express $h_{\mu\nu} = e_\mu e_\nu$ in terms of a single spatial vielbein $e_\mu$,
and likewise $h^{\mu\nu}=e^\mu e^\nu$.
We effectively already used the counterterm with coefficient $c$ in~\eqref{eq:finite-d=1-counterterm} above to achieve Weyl invariance,
and now having $c\neq0$ would break this.
However, it could be possible that one can swap Carroll boost invariance for Weyl invariance,%
\footnote{%
  We will find out that this is, in fact, not possible.
}
so we will allow for this counterterm, too.

Up to total derivative terms, we have for any function $f$ that
\begin{equation}
    e f\delta_\lambda\left(e^\mu a_\mu\right)=-e\lambda\mathcal{L}_v f\,,
\end{equation}
where we now denote the $d=1$ Carroll boost parameter using $\lambda_\mu=\lambda e_\mu$. 
Furthermore, we know that $K$ is Carroll boost-invariant.
Hence, we obtain
\begin{align}
  &\delta_\lambda
  \int_{r = \Lambda} d^{2}x\, e\left(aK^2+bK e^\mu a_\mu+ch^{\mu\nu}a_\mu a_\nu\right)
  \nonumber
  \\
  &{}\qquad
  =-\int_{r = \Lambda} d^{2}x\, e\lambda\left(
    b\mathcal{L}_vK+2c\mathcal{L}_v\left(e^\mu a_\mu\right)
  \right)\,.
\end{align}
From this, we can see that there are no values we can choose for $(a,b,c)$ such that we would cancel the non-invariance in~\eqref{eq:Cboostnoninva}.
We conclude that we cannot remove the nonzero terms appearing in $h_{\mu\nu}T^\nu$,
which violate the $d=1$ Carroll boost Ward identity,
whilst at the same time respecting the diffeomorphism Ward identity. 

Therefore, we conclude that~\eqref{eq:Cboostnoninva} leads to a Carroll boost anomaly $\mathcal{A}^\mu_B$,
\begin{equation}\label{eq:d=1anomaly}
    \mathcal{A}^\mu_B=v^\rho h^{\mu\sigma}\left(\pd_\rho\tilde b_\sigma-\pd_\sigma \tilde b_\rho\right)\,,
\end{equation}
so that, under a Carroll boost,
\begin{equation} \delta_\lambda\left(S_{\text{EH}}+S_{\text{ext}}+S_\text{norm}+S_\text{fin}\right)\big\vert_{\text{os}}
  =\cdots+2\int_{r = \Lambda} d^{2}x\, e \lambda e_\mu \mathcal{A}^\mu_B\,.
\end{equation}
It is straightforward to verify that 
the anomaly \eqref{eq:d=1anomaly} transforms under boundary diffeomorphisms, Weyl transformations and Carroll boosts as required for consistency in~\eqref{eq:trafoanomaly}.
If we define 
\begin{equation}
    \mathcal{A}_B=e_\mu\mathcal{A}^\mu_B\,,
\end{equation}
then we find
\begin{equation}
    \delta \mathcal{A}_B=\mathcal{L}_\chi\mathcal{A}_B-2\Lambda_D\mathcal{A}_B+e^{-1}\partial_\mu\left(e v^\mu\mathcal{L}_v\lambda\right)\,.
\end{equation}
By using the above transformation properties of the Carroll boost anomaly,
we can also directly verify that \eqref{eq:d=1WZ} holds, so that Wess--Zumino consistency is obeyed.
The existence of a $d=1$ Carroll boost anomaly was also observed in \cite{Campoleoni:2022wmf},
and see also~\cite{Jensen:2017tnb,Copetti:2019ooe} for similar observations in the context of warped CFTs.

\paragraph{Relation to central charges.}
Consider the following identity,
\begin{align}
  &e\left(T^\mu\mathcal{L}_K\tau_\mu+\frac{1}{2}T^{\mu\nu}\mathcal{L}_K h_{\mu\nu}\right)
  \nonumber
  \\
  &{}\qquad
  =\partial_\mu\left(eT^\mu{}_\rho K^\rho\right)
  +K^\rho\left[-\partial_\mu\left(e T^\mu{}_\rho\right)+eT^\mu\partial_\rho\tau_\mu+\frac{e}{2}T^{\mu\nu}\partial_\rho h_{\mu\nu}\right]\,,
\end{align}
where, at this stage, $K^\mu$ is any vector.
We remind the reader that the energy-momentum tensor is $T^\mu{}_\nu=T^\mu\tau_\nu+T^{\mu\rho}h_{\rho\nu}$.
If we invoke the diffeomorphism Ward identity \eqref{eq:boundary-diffeo-WI-not-covariant}, 
with vanishing shear, then we can write this as
\begin{equation}
    \partial_\mu\left(eT^\mu{}_\rho K^\rho\right)=e\left(T^\mu\mathcal{L}_K\tau_\mu+\frac{1}{2}T^{\mu\nu}\mathcal{L}_K h_{\mu\nu}\right)\,.
\end{equation}
We then also demand that $K^\mu$ corresponds to a conformal Carroll Killing vector,
which means that it is a solution to  
\begin{subequations}
    \begin{align}
    0&=\delta\tau_\mu  =  \mathcal{L}_K\tau_\mu+\Lambda_D\tau_\mu+\lambda_\mu\,,\\
    0&=\delta h_{\mu\nu}  = \mathcal{L}_K h_{\mu\nu}+2\Lambda_D h_{\mu\nu}\,,
\end{align}
\end{subequations}
for appropriate values of the Weyl and boost parameters $\Lambda_D$ and $\lambda_\mu=\lambda e_\mu$.
With that assumption, we can now write
\begin{equation}
  \label{eq:charge-eqn-d1}
  \partial_\mu\left(eT^\mu{}_\rho K^\rho\right)=-eT^\mu\lambda_\mu=-e\lambda\mathcal{A}_B\,,
\end{equation}
where we used the Weyl Ward identity $T^\mu\tau_\mu+T^{\mu\nu}h_{\mu\nu}=0$
as well as the anomalous Carroll boost Ward identity
$T^\rho h^\mu_\rho=\mathcal{A}_B^\mu$,
which follows for example from~\eqref{eq:imprcurrentsd=1}. 

It is well known that the vectors $K^\mu$ that obey the conformal Carroll Killing equations form the BMS$_3$ algebra under the Lie bracket,
without any central extensions
(see for example~\cite{Duval:2014uva,Hartong:2015usd,Ciambelli:2019lap}).
In a flat two-dimensional Carroll space, we have $a_\mu=0$ and $K=0$.
Then, the anomaly~\eqref{eq:d=1anomaly} vanishes, so that the BMS$_3$ charges we construct above are conserved following~\eqref{eq:charge-eqn-d1}.
The anomaly then shows up in the charge algebra~\cite{Barnich:2006av} with respect to the Poisson bracket.
In other words, the Carroll boost anomaly is responsible for the $c_M$ central charge in the BMS$_3$ symmetry algebra (see also \cite{Campoleoni:2022wmf}).

The BMS$_3$ algebra also admits another central extension, which is often denoted as $c_L$.
It would be interesting to see if this second central charge is related to a diffeomorphism anomaly.
We can view the BMS$_3$ algebra as the In\"on\"u--Wigner contraction of two copies of the Virasoro algebra,
with left- and right-moving central charges $C_L$ and $C_R$.
In the limit, $c_L$ then arises from $C_L-C_R$ and $c_M$ from $C_L + C_R$~\cite{Bagchi:2012cy,Barnich:2012aw}.
Since $C_L - C_R$ is related to a diffeomorphism (or gravitational) anomaly in a two-dimensional relativistic CFT~\cite{Alvarez-Gaume:1983ihn,Kraus:2005zm}, it would be natural to expect an analogous interpretation of $c_L$ in the Carrollian theory. 

\subsection{Boundary energy-momentum tensor for \texorpdfstring{$d=2$}{d=2}}
\label{ssub:EMT-news-for-d=2}
For $d=2$, the analysis in~\eqref{eq:integrandorder} tells us that the variational result in~\eqref{eq:varSEH+ext+norm} is order $r$, and hence it is still divergent.
Equations \eqref{eq:calT1mu} and \eqref{eq:calT3munu} now give
\begin{align}
  \overset{(1)}{\mathcal{T}}{}^\mu
  =& 2v^\mu\left(\overset{(1)}{\mathcal{K}}-\overset{(0)}{S}\right)
  +h^{\mu\nu}\partial_\nu K\nonumber
  \\
  =& -v^\mu\mathcal{Q}
  +h^{\mu\nu}\partial_\nu K\,,\label{eq:calT(1)}
  \\
  \overset{(3)}{\mathcal{T}}{}^{\mu\nu}
  =& -h^{\mu\nu}
    \os{1}{\mathcal{K}}
  -2\overset{(1)}{\mathcal{Z}}{}^{(\mu}v^{\nu)}
  +h^{\mu\rho}h^{\nu\sigma}\left(
    -\mathcal{L}_v C_{\rho\sigma}
    +2A_{\rho\sigma}\right)
  \nonumber\\
  =& -h^{\mu\nu}
    \mathcal{D}_\rho a^\rho
  -2v^{(\mu}h^{\nu)\sigma}\left(\mathcal{D}_\rho+a_\rho\right)F^\rho{}_\sigma +2A^{\mu\nu}\nonumber\\
  &+2v^{(\mu}h^{\nu)\sigma}\left(\mathcal{D}_\rho-a_\rho\right)C^{\rho}{}_\sigma-h^{\mu\rho}h^{\nu\sigma}
    \mathcal{L}_v C_{\rho\sigma}
  \,,\label{eq:calT(3)}
\end{align}
where we used~\eqref{eq:S0d=2} to relate~$\os{0}{S}$ to~$\mathcal{Q}$,
and where we used the expression for~$\os{1}{\mathcal{Z}}^\mu$ from Equation~\eqref{eq:app-curly-Z-expansion-d2-r1-with-U2-or-hvPi0} in the third equality above.

\paragraph{Total derivative from overleading shear terms.}
Note that the shear terms in the last line of~\eqref{eq:calT(3)} are of the form of the EMT ambiguity~$t^{\mu\nu}$
in~\eqref{eq:constraint-response-term} that arises due to the $K_{\langle\mu\nu\rangle}=0$ constraint on $h_{\mu\nu}$, with $\zeta^{\mu\nu}=-2C^{\mu\nu}$ as its parameter.
On shell, the variations respect this constraint,
so we can invoke the identity~\eqref{eq:idtmunuosvar} involving $t^{\mu\nu}\delta h_{\mu\nu}$.
This now tells us that the aforementioned shear terms,
which appear in \eqref{eq:varSEH+ext+norm} at order $r$ as
\begin{equation}
    \label{eq:hol-ren-d=2-overleading-total-derivative-shear-term}
    \int_{r=\Lambda} d^3x e \overset{(3)}{\mathcal{T}}{}^{\mu\nu}\delta h_{\mu\nu}\,,
\end{equation}
form a total derivative term.
Therefore, if we were left with only these terms,
the divergent contributions would form a total derivative and,
since we work up to total derivatives,
we can discard this term.

\paragraph{Intrinsic counterterm.}
For now, we still need to deal with the shear-independent terms in~\eqref{eq:calT(1)} and~\eqref{eq:calT(3)}.
In order to cancel these terms, we will add the following intrinsic counterterm,
\begin{equation}\label{eq:intct}
  S_\text{int}
  =-c\int_{r = \Lambda} d^{3}x\,E\, r Q\,,
\end{equation}
where $Q$ is defined in terms of $\Pi^{\mu\rho}\hat R_{\mu\rho}$ in \eqref{eq:spatialRicscalard=2}. We remind the reader that $\hat C^\rho_{\mu\nu}$~is our connection on the $r=\Lambda$ hypersurface,
as discussed in Section~\ref{ssec:connection-choice}.
Next, we show that the variation of $S_\text{int}$ can be written in the form
\begin{equation}
  \label{eq:int-var-d2}
  \delta S_{\text{int}}
  = c\int d^{3}x E r\left(
    \mathcal{T}^\mu_{\text{int}}\delta V_\mu
    +\frac{1}{2}\mathcal{T}_{\text{int}}^{\mu\nu}\delta\Pi_{\mu\nu}-E^{-1}\partial_\mu\left(E Y^\mu\right)
  \right)\,.
\end{equation}
In the following, we will take $c=1$.
It can be shown that $Q$ varies for $d=2$ as
\begin{align}
    \delta Q & =  -QE^{-1}\delta E+\left(-QU^\mu+\Pi^{\mu\nu}\partial_\nu\mathcal{K}-2\Pi^{\mu\sigma}\left(\hat D_\rho-\mathcal{A}_\rho\right)\mathcal{K}^T{}^\rho{}_\sigma\right)\delta V_\mu\nonumber\\
    &\hspace{-.5cm}+\frac{1}{2}\left(-\Pi^{\mu\nu}\hat D_\rho \mathcal{A}^\rho+2\Pi^{\rho\langle\mu}\Pi^{\nu\rangle\sigma}\left(\hat D_\rho \mathcal{A}_\sigma+\mathcal{A}_\rho \mathcal{A}_\sigma\right)+2\mathcal{K}^T{}^\mu{}_\rho\mathcal{F}^{\rho\nu}-2U^\mu\left(\hat D_\rho+\mathcal{A}_\rho\right)\mathcal{F}^{\rho\nu}\right)\delta\Pi_{\mu\nu}\nonumber\\
    &\hspace{-.5cm}+\hat D_\rho\left(-U^\rho\Pi^{\mu\nu}\hat D_\mu\delta V_\nu+\Pi^\rho_\sigma Y^\sigma\right)\,.
\end{align}
This follows from the fact that,
after the obvious substitutions,
the variation of~$Q$ is the same as the variation of~$\mathcal{Q}$ given in~\eqref{eq:varcalQ}. The vector in the total derivative on the last line is split into a term that is proportional to $U^\rho$ and a spatial piece that 
is given by
\begin{align}
  \Pi^\rho_\sigma Y^\sigma
  = \Pi^\rho_\sigma
  &\left(
    \delta \mathcal{A}^\sigma
    +\mathcal{A}^\sigma \Pi^{\mu\nu}\delta \Pi_{\mu\nu}
    +\Pi^{\mu\nu}\delta\hat C^\sigma_{\mu\nu}
  \right.
  \\
  &{}\qquad\nonumber
  \left.
    -\Pi^{\sigma\nu}\partial_\nu\left(E^{-1}\delta E\right)
    +U^\mu \mathcal{F}^{\sigma\nu}\delta \Pi_{\mu\nu}
  \right)\,.
\end{align}
This variation tells us that 
\begin{subequations}
\label{eq:int-curly-Ts}
\begin{align}
    \mathcal{T}^\mu_{\text{int}} & = QU^\mu-\Pi^{\mu\nu}\partial_\nu\mathcal{K}+2\Pi^{\mu\sigma}\left(\hat D_\rho-\mathcal{A}_\rho\right)\mathcal{K}^T{}^\rho{}_\sigma\,,\label{eq:mathcalTintmu}\\
    \mathcal{T}_{\text{int}}^{\mu\nu} & =  \Pi^{\mu\nu}\hat D_\rho \mathcal{A}^\rho-2\Pi^{\rho\langle\mu}\Pi^{\nu\rangle\sigma}\left(\hat D_\rho \mathcal{A}_\sigma+\mathcal{A}_\rho \mathcal{A}_\sigma\right)-2\mathcal{K}^T{}^\mu{}_\rho\mathcal{F}^{\rho\nu}\nonumber\\   &\qquad+2U^{(\mu}\Pi^{\nu)\sigma}\left(\hat D_\rho+\mathcal{A}_\rho\right)\mathcal{F}^{\rho}{}_{\sigma}\,.\label{eq:mathcalTintmunu}
\end{align}
\end{subequations}
In the last term in $\mathcal{T}_{\text{int}}^{\mu\nu}$,
we dropped any term proportional to $U^\mu U^\nu$,
which would cancel in~\eqref{eq:int-var-d2} since we have $U^\mu U^\nu\delta\Pi_{\mu\nu}=0$.

After adding this intrinsic counterterm,
and recalling $a=2$, $b=1$ and $c=1$,
we can now replace the total variation in \eqref{eq:varSEH+ext+norm} with
\begin{eqnarray} \delta\left(S_{\text{EH}}+S_{\text{ext}}+S_{\text{norm}}+S_{\text{int}}\right)\Big\vert_{\text{Car-cov BS gauge}} & = & \cdots\label{eq:finitevard=2}\\
&&\hspace{-8cm}+\int_{r=\Lambda}d^{3}x E\left[-E^{-1}\partial_\mu\left(E \left(X^\mu+rY^\mu\right)\right)+\left(\mathcal{T}^\mu+r\mathcal{T}^\mu_{\text{int}}\right)\delta V_\mu+\frac{1}{2}\left(\mathcal{T}^{\mu\nu}+r\mathcal{T}^{\mu\nu}_{\text{int}}\right)\delta\Pi_{\mu\nu}\right]\,,\nonumber
\end{eqnarray}
where $\mathcal{T}^\mu$ is given in \eqref{eq:calTmu} and $\mathcal{T}^{\mu\nu}$ is given in \eqref{eq:calTmunu}.
For $d=2$, they read
\begin{subequations}
\label{eq:curly-Ts}
  \begin{align}
    \mathcal{T}^\mu
    &= 2U^\mu\bar{\mathcal{K}}-S\mathcal{Z}^\mu+\Pi^{\mu\nu}\partial_\nu S\,,
    \\
    \mathcal{T}^{\mu\nu}
    &= \Pi^{\mu\nu}\left(-\bar{\mathcal{K}}+\partial_r S-r^{-1}S+2S\partial_r\beta\right)
    \\
    &{}\qquad\nonumber
    +2\mathcal{K}^{T\,\mu\nu}-S\mathcal{G}^{\mu\nu}-2\mathcal{Z}^{(\mu}U^{\nu)}\,.
  \end{align}
\end{subequations}
At leading order, the additional contributions from~\eqref{eq:int-curly-Ts} are equal to 
\begin{equation}
\begin{split}
        \os{2}{\mathcal{T}}{}^\mu_{\text{int}} & =  \mathcal{Q}v^\mu-h^{\mu\nu}\partial_\nu K\,,\\
    \os{4}{\mathcal{T}}{}_{\text{int}}^{\mu\nu} & =  h^{\mu\nu}\mathcal{D}_\rho a^\rho-2A^{\mu\nu}+2v^{(\mu} h^{\nu)\sigma}\left(\mathcal{D}_\rho+a_\rho\right)F^{\rho}{}_{\sigma}\,.
\end{split}
\end{equation}
Comparing this with the overleading terms in~\eqref{eq:calT(1)} and~\eqref{eq:calT(3)},
we see that these precisely cancel the shear-independent terms.
We thus obtain
\begin{align}
    &E\left(\left(\mathcal{T}^\mu+r\mathcal{T}^\mu_{\text{int}}\right)\delta V_\mu+\frac{1}{2}\left(\mathcal{T}^{\mu\nu}+r\mathcal{T}^{\mu\nu}_{\text{int}}\right)\delta\Pi_{\mu\nu}\right)\nonumber\\
    &{}\qquad
    =e\frac{r}{2}\left(\mathcal{T}^{\mu\nu}+r \mathcal{T}^{\mu\nu}_{\text{int}}\right)\Big\vert_{@r^{-3}}\delta h_{\mu\nu}+\mathcal{O}(1)\nonumber\\
    &{}\qquad
    =e\frac{r}{2}\left(2v^{(\mu}h^{\nu)\sigma}\left(\mathcal{D}_\rho-a_\rho\right)C^{\rho}{}_\sigma-h^{\mu\rho}h^{\nu\sigma}
    \mathcal{L}_v C_{\rho\sigma}\right)\delta h_{\mu\nu}+\mathcal{O}(1)\nonumber\\
    &{}\qquad
    =-\frac{r}{2}\partial_\rho\left(e\left[v^\rho C^{\mu\nu}-C^{\rho\mu}v^\nu-C^{\rho\nu}v^\mu\right]\delta h_{\mu\nu}\right)+\mathcal{O}(1)\,,
\end{align}
where we recall that the total derivative term in the last line arises due to the arguments around~\eqref{eq:hol-ren-d=2-overleading-total-derivative-shear-term} above.
For later purposes, we note this EMT-ambiguity-type term arises from
\begin{equation}
\label{eq:identity-with-zeta=C}
    \left(\mathcal{T}^{\mu\nu}+r \mathcal{T}^{\mu\nu}_{\text{int}}\right)\Big\vert_{@r^{-3}}=2v^{(\mu}h^{\nu)\sigma}\left(\mathcal{D}_\rho-a_\rho\right)C^{\rho}{}_\sigma-h^{\mu\rho}h^{\nu\sigma}
    \mathcal{L}_v C_{\rho\sigma}\,.
\end{equation}
This shows that~\eqref{eq:finitevard=2} is finite if we ignore total derivative terms.

\paragraph{Finite currents.}
Subsequently, the renormalised boundary EMT-news complex is defined in terms of the following finite part of the integrand,
\begin{align}
  &E\left(\left(\mathcal{T}^\mu+r \mathcal{T}^\mu_{\text{int}}\right)\delta V_\mu+\frac{1}{2}\left(\mathcal{T}^{\mu\nu}+r \mathcal{T}^{\mu\nu}_{\text{int}}\right)\delta\Pi_{\mu\nu}\right)\big\vert_{@r^0}\nonumber
  \\
  &{}\qquad
  = e\left(\mathcal{T}^\mu+r \mathcal{T}^\mu_{\text{int}}\right)\Big\vert_{@r^{-2}}\delta\tau_\mu
  +\frac{e}{2}\left(\mathcal{T}^{\mu\nu}+r \mathcal{T}^{\mu\nu}_{\text{int}}\right)\big\vert_{@r^{-4}}\delta h_{\mu\nu}\nonumber
  \\
  &{}\qquad\qquad\nonumber
  +\frac{e}{2}\left(\mathcal{T}^{\mu\nu}+r \mathcal{T}^{\mu\nu}_{\text{int}}\right)\big\vert_{@r^{-3}}\delta\overset{(-1)}{\Pi}_{\mu\nu}
  \\
  &{}\qquad
  =2e\left(T_{\text{ren}}^\mu\delta\tau_\mu+\frac{1}{2}T_{\text{ren}}^{\mu\nu}\delta h_{\mu\nu}+\frac{1}{2} S_{\text{ren}}^{\mu\nu}\delta C_{\mu\nu}\right)\,,\label{eq:ren-EMT-N-def}
\end{align}
where we remind the reader that $S_{\text{ren}}^{\mu\nu}$ is taken to be STF.
In the last line, we defined the renormalised EMT-news complex,
and our goal is now to obtain it explicitly.
Crucially, the expression again contains the overleading current contributions which,
when paired with the leading metric variations, gave rise to a total derivative term.
On the other hand, at this order, they pair with the subleading metric variations,
which gives rise to
\begin{align}
    &\frac{1}{2}\left(\mathcal{T}^{\mu\nu}+r \mathcal{T}^{\mu\nu}_{\text{int}}\right)\big\vert_{@r^{-3}}\delta\overset{(-1)}{\Pi}_{\mu\nu}\nonumber\\
    &=\mathcal{D}_\rho\left[\left(v^\rho\left(\mathcal{D}_\mu-a_\mu\right)C^{\mu\nu}-v^\nu\left(\mathcal{D}_\mu-a_\mu\right)C^{\mu\rho}\right)\delta\tau_\nu\right]\nonumber\\
    &{}\qquad
    +\left[-\left(\mathcal{L}_v-K\right)\left(\mathcal{D}_\rho-a_\rho\right)C^{\rho\mu}+v^\mu\mathcal{D}_\sigma\left(\mathcal{D}_\rho-a_\rho\right)C^{\rho\sigma}+h^{\mu\rho}a^\sigma\mathcal{L}_v C_{\rho\sigma}\right]\delta\tau_\mu\nonumber\\
    &{}\qquad
    +\left[F_\sigma{}^\mu v^\nu\left(\mathcal{D}_\rho-a_\rho\right)C^{\rho\sigma}+v^\mu C^{\nu\sigma}\left(\mathcal{D}_\rho-a_\rho\right)C^{\rho}{}_\sigma\right]\delta h_{\mu\nu}\nonumber\\
    &{}\qquad
    -\frac{1}{2}h^{\mu\rho}h^{\nu\sigma}\mathcal{L}_v C_{\rho\sigma}\delta C_{\mu\nu}\,.
\end{align}
This is the mechanism by which variations of the shear arise on the boundary through the variational procedure, even though it appears as a subleading field in the metric (though see footnote~\ref{fn:shear-subleading}).
This leads to the intermediate result
\begin{subequations}
\begin{align}
    T_{\text{ren}}^\mu & = \frac{1}{2}\left(\mathcal{T}^\mu+r \mathcal{T}^\mu_{\text{int}}\right)\Big\vert_{@r^{-2}}-\frac{1}{2}\left(\mathcal{L}_v-K\right)\left(\mathcal{D}_\rho-a_\rho\right)C^{\rho\mu}\nonumber\\
    &\qquad+\frac{1}{2}v^\mu\mathcal{D}_\sigma\left(\mathcal{D}_\rho-a_\rho\right)C^{\rho\sigma}+\frac{1}{2}h^{\mu\rho}a^\sigma\mathcal{L}_v C_{\rho\sigma}\,,\\
    T_{\text{ren}}^{\mu\nu} & =  \frac{1}{2}\left(\mathcal{T}^{\mu\nu}+r \mathcal{T}^{\mu\nu}_{\text{int}}\right)\Big\vert_{@r^{-4}}+F_\sigma{}^{(\mu} v^{\nu)}\left(\mathcal{D}_\rho-a_\rho\right)C^{\rho\sigma}\nonumber\\
    &\qquad+v^{(\mu} C^{\nu)\sigma}\left(\mathcal{D}_\rho-a_\rho\right)C^{\rho}{}_\sigma\,,\\
    S_{\text{ren}}^{\mu\nu} & =  -\frac{1}{2}h^{\mu\rho}h^{\nu\sigma}\mathcal{L}_v C_{\rho\sigma}\,.
\end{align}
\end{subequations}
The resulting news current is already reasonable,
but it is not yet Weyl-covariant.

At this point, let us therefore consider the effect of adding the `finite norm' counterterm we previously considered in~\eqref{eq:finnormct} to the action.
Its on-shell value in the limit $r\to\infty$ is 
\begin{equation}
S_{\text{fin,norm}}\big\vert_{@r^0}=-\frac{\tilde b}{8}\int_{r=\Lambda} d^3x eK (C^2-F^2)\,.
\end{equation}
Furthermore, using~\eqref{eq:varSnormfin}, it can be checked that its variation evaluated on shell is the variation of the on-shell value,
and so we have
\begin{equation}
    \delta S_{\text{fin,norm}}\big\vert_{@r^0}=-\frac{\tilde b}{8}\delta\int_{r=\Lambda} d^3x eK (C^2-F^2)\,.
\end{equation}
We can recognize this as a combination of the improvement terms~\eqref{eq:fourdim-weyl-improvement-term-1}
and~\eqref{eq:fourdim-weyl-improvement-term-4}
we already considered in Section~\ref{ssec:bulk-improvements-weyl-4d}.
This means that its contribution to the currents can be read off from~\eqref{eq:fourdim-weyl-improvement-term-23-currents} after setting $a_1=-\tilde b/8=-a_4$ and $a_2=0$ there.
Adding this term too, we set $\tilde b=b=1$ (as well as the earlier values $a=2$ and $c=1$) and define (up to boundary terms)
\begin{align}
  &\lim_{\Lambda\to \infty}\delta\left(S_{\text{EH}}+S_{\text{ext}}+S_{\text{norm}}+S_{\text{fin,norm}}+S_{\text{int}}\right)\big\vert_{\text{os}}
  \nonumber
  \\
  &{}\qquad\qquad
  =\cdots+2\int_{\mathcal{I}^+}d^{3}x e\left[T'^\mu_{\text{ren}}\delta\tau_\mu+\frac{1}{2}T'^{\mu\nu}_{\text{ren}}\delta h_{\mu\nu}+\frac{1}{2}S'^{\mu\nu}_{\text{ren}}\delta h_{\mu\nu}\right]\,.\label{eq:finitevard=2order1}
\end{align}
Together with these additional contributions,
we then obtain
\begin{subequations}
\begin{align}
    T'^\mu_{\text{ren}}
    &=  \frac{1}{2}\left(\mathcal{T}^\mu+r \mathcal{T}^\mu_{\text{int}}\right)\big\vert_{@r^{-2}}
    -\frac{1}{2}\left(\mathcal{L}_v-K\right)\left(\mathcal{D}_\rho-a_\rho\right)C^{\rho\mu}
    \nonumber
    \\
    &\qquad
    +\frac{1}{2}v^\mu\mathcal{D}_\sigma\left(\mathcal{D}_\rho-a_\rho\right)C^{\rho\sigma}
    +\frac{1}{2}h^{\mu\rho}a^\sigma\mathcal{L}_v C_{\rho\sigma}
    -\frac{1}{8}KF^2 v^\mu
    \nonumber
    \\
    &\qquad
    -\frac{1}{4}\partial_\rho K F^{\rho\mu}
    -\frac{1}{4}K h^\mu_\sigma\mathcal{D}_\rho F^{\rho\sigma}
    +\frac{1}{4}K a_\nu F^{\mu\nu}\,,
    \\
    T'^{\mu\nu}_{\text{ren}} & = \frac{1}{2}\left(\mathcal{T}^{\mu\nu}+r \mathcal{T}^{\mu\nu}_{\text{int}}\right)\big\vert_{@r^{-4}}+F_\sigma{}^{(\mu} v^{\nu)}\left(\mathcal{D}_\rho-a_\rho\right)C^{\rho\sigma}\nonumber\\
    &\qquad+v^{(\mu} C^{\nu)\sigma}\left(\mathcal{D}_\rho-a_\rho\right)C^{\rho}{}_\sigma+\frac{1}{16}KC^2h^{\mu\nu}+\frac{1}{8}C\cdot N h^{\mu\nu}\nonumber\\
    &\qquad+\frac{1}{16}\left(v^\mu h^{\nu\rho}+v^\nu h^{\mu\rho}\right)\partial_\rho \left(C^2-F^2\right)-\frac{1}{8}KF^2 h^{\mu\nu}+\frac{1}{16}h^{\mu\nu}\mathcal{L}_v F^2\,,\\
    S'^{\mu\nu}_{\text{ren}} & =  -\frac{1}{2}h^{\mu\rho}h^{\nu\sigma}\left(\mathcal{L}_v+\frac{1}{2}K\right) C_{\rho\sigma}=\frac{1}{2}N^{\mu\nu}\,.
    \label{eq:hol-ren-d=2-weyl-improved-news}
\end{align}
\end{subequations}
In particular, we see that we have recovered the Weyl-covariant news tensor~$N_{\mu\nu}$ from~\eqref{eq:NewsDefn-repeat} as the news current,
just as in the Weyl-covariant EMT-news complex we obtained in~\eqref{eq:WeylImprovedEMTNewsComplex} in Section~\ref{ssec:bulk-improvements-weyl-4d}.
In terms of the various projections of the EMT, we get
\begin{subequations}
\label{eq:hol-ren-d2-emtn-shear-terms-news}
\begin{align}
    \tau_\mu T'^\mu_{\text{ren}} & = \frac{1}{2}\tau_\mu \left(\mathcal{T}^\mu+r \mathcal{T}^\mu_{\text{int}}\right)\big\vert_{@r^{-2}}\nonumber\\
    &\qquad-\frac{1}{2}\mathcal{D}_\rho\mathcal{D}_\sigma C^{\rho\sigma}+\frac{1}{2}C\cdot A+a_\sigma\mathcal{D}_\rho C^{\rho\sigma}-a_\rho a_\sigma C^{\rho\sigma}+\frac{1}{8}KF^2\,,\\
    h^\mu_\rho T'^\rho_{\text{ren}} & = \frac{1}{2}h^\mu_\rho\left(\mathcal{T}^\rho+r \mathcal{T}^\rho_{\text{int}}\right)\big\vert_{@r^{-2}}\nonumber\\
    &\qquad+\frac{1}{2}\left(\mathcal{D}_\rho-a_\rho\right)N^{\rho\mu}+\frac{1}{4}\partial_\rho K C^{\rho\mu}-\frac{1}{4}K\mathcal{D}_\rho C^{\rho\mu}+\frac{1}{4}Ka_\rho C^{\rho\mu}\nonumber\\
    &\qquad-\frac{1}{4}\partial_\rho K F^{\rho\mu}-\frac{1}{4}K a_\rho F^{\rho\mu}-\frac{1}{4}Kh^\mu_\sigma\mathcal{D}_\rho F^{\rho\sigma}\,,\\
    h_{\mu\nu}T'^{\mu\nu}_{\text{ren}} & = \frac{1}{2}h_{\mu\nu}\left(\mathcal{T}^{\mu\nu}+r \mathcal{T}^{\mu\nu}_{\text{int}}\right)\big\vert_{@r^{-4}}+\frac{1}{8}KC^2+\frac{1}{4}C\cdot N\nonumber\\
    &\qquad-\frac{1}{4}KF^2+\frac{1}{8}\mathcal{L}_v F^2\,,\\
    \tau_{\mu}h_{\nu\kappa}T'^{\mu\nu}_{\text{ren}} & = \frac{1}{2}\tau_{\mu}h_{\nu\kappa}\left(\mathcal{T}^{\mu\nu}+r \mathcal{T}^{\mu\nu}_{\text{int}}\right)\big\vert_{@r^{-4}}+\frac{1}{2}\os{1}{\mathcal{Z}}^\rho \left(F_{\rho\kappa}+C_{\rho\kappa}\right)-\frac{1}{2}F_{\sigma\kappa}\left(\mathcal{D}_\rho+a_\rho\right)F^{\rho\sigma}\nonumber\\
    &\qquad-\frac{1}{2}C_{\sigma\kappa}\left(\mathcal{D}_\rho+a_\rho\right)F^{\rho\sigma}-\frac{1}{16}h^\rho_\kappa\partial_\rho\left(C^2-F^2\right)\,,\\
    h^{\langle\mu}_\rho h^{\nu\rangle}_\sigma T'^{\rho\sigma}_{\text{ren}} & = \frac{1}{2}h^{\langle\mu}_\rho h^{\nu\rangle}_\sigma\left(\mathcal{T}^{\rho\sigma}+r \mathcal{T}^{\rho\sigma}_{\text{int}}\right)\big\vert_{@r^{-4}}\,,
\end{align}
\end{subequations}
where we used Equation~\eqref{eq:Z1Simp} for $\os{1}{\mathcal{Z}}^\mu$ in the expression for $\tau_{\mu}h_{\nu\kappa}T'^{\mu\nu}_{\text{ren}}$.
Additionally, in the result for $h^\mu_\rho T'^\rho_{\text{ren}}$,
we used~\eqref{eq:comLiecovSTF} to show that
\begin{equation}
    h^\mu_\sigma\left(\mathcal{L}_v-K\right)\left(\mathcal{D}_\rho-a_\rho\right)C^{\rho\sigma}=-\mathcal{D}_\rho N^{\rho\mu}-Ka_\rho C^{\rho\mu}-\frac{1}{2}\partial_\rho K C^{\rho\mu}+\frac{1}{2}K\mathcal{D}_\rho C^{\rho\mu}\,.
\end{equation}

We will now proceed to compute $\left(\mathcal{T}^\mu+r \mathcal{T}^\mu_{\text{int}}\right)\big\vert_{@r^{-2}}$ and $\left(\mathcal{T}^{\mu\nu}+r \mathcal{T}^{\mu\nu}_{\text{int}}\right)\big\vert_{@r^{-4}}$.
Using the unexpanded expressions in \eqref{eq:calTmu} and \eqref{eq:mathcalTintmu}, we can write
\begin{align} 
\mathcal{T}^\mu+r\mathcal{T}^\mu_{\text{int}}
   & = U^\mu\left(2\bar{\mathcal{K}}+rQ\right)-S\mathcal{Z}^\mu+\Pi^{\mu\nu}\partial_\nu \left(S-r\mathcal{K}\right)\nonumber\\
   &\qquad+2r\Pi^{\mu\sigma}\left(\hat D_\rho-\mathcal{A}_\rho\right)\mathcal{K}^T{}^\rho{}_\sigma\,,
\end{align}
and, using~\eqref{eq:calTmunu} and~\eqref{eq:mathcalTintmunu},
we similarly obtain
\begin{align}
    \mathcal{T}^{\mu\nu}+r\mathcal{T}^{\mu\nu}_{\text{int}}
    &= \Pi^{\mu\nu}\left(-\bar{\mathcal{K}}+r\hat D_\rho \mathcal{A}^\rho+\partial_r S-r^{-1}S+2S\partial_r\beta\right)\nonumber
    \\
    &\qquad\nonumber
    +2\mathcal{K}^{T\,\mu\nu}-S\mathcal{G}^{\mu\nu}-2\mathcal{Z}^{(\mu}U^{\nu)}\nonumber\\
&\qquad-2r\Pi^{\rho\langle\mu}\Pi^{\nu\rangle\sigma}\left(\hat D_\rho \mathcal{A}_\sigma+\mathcal{A}_\rho \mathcal{A}_\sigma\right)-2r\mathcal{K}^T{}^\mu{}_\rho\mathcal{F}^{\rho\nu}\nonumber\\   &\qquad+2rU^{(\mu}\Pi^{\nu)\sigma}\left(\hat D_\rho+\mathcal{A}_\rho\right)\mathcal{F}^{\rho}{}_{\sigma}\,.
  \end{align}
Using that we have $Q=r^{-2}\mathcal{Q}+\mathcal{O}(r^{-3})$,
along with the identity~\eqref{eq:S0d=2} linking~$\os{0}{S}$ to~$\mathcal{Q}$,
as well as the identity~\eqref{eq:spatialRicscalard=2} for the spatial trace of the Ricci curvature,
\begin{subequations}
  \begin{align} 
    \left(\mathcal{T}^\mu+r\mathcal{T}^\mu_{\text{int}}\right)\big\vert_{@r^{-2}}
    & =  v^\mu\left(2\os{2}{\mathcal{K}}-2\os{1}{S}+\hat D_\rho \mathcal{A}^\rho\big\vert_{@r^{-3}}+\Pi^{\rho\sigma}\hat R_{\rho\sigma}\big\vert_{@r^{-3}}\right)
    \\
    \nonumber
    &\qquad-K\os{1}{\mathcal{Z}}^\mu+\frac{1}{2}h^{\mu\nu}\partial_\nu \mathcal{Q}+2h^{\mu\sigma}\left(\mathcal{D}_\rho-a_\rho\right)\left(h^{\rho\alpha}\os{-1}{\mathcal{K}}^T_{\alpha\sigma}\right)\,,\\
    \left(\mathcal{T}^{\mu\nu}+r\mathcal{T}^{\mu\nu}_{\text{int}}\right)\big\vert_{@r^{-4}}
    &= h^{\mu\nu}\left(-\os{2}{\mathcal{K}}-\os{1}{S}+\hat D_\rho \mathcal{A}^\rho\big\vert_{@r^{-3}}-\frac{1}{4}K\left(F^2-C^2\right)\right)\nonumber
    \\
    &\qquad\nonumber
    +2\mathcal{K}^{T\,\mu\nu}\big\vert_{@r^{-4}}+\os{0}{S}C^{\mu\nu}-K\mathcal{G}^{\mu\nu}\big\vert_{@r^{-5}}+\os{1}{\mathcal{Z}}^\mu a^\nu+\os{1}{\mathcal{Z}}^\nu a^\mu\nonumber\\
    &\qquad-2v^{(\mu}h^{\nu)\rho}\os{2}{\mathcal{Z}}_\rho+2v^{(\mu}C^{\nu)\rho}\os{1}{\mathcal{Z}}_\rho-N^\mu{}_\rho F^{\rho\nu}+\frac{1}{2}KC^\mu{}_\rho F^{\rho\nu}\nonumber\\
    &\qquad-2A^\mu{}_\rho F^{\rho\nu}-2\Pi^{\rho\langle\mu}\Pi^{\nu\rangle\sigma}\left(\hat D_\rho \mathcal{A}_\sigma+\mathcal{A}_\rho \mathcal{A}_\sigma\right)\big\vert_{@r^{-5}}\nonumber\\   
    &\qquad+2U^{(\mu}\Pi^{\nu)\sigma}\left(\hat D_\rho+\mathcal{A}_\rho\right)\mathcal{F}^{\rho}{}_{\sigma}\big\vert_{@r^{-5}}\,,\label{eq:full-curly-T-at-order-minus-4}
  \end{align}
\end{subequations}
where we used the expression for $\os{-1}{\mathcal{K}}^T_{\mu\nu}$ in \eqref{eq:app-curly-KT-expansion-d2-r1} and where $\os{2}{\mathcal{K}}$ is given by~\eqref{eq:app-curly-K-expansion-d2-r-2}.
In order to make progress, we compute the following expansions
\begin{subequations}
\begin{align}
    \hat D_\rho \mathcal{A}^\rho\big\vert_{@r^{-3}} & = \mathcal{D}_\rho\left(F^{\rho\sigma}a_\sigma-C^{\rho\sigma}a_\sigma\right)\\
    & = a_\sigma\mathcal{D}_\rho F^{\rho\sigma}+\frac{1}{4}\left(\mathcal{L}_v-2K\right)F^2-a_\sigma\left(\mathcal{D}_\rho-a_\rho\right)C^{\rho\sigma}-C\cdot A\,,\nonumber\\
    \mathcal{K}^{T\,\mu\nu}\big\vert_{@r^{-4}} & = h^{\mu\rho}h^{
    \nu\sigma}\os{0}{\mathcal{K}}^{T}_{\rho\sigma}-h^{\mu\nu}\os{-1}{\mathcal{K}}\cdot C\,,\\
    &= h^{\rho\langle\mu}h^{\nu\rangle\sigma}\os{0}{\mathcal{K}}_{\rho\sigma}-\frac{1}{2}h^{\mu\nu}C\cdot\os{-1}{\mathcal{K}}^T-\frac{1}{2}C^{\mu\nu}\mathcal{D}_\rho a^\rho-\frac{1}{2}Kh^{\rho\langle\mu}h^{\nu\rangle\sigma}\os{0}{\Pi}_{\rho\sigma}\,,\nonumber\\
    \mathcal{G}^{\mu\nu}\big\vert_{@r^{-5}} & =  h^{\mu\rho}h^{\nu\sigma}\os{1}{\mathcal{G}}_{\rho\sigma}+C^2 h^{\mu\nu}=-2h^{\mu\rho}h^{\nu\sigma}\os{0}{\Pi}_{\rho\sigma}+C^2 h^{\mu\nu}\,,\\
    \Pi^{\mu\nu}\hat R_{\mu\nu}\big\vert_{@r^{-3}} & = h^{\mu\nu}\mathcal{D}_\rho \hat C^\rho_{\mu\nu}\big\vert_{@r^{-1}}+C\cdot A\nonumber\\
    & = \mathcal{D}_\rho \left(h^{\mu\nu}\hat C^\rho_{\mu\nu}\big\vert_{@r^{-1}}\right)+\frac{1}{4}\left(\mathcal{L}_v-2K\right)F^2+C\cdot A\nonumber\\
    & = \frac{1}{2}\left(\mathcal{L}_v-2K\right)F^2+a_\alpha\mathcal{D}_\rho F^{\rho\alpha}+\mathcal{D}_\rho\left(\mathcal{D}_\alpha-a_\alpha\right)C^{\alpha\rho}+C\cdot A\,,
\end{align}
\end{subequations}
where $\os{1}{\mathcal{G}}_{\rho\sigma}$ is in~\eqref{eq:app-CalG1} and we used $h^{\mu\rho}h^{\nu\sigma}\os{1}{\mathcal{G}}_{\rho\sigma}=-2h^{\mu\rho}h^{\nu\sigma}\os{0}{\Pi}_{\rho\sigma}$, as well as
\begin{align}
-2\Pi^{\rho\langle\mu}\Pi^{\nu\rangle\sigma}\left(\hat D_\rho \mathcal{A}_\sigma+\mathcal{A}_\rho \mathcal{A}_\sigma\right)\big\vert_{@r^{-5}} & =  h^{\mu\nu}A\cdot C+2A^{\mu\alpha}F_\alpha{}^\nu\\
&\qquad+h^{\rho\langle\mu}h^{\nu\rangle\sigma}\left(\mathcal{L}_a C_{\rho\sigma}+2a_\rho\left(\mathcal{D}_\alpha+a_\alpha\right)F^\alpha{}_\sigma\right.\nonumber\\
&\qquad\left.-2a_\rho\left(\mathcal{D}_\alpha-2a_\alpha\right)C^\alpha{}_\sigma-C^\alpha{}_\sigma\mathcal{L}_v F_{\rho\alpha}\right)\,,\nonumber
\end{align}
and
\begin{eqnarray}
    2U^{(\mu}\Pi^{\nu)\sigma}\left(\hat D_\rho+\mathcal{A}_\rho\right)\mathcal{F}^\rho{}_\sigma\big\vert_{@r^{-5}}=2v^{(\mu}a^{\nu)}F^2-2a^{(\mu}h^{\nu)\sigma}\left(\mathcal{D}_\alpha+a_\alpha\right)F^\alpha{}_\sigma\,.
\end{eqnarray}
In deriving these results, we made frequent use of \eqref{eq:app-d2-spatial-STF-STF-product} and \eqref{eq:app-d2-spatial-STF-antisym-product},
as well as results from Appendix~\ref{subsec:speciald=2id},
but results such as \eqref{eq:C1} and \eqref{eq:CDa} have also been important.

With the help of these results, it can be shown that we are getting close to the expressions for the EMT components that we derived previously from the bulk equations of motion in~\eqref{eq:EMTNewsComplexSec7},
\begin{align}
\label{eq:energyflux}
\begin{split}
        \tau_\mu \left(\mathcal{T}^\mu+r \mathcal{T}^\mu_{\text{int}}\right)\big\vert_{@r^{-2}} & =  2\tau_\mu T^\mu_{\text{EOM}}+KF^2\,,\\
    h^\mu_\rho\left(\mathcal{T}^\rho+r \mathcal{T}^\rho_{\text{int}}\right)\big\vert_{@r^{-2}} & =  2h^\mu_\rho T^\rho_{\text{EOM}}-2Ka_\rho F^{\rho\mu}-2Kh^\mu_\sigma\mathcal{D}_\rho F^{\rho\sigma}\\
    &\qquad-2F^{\rho\mu}\partial_\rho K\,.
\end{split}
\end{align}
In deriving \eqref{eq:energyflux},
we used Equation~\eqref{eq:Z1Simp} for $\os{1}{\mathcal{Z}}^\mu$ and $\os{-1}{\mathcal{K}}^T_{\mu\nu}$ as given in \eqref{eq:app-curly-KT-expansion-d2-r1},
and we subsequently used \eqref{eq:idLievP0} to eliminate $\left(\mathcal{D}_\rho-a_\rho\right)N^{\rho\mu}$.
We also used the Bianchi identity \eqref{eq:BItwo} and \eqref{eq:aA} to rewrite $\left(\mathcal{D}_\rho-a_\rho\right)A^{\rho\mu}$. We can furthermore show that
\begin{subequations}
\begin{align}
    h_{\mu\nu}\left(\mathcal{T}^{\mu\nu}+r\mathcal{T}^{\mu\nu}_{\text{int}}\right)\big\vert_{@r^{-4}} & =  2h_{\mu\nu}T^{\mu\nu}_{\text{EOM}}-2a_\sigma\mathcal{D}_\rho C^{\rho\sigma}+2a_\rho a_\sigma C^{\rho\sigma}
    \\
    \nonumber
    &\qquad-C\cdot N-\frac{1}{2}KC^2+\mathcal{L}_v F^2-2KF^2\,,\\
    \tau_\mu h_{\nu\kappa}\left(\mathcal{T}^{\mu\nu}+r\mathcal{T}^{\mu\nu}_{\text{int}}\right)\big\vert_{@r^{-4}} & =  h^\rho_\kappa\os{2}{\mathcal{Z}}_\rho-C^\rho{}_\kappa\os{1}{\mathcal{Z}}_\rho-a_\kappa F^2\\
    & =  2P_\kappa^{\text{EOM}}
    -\overset{(1)}{\mathcal{Z}}{}^\rho\left(F_{\rho\kappa}+C_{\rho\kappa}\right)
    +\frac{1}{4}h^\mu_\kappa\partial_\mu (C^2+F^2) -a_\kappa F^2\,,\nonumber\\
    h^{\langle\mu}_\rho h^{\nu\rangle}_\sigma\left(\mathcal{T}^{\rho\sigma}+r \mathcal{T}^{\rho\sigma}_{\text{int}}\right)\Big\vert_{@r^{-4}} & =  2\tilde T^{\mu\nu}_{\text{EOM}}-h^{\mu\langle\rho}h^{\sigma\rangle\nu}\mathcal{L}_v\left(F_{\rho\alpha}C^\alpha{}_\sigma\right)
    \\
    \nonumber
    &\qquad-h^{\langle\mu}_\rho h^{\nu\rangle}_\sigma\left(-\mathcal{L}_a C^{\rho\sigma}-2C^{\rho\sigma}\mathcal{D}_\alpha a^\alpha+2a^\rho\mathcal{D}_\alpha C^{\alpha\sigma}\right)\,.
\end{align}
\end{subequations}
In the last expression, we used the identity
\begin{align}
  &h^{\rho\langle\mu}h^{\nu\rangle\sigma}\left(\mathcal{L}_a C_{\rho\sigma}-2a_\rho\left(\mathcal{D}_\alpha-2a_\alpha\right)C^\alpha{}_\sigma\right)
  \nonumber
  \\
  &{}\qquad
  =-h^{\langle\mu}_\rho h^{\nu\rangle}_\sigma\left(-\mathcal{L}_a C^{\rho\sigma}-2C^{\rho\sigma}\mathcal{D}_\alpha a^\alpha+2a^\rho\mathcal{D}_\alpha C^{\alpha\sigma}\right)\,.
\end{align}
Together with the contributions we already found in~\eqref{eq:hol-ren-d2-emtn-shear-terms-news}, we therefore have
\begin{subequations}
\begin{align}
    \tau_\mu T'^\mu_{\text{ren}} & = \tau_\mu T^\mu_{\text{EOM}}\nonumber\\
    &\qquad-\frac{1}{2}\mathcal{D}_\rho\mathcal{D}_\sigma C^{\rho\sigma}+\frac{1}{2}C\cdot A+a_\sigma\mathcal{D}_\rho C^{\rho\sigma}-a_\rho a_\sigma C^{\rho\sigma}+\frac{5}{8}KF^2\,,\\
    h^\mu_\rho T'^\rho_{\text{ren}} & =  h^\mu_\rho T^\rho_{\text{EOM}}\nonumber\\
    &\qquad+\frac{1}{2}\left(\mathcal{D}_\rho-a_\rho\right)N^{\rho\mu}+\frac{1}{4}\partial_\rho K C^{\rho\mu}-\frac{1}{4}K\mathcal{D}_\rho C^{\rho\mu}+\frac{1}{4}Ka_\rho C^{\rho\mu}\nonumber\\
    &\qquad-\frac{5}{4}\partial_\rho K F^{\rho\mu}-\frac{5}{4}K a_\rho F^{\rho\mu}-\frac{5}{4}Kh^\mu_\sigma\mathcal{D}_\rho F^{\rho\sigma}\,,\\
    h_{\mu\nu}T'^{\mu\nu}_{\text{ren}} & = h_{\mu\nu}T^{\mu\nu}_{\text{EOM}}-a_\sigma\mathcal{D}_\rho C^{\rho\sigma}+a_\rho a_\sigma C^{\rho\sigma}-\frac{1}{8}KC^2-\frac{1}{4}C\cdot N\nonumber\\
    &\qquad-\frac{5}{4}KF^2+\frac{5}{8}\mathcal{L}_v F^2\,,\\
    \tau_{\mu}h_{\nu\kappa}T'^{\mu\nu}_{\text{ren}} & = P^{\text{EOM}}_\kappa+\frac{1}{16}h^\rho_\kappa\partial_\rho C^2+\frac{5}{16}h^\rho_\kappa\partial_\rho F^2+\frac{1}{2}F_{\kappa\sigma}\left(\mathcal{D}_\rho-2a_\rho\right)C^{\rho\sigma}\nonumber\\
    &\qquad+\frac{1}{2}a_\rho C^{\rho\sigma}F_{\sigma\kappa}-\frac{1}{2}h_{\kappa\alpha}\left(\mathcal{D}_\rho-a_\rho\right)\left(F^{\rho\sigma}C_\sigma{}^\alpha\right)\,,
\end{align}
\end{subequations}
where, in the expression for $\tau_{\mu}h_{\nu\kappa}T'^{\mu\nu}_{\text{ren}}$,
we used \eqref{eq:FDC} and \eqref{eq:FDFid}.
The last term in the expression for $\tau_{\mu}h_{\nu\kappa}T'^{\mu\nu}_{\text{ren}}$ is of the form of the $\zeta^{\mu\nu}$ term in \eqref{eq:finalWeylP}.
Finally,
\begin{equation}
    \label{eq:hol-ren-d=2-news-ren-to-news-eom}
    S'^{\mu\nu}_{\text{ren}}
    = \frac{1}{2}N_{\mu\nu}
    =S^{\mu\nu}_{\text{EOM}}+\frac{1}{4}KC^{\mu\nu}-A^{\mu\nu}\,.
\end{equation}
Again, we see these results are closely related to the form of the EMT-news complex we derived from the bulk equations of motion in~\eqref{eq:EMTNewsComplexSec7}.

In fact, we can see that the additional terms in $T'^\mu_{\text{ren}}$, $T'^{\mu\nu}_{\text{ren}}$ and $S'^{\mu\nu}_{\text{ren}}$ above
take the form of the general improvement transformation we considered in~\eqref{eq:improve-eom-to-weyl-d=2} with $a_1=1/16$, $a_2=-1/2$ and $a_4=5/16$.  
With the exception of~$a_4$, these are the values that improved the EOM energy-momentum-news complex to the Weyl energy-momentum-news complex in Section~\ref{ssec:bulk-improvements-weyl-4d}.
In order to change the value of $a_4$, we need to add a finite counterterm that is equal to $\int d^3x e KF^2$ on shell.
Adding the following finite intrinsic counterterm 
to~\eqref{eq:intct}
will do the job,
\begin{equation}
  \label{eq:hol-ren-d=2-ctilde-counterterm}
    S_{\text{int, fin}}=\tilde c\int_{r=\Lambda}d^3x ErS\mathcal{F}^2\,.
\end{equation}
The on-shell value of this counterterm is $\tilde c\int_{\mathcal{I}^+}d^3x e KF^2$,
and so $\tilde c$ plays the same role as $a_4$ in the improvements generated by \eqref{eq:improvgen}. 
If we take~$\tilde c=-1/4$, we therefore obtain the Weyl-improved energy-momentum-news complex of Section~\ref{ssec:bulk-improvements-weyl-4d}.
Another way of saying this is that we replace $Q$ in the intrinsic counterterm~\eqref{eq:intct} with $Q+\frac{1}{4}S\mathcal{F}^2$.

To summarize,
for $d=2$,
we have shown that if we define
\begin{align}
\begin{split}
        S'_{\text{norm}} & =  \int_\Sigma d^3\xi\sqrt{-g}N^2\nabla_M V^M=-\int_{r=\Lambda} d^{3}x E\left(2r^{-1} S +S\partial_r\beta\right)\,,\\
    S'_\text{int}
  & =  -\int_{r = \Lambda} d^{3}x\,E\, r\left(Q+\frac{1}{4}S\mathcal{F}^2\right)\,,
\end{split}
\end{align}
then the following total variation,
evaluated on shell,
\begin{align} &\delta\left(S_{\text{EH}}+S_{\text{ext}}+S'_{\text{norm}}+S'_{\text{int}}\right)\big\vert_{\text{os}}
  \nonumber
  \\
  &{}\qquad
  =\cdots+2\int_{\mathcal{I}^+}d^3 xe\left(T_{\text{W}}^\mu\delta\tau_\mu+\frac{1}{2}T_{\text{W}}^{\mu\nu}\delta h_{\mu\nu}+\frac{1}{2}S_{\text{W}}^{\mu\nu}\delta C_{\mu\nu}\right)\,,
  \label{eq:variationosd=2}
\end{align}
reproduces the Weyl-covariant EMT-news complex as given in~\eqref{eq:WeylImprovedEMTNewsComplex}.

\subsection{Including logarithmic contributions}
\label{ssec:on-shell-action-logs}

So far, the analysis in this section has been performed for a radial expansion without logs. As we have done previously, here we will separately discuss what changes when we include the $\log r$ terms of Section~\ref{ssec:radial-expansion-logs}. In Section~\ref{ssec:log-emt-news}, we showed that including log terms does not lead to new terms in the EMT-news complex obtained from the equations of motion. Here, we show the same is true for the EMT-news complex extracted from the renormalised on-shell action. 

We saw in Equation~\eqref{eq:ren-EMT-N-def} that the finite part of the on-shell renormalised variation gives rise to the EMT-news complex.
In what follows, we will explore how these expressions change in the presence of logarithms.
From Section~\ref{ssec:radial-expansion-logs}, we know that 
\begin{equation}
    \os{3,1}{\beta} = 0\,,\qquad \os{1,1}{S} = 0\,. 
\end{equation}
Using these along with the log-inclusive expansions from Equations~\eqref{eq:first-log-block}, \eqref{eq:second-log-block} and~\eqref{eq:logexpansionsNNLO}
as well as the explicit expressions for the responses in Equations~\eqref{eq:int-curly-Ts} and~\eqref{eq:curly-Ts}, we see that only the combination 
\begin{equation}
(\mathcal{T}^{\mu\nu} + r \mathcal{T}^{\mu\nu}_{\text{int}})    
\end{equation}
receives a relevant log correction through the term
\begin{equation}
    -2 \mathcal{Z}^{(\mu} U^{\nu)}\,.
\end{equation}
The expansion of $\mathcal{Z}_\mu$ in~\eqref{eq:NNLOinZ} is done only up to $\OO(r^{-2}\log r)$, but, as we saw in~\eqref{eq:full-curly-T-at-order-minus-4}, we also require the order-$r^{-2}$ contribution when calculating $(\mathcal{T}^{\mu\nu} + r \mathcal{T}^{\mu\nu}_{\text{int}})\big\vert_{@r^{-4}}$, and the presence of logs leads to a modification at order $r^{-2}$:
\begin{equation}
    \mathcal{Z}_\mu = \cdots + r^{-2}\log r\,\os{2,1}{\mathcal{Z}}_\mu - \frac{1}{3}r^{-2} \os{2,1}{\mathcal{Z}}_\mu + \cdots\,,
\end{equation}
where
\begin{equation}
    \os{2,1}{\mathcal{Z}}_\mu = -2h^{\nu\rho}\mathcal{D}_\nu D_{\rho\mu} = -2h^\nu_\mu(\mathcal{D}_\rho - a_\rho)D^\rho{_\nu}\,.
\end{equation}
Raising the index with $\Pi^{\mu\nu} = r^{-2}h^{\mu\nu} + \cdots$, we get
\begin{equation}
    \mathcal{Z}^\mu = \cdots + r^{-4}\log r\,\os{2,1}{\mathcal{Z}}^\mu - \frac{1}{3}r^{-4} \os{2,1}{\mathcal{Z}}^\mu + \cdots\,,
\end{equation}
so that the contribution due to logs is
\begin{equation}
\label{eq:log-cont}
    -2 \mathcal{Z}^{(\mu} U^{\nu)} = \cdots + 4r^{-4}\left(\log r-\frac{1}{3}\right) v^{(\mu}h^{\nu)\rho}(\mathcal{D}_\sigma - a_\sigma )D^\sigma{_\rho} + \cdots\,.
\end{equation}
Using the fact that, following~\eqref{eq:STF1/rSimple}, we have
\begin{equation}
    \mathcal{L}_v D_{\mu\nu} = 0\,,
\end{equation}
we can rewrite the logarithmic contribution~\eqref{eq:log-cont} in terms of the ambiguity tensor~$t^{\mu\nu}$ introduced in~\eqref{eq:constraint-response-term} with $\zeta_{\mu\nu} = -2D_{\mu\nu}$ as its parameter,
\begin{equation}
    -2 \mathcal{Z}^{(\mu} U^{\nu)} = \cdots + 2r^{-4}\left(\log r-\frac{1}{3}\right) t^{\mu\nu}\big\vert_{\zeta_{\mu\nu} = -2D_{\mu\nu}} + \cdots\,.
\end{equation}
With that, the log contributions to
\begin{equation}
    (\mathcal{T}^{\mu\nu} + r \mathcal{T}^{\mu\nu}_{\text{int}}) \delta h_{\mu\nu}\,,
\end{equation}
are total derivatives by the identity~\eqref{eq:idtmunuosvar}.
This precisely mirrors what happened in Equation~\eqref{eq:identity-with-zeta=C} above,
where the ostensibly superleading term satisfied
\begin{equation}
    (\mathcal{T}^{\mu\nu} + r\mathcal{T}^{\mu\nu}_{\text{int}})\big\vert_{@r^{-3}}\delta h_{\mu\nu} = t^{\mu\nu}\big\vert_{\zeta_{\mu\nu} = -2C_{\mu\nu}}\delta h_{\mu\nu} \,,
\end{equation}
and therefore gave a total derivative.
Thus, just as we saw in Section~\ref{ssec:log-emt-news} for the EMT-news complex derived from the Einstein equation,
adding the log terms discussed in Section~\ref{ssec:radial-expansion-logs} does not modify the EMT-news complex derived from the on-shell variation of the action.

\subsection{Carroll boost anomaly for \texorpdfstring{$d=2$}{d=2}}\label{ssec:hol-ren-fourdim-anomalies-improvements}
We have managed to obtain the Weyl improved EMT-news complex from the Einstein--Hilbert action, supplemented with a set of appropriate counterterms.
We know from the discussion of Section \ref{ssec:bulk-improvements-weyl-covariant-currents} that,
while it does satisfy the Weyl Ward identity~\eqref{eq:boundary-Weyl-WI},
this EMT-news complex does not obey the Carroll boost Ward identity as derived in~\eqref{eq:boundary-boost-WI}.
Instead, we obtain the expression in~\eqref{eq:anomCboostWI},
\begin{equation}\label{eq:Carbanom}
    h^\mu_\rho T^\rho - \left(\mathcal{D}_\rho-a_\rho\right)S^{\rho\mu} =\mathcal{A}_{\text{B}}^\mu\,,
\end{equation}
where, instead of zero, we obtain~$\mathcal{A}_{\text{B}}^\mu$ on the right-hand side,
which is given in~\eqref{eq:Cboostanom} and~\eqref{eq:Cboostanomv2} by the following expressions,
\begin{align}
\label{eq:anominanomSec}
\begin{split}
    \mathcal{A}_\text{B}^\mu & =  \frac{1}{2}h^{\mu\nu}\left(\partial_\nu+2a_\nu\right)\left(\overset{(0)}{S}-a^2\right)
    +h^\mu_\rho\left(\mathcal{L}_v-\frac{3}{2}K\right)\os{0}{P}^\rho\\
    &\qquad
    -\frac{1}{2}v^\rho\left(\partial_\rho\tilde b_\sigma-\partial_\sigma\tilde b_\rho\right)\left(C^{\sigma\mu}-F^{\sigma\mu}\right)\\
     & =  \frac{1}{2}h^{\mu\nu}\left(\partial_\nu+2a_\nu\right)\left(\overset{(0)}{S}-a^2\right)
    +\frac{1}{2}\left(\mathcal{D}_\rho-a_\rho\right)N^{\rho\mu}\\
    &\qquad
    +\frac{1}{2}v^\rho\left(\partial_\rho\tilde b_\sigma-\partial_\sigma\tilde b_\rho\right)F^{\sigma\mu}+\frac{1}{2}h^\mu_\rho\left(\mathcal{L}_v-\frac{3}{2}K\right)\mathcal{D}_\sigma F^{\sigma\rho}\,.
    \end{split}
\end{align}
A natural question is if we can perform a further improvement such that the Carroll boost Ward identity is obeyed,
in the sense that the resulting improved EMT-news complex would satisfy~\eqref{eq:Carbanom} with zero on the right-hand side.
In the following, we will not require that the Weyl Ward identity still needs to be obeyed,
but we do demand that the diffeomorphism Ward identity is obeyed.
This latter condition tells us that we can only add counterterms that are boundary covariant. 
Our insistence on diffeomorphism-invariance of the boundary theory is strongly motivated by the fact that the relevant bulk equations of motion can be written into the form of the Carroll diffeomorphism Ward identity, as we showed in Section~\ref{sec:bulk-conservation-equations}.

The object $\mathcal{A}_\text{B}^\mu$ contains shear-independent terms that are cubic in derivatives and one term that is linear in shear and second order in derivatives.
Let us focus on the former.
In order to remove these terms from~$\mathcal{A}_\text{B}^\mu$ we thus need a finite counterterm that is cubic in derivatives.
Working up to boundary terms, it turns out that all such terms are proportional to $K$,
such as $Ka^2$ or $K\mathcal{Q}$.

This can be shown as follows.
Terms constructed from the boundary geometry that are first order in derivatives are $K$, $a_\mu$ and $F_{\mu\nu}$.
There is only one term that is genuinely second order, meaning that it is not a product of first order terms, and that is $\mathcal{Q}$, which is constructed from the curvature following the discussion in Appendix~\ref{app:curvten}.
There are no such terms at cubic order, in the sense that a cubic-order term is either
a triple product of first-order derivative terms or a product of second- and first-order derivative terms.
More specifically, at first order in derivatives, we have the following tensors:
\begin{equation}
    \mathcal{O}(\partial)\quad :\quad K\,,\qquad a_\mu\,,\qquad F_{\mu\nu}\,.
\end{equation}
At second order in derivatives, we have
\begin{equation}
\begin{array}{ccccccccc}
    \mathcal{O}(\partial^2) 
    & : & Q\,,& a^2\,,& h^\rho_{\langle\mu}h^\sigma_{\nu\rangle}a_\rho a_\sigma\,,& Ka_\mu\,,& K^2\,,& KF_{\mu\nu}\,,& F^2\,,\nonumber\\
    && a^\rho F_{\rho\mu}\,,&\mathcal{L}_v K\,,& \mathcal{L}_v a_\mu\,,& \mathcal{L}_v F_{\mu\nu}\,,& \mathcal{D}_\rho a^\rho\,,& h^\rho_\mu\partial_\rho K\,,& A_{\mu\nu}\,,
\end{array}
\end{equation}
where we organised the terms in irreducible spatial tensors. 
At second order, we did not consider objects with three free spatial indices.
This is because these are not needed to construct scalars at cubic order.
At cubic order,
up to total derivatives,
we can construct the following scalars,
\begin{equation}\label{eq:cubiccts}
    \mathcal{O}(\partial^3)
    \quad:\quad
    K\mathcal{Q}\,,\qquad Ka^2\,,\qquad K^3\,,\qquad KF^2\,,\qquad K\mathcal{D}_\rho a^\rho\,.
\end{equation}
These are the cubic local covariant counterterms we can add to the on-shell action.
As we claimed above, we see that they are all proportional to $K$.

The energy current $T^\mu$ is the response to $\tau_\mu$ variations.
As we can see from~\eqref{eq:anominanomSec}, the expression for $\mathcal{A}_B^\mu$ contains some terms that are independent of $K$.
If we compute the variation of a putative improvement-generating term of the form
\begin{equation}
  \int d^3x eKF\,,
\end{equation}
where $F$ is some scalar function,
we find that all terms in the response to $\tau_\mu$ variations are proportional to $K$.
This follows from the expression for~$\delta K$ in~\eqref{eq:varK} and the variation of $e$ given in~\eqref{eq:app-integration-measure-variation}.
As a result, we conclude that none of the counterterms in~\eqref{eq:cubiccts} can remove the right-hand side of~\eqref{eq:Carbanom}.
Since it cannot be removed with improvement transformations,
we therefore call $\mathcal{A}_B^\mu$ a Carroll boost anomaly.

Using our variational definition of the EMT-news complex in~\eqref{eq:variationosd=2} we thus have that, under a local Carroll boost with parameter $\lambda_\mu=\lambda_a e^a_\mu$,
the variation is
\begin{equation}
\delta_\lambda\left(S_{\text{EH}}+S_{\text{ext}}+S'_{\text{norm}}+S'_{\text{int}}\right)\big\vert_{\text{os}}=\cdots+\int_{\mathcal{I}^+}d^3 xe\lambda_\mu\mathcal{A}_B^\mu\,.
\end{equation}
Since the variation evaluated on shell at $\mathcal{I}^+$ does give zero for boundary Weyl transformations and for boundary diffeomorphisms,
this result is of the form we anticipated in~\eqref{eq:gaugetrafoSos} when discussing consistency conditions.

The form of the anomaly in \eqref{eq:anominanomSec} is boundary diffeomorphism-covariant and,
using the results of Section~\ref{ssec:bulk-improvements-weyl-covariant-currents},
it is easy to see that it has Weyl weight $-4$.
Under the boundary gauge transformations, we thus have  
\begin{equation}
    \label{eq:anomtrafo-repeat}
    \delta \mathcal{A}_B^\mu=\mathcal{L}_\chi \mathcal{A}_B^\mu-4\Lambda_D \mathcal{A}_B^\mu+\delta_\lambda \mathcal{A}_B^\mu\,,
\end{equation}
where $\delta_\lambda \mathcal{A}_B^\mu$ is the transformation under local Carroll boosts.
This agrees with our expectations in Equation~\eqref{eq:anomtrafo}.
We can now verify that the Carroll boost anomaly obeys the Wess--Zumino consistency condition \eqref{eq:WZcon} by a direct calculation.
If we use $\delta_1$ to denote a gauge transformation with parameters $\chi_1^\mu\,,\lambda_1^a\,,\lambda_1^{ab}=-\lambda_a^{ba}$ and $\Lambda_D^1$,
and likewise for $\delta_2$,
then we can show that 
\begin{equation}
    \left[\delta_1\,,\delta_2\right]\left(S_{\text{EH}}+S_{\text{ext}}+S'_{\text{norm}}+S'_{\text{int}}\right)\big\vert_{\text{os}}=\cdots+\int_{\mathcal{I}^+}d^3 xe\lambda^{[1,2]}_\mu\mathcal{A}_B^\mu\,,
\end{equation}
where $\lambda^{[1,2]}_\mu$ is the commutator of the boost transformation parameters we computed in~\eqref{eq:lambda3}.
This shows that Wess--Zumino consistency~\eqref{eq:WZcon} is obeyed.
In deriving this result, we used that
\begin{equation}\label{eq:propCboostanom}
    \int d^3x e e^a_\mu\left(\lambda^a_1\delta_{\lambda_2}\mathcal{A}^\mu_B-\left(1\leftrightarrow 2\right)\right)=0\,.
\end{equation}
This can be checked directly, but it is a rather lengthy calculation.
The details can be found in Appendix~\ref{app:Carrollboosts}, where we show that
\begin{equation}\label{eq:boosttrafoanom}
    \delta_\lambda\mathcal{A}_{B\,\mu}=\frac{1}{2}\lambda^\nu\left[\mathcal{L}_v N_{\mu\nu}+2h_{\langle\mu}^\rho h^\sigma_{\nu\rangle}\left(\mathcal{D}_\rho+3a_\rho\right)G_\sigma\right]\,,
\end{equation}
where we defined 
\begin{equation}
    G_\sigma=v^\alpha\left(\partial_\alpha\tilde b_\sigma-\partial_\sigma\tilde b_\alpha\right)=\mathcal{L}_v a_\sigma+\frac{1}{2}h_\sigma^\alpha\partial_\alpha K+\frac{1}{2}Ka_\sigma\,.
\end{equation}
Using this result, we have
\begin{equation}
    e^a_\mu\lambda^a_1\delta_{\lambda_2}\mathcal{A}^\mu_B=\frac{1}{2}\lambda^\mu_1\lambda^\nu_2\left[\mathcal{L}_v N_{\mu\nu}+2h_{\langle\mu}^\rho h^\sigma_{\nu\rangle}\left(\mathcal{D}_\rho+3a_\rho\right)G_\sigma\right]\,,
\end{equation}
where the term in square brackets is STF. This shows that \eqref{eq:propCboostanom} is obeyed.
For an alternative derivation of this, we refer the reader to \cite{Hartong:2025WIP2}.

\section{Discussion}
\label{sec:discussion}
In this work, we have constructed a boundary energy-momentum tensor at future null infinity in asymptotically flat spacetimes of dimensions $3$ and $4$, completing the programme laid out in~\cite{Hartong:2025jpp}.
When the spacetime is four-dimensional, the boundary energy-momentum tensor is augmented by the news, and together they form an energy-momentum-news complex.
By keeping the most general allowed boundary structure on $\scri^+$,
which is made up of a conformal Carrollian structure and the shear,
we showed that the energy-momentum-news complex is the response to the variation of the on-shell Einstein--Hilbert action with respect to the boundary structure.
Here, divergences were subtracted using a procedure that is very close in spirit to the holographic renormalisation procedure in the AdS/CFT correspondence.
The diffeomorphism Ward identity for a boundary functional turns into the Bondi loss equations for the energy-momentum-news complex, and is non-anomalous,
as is the Weyl Ward identity.
The Carroll boost Ward identity, in contrast, is anomalous in both three and four bulk spacetime dimensions.
The energy-momentum-news complex can also be obtained by recasting the $U^\mu R_{\mu\nu} = 0 $ Einstein equation at order $r^{-d}$ so that it takes the form of the boundary diffeomorphism Ward identity.
The energy-momentum-news complex is defined up to improvements, and we have explicitly exhibited the improvement transformation that relates the answers obtained from these two distinct approaches.

Our results suggest many interesting directions for further study, including:
\begin{description}
 \item[BMS algebra from EMT-news complex] 
         Now that we have an energy-momentum tensor, a notion of a conformal Carroll Killing vector and a diffeomorphism Ward identity, it should be possible to generalise equations such as \eqref{eq:charge-eqn-d1} to $d=2$ and to use these to define the BMS charges and rederive their algebra.
    \item[Carroll-covariant Newman--Unti gauge] 
        In this work,
        we introduced a Carroll-covariant generalisation of the Bondi gauge to fix all the subleading gauge redundancies
        while keeping the most general boundary conformal Carrollian structure.
        As briefly discussed in Subsection~\ref{ssec:Newman-Unti},
        choosing Newman--Unti gauge replaces the `determinant condition'~\eqref{eq:bondi-gauge-condition} with the requirement that $\beta = \mathcal{O}(r^{-1})$,
        which leads to a Carroll-covariant Newman--Unti gauge.
        On shell, in the absence of matter,
        we get $\beta=0$.
        In particular, as we demonstrated in~\eqref{eq:NU-affine},
        this implies that integral curves of $V^M$ are affinely parametrised null geodesics.
        Letting $x^M(\lambda)$ denote such an integral curve, we have
        \begin{equation}
            \frac{d x^M}{d\lambda} = V^M = -(\D_r)^M\Longrightarrow \frac{d r}{d\lambda} = -1\,,\quad\text{and}\quad \frac{d x^\mu}{d\lambda} = 0\,,         
        \end{equation}
        so that $\lambda = -r + \text{constant}$.
        In other words, $-r$ is an affine parameter for the null geodesics generated by $V^M$.
        Crucially, as argued in~\cite{Geiller:2022vto,Geiller:2024amx},
        and as discussed in Subsection~\ref{ssec:Newman-Unti},
        the trace $\os{-1}{g}_{\mu\nu}$ with respect to $h^{\mu\nu}$ does not automatically vanish in the Carroll-covariant Newman--Unti gauge,
        although there is enough residual gauge symmetry to remove it.
        In~\cite{Geiller:2022vto}, it has been argued that if we do not fix the radial diffeomorphisms
        (which leads to the partial Bondi gauge),
        we can end up with a larger  asymptotic charge algebra.
        It would be interesting to repeat our construction including the trace of~$\os{-1}{g}_{\mu\nu}$ and to see if the findings of \cite{Geiller:2022vto} can be understood from the boundary EMT-news complex and its associated charges. 
        From a computational perspective,
        because $-r$ is affine along $V^M$,
        the Newman--Unti gauge is often better suited for a treatment using the Newman--Penrose formalism~\cite{Newman:1962cia,Barnich:2011ty}, which we discuss next.

    \item[Relation to the Newman--Penrose formalism]
        As first devised in~\cite{Newman:1961qr} (see~\cite{Newman:2009} for a review), the Newman--Penrose formalism for a four-dimensional spacetime is based on a complex `doubly-null' tetrad $(U^M,V^M,\texttt{m}^M,\bar{\texttt{m}}^M)$, where the inverse nullbeine $(U^M,V^M)$ appearing in the metric~\eqref{eq:gen-metric-null-frame-decomposition} are augmented by the complex null dyad
        \begin{equation}
            \texttt{m}^M = \frac{1}{\sqrt{2}}(E_1^M + i E_2^M)\,,\qquad \bar{\texttt{m}}^M = \frac{1}{\sqrt{2}}(E_1^M - i E_2^M)\,,
        \end{equation}
        satisfying the conditions
        \begin{equation}
            g(\texttt{m},\texttt{m}) = g(\bar{\texttt{m}},\bar{\texttt{m}}) = 0\,,\qquad g(\texttt{m},\bar{\texttt{m}}) = +1\,.
        \end{equation}
        In addition to $12$ spin coefficients, which encode connection components, curvature is encoded in five complex Weyl scalars~$\Psi_0,\dots,\Psi_4$ and Ricci scalars~$\Phi_{ij}$ and~$\Lambda$.
        Rather than the tensorial equations we considered in Sections~\ref{sec:rewriting-EE} and~\ref{sec:radial-expansion}, Einstein's equations (and the Bianchi identities) turn into a set of $30$ scalar equations in Newman--Penrose variables.
        For our purposes, the Newman--Penrose formalism is particularly well suited to describe the Bondi loss equations at $\scri^+$,
        with the radiative degrees of freedom captured by the leading piece of one of the spin coefficients (known as $\sigma$) and $\Psi_4$.
        For this reason, it would be interesting to recast our results using Newman--Penrose variables, which would also facilitate comparison with recent works such as~\cite{Ruzziconi:2025fuy,Geiller:2025dqe}.
        
    \item[Variational principle and the gravitational $S$-matrix]
        In the AdS/CFT correspondence,
        the renormalised on-shell boundary action with Dirichlet boundary conditions becomes a generating functional for CFT correlation functions,
        see for example~\eqref{eq:os-variation-intro}. 
        To properly define scattering using a Carrollian partition function, Refs.~\cite{Kim:2023qbl,Kraus:2024gso,Isen:2026xoc} used the Arefeva--Faddeev--Slavnov (AFS) formalism
        to construct generating functionals
        that depend on data on both $\scri^+$ and $\scri^-$
        for massless scalar fields, scalar QED and gravity.
        The boundary conditions that allow for the computation of $S$-matrix elements fix the positive frequency part of the massless propagating field on $\scri^-$
        and the negative frequency part on $\scri^+$.
        Very recently, a Hamilton--Jacobi approach to holographic renormalisation for scalar fields was developed in~\cite{Ammon:2025avo} and was shown to match the AFS generating functional.
        It would be very interesting to use these methods in the context of gravity, and combine the AFS formalism with the renormalisation procedure devised in this work.

    \item[Other boundary components of asymptotically flat spacetimes] 
    Finally, as we already remarked above, asymptotically flat spacetimes have multiple boundary components, namely $\scri^\pm$ as well as $i^\pm$ and $i^0$.
    In generic asymptotically flat spacetimes, the relevant asymptotic fields typically have direction-dependent limits at $i^0$.
    Ashtekar and Hansen resolved this by `blowing up' the point $i^0$ to construct a four-dimensional space they named Spi~\cite{Ashtekar:1978zz}. The space of spatial directions at $i^0$ is the three-dimensional unit hyperboloid $dS_3$, and the space Spi is a line bundle over this space of directions. 
    Gibbons later demonstrated that the space Spi has a type of Carrollian structure~\cite{Gibbons:2019zfs},
    while it was shown in~\cite{Figueroa-OFarrill:2021sxz} that the blow-up of $i^\pm$ is the Carrollian analogue of AdS (see also~\cite{Have:2024dff,Borthwick:2024skd}).
    In particular, the paper~\cite{Figueroa-OFarrill:2021sxz} showed that all the boundaries of flat spacetime have a Carrollian structure: the Carrollian analogue of anti-de Sitter space for $i^\pm$, and a pseudo-Carrollian geometry that looks like $\RR\times dS_3$ in the case of $i^0$ (see~\cite{Compere:2011ve,Capone:2022gme,Fuentealba:2022xsz,Girelli:2026gbr} for recent work generalising the original Beig--Schmidt expansion~\cite{Beig:1982ifu,Beig:1984bla} near $i^0$).
    A full holographic treatment of asymptotically flat spacetimes must also take these other boundary components into account,
    and, while this is beyond the scope of this work,
    we expect that Carrollian boundary geometry again will play a crucial role.
    In particular, for a globally well-defined variational principle,
    we would need to consider all boundary components with appropriate boundary conditions, boundary terms and corner terms~\cite{Hayward:1993my,Lehner:2016vdi}.

    \item[Limit from AdS]
        Considering flat limits of the AdS/CFT correspondence has a long and illustrious history, starting with~\cite{Susskind:1998vk,Polchinski:1999ry,Giddings:1999jq}.
        An obvious question is whether there exists a limit where the cosmological constant goes to zero such that,
        on the one hand,
        the solution space becomes the one we considered in this paper
        and, on the other, the holographically renormalised action we derived arises from the limiting procedure.
        A natural starting point would be the Bondi gauge for (A)dS space developed in~\cite{Poole:2018koa,Compere:2019bua}. 
\end{description}

\subsection*{Acknowledgements}
We are grateful to Federico Capone, Luca Ciambelli, Jos\'e Figueroa-O'Farrill, Laurent Freidel, Marc Geiller, Lionel Mason, Kévin Nguyen, Niels Obers, Romain Ruzziconi and C\'eline Zwikel for useful discussions.
The work of JH is and the work of GO was supported by the Royal Society URF Research Fellows Enhanced Research Expenses 2022 (RF\textbackslash ERE\textbackslash 221013).
The work of EH is supported by Carlsberg Foundation grant CF24-1656.
The Center of Gravity is a Center of Excellence funded by the Danish National Research Foundation under grant No.~184.
The work of GO is supported by the European Research Council (ERC) grant TraBHolo/101222495.
Views and opinions expressed are however those of the authors only and do not necessarily reflect those of the European Union or the European Research Council Executive Agency. Neither the European Union nor the granting authority can be held responsible for them.

\appendix

\newpage
\section{Conventions and useful identities}
\label{app:conventions}

Our starting point is a $(d+2)$-dimensional bulk geometry,
which we describe using coordinates $x^M$ labelled by capital indices.
Early on in Section~\ref{sec:bulk-geom},
we single out a radial coordinate $r$ and split
\begin{equation}
  x^M
  = (r,x^\mu)\,.
\end{equation}
The remaining $d+1$ coordinates $x^\mu$ describe equal $r$ surfaces and are labelled by Greek indices.
In particular, as $r\to\infty$ for fixed $x^\mu$, they describe future null infinity.
We use
\begin{equation}
  X_{(\mu\nu)}
  = \frac{1}{2} (X_{\mu\nu}+X_{\nu\mu})\,,
  \qquad
  X_{[\mu\nu]}
  = \frac{1}{2} (X_{\mu\nu}-X_{\nu\mu})\,,
\end{equation}
to denote symmetrisation and antisymmetrisation, respectively.

\subsection{General curvature conventions and identities}
\label{sapp:conventions-curvature}
We use the mostly plus sign convention for the bulk spacetime metric.
We use both $(d+2)$-dimensional and $(d+1)$-dimensional curvature tensors.
For ease of notation, we will discuss them only in the latter case,
which is described using $x^\mu$ coordinates.
Our sign conventions are of course the same in both cases.

In this subsection,
we use $\Gamma^\rho_{\mu\nu}$ and $D_\mu$
to denote the components of a fully general affine connection
and its associated covariant derivative.
Its Riemann tensor $R_{\mu\nu\rho}{}^\sigma$ and torsion $T^\sigma{}_{\mu\nu}$ are defined by the following Ricci identities
\begin{align}
  \label{eq:app-general-torsion-def}
  \left[D_\mu, D_\nu\right] f
  &= - T^\rho{}_{\mu\nu} \pd_\rho f\,,
  \\
  \label{eq:app-general-riemann-vector-def}
  \left[D_\mu, D_\nu\right] W^\rho
  &= - R_{\mu\nu\sigma}{^\rho}W^\sigma
  - T^\sigma{}_{\mu\nu} D_\sigma W^\rho\,,
  \\
  \label{eq:app-general-riemann-form-def}
  \left[D_\mu, D_\nu\right] X_\rho
  &=  R_{\mu\nu\rho}{}^\sigma X_\sigma
  - T^\sigma{}_{\mu\nu}D_\sigma X_\rho\,,
\end{align}
where $f$, $W^\mu$ and $X_\mu$ are arbitrary functions,
vector fields and one-forms.
These conventions correspond to
\begin{align}
  \label{eq:app-general-torsion-cpts-def}
  T^\rho{ }_{\mu \nu}
  &= 2 \Gamma_{[\mu \nu]}^\rho\,,
  \\
  \label{eq:app-general-riemann-cpts-def}
  R_{\mu \nu \sigma}{ }^\rho
  &=-\partial_\mu \Gamma_{\nu \sigma}^\rho
  +\partial_\nu \Gamma_{\mu \sigma}^\rho
  -\Gamma_{\mu \lambda}^\rho \Gamma_{\nu \sigma}^\lambda
  +\Gamma_{\nu \lambda}^\rho \Gamma_{\mu \sigma}^\lambda\,.
\end{align}
By definition, the Riemann tensor 
$R_{\mu\nu\sigma}{}^\rho$ is antisymmetric in its first two indices.
It also obeys the algebraic and differential Bianchi identities,
\begin{align}
  \label{eq:app-general-riemann-alg-bianchi}
  R_{[\mu \nu \sigma]}{ }^\rho
  &= T^\lambda{ }_{[\mu \nu} T_{\sigma] \lambda}^\rho
  - D_{[\mu} T_{\nu \sigma]}^\rho\,,
  \\
  \label{eq:app-general-riemann-dif-bianchi}
  D_{[\lambda} R_{\mu \nu] \sigma}{ }^\rho
  &= T^\kappa{ }_{[\lambda \mu} R_{\nu] \kappa \sigma}{ }^\rho\,.
\end{align}
For the purposes of this subsection, we let $E$ denote a general integration measure
(corresponding for example to the square root of the determinant of a non-degenerate Riemannian or Lorentzian metric).
We can always write
\begin{equation}
  \Gamma^\rho_{\rho\mu}
  \label{eq:app-general-affine-connection-trace}
  = E\inv\pd_\mu E + B_\mu\,,
\end{equation}
where $B_\mu$ is a one-form.
The covariant derivative of a weight~$w$ tensor density is
\begin{equation}
  \label{eq:app-general-cov-deriv-of-tensor-density}
  D_\rho X^\mu{}_\nu
  = \pd_\rho X^\mu{}_\nu
  + \Gamma^\mu_{\rho\sigma} X^\sigma{}_\nu
  - \Gamma^\sigma_{\rho\nu} X^\mu{}_\sigma
  - w \Gamma^\sigma_{\sigma\rho} X^\mu{}_\nu\,,
\end{equation}
which generalises linearly to other tensor densities.
Since $E$ transforms as a scalar density of weight $+1$,
its covariant derivative is
\begin{equation}
  \label{eq:app-general-cov-deriv-of-measure}
  D_\rho E
  = \pd_\rho E
  - \Gamma^\sigma_{\sigma\rho} E
  = - B_\rho E\,.
\end{equation}
If $B_\rho$ is zero for a given connection,
we therefore say this connection is volume-form compatible.
The covariant divergence of a (weight zero) vector field $X^\mu$ is then
\begin{equation}
  \label{eq:app-general-cov-divergence}
  D_\rho X^\rho
  = \pd_\rho X^\rho
  + \Gamma^\rho_{\rho\sigma} X^\sigma
  = E\inv \pd_\rho \left(E X^\rho\right)
  + B_\rho X^\rho\,.
\end{equation}
Hence, when $B_\rho=0$, we can perform partial integrations more easily.

The Ricci tensor is defined as follows,
\begin{equation}
  \label{eq:app-general-ricci-def}
  R_{\mu\nu}
  = R_{\mu \rho \nu}{ }^\rho.
\end{equation}
Note that the Ricci tensor is generically not symmetric.
Instead, equation \eqref{eq:app-general-riemann-alg-bianchi} tells us that
\begin{align}
  \label{eq:app-general-ricci-antisymmetric-part}
  2 R_{[\mu\nu]}
  &= R_{\mu\nu\rho}{}^\rho
  + D_\rho T^\rho{}_{\mu\nu}
  + 2D_{[\mu} T^\rho{}_{\nu]\rho}
  - T^\rho{}_{\mu\nu} T^\sigma{}_{\sigma\rho}\nonumber\\
  &=- \pd_\mu \Gamma_{\rho\nu}^\rho
  + \pd_\nu \Gamma_{\rho\mu}^\rho+ D_\rho T^\rho{}_{\mu\nu}\,,
\end{align}
where we used
\begin{equation}
  \label{eq:app-general-riemann-three-four-trace}
  R_{\mu\nu\rho}{}^\rho
  = - \pd_\mu \Gamma_{\nu\rho}^\rho
  + \pd_\nu \Gamma_{\mu\rho}^\rho
  = - \pd_\mu \Gamma_{\rho\nu}^\rho
  + \pd_\nu \Gamma_{\rho\mu}^\rho-2D_{[\mu}T^\rho_{\nu]\rho}+T^\sigma_{\mu\nu}T^\rho_{\rho\sigma}\,.
\end{equation}
The second equality for $R_{[\mu\nu]}$ also follows directly from \eqref{eq:app-general-riemann-cpts-def}. We thus see that there are two contributions to the antisymmetric part. One comes from torsion and the other from volume-form incompatibility.
We will always use torsion-free and volume-form compatible connections,
so the corresponding Ricci tensor will be symmetric and the contraction $R_{\mu\nu\rho}{}^\rho$ vanishes.

\subsection{Lie derivatives}
\label{sapp:conventions-lie-derivative}
We will consider Lie derivatives both with respect to bulk vector fields $\xi^M$
as well as vector fields $\xi^\mu$ on equal-$r$ surfaces.
To avoid any possible confusion,
we emphasise that for a bulk tensor $X^M{}_N$ which is expressed in terms of the $(d+2)$-dimensional bulk coordinates $x^M$, the Lie derivative with respect to $\xi^M$ is given by
\begin{equation}
  \label{eq:app-lie-derivative-bulk}
  \LL_\xi X^M{}_N
  = \xi^R \pd_R X^M{}_N
  + X^M{}_R \pd_N \xi^R
  - X^R{}_N \pd_R \xi^M\,.
\end{equation}
On the other hand,
the Lie derivative of an equal-$r$ hypersurface tensor $X^\mu{}_\nu$
with respect to a hypersurface vector field $\xi^\mu$ is
\begin{equation}
  \label{eq:app-lie-derivative-boundary}
  \LL_\xi X^\mu{}_\nu
  = \xi^\rho \pd_\rho X^\mu{}_\nu
  + X^\mu{}_\rho \pd_\nu \xi^\rho
  - X^\rho{}_\nu \pd_\rho \xi^\mu\,.
\end{equation}
The former preserves $(d+2)$-dimensional bulk covariance,
while the latter is covariant on $(d+1)$-dimensional equal-$r$ hypersurfaces.
After splitting $x^M=(r,x^\mu)$,
the former includes $\xi^r$ terms,
while the latter does not.
We often need to rewrite Lie derivatives in terms of covariant derivatives and vice versa. Here, it is again helpful to use a connection without torsion as we can then simply replace the partial derivatives with covariant derivatives.

We will use the convention that the notation $\mathcal{L}_\xi X^\mu{}_\nu(r,x)$ denotes a $(d+1)$-dimensional Lie derivative 
along $\xi^\mu$ of a hypersurface tensor $X^\mu{}_\nu(r,x)$. In other words, it is not to be read  as the hypersurface components of $\mathcal{L}_\xi X^M{}_N$ which is taken along $\xi^P$.
Said differently, when writing $\mathcal{L}_\xi X^M{}_N$ and $\mathcal{L}_\xi X^\mu{}_\nu$, the former is a Lie derivative along $\xi^P$ and the latter along $\xi^\rho$.

\subsection{Spatial tensors}
\label{sapp:spatial-tensors}
As summarised in Appendix~\ref{app:carroll-geometry} below,
the $(d+1)$-dimensional boundary geometry
is described by a vector field (loosely speaking, the covariant version of the direction of retarded time along $\mathcal{I}^+$), a clock one-form (subject to a gauge ambiguity)
and a degenerate rank~$d$ metric $h_{\mu\nu}$ (so that its signature is $(0,1,\ldots,1)$) and its `inverse'~$h^{\mu\nu}$,
which correspond to the spatial directions.
They can be used to construct a spatial projection operator,
\begin{equation}
  \label{eq:app-spatial-projection-operator-boundary}
  h^\mu_\nu
  = h^{\mu\rho} h_{\rho\nu}\,,
\end{equation}
and any tensor that is invariant under this projection is said to be spatial.
In this paper, we will mainly be interested in $d=1$ and $d=2$,
corresponding to a three-dimensional and four-dimensional bulk.
In these cases, some useful identities hold for spatial tensors,
which we collect in the following.

Note that we can alternatively describe the geometry of general equal-$r$ surfaces
using the $r$-dependent degenerate rank $d$ tensors $\Pi^{\mu\nu}$ and $\Pi_{\mu\nu}$,
again complemented with an ($r$-dependent) time vector field and a clock one-form.
They can similarly be used to define a
$\Pi^\mu_\nu = \Pi^{\mu\rho} \Pi_{\rho\nu}$
spatial projection operator.
At leading order, these tensors reduce to the boundary objects above.
At general $r$, the resulting notion of spatial tensors is different,
but it should be clear from the context which one is meant.
Regardless, this $r$-dependent notion of spatial tensors likewise leads to equivalent $d=1,2$ properties after appropriate substitutions have been made.
In the following discussion, we will stick to boundary variables.

\paragraph{Raising, lowering and contractions.}
Consider a spatial tensor $X^\mu{}_\nu$,
which is invariant under projections by~\eqref{eq:app-spatial-projection-operator-boundary},
so that
\begin{equation}
  \label{eq:app-spatial-tensor-def}
  X^\mu{}_\nu
  = h^\mu_\rho h_\nu^\sigma
  X^\rho{}_\sigma
  = h^\mu_\rho
  X^\rho{}_\nu
  = h_\nu^\sigma
  X^\mu{}_\sigma\,.
\end{equation}
Since there is no non-degenerate metric on $\mathcal{I}^+$,
we are not able to raise and lower indices on generic tensors.
However, we can unambiguously define
\begin{equation}
  \label{eq:app-spatial-tensor-raise-lower-def}
  X^{\mu\nu}
  = X^\mu{}_\sigma h^{\sigma\nu}\,,
  \qquad
  X_{\mu\nu}
  = h_{\mu\rho} X^\rho{}_\nu\,,
\end{equation}
for spatial tensors.
This convention is almost universally implemented in the main text
since it leads to more compact expressions.
Additionally,
if $Y^\mu_\nu$ is another spatial tensor,
we will use the abbreviations
\begin{equation}
  \label{eq:app-spatial-tensor-contractions-def}
  X^2
  = X^{\mu\nu} X_{\mu\nu}\,,
  \qquad
  X \cdot Y
  = X^{\mu\nu} Y_{\mu\nu}\,,
\end{equation}
for contractions of spatial tensors.
Of course, these expressions generalise linearly to higher-rank spatial tensors.

\paragraph{STF tensors.}
The spatial trace of a general tensor~$X^\mu{}_\nu$ is
\begin{equation}
  \label{eq:app-spatial-trace-def}
  h^{\mu\nu} X_{\mu\nu}\,.
\end{equation}
If $X^\mu{}_\nu$ is spatial,
we will often also just call this the trace of the tensor.
Even if it is not spatial,
we can define its spatial symmetric trace-free (STF) part as
\begin{equation}
  \label{eq:app-spatial-STF-def}
  X_{\langle \mu\nu \rangle}
  = h_{(\mu}^\rho h_{\nu)}^\sigma X_{\rho\sigma}
  - \frac{1}{d} h_{\mu\nu} h^{\rho\sigma} X_{\rho\sigma}\,.
\end{equation}
The resulting expression is spatial by construction,
and its (spatial) trace vanishes.

\paragraph{Special properties in three bulk dimensions.}
With $d=1$,
corresponding to three bulk dimensions and two boundary dimensions,
we have only a single spatial boundary direction.
As a result, all antisymmetric spatial tensors
and spatial symmetric trace-free tensors are zero.
For example,
in $d=1$, the (antisymmetric) twist tensor~$F_{\mu\nu}$ and the (STF) shear tensor~$C_{\mu\nu}$ both vanish,
\begin{equation}
  \label{eq:app-d1-twist-shear-zero}
  F_{\mu\nu}
  = 0\,,
  \qquad
  C_{\mu\nu}
  = 0\,.
\end{equation}
This greatly simplifies many computations in three bulk spacetime dimensions.

\paragraph{Special properties in four bulk dimensions.}
On the other hand, with $d=2$,
the twist $F_{\mu\nu}$ and the shear $C_{\mu\nu}$ are of course nonzero.
However, products of such tensors do satisfy useful identities.
Let $X_{\mu\nu}$ and $Y_{\mu\nu}$ be spatial symmetric tensors,
and let $Z_{\mu\nu}$ be a spatial antisymmetric tensor.
Using a vielbein decomposition
\begin{equation}
  \label{eq:app-d2-spatial-frame-decomposition}
  h_{\mu\nu}
  = \delta_{ab} e^a_\mu e^b_\nu\,,
  \qquad
  X_{\mu\nu}
  = X_{ab} e^a_\mu e^b_\nu\,,
  \qquad
  a=1,2\,,
\end{equation}
following~\eqref{eq:app-spatial-vielbein-def} below,
and similar for $Y_{\mu\nu}$ and $Z_{\mu\nu}$,
we have
\begin{equation}
  \label{eq:app-d2-spatial-STF-antisym-Pauli-decomposition}
  X_{\mu\nu}
  = \left(
    X_{11} \sigma^3_{ab}
    + X_{12} \sigma^1_{ab}
  \right) e^a_\mu e^b_\nu\,,
  \qquad
  Z_{\mu\nu}
  = Z_{12} i\sigma^2_{ab} e^a_\mu e^b_\nu\,,
\end{equation}
where $\sigma^i_{ab}$ are the Pauli matrices.
From this, it is easy to see that
\begin{subequations}
  \label{eq:app-d2-spatial-STF-and-antisym-product-rules}
  \begin{align}
    \label{eq:app-d2-spatial-STF-STF-product}
    X_\mu{}^\rho Y_{\rho\nu}
    + Y_\mu{}^\rho X_{\rho\nu}
    &= h_{\mu\nu} X \cdot Y\,,
    \\
    \label{eq:app-d2-spatial-STF-antisym-product}
    X_\mu{}^\rho Z_{\rho\nu}
    -  X_{\nu}{}^\rho Z_{\rho\mu}
    &= 0\,,
    \\
    \label{eq:app-d2-spatial-antisym-antisym-product}
    Z_\mu{}^\rho Z_{\rho\nu}
    &= - \frac{1}{2} h_{\mu\nu} Z^2\,.
  \end{align}
\end{subequations}
Here, we have used the conventions~\eqref{eq:app-spatial-tensor-contractions-def} above for multiplying spatial tensors.

Additionally,
it is useful to note that we have
\begin{equation}
  \label{eq:app-d2-three-spatial-metrics-triply-antisymmetrized-vanishes}
  h_{\mu[\rho} h_{\sigma][\kappa} h_{\lambda]\nu}
  = h_{\nu[\rho} h_{\sigma][\kappa} h_{\lambda]\mu}\,,
\end{equation}
which means that the antisymmetrisation of a product of three degenerate spatial metric tensors in $d=2$ vanishes if any three pairs of indices are antisymmetrised.
This follows from the standard identities for Levi-Civita symbols as follows,
\begin{equation}
  \eps_{ab}\eps_{cd}
  = 2 \delta_{a[c} \delta_{d]b}
  \qiq
  2 h_{\mu[\rho} h_{\sigma]\nu}
  = \eps_{ab} \eps_{cd}
  \,
  e^a_\mu e^b_\nu e^c_\rho e^d_\sigma\,.
\end{equation}
Antisymmetrising the left-hand side of~\eqref{eq:app-d2-three-spatial-metrics-triply-antisymmetrized-vanishes} in its outermost indices then gives
\begin{align}
  \eps^{ab} e_a^\mu e_b^\nu
  h_{\mu[\rho} h_{\sigma][\kappa} h_{\lambda]\nu}
  &= 2 \eps_{cd} \eps^{ab} \eps_{a[k} \delta_{l]b}
  \,
  e^a_\mu e^b_\nu e^c_\rho e^d_\sigma e^k_\kappa e^l_\lambda
  \\
  &= \eps_{cd} \left(\delta_{kl} - \delta_{lk}\right)
  e^a_\mu e^b_\nu e^c_\rho e^d_\sigma e^k_\kappa e^l_\lambda
  = 0\,.
\end{align}
We will frequently use the identity~\eqref{eq:app-d2-three-spatial-metrics-triply-antisymmetrized-vanishes} with three of its indices raised,
\begin{equation}
  \label{eq:d=2projid}
  h^\mu_{[\alpha}h^\nu_{\beta]}h^\rho_\gamma
  +h^{\mu\nu}h^\rho_{[\alpha}h_{\beta]\gamma}
  =h^{\nu\rho}h^\mu_{[\alpha}h_{\beta]\gamma}
  +h^\mu_\gamma h^\rho_{[\alpha}h^\nu_{\beta]}\,,
\end{equation}
which is often useful when contracted with particular spatial tensors.
We will give specific examples involving covariant derivatives in Section \ref{subsec:speciald=2id}.

\section{Boundary conformal Carrollian geometry}
\label{app:carroll-geometry}

In this appendix, we give a self-contained overview of Carrollian geometry in arbitrary dimensions.
We also collect useful properties of the connection we choose to use to describe the conformal Carrollian geometry at null infinity, including many identities involving the associated curvature and covariant derivative.
Finally, we review the construction of Weyl-covariant objects and list several variational results.

\subsection{Metric data and gauge transformations}\label{subsec:Carrolldata}

The metric data (or structure) of a $(d+1)$-dimensional Carrollian geometry is defined by a degenerate symmetric tensor $h_{\mu\nu}$ of rank $d$ and a vector field $v^\mu$ which spans the kernel of the degenerate metric.
The former is a degenerate spatial metric,
while the latter parametrises the time direction.
These tensors can be complemented by a clock one-form $\tau_\mu$ and an inverse spatial metric $h^{\mu\nu}$,
which transform under local Carroll boosts.
Since we consider a conformal class of boundary Carrollian geometry at null infinity,
they also transform under boundary Weyl transformations,
\begin{subequations}
  \label{eq:app-v-tau-h-h-transformations}
  \begin{align}
    \delta v^\mu
    &= \LL_\chi v^\mu
    - \Lambda_D v^\mu\,,
    \\
    \delta \tau_\mu
    &= \LL_\chi \tau_\mu
    + \Lambda_D \tau_\mu
    + \lambda_a e^a_\mu\,,
    \\
    \delta h_{\mu\nu}
    &= \LL_\chi h_{\mu\nu}
    + 2 \Lambda_D h_{\mu\nu}\,,
    \\
    \delta h^{\mu\nu}
    &= \LL_\chi h^{\mu\nu}
    - 2 \Lambda_D h^{\mu\nu}
    + 2 \lambda^{(\mu} v^{\nu)}\,.
  \end{align}
\end{subequations}
We derived these objects and their gauge transformations from a partial gauge fixing of $(d+2)$-dimensional bulk metric degrees of freedom in Section~\ref{sec:bulk-geom}.
Here, we use $\chi^\mu(x)$ to parametrise diffeomorphisms, and
$\Lambda_D(x)$ parametrise Weyl transformations.
Furthermore, the local Carroll boost parameter $\lambda_\mu(x)$ is spatial
in the sense that $v^\mu\lambda_\mu$ vanishes.
We can therefore consistently raise and lower its indices
using
$\lambda^\mu= h^{\mu\nu} \lambda_\nu$,
as discussed in Appendix~\ref{sapp:spatial-tensors}.

\paragraph{Spatial vielbeine.}
Additionally, we occasionally use spatial vielbeine~$e^a_\mu$
and inverse spatial vielbeine~$e_a^\mu$
to factorise the degenerate rank $d$ spatial metrics,
\begin{equation}
  \label{eq:app-spatial-vielbein-def}
  h_{\mu\nu}
  = \delta_{ab} e^a_\mu e^b_\nu\,,
  \qquad
  h^{\mu\nu}
  = \delta^{ab} e_a^\mu e_b^\nu\,.
\end{equation}
This allows us to write the boost parameter $\lambda_\mu = \lambda_a e^a_\mu$ with local frame indices.
In addition to the above,
these fields transform under local $\mathfrak{so}(d)$ rotations with parameters $\lambda^{ab}=-\lambda^{ba}$,
so that
\begin{subequations}
  \label{eq:app-spatial-vielbein-transformations}
  \begin{align}
    \delta e^a_\mu
    &= \LL_\chi e^a_\mu
    + \Lambda_D e^a_\mu
    + \lambda^a{}_b e^b_\mu\,,
    \\
    \delta e_a^\mu
    &= \LL_\chi e_a^\mu
    - \Lambda_D e^\mu_a
    + \lambda_a v^\mu
    - \lambda^b{}_a e^\mu_b\,.
  \end{align}
\end{subequations}
We raise and lower tangent space indices $a,b,\ldots$ with the Kronecker delta.
Throughout most of the main text,
we work with the spatial metric instead of the spatial vielbeine,
so that all objects are manifestly invariant under local rotations.

\paragraph{Orthogonality and variations.}
The set of variables $(\tau_\mu, h_{\mu\nu})$
fully specify the Carroll metric data,
and the inverse variables $(v^\mu, h^{\mu\nu})$
are determined from them by the orthogonality conditions
\begin{equation}
  \label{eq:app-car-boundary-orthogonality-completeness}
  \begin{gathered}
    -1
    = v^\mu \tau_\mu\,,
    \qquad
    0
    = v^\mu h_{\mu\nu}\,,
    \qquad
    0
    = \tau_\mu h^{\mu\nu}\,,
    \qquad
    \delta^\mu_\nu
    = - v^\mu \tau_\nu + h^\mu_\nu\,.
  \end{gathered}
\end{equation}
Here,
$h^\mu_\nu = h^{\mu\rho} h_{\rho\nu}$
is the spatial projection operator,
and
$-v^\mu \tau_\nu$
is the projection operator on the time directions.
Using a general variation of the relations~\eqref{eq:app-car-boundary-orthogonality-completeness},
we see that we can translate between the variations of either set of variables,
\begin{subequations}
  \label{eq:app-variation-relations}
  \begin{align}
    \label{eq:app-variation-relations-up-to-down}
    \delta h^{\mu\nu}
    &= 2 v^{(\mu} h^{\nu)\rho} \delta \tau_\rho
    - h^{\mu\rho} h^{\nu\sigma} \delta h_{\rho\sigma}\,,
    &
    \delta v^\mu
    &= v^\mu v^\rho \delta \tau_\rho
    - h^{\mu\rho} v^\sigma \delta h_{\rho\sigma}\,,
    \\
    \label{eq:app-variation-relations-down-to-up}
    \delta h_{\mu\nu}
    &= 2 \tau_{(\mu} h_{\nu)\rho} \delta v^\rho
    - h_{\mu\rho} h_{\nu\sigma} \delta h^{\rho\sigma}\,,
    &
    \delta \tau_\mu
    &= \tau_\mu \tau_\rho \delta v^\rho
    - h_{\mu\rho} \tau_\sigma \delta h^{\rho\sigma}\,.
  \end{align}
\end{subequations}
We can also use the Carroll metric data to define an integration measure,
\begin{equation}
  \label{eq:app-integration-measure-def}
  e
  = \sqrt{
    \det\left(\tau_\mu \tau_\nu + h_{\mu\nu}\right)
  }
  = \det(\tau_\mu \,, e^a_\mu)\,.
\end{equation}
Its variation is given by
\begin{equation}
  \label{eq:app-integration-measure-variation}
  \delta e
  = e \left(
    - v^\mu \delta \tau_\mu
    + \frac{1}{2} h^{\mu\nu} \delta h_{\mu\nu}
  \right)
  = e \left(
    \tau_\mu \delta v^\mu
    - \frac{1}{2} h_{\mu\nu} \delta h^{\mu\nu}
  \right).
\end{equation}

\paragraph{Extrinsic curvature and acceleration.}
From the fundamental variables above,
we can define further Carroll-geometric objects,
including the boundary extrinsic curvature~$K_{\mu\nu}$,
its trace $K$ and the acceleration $a_\mu$,
\begin{align}
  \label{eq:app-extrinsic-curvature-trace-acceleration-def}
  K_{\mu\nu}
  = - \frac{1}{2} \LL_v h_{\mu\nu}\,,
  \qquad
  K
  = h^{\mu\nu} K_{\mu\nu}\,,
  \qquad
  a_\mu
  = \LL_v \tau_\mu\,.
\end{align}
Note that both the extrinsic curvature and the acceleration are spatial so that their $v^\mu$ contraction vanishes. Hence, we raise and lower their indices using the degenerate spatial metric and its inverse,
as discussed in Appendix~\ref{sapp:spatial-tensors}.
The trace $K$ of the extrinsic curvature is the divergence of $v^\rho$, since
\begin{equation}
  \label{eq:app-K-is-total-derivative}
  K
  = - \frac{1}{2} h^{\mu\nu} \LL_v h_{\mu\nu}
  = - e\inv \pd_\rho \left(e v^\rho\right)\,.
\end{equation}
The simple expression for $K$ in~\eqref{eq:app-K-is-total-derivative} therefore often helps when considering for example variations of the extrinsic curvature.

By replacing the general variation of $h^{\mu\nu}$ in~\eqref{eq:app-variation-relations} with a Lie derivative with respect to $v^\mu$,
and by subsequently applying the resulting expression to the projection
$h^\mu_\nu = h^{\mu\rho} h_{\rho\nu}$,
we obtain the identities
\begin{equation}
  \label{eq:app-v-lie-derivs-of-h-up-and-h-proj}
  \LL_v h^{\mu\nu}
  = 2 K^{\mu\nu}
  + 2 v^{(\mu} a^{\nu)}\,,
  \qquad
  \LL_v h^\mu_\nu
  = v^\mu a_\nu\,.
\end{equation}

\paragraph{Exterior derivatives.}
We occasionally denote the exterior derivative of $\tau_\mu$ using
\begin{equation}
  \label{eq:app-tau-mu-nu-def}
  \tau_{\mu\nu}
  = (d\tau)_{\mu\nu}
  = 2 \pd_{[\mu} \tau_{\nu]}.
\end{equation}
It is easy to see that its time projection
$v^\rho \tau_{\rho\mu}=a_\mu$
is equal to the acceleration.
The double spatial projection gives
\begin{equation}
  \label{eq:app-twist-def}
  F_{\mu\nu}
  = h_\mu^\rho h_\nu^\sigma \tau_{\rho\sigma}\,,
\end{equation}
which is also known as the twist tensor.
We have
$\tau\wedge d\tau = \tau \wedge F$,
so that the twist tensor controls to what extent the Frobenius integrability condition on $\tau$ is violated.
We can decompose $\tau_{\mu\nu}$ in terms of the twist and the acceleration as follows,
\begin{equation}
  \label{eq:app-dtau-decomposition}
  \tau_{\mu\nu}
  = F_{\mu\nu}
  - 2 \tau_{[\mu} a_{\nu]}\,.
\end{equation}
Since exterior derivatives and Lie derivatives commute,
we also have
\begin{equation}
  \label{eq:app-da-decomposition}
  2 \pd_{[\mu} a_{\nu]}
  = \LL_v \tau_{\mu\nu}
  = \LL_v F_{\mu\nu}
  - 2 \tau_{[\mu} \LL_v a_{\nu]}\,,
\end{equation}
where the last equality follows from~\eqref{eq:app-dtau-decomposition}.

\subsection{Boundary connection}
\label{sapp:carroll-geometry-connection}
We now collect some properties of the affine connection~\eqref{eq:bdy-Ccon} that we use throughout most of this paper,
which is given by
\begin{equation}
  \label{eq:app-calCcon}
  \begin{split}
    \mathcal{C}^\rho_{\mu\nu}
    & =  -\frac{1}{2}v^\rho\left(\partial_\mu \tau_\nu+\partial_\nu \tau_\mu\right)
    -\frac{1}{2}v^\rho\left(\tau_\mu a_\nu+\tau_\nu a_\mu\right)\\
    &{}\qquad
    +\frac{1}{2}h^{\rho\sigma}\left(
      \partial_\mu h_{\nu\sigma}+\partial_\nu h_{\mu\sigma}-\partial_\sigma h_{\mu\nu}
    \right)\,.
  \end{split}
\end{equation}
This connection is the leading-order term in the large $r$ expansion of the $r$-dependent connection $\hat{C}^\rho_{\mu\nu}$ in~\eqref{eq:hatCcon} and it satisfies similar properties.
As discussed in Section~\ref{ssec:connection-choice},
these connections are not unique, but their covariant derivatives have several desirable properties.
We use $\mathcal{D}_\mu$ to denote the covariant derivative of the boundary connection $\mathcal{C}^\rho_{\mu\nu}$ in~\eqref{eq:app-calCcon}.
First, we see that the connection is symmetric and therefore has zero torsion,
and we also have
\begin{equation}
  \label{eq:app-boundary-connection-trace}
  \mathcal{C}^\rho_{\rho\mu}
  = e\inv \pd_\mu e\,.
\end{equation}
Following~\eqref{eq:app-general-cov-divergence}, this means that the covariant derivative is volume-form compatible
and that the covariant and ordinary divergence of $eX^\rho$ are the same,
\begin{equation}
  \label{eq:app-boundary-vf-compatibility}
  \mathcal{D}_\rho e
  = 0\,,
  \qquad
  \mathcal{D}_\rho X^\rho
  = e\inv \pd_\rho \left(e X^\rho\right)\,,
\end{equation}
where $X^\mu$ is a vector field.
These properties are very convenient when integrating by parts.
However, none of the fundamental Carroll metric variables are covariantly constant,
and we have
\begin{subequations}
  \label{eq:app-boundary-metricity-properties}
  \begin{gather}
    \mathcal{D}_\rho v^\mu
    = - K^\mu{}_{\rho}\,,
    \qquad
    \mathcal{D}_\rho h_{\mu\nu}
    = - 2 \tau_{(\mu} K_{\nu)\rho}\,,
    \\
    \mathcal{D}_\rho \tau_\mu
    = \frac{1}{2} F_{\rho\mu}-\tau_\rho a_\mu\,,
    \qquad
    \mathcal{D}_\rho h^{\mu\nu}
    = - v^{(\mu} h^{\nu)\alpha} F_{\alpha\rho}-2\tau_\rho v^{(\mu}a^{\nu)}\,.
  \end{gather}
\end{subequations}
Note that the leading-order bulk equations of motion imply that $K_{\mu\nu}=Kh_{\mu\nu}/d$ is pure trace.
We often already implement this constraint ``off-shell'' to simplify our expressions at subleading orders. However, we decided to keep this appendix fully general and so here we will not make this assumption.
Some useful consequences of the metricity properties~\eqref{eq:app-boundary-metricity-properties} include
\begin{gather}
  \label{eq:app-boundary-cov-der-h-up-v-divergence}
  \mathcal{D}_\rho h^{\rho\nu}
  = a^\nu\,,
  \qquad
  \mathcal{D}_\rho v^\rho
  = - K
  =  e\inv \pd_\rho \left(e v^\rho\right)\,.
\end{gather}

\paragraph{Lie derivatives.}
Since the torsion vanishes, we can convert Lie derivatives into covariant derivatives simply using
\begin{equation}
  \label{eq:app-lie-to-boundary-covariant-derivative}
  \LL_\xi X^\mu{}_\nu
  = \xi^\rho \mathcal{D}_\rho X^\mu{}_\nu
  - X^\rho{}_\nu \mathcal{D}_\rho \xi^\mu
  + X^\mu{}_\rho \mathcal{D}_\nu \xi^\rho\,,
\end{equation}
where $\xi^\mu$ is a vector field and $X^\mu{}_\nu$ is a tensor.

\paragraph{Exterior derivatives.}
Since the connection is torsion-free,
it reduces to the exterior derivative on $p$-forms $X_{\mu_1 \cdots \mu_p}$ after antisymmetrisation,
\begin{equation}
  (p+1)!\,
  \mathcal{D}_{[\rho} X_{\mu_1 \cdots \mu_p]}
  = (p+1)!\,
  \pd_{[\rho} X_{\mu_1 \cdots \mu_p]}\,.
\end{equation}
In particular,
we see that the antisymmetrisation of $\mathcal{D}_\mu \tau_\nu$ reduces to~\eqref{eq:app-dtau-decomposition} above.

\paragraph{Derivatives of acceleration.}
It is also useful to decompose the covariant derivative $\mathcal{D}_\mu a_\nu$ of the acceleration.
The antisymmetric part of $\mathcal{D}_\mu a_\nu$ follows from~\eqref{eq:app-da-decomposition}.
The $v^\mu$ projections are given by
\begin{subequations}
  \label{eq:app-boundary-cov-der-of-a-time-projections}
  \begin{gather}
    v^\rho 
    \mathcal{D}_\rho a_\nu
    = \LL_v a_\nu
    + K_{\nu\rho} a^\rho
    \\
     v^\sigma
    \mathcal{D}_\mu a_\sigma
    = K_{\mu\rho} a^\rho\,.
  \end{gather}
\end{subequations}
For the spatial projection of $\mathcal{D}_\rho a_\sigma$ we can write
\begin{align}
  \label{eq:app-boundary-cov-der-of-a-decomposition}
  h^\rho_\mu h^\sigma_\nu \mathcal{D}_\rho a_\sigma
  = 
  \frac{1}{2} \LL_v F_{\mu\nu} + A_{\mu\nu}
  + \frac{1}{d} h_{\mu\nu}  \mathcal{D}_\rho a^\rho
  - a_{\mu} a_{\nu}\,,
\end{align}
where we defined the STF tensor
\begin{equation}
  \label{eq:app-Amunu-def}
  A_{\mu\nu}
  = h^\rho_{\langle\mu} h^\sigma_{\nu\rangle} \left(
    \mathcal{D}_\rho a_\sigma
    + a_\rho a_\sigma
  \right)\,.
\end{equation}
It is convenient to write it like this because the objects $A_{\mu\nu}$ and $\mathcal{D}_\rho a^\rho$ occur frequently often allowing us to use identities such as \eqref{eq:app-d2-spatial-STF-STF-product} and \eqref{eq:app-d2-spatial-STF-antisym-product}.

\subsection{Properties of the boundary curvature tensor}\label{app:curvten}
We now collect some useful properties of the curvature tensor~$\mathcal{R}_{\mu\nu\sigma}{}^\rho$ associated to the boundary connection~\eqref{eq:app-calCcon}.
Since the latter has zero torsion,
the Bianchi identities~\eqref{eq:app-general-riemann-alg-bianchi} and~\eqref{eq:app-general-riemann-dif-bianchi} reduce to
\begin{align}
  \label{eq:app-boundary-riemann-alg-bianchi}
  \mathcal{R}_{[\mu \nu \sigma]}{ }^\rho
  &= 0\,,
  \\
  \label{eq:app-boundary-riemann-dif-bianchi}
  \mathcal{D}_{[\lambda} \mathcal{R}_{\mu \nu] \sigma}{ }^\rho
  &= 0\,.
\end{align}
Furthermore, because $\mathcal{C}^\rho_{\mu\nu}$ is volume-form compatible the Ricci tensor $\mathcal{R}_{\mu\sigma}=\mathcal{R}_{\mu \rho \sigma}{ }^\rho$ is symmetric. 

We will use the same construction as for the curvature tensor of the constant-$r$ hypersurfaces applied to the connection \eqref{eq:app-calCcon}. This is the boundary value of what is called $\hat R_{\mu\nu\rho}{}^\sigma$ in Section \ref{subsec:hypercurv}. Let us define
\begin{equation}
    \mathcal{Q}_{\mu\nu\rho\alpha}=\mathcal{R}_{\mu\nu\rho}{}^\sigma h_{\sigma\alpha}+\mathcal{T}_{\mu\nu\rho\alpha}\,,
\end{equation}
where 
\begin{equation}\label{eq:defcalT}
    \mathcal{T}_{\mu\nu\rho\alpha}=-\frac{1}{2}\mathcal{D}_\mu\left(\mathcal{D}_\nu h_{\rho\alpha}+\mathcal{D}_\rho h_{\nu\alpha}-\mathcal{D}_\alpha h_{\nu\rho}\right)-\left(\mu\leftrightarrow\nu\right)\,.
\end{equation}
Following an analogous argument as in Section \ref{subsec:hypercurv}, we conclude that $\mathcal{Q}_{\mu\nu\rho\alpha}$ has the same symmetries as the standard Levi-Civita Riemann tensor,
so we call it a `Riemann-like' tensor. Using \eqref{eq:app-boundary-metricity-properties}, we can show that
\begin{equation}\label{eq:calTexplicit}
    \mathcal{T}_{\mu\nu\rho\alpha}=\frac{1}{2}K_{\nu\rho}F_{\mu\alpha}-K_{\nu\rho}\tau_\mu a_\alpha+\tau_\alpha\mathcal{D}_\mu K_{\nu\rho}-\left(\mu\leftrightarrow\nu\right)\,.
\end{equation}

We can decompose the curvature tensor as
\begin{equation}\label{eq:relcalRandcalQ}
    \mathcal{R}_{\mu\nu\rho}{}^\kappa=h^{\kappa\alpha}\mathcal{Q}_{\mu\nu\rho\alpha}-h^{\kappa\alpha}\mathcal{T}_{\mu\nu\rho\alpha}-v^\kappa\left[\mathcal{D}_\mu\,,\mathcal{D}_\nu\right]\tau_\rho\,,
\end{equation}
where we used the Ricci identity for $\mathcal{R}_{\mu\nu\rho}{}^\kappa\tau_\kappa$. The Ricci tensor is thus 
\begin{equation}\label{eq:relcalRicandcalQ}
    \mathcal{R}_{\mu\rho}=h^{\nu\alpha}\mathcal{Q}_{\mu\nu\rho\alpha}-h^{\nu\alpha}\mathcal{T}_{\mu\nu\rho\alpha}-v^\nu\left[\mathcal{D}_\mu\,,\mathcal{D}_\nu\right]\tau_\rho\,.
\end{equation}

All contractions of $\mathcal{R}_{\mu\nu\rho}{}^\kappa$ with either $v^\lambda$ or $\tau_\kappa$ are determined by the non-metricity. For $\mathcal{R}_{\mu\nu\rho}{}^\kappa \tau_\kappa$ we have, 
after using various identities derived in this appendix,
\begin{eqnarray}
    \mathcal{R}_{\mu\nu\rho}{}^\sigma\tau_\sigma=\left[\mathcal{D}_\mu\,,\mathcal{D}_\nu\right]\tau_\rho & = & \left[-\frac{1}{2}\tau_\rho F_{\sigma\mu}K^\sigma{}_\nu-\frac{1}{2}\tau_\nu F_{\sigma\mu}K^\sigma{}_\rho+\frac{1}{2}F_{\rho\mu}a_\nu-\tau_\mu\tau_\rho K_\nu{}^\sigma a_\sigma\right.\nonumber\\
    &&\left.+\tau_\mu\left(A_{\nu\rho}+\frac{1}{d}h_{\nu\rho}\mathcal{D}_\sigma a^\sigma\right)-\left(\mu\leftrightarrow\nu\right)\right]\label{eq:comDtau}\\
    &&-\frac{1}{2}h^\sigma_\rho h^\alpha_\mu h^\beta_\nu\left(\mathcal{D}_\sigma+a_\sigma\right)F_{\alpha\beta}\,,\nonumber
\end{eqnarray}
where $A_{\nu\rho}$ is defined in \eqref{eq:app-Amunu-def}. The antisymmetrisation in $\mu$ and $\nu$ only applies to the terms inside the square brackets.
The expression for $\mathcal{R}_{\mu\nu\rho}{}^\sigma v^\rho$ follows likewise from a Ricci identity, leading to
\begin{equation}
    \mathcal{R}_{\mu\nu\rho}{}^\sigma v^\rho=\mathcal{D}_\mu K_{\nu}{}^\sigma-\mathcal{D}_\nu K_{\mu}{}^\sigma\,.
\end{equation}
To derive an expression for $\mathcal{R}_{\mu\nu\rho}{}^\kappa v^\mu$ we use the symmetry properties of $\mathcal{Q}_{\mu\nu\rho\alpha}$. Again, the argument is entirely analogous to the derivation of \eqref{eq:RUfirstindex}, and so we just quote the result
\begin{align}
\begin{split}
    \mathcal{R}_{\mu\nu\rho}{}^\kappa v^\mu & =  -h^{\kappa\alpha}h_{\nu\gamma}\left[\mathcal{D}_\rho\,,\mathcal{D}_\alpha\right]v^\gamma+h^{\kappa\alpha}h^\lambda_\nu v^\mu\left(\mathcal{T}_{\rho\alpha\mu\lambda}-\mathcal{T}_{\mu\lambda\rho\alpha}\right)\\
    &\quad\,-v^\kappa v^\mu\left[\mathcal{D}_\mu\,,\mathcal{D}_\nu\right]\tau_\rho\,.
\end{split}
\end{align}
In here, we have
\begin{equation}\label{eq:vconcomDDtau}
    v^\mu\left[\mathcal{D}_\mu\,,\mathcal{D}_\nu\right]\tau_\rho = -\frac{1}{2} F_{\sigma\nu}K^\sigma{}_\rho+\tau_\rho K_\nu{}^\sigma a_\sigma-A_{\nu\rho}-\frac{1}{d}h_{\nu\rho}\mathcal{D}_\sigma a^\sigma\,,
\end{equation}
which follows from \eqref{eq:comDtau}. We furthermore have
\begin{eqnarray}
    h^{\kappa\alpha}h^\lambda_\nu v^\mu\left(\mathcal{T}_{\rho\alpha\mu\lambda}-\mathcal{T}_{\mu\lambda\rho\alpha}\right) & = & -K_{\nu\rho}a^\kappa\,,\\
    -h^{\kappa\alpha}h_{\nu\gamma}\left[\mathcal{D}_\rho\,,\mathcal{D}_\alpha\right]v^\gamma & = & h^{\alpha\kappa}\left(\mathcal{D}_\rho K_{\alpha\nu}-\mathcal{D}_\alpha K_{\rho\nu}\right)\,.
\end{eqnarray}
Hence, in the end we find
\begin{eqnarray}
    \mathcal{R}_{\mu\nu\rho}{}^\kappa v^\mu & = & h^{\alpha\kappa}\left(\mathcal{D}_\rho K_{\alpha\nu}-\mathcal{D}_\alpha K_{\rho\nu}\right)-K_{\nu\rho}a^\kappa\nonumber\\
    &&+v^\kappa \left[\frac{1}{2} F_{\sigma\nu}K^\sigma{}_\rho-\tau_\rho K_\nu{}^\sigma a_\sigma+A_{\nu\rho}+\frac{1}{d}h_{\nu\rho}\mathcal{D}_\sigma a^\sigma\right]\,.\label{eq:vfirstindexcalR}
\end{eqnarray}
The purely spatial projection of the curvature tensor is
\begin{equation}\label{eq:spatialprojRiem}
    h^\mu_\alpha h^\nu_\beta h^\rho_\gamma h_{\delta\kappa}\mathcal{R}_{\mu\nu\rho}{}^\kappa=h^\mu_\alpha h^\nu_\beta h^\rho_\gamma h_{\delta}^{\sigma}\left(\mathcal{Q}_{\mu\nu\rho\sigma}-\frac{1}{2}K_{\nu\rho}F_{\mu\sigma}+\frac{1}{2}K_{\mu\rho}F_{\nu\sigma}\right)\,,
\end{equation}
where we used \eqref{eq:calTexplicit}. In this expression, we can write
\begin{equation}
    h^\mu_\alpha h^\nu_\beta h^\rho_\gamma h_{\delta}^{\sigma}\mathcal{Q}_{\mu\nu\rho\sigma}=e^a_\alpha e^b_\beta e^c_\gamma e^d_{\delta}\mathcal{Q}_{abcd}\,,
\end{equation}
where $\mathcal{Q}_{abcd}$ is a $d$-dimensional tensor with `Riemann-like' symmetries.

We can write the Ricci tensor \eqref{eq:relcalRicandcalQ} as
\begin{equation}
    \mathcal{R}_{\mu\rho}=h^{\nu\alpha}\mathcal{Q}_{\mu\nu\rho\alpha}+\tau_\mu K^\sigma{}_{\rho}a_\sigma+\tau_\rho K_\mu{}^\sigma a_\sigma-A_{\mu\rho}-\frac{1}{d}h_{\mu\rho}\mathcal{D}_\sigma a^\sigma\,,
\end{equation}
where we used \eqref{eq:vconcomDDtau} and \eqref{eq:calTexplicit}. The $v^\mu$ projection of the Ricci tensor can be obtained by contracting $\kappa$ and $\nu$ in \eqref{eq:vfirstindexcalR} leading to
\begin{equation}\label{eq:vRic}
     \mathcal{R}_{\mu\rho} v^\mu =
     \partial_\rho K-\mathcal{D}_\alpha K^\alpha{}_{\rho}\,.
\end{equation}
The spatial projection of the Ricci tensor is 
\begin{equation}
    h^\mu_\alpha h^\rho_\beta\mathcal{R}_{\mu\rho}=h^\mu_\alpha h^\rho_\beta h^{\nu\alpha}\mathcal{Q}_{\mu\nu\rho\alpha}-A_{\alpha\beta}-\frac{1}{d}h_{\alpha\beta}\mathcal{D}_\sigma a^\sigma\,.
\end{equation}

For $d=2$ we can write
\begin{equation}
  \label{eq:Qscalar-boundary-d=2}
    \mathcal{Q}_{abcd}=\frac{1}{2}\mathcal{Q}\left(\delta_{ac}\delta_{bd}-\delta_{bc}\delta_{ad}\right)\,,
\end{equation}
so that 
\begin{equation}\label{eq:traceQ}
    h^\mu_\alpha h^\rho_\beta h^{\nu\alpha}\mathcal{Q}_{\mu\nu\rho\alpha}=\frac{1}{2}\mathcal{Q}h_{\alpha\beta}\,.
\end{equation}
So, in the end, for $d=2$ we obtain 
\begin{equation}
    h^\mu_\alpha h^\rho_\beta\mathcal{R}_{\mu\rho}=\frac{1}{2}\left(\mathcal{Q}-\mathcal{D}_\sigma a^\sigma\right)h_{\alpha\beta}-A_{\alpha\beta}\,.\label{d=2RicCar}
\end{equation}
Hence, we have
\begin{equation}\label{eq:spatRicandQ}
    h^{\mu\nu}\mathcal{R}_{\mu\nu}=\mathcal{Q}-\mathcal{D}_\sigma a^\sigma\,.
\end{equation}
We conclude that $\mathcal{Q}$, or equivalently $h^{\mu\rho}\mathcal{R}_{\mu\rho}$, is the only part of the curvature tensor of a general 3-dimensional Carrollian manifold that is not expressible in terms of the non-metricity objects that appear in \eqref{eq:app-boundary-metricity-properties}.

Armed with these results we can now derive Bianchi identities for the curvature scalar. 
If we contract \eqref{eq:app-boundary-riemann-dif-bianchi} once, we obtain\footnote{If we contract \eqref{eq:oncecontracteddiffBI} with $h^{\nu\sigma}$ then we can obtain the resulting equation from a variational argument using that for $d=2$ the integral $\int d^3x e \mathcal{Q}$ is diffeomorphism-invariant (see Section~\ref{app:variations} for more details).}
\begin{equation}\label{eq:oncecontracteddiffBI}
    \mathcal{D}_\mu\mathcal{R}_{\nu\sigma}-\mathcal{D}_\nu\mathcal{R}_{\mu\sigma}+\mathcal{D}_\rho\mathcal{R}_{\mu\nu\sigma}{}^\rho=0\,.
\end{equation}
If we next contract all indices with $v^\mu h^{\nu\sigma}$ and we use the results derived in this subsection, we obtain
\begin{equation}\label{eq:BianchiQ}
    \left(\mathcal{L}_v-K\right)\mathcal{Q}=\mathcal{D}_\mu\left(h^{\mu\nu}\partial_\nu K+Ka^\mu\right)\,.
\end{equation}
This last identity has been derived only for $d=2$ and the case where $K_{\mu\nu}=\frac{1}{2}Kh_{\mu\nu}$. Under the same conditions, we can derive a second Bianchi identity by contracting~\eqref{eq:oncecontracteddiffBI} with $h_\kappa^\mu h^{\nu\sigma}$. This leads to 
\begin{align}
\label{eq:BItwo}
0 & =  \frac{1}{2}(\mathcal{L}_v-K)\left(\mathcal{D}_\rho F^\rho{}_\kappa+a_\rho F^\rho{}_\kappa\right)-\frac{1}{2}\partial_\rho K F^\rho{}_\kappa\nonumber\\
&\quad\,+\frac{1}{2}\mathcal{Q}a_\kappa-a_\kappa \mathcal{D}_\sigma a^\sigma+\frac{1}{2}h^\rho_\kappa(\partial_\rho+3a_\rho)\mathcal{D}_\sigma a^\sigma-\mathcal{D}_\rho A^\rho{}_\kappa\,.
\end{align}
In deriving this, we used many of the $d=2$ curvature tensor properties derived in this subsection. Most of the work comes from the last term in \eqref{eq:oncecontracteddiffBI}. Here, it is useful to note that 
\begin{subequations}
\begin{align}
    h^\sigma_\rho h^{\mu\nu}\mathcal{R}_{\sigma\mu\nu}{}^\rho & =  -\mathcal{Q}\,,\\
    h^\sigma_\kappa h^{\mu\nu}\tau_\gamma\mathcal{R}_{\sigma\mu\nu}{}^\gamma & = \frac{1}{2}\mathcal{D}_\alpha F^\alpha{}_\kappa+\frac{1}{2}a_\alpha F^\alpha{}_\kappa\,,\\
    v^\sigma h^{\mu\nu}\mathcal{R}_{\sigma\mu\nu}{}^\rho & =  -\frac{1}{2}h^{\rho\sigma}\partial_\sigma K-Ka^\rho+v^\rho\mathcal{D}_\sigma a^\sigma\,,\label{eq:BItwospatial}
\end{align}
\end{subequations}
as well as
\begin{equation}
    h^\sigma_\kappa\mathcal{R}_{\sigma\mu\nu}{}^\rho\mathcal{D}_\rho h^{\mu\nu} = -\frac{1}{2}K a_\rho F^\rho{}_\kappa+\frac{3}{4}\partial_\rho K F^\rho{}_\kappa-\frac{1}{2}a_\kappa\mathcal{D}_\sigma a^\sigma-a_\rho A^\rho{}_\kappa\,.
\end{equation}

\subsection{Special \texorpdfstring{$d=2$}{d=2} identities}\label{subsec:speciald=2id}

Certain index identities are dimension-specific. To illustrate what sort of identities we will be discussing, we remind the reader of the well-known property that, for $i,j,k=1,2$, the following two equations are identical
\begin{equation}
    \partial_i X_{jk}-\partial_{j}X_{ik}=0\qquad\Longleftrightarrow\qquad\partial_i X_{ij}=0\,,
\end{equation}
whenever $X_{ij}$ is STF. We will be interested in the curved space version of this statement. Contracting~\eqref{eq:d=2projid}
with $\mathcal{D}_\mu X_{\nu\rho}$,
where $X_{\mu\nu}$ is any spatial STF tensor,
the right-hand side vanishes,
and we obtain
\begin{equation}
\label{eq:DCident}
    h^\mu_{[\alpha}h^\nu_{\beta]}h^\rho_\gamma\mathcal{D}_\mu X_{\nu\rho}
    +h^{\mu\nu}h^\rho_{[\alpha}h_{\beta]\gamma}\mathcal{D}_\mu X_{\nu\rho}
    =0\,.
\end{equation}
This implies in particular the following identities for the shear,
\begin{eqnarray}
    h^{\rho\langle\mu}h^{\nu\rangle\sigma}a^\alpha\left(\mathcal{D}_\alpha C_{\rho\sigma}-\mathcal{D}_\rho C_{\alpha\sigma}\right) & = & h^{\rho\langle\mu}h^{\nu\rangle\sigma}a_\sigma\left(\mathcal{D}_\alpha-a_\alpha\right)C^\alpha{}_\rho\,,\\
    F^{\rho\sigma}\mathcal{D}_\rho C_{\sigma\alpha} & = & -F^\sigma{}_\alpha\left(\mathcal{D}_\rho-a_\rho\right)C^\rho{}_\sigma\,.\label{eq:FDC}
\end{eqnarray}

We can also use \eqref{eq:d=2projid} to obtain identities involving the twist tensor. If we contract it with $\mathcal{D}_\mu F_{\nu\rho}$, we obtain
\begin{eqnarray}
    0 & = & h^\mu_\alpha h^\nu_\beta h^\rho_\gamma\left(\frac{1}{2}\mathcal{D}_\mu F_{\nu\rho}+\frac{1}{2}\mathcal{D}_\nu F_{\rho\mu}+\mathcal{D}_\rho F_{\mu\nu}+\frac{1}{2}h_{\nu\rho}h^{\kappa\lambda}\mathcal{D}_\kappa F_{\lambda\mu}-\frac{1}{2}h_{\mu\rho}h^{\kappa\lambda}\mathcal{D}_\kappa F_{\lambda\nu}\right)\,.\nonumber\\
    &&\label{eq:DFd=2id}
\end{eqnarray}
The twist tensor satisfies the following Bianchi identity,
\begin{equation}\label{eq:BianchiideF}
    h^\alpha_\mu h^\beta_\nu h^\gamma_\rho\left(\left(\mathcal{D}_\alpha-a_\alpha\right)F_{\beta\gamma}+\left(\mathcal{D}_\gamma-a_\gamma\right)F_{\alpha\beta}+\left(\mathcal{D}_\beta-a_\beta\right)F_{\gamma\alpha}\right)=0\,.
\end{equation}
By contracting \eqref{eq:d=2projid} with $a_\rho F_{\mu\nu}$, we learn that
\begin{equation}
    a_\alpha F_{\beta\gamma}+a_\beta F_{\gamma\alpha}+2a_\gamma F_{\alpha\beta}=h_{\alpha\gamma}a^\sigma F_{\sigma\beta}-h_{\beta\gamma}a^\sigma F_{\sigma\alpha}\,.
\end{equation}
Combining the last two results, we can write \eqref{eq:DFd=2id} as
\begin{eqnarray}
    \hspace{-0.5cm}h^\mu_\alpha h^\nu_\beta h^\rho_\gamma\left((\mathcal{D}_\rho-a_\rho)F_{\mu\nu}-h_{\mu\rho}h^{\kappa\lambda}(\mathcal{D}_\kappa-a_\kappa)F_{\lambda\nu}+h_{\nu\rho}h^{\kappa\lambda}(\mathcal{D}_\kappa-a_\kappa)F_{\lambda\mu}\right)=0\,.
\end{eqnarray}
This covariantises the well-known fact that, in $d=2$, we have the identity
\begin{equation}
    \partial_k F_{ij}=\delta_{ki}\partial_l F_{lj}-\delta_{kj}\partial_l F_{li}\,.
\end{equation}

If we contract \eqref{eq:DFd=2id} with $a^\alpha$, use the Bianchi identity
\eqref{eq:BianchiideF} and take the STF part in $\beta$ and $\gamma$, we obtain 
\begin{equation}\label{eq:calDFid}
    h^{\rho\langle\mu}h^{\nu\rangle\sigma}\left[a_\rho\mathcal{D}_\kappa F^\kappa{}_\sigma-a_\rho a_\kappa F^\kappa{}_\sigma-a^\kappa\mathcal{D}_\rho F_{\kappa\sigma}\right]=0\,.
\end{equation}
The Bianchi identity \eqref{eq:BianchiideF} can also be used to derive various other identities involving the twist tensor. For example, if we contract this with $F^{\mu\nu}$ and use that $F^{\alpha\beta}F_{\beta\gamma}=-\frac{1}{2}h^\alpha_\gamma F^2$, we can derive the following (Weyl-covariant) identity,
\begin{equation}\label{eq:FDFid}
    F_{\sigma\alpha}\mathcal{D}_\rho F^{\rho\sigma}=-\frac{1}{4}h^\rho_\alpha\left(\partial_\rho+2a_\rho\right) F^2\,.
\end{equation}
The last two identities have been used in the derivation of \eqref{eq:STF1/rSimple},
and \eqref{eq:FDFid} has furthermore been used in several other places in the main text.

\subsection{Commuting Lie and covariant derivatives}
\label{sapp:carroll-lie-covariant-commutators}
We will often have to commute a Lie derivative along $v^\mu$ with a covariant derivative. This leads to curvature tensor components that we discussed earlier in Section \ref{app:curvten}. The details depend on the tensor on which these derivatives act. In this appendix, we will derive a few useful results that are used in the main text.

Our first example is the following identity,
\begin{equation}\label{eq:vectorcase}
    \left(\mathcal{L}_v-K\right)\mathcal{D}_\mu X^\mu=\mathcal{D}_\mu\left(\mathcal{L}_v-K\right) X^\mu\,,
\end{equation}
where $X^\mu$ is any vector. This follows directly from $\mathcal{L}_v\mathcal{D}_\mu X^\mu=v^\rho\mathcal{D}_\rho\mathcal{D}_\mu X^\mu$ and commuting the covariant derivatives. This leads to a Ricci tensor contracted with $v^\mu X^\nu$. At this point we use the result \eqref{eq:vRic}.

Case number two concerns $\left[\mathcal{L}_v\,,\mathcal{D}_\mu\right]X^{\mu\nu}$. Using standard manipulations with covariant derivatives, we can write for any $X^{\mu\nu}$
\begin{eqnarray}
    \left[\mathcal{L}_v\,,\mathcal{D}_\mu\right]X^{\mu\nu} & = & -v^\rho\mathcal{R}_{\rho\sigma}X^{\sigma\nu}-v^\rho\mathcal{R}_{\rho\mu\sigma}{}^\nu X^{\mu\sigma}\nonumber\\
    &&+X^{\mu\rho}\mathcal{D}_\mu\mathcal{D}_\rho v^\nu+X^{\sigma\nu}\mathcal{D}_\mu\mathcal{D}_\sigma v^\mu\,.
\end{eqnarray}
To make progress, we need to use the results from Section \ref{app:curvten}. First, using \eqref{eq:vRic}, we obtain
\begin{eqnarray}
    \left[\mathcal{L}_v\,,\mathcal{D}_\mu\right]X^{\mu\nu} & = & -X^{\sigma\nu}\partial_\sigma K-X^{\mu\sigma}\left(v^\rho\mathcal{R}_{\rho\mu\sigma}{}^\nu-\mathcal{D}_\mu\mathcal{D}_\sigma v^\nu\right)\,.
\end{eqnarray}
Equation \eqref{eq:vfirstindexcalR} tells us that 
\begin{equation}
    h_{\nu\gamma}\left(v^\rho\mathcal{R}_{\rho\mu\sigma}{}^\nu-\mathcal{D}_\mu\mathcal{D}_\sigma v^\nu\right)=h_\gamma^{\alpha}\left(\mathcal{D}_\mu K_{\sigma\alpha}+\mathcal{D}_\sigma K_{\mu\alpha}-\mathcal{D}_\alpha K_{\mu\sigma}-a_\alpha K_{\mu\sigma}\right)\,.
\end{equation}
Notice that this is symmetric in $\mu$ and $\sigma$. 

We will now assume that $d=2$, $X^{\mu\nu}$ is spatial and that $K_{\mu\nu}=\frac{1}{2}Kh_{\mu\nu}$. We now have
\begin{eqnarray}
    h_{\nu\gamma}\left[\mathcal{L}_v\,,\mathcal{D}_\mu\right]X^{\mu\nu} & = &  -h_{\nu\gamma}X^{\sigma\nu}\partial_\sigma K\label{eq:comLieDXmunu}\\
    &&-\frac{1}{2}X^{\mu\sigma}h_\gamma^{\alpha}\left( h_{\sigma\alpha}\partial_\mu K+ h_{\mu\alpha}\partial_\sigma K-h_{\mu\sigma}\partial_\alpha K- Ka_\alpha h_{\mu\sigma}\right)\,.\nonumber
\end{eqnarray}
We see that, if $X^{\mu\nu}$ is antisymmetric, we find
\begin{equation}
    h_{\nu\gamma}\left[\mathcal{L}_v\,,\mathcal{D}_\mu\right]X^{\mu\nu} = -h_{\nu\gamma}X^{\sigma\nu}\partial_\sigma K\,,
\end{equation}
which can also be written as 
\begin{equation}
    h_{\nu\gamma}\left(\mathcal{L}_v-K\right)\mathcal{D}_\mu X^{\mu\nu}=h_{\nu\gamma}\mathcal{D}_\mu \left(\mathcal{L}_v-K\right)X^{\mu\nu}\,,
\end{equation}
for any spatial and antisymmetric $X^{\mu\nu}$.
Now, if we assume that $X^{\mu\nu}$ is STF, then \eqref{eq:comLieDXmunu} becomes
\begin{eqnarray}\label{eq:hprojcomLievDSTF}
    h_{\nu\gamma}\left[\mathcal{L}_v\,,\mathcal{D}_\mu\right]X^{\mu\nu} & = & -2h_{\nu\gamma}X^{\sigma\nu}\partial_\sigma K\,,
\end{eqnarray}
which we can rewrite as
\begin{equation}\label{eq:comLiecovSTF}
    h_{\nu\gamma}\left(\mathcal{L}_v-2K\right)\mathcal{D}_\mu X^{\mu\nu}=h_{\nu\gamma}\mathcal{D}_\mu \left(\mathcal{L}_v-2K\right)X^{\mu\nu}\,,
\end{equation}
for any spatial STF tensor $X^{\mu\nu}$.

Finally, using $d=2$, $K_{\mu\nu}=\frac{1}{2}Kh_{\mu\nu}$ and assuming that $X_\nu$ is spatial we can similarly show that
\begin{equation}
    \mathcal{D}_\mu\mathcal{L}_v X_\nu=\mathcal{L}_v\mathcal{D}_\mu X_\nu-\frac{1}{2}\lambda_\mu\partial_\nu K-\frac{1}{2}\lambda_\nu\partial_\mu K+\frac{1}{2}h_{\mu\nu}\lambda^\rho\partial_\rho K+\frac{1}{2}K h_{\mu\nu}\lambda^\rho a_\rho\,.
\end{equation}

\subsection{Weyl covariance}
\label{ssec:bulk-improvements-weyl-cov-derivs}
We collect the Weyl transformations of our boundary variables,
and we also construct derivatives that transform homogeneously under Weyl transformations.
Our main application for these Weyl-covariant derivatives is that they suggest natural `building blocks' which are used to obtain a more manifestly Weyl-covariant formulation of the Weyl-improved EMT-news currents in Section~\ref{sec:bulk-improvements}.

\paragraph{Weyl transformations.}
We say that a general tensor
$X^{\mu_1 \cdots \mu_p}{}_{\nu_1 \cdots \nu_q}$
transforms homogeneously under Weyl transformations~$\Lambda_D$
with weight~$w$
if we have
\begin{equation}
  \label{eq:weyl-tr-weight-def}
  \delta_{\Lambda_D}
  X^{\mu_1 \cdots \mu_p}{}_{\nu_1 \cdots \nu_q}
  = w \Lambda_D
  X^{\mu_1 \cdots \mu_p}{}_{\nu_1 \cdots \nu_q}\,.
\end{equation}
Following the expressions in for example~\eqref{eq:app-v-tau-h-h-transformations},
recall that
the boundary Carroll metric variables
transform under boundary Weyl transformations
with weights $\pm1$ and $\pm2$,
\begin{subequations}
  \label{eq:weyl-tr-metric-variables}
  \begin{align}
    \delta_{\Lambda_D}
    \tau_\mu
    &= \Lambda_D \tau_\mu\,,
    &&
    \delta_{\Lambda_D}
    h_{\mu\nu}
    = 2 \Lambda_D h_{\mu\nu}\,,
    \\
    \delta_{\Lambda_D}
    v^\mu
    &= - \Lambda_D v^\mu\,,
    &&
    \delta_{\Lambda_D}
    h^{\mu\nu}
    = - 2 \Lambda_D h^{\mu\nu}\,.
  \end{align}
\end{subequations}
This means that the spatial and temporal projection operators
$h^\mu_\nu$
and
$v^\mu \tau_\nu$
are both Weyl-invariant.
Using
the general expression~\eqref{eq:app-integration-measure-variation} for the variation
of the integration measure $e$,
we also have
\begin{equation}
  \label{eq:weyl-tr-measure}
  \delta_{\Lambda_D}
  e
  = (d+1) \Lambda_D e\,.
\end{equation}
It is easy to see that the twist tensor $F_{\mu\nu}$ defined in~\eqref{eq:app-twist-def}
has Weyl weight one,
and
the same holds for the shear tensor $C_{\mu\nu}$
following~\eqref{eq:general-gauge-tr-of-shear-LO-on-shell},
so that
\begin{equation}
  \label{eq:weyl-tr-shear-twist}
  \delta_{\Lambda_D}
  C_{\mu\nu}
  = \Lambda_D C_{\mu\nu}\,,
  \qquad
  \delta_{\Lambda_D}
  F_{\mu\nu}
  = \Lambda_D F_{\mu\nu}\,.
\end{equation}
On the other hand,
the acceleration $a_\mu$
and the trace $K$ of the boundary extrinsic curvature
do not have a definite Weyl weight,
and they transform as follows,
\begin{equation}
  \label{eq:weyl-tr-acceleration-K}
  \delta_{\Lambda_D}
  a_\mu
  = h_\mu^\rho \pd_\rho \Lambda_D\,,
  \qquad
  \delta_{\Lambda_D}
  K
  = - \Lambda_D K
  - d \LL_v \Lambda_D\,.
\end{equation}
Note that these transformations give the spatial and temporal projections of the inhomogeneous $\pd_\mu\Lambda_D$ term.
We can combine them into
\begin{equation}
  \label{eq:b-tilde-def-and-weyl-tr}
  \tilde{b}_\mu
  = a_\mu
  + \frac{1}{d} K \tau_\mu\,,
  \qquad
  \delta_{\Lambda_D}
  \tilde{b}_\mu
  = \pd_\mu \Lambda_D\,,
\end{equation}
which transforms as a Weyl connection.%
\footnote{%
  Note that
  $\tilde{b}_\mu$
  is not equal to the gauge fixing of
  $b_\mu = (1/2)\os{-1}{S} \tau_\mu = (1/d) K \tau_\mu$
  we obtained in Section~\ref{ssec:car-cov-BS-gauge-fix-Sigma}.
  This would be true if had used the  alternative gauge fixing
  $U^\mu \pd_\rho E^a_\mu=0$
  which we discussed
  in that section
  in footnote~\ref{fn:alternative-sigma-gauge}. What is true in both cases is that we can write
  \begin{equation}
      \os{-1}{g}_{\mu\nu}=-\tilde b_\mu\tau_\nu-\tilde b_\nu\tau_\nu+C_{\mu\nu}\,.
  \end{equation}
}
Its curvature $d\tilde{b}$ is Weyl-invariant.

\paragraph{Weyl-covariant derivatives.}
We can use the inhomogeneous transformations in~\eqref{eq:weyl-tr-acceleration-K} and~\eqref{eq:b-tilde-def-and-weyl-tr} to construct derivatives that transform covariantly under Weyl transformations.
If $X^\mu{}_\nu$ is a tensor with Weyl weight $w$,
we see that
\begin{equation}
  \delta_{\Lambda_D}
  \left(\LL_v X^\mu{}_\nu\right)
  = (w-1) \Lambda_D \LL_v X^\mu{}_\nu
  + w \LL_v \Lambda_D X^\mu{}_\nu
  + v^\mu X^\rho{}_\nu \pd_\rho \Lambda_D
  - X^\mu{}_\rho v^\rho \pd_\nu \Lambda_D\,.
\end{equation}
Using the transformations of $K$ and $\tilde{b}_\mu$ in~\eqref{eq:weyl-tr-acceleration-K} and~\eqref{eq:b-tilde-def-and-weyl-tr},
we see that
\begin{align}
  \label{eq:weyl-homog-lie-deriv-general}
  \left(\LL_v + \frac{w}{d}K\right) X^\mu{}_\nu
  - v^\mu X^\rho{}_\nu \tilde{b}_\rho
  + X^\mu{}_\rho v^\rho \tilde{b}_\nu\,,
\end{align}
transforms homogeneously with weight $w-1$ under Weyl transformations.
This expression generalises to other tensors of weight $w$ in the obvious way.
The latter two terms can be removed using a spatial projection and by considering spatial tensors.
For example, this means that
for a general weight~$w$ vector $X^\mu$
and a weight~$w$ spatial tensor $X_{\mu\nu}$,
we see that
\begin{equation}
  \label{eq:weyl-homog-lie-deriv-special-cases}
  h^\mu_\rho \left(
    \LL_v
    + \frac{w}{d} K
  \right) X^\rho\,,
  \qquad
  \left(
    \LL_v
    + \frac{w}{d} K
  \right)
  X_{\mu\nu}\,,
\end{equation}
are both homogeneous with weight $w-1$. The latter generalises straightforwardly to any $(0,q)$ tensor.

Now let us consider covariant derivatives.
We can write the variation of the hypersurface connection~\eqref{eq:app-calCcon} as follows,
\begin{equation}
    \delta_{\Lambda_D}\mathcal{C}^\rho_{\mu\nu}=-v^\rho\tau_\mu\tau_\nu\mathcal{L}_v\Lambda_D-h_{\mu\nu}h^{\rho\sigma}\partial_\sigma\Lambda_D+\delta^\rho_\mu\partial_\nu\Lambda_D+\delta^\rho_\nu\partial_\mu\Lambda_D\,,
\end{equation}
which, using~\eqref{eq:weyl-tr-metric-variables} and~\eqref{eq:b-tilde-def-and-weyl-tr},
we identify as the total variation of the tensor
\begin{gather}
  \mathcal{W}^\rho{}_{\mu\nu} 
  =
  dK  \tau_{\mu}\tau_\nu v^\rho
  -h_{\mu\nu} a^\rho+ 2\delta^\rho_{(\mu} \tilde{b}_{\nu)}\,.
\end{gather}
Using this observation, we can define a new connection
$\check{\mathcal{C}}^\rho_{\mu\nu} = \mathcal{C}^\rho_{\mu\nu} - \mathcal{W}^\rho{}_{\mu\nu}$
with a covariant derivative~$\check{\mathcal{D}}_\rho$
that acts on a tensor $X^\mu{}_\nu$ with Weyl weight $w$ as follows,
\begin{gather}
  \label{eq:weyl-homog-covariant-deriv}
  \check{\mathcal{D}}_\rho X^\mu{}_\nu
  = \left(\mathcal{D}_\rho - w \tilde{b}_\rho\right) X^\mu{}_\nu
  - \mathcal{W}^\mu{}_{\rho\alpha} X^\alpha{}_\nu
  + \mathcal{W}^\alpha{}_{\rho\nu} X^\mu{}_\alpha\,.
\end{gather}
This expression is homogeneous with the same weight $w$ under Weyl transformations.
As always,
this definition extends linearly to general homogeneous tensors. In Section \ref{subsec:Weylcov} we work out a few important examples.

\subsection{Variations}\label{app:variations}

In this section we collect some important identities involving the variation of various geometric objects. For general $K_{\mu\nu}$ and any $d$, we have
\begin{align}
    \delta\mathcal{C}^\rho_{\mu\nu} & =  -\frac{1}{2}v^\rho\left(\left(\mathcal{D}_\mu+a_\mu\right)\delta\tau_\nu+\left(\mathcal{D}_\nu+a_\nu\right)\delta\tau_\mu+\tau_\mu\delta a_\nu+\tau_\nu\delta a_\mu\right)+K_{\mu\nu}h^{\rho\sigma}\delta\tau_\sigma\nonumber\\
    &\quad\,+\frac{1}{2}h^{\rho\sigma}\left(\mathcal{D}_\mu\delta h_{\nu\sigma}+\mathcal{D}_\nu\delta h_{\mu\sigma}-\mathcal{D}_\sigma\delta h_{\mu\nu}\right)\,,\label{eq:varconn}
\end{align}
as well as
\begin{equation}\label{eq:vara}
    \delta a_\mu=\mathcal{L}_v\delta\tau_\mu-\left(\partial_\mu-a_\mu\right)\left(v^\rho\delta\tau_\rho\right)+F_\mu{}^{(\rho}v^{\sigma)}\delta h_{\rho\sigma}-\tau_\mu a^{(\rho} v^{\sigma)}\delta h_{\rho\sigma}\,,
\end{equation}
and
\begin{align}
    \delta K & =  Kv^\mu\delta\tau_\mu+\mathcal{D}_\rho\left(h^{\mu\rho}v^\nu\delta h_{\mu\nu}\right)-\frac{1}{2}\mathcal{L}_v\left(h^{\mu\nu}\delta h_{\mu\nu}\right)\nonumber\\
    & =  Kv^\mu\delta\tau_\mu-2K^{\mu\nu}\delta h_{\mu\nu}-\frac{1}{2}h^{\mu\nu}\mathcal{L}_v\delta h_{\mu\nu}+h^{\rho\mu}v^\nu\mathcal{D}_\rho\delta h_{\mu\nu}\,.\label{eq:varK}
\end{align}
The first line in $\delta K$ is useful for doing integration by parts.

Next, moving on to the variation of curvatures we first observe that
the general variation of the boundary Ricci tensor is given by
\begin{equation}
    \delta\mathcal{R}_{\mu\nu}=\mathcal{D}_\rho\delta \mathcal{C}^\rho_{\mu\nu}-\mathcal{D}_\mu\delta\mathcal{C}^\rho_{\rho\nu}\,.
\end{equation}
However, for $d=2$, we know from the discussion in Appendix~\ref{app:curvten} that we can use $\mathcal{Q}$ to parametrise the only component of the curvature tensor that is not determined by its metricity properties. Using~\eqref{eq:spatRicandQ}, and keeping $K_{\mu\nu}$ general for now, the variation of this scalar is given by
\begin{align}
\delta \mathcal{Q} & =  -2(\mathcal{D}_\rho a^\rho) e^{-1}\delta e-(\mathcal{D}_\rho h^{\mu\nu})\delta C^\rho_{\mu\nu}-h^{\mu\rho}h^{\nu\sigma}\mathcal{R}_{\rho\sigma}\delta h_{\mu\nu}+2v^\rho h^{\sigma\mu}\mathcal{R}_{\rho\sigma}\delta\tau_\mu\nonumber\\
&\quad\,+\mathcal{D}_\rho\left(v^\rho a^\mu\delta\tau_\mu-v^\rho\tau_\sigma h^{\mu\nu}\delta C^\sigma_{\mu\nu}+h^\rho_\sigma X^\sigma\right)\,,\label{eq:deltaQ}
\end{align}
where we abbreviated
\begin{equation}
    h^\rho_\sigma X^\sigma=h^\rho_\sigma\left(\delta a^\sigma+2a^\sigma e^{-1}\delta e+h^{\mu\nu}\delta \mathcal{C}^\sigma_{\mu\nu}-h^{\sigma\nu}\partial_\nu\left(e^{-1}\delta e\right)\right)\,.
\end{equation}
Using \eqref{eq:varconn} and \eqref{eq:vara}, we can write this as
\begin{eqnarray}
    \delta\mathcal{Q} & = & -\mathcal{Q}e^{-1}\delta e+\left(-\mathcal{Q}v^\mu+h^{\mu\nu}\partial_\nu K-2h^{\mu\sigma}\left(\mathcal{D}_\rho-a_\rho\right)K^T{}^\rho{}_\sigma\right)\delta \tau_\mu\nonumber\\
    &&+\frac{1}{2}\left(-h^{\mu\nu}\mathcal{D}_\rho a^\rho+2A^{\mu\nu}+2K^T{}^\mu{}_\rho F^{\rho\nu}-2v^\mu\left(\mathcal{D}_\rho+a_\rho\right)F^{\rho\nu}\right)\delta h_{\mu\nu}\nonumber\\
    &&+\mathcal{D}_\rho\left(-v^\rho h^{\mu\nu}\mathcal{D}_\mu\delta \tau_\nu+h^\rho_\sigma Y^\sigma\right)\,,\label{eq:varcalQ}
\end{eqnarray}
where we furthermore used Equations \eqref{eq:vRic} and \eqref{d=2RicCar} and where $h^\rho_\sigma Y^\sigma$ is
\begin{eqnarray}
   h^\rho_\sigma Y^\sigma & = & h^\rho_\sigma X^\sigma + v^\mu F^{\rho\nu}\delta h_{\mu\nu}+2a^\rho v^\mu\delta\tau_\mu\\
   & = & h^\rho_\sigma\left(\delta a^\sigma+a^\sigma h^{\mu\nu}\delta h_{\mu\nu}+h^{\mu\nu}\delta\mathcal{C}^\sigma_{\mu\nu}-h^{\sigma\nu}\partial_\nu\left(e^{-1}\delta e\right)+v^\mu F^{\sigma\nu}\delta h_{\mu\nu}\right)\,.\nonumber
\end{eqnarray}
This result for $\delta\mathcal{Q}$ is valid for $d=2$ and any $K_{\mu\nu}$,
hence the appearance of $K^T_{\mu\nu}=K_{\mu\nu}-\frac{1}{2}K h_{\mu\nu}$
which would be zero in most of the main text.

Let us consider $\int d^3x e \mathcal{Q}$. Using the fact that this is diffeomorphism-invariant, we can show that 
\begin{equation}\label{eq:twicecontracedBI}
    0=\int d^3x e\chi^\alpha h^{\mu\nu}\left[\mathcal{D}_\mu\mathcal{R}_{\nu\alpha}-\mathcal{D}_\alpha\mathcal{R}_{\mu\nu}-\mathcal{D}_\rho\mathcal{R}_{\alpha\mu\nu}{}^\rho\right]\,,
\end{equation}
where we dropped boundary terms. In deriving this we used \eqref{eq:deltaQ} as well as that under a diffeomorphism we have
\begin{align}
    \delta_\chi \mathcal{C}^\rho_{\mu\nu} & =  \mathcal{D}_\mu\mathcal{D}_\nu\chi^\rho-\chi^\alpha\mathcal{R}_{\alpha\mu\nu}{}^\rho\,,\\
    \delta e & =  \partial_\alpha\left(e\chi^\alpha\right)\,.
\end{align}
The result \eqref{eq:twicecontracedBI} leads to the twice-contracted differential Bianchi identity
\begin{equation}\label{eq:twicecontracedBI2}
    h^{\mu\nu}\left[\mathcal{D}_\mu\mathcal{R}_{\nu\alpha}-\mathcal{D}_\alpha\mathcal{R}_{\mu\nu}-\mathcal{D}_\rho\mathcal{R}_{\alpha\mu\nu}{}^\rho\right
    ]=0\,,
\end{equation}
which holds by virtue of \eqref{eq:oncecontracteddiffBI}.

\subsection{Carroll boosts}\label{app:Carrollboosts}

Here, we will collect some results for the Carroll boost transformations of various geometric objects and derived quantities including the shear tensor. We know that $v^\mu$ and $h_{\mu\nu}$ are Carroll boost-invariant, while $\tau_\mu$ and $h^{\mu\nu}$ transform as
\begin{align}
\begin{split}
        \delta_\lambda\tau_\mu & =  \lambda_\mu\,,\\
    \delta_\lambda h^{\mu\nu} & =  v^\mu\lambda^\nu+v^\nu\lambda^\mu\,,
\end{split}
\end{align}
where $\lambda^\mu=h^{\mu\rho}\lambda_\rho$ and $v^\rho\lambda_\rho=0$. The integration measure $e$ is Carroll boost-invariant as follows from $e^{-1}\delta e=-v^\mu\delta\tau_\mu+\frac{1}{2}h^{\mu\nu}\delta h_{\mu\nu}$ and taking $\delta$ to be $\delta_\lambda$. Using that $K=-e^{-1}\partial_\mu\left(e v^\mu\right)$,  $a_\mu=\mathcal{L}_v\tau_\mu$ and $F_{\mu\nu}=h^\rho_\mu h^\sigma_\nu\left(\partial_\rho\tau_\sigma-\partial_\sigma\tau_\rho\right)$, we see that 
\begin{align}
\begin{split}
    \delta_\lambda K & =  0\,,\\
\delta_\lambda a_\mu & = \mathcal{L}_v\lambda_\mu\,,\\
\delta_\lambda F_{\mu\nu} & =  h^\rho_\mu h^\sigma_\nu\left[\left(\partial_\rho-a_\rho\right)\lambda_\sigma-\left(\partial_\sigma-a_\sigma\right)\lambda_\rho\right]\,.
\end{split}
\end{align}
Using the general variational results of Section \ref{app:variations}, we can obtain the Carroll boost transformations of derived objects. In particular, we have
\begin{equation}
    \delta_\lambda \mathcal{C}^\rho_{\mu\nu}=-\frac{1}{2}v^\rho\left(\left(\mathcal{D}_\mu+a_\mu\right)\lambda_\nu+\left(\mathcal{D}_\nu+a_\nu\right)\lambda_\mu+\tau_\mu\delta a_\nu+\tau_\nu\delta a_\mu\right)+K_{\mu\nu}\lambda^\rho\,.
\end{equation}

In the remainder of this subsection, we will use $d=2$ and $K_{\mu\nu}=\frac{1}{2}K h_{\mu\nu}$. We will frequently use results from Section \ref{sapp:carroll-lie-covariant-commutators}. Using the variations of the Ricci tensor and the connection in Section \ref{app:variations}, we can show that 
\begin{equation}
\delta_\lambda\left(h^{\mu\nu}\mathcal{R}_{\mu\nu}\right)=-\left(\mathcal{D}_\mu+a_\mu\right)\left(h^{\mu\nu}\mathcal{L}_v\lambda_\nu\right)-\lambda^\mu\mathcal{L}_v a_\mu+K\mathcal{D}_\mu\lambda^\mu+2\lambda^\mu\partial_\mu K\,.
\end{equation}
Using that
\begin{equation}
    \delta_\lambda\left(\mathcal{D}_\mu a^\mu\right)=\left(\mathcal{D}_\mu+a_\mu\right)\left(h^{\mu\nu}\mathcal{L}_v\lambda_\nu\right)+\lambda^\mu\mathcal{L}_v a_\mu\,,
\end{equation}
we can show that
\begin{equation}\label{eq:varS0-a^2}
    \delta_\lambda\left(\os{0}{S}-a^2\right)=\left(\mathcal{D}_\mu-a_\mu\right)\left(h^{\mu\nu}\left(\mathcal{L}_v+\frac{1}{2}K\right)\lambda_\nu\right)+\lambda^\sigma G_\sigma\,,
\end{equation}
where we defined 
\begin{equation}
    G_\sigma=v^\alpha\left(\partial_\alpha\tilde b_\sigma-\partial_\sigma\tilde b_\alpha\right)\,.
\end{equation}
In here, we used \eqref{eq:S0d=2}. Equation \eqref{eq:varS0-a^2} is Weyl-covariant. The parameter $\lambda_\mu=e^a_\mu\lambda^a$ has Weyl weight $+1$.

Next, we consider the shear tensor, which transforms under Carroll boosts as 
\begin{equation}
    \delta_\lambda C_{\mu\nu}=2h^\rho_{\langle\mu}h^\sigma_{\nu\rangle}\left(\mathcal{D}_\rho+a_\rho\right)\lambda_\sigma\,.
\end{equation}
We will be interested in the Carroll boost transformation of the anomaly \eqref{eq:anominanomSec}. We will use the first equality in \eqref{eq:anominanomSec} to compute this. This involves $\os{0}P^\mu$, so we will look at that first. We remind the reader that we have
\begin{equation}
    \os{0}{P}^\mu=-\frac{1}{2}\left(\mathcal{D}_\rho-2a_\rho\right)C^{\rho\mu}+\frac{1}{2}h^\mu_\sigma\mathcal{D}_\rho F^{\rho\sigma}\,.
\end{equation}
As an intermediate result, we can show that
\begin{align}
    \delta\os{0}{P}^\mu & =  v^\mu\lambda_\sigma\os{0}{P}^\sigma+\frac{1}{2}C^{\mu\alpha}\left(\mathcal{L}_v+\frac{1}{2}K\right)\lambda_\alpha+\frac{1}{2}\lambda_\alpha N^{\mu\alpha}+\frac{1}{2}h^\mu_\sigma\mathcal{D}_\rho\delta F^{\rho\sigma}\nonumber\\
    &\qquad-\frac{1}{2}\left(\mathcal{D}_\rho-2a_\rho\right)\left(h^{\rho\alpha}h^{\mu\beta}\delta C_{\alpha\beta}\right)\,.
\end{align}
Next, we show that 
\begin{align}
    &\frac{1}{2}h^\mu_\sigma\mathcal{D}_\rho\delta F^{\rho\sigma}-\frac{1}{2}\left(\mathcal{D}_\rho-2a_\rho\right)\left(h^{\rho\alpha}h^{\mu\beta}\delta C_{\alpha\beta}\right)\nonumber\\
    =&\lambda_\rho h^\mu_\sigma\left(\mathcal{L}_v-\frac{3}{2}K\right)F^{\rho\sigma}-h^{\rho\mu}\left(\mathcal{D}_\rho+a_\rho\right)\left[\left(\mathcal{D}_\alpha-2a_\alpha\right)\lambda^\alpha\right]\nonumber\\
    &-F^{\alpha\mu}\left(\mathcal{L}_v+\frac{1}{2}K\right)\lambda_\alpha-2\lambda^\mu\left(\os{0}{S}-a^2\right)\,,
\end{align}
where we used \eqref{eq:spatialprojRiem} and \eqref{eq:traceQ}.

The first equality in \eqref{eq:anominanomSec} gives an expression for $\mathcal{A}_B^\mu$ whose transformations under boosts we want to compute,
but it will be slightly easier to work with $\mathcal{A}_{B\,\mu}=h_{\mu\nu}\mathcal{A}_B^\nu$. It is straightforward to derive the following result
\begin{align}
    \delta \mathcal{A}_{B\,\mu}  = & \frac{1}{2}\lambda_\mu\left(\mathcal{L}_v-K\right)\left(\os{0}{S}-a^2\right)+\left(\os{0}{S}-a^2\right)\left(\mathcal{L}_v+\frac{1}{2}K\right)\lambda_\mu\nonumber\\
    &+\frac{1}{2}h^\nu_\mu\left(\partial_\nu+2a_\nu\right)\delta_\lambda\left(\os{0}{S}-a^2\right)+\left(\mathcal{L}_v-\frac{1}{2}K\right)\delta_\lambda\os{0}{P}_\mu\\
    &-\frac{1}{2}\left(C^\sigma{}_\mu-F^\sigma{}_\mu\right)\mathcal{L}_v\left(\mathcal{L}_v+\frac{1}{2}K\right)\lambda_\sigma-\frac{1}{2}G_\sigma\left(\delta_\lambda C^\sigma{}_\mu-\delta_\lambda F^\sigma{}_\mu\right)\nonumber\,.
\end{align}
We can show
\begin{align}
    \left(\mathcal{L}_v-\frac{1}{2}K\right)\delta_\lambda\os{0}{P}_\mu  = & \frac{1}{2}\left(C^\sigma{}_\mu-F^\sigma{}_\mu\right)\mathcal{L}_v\left(\mathcal{L}_v+\frac{1}{2}K\right)\lambda_\sigma-\left(\os{0}{S}-a^2\right)\left(\mathcal{L}_v+\frac{1}{2}K\right)\lambda_\mu\nonumber\\
    &\hspace{-1cm}+\frac{1}{2}\lambda^\alpha\mathcal{L}_v N_{\alpha\mu}+\frac{1}{2}\lambda^\alpha\mathcal{L}_v\left(\mathcal{L}_v+\frac{1}{2}K\right)F_{\alpha\mu}-\lambda_\mu\left(\mathcal{L}_v-K\right)\left(\os{0}{S}-a^2\right)\nonumber\\
    &\hspace{-1cm}-\frac{1}{2}\left(\mathcal{L}_v-\frac{1}{2}K\right)\left(h^\rho_\mu\left(\partial_\rho+a_\rho\right)\left[\left(\mathcal{D}_\alpha-2a_\alpha\right)\lambda^\alpha\right]\right)\,.
\end{align}
Using results from Section \ref{sapp:carroll-lie-covariant-commutators}, we can show that the last line can be rewritten as
\begin{equation}
    \left(\mathcal{L}_v-\frac{1}{2}K\right)\left(h^\rho_\mu\left(\partial_\rho+a_\rho\right)f\right)=h^\rho_\mu\left(\partial_\rho+2a_\rho\right)\left(\mathcal{L}_v-\frac{1}{2}K\right)f+f G_\mu\,,
\end{equation}
for any scalar $f$. Here, we take $f=\left(\mathcal{D}_\alpha-2a_\alpha\right)\lambda^\alpha$.
Similarly, we can derive
\begin{equation}
    \left(\mathcal{L}_v-\frac{1}{2}K\right)\left[\left(\mathcal{D}_\alpha-2a_\alpha\right)\lambda^\alpha\right]=\left(\mathcal{D}_\alpha-a_\alpha\right)\left[h^{\alpha\beta}\left(\mathcal{L}_v+\frac{1}{2}K\right)\lambda_\beta\right]-\lambda^\sigma G_\sigma\,.
\end{equation}
These results lead to 
\begin{align}
    \delta\mathcal{A}_{B\,\mu}  = & -\frac{1}{2}G_\sigma\left[\delta_\lambda C^\sigma{}_\mu-\delta_\lambda F^\sigma{}_\mu+h^\sigma_\mu\left(\mathcal{D}_\alpha-2a_\alpha\right)\lambda^\alpha-2h^\nu_\mu\mathcal{D}_\nu\lambda^\sigma-4a_\mu\lambda^\sigma\right]\\
    &\hspace{-1cm}+h^\nu_\mu\lambda^\sigma\mathcal{D}_\nu G_\sigma+\frac{1}{2}\lambda^\alpha\left(\mathcal{L}_v N_{\alpha\mu}+\mathcal{L}_v\left(\mathcal{L}_v+\frac{1}{2}K\right)F_{\alpha\mu}-h_{\alpha\mu}\left(\mathcal{L}_v-K\right)\left(\os{0}{S}-a^2\right)\right)\,,\nonumber
\end{align}
where we used 
\begin{align}
    &\frac{1}{2}h^\nu_\mu\left(\partial_\nu+2a_\nu\right)\left[\delta_\lambda\left(\os{0}{S}-a^2\right)-\left(\mathcal{L}_v-\frac{1}{2}K\right)\left[\left(\mathcal{D}_\alpha-2a_\alpha\right)\lambda^\alpha\right]\right]\nonumber\\
    =&\left(h^\nu_\mu\mathcal{D}_\nu\lambda^\sigma\right)G_\sigma+h^\nu_\mu\lambda^\sigma\mathcal{D}_\nu G_\sigma+2a_\mu\lambda^\sigma G_\sigma\,.
\end{align}
Using Equation \eqref{eq:S0d=2} and the Bianchi identity \eqref{eq:BianchiQ}, we obtain
\begin{equation}
    \left(\mathcal{L}_v-K\right)\left(\os{0}{S}-a^2\right)=\mathcal{D}_\mu G^\mu\,.
\end{equation}
Finally, by splitting the spatial projection of $\mathcal{D}_\nu G_\sigma$ into an antisymmetric, a STF and trace part and using the identity
\begin{equation}
    2h^\nu_\alpha h^\sigma_\beta\left(\partial_{[\nu}+a_{[\nu}\right)G_{\sigma]}=h^\nu_\alpha h^\sigma_\beta\mathcal{L}_v\left(\mathcal{L}_v+\frac{1}{2}K\right)F_{\nu\sigma}\,,
\end{equation}
we get the final result 
\begin{equation}
    \delta_\lambda\mathcal{A}_{B\,\mu}=\frac{1}{2}\lambda^\nu\left[\mathcal{L}_v N_{\mu\nu}+2h_{\langle\mu}^\rho h^\sigma_{\nu\rangle}\left(\mathcal{D}_\rho+3a_\rho\right)G_\sigma\right]\,,
\end{equation}
which is Equation~\eqref{eq:boosttrafoanom}.

\section{Overview of expansion results}
\label{app:overview-of-expansion-results}
In this appendix, we collect our main results regarding the large $r$ expansion of the bulk metric components.
This appendix exclusively uses the Carroll-covariant Bondi--Sachs gauge introduced in Section~\ref{sec:car-cov-BS-gauge},
which includes the Bondi determinant condition.
Section~\ref{sapp:on-shell-expansion-fundamental-variables} focuses on the expansion of the metric components in the Carroll-covariant Bondi--Sachs gauge,
while Section~\ref{eq:app-car-cov-S-beta-Pi-dd-expansion} contains the composite objects that we encountered in Section~\ref{sec:rewriting-EE}.

\subsection{On-shell expansion of Bondi--Sachs variables}
\label{sapp:on-shell-expansion-fundamental-variables}
As summarised in Section~\ref{sec:BS-summary},
our Carroll-covariant Bondi gauge parametrises a general $(d+2)$-dimensional bulk metric $g_{MN}$ as follows,
\begin{equation}
  \label{eq:app-car-cov-bondi-metric}
  ds^2
  = - 2 e^\beta \tau_\mu dr dx^\mu
  + \left(
    - e^{2\beta} S \tau_\mu \tau_\nu
    + \Pi_{\mu\nu}
  \right) dx^\mu dx^\nu\,.
\end{equation}
The functions $S$ and $\beta$ and the rank $d$ degenerate symmetric tensor $\Pi_{\mu\nu}$
depend on all of the $d+2$ coordinates $x^M=(r,x^\mu$).
Their radial expansion is given by\footnote{In this appendix we do not consider $\log r$ terms. These are treated separately in Section \ref{ssec:radial-expansion-logs}.}
\begin{subequations}
  \label{eq:app-car-cov-S-beta-Pi-dd-expansion}
  \begin{align}
    \beta
    &= r^{-2} \os{2}{\beta}
    + \OO(r^{-3})\,,
    \\
    S
    &= \frac{2}{d}r K
    + \os{0}{S}+r^{-1}\os{1}{S}
    + \OO(r^{-2})\,,
    \\
    \Pi_{\mu\nu}
    &= r^2 h_{\mu\nu}
    + r \left(
      C_{\mu\nu}
      - 2 \tau_{(\mu} a_{\nu)}
    \right)
    + \os{0}{\Pi}_{\mu\nu}
    +r^{-1}\os{1}{\Pi}_{\mu\nu}+\mathcal{O}(r^{-2})\,.
  \end{align}
\end{subequations}
Here we have already implemented the leading-order equations of motion derived in Section~\ref{ssec:bulk-geom-LO-EE-in-partial-gauge},
which imply that
\begin{equation}
  \label{eq:app-car-cov-bondi-LO-eom-results}
  \os{1}{\beta}
  = 0\,,
  \qquad
  h_\mu^{\rho} v^{\sigma} \os{-1}{\Pi}_{\rho\sigma}
  = a_\mu\,,
  \qquad
  \os{-1}{S}
  = \frac{2}{d} K\,,
  \qquad
  K_{\mu\nu}
  = \frac{1}{d} h_{\mu\nu} K\,.
\end{equation}
Additionally,
following~\eqref{eq:shear-tensor-initial-definition},
we have introduced the shear tensor,
\begin{equation}
  C_{\mu\nu}
  = h^\rho_{\langle \mu} h^\sigma_{\nu\rangle} \os{-1}{\Pi}_{\rho\sigma}\,,
\end{equation}
which is spatial and STF.
The Bondi determinant condition sets $h^{\mu\nu}\os{-1}{\Pi}_{\mu\nu}=0$.

\paragraph{Integration measure.}
As we showed in~\eqref{eq:gauge-fixed-det-met} and~\eqref{eq:bond-gauge-condition-as-bulk-vielbein-determinant},
the square root of the bulk metric determinant reduces to
\begin{equation}
  \label{eq:app-E-in-Bondi-gauge}
  E
  =\sqrt{\det\left(V_\mu V_\nu+\Pi_{\mu\nu}\right)}
  = r^d e^\beta e\,.
\end{equation}
This is the integration measure we use on equal-$r$ surfaces.
Here, $e$
is the integration measure on the asymptotic null boundary
defined in~\eqref{eq:app-integration-measure-def}.

\paragraph{Expansion of inverse variables.}
The inverse bulk metric is parametrised by
\begin{subequations}
  \label{eq:app-car-cov-bondi-metric-inverse}
  \begin{alignat}{2}
    g^{rr}
    &= S
    \\
    g^{r\mu}
    &= U^\mu
    &&= v^\mu
    + r\inv \os{1}{U}^\mu
    + \OO(r^{-2})\,,
    \\
    g^{\mu\nu}
    &= \Pi^{\mu\nu}
    &&= r^{-2} h^{\mu\nu}
    + r^{-3} \os{3}{\Pi}^{\mu\nu}
    + \OO(r^{-4})\,.
  \end{alignat}
\end{subequations}
The function~$S$ already appears in~\eqref{eq:app-car-cov-S-beta-Pi-dd-expansion}.
Geometrically, it determines the signature of equal-$r$ surfaces.
The remaining variables $U^\mu$ and $\Pi^{\mu\nu}$
are the inverse of the (degenerate) metric variables $V_\mu = e^\beta \tau_\mu$ and $\Pi_{\mu\nu}$ on equal-$r$ surfaces.
At leading order,
these variables correspond to the independent boundary Carroll metric variables
$(\tau_\mu, h_{\mu\nu})$
and the corresponding inverse variables
$(v^\mu, h^{\mu\nu})$.

The aforementioned $r$-dependent hypersurface metric variables satisfy the following orthogonality conditions,
\begin{equation}
  \label{eq:app-car-cov-bondi-bulk-orthogonality-completeness}
  \begin{gathered}
    -1
    = e^\beta U^\mu \tau_\mu\,,
    \qquad
    0
    = U^\mu \Pi_{\mu\nu}\,,
    \qquad
    0
    = \tau_\mu \Pi^{\mu\nu}\,,
    \qquad
    \delta^\mu_\nu
    = - e^\beta U^\mu \tau_\nu + \Pi^\mu_\nu\,.
  \end{gathered}
\end{equation}
Here, $\Pi^\mu_\nu = \Pi^{\mu\rho} \Pi_{\rho\nu}$ is the spatial projection operator
on equal-$r$ surfaces.
At leading order in the radial expansion, these equations reduce
to the corresponding boundary Carroll orthogonality relations~\eqref{eq:app-car-boundary-orthogonality-completeness},
\begin{equation}
  \label{eq:app-car-boundary-orthogonality-completeness-repeat}
  \begin{gathered}
    -1
    = v^\mu \tau_\mu\,,
    \qquad
    0
    = v^\mu h_{\mu\nu}\,,
    \qquad
    0
    = \tau_\mu h^{\mu\nu}\,,
    \qquad
    \delta^\mu_\nu
    = - v^\mu \tau_\nu + h^\mu_\nu\,.
  \end{gathered}
\end{equation}
The relations~\eqref{eq:app-car-cov-bondi-bulk-orthogonality-completeness} allow us to solve for the terms in the expansion of $U^\mu$ and $\Pi^{\mu\nu}$
given the expansion of $\Pi_{\mu\nu}$ and $\beta$.
Using the first two equations of~\eqref{eq:app-car-cov-bondi-bulk-orthogonality-completeness},
we get
\begin{equation}
  \label{eq:app-expansion-U}
  U^\mu
  = v^\mu
  - r^{-1} a^\mu
  + r^{-2} \left[
    - \os{2}{\beta} v^\mu
    - h^{\mu\rho} v^\sigma \os{0}{\Pi}_{\rho\sigma}
    + C^{\mu\rho} a_\rho
  \right]
  + \OO(r^{-3})\,.
\end{equation}
We can then expand $U^\mu U^\nu \Pi_{\mu\nu}=0$ to fix the $v^\mu v^\nu$ projections of $\Pi_{\mu\nu}$,
so that
\begin{subequations}
  \label{eq:app-expansion-vvPi}
  \begin{align}
    v^\mu v^\nu \os{-1}{\Pi}_{\mu\nu}
    &= 0\,,
    \\
    v^\mu v^\nu \os{0}{\Pi}_{\mu\nu}
    &= a^2\,,
    \\
    v^\mu v^\nu \os{1}{\Pi}_{\mu\nu}
    &= 2 a^\mu v^\nu \os{0}{\Pi}_{\mu\nu}
    - a^\mu a^\nu C_{\mu\nu}\,,\label{eq:vvPione}
  \end{align}
\end{subequations}
at the first three subleading orders.
From $\tau_\mu \Pi^{\mu\nu}=0$ we already know that only the purely spatial projections of the terms in the expansion of $\Pi^{\mu\nu}$ are nonzero.
Using the above, we can then expand
$\Pi_{\mu\rho}\Pi_{\nu\sigma}\Pi^{\rho\sigma}=\Pi_{\mu\nu}$
to obtain
\begin{align}
  \label{eq:app-expansion-PiMuNu}
  \Pi^{\mu\nu}
  &= r^{-2} h^{\mu\nu}
  - r^{-3} C^{\mu\nu}
  - r^{-4} \left[
    h^{\mu\rho} h^{\nu\sigma} \os{0}{\Pi}_{\rho\sigma}
    - C_{\mu}{}^\sigma C_{\sigma\nu}
  \right]
  + \OO(r^{-5})\,.
\end{align}
In this way, the expansion of $\beta$ and $\Pi_{\mu\nu}$ determine the subleading terms in the expansion of the inverse variables $U^\mu$ and $\Pi^{\mu\nu}$ to any order.

\paragraph{Bondi determinant condition.}
Recall from~\eqref{eq:bondi-gauge-condition} that we implemented the Carroll-covariant equivalent of the Bondi determinant condition using
\begin{equation}
  \label{eq:app-bondi-gauge-condition}
  \frac{1}{2} \Pi^{\mu\nu} \pd_r \Pi_{\mu\nu}
  = r\inv d\,.
\end{equation}
Expanding this equation allows us to fix the trace
$h^{\mu\nu}{\Pi}_{\mu\nu}$
at $r^{-n}$
for $n\geq-1$,
so that
\begin{subequations}
  \label{eq:app-bondi-gauge-condition-solving-Pi-traces}
  \begin{align}
    \label{eq:app-bondi-gauge-condition-solving-Pi-traces-m1}
    h^{\mu\nu} \os{-1}{\Pi}_{\mu\nu}
    &= 0\,,
    \\
    \label{eq:app-bondi-gauge-condition-solving-Pi-traces-0}
    h^{\mu\nu} \os{0}{\Pi}_{\mu\nu}
    &= \frac{1}{2} C^{\mu\nu} C_{\mu\nu}\,,
    \\
\label{eq:app-bondi-gauge-condition-solving-Pi-traces--1}
     h^{\mu\nu} \os{1}{\Pi}_{\mu\nu}&=  C^{\mu\nu}D_{\mu\nu}\,.
  \end{align}
\end{subequations}
Since we also fixed the double-time projections~\eqref{eq:app-expansion-vvPi} by expanding
$U^\mu U^\nu \Pi_{\mu\nu}=0$,
the remaining undetermined components in the subleading terms in the expansion of
${\Pi}_{\mu\nu}$
are therefore their mixed space-time projection and their STF part,
\begin{equation}
  \label{eq:app-unfixed-Pi-components-pre-EOM}
  v^\rho h^\sigma_\mu
  \os{n}{\Pi}_{\rho\sigma}\,,
  \qquad
  h^\rho_{\langle\mu} h^\sigma_{\nu\rangle}
  \os{n}{\Pi}_{\rho\sigma}\,.
\end{equation}

\paragraph{Consequences of bulk equations of motion.}
At this point,
our remaining independent degrees of freedom are given by the functions $\beta$ and $S$,
together with the space-time and STF components of $\Pi_{\mu\nu}$ in~\eqref{eq:app-unfixed-Pi-components-pre-EOM}.
It turns out that the function $\beta$ is fixed to all orders by the $R_{rr}=0$ bulk equation of motion, as we discussed in Section~\ref{ssec:radial-expansion-rr}.
We also saw in Section~\ref{ssec:radial-expansion-rmu-trace} that
the trace equation of motion $h^{\mu\nu} R_{\mu\nu}=0$
allows us to solve for the $r^0$~contribution to~$S$ unless $d=1$.
Finally, the $R_{\mu r}=0$ equations of motion allow us to solve for the remaining coefficients in the expansion of
$S$
and
$h_\mu^\rho v^\sigma \Pi_{\rho\sigma}$,
with the exception of
\begin{equation}
  \label{eq:app-car-cov-bondi-eom-unfixed-S-hvPi-coeffs}
  \qquad
  \os{d-1}{S}\,,
  \qquad
  h^\rho_\mu v^\sigma \os{d-1}{\Pi}_{\rho\sigma}\,.
\end{equation}
For $d=1$, the former corresponds to $\os{0}{S}$, which is then likewise not fixed by these equations of motion.
These variables parametrise the Bondi mass aspect and the Bondi angular momentum aspect.
As we show in Section~\ref{sec:bulk-conservation-equations},
their evolution along the Carroll time vector field $v^\mu$ is fixed
by the boundary diffeomorphism Ward identities,
which give a Carroll-covariant generalisation of the Bondi loss equations.

Following Section~\ref{ssec:radial-expansion-stf},
the STF part of the $\Pi^\rho_\mu \Pi^\sigma_\nu R_{\rho\sigma}=0$ bulk equations of motion
fixes the evolution of the STF part of the subleading components of $h^\rho_\mu h^\sigma_\nu\Pi_{\rho\sigma}$
along the $v^\mu$ direction,
\begin{equation}
  \label{eq:app-Pimunu-stf-evolution}
  \LL_v \left(
    h^\rho_{\langle\mu} h^\sigma_{\nu\rangle}
    \os{n}{\Pi}_{\rho\sigma}
  \right)
  = \cdots,
  \qquad
  n \geq 0\,.
\end{equation}
The STF constraint on the extrinsic curvature from the leading-order equations of motion~\eqref{eq:app-car-cov-bondi-LO-eom-results}
can be seen as the order $r^2$ version of this statement.
On the other hand, the STF equation vanishes identically at order $r$,
and the evolution of the shear $C_{\mu\nu}$ is not fixed.
For $d=2$
the evolution equation~\eqref{eq:app-Pimunu-stf-evolution} at $r^0$ leads to\footnote{Since we cannot algebraically solve for $h^\rho_{\langle\mu} h^\sigma_{\nu\rangle}
    \os{n}{\Pi}_{\rho\sigma}$, obtaining a unique solution near future null infinity requires specifying initial data for $h^\rho_{\langle\mu} h^\sigma_{\nu\rangle}
    \os{n}{\Pi}_{\rho\sigma}$ at some instant of boundary time. This is an all order in $r$ set of initial conditions.}
\begin{equation}
  \label{eq:app-Dmunu-def-and-evolution}
  \LL_v D_{\mu\nu}
  = 0\,,
  \qquad
  D_{\mu\nu}
  =h^{\rho}_\mu h^\sigma_\nu\left(\os{0}{\Pi}_{\rho\sigma}-\frac{1}{4}C^2h_{\rho\sigma}\right)
  -\frac{1}{2}F_{\mu\sigma} C^\sigma{}_\nu\,.
\end{equation}
Note that $D_{\mu\nu}$ is STF due to~\eqref{eq:app-bondi-gauge-condition-solving-Pi-traces-0} above.
This tensor also arises at order $r^{-3}$ in the $\Pi_\kappa^\mu R_{\mu r}=0$ equations of motion in $d=2$,
which imposes the constraint
\begin{equation}
  \label{eq:app-cov-der-of-Dmn-constraint}
   h^{\mu\nu}\mathcal{D}_\mu D_{\nu\kappa}=0\,,
\end{equation}
following~\eqref{eq:PDZ}.
As we discussed in Section~\ref{ssec:radial-expansion-logs},
this equation can be removed by adding logarithmic terms to the radial expansion in~\eqref{eq:app-car-cov-S-beta-Pi-dd-expansion}.

\paragraph{Three-dimensional expansion.}
For $d=1$,
which corresponds to a three-dimensional bulk,
the situation simplifies significantly.
First, all antisymmetric and all STF tensors vanish,
so that $F_{\mu\nu}=0$
and $h_{\langle \mu}^\rho h_{\nu\rangle}^\sigma \Pi_{\rho\sigma} = 0$
at all orders.
In particular,
there is no shear and no $D_{\mu\nu}$ tensor.
We also get $\beta=0$ at all orders, as discussed in Section~\ref{ssec:rewriting-EE-3d-simplifications}.
Furthermore, we saw in Section~\ref{ssec:radial-expansion-threedim}
that the three-dimensional radial expansion terminates.
Specifically, from~\eqref{eq:three-dim-radial-expansion-terminates-preview} we have
\begin{equation}
  \pd_r^3 g_{MN}
  = 0.
\end{equation}
Applying this condition to the parametrisation~\eqref{eq:app-car-cov-bondi-metric},
and taking into account~\eqref{eq:app-expansion-vvPi} and~\eqref{eq:app-bondi-gauge-condition-solving-Pi-traces} above,
we see that
\begin{align}
  ds^2
  &= - 2\tau_\mu dr dx^\mu
  + \left(- S\tau_\mu \tau_\nu + \Pi_{\mu\nu}\right) dx^\mu dx^\nu
  \\
  \label{eq:app-3Dsol}
  &= - 2\tau_\mu dr dx^\mu
  + r^2 h_{\mu\nu}dx^\mu dx^\nu
  -2r \left(K \tau_\mu \tau_\nu + \tau_{(\mu} a_{\nu)}\right) dx^\mu dx^\nu
  \\
  &{}\qquad\nonumber
  - \left(\os{0}{S} - a^2\right)\tau_\mu \tau_\nu dx^\mu dx^\nu
  - 2 \tau_{(\mu} v^\rho h^\sigma_{\nu)} \os{0}{\Pi}_{\rho\sigma} dx^\mu dx^\nu\,.
\end{align}
At subleading orders, this only leaves the following components,
\begin{equation}
  \label{eq:app-3Dsol-S0-hvPi0-arbitrary}
  \os{0}{S}\,,
  \qquad
  P_\mu
  = h_\mu^\rho v^\sigma \os{0}{\Pi}_{\rho\sigma}\,,
\end{equation}
in accordance with~\eqref{eq:app-car-cov-bondi-eom-unfixed-S-hvPi-coeffs}.
They parametrise the Bondi mass and the Bondi angular momentum aspects.
For reference,
in terms of our Carroll-covariant Bondi--Sachs variables,
this bulk metric corresponds to
\begin{subequations}
  \label{eq:app-3Dsol-components}
  \begin{align}
    \label{eq:app-3Dsol-components-beta}
    \beta
    &= 0\,,
    \\
    \label{eq:app-3Dsol-components-S}
    S
    &= 2rK
    + \os{0}{S}
    + 2r\inv a^\mu P_\mu
    + r^{-2} P^2
    \\
    \label{eq:app-app-3Dsol-components-Pi}
    \Pi_{\mu\nu}
    &= r^2 h_{\mu\nu}
    - 2r \tau_{(\mu}a_{\nu)}
    + a^2 \tau_\mu \tau_\mu
    - 2 \tau_{(\mu}P_{\nu)}
    \\
    &{}\qquad\nonumber
    + 2r\inv a^\rho P_\rho \tau_\mu \tau_\nu
    + r^{-2} P^2 \tau_\mu \tau_\nu\,,
    \\
    \label{eq:app-3Dsol-components-U}
    U^\mu
    &= v^\mu - r\inv a^\mu - r^{-2} P^\mu\,,
    \\
    \label{eq:app-3Dsol-components-inverse-Pi}
    \Pi^{\mu\nu}
    &= r^{-2} h^{\mu\nu}\,.
  \end{align}
\end{subequations}
corresponding to~\eqref{eq:3Dsol-inverse-components} and~\eqref{eq:3Dsol-components-Pi}
in the main text.

\paragraph{Four-dimensional expansion.}
It is also useful to collect some explicit expressions for the subleading terms in the $d=2$ expansion,
corresponding to four bulk spacetime dimensions.
In addition to the leading-order terms in~\eqref{eq:app-car-cov-S-beta-Pi-dd-expansion},
we obtain
\begin{align}
  \label{eq:app-4Dsol-Pi0}
  \os{0}{\Pi}_{\mu\nu}
  &= a^2\tau_\mu\tau_\nu
  -2\tau_{(\mu}\os{0}{P}_{\nu)}
  +D_{\mu\nu}
  +\frac{1}{2}F_{\mu\rho}C^\rho{}_\nu
  +\frac{1}{4}h_{\mu\nu}C^2\,,
  \\
  \label{eq:app-4Dsol-Pi1}
  \os{1}{\Pi}_{\mu\nu}
  &= \left(2a^\rho \os{0}{P}_\rho
  -C^{\rho\sigma}a_\rho a_\sigma\right)\tau_\mu\tau_\nu
  -2\tau_{(\mu}\os{1}{P}_{\nu)}
  +h^\rho_\mu h^\sigma_\nu\os{1}{\Pi}_{\rho\sigma}\,,
  \\
  \label{eq:app-4Dsol-S0}
  \os{0}{S}
  &=  \frac{1}{2}h^{\mu\nu}\mathcal{R}_{\mu\nu}+\frac{3}{2}\mathcal{D}_\mu a^\mu=\frac{1}{2}\mathcal{Q}+\mathcal{D}_\mu a^\mu\,,
  \\
  \label{eq:app-4Dsol-beta2}
  \os{2}{\beta}
  &= \frac{1}{16}\left(F^2-C^2\right)\,.
\end{align}
We do not need $h^\rho_\mu h^\sigma_\nu\Pi_{\rho\sigma}$ at order $r^{-1}$ for the purposes of this paper.
In the above,
we used the following abbreviations,
\begin{align}
  \label{eq:app-4Dsol-P0-def}
  \os{0}{P}_\mu
  &= h_\mu^\rho v^\sigma \os{0}{\Pi}_{\rho\sigma}
  = -\frac{1}{2}\left(\mathcal{D}_\rho-2a_\rho\right)C^\rho{}_\mu
  +\frac{1}{2}\mathcal{D}_\rho F^\rho{}_\mu\,,
  \\
  \label{eq:app-4Dsol-P1-def}
  \os{1}{P}_\mu
  &= h_\mu^\rho v^\sigma \os{1}{\Pi}_{\rho\sigma}\,.
\end{align}
The latter parametrises the Bondi angular momentum,
and $\os{1}{S}$ parametrises the Bondi mass. We refer to Appendix \ref{app:curvten} for the definition of the boundary Ricci tensor
$\mathcal{R}_{\mu\nu}$ and the associated scalar $\mathcal{Q}$.

\subsection{Expansions of composite objects}
\label{sapp:intermediate-results-expansions}
In Section~\ref{sec:rewriting-EE},
we rewrote the bulk Einstein equations in terms of the metric variables
of our Carroll-covariant Bondi gauge.
There, it was useful to introduce $r$-dependent generalisations
of some of the natural objects that appear in the asymptotic Carrollian geometry.
Following~\eqref{eq:curly-A-K-F-def-repeat},
we have
\begin{subequations}
  \label{eq:app-curly-A-K-F-def-repeat}
  \begin{gather}
    \mathcal{K}_{\mu\nu}
    = -\frac{1}{2} \LL_U \Pi_{\mu\nu}
    = r^2 K_{\mu\nu}
    + \OO(r)\,,
    \qquad
    \mathcal{K}
    = \Pi^{\mu\nu} \mathcal{K}_{\mu\nu}
    = K
    + \OO(r\inv)\,,
    \\
    \mathcal{A}_\mu
    = \LL_U V_\mu
    = a_\mu
    + \OO(r\inv)\,,
    \qquad
    \mathcal{F}_{\mu\nu}
    = 2\Pi_\mu^\rho \Pi_\nu^\sigma \pd_{[\rho} V_{\sigma]}
    = F_{\mu\nu}
    + \OO(r\inv)\,,
  \end{gather}
\end{subequations}
where
$K_{\mu\nu}$,
$K$,
$a_\mu$
and
$F_{\mu\nu}$
are the boundary extrinsic curvature,
its trace,
the acceleration
and the twist tensor,
as defined in~\eqref{eq:app-extrinsic-curvature-trace-acceleration-def}
and~\eqref{eq:app-twist-def} above.

Motivated by the constraints~\eqref{eq:app-car-cov-bondi-LO-eom-results} imposed by the leading-order equations of motion,
we also introduced the following `composite' objects
in~\eqref{eq:KTDefn}
and~\eqref{eq:BarCalK-def},
\begin{equation}
  \label{eq:app-KT-BarCalK-def}
  \mathcal{K}^T_{\mu\nu}
  = \mathcal{K}_{\mu\nu}
  - \frac{1}{d} \Pi_{\mu\nu} \mathcal{K}
  = \OO(r)\,,
  \qquad
  \bar{\mathcal{K}}
  = \mathcal{K}
  - \frac{d}{2r} S
  = \OO(r\inv)\,.
\end{equation}
These objects are defined such that the leading-order terms of their individual terms cancel
due to the equations of motion~\eqref{eq:app-car-cov-bondi-LO-eom-results}.
Note that $\mathcal{K}^T_{\mu\nu}$
is symmetric trace-free (STF) with respect to the $r$-dependent spatial tensor $\Pi^{\mu\nu}$,
but the terms in its radial expansion will generically not be STF with respect to the $h^{\mu\nu}$ boundary spatial tensor except at leading order.

Using similar arguments as in~\eqref{eq:app-K-is-total-derivative},
we can write the trace of the $r$-dependent extrinsic curvature as follows,
\begin{equation}
  \label{eq:app-calK-is-total-derivative}
  \mathcal{K}
  = - E\inv \pd_\rho \left(E U^\rho\right)
  = - e\inv \pd_\rho\left(e U^\rho\right)
  - U^\rho \pd_\rho \beta\,,
\end{equation}
where we used the expression for $E$ in~\eqref{eq:app-E-in-Bondi-gauge} in the second equality.
This expression for $\mathcal{K}$ is very convenient when computing its expansion.

In the course of our rewriting
of the bulk equations of motion,
we came across several other convenient composite variables.
In~\eqref{eq:curly-G-def} and~\eqref{eq:calZ}
we introduced
\begin{subequations}
  \begin{gather}
    \label{eq:app-curly-G-def}
    \mathcal{G}_{\mu\nu}
    = \Pi_\mu^\rho \Pi_\nu^\sigma \left(
      \pd_r \Pi_{\rho\sigma}
      - 2 r\inv \Pi_{\rho\sigma}
    \right)
    = - C_{\mu\nu}
    + \OO(r\inv)\,,
    \\
    \label{eq:app-calZ}
    \mathcal{Z}_\mu
    = \Pi_{\mu\nu}\partial_r U^\nu
    - \mathcal{A}_\mu
    = \OO(r^{-1})\,.
  \end{gather}
\end{subequations}
The first tensor $\mathcal{G}_{\mu\nu}$ can be interpreted as a generalisation of the asymptotic shear tensor $C_{\mu\nu}$ to arbitrary equal-$r$ surfaces,
while $\mathcal{Z}_\mu$ parametrises the $r$-dependence of the $U^\mu$ vector field.
Again, the leading-order contributions of their constituent terms cancel.
Both terms are spatial
with respect to the $r$-dependent $\Pi_{\mu\nu}$ spatial tensors,
and $\mathcal{G}_{\mu\nu}$ is STF.
The first nonzero contribution to the expansion of~$\mathcal{Z}_\mu$ is
\begin{eqnarray}
  \os{1}{\mathcal{Z}}_{\mu}
  & = & -2h_{\mu\rho}\os{2}{U}^\rho+a_\rho C^\rho{}_\mu+a_\rho F^\rho{}_\mu\nonumber\\
  & = & 2h^\rho_\mu v^\sigma \os{0}{\Pi}_{\rho\sigma}
  + a_\rho F^\rho{}_\mu
  - a_\rho C^\rho{}_\mu\,,\label{eq:app-curly-Z-expansion-d2-r1-with-U2-or-hvPi0}
\end{eqnarray}
using the expansion of $U^\mu$ in~\eqref{eq:app-expansion-U} as well as
\begin{equation}
    \overset{(1)}{\mathcal{A}}_\mu =-a^{\rho}\tau_{\rho\mu}=-a^{\rho}F_{\rho\mu}-a^2\tau_\mu\,.
\end{equation}
Likewise, at the next order, we have
\begin{align}
  \os{2}{\mathcal{Z}}_{\mu}
  &= -3h_{\mu\nu}\os{3}{U}{}^\nu
  -2\os{-1}{\Pi}_{\mu\nu}\os{2}{U}{}^\nu
  +\os{0}{\Pi}_{\mu\nu}a^\nu
  -\os{2}{\mathcal{A}}{}_\mu\nonumber
  \\
  &= 3\os{1}{\Pi}_{\mu\nu} v^\nu
  - 2\os{0}{\Pi}_{\mu\nu} a^\nu
  + \os{-1}{\Pi}_{\mu\nu}\os{2}{U}^\nu
  - \os{2}{\mathcal A}_\mu\,,\label{eq:calZ2}
\end{align}
where we used $U^\mu\Pi_{\mu\nu}=0$ to solve for $h_{\mu\nu}\os{3}{U}{}^\nu$ in terms of $\os{1}{\Pi}_{\mu\nu} v^\nu$.
In particular, note that
$h_\mu^\rho v^\sigma \os{1}{\Pi}_{\rho\sigma}$
enters in
$\os{2}{\mathcal{Z}}_\mu$
through the first term on the second line above.
We will give the explicit on-shell $d=2$ expressions below.

\paragraph{Subleading expressions in four dimensions.}
We now collect some explicit results for the subleading terms in the expansion of the composite objects defined above.
Here we consider the case of $d=2$ with no logs.

We start with $\mathcal{G}_{\mu\nu}$ which is given in \eqref{eq:app-curly-G-def}. At order $r^{-1}$ we find
\begin{eqnarray}
    \os{1}{\mathcal{G}}_{\mu\nu} &=& -2D_{\mu\nu}-\frac{1}{2}C^2 h_{\mu\nu}-F_{\mu\rho}C^\rho{}_\nu + 2a^\rho C_{\rho(\mu}\tau_{\nu)}
    \label{eq:app-CalG1}\,,
\end{eqnarray}
where we used \eqref{eq:app-4Dsol-Pi0}.
Next we consider $\mathcal{Z}_\mu$ which we will need at orders $r^{-1}$ and $r^{-2}$. Using \eqref{eq:app-curly-Z-expansion-d2-r1-with-U2-or-hvPi0} and \eqref{eq:app-4Dsol-P0-def} we have
\begin{equation}\label{eq:CalZ1new}
    \os{1}{\mathcal{Z}}_\mu =  -(\mathcal{D}_\rho -a_\rho)C^{\rho}{}_\mu+\mathcal{D}_\rho F^{\rho}{}_\mu+a_\rho F^{\rho}{}_\mu\,,
\end{equation}
which agrees with \eqref{eq:Z-eq}.
At the next subleading order,
we have $\os{2}{\mathcal{Z}}_{\mu}$ given by \eqref{eq:calZ2}.
We will not need the temporal component of this expression.
Note that this expression contains
$h_\mu^\rho v^\sigma \os{1}{\Pi}_{\rho\sigma}$,
which we identified in~\eqref{eq:app-4Dsol-P1-def} as the term parametrising the Bondi angular momentum.
Following~\eqref{eq:app-expansion-U},
we have
\begin{align}
  \os{2}{U}^\mu
  &= - \os{2}{\beta} v^\mu
  - h^{\mu\rho} v^\sigma \os{0}{\Pi}_{\rho\sigma}
  + C^{\mu\rho} a_\rho\nonumber
  \\
  &=
  \frac{1}{16}v^\mu \left(C^2 - F^2\right)
  + \frac{1}{2}\mathcal{D}_\rho C^{\rho\mu}
  - \frac{1}{2}h^\mu_\sigma\mathcal{D}_\rho F^{\rho\sigma}\,,\label{eq:app-expansion-U-d2-r2}
\end{align}
so that \eqref{eq:calZ2} leads to
\begin{equation}\label{eq:genZ2}
    h^\mu_\kappa\os{2}{\mathcal{Z}}_\mu=3 h^\mu_\kappa v^\nu\os{1}{\Pi}_{\mu\nu}-\frac{1}{16}h^\mu_\kappa\left(\partial_\mu+a_\mu\right)\left(F^2-C^2\right)-2a^\mu D_{\mu\kappa}+\os{0}{P}^\mu\left(F_{\mu\kappa}-C_{\mu\kappa}\right)\,,
\end{equation}
where we used
\begin{equation}
    h^\mu_\kappa\overset{(2)}{\mathcal{A}}{}_\mu=\overset{(2)}{U}{}^\sigma F_{\sigma\kappa}+h^\mu_\kappa\partial_\mu\overset{(2)}{\beta}\,,
\end{equation}
as well as \eqref{eq:app-4Dsol-Pi0}.

Finally, we consider the expansion of $\mathcal{K}_{\mu\nu}$ and related objects such as $\mathcal{K}^T_{\mu\nu}$ and $\mathcal{K}$. We start with the trace $\mathcal{K}$.
Using the expression for~$\mathcal{K}$ in~\eqref{eq:app-calK-is-total-derivative}
together with the expansion of $U^\mu$ in~\eqref{eq:app-expansion-U},
\begin{equation}
  \label{eq:app-curly-K-expansion-d2-r-1}
  \os{1}{\mathcal{K}}
  =-e^{-1}\partial_\mu\left(e\overset{(1)}{U}{}^\mu\right)
  = e\inv \pd_\mu \left(e a^\mu\right)
  = \mathcal{D}_\mu a^\mu
  =h^{\mu\nu}\left(\mathcal{D}_{\mu}a_{\nu}+a_{\mu}a_{\nu}\right),
\end{equation}
at leading order.
Here, we have included various alternative forms of the same expression that are used throughout the main text.
Next,
using~\eqref{eq:app-expansion-U-d2-r2} and 
\eqref{eq:app-4Dsol-beta2}
we get
\begin{align}
  \os{2}{\mathcal{K}}
  &=-e^{-1}\partial_\mu\left(e\overset{(2)}{U}{}^\mu\right)
  -\mathcal{L}_v\overset{(2)}{\beta}\,,\nonumber
  \\
  & =  \frac{1}{4}\mathcal{L}_v F^2-K\os{2}{\beta}-\frac{1}{2}KF^2-\frac{1}{2}a^\rho\mathcal{D}_\mu C^\mu{}_\rho\nonumber\\
  &\qquad+\frac{1}{2}a^\rho\mathcal{D}_\mu F^\mu{}_\rho-\frac{1}{2}C^{\mu\rho}A_{\mu\rho}+\frac{1}{2}\mathcal{D}_\mu\left(h^{\mu\nu}\os{1}{\mathcal{Z}}_\nu\right)\nonumber\\
  &= -\frac{1}{2}\mathcal{D}_\mu\mathcal{D}_\nu C^{\mu\nu}+\frac{1}{16}KC^2+\frac{3}{16}KF^2-\frac{1}{4}\mathcal{L}_v F^2\,,\label{eq:app-curly-K-expansion-d2-r-2}
\end{align}
at the first subleading order. Finally, we consider 
$\mathcal{K}_{\mu\nu}^T=\mathcal{K}_{\mu\nu}-\frac{1}{2}\mathcal{K}\Pi_{\mu\nu}$
where
$\mathcal{K}_{\mu\nu}=-\frac{1}{2}\mathcal{L}_U\Pi_{\mu\nu}$. At leading order which is order $r$, we have that $\overset{(-1)}{\mathcal{K}}{}^T_{\mu\nu}$ is given by
\begin{eqnarray}
  \overset{(-1)}{\mathcal{K}}{}^T_{\mu\nu}
  & = & \os{-1}{\mathcal{K}}_{\mu\nu}-\frac{1}{2}\os{1}{\mathcal{K}}h_{\mu\nu}-\frac{1}{2}K\os{-1}{\Pi}_{\mu\nu}\nonumber\\
  & = & -\frac{1}{2}\mathcal{L}_v\os{-1}{\Pi}_{\mu\nu}+\frac{1}{2}\mathcal{L}_a h_{\mu\nu}-\frac{1}{2}\os{1}{\mathcal{K}}h_{\mu\nu}-\frac{1}{2}K\os{-1}{\Pi}_{\mu\nu}\nonumber\\
  &=&  \frac{1}{2} N_{\mu\nu}
  - \frac{1}{4}K C_{\mu\nu}
  + A_{\mu\nu}\,.\label{eq:app-curly-KT-expansion-d2-r1}
\end{eqnarray}
This object is spatial and STF. At the next order we have $\overset{(0)}{\mathcal{K}}{}^T_{\mu\nu}$. This object was written in terms of $\overset{(0)}{\mathcal{K}}{}_{\langle\mu\nu\rangle}$ in equation \eqref{eq:calKT0v2} where
\begin{equation}
    \os{0}{\mathcal{K}}_{\langle\mu\nu\rangle}=h^\rho_\mu h^\sigma_\nu\os{0}{\mathcal{K}}_{\rho\sigma}-\frac{1}{2}h_{\mu\nu}h^{\rho\sigma}\os{0}{\mathcal{K}}_{\rho\sigma}\,.
\end{equation}
Hence, it remains to compute $\os{0}{\mathcal{K}}_{\langle\mu\nu\rangle}$ where
\begin{equation}
    \overset{(0)}{\mathcal{K}}{}_{\mu\nu}=-\frac{1}{2}\mathcal{L}_v\overset{(0)}{\Pi}_{\mu\nu}+\frac{1}{2}\mathcal{L}_a\overset{(-1)}{\Pi}_{\mu\nu}-\frac{1}{2}\mathcal{L}_{\overset{(2)}{U}}h_{\mu\nu}\,.
\end{equation}
Using \eqref{eq:app-4Dsol-Pi0} and \eqref{eq:app-expansion-U-d2-r2} we obtain
\begin{eqnarray}
    \os{0}{\mathcal{K}}_{\langle\mu\nu\rangle} & = & h^\rho_{\langle\mu}h^\sigma_{\nu\rangle}\os{0}{\mathcal{K}}_{\rho\sigma} \nonumber\\
    & = &  h^\rho_{\langle\mu}h^\sigma_{\nu\rangle}\left(\left(\mathcal{D}_\rho+a_\rho\right)\os{0}{P}_\sigma-\frac{1}{4}\mathcal{L}_v\left(F_{\rho\alpha}C^\alpha{}_\sigma\right)-a^\alpha F_{\alpha\rho}a_\sigma\right.\nonumber\\
    &&\left.+\frac{1}{2}a^\alpha\left(\mathcal{D}_\alpha C_{\rho\sigma}-\mathcal{D}_\rho C_{\sigma\alpha}-\mathcal{D}_\sigma C_{\rho\alpha}\right)\right)\,.\label{eq:STFpartofcalK0}
\end{eqnarray}

\section{Further details on EOM expansions}
\label{app:intermediate-results}
In Section~\ref{sec:bulk-conservation-equations} we suppressed details on relevant intermediate calculations in order to more quickly get to the main result. This concerns the $d=2$ computation of the Bondi loss equations by expanding $\Pi_\kappa^\mu U^\nu R_{\mu\nu}=0$ and $U^\mu U^\nu R_{\mu\nu}=0$ to order~$r^{-2}$. In this appendix we provide the necessary details. 

\subsection{Boundary Bianchi identities and the loss equations at \texorpdfstring{$\mathcal{O}(r^{-1})$}{order r(-1)}}
\label{ssapp:order-r-m1-EOM}
As outlined in Section~\ref{sec:radial-expansion}, it is redundant to verify that the  projections $U^\mu U^\nu R_{\mu\nu}=0$ and $\Pi^\mu_\kappa U^\nu R_{\mu\nu}=0$ vanish at any order save $\mathcal{O}(r^{-2})$. The bulk Bianchi identities allow us to construct a minimal list of projections, the vanishing of which automatically ensure the vanishing of the bulk Ricci tensor and, by extension, the above two projections save at $\mathcal{O}(r^{-2})$.

This is not to say that solving them is not useful. The automatic vanishing of these equations order by order involves complicated combinations of the boundary geometry (among other things). Actually proving that these terms sum to zero requires using boundary identities such as Bianchi identities for the boundary curvature tensor. 
Turning this around, we see that these equations can therefore serve as an interesting source for Bianchi identities or other non-trivial identities based on the boundary geometry. This in itself sounds like a good reason to study these equations. Additionally, such identities at lower order can also be very useful for simplifying equations at higher order in the radial expansion.

Following this thread, we now explicitly demonstrate the vanishing of both loss equations in $d=2$ at $\mathcal{O}(r^{-1})$. Doing so will yield a pair of Bianchi identities that will be of particular relevance for the loss equations one order lower.

\subsubsection{Evaluating \texorpdfstring{$U^\mu U^\nu R_{\mu\nu}=0$}{Umu Unu Rmunu} at \texorpdfstring{$\mathcal{O}(r^{-1})$}{order r(-1)}} 
\label{ssapp:order-r-m1-U-U-R-EOM}
As we have seen from the trace equation, the object $\os{0}S$ in $d=2$ encodes properties of the boundary geometry. In particular, $\os{0}S$ depends on the Ricci scalar of the boundary curvature as in Equation \eqref{eq:S0d=2}, which we repeat here for convenience,
\begin{equation}
    2\os{0}{S}=h^{\mu\nu}\mathcal{R}_{\mu\nu}+3\mathcal{D}_\mu a^\mu=\mathcal{Q}+2\mathcal{D}_\mu a^\mu\,.
\end{equation}
We can expand out $U^\mu U^\nu R_{\mu\nu} = 0$ at $\mathcal{O}(r^{-1})$. For $d=2$, using \eqref{eq:newformUUR} (where only the terms on the first line contribute), this works out to be 
\begin{equation}\label{eq:UURr-1}
     e^{-1}\partial_\mu\left(e \left[h^{\mu\nu}\partial_\nu K+K a^\mu\right]\right)-2K\os{1}{\bar{\mathcal{K}}}+2\mathcal{L}_v\os{1}{\bar{\mathcal{K}}}=0\,.
\end{equation}
Next, we observe that
\begin{equation}
    \os{1}{\bar{\mathcal{K}}}=\os{1}{\mathcal{K}}-\os{0}{S}=\mathcal{D}_\mu a^\mu-\os{0}{S}=-\frac{1}{2}\mathcal{Q}\,.
\end{equation}
We thus see that \eqref{eq:UURr-1} is obeyed by virtue of the Bianchi identity \eqref{eq:BianchiQ}.

\subsubsection{Evaluating \texorpdfstring{$\Pi^\mu_\kappa U^\nu R_{\mu\nu}=0$}{Pi-U-Rmn} at \texorpdfstring{$\mathcal{O}(r^{-1})$}{order -1}} 
\label{ssapp:order-r-m1-Pi-U-R-EOM}

We now turn to the second loss equation at $\mathcal{O}(r^{-1})$.
The $U^\mu \Pi^\nu_\kappa R_{\mu\nu}=0$ equation at this order, for $d=2$, reads
\begin{align}
0 & =  \overset{(0)}{S}a_\kappa-\frac{1}{2}\partial_\rho K F^\rho{}_\kappa-\frac{1}{2}K\left(\mathcal{D}_\rho F^\rho{}_\kappa+a_\rho F^\rho{}_\kappa\right)-\frac{1}{2}K a_\rho C^\rho{}_\kappa-\frac{1}{2}\partial_\rho K C^\rho{}_\kappa\nonumber\\
&\quad\,+\frac{1}{2}h^\rho_\kappa\partial_\rho\overset{(1)}{\mathcal{K}}-\frac{1}{2}a_\kappa \overset{(1)}{\mathcal{K}}+\frac{1}{2}\mathcal{L}_v\overset{(1)}{\mathcal{Z}}_\kappa-\mathcal{D}_\rho\left(h^{\rho\sigma}\overset{(-1)}{\mathcal{K}}{}^T_{\sigma\kappa}\right)\,,\label{eq:UPiRr-1}
\end{align}
and it is helpful to recall that $\os{1}{\mathcal{K}}=\mathcal{D}_\sigma a^\sigma$ and 
\begin{equation}
\begin{split}
     \os{1}{\mathcal{Z}}_\kappa 
    &= -  \mathcal{D}_\rho C^\rho{_\kappa}  + a_\rho C^\rho{_\kappa} +   \mathcal{D}_\rho F^\rho{_\kappa}+ a_\rho F^\rho{_\kappa}\,,
\end{split}
\end{equation}
as well as
\begin{equation}
  \overset{(-1)}{\mathcal{K}}{}^{T}_{\mu\nu}
  =\frac{1}{2}N_{\mu\nu}
  -\frac{1}{4}KC_{\mu\nu}
  +A_{\mu\nu}\,.
\end{equation}
Unlike our previous calculation, $U^\mu \Pi^\nu_\kappa R_{\mu\nu}=0$ at $\mathcal{O}(r^{-1})$ contains both terms that do not contain the shear tensor and terms that are linear in the shear tensor. Naturally, the two sets of terms must be separated and individually shown to vanish. We will be brief with our exposition of the latter, since this does not require any new non-trivial identities. Indeed, the terms linear in the shear can be seen to cancel by virtue of the identity \eqref{eq:comLiecovSTF} that involves commuting a covariant divergence with a Lie derivative along $v$ when acting on the shear. 
This leaves us with the shear-independent terms, which amount to
\begin{align}
\label{eq:BId2R1}
0 & =  \frac{1}{2}(\mathcal{L}_v-K)\left(\mathcal{D}_\rho F^\rho{}_\kappa+a_\rho F^\rho{}_\kappa\right)-\frac{1}{2}\partial_\rho K F^\rho{}_\kappa\nonumber\\
&\qquad+\frac{1}{2}\mathcal{Q}a_\kappa-a_\kappa \mathcal{D}_\sigma a^\sigma+\frac{1}{2}h^\rho_\kappa(\partial_\rho+3a_\rho)\mathcal{D}_\sigma a^\sigma-\mathcal{D}_\rho A^\rho{}_\kappa\,,
\end{align}
where we used \eqref{eq:S0d=2}. This equation is identically satisfied due to the Bianchi identity \eqref{eq:BItwo}. We conclude that the $\Pi^\mu_\kappa U^\nu R_{\mu\nu}=0$ projection vanishes identically at $\mathcal{O}(r^{-1})$ as well.

We note that \eqref{eq:BId2R1} takes the form of the spatial projection of the diffeomorphism Ward identity (see for example Equation \eqref{eq:diffeoWIspatialproj-repeat}).\footnote{Equation \eqref{eq:BianchiQ} can likewise be recast into the form of the temporal projection of the diffeomorphism Ward identity.} Since this equation is identically satisfied it takes the form of an improvement transformation. If this improvement transformation is variational, we only need to guess the term whose variation under diffeomorphisms leads to the identity above. We saw below equation \eqref{eq:twicecontracedBI} that the diffeomorphism-invariance of $\int d^3x e \mathcal{Q}$ leads to the Bianchi identity \eqref{eq:twicecontracedBI2} whose temporal and spatial projections give the Bianchi identities needed to check that $U^\mu U^\nu R_{\mu\nu} = 0$ and $\Pi_\kappa^\mu U^\nu R_{\mu\nu} = 0$ are satisfied at $\mathcal{O}(r^{-1})$.

\subsection{Evaluating the loss equations at \texorpdfstring{$\mathcal{O}(r^{-2})$}{order -2}}
\label{app:LossEqnLists}
In this appendix, we briefly sketch the calculations that allow us to go from the all-orders loss equations~\eqref{eq:U-U-R-d2} and~\eqref{eq:Pi-U-R-d2} to the evolution equations~\eqref{eq:R1UUR2} and~\eqref{eq:UPiRinterm} at order $r^{-2}$ in the radial expansion of the former.
We begin with the latter.

\subsubsection{Evaluating \texorpdfstring{$\Pi^\mu_\kappa U^\nu R_{\mu\nu}=0$}{Pi-U-Rmn} at \texorpdfstring{$\mathcal{O}(r^{-2})$}{order -2}}
\label{app:LossEqnLists-momentum}

We begin with~\eqref{eq:Pi-U-R-d2} which reads
\begin{align}
  0
  = \Pi_\kappa^\mu U^\nu R_{\mu\nu}
  &= -\frac{1}{2}\mathcal{F}^\rho{}_\kappa\left(\partial_\rho +\mathcal{A}_\rho \right)S+\frac{1}{2}\mathcal{G}^\rho{}_\kappa\left(\partial_\rho +\mathcal{A}_\rho \right)S\label{eq:appDUPiR}
  \\
  &{}\qquad\nonumber 
  -\partial_r\beta\Pi^\alpha_\kappa\left(\partial_\alpha +\mathcal{A}_\alpha \right)S-\frac{1}{2}\Pi^\alpha_\kappa\left(\partial_\alpha+\mathcal{A}_\alpha\right)\left(\partial_r S-r^{-1} S\right)
  \\
  &{}\qquad\nonumber
  -\frac{1}{2}S\Pi^\mu_\kappa\hat D_\rho \mathcal{F}^\rho{}_\mu+\frac{1}{2}S\Pi^\rho_\kappa\left(
    \partial_r\mathcal{Z}_\rho+r^{-1}\mathcal{Z}_\rho
  \right)
  \\
  &{}\qquad\nonumber
  +\frac{1}{2}\left(\mathcal{L}_U-\bar{\mathcal{K}}\right)\mathcal{Z}_\kappa
  +\frac{1}{2}\Pi^\mu_\kappa \left(\partial_\mu-\mathcal{A}_\mu\right)\bar{\mathcal{K}}
  -\Pi^\mu_\kappa\hat D_\rho \mathcal{K}^{T\rho}{}_{\mu}\,.
\end{align}

The first step is to evaluate each term above at the order of interest. 
Because the $U^\kappa$ projection vanishes, we can restrict our attention to the $h^\kappa_\alpha$ projection of $\Pi_\kappa^\mu U^\nu R_{\mu\nu}=0$. This is because the $v^\kappa$ projection is guaranteed to be obeyed once the equation is obeyed at order $r^{-1}$. Using the expansion results of Appendix \ref{app:overview-of-expansion-results}, we then obtain the following list of terms:
\begin{subequations}
\begin{align}
    -\frac{1}{2}h^\kappa_\alpha\mathcal{F}^\rho{}_\kappa\left(\partial_\rho +\mathcal{A}_\rho \right)S\Big\vert_{@r^{-2}} & =  \frac{1}{2}\partial_\rho K C^{\rho\mu}F_{\mu\alpha}-\frac{1}{2}\partial_\rho\overset{(0)}{S}F^\rho{}_\alpha\\
    & \quad\, +\frac{1}{2}Ka_\rho C^{\rho\mu}F_{\mu\alpha}-\frac{1}{4}Ka_\alpha F^2-\frac{1}{2}\overset{(0)}{S}a_\rho F^\rho{}_{\alpha}\,,\nonumber\\
    +\frac{1}{2}h^\kappa_\alpha\mathcal{G}^\rho{}_\kappa\left(\partial_\rho +\mathcal{A}_\rho \right)S\Big\vert_{@r^{-2}} & =  -\frac{1}{2}\overset{(0)}{S}a_\rho C^\rho{}_\alpha-\frac{1}{2}C^\rho{}_\alpha\partial_\rho\overset{(0)}{S}-Ka_\rho D^\rho{}_\alpha\nonumber\\
    &\quad\, -D^\rho{}_\alpha\partial_\rho K-\frac{1}{2}F^{\rho\kappa}C_{\kappa\alpha}\partial_\rho K\,,\\
    -h^\kappa_\alpha\partial_r\beta\Pi^\alpha_\kappa\left(\partial_\alpha +\mathcal{A}_\alpha\right)S\Big\vert_{@r^{-2}} & =  \frac{1}{8}\left(F^2-C^2\right)\left(h^\gamma_\alpha\partial_\gamma K+Ka_\kappa\right)\,,\\
    -\frac{1}{2}h^\kappa_\alpha\Pi^\alpha_\kappa\left(\partial_\alpha+\mathcal{A}_\alpha\right)\left(\partial_r S-r^{-1}S\right)\Big\vert_{@r^{-2}} & =  h^\gamma_\alpha\partial_\gamma \overset{(1)}{S}+a_\alpha\overset{(1)}{S}-\frac{1}{2}a^\rho F_{\rho\alpha}\overset{(0)}{S}\,,\\
    -\frac{1}{2}h^\kappa_\alpha S\Pi^\mu_\kappa\hat D_\rho \mathcal{F}^\rho{}_\mu\Big\vert_{@r^{-2}} & =  -\frac{1}{2}\overset{(0)}{S}h_{\alpha\sigma}\mathcal{D}_\rho F^{\rho\sigma}\nonumber\\
    &\quad\,-\frac{1}{2}KC_{\alpha\sigma}\mathcal{D}_\rho F^{\rho\sigma}-\frac{1}{4}Ka_\alpha F^2\,,\\
    +\frac{1}{2}h^\kappa_\alpha S\Pi^\rho_\kappa\left(\partial_r\mathcal{Z}_\rho+r^{-1}\mathcal{Z}_\rho\right)\Big\vert_{@r^{-2}} & =  -\frac{1}{2}Kh^\rho_\alpha\overset{(2)}{\mathcal{Z}}_\rho\,,\\
    +\frac{1}{2}h^\kappa_\alpha\left(\mathcal{L}_U-\bar{\mathcal{K}}\right)\mathcal{Z}_\kappa\Big\vert_{@r^{-2}} & =  \frac{1}{2}\mathcal{L}_v\left(h^\kappa_\alpha\overset{(2)}{\mathcal{Z}}_\kappa\right)-\frac{1}{2}a_\alpha a^\kappa\overset{(1)}{\mathcal{Z}}_\kappa-\frac{1}{2}h^\kappa_\alpha\mathcal{L}_a\overset{(1)}{\mathcal{Z}}_\kappa\nonumber\\
    &\quad\,-\frac{1}{2}\left(\overset{(1)}{\mathcal{K}}-\overset{(0)}{S}\right)\overset{(1)}{\mathcal{Z}}_\alpha\,,\label{eq:PIUR11thterm}\\
    +\frac{1}{2}h^\kappa_\alpha\Pi^\mu_\kappa \left(\partial_\mu-\mathcal{A}_\mu\right)\bar{\mathcal{K}}\Big\vert_{@r^{-2}} & = \frac{1}{2}h^\mu_\alpha\partial_\mu\left(\overset{(2)}{\mathcal{K}}-\overset{(1)}{S}\right)-\frac{1}{2}a_\alpha\left(\overset{(2)}{\mathcal{K}}-\overset{(1)}{S}\right)\nonumber\\
    &\quad\, +\frac{1}{2}a^\rho F_{\rho\alpha}\left(\overset{(1)}{\mathcal{K}}-\overset{(0)}{S}\right)\,,\\
    \intertext{}
    -h^\kappa_\alpha\Pi^\mu_\kappa\hat D_\rho \left(\Pi^{\rho\sigma}\mathcal{K}^T_{\sigma\mu}\right)\Big\vert_{@r^{-2}} & =  -h^\nu_\alpha\mathcal{D}_\rho\left(h^{\rho\sigma}h^\mu_\nu\overset{(0)}{\mathcal{K}}{}^T_{\sigma\mu}-C^{\rho\sigma}h^\mu_\nu\overset{(-1)}{\mathcal{K}}{}^T_{\sigma\mu}\right)\nonumber\\
    &\quad\,+F^\nu{}_\alpha a^\mu \overset{(-1)}{\mathcal{K}}{}^T_{\mu\nu}+\frac{1}{2}h^\mu_\alpha\overset{(-1)}{\mathcal{K}}{}^{T}{}^{\nu\rho}\mathcal{D}_\mu C_{\nu\rho}\,.
\end{align}
\end{subequations}
In deriving these expressions, we used that the expansions of $U^\mu\mathcal{K}^T_{\mu\nu}=0$
and $U^\mu\mathcal{Z}_\mu=0$ lead to $v^\mu\overset{(0)}{\mathcal{K}}{}^T_{\mu\nu}=a^\mu\overset{(-1)}{\mathcal{K}}{}^T_{\mu\nu}$ and $v^\mu\overset{(2)}{\mathcal{Z}}_\mu=a^\mu\overset{(1)}{\mathcal{Z}}_\mu$.

Using these results, Equation \eqref{eq:appDUPiR} at order $r^{-2}$ leads to
\begin{align}
    0 & = a_\alpha\overset{(1)}{S}-\frac{3}{2}a^\rho F_{\rho\alpha}\overset{(0)}{S}- \frac{1}{8}K\left(3F^2+C^2\right)a_\alpha+ \partial_\rho K C^{\rho\mu}F_{\mu\alpha}-\frac{1}{2}\partial_\rho\overset{(0)}{S}F^\rho{}_\alpha\nonumber\\
    &\quad\, -\frac{1}{2}\overset{(0)}{S}h_{\alpha\sigma}\mathcal{D}_\rho F^{\rho\sigma}-\frac{1}{2}KC_{\alpha}{}^{\sigma}\left(\mathcal{D}_\rho F^{\rho}{}_{\sigma}+a_\rho F^{\rho}{}_{\sigma}\right) -\frac{1}{2}\overset{(0)}{S}a_\rho C^\rho{}_\alpha\nonumber\\
    &\quad\,-\frac{1}{2}C^\rho{}_\alpha\partial_\rho\overset{(0)}{S}-Ka_\rho D^\rho{}_\alpha-D^\rho{}_\alpha\partial_\rho K\nonumber\\
    &\quad\,+ \frac{1}{8}\left(F^2-C^2\right)h^\gamma_\alpha\partial_\gamma K+\frac{1}{2}\left(\mathcal{L}_v-K\right)\left(h^\kappa_\alpha\overset{(2)}{\mathcal{Z}}_\kappa\right)  \nonumber\\
   &\quad\,+\frac{1}{2}\overset{(0)}{S}\overset{(1)}{\mathcal{Z}}_\alpha -\frac{1}{2}\overset{(1)}{\mathcal{K}}\overset{(1)}{\mathcal{Z}}_\alpha-\frac{1}{2}a_\alpha a^\kappa\overset{(1)}{\mathcal{Z}}_\kappa-\frac{1}{2}h^\kappa_\alpha\mathcal{L}_a\overset{(1)}{\mathcal{Z}}_\kappa\nonumber\\
   &\quad\, + \frac{1}{2}h^\mu_\alpha\partial_\mu\left(\overset{(2)}{\mathcal{K}}+\overset{(1)}{S}\right)-\frac{1}{2}a_\alpha\left(\overset{(2)}{\mathcal{K}}-\overset{(1)}{S}\right)+\frac{1}{2}a^\rho F_{\rho\alpha}\overset{(1)}{\mathcal{K}}\nonumber\\
    &\quad\, -h^\nu_\alpha\mathcal{D}_\rho\left(h^{\rho\sigma}h^\mu_\nu\overset{(0)}{\mathcal{K}}{}^T_{\sigma\mu}-C^{\rho\sigma}\overset{(-1)}{\mathcal{K}}{}^T_{\sigma\nu}\right)\nonumber\\
    &\quad\,+F^\nu{}_\alpha a^\mu \overset{(-1)}{\mathcal{K}}{}^T_{\mu\nu}+\frac{1}{2}h^\mu_\alpha\overset{(-1)}{\mathcal{K}}{}^{T}{}^{\nu\rho}\mathcal{D}_\mu C_{\nu\rho}\,.\label{eq:UPiRinterm2}
\end{align}
which is precisely~\eqref{eq:UPiRinterm}. This serves as our starting point for Section~\ref{sssec:d2AngLoss}.

\subsubsection{Evaluating \texorpdfstring{$U^\mu U^\nu R_{\mu\nu}=0$}{U-U-Rmn} at \texorpdfstring{$\mathcal{O}(r^{-2})$}{order -2}}
\label{app:LossEqnLists-mass}

In precisely the same fashion as before, we start off with the full equation $U^\mu U^\nu R_{\mu\nu}=0$ which is given in 
\eqref{eq:U-U-R-d2}, or in a slightly alternative form in \eqref{eq:newformUUR}. Here, we will use the form in~\eqref{eq:U-U-R-d2}, which we repeat here,
\begin{align}
  0
  = U^\mu U^\nu R_{\mu\nu}
  &= \hat D_\rho\left(S\mathcal{A}^\rho+\frac{1}{2}\Pi^{\rho\sigma}\partial_\sigma S\right)-\frac{1}{2}\mathcal{A}^\rho\partial_\rho S-\mathcal{Z}^\rho\left(S\mathcal{A}_\rho+\frac{1}{2}\partial_\rho S\right)
  \\
  &{}\qquad\nonumber
  +\frac{1}{4}S^2\mathcal{F}^2 -\frac{1}{2}S\mathcal{Z}_\mu\mathcal{Z}^\mu+S\mathcal{L}_U\partial_r\beta
  +S\left(\partial_r+r^{-1}\right)\left(S\partial_r\beta\right)
  \\
  &{}\qquad\nonumber
  +\frac{1}{2}S\left(\partial_r^2S+r^{-1}\partial_r S-r^{-2}S\right)
  +\frac{1}{2}S\partial_r\beta\left(\partial_r S+2S\partial_r\beta\right)
  \\
  &{}\qquad\nonumber
  -\frac{1}{2}\bar{\mathcal{K}}\left(
    \partial_r S+2r^{-1}S+2S\partial_r\beta
  \right)
  -\mathcal{K}^{T\rho\sigma}\mathcal{K}^T_{\rho\sigma}
  -\frac{1}{2}\bar{\mathcal{K}}^2
  +\mathcal{L}_U\bar{\mathcal{K}}\,.
\end{align}
We expand each term to order $r^{-2}$. This results in the list below:
\begin{subequations}
\begin{align}
    \bar D_\rho\left(S\Pi^{\rho\sigma}\mathcal{A}_\sigma+\frac{1}{2}\Pi^{\rho\sigma}\partial_\sigma S\right)\Big\vert_{@r^{-2}} & =  e^{-1}\partial_\rho\left[e\left(\frac{1}{2}h^{\rho\sigma}\partial_\sigma\overset{(0)}{S}+\overset{(0)}{S}a^\rho-\frac{1}{2}C^{\rho\sigma}\partial_\sigma K\right.\right.\nonumber\\
    &\quad\,\left.\left.-KC^{\rho\sigma}a_\sigma+Ka^\sigma F^\rho{}_\sigma\right)\right]\,, \\
    -\mathcal{Z}_\rho\left(S\Pi^{\rho\sigma}\mathcal{A}_\sigma+\frac{1}{2}\Pi^{\rho\sigma}\partial_\sigma S\right)\Big\vert_{@r^{-2}} & =  -\overset{(1)}{\mathcal{Z}}_\rho\left(K a^\rho+\frac{1}{2}h^{\rho\sigma}\partial_\sigma K\right)\,, \\
    +\frac{1}{4}S^2\mathcal{F}^2\Big\vert_{@r^{-2}} & =  +\frac{1}{4}K^2 F^2\,,\\
    -\frac{1}{2}\Pi^{\rho\sigma}\partial_\rho S\mathcal{A}_\sigma \Big\vert_{@r^{-2}}& =  +\frac{1}{2}C^{\rho\sigma}a_\rho\partial_\sigma K-\frac{1}{2}a^\rho\partial_\rho\overset{(0)}{S}\nonumber\\
    &\quad\,-\frac{1}{2}F^\rho{}_\sigma a^\sigma\partial_\rho K\,,\\
    -\frac{1}{2}S\Pi^{\rho\sigma}\mathcal{Z}_\rho\mathcal{Z}_\sigma \Big\vert_{@r^{-2}}& =  0\,,\\
    +S\mathcal{L}_U\partial_r\beta \Big\vert_{@r^{-2}}& =  +\frac{1}{8}K\mathcal{L}_v C^2-\frac{1}{8}K\mathcal{L}_v F^2\,,\\
    +\frac{1}{2}S\left(\partial_r^2S+r^{-1}\partial_r S-r^{-2}S\right)\Big\vert_{@r^{-2}} & =  -\frac{1}{2}\left(\overset{(0)}{S}\right)^2\,,\\
+S\left(\partial_r+r^{-1}\right)\left(S\partial_r\beta\right) \Big\vert_{@r^{-2}}& =  +\frac{1}{8}K^2\left(F^2-C^2\right)\,,\\
    +\frac{1}{2}S\partial_r\beta\left(\partial_r S+2S\partial_r\beta\right)\Big\vert_{@r^{-2}} & =  +\frac{1}{16}K^2\left(C^2-F^2\right)\,,\\
    -\frac{1}{2}\bar{\mathcal{K}}\left(\partial_r S+2r^{-1}S+2S\partial_r\beta\right)\Big\vert_{@r^{-2}}& =  +\left(\overset{(0)}{S}\right)^2-\overset{(0)}{S}\overset{(1)}{\mathcal{K}}+\frac{3}{2}K\left(\overset{(1)}{S}-\overset{(2)}{\mathcal{K}}\right)\,,\\
    -\mathcal{K}^{T\,\rho\sigma}\mathcal{K}^T_{\rho\sigma} \Big\vert_{@r^{-2}}& =  -\overset{(-1)}{\mathcal{K}}{}^{T\,\mu\nu}\overset{(-1)}{\mathcal{K}}{}^{T}_{\mu\nu}\,,\\
    -\frac{1}{2}\bar{\mathcal{K}}^2 \Big\vert_{@r^{-2}}& =  -\frac{1}{2}\left(\overset{(1)}{\mathcal{K}}-\overset{(0)}{S}\right)^2\,,\\
    +\mathcal{L}_U\bar{\mathcal{K}} \Big\vert_{@r^{-2}}& =  +\mathcal{L}_v\left(\overset{(2)}{\mathcal{K}}-\overset{(1)}{S}\right)-\mathcal{L}_a\left(\overset{(1)}{\mathcal{K}}-\overset{(0)}{S}\right)\,.
\end{align}
\end{subequations}
Summing the terms in this list leads to the following result for $U^\mu U^\nu R_{\mu\nu}=0$ at order $r^{-2}$,
\begin{align}
\label{eq:UUR2R1}
    0 &=  \mathcal{D}_\rho\left[\frac{1}{2}h^{\rho\sigma}\partial_\sigma\overset{(0)}{S}+\overset{(0)}{S}a^\rho-\frac{1}{2}C^{\rho\sigma}\partial_\sigma K-KC^{\rho\sigma}a_\sigma+Ka^\sigma F^\rho{}_\sigma\right]\nonumber\\
    &\quad\,-\overset{(1)}{\mathcal{Z}}_\rho\left(K a^\rho+\frac{1}{2}h^{\rho\sigma}\partial_\sigma K\right)+\frac{1}{4}K^2 F^2+\frac{1}{2}C^{\rho\sigma}a_\rho\partial_\sigma K+\frac{1}{2}a^\rho\partial_\rho\overset{(0)}{S}-\frac{1}{2}F^\rho{}_\sigma a^\sigma\partial_\rho K\nonumber\\
    &\quad\,+\frac{1}{8}K\mathcal{L}_v C^2-\frac{1}{8}K\mathcal{L}_v F^2+\frac{1}{16}K^2\left(F^2-C^2\right)\nonumber\\
    &\quad\,+\frac{3}{2}K\left(\overset{(1)}{S}-\overset{(2)}{\mathcal{K}}\right)-\mathcal{L}_v\left(\overset{(1)}{S}-\overset{(2)}{\mathcal{K}}\right)-\mathcal{L}_a\overset{(1)}{\mathcal{K}}-\frac{1}{2}\left(\overset{(1)}{\mathcal{K}}\right)^2\nonumber\\
    &\quad\,-\overset{(-1)}{\mathcal{K}}{}^{T\,\mu\nu}\overset{(-1)}{\mathcal{K}}{}^{T}_{\mu\nu}\,,
\end{align}
which is~\eqref{eq:R1UUR2}.

\section{Reduction to standard Bondi loss equations}
\label{app:reduction-to-standard-BS}

In this appendix, we demonstrate how the diffeomorphism Ward identity for the Carroll-covariant currents obtained in Section~\ref{ssec:bondi-loss-4d-simplified} reduces to the standard $d=2$ Bondi loss equations when working in Bondi--Sachs gauge as discussed in Section~\ref{ssec:reduction-to-standard-Bondi--Sachs-gauge}.
Since the results of Section~\ref{ssec:bondi-loss-4d-simplified} are a special case of the general results of Sections~\ref{sec:bulk-improvements} and~\ref{sec:HoloRenormAndOn-ShellActions}, we can also view this appendix as embedding the standard BS phase space results into the general framework developed in this work. 

As we saw in Section \ref{ssec:reduction-to-standard-Bondi--Sachs-gauge}, the boundary structures in Bondi coordinates $x^\mu = (u,x^A)$ simplify to
\begin{align}
\label{eq:Carroll-structure-BS}
\begin{aligned}
    \tau_\mu dx^\mu &= du\,,& v^\mu \D_\mu &= -\D_u\,,\\
    h_{\mu\nu}dx^\mu dx^\nu &= \gamma_{AB}(x)dx^A dx^B\,,& h^{\mu\nu}\D_\mu \D_\nu &= \gamma^{AB}(x)\D_A \D_B\,,
\end{aligned}
\end{align}
where $\gamma_{AB}$ and $\gamma^{AB}$ are independent of $u$. These choices imply that
\begin{equation}
    a_\mu = 0\,,\qquad K = 0~(\implies K_{\mu\nu} = 0)\,,\qquad F_{\mu\nu} = 0\,.
\end{equation}
Furthermore, imposing the relations~\eqref{eq:Carroll-structure-BS} in the expression for the connection coefficients $\mathcal{C}^\rho_{\mu\nu}$ (cf.~\eqref{eq:app-calCcon}) reveals that all components involving the $u$-direction vanish, and so, for example, if $X^\mu = (0,X^A)$ is a spatial tensor, which may depend on $u$, we get simplifications of the form
\begin{equation}
    \mathcal{D}_\mu X^\mu = D_A X^A\,,
\end{equation}
where $D_A$ is the Levi-Civita connection of the two-dimensional metric $\gamma_{AB}$. 

We begin with the Bondi mass loss equation~\eqref{eq:energyeq} for the EMT-news complex derived in Section \ref{ssec:bulk-improvements-weyl-covariant-currents}. In Bondi--Sachs gauge this reduces to
\begin{equation}
    0 = \D_u T_{\text{W}}^u - \frac{1}{4}N_{AB}N^{AB} + D_A T_{\text{W}}^A\,,
\end{equation}
where $N_{AB} = \D_u C_{AB}$. In writing the above, we used that
\begin{equation}
    S_{\text{W}}^{AB} = \frac{1}{2}N^{AB}\,.
\end{equation}
The energy flux $T^A_{\text{W}}$ is obtained from the Carroll boost Ward identity~\eqref{eq:anomCboostWI}
\begin{equation}
\label{eq:boost-WI-app}
    h^\mu_\rho T^\rho_{\text{W}} = \mathcal{A}^\mu_{(\text{B})} + (\mathcal{D}_\rho - a_\rho)S_{\text{W}}^{\rho\mu} \implies T^A_{\text{W}} = \mathcal{A}^A_{(\text{B})} + \frac{1}{2}D_B N^{AB}\,,
\end{equation}
where we put parentheses around the `B' that indicates `boost anomaly' to avoid confusing it with an index. 
The boost anomaly~\eqref{eq:Cboostanom}, which has no $u$-component, reduces to
\begin{equation}
    \mathcal{A}^A_{(\text{B})} = \frac{1}{4}\D^A R[\gamma] + \frac{1}{2}D_B N^{BA}\,,
\end{equation}
where we used that, in Bondi--Sachs gauge, 
\begin{equation}
    \os{0}{S} = \frac{1}{2}R[\gamma]\,,\qquad \os{0}{P}_A = - \frac{1}{2}D^B C_{BA}\,,
\end{equation}
where $R[\gamma]$ is the Ricci scalar of the Levi-Civita connection for $\gamma_{AB}$. Combining these findings allows us to conclude that
\begin{equation}
    T^A_{\text{W}} = \frac{1}{4}\D^A R[\gamma] + D_B N^{BA}\,.
\end{equation}
The next ingredient in the mass loss equation is the $u$-component of the energy current $T^u_{\text{W}} = \tau_\mu T^\mu_{\text{W}}$ (cf.~\eqref{eq:WeylImprovedEMTNewsComplex-energy-density}), which in Bondi--Sachs gauge reduces to 
\begin{equation}
\label{eq:T-u}
     T^u_{\text{W}} = \os{1}{S} - \frac{1}{2}D^A D^B C_{AB}\,,
\end{equation}
so that, defining the Bondi mass to be
\begin{equation}
\label{eq:Bondi-mass-d=2}
    M := -\frac{1}{2}\os{1}{S}\,,
\end{equation}
the Bondi mass loss equation becomes
\begin{equation}
    \label{eq:app-standard-bondi-mass-loss}
    \D_u M = \frac{1}{4} D^A D^B N_{AB} + \frac{1}{8} D_A D^A R[\gamma] - \frac{1}{8} N^{AB}N_{AB}\,,
\end{equation}
which is the standard form of the Bondi mass loss equation (see, for example Equation (2.6) in~\cite{Donnay:2023mrd}). Using \eqref{eq:S-1-Weyl-covariant-combination} we can relate $M$ to the metric via
\begin{align}
  \os{1}{g}_{uu}
  = - \os{1}{S}=2M\,.
\end{align}

Next, we consider the angular momentum aspect equation~\eqref{eq:diffeoWIspatialproj} for the EMT-news complex derived in Section \ref{ssec:bulk-improvements-weyl-covariant-currents}, which in Bondi--Sachs gauge simplifies to
\begin{equation}
\begin{split}
\label{eq:angular-mom-eq}
    0 &= \gamma_{AB}\D_u T^{uB}_{\text{W}} + D_B \tilde T_{\text{W}}^{BC}\gamma_{CA} + \frac{1}{2} \D_A\left(\gamma_{BC}T_{\text{W}}^{BC} + \frac{1}{2}C_{BC}N^{BC}\right)\\
    &\quad\,+ \frac{1}{4} D_B(N^{BC}C_{AC} - N_{AC}C^{CB}) - \frac{1}{4} N^{BC} D_A C_{BC}\,.
\end{split}
\end{equation}
The STF stress tensor $ \tilde T_{\text{W}}^{AB}$ becomes
\begin{equation}
    \tilde T^{AB}_{\text{W}} = \frac{1}{4} R[\gamma] C^{AB} - \frac{1}{2}D^{\langle A} D_C C^{B\rangle C}\,,
\end{equation}
where we set the Lagrange multiplier $\zeta^{\mu\nu}$ introduced in~\eqref{eq:on-shell-action-leading-order-with-constraint} to zero. Similarly to the $u$-component of the energy current, which was obtained by solving the boost Ward identity~\eqref{eq:boost-WI-app}, the trace $\gamma_{AB} T^{BC}_{\text{W}}$ is obtained by solving the Weyl Ward identity~\eqref{eq:boundary-Weyl-WI}, which tells us that
\begin{equation}
\begin{split}
    h_{\mu\nu}T^{\mu\nu}_{\text{W}} &= -\frac{1}{2}S^{\mu\nu}_{\text{W}} C_{\mu\nu} - T^\mu_{\text{W}}\tau_\mu\,,\\
    \implies \gamma_{AB}T_{\text{W}}^{AB} &= - \frac{1}{4} N^{AB} C_{AB} + 2M + \frac{1}{2}D_A D_B C^{AB}\,,
\end{split}
\end{equation}
where we used~\eqref{eq:T-u} and~\eqref{eq:Bondi-mass-d=2}. The last ingredient is the momentum density~\eqref{eq:WeylImprovedEMTNewsComplex-momentum-density}, which in Bondi--Sachs gauge reads
\begin{eqnarray}
    \gamma_{AB} T^{uB}_{\text{W}} = \frac{3}{2}\os{1}{P}_A - \frac{1}{32}\D_A C^2 + \frac{1}{4}C_A{^B}D_C C^C{}_{B}\,,
\end{eqnarray}
where we again set $\zeta^{\mu\nu} = 0$. The quantity $\os{1}{P}_A$ contains the angular momentum aspect. To relate it to the standard appearance of the angular momentum aspect in Bondi--Sachs gauge, note that, by definition,
\begin{equation}
    \os{1}{P}_A = - \os{1}{\Pi}_{Au} =-\os{1}{g}_{Au}= \Pi_{AB}\tilde U^B\Big\vert_{@r^{-1}}\,,
\end{equation}
where we used~\eqref{eq:Bondi-rels} and \eqref{eq:gauge-fixed-dictionary-to-standard-Bondi--Sachs-metric-form}.

The radial expansion of $\Pi_{AB}$ is given in~\eqref{eq:spatial-metric-exp}, while $\tilde U^A$ is $\OO(r^{-2})$ with the radial expansion (cf.~Eq.~(2.5) in~\cite{Donnay:2023mrd}, which uses the conventions of~\cite{Barnich:2010eb})
\begin{equation}
    \tilde U^A = -\frac{1}{2r^2}D_B C^{AB} - \frac{2}{3r^3}\left( N^A - \frac{1}{2}C^{AB} D_C C^C{_B}  \right) + \OO(r^{-4})\,,
\end{equation}
where $N^A$ is the angular momentum aspect. This implies that
\begin{equation}
    \os{1}{P}_A = \gamma_{AB}\os{3}{\tilde U}^B + C_{AB}\os{2}{\tilde U}^B = -\frac{2}{3}N_A - \frac{1}{6}C_{AB} D_C C^{CB}\,,
\end{equation}
and therefore that
\begin{equation}
       \gamma_{AB} T^{uB}_{\text{W}} = -N_A - \frac{1}{32}\D_A C^2\,.
\end{equation}
With this, the angular momentum aspect equation~\eqref{eq:angular-mom-eq} reads
\begin{equation}
\label{eq:mom-eq-precursor}
\begin{split}
    \D_u N_A &= \D_A M + \frac{1}{16}\D_A(N^{BC}C_{BC}) - \frac{1}{4}N^{BC}D_A C_{BC}\\
    &\quad\,+ \frac{1}{4}D_B(N^{BC}C_{AC} - N_{AC}C^{BC}) - \frac{1}{4}D_B D^B D_C C^C{_A}\\
    &\quad\, + \frac{1}{4} D_A D^B D^C C_{BC} + \frac{1}{4}C^B{_A}D_B R[\gamma]\\
    &\quad\, + \frac{1}{4}R[\gamma]D_B C^B{_A} + \frac{1}{4}D_A D_BD_CC^{BC} - \frac{1}{4}D_B D_AD_C C^{BC}\,.
\end{split}
\end{equation}
The Ricci identity~\eqref{eq:app-general-riemann-vector-def} for the two-dimensional metric $\gamma_{AB}$ implies that
\begin{equation}
[D_A,D_B]V^B = -R_{AB}[\gamma]V^B = - \frac{1}{2}R[\gamma]V_A\,,
\end{equation}
for any $V^B$, where we used the two-dimensional identity $R_{AB}[\gamma] = \frac{1}{2}R[\gamma]\gamma_{AB}$. Setting $V^B = D_C C^{CB}$, this implies that the last line can be simplified to give
\begin{equation}
    \frac{1}{4}R[\gamma]D_B C^B{_A} + \frac{1}{4}D_A D_BD_CC^{BC} - \frac{1}{4}D_B D_AD_C C^{BC} = \frac{1}{8}R[\gamma]V_A\,.
\end{equation}
Combining this with the first term on the penultimate line of~\eqref{eq:mom-eq-precursor} gives
\begin{equation}
    \frac{1}{4} D_A D^B D^C C_{BC} +  \frac{1}{8}R[\gamma]V_A = \frac{1}{4}D_B D_A D_C C^{CB}\,,
\end{equation}
so that the angular momentum aspect equation takes the final form
\begin{equation}
    \begin{split}
    \D_u N_A &= \D_A M + \frac{1}{16}\D_A(N^{BC}C_{BC}) - \frac{1}{4}N^{BC}D_A C_{BC}\\
    &\quad\,+ \frac{1}{4}D_B(N^{BC}C_{AC} - N_{AC}C^{BC}) - \frac{1}{4}D_B D^B D_C C^C{_A}\\
    &\quad\, + \frac{1}{4} D_B D_A D_C C^{BC} + \frac{1}{4}C^B{_A}D_B R[\gamma]\,,
\end{split}
\end{equation}
which is the standard Bondi angular momentum aspect equation (see, for example, the second equation in~(2.6) in~\cite{Donnay:2023mrd}). 

\newpage

\newpage
\bibliographystyle{JHEP}
\bibliography{Masterbibliography}

\end{document}